\documentclass[15pt,french,epsfig,graphics,twoside]{book}

\usepackage{graphicx}
\usepackage{amssymb}
\usepackage{textcomp}

\usepackage{dingbat}

\usepackage{pifont}

\usepackage{stmaryrd}
\usepackage{mathabx}

\usepackage{fancyhdr}
\usepackage[francais]{babel}
\usepackage[latin1]{inputenc}
\usepackage{tipa}

\usepackage{slashed}

\usepackage[hang,small]{caption}

\begin{document}

~~~\vskip 4 cm
\begin{center}
~~~\\
\vskip 0.3cm
{\Huge CALCUL SPINORIEL} \\
~~~\\
{\Large en Physique des Particules}
\end{center}

\vskip 5 cm

\centerline{\Large \bf Christian Carimalo} \vskip 0.75 cm

\pagestyle{fancy}

\thispagestyle{empty}

\fancyhf{}


\lhead{} \chead{\nouppercase{\leftmark} }

\lfoot{\em Christian Carimalo}

\cfoot{\thepage}

\rfoot{\em Calcul spinoriel}

\renewcommand{\footrulewidth}{0.4pt}

\newpage
\thispagestyle{empty}

\newpage

\setcounter{page}{0}

\setcounter{equation}{0}

\setcounter{section}{0}
\renewcommand{\theequation}{\mbox{1.}\arabic{equation}}

\newcommand{\beq}{\begin{equation}}
\newcommand{\enq}{\end{equation}}
\newcommand{\und}{\underline}
\newcommand{\biz}{\begin{itemize}}
\newcommand{\eiz}{\end{itemize}}
\newcommand{\di}{\displaystyle}

\newcommand{\Vec}{\stackrel{\longrightarrow}}
\newcommand{\Bip}{\stackrel{\longmapsto}}
\newcommand{\Alg}{\overline}
\newcommand{\arc}{\stackrel{\frown}}

\newcommand{\nin}{\noindent}
\newcommand{\vv}{\vskip 0.25 cm}
\newcommand{\vvv}{\vskip 0.3 cm}
\newcommand{\ov}{\overline}

\newcommand{\ro}{\mathring}

\newcommand{\uta}{\begin{array}[t]{c}
{A'} \\
\stackrel{\sim}{\stackrel{~~}{~~}}
\end{array}}

\newcommand{\utx}{\begin{array}[t]{c}
{x} \\
\stackrel{\sim}{\stackrel{~~}{~~}}
\end{array}}

\newcommand{\uthp}{\begin{array}[t]{c}
{\hat{p}} \\
\stackrel{\sim}{\stackrel{~~}{~~}}
\end{array}}

\newcommand{\uthpm}{\begin{array}[t]{c}
{\hat{p}^{\,-1}} \\
\stackrel{\hskip -0.5 cm\sim}{\stackrel{~}{~}}
\end{array}}

\newcommand{\utxi}{\begin{array}[t]{c}
{\xi} \\
\stackrel{\sim}{\stackrel{~~}{~~}}
\end{array}}

\newcommand{\utnz}{\begin{array}[t]{c}
{n_\alpha(\ro{p})} \\
\stackrel{\thicksim}{\stackrel{~~}{~~}}
\end{array}}

\newcommand{\utnzz}{\begin{array}[t]{c}
{n_\alpha} \\
\stackrel{\thicksim}{\stackrel{~~}{~~}}
\end{array}}

\newcommand{\utn}{\begin{array}[t]{c}
{n_\alpha(p)} \\
\stackrel{\thicksim}{\stackrel{~~}{~~}}
\end{array}}

\newcommand{\td}{\textsubtilde}

\renewcommand{\footrulewidth}{0.4pt}

\chapter{Formalisme d'h\'elicit\'e}

\section{Spin-H\'elicit\'e }

\vv \nin Le spin, ou l'h\'elicit\'e selon le cas, d'une particule
est d\'efini \`a l'aide de l'op\'erateur de Pauli-Lubanski

\beq W_{\mu}(p) = \di{1\over 2} \epsilon_{\mu \nu \rho \sigma}
p^{\nu} J^{\rho \sigma}  \label{PL} \enq

\nin o\`u $p$ est la 4-impulsion\footnote{Dans la suite, sauf
indication contraire, les termes de ``impulsion" et de ``vecteur"
seront utilis\'es pour ``4-impulsion" et ``4-vecteur",
respectivement.} de la particule et $J^{\rho \sigma}$ la
repr\'esentation hermitique des g\'en\'erateurs du groupe de Lorentz
$L^{\uparrow}_{+}$ dans l'espace des \'etats de la particule. Les
op\'erateurs $W_{\mu}(p)$ sont simplement les g\'en\'erateurs du
{\em petit groupe} de $p$, qui laisse ce vecteur invariant. Sous
$L^{\uparrow}_{+}$, ils se transforment selon

\beq U(\Lambda) ~W_{\mu}(p) ~U(\Lambda)^{-1} = W_{\mu}(\Lambda p)
\enq

\vv \nin $\Lambda$ \'etant une tranformation de Lorentz, $U(
\Lambda)$ sa repr\'esentation (unitaire) pour la particule
consid\'e-r\'ee.

\vv \nin Pour une particule massive r\'eelle de masse $m$, son
impulsion est du genre temps - futur. On a en effet $p_{\mu} p^{\mu}
= p^2_0 ~- \Vec{~p~}^2 = p^2 = m^2
> 0$, $p^0 > 0$. On peut alors attacher \`a chacune des impulsions
$p$ de la particule une triade de vecteurs unitaires du genre
espace, soit $n_i(p)$ avec $ i = 1, 2, 3$, formant avec $n_0(p) =
\hat{p} = p/m$ une base de l'espace de Minkowski :

\begin{eqnarray}
&n_i(p) \cdot n_j (p) = - \delta_{i j}~,~~~n_0(p) \cdot n_i(p) =
0~,~~n_0(p)^2 = 1~ & \nonumber \\
& \epsilon_{\mu \nu \rho \sigma} \,n_0(p)^{\mu} \,n_1(p)^{\nu}
n_2(p)^{\rho}\, n_3(p)^{\sigma}~=~1 &
\end{eqnarray}

\nin Les op\'erateurs de spin sont alors d\'efinis comme

\beq S_i(p) = - n_i(p) \cdot W(p) \enq

\nin Ils engendrent le groupe $SO(3)(p)$ des rotations dans l'hyperplan
orthogonal \`a $p$ et satisfont aux relations de
commutation de l'alg\`ebre de Lie de $SU(2)$. Comme

\beq  W_\mu \,p^\mu = 0 \enq

\vv \nin l'op\'erateur de Pauli-Lubanski peut \^etre \'ecrit sous la
forme

\beq W(p) = \di{\sum^3_{i = 1}}~ n_i(p) S_i(p) \enq

\nin et l'invariant $W^2 = W^{\mu} W_{\mu}$ s'exprime en fonction du
spin $s$ de la particule comme

\beq W^2  = - m^2 \Vec{\,S\,}^2\hskip -0.2cm(p) ~=~-m^2\,s(s+1) \enq

\nin Les divers petits groupe $SO(3)(p)$ se d\'eduisent les uns des
autres par des transformations de $L^{\uparrow}_{+}$, soit,
symboliquement,

\beq U(\Lambda) ~SO(3)(p) ~U(\Lambda)^{-1} = SO(3)(\Lambda p)\enq

\nin mais les op\'erateurs $S_i(p)$ ont, en g\'en\'eral, une loi de
transformation compliqu\'ee :

\beq U(\Lambda) ~S_i(p) ~U(\Lambda)^{-1} = - \di{\sum_j}~ \left[~
n_j(\Lambda p) \cdot \Lambda n_i(p) ~\right] ~S_j(\Lambda p)
\label{noninvar}  \enq

\nin du fait qu'il n'y a pas a priori de relation simple entre les
vecteurs $n_j(\Lambda p)$ de la triade associ\'ee \`a $\Lambda p$ et
les transform\'es $\Lambda n_i(p)$ des vecteurs de la triade
associ\'ee \`a $p$. Ici, les $S_j(\Lambda p)$ sont les op\'erateurs
de spin associ\'es \`a $\Lambda p$, d\'efinis par

\beq S_j(\Lambda p) = - n_j(\Lambda p) \cdot W(\Lambda p) \enq

\vv \nin Lorsque la masse de la particule est nulle ($p^2 = 0,~p_0 >
0$), son impulsion $p$ ne peut plus \^etre utilis\'ee pour d\'efinir
elle-m\^eme un vecteur unitaire de base et la d\'emarche
pr\'ec\'edente ne convient plus. On associe alors \`a $p$, d'une
part, deux vecteurs unitaires du genre espace $n_i(p)$ ($i=1,2$)
orthogonaux entre eux et orthogonaux \`a $p$ et, d'autre part, un
vecteur unitaire du genre temps $n_0(p)$ orthogonal aux $n_i(p)$ :

$$ n_i(p) \cdot n_j(p) = - \delta_{ij}~,~~~n_i(p) \cdot p =
0~,~~n_i(p) \cdot n_0(p) = 0~,~~n^2_0(p)=1 $$

\vv \nin Pour une telle particule, on a maintenant

\beq W_\mu (p) = \lambda(p) p_\mu + \di{\sum^2_{i = 1}}~ n_i(p)_\mu
W_i(p) ~~~{\rm avec}~~~ W_i(p) = - n_i(p) \cdot W(p) \enq

\vv \nin On montre que les trois op\'erateurs $\lambda(p)$, $W_1(p)$
et $W_2(p)$ constituent un ensemble isomorphe \`a l'alg\`ebre de Lie
de $P(2)$. On a en effet

$$ \left[W_1, W_2 \right] = 0~,~~~\left[ \lambda, W_1 \right] = i
W_2~,~~\left[\lambda , W_2 \right] = - i W_1 $$

\vv \nin Le sous-groupe ab\'elien invariant et engendr\'e par $W_1$
et $W_2$ s'appelle le ``groupe de jauge" de $p$. Ses
repr\'esentations physiques sont telles que $W_1$ et
$W_2$ y prennent la valeur z\'ero (jauge invariante). On a alors

\beq W(p) = \lambda(p)~p~,~~~W^2(p) = 0~,~~~U(\Lambda) \lambda(p)
U^{-1}(\Lambda) = \lambda(\Lambda p) \label{hel0} \enq

\vv \nin o\`u $\Lambda$ est une transformation de $L^{\uparrow}_{+}$
et $U(\Lambda)$ sa repr\'esentation unitaire dans l'espace des
\'etats de la particule. La derni\`ere relation dans \ref{hel0}
montre que $\lambda(p)$ est un op\'erateur {\it scalaire} qui
repr\'esente donc un invariant relativiste.

\vv \nin Une particule de masse nulle est associ\'ee \`a une valeur
propre particuli\`ere $\lambda$ de cet op\'erateur ou, dans le cas
du photon, \`a deux valeurs propres oppos\'ees, \`a savoir $+1$ et
$-1$. Le nombre $\lambda$, appel\'e {\it h\'elicit\'e} ou encore,
polarisation circulaire de la particule, est conserv\'e dans les
transformations de $L^{\uparrow}_{+}$, celles-ci se r\'eduisant
alors \`a une simple multiplication des \'etats de la particule par
des facteurs de phase $e^{-i \lambda \psi}$ et \`a la transformation
de $p$ en $\Lambda p$. Il repr\'esente en quelque sorte le ``spin"
de la particule et caract\'erise compl\`etement celle-ci du point de
vue du groupe de Lorentz.

\vv \nin Alors que l'h\'elicit\'e d'une particule de masse nulle est
un invariant relativiste, une composante de spin $S_i(p)$ pour une
particule massive seule n'est jamais invariante. Pour qu'elle le
f\^ut, il aurait fallu, d'apr\`es \ref{noninvar}, que l'on e\^ut

\beq \Lambda \left\{n_i(p) \right\} = n_i(\Lambda p) \label{covar0}
\enq

\vv \nin relation qui impose au vecteur $n_i(p)$ de satisfaire \`a
une certaine covariance sous $L^{\uparrow}_{+}$.

\vv \nin Pour une particule unique, c'est-\`a-dire lorsqu'on ne
dispose que d'une seule impulsion, celle de la particule, on ne peut
satisfaire cette relation. En revanche, lorsqu'on effectue, au
minimum, un produit tensoriel de deux espaces d'\'etats \`a une
particule chacun et que l'on peut alors disposer de deux impulsions,
il est possible de construire \`a partir de celles-ci et pour
chacune des particules un vecteur ayant la covariance requise. C'est
le principe du {\it couplage d'h\'elicit\'e}, pour deux particules
r\'eelles\footnote{Voir : P. Moussa, R. Stora, Methods in subnuclear
Physics, Hercegnovi 1966, vol II, p. 265 (Gordon and Breach) ; P.
Moussa, Th\`ese d'\'etat, Orsay (1968) ; G. Mahoux, cours de 2nde ann\'ee de troisi\`eme cycle de Physique th\'eorique, facult\'e d'Orsay, 1970-1971.}. Les
composantes de spin suivant ces vecteurs particuliers, appel\'ees
aussi {\it h\'elicit\'es}, sont des invariants relativistes dans
l'espace produit tensoriel. Cette circonstance se rencontre tout
particuli\`erement en th\'eorie quantique des champs o\`u la
description de processus \'el\'ementaires fait intervenir des {\it
vertex} \`a trois particules impliquant de tels produits tensoriels.

\section{Vertex \`a trois particules - T\'etrades d'h\'elicit\'e}

\begin{figure}[hbt]
\centering
\includegraphics[scale=0.3, width=3cm, height=3cm]{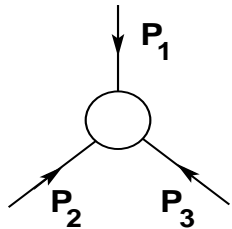}
\vskip 0.25cm

\caption{Vertex \`a trois lignes entrantes} \label{vertex3}
\end{figure}

\vv \nin Consid\'erons d'une mani\`ere g\'en\'erale un vertex
comportant trois ``lignes entrantes" d'impulsions $P_1$, $P_2$ et
$P_3$, correspondant chacune \`a une particule, r\'eelle ou
virtuelle (fig \ref{vertex3}).

\vvv

\vv \nin La conservation de l'impulsion \`a ce vertex sera
exprim\'ee sous la forme

\beq P_1 + P_2 + P_3 = 0 \enq

\vv \nin Pour simplifier, nous supposerons que les normes des
impulsions sont diff\'erentes de z\'ero

\beq P^2_i = (P_{i 0})^2 - (\Vec{P_i})^2 = s_i \neq 0 \enq

\vv \nin Notons $P_{ij} = - P_k$ ($ i \neq j \neq k$) la somme des
impulsions $P_i$ et $P_j$ et par $\hat{P}_{ij}$ le vecteur unitaire
correspondant. En introduisant la ``fonction signe"
$\varepsilon(x)$, on a

\beq \hat{P}_{ij} = - \hat{P}_k = -
\di{P_k\over{\sqrt{|s_k|}}}~,~~~\hat{P}^2_{ij} =
\varepsilon(s_k)~,~~~\hat{P}^2_i = \varepsilon(s_i) \enq

\vv \nin Nous d\'efinirons ensuite ``l'impulsion relative" de $P_i$
et $P_j$ de la mani\`ere habituelle :

\beq Q_{ij} = \di{1\over 2} \left\{P_i - P_j + \di{{s_i - s_j}\over
s_k} P_k \right\} = P_i - \di{{P_i \cdot P_k}\over s_k} = - P_j +
\di{{P_j \cdot P_k}\over s_k} = - Q_{ji} \label{vecelic}\enq

\vv \nin Ce vecteur, que nous appellerons aussi ``vecteur
d'h\'elicit\'e" de $P_k$, est orthogonal \`a $P_k$ et sa norme est

\beq Q^2_{ij} = - \di{1 \over {4 s_k}} \Lambda(s_1, s_2, s_3) \enq

\vv \nin o\`u $\Lambda(x,y,z)$ est la fonction sym\'etrique en ses
trois arguments $x$, $y$ et $z$, d\'efinie par

\beq \Lambda(x,y,z) = x^2 + y^2 + z^2 - 2 xy -2 yz -2 zx \label{fctlambda} \enq

\vv \nin On notera que

\beq \Lambda(s_1, s_2, s_3) = 4\left\{ (P_1 \cdot P_j)^2 - s_i s_j
\right\} \enq

\vv \nin et que, de l'orthogonalit\'e de $Q_{ij}$ et $P_k$ on
d\'eduit les relations

\beq Q_{ij} \cdot P_i =  Q_{ij} \cdot P_j = Q^2_{ij}~,~~~Q_{ij}
\cdot Q_{ik} = \di{{(P_i \cdot P_j)}\over{4 s_j s_k}} \Lambda(s_1 ,
s_2, s_3) \enq

\vv \nin Nous ne consid\`ererons que les cas o\`u la fonction $\Lambda$ (eq. \ref{fctlambda})
prend une valeur positive ou \`a la limite nulle. Le vecteur unitaire port\'e par le vecteur d'h\'elicit\'e  \ref{vecelic} aura alors pour norme $- \varepsilon(s_k)$. Par suite, l'un des deux vecteurs $\hat{Q}_{ij}$ ou $\hat{P}_k$ sera du genre temps et l'autre du genre espace. Au besoin, nous changerons le signe du vecteur du genre temps afin que celui-ci pointe vers le futur et nous le noterons provisoirement $e_{t,k}$ tandis que le vecteur du genre espace sera not\'e $e_{s,k}$.

\vv \nin Introduisons ensuite deux nouveaux vecteurs r\'eels, $e_1$ et $e_2$, unitaires et du genre espace, orthogonaux entre eux et orthogonaux \`a $e_{t,k}$ et $e_{s,k}$ et formant avec ceux-ci une base orthogonale, norm\'ee et directe de l'espace-temps. Par la suite, nous utiliserons plut\^ot les vecteurs de ``polarisation circulaire"

\beq e^{(\pm)} = \mp \di{1 \over{\sqrt{2}}} \left[ e_1 \pm i e_2 \right] \enq

\vv \nin Pour d\'esigner la t\'etrade ainsi form\'ee, nous adopterons \'egalement la notation g\'en\'erale $e^{(\lambda)}_k$ et \'ecrirons alors les relations d'orthogonalit\'e et de fermeture de cette base sous la forme

\beq e^{(\lambda') \star}_k  \cdot e^{(\lambda)}_k = \eta_\lambda \delta_{\lambda' \lambda}~,~~~
\di{\sum_\lambda} \eta_\lambda ~e^{(\lambda) \star \mu}_k ~
e^{(\lambda) \nu}_k = g^{\mu \nu} \enq

\vv \nin o\`u $\eta_\lambda = 1$ pour $\lambda = 0_t$ et $\eta_\lambda = -1$ pour $\lambda = 0_s, \pm1$, $\lambda = 0_t$ ou $= 0_s$ signifiant que $\lambda$ prend effectivement la valeur 0 lorsque le vecteur correspondant est du genre temps (t) ou du genre espace (s), respectivement.  La raison de cette notation vient du fait bien connu que la t\'etrade $e^{(\lambda)}_k$ est associ\'ee \`a la repr\'esentation quadri-vectorielle du groupe de Lorentz. L'indice $\lambda$ correspond aux valeurs propres de l'op\'erateur $W(e_{t,k}) \cdot e_{s,k}$, o\`u $W(e_{t,k})$ est la repr\'esentation quadri-vectorielle de l'op\'erateur de Pauli-Lubansky correspondant au vecteur $e_{t,k}$, soit

\beq \left\{W_\mu (e_{t,k}) \right\}_{\alpha \beta} = i \varepsilon_{\mu \nu \alpha \beta} ~e^\nu_{t,k} \enq

\vv \nin On a en effet

$$ W_{\alpha \beta} = \left\{ W (e_{t,k})  \cdot e_{s,k} \right\}_{\alpha \beta} = i
\varepsilon_{\mu \nu \alpha \beta} ~e^\mu_{s,k}~e^\nu_{t,k}~,~~~{\rm et}~~~W(e^{(\lambda)}_k) = \lambda~e^{(\lambda)}_k $$

\vv \nin La triade $\left( e_{s,k}~,e^(\pm)_k \right)$ est quant \`a elle associ\'ee \`a une repr\'esentation unitaire de spin 1 du petit groupe de $e_{t,k}$, tandis que ce dernier vecteur est associ\'e \`a une repr\'esentation scalaire de ce m\^eme petit groupe.

\vv \nin Remarquons que puisque les vecteurs $e^{(\pm)}$ sont orthogonaux aux vecteurs $P_k$ et $Q_{ij}$, ils sont \'egalement orthogonaux aux deux autres impulsions $P_i$ et $P_j$ ainsi qu'aux impulsions relatives. Par suite, tous les vecteurs de polarisations circulaires $e^{(\pm)}_i$ ($i = 1, 2, 3$) des diverses impulsions sont situ\'es dans le biplan orthogonal au plan d\'efini par les impulsions, appel\'e ``plan du vertex", et ne diff\`erent donc entre eux que par des rotations qui se r\'eduisent ici \`a de simples multiplications par des facteurs de phase $e^{-i \lambda \psi}$.

\vv \nin Nous appellerons de mani\`ere g\'en\'erale ``t\'etrades d'h\'elicit\'e" les t\'etrades ainsi d\'efinies. Nous verrons par la suite la signification de cette d\'enomination.

\section{Les couplages d'h\'elicit\'e}

\vv \nin Revenons au vertex \`a trois ``particules", r\'eelles ou virtuelles de la figure \ref{vertex3}, chacune ayant une impulsion donn\'ee et correspondant \`a une repr\'esentation particuli\`ere du groupe de Lorentz.

\vv \nin Nous dirons que nous effectuons un {\it couplage d'h\'elicit\'e} entre ces trois particules si nous attribuons \`a chacune d'elles sa propre t\'etrade d'h\'elicit\'e, o\`u les vecteurs de polarisation circulaire ne pourrons diff\'erer que par un signe global d'une t\'etrade \`a une autre. Nous examinerons ici les trois situations suivantes :

\vv \nin $\clubsuit$ celle o\`u les trois impulsions $P_1$, $P_2$ et $-P_3$ sont du genre temps futur ;

\vv \nin $\clubsuit$ celle o\`u les les deux impulsions $P_1$ et $-P_2$ sont du genre temps futur, $P_3$ \'etant du genre espace ;

\vv \nin $\clubsuit$  enfin, celle o\`u $P_1$ et $P_2$ sont du genre espace, $-P_3$ \'etant du genre temps.

\begin{figure}[hbt]
\centering
\includegraphics[scale=0.3, width=3cm, height=3cm]{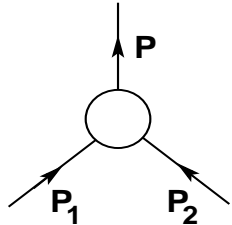}
\vskip 0.25cm

\caption{Vertex \`a deux particules entrantes r\'eelles} \label{vertex31}
\end{figure}

\subsection{Premier cas (fig \ref{vertex31})}

\vvv

\vv \nin Nous supposerons que les lignes entrantes 1 et 2 repr\'esentent ici deux particules r\'eelles, d'impulsions donn\'ees, chacune \'etant caract\'eris\'ee par un espace de repr\'esentation irr\'eductible $[m, s, \eta]$ du groupe de Poincar\'e, $m$ \'etant la masse, suppos\'ee non nulle, de la particule consid\'er\'ee, $s$ son spin et $\eta$ sa parit\'e. Nous poserons

\beq -P_3 = p = p_1 + p_2~,~~~p^2_i = m^2_i~~(i=1,2)~,~~~p^2=s >0 \enq

\vv \nin Le couplage envisag\'e ici est le couplage usuel entre deux particules r\'eelles 1 et 2 dans l'espace $E$, produit tensoriel des espaces d'\'etats associ\'es respectivement  \`a chacune des particules. On peut \'egalement le qualifier de ``couplage d'h\'elicit\'e dans la voie s". Nous allons l'expliciter bri\`evement, en renvoyant, pour plus de d\'etails,  le lecteur aux travaux
d\'ej\`a cit\'es de P. Moussa et R. Stora, dont nous adoptons les notations.

\vv \nin Un \'etat d'une particule d'impulsion $p$ et de composante de spin $\sigma$ , valeur propre de l'op\'erateur $S_3(p)$, d\'epend du choix de l'axe $n_3(p)$ sur lequel est mesur\'e le spin, c'est-\`a-dire, en fait, de la triade de vecteurs que l'on associe \`a $p$. Pour cette raison, cet \'etat est not\'e $ | [p], \sigma> $, o\`u $[p]$ est, modulo $\pm$, la matrice 2$\times$2 de SL(2,C) repr\'esentant la transformation de Lorentz $\Lambda_{\ro{p} \rightarrow p}$ qui permet de passer d'une t\'etrade de r\'ef\'erence, choisie une fois pour toutes et pour laquelle $p$ est orient\'ee suivant l'axe des temps :

\beq  p \equiv ~\ro{p} ~= (m,0,0,0) = n_0(\ro{p})~,~~n_i(\ro{p})~,~~i=1,2,3 \enq

\vv \nin en une t\'etrade particuli\`ere associ\'ee \`a $p$, d\'efinie comme

\beq  \Lambda_{\ro{p} \rightarrow p} \left(n_\alpha (\ro{p}) \right) = n_\alpha(p)~,~~~\alpha = 0,1,2,3 \enq

\vv \nin Dans SL(2,C), cette derni\`ere relation est traduite sous la forme

\beq [\,p\,]\hskip -0.1cm  \utnz \hskip -0.1cm[\,p\,]^\dagger = \hskip -0.1cm \utn \enq

\vv \nin Dans cette notation, $\utx$ est une matrice 2$\times$2 associ\'ee \`a un vecteur $x$ d\'efinie par

\beq \utx = x_0 \,\tau_0 + \Vec{\,x\,} \cdot \Vec{\,\tau\,} \enq

\nin $\tau_0$ \'etant la matrice 2$\times$2 unit\'e et $\Vec{~\tau~}$ l'ensemble des trois matrices de Pauli :

\beq \tau_1 =  \left( \begin{array}{cc} 0 & 1 \\ 1 & 0 \end{array} \right) ~,~~
\tau_2 =  \left( \begin{array}{cc} 0 & -i \\ i & 0 \end{array} \right) ~,~~
\tau_3 = \left( \begin{array}{cc} 1 & 0 \\ 0 & -1 \end{array} \right)
\enq

\vv \nin Par d\'efinition, on a $[\,\ro{p}\,] = \tau_0$. Par extension, la matrice $[\,p\,]$ sera aussi appel\'ee {\it t\'etrade de $p$}. Les \'etats de la particule sont d\'efinis par

\beq  | [\,p\,], \sigma> ~=\, U([\,p\,]) ~ | [\,\ro{p}\,], \sigma> \enq

\vv \nin o\`u $U([\,p\,])$ est, dans l'espace des \'etats de la particule, la repr\'esentation unitaire de $[\,p\,]$. Pour simplifier les notations, nous identifierons par la suite une transformation de Lorentz $\Lambda(A)$ avec sa repr\'esentation $A$ dans SL(2,C).

\vv \nin Soit $U(a,A)$ la repr\'esentation unitaire de la transformation $(a, A)$ du groupe de Poincar\'e restreint, $a$ \'etant une translation des vecteurs et $A$ une transformation de Lorentz. On a

\beq U(a,A)~| [\,p\,], \sigma> ~=~e^{i a \cdot Ap}~{\cal D}^s(R_A)_{\sigma' \sigma}~| [\,Ap\,], \sigma'> \enq

\nin o\`u

\beq R_A = [\,Ap\,]^{-1} ~A~[\,p\,] \label{rotW1} \enq

\nin est la {\it rotation de Wigner} associ\'ee \`a $A$, et qui appartient au petit groupe de $p$, et ${\cal D}^s$ sa repr\'esentation agissant sur les variables de spin de la particule.

\vv \nin Dans l'espace produit tensoriel $E$ des espaces des \'etats des particules 1 et 2, l'\'etat

\beq |[\,p_1\,], \sigma_1 ; [p_2], \sigma_2 > ~= |[\,p_1\,], \sigma_1> \otimes |[\,p_2\,], \sigma_2> \label{etatE}
\enq

\nin se transforme donc selon la loi

\begin{eqnarray} & U(a, A) ~|[\,p_1\,], \sigma_1 ; [\,p_2\,], \sigma_2 >~=& \nonumber \\
 &e^{ia \cdot A(p_1+p_2)} ~{\cal D}^{s_1} (R_{A1})_{\sigma'_1 \sigma_1}~{\cal D}^{s_2}(R_{A2})_{\sigma'_2 \sigma_2} ~|[Ap_1], \sigma'_1 ; [Ap_2], \sigma'_2 >& \end{eqnarray}

\nin avec

\beq R_{Ai}  = [A p_i]^{-1}\,A\,[\,p_i\,] ~,~~~i = 1,2 \label{rotW2} \enq

\vv \nin En g\'en\'eral, les deux rotations de Wigner \ref{rotW2} sont diff\'erentes. Le probl\`eme de la r\'eduction de l'espace $E$ en sous-espaces de repr\'esentations irr\'eductibles du type $[\sqrt{s}, J, \eta]$ consiste alors \`a rechercher, pour chacune des particules, des t\'etrades particuli\`eres telles que ces rotations soient identiques, ou, au moins, reli\'ees simplement l'une \`a l'autre, de telle sorte que l'on puisse effectuer le plus simplement possible la r\'eduction du produit tensoriel ${\cal D}^{s_1} \otimes {\cal D}^{s_2}$.

\vv \nin Il s'av\`ere que l'on peut effectivement construire des t\'etrades de ce type. Remarquons tout d'abord qu'une identit\'e entre les deux rotations de Wigner indique que celles-ci ne doivent plus d\'ependre ni de $p_1$ ou de $p_2$, ni des t\'etrades qui leur sont attach\'ees. Cette constatation sugg\`ere de construire les t\'etrades \`a partir d'une t\'etrade arbitraire $[p]$ associ\'ee \`a l'impulsion totale $p = p_1 + p_2$ de l'\'etat \ref{etatE} en posant

\beq R_{A1} = R_{A2} = R_A = [Ap]^{-1} \,A\,[\,p\,]  \label{rotid} \enq

\vv \nin Cette derni\`ere \'equation impose \`a la transformation $[\,p_i\,]~[\,p\,]^{-1}$, qui permet de passer de la t\'etrade associ\'ee \`a $p$ \`a celle associ\'ee \`a $p_i$, de satisfaire \`a une certaine
covariance sous $L^{\uparrow}_{+}$, \`a savoir :

\beq A ~[\,p_i\,]~[\,p\,]^{-1} \,A^{-1}~=~[Ap_i]~[Ap]^{-1}~,~~~{\rm ou}~~~A ~A_{p \rightarrow p_i}~A^{-1} = A_{Ap \rightarrow Ap_i} \enq

\vv \nin Or, la transformation de Lorentz pure $[ \hat{p} \rightarrow \hat{p}_i]$, transformant $\hat{p}$ en $\hat{p_i}$, poss\`ede cette propri\'et\'e. On trouve alors deux solutions au probl\`eme de la r\'eduction.

\vv \nin \ding{52} La premi\`ere consiste \`a poser

\beq [\,p_i\,] = [ \,\hat{p} \rightarrow \hat{p}_i \,]\, [\,p\,]~~~~i=1,2 \enq

\vv \nin et conduit au couplage appel\'e {\it couplage $\ell-s$}, avec une d\'efinition relativiste du moment orbital ; elle satisfait la relation \ref{rotid}.

\vv \nin \ding{52} La seconde solution, donnant des formules de r\'eduction plus simples, consiste \`a poser

\beq [\,p_i\,] \equiv [\,p_i\,]_h = [ \,\hat{p} \rightarrow \hat{p}_i\, ] \,R(n_3(p) \rightarrow \hat{q}_{12})~ [\,p\,] \enq

\vv \nin $q_{12}$ \'etant l'impulsion relative de $p_1$ et $p_2$ et $R(n_3(p) \rightarrow \hat{q}_{12})$
la rotation du petit groupe de $p$ amenant $n_3(p)$ sur $\hat{q}_{12}$ (vecteur unitaire port\'e par $q_{12}$). Posant

\beq \hat{q} = [\,p\,]^{-1}~ \hat{q}_{12}~,~~~n_3 = [\,p\,]^{-1}\,n_3(p) \enq

\vv \nin on a aussi

\beq [\,p\,]^{-1}~R(n_3(p) \rightarrow \hat{q}_{12})~ [\,p\,] = R(n_3 \rightarrow \hat{q}) \enq

\vv \nin On voit que la transformation

\beq [\,p\,]_h = R(n_3(p) \rightarrow \hat{q}_{12})~ [\,p\,] \enq

\nin associe \`a $p$ une t\'etrade d'h\'elicit\'e, puisque

\beq e_s(p) = [\,p\,]_h~n_3 = \hat{q}_{12} ~,~~~e_t (p) = [\,p\,]_h~n_0 = \hat{p} \enq

\vv \nin En outre, il est facile de v\'erifier (voir Appendice A) que la transformation de Lorentz\, pure
$[ \hat{p} \rightarrow \hat{p}_1]$ transforme $\hat{q}_{12}$ en l'impulsion relative unitaire de $-p$ et $p_2$, c'est-\`a-dire, en vecteur d'h\'elicit\'e de $p_1$ :

\begin{eqnarray} & e_s(p_1) = h(p_1,p) = \di{{2 m_1}\over \sqrt{\Lambda}}~\left(-p + p_1 \di{{p_1 \cdot p}\over m^2_1}\right)  =    \di{{2 m_1}\over \sqrt{\Lambda}}~\left(-p_2 + p_1 \di{{p_1 \cdot p_2}\over m^2_1}\right) & \nonumber \\
& {\rm avec}~~~\Lambda = \Lambda(s, m^2_1, m^2_2)~ & \end{eqnarray}

\vv \nin Ainsi, $[\,p_1\,]_h$ associe bien \`a $p_1$ une t\'etrade d'h\'elicit\'e. Pour ce qui concerne la particule 2, comme $\hat{q}_{12} = - \hat{q}_{21}$, la transformation de Lorentz pure $[ \,\hat{p} \rightarrow \hat{p}_2\,]$ transforme $\hat{q}_{12}$ en l'oppos\'e de l'impulsion relative unitaire de $-p$ et $p_2$, c'est-\`a-dire, en vecteur d'h\'elicit\'e de $p_2$ :

\begin{eqnarray} & e_s(p_2) = [ \,\hat{p} \rightarrow \hat{p}_2\,]~\hat{q}_{21} = h(p_2, p) & \nonumber \\
& = \di{{2 m_2}\over \sqrt{\Lambda}}~\left(-p + p_2 \di{{p_2 \cdot p}\over m^2_2}\right)  =    \di{{2 m_2}\over \sqrt{\Lambda}}~\left(-p_1 + p_2 \di{{p_1 \cdot p_2}\over m^2_2}\right) & \end{eqnarray}

\vv \nin Aussi, d'apr\`es Jacob et Wick\footnote{M. Jacob, G. C. Wick, Ann. Phys. 7, 404 (1959).},  la t\'etrade associ\'ee \`a $p_2$ est-elle plut\^ot d\'efinie comme

\beq [\,p_2\,] = [\,p_2\,]_h = [\, \hat{p} \rightarrow \hat{p}_2\,]~[\,p\,]_h~ Y \enq

\vv \nin Y \'etant la rotation d'angle $\pi$ autour de l'axe $n_2(\ro{p}) = n_2$, qui renverse donc $n_3$ et permet d'obtenir ainsi $n_3(p_2) = h(p_2,p)$. En outre, les vecteurs de polarisation circulaire de la t\'etrade de $p$ sont conserv\'es dans les deux transformations  $[\, \hat{p} \rightarrow \hat{p_1}\,]$ et  $[ \,\hat{p} \rightarrow \hat{p}_2\,]$, ce qui fait que :

\beq e^{(\pm)}(p_1) = e^{(\pm)}(p) = e^{(\mp)}(p_2) \enq

\vv \nin Ainsi, cette d\'efinition des t\'etrades correspond bien \`a un couplage d'h\'elicit\'e entre les trois lignes du vertex. L'int\'er\^et majeur de celui-ci r\'eside dans les deux faits suivants :

\vv \nin \leftthumbsup Bien que les deux rotations de Wigner \ref{rotW2} ne soient pas identiques, elles sont n\'eanmoins reli\'ees simplement puisqu'elles sont inverses l'une de l'autre :

\begin{eqnarray}   & R_{A1} = R^{-1}_{A2} = R^{-1}(\hat{q}_A) ~R_A~R(\hat{q}) = [Ap]^{-1}_h ~A~[p]_h & \nonumber \\
& {\rm avec}~~~R_A = [Ap]^{-1}~A~[p]~,~~~\hat{q}_A = R_A \hat{q} & \\
&R(\hat{q}) = R(n_3 \rightarrow \hat{q})~,~~~R(\hat{q}_A) = R(n_3 \rightarrow \hat{q}_A) & \nonumber
\end{eqnarray}

\vv \nin $\spadesuit$ $R_{A1}$ est une rotation d'angle $\psi_A$ dans le bi-plan $(n_1, n_2)$ et conserve donc $n_3$. Par cons\'equent, on a

\beq {\cal D}^{s_1}_{\lambda'_1 \lambda_1} (R_{A1}) = \delta_{\lambda_1 \lambda'_1}~e^{-i \lambda_1 \psi_A}~,~~~{\cal D}^{s_2}_{\lambda'_2 \lambda_2} (R_{A2}) = \delta_{\lambda_2 \lambda'_2}~e^{i \lambda_2 \psi_A} \enq

\vv \nin ce qui montre que les indices respectifs $\lambda_1$ et $\lambda_2$ des deux particules sont ici des invariants relativistes. Nous pouvons retrouver ce r\'esultat en remarquant que les vecteurs
d'h\'elicit\'e $h_(p_1, p)$ et $h(p_2, p)$ satisfont la relation de covariance \ref{covar0} :

\beq \Lambda\{ h(p_i,p) \} = h(\Lambda p_i, \Lambda p) ~~~i=1,2\enq

\vv \nin et que, comme nous l'avons indiqu\'e plus haut, les composantes de spin suivant ces vecteurs sont des invariants relativistes dans l'espace $E$. C'est par analogie avec l'invariance relativiste de l'h\'elicit\'e d'une particule de masse nulle que ces indices de spin particuliers sont appel\'es aussi {\it h\'elicit\'es} et le couplage obtenu {\it couplage d'h\'elicit\'e}. Par rapport au couplage $\ell-s$ standard, ce couplage pr\'esente l'immense avantage de ramener chacune des rotations de Wigner \`a une rotation autour de l'axe $n_3$ dans le r\'ef\'erentiel o\`u la particule consid\'er\'ee est au repos, auquel cas les matrices ${\cal D}^{s_i}$ sont diagonales et les indices de spin correspondants de la particule apparaissent alors comme des invariants relativistes.  Devant l'extr\^eme simplification apport\'e par ce dernier r\'esultat, le probl\`eme de la r\'eduction devient secondaire et pour \'etudier les deux autres cas annonc\'es nous nous laisserons guider par ce qui vient d'\^etre fait ici.

\vv \nin Sous les transformations du groupe de Poincar\'e restreint, les \'etats \`a deux particules en couplage d'h\'elicit\'e se transforment donc de la mani\`ere suivante :

\beq U(a, A) ~|[\,p_1\,]_h, \lambda_1 ; [\,p_2\,]_h , \lambda_2 > = e^{i a \dot A p} ~e^{i(\lambda_2 - \lambda_1) \psi_A} ~|[Ap_1]_h, \lambda_1 ; [Ap_2]_h , \lambda_2 > \enq

\vv \nin En particulier, si $A$ est une rotation dans le plan orthogonal au ``plan de vertex", plan d\'efini par les trois impulsions du vertex, les impulsions restent inchang\'ees et, dans cette transformation, l'\'etat des deux particules est simplement multipli\'e par un facteur de phase :

\beq U(A) ~|[\,p_1\,]_h, \lambda_1 ; [\,p_2\,]_h , \lambda_2 > = e^{i(\lambda_2 - \lambda_1) \psi} ~|[\,p_1\,]_h, \lambda_1 ; [\,p_2\,]_h , \lambda_2 > \enq

\vv \nin Il s'agit ici d'une rotation des vecteurs de polarisation circulaire $e^{(\pm)}(p)$, ou encore, d'un changement de t\'etrade associ\'ee \`a $p$ qui se traduit, dans le syst\`eme du centre de masse des deux particules o\`u $\Vec{p_1}$ et $\Vec{p_2}$ sont colin\'eaires \`a $n_3$, par une rotation d'angle $\psi$ dans le plan $(n_1, n_2)$.

\vv \nin Pour ce qui concerne les vecteurs constituant les t\'etrades d'h\'elicit\'e, on a aussi

\beq A ~e^{(\lambda)}(p_i) = [Ap_1]_h ~R_{Ai}~e^{(\lambda)}(\ro{p}) = e^{-i \lambda \psi_i}~e^{(\lambda)}(Ap_i) \enq

\vv \nin avec $\psi_p = \psi_A = \psi_{p_i} = - \psi_{p_2} $. Par cons\'equent, l'indice $\lambda$ lui-m\^eme est un invariant relativiste. Ce r\'esultat peut \^etre retrouv\'e simplement en consid\'erant \`a nouveau l'\'equation aux valeurs propres de l'op\'erateur d'h\'elicit\'e

\beq \lambda_{\alpha \beta}(p_i) = i ~\varepsilon_{\alpha \beta \mu \nu}~e_s(p_i)^\mu~e_t(p_i)^\nu \enq

\vv \nin op\'erateur qui, les deux vecteurs $e_s(p_i)$ et $e_t(p_i)$ ayant ici la covariance requise, poss\`ede lui aussi des valeurs propres invariantes relativistes. On a en particulier

\beq  \lambda_{\alpha \beta}(p) = i ~\varepsilon_{\alpha \beta \mu \nu}~\hat{q}^\mu_{12}~\hat{p}^\nu \label{ophelcas1} \enq

\begin{figure}[hbt]
\centering
\includegraphics[scale=0.3, width=3cm, height=3cm]{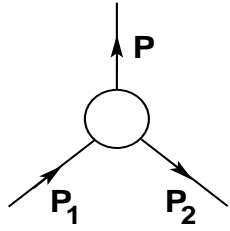}
\vskip 0.25cm

\caption{Vertex \`a deux particules r\'eelles, l'une entrante, l'autre sortante} \label{vertex32}
\end{figure}

\subsection{Deuxi\`eme cas (fig \ref{vertex32})}

\vv \nin Nous supposerons ici que la ligne entrante 1 et la ligne \und{sortante} 2 du vertex repr\'esentent encore deux particules r\'eelles d'impulsions $p_1$ et $p_2$ respectivement, tandis que la troisi\`eme ligne sortante repr\'esente une particule virtuelle \'emise \`a ce vertex et dont
l'impulsion $p=p_1 - p_2$, du genre espace : $p^2 = -t < 0$, est le quadri-transfert d'impulsion entre les deux particules 1 et 2.

\vv \nin Le cas envisag\'e ici correspond, en quelque sorte, \`a un ``couplage d'h\'elicit\'e dans la voie $t$" pour les deux particules 1 et 2. Pour d\'efinir ce couplage, nous proc\`ederons par analogie avec le cas pr\'ec\'edent.

\vv \nin Ici, $p$ est l'impulsion totale de $p_1$ et $-p_2$ et son vecteur unitaire $\hat{p}$ est du genre espace

$$ \hat{p} = \di{p \over \sqrt{t}} ~,~~~\hat{p}^2 = -1 $$

\vv \nin Son vecteur d'h\'elicit\'e, que nous noterons encore $q_{12}$, est l'impulsion relative de $p_1$ et $-p_2$, soit

\beq q_{12} = q_{21} = \di{1\over 2} \left( p_1 + p_2 + \di{{m^2_1 - m^2_2}\over t} ~p \right) 
= p_1 + \di{{p_1 \cdot p}\over t}\,p \label{relcas2} \enq

 \vv \nin Il s'agit d'un vecteur du genre temps car

 \beq q^2_{12} = \di{{\Lambda(-t, m^2_1 , m^2_2)}\over {4 t}} > 0   \enq

\vv \nin et qui pointe vers le futur, puisque ($q_{12} \cdot (p_1-p_2) =0$) \footnote{Rappelons que si un vecteur du genre temps pointe vers le futur dans un r\'ef\'erentiel donn\'e, alors il pointe aussi vers le futur dans tout autre r\'ef\'erentiel qui s'en d\'eduit par une transformation de $L^{\uparrow}_{+}$. Or, la relation \ref{q12futur} montre que dans le r\'ef\'erentiel o\`u la particule 1 est au repos, $q^0_{12} = q_{12}\cdot p_1/m_1 >0$. Par suite, $q_{12}$ est bien du genre temps futur.}

\beq q_{12}\cdot p_1 = q_{12} \cdot p_2 = \di{1\over 2} q_{12} \cdot (p_1 + p_2) = q^2_{12} > 0 \label{q12futur} \enq

\vv \nin Les r\^oles de $p$ et $q_{12}$ \'etant \'echang\'es par rapport au cas pr\'ec\'edent, nous d\'efinirons une t\'etrade d'h\'elicit\'e de $p$ de la mani\`ere suivante. Choisissons en premier lieu une t\'etrade quelconque $[q_{12}]$ associ\'ee \`a $q_{12}$. Effectuons ensuite la rotation du petit groupe de ce vecteur qui am\`ene $n_3(q_{12})$ sur $\hat{p}$. La t\'etrade

\beq [\,p\,]_h = R(n_3(q_{12}) \rightarrow \hat{p}) ~[q_{12}] \enq

\vv \nin donne

\beq [\,p\,]_h n_0 = e_t(p) = \hat{q}_{12}~,~~[\,p\,]_h~n_3 = e_s(p) = \hat{p} \enq

\vv \nin D'apr\`es une remarque pr\'ec\'edente, pour satisfaire l'\'egalit\'e entre les trois rotations de Wigner associ\'ees respectivement aux trois particules du vertex, il suffit d'utiliser des transformations de Lorentz pures appropri\'ees. Il est facile de v\'erifier (voir appendice A) que celle, $[\,\hat{q}_{12} \rightarrow \hat{p}_1\,]$, transformant $\hat{q}_{12}$ en $\hat{p}_1$ transforme $\hat{p}$ en l'impulsion relative de $p_2$ et $p$, c'est-\`a-dire, en vecteur d'h\'elicit\'e de $p_1$ :

\begin{eqnarray} &  [\hat{q}_{12} \rightarrow \hat{p}_1]~\hat{p} = e_s(p_1) = h(p_1,p) = \di{{2 m_1}\over{\sqrt{\Lambda}}}~\left(p - \di{{p_1\cdot p}\over m^2_1}~p_1 \right) & \\
&{\rm avec}~~~\Lambda = \Lambda (-t,m^2_1, m^2_2) & \nonumber \end{eqnarray}

\vv \nin Par ailleurs, puisque $h(p_2, -p) = - h(p_2, p)$, on a

\beq  [\hat{q}_{12} \rightarrow \hat{p}_2]~\hat{p} = e_s(p_2) = -h(p_2, p_2 - p_1) = h(p_2, p) \label{hel2} \enq

\vv \nin c'est-\`a-dire, la transformation de Lorentz pure $[\,\hat{q}_{12} \rightarrow \hat{p}_2\,]$ transforme $\hat{p}$ en l'impulsion relative unitaire de $-p_1$ et $p$, vecteur d'h\'elicit\'e de $p_2$.
D'apr\`es \ref{hel2}, il ne sera pas n\'ecessaire, pour construire la t\'etrade d'h\'elicit\'e de la particule 2, de renverser les axes au d\'epart. Nous d\'efinirons donc les t\'etrades d'h\'elicit\'e des deux particules 1 et 2 par

\beq  [\,p_i\,]_h = [\,\hat{q}_{12} \rightarrow \hat{p}_i\,]~[\,p\,]_h ~~~~i = 1,2 \enq

\vv \nin Cette d\'efinition conduit bien \`a un couplage d'h\'elicit\'e entre les trois particules, pour lequel les trois rotations de Wigner sont identiques et repr\'esentent, chacune pour sa part, une rotation autour de $n_3$ (donc effectu\'ee dans le bi-plan $(n_1,n_2)$), dans un r\'ef\'erentiel particulier. Pour l'une et l'autre des deux particules 1 et 2, ce r\'ef\'erentiel est celui o\`u la particule correspondante est au repos, tandis que pour la particule virtuelle d'impulsion $p$, il s'agit du r\'ef\'erentiel o\`u $q_{12}$ est orient\'e suivant l'axe des temps. Ce dernier est l'analogue du r\'ef\'erentiel du centre de masse dans le cas pr\'ec\'edent. Nous poserons ici

\begin{eqnarray} & \hat{q} = [q_{12}]^{-1}~\hat{p}~,~~~R_A = [Aq_{12}]^{-1}~A~[q_{12}]  \nonumber \\
& R(\hat{q}) = R(n_3 \rightarrow \hat{q})~,~~~\hat{q}_A = R_A ~\hat{q} &
\end{eqnarray}

\vv \nin On obtient donc

\begin{eqnarray} & R_{A1} = R_{A2} = R_A(p) = [Ap]^{-1}_h~A~[p]_h = R^{-1}(\hat{q}_A)~R_A~R(\hat{q}) & \nonumber \\
& {\cal D}^{s_i}_{\lambda'_i \lambda_i} (R_{Ai}) = \delta_{\lambda_i \lambda'_i}~e^{-i \lambda_i \psi_A} & \label{transfovoiet}
\end{eqnarray}

\vv \nin la toute derni\`ere relation traduisant le fait que les composantes de spin des particules 1 et 2 suivant leur propre vecteur d'h\'elicit\'e, $\lambda(p_i) = W_i(\hat{p}_i) \cdot h(p_i,p)$, sont des invariants relativistes. Lorsque la transformation $A$ est une rotation dans le plan perpendiculaire au plan de vertex, les impulsions sont inchang\'ees et les \'etats d'h\'elicit\'e des deux particules sont simplement multipli\'ees par un facteur de phase :

\beq U(A)~|[\,p_i\,]_h, \lambda_i >  = e^{-i \lambda_i \psi}~|[\,p_i\,]_h, \lambda_i > \enq

\vv \nin Dans le r\'ef\'erentiel o\`u $q_{12}$ est orient\'e suivant l'axe des temps et $\hat{q}$ orient\'e suivant $n_3$ ($\Vec{p_1}$ et $\Vec{p_2}$ sont alors \'egalement colin\'eaires \`a $n_3$), cette transformation est une rotation d'angle $\psi$, effectu\'ee dans le plan $(n_1,n_2)$, des vecteurs de polarisation circulaire, lesquels sont identiques pour les trois t\'etrades :

\beq e^{\pm}(p_1) =  e^{\pm}(p_2) = e^{\pm}(p) \enq

\vv \nin Soulignons ici que le couplage dont il vient d'\^etre question n'a de sens qu'\`a condition d'envisager les transformations de Lorentz simultan\'ement dans les deux espaces d'\'etats associ\'es respectivement \`a chacune des particules 1 et 2 r\'eelles. Cette circonstance intervient lorsqu'on \'etudie des \'el\'ements de matrice  de la forme

\beq <[\,p_2\,], \sigma_2 | {\cal O} |[\,p_1\,], \sigma_1> \label{ampliver} \enq

\begin{figure}[hbt]
\centering
\includegraphics[scale=0.3, width=9cm, height=5cm]{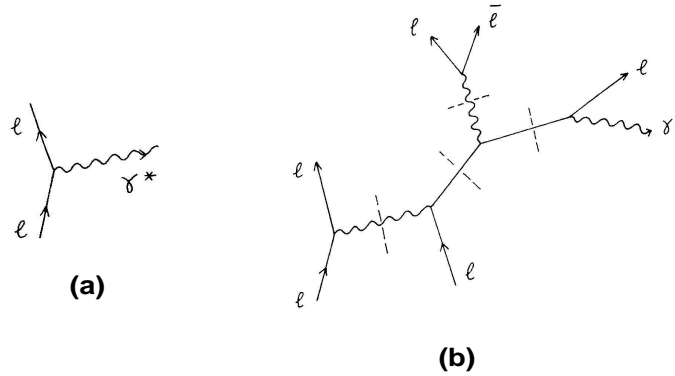}
\vskip 0.25cm

\caption{Vertex (a) dans le diagramme en arbre (b)} \label{vertex33}
\end{figure}

\vv \nin qui repr\'esentent des {\it amplitudes de vertex} pour des vertex tel celui de la figure \ref{vertex33} (a)  intervenant dans un diagramme en arbre tel celui de la figure \ref{vertex33} (b)  d\'ecrivant un certain processus \'el\'ementaire.
L'op\'erateur ${\cal O}$ qui y figure repr\'esente le {\it courant} associ\'e \`a ce vertex, {\it coupl\'e} \`a une particule virtuelle que nous avons sch\'ematis\'ee par la ligne sortante d'impulsion $p$ dans la figure \ref{vertex32}.

\vvv

\vv \nin L'introduction du couplage d'h\'elicit\'e dans la voie t entre les particules 1 et 2 permet d'obtenir de nouvelles amplitudes de vertex dont la loi de transformation est, d'apr\`es  \ref{transfovoiet},  beaucoup plus simple que celle de \ref{ampliver} o\`u les t\'etrades seraient choisies arbitrairement. En effet, $A$ \'etant une transformation de Lorentz, on a d\`es lors

\begin{eqnarray} & <[\,p_2\,]_h, \lambda_2 | U^{-1}(A) U(A) ~J~ U^{-1}(A) U(A) |[\,p_1\,]_h, \lambda_1> = &
\nonumber \\
& e^{i(\lambda_2 - \lambda_1)\psi_A}~<[Ap_2]_h, \lambda_2 |
~J_A~|[Ap_1]_h, \lambda_1>&
\end{eqnarray}

\vv \nin o\`u $J$ est le courant du vertex et $J_A = U(A) ~J~U^{-1}(A)$ le courant transform\'e dont l'expression d\'epend de la repr\'esentation de Lorentz associ\'ee \`a ce courant, c'est-\`a-dire, en fait, de celle de la particule virtuelle \`a laquelle il est coupl\'e, puisque, d'apr\`es les principes g\'en\'eraux de la th\'eorie des champs, lesdites repr\'esentations doivent \^etre identiques.

\vv \nin C'est en projetant ces amplitudes de vertex avec couplage d'h\'elicit\'e sur les fonctions d'onde d'h\'elicit\'e de la particule virtuelle que l'on obtient ce qu'on appelle des {\it amplitudes d'h\'elicit\'e}, caract\'eristiques du vertex consid\'er\'e et qui sont {\it invariantes relativistes}.

\vv \nin On notera que chacun des vecteurs $e^{(\lambda)}(p)$ de la t\'etrade d'h\'elicit\'e de $p$ est vecteur propre, avec la valeur propre $\lambda$, de l'op\'erateur d'h\'elicit\'e quadri-vectorielle

\beq \lambda_{\alpha \beta }(p) = i~\varepsilon_{\alpha \beta \mu \nu}~\hat{p}^\mu~\hat{q}^\nu_{12} \label{ophelcas2} \enq

\vv \nin qui peut \^etre consid\'er\'e comme le {\it prolongement analytique} de l'op\'erateur similaire \ref{ophelcas1} du cas pr\'ec\'edent

\beq \lambda_{{\rm cas}\, 1} (p_1+p_2) ~\rightarrow ~ - \lambda_{{\rm cas} \, 2} (p_1 - p_2) \enq

\vv \nin On peut d'ailleurs relier facilement les vecteurs $e_t$ et $e_s$ obtenus dans l'un et l'autre cas
par la substitution $p_2 \rightarrow - p_2$

\beq e^{{\rm cas}\, 1}_t (p_1 + p_2) \rightarrow  e^{{\rm cas}\, 2}_t (p_1 - p_2) ~,~~
e^{{\rm cas}\, 1}_s (p_1 + p_2) \rightarrow  e^{{\rm cas}\, 2}_s (p_1 - p_2) \enq

\vv \nin Rappelons ici que les g\'en\'erateurs du petit groupe d'un vecteur $p$ du genre espace sont donn\'es par

\beq J(p) = n_0(p) \cdot W(p)~,~~W_1(p) = - n_1(p) \cdot W(p)~,~~W_2(p) = - n_2(p) \cdot W(p) \enq

\vv \nin $W(p)$ \'etant l'op\'erateur de Pauli-Lubansky \ref{PL} et $(n_0, n_1, n_2, \hat{p})$ une base orthonormale de l'espace-temps ($n^2_0(p) = 1~,~n^2_1(p) = n^2_2(p) = -1$). L'ensemble de ces trois op\'erateurs satisfait aux relations de commutation de $L(2,1)$ et $J(p)$ engendre le groupe des rotations dans le plan $(n_1, n_2)$. C'est pr\'ecis\'ement ce groupe de rotations que l'on retrouve aussi bien lorsque $p$ est du genre temps que lorsque $p$ est du genre espace : dans le premier cas, $J(p)$ est la composante du spin suivant $n_3(p)$, tandis que dans le second cas, c'est la composante du spin suivant $n_0(p)$. Dans les deux cas, $J(p)$ a le m\^eme ensemble de valeurs propres.  Si la projection se fait suivant un vecteur d'h\'elicit\'e de $p$, la composante $J(p)$ obtenue est l'op\'erateur d'h\'elicit\'e de la particule, dont les valeurs propres poss\`edent alors l'invariance relativiste.

\vv \nin En remarque finale, notons aussi que les \'etats de spin d'un photon virtuel  sont d\'ecrits dans le cadre de la repr\'esentation quadri-vectorielle du groupe de Lorentz. Aussi, lorsque la particule virtuelle est un photon, l'h\'elicit\'e quadri-vectorielle \ref{ophelcas1}  ou \ref{ophelcas2} repr\'esente effectivement l'h\'elicit\'e du photon. Les vecteurs propres $e^{(\lambda)}(p)$ de cet op\'erateur, qui constituent la t\'etrade de $p$, repr\'esentent quant \`a eux les {\it fonctions d'onde d'h\'elicit\'e} de ce photon.

\begin{figure}[hbt]
\centering
\includegraphics[scale=0.3, width=14cm, height=5cm]{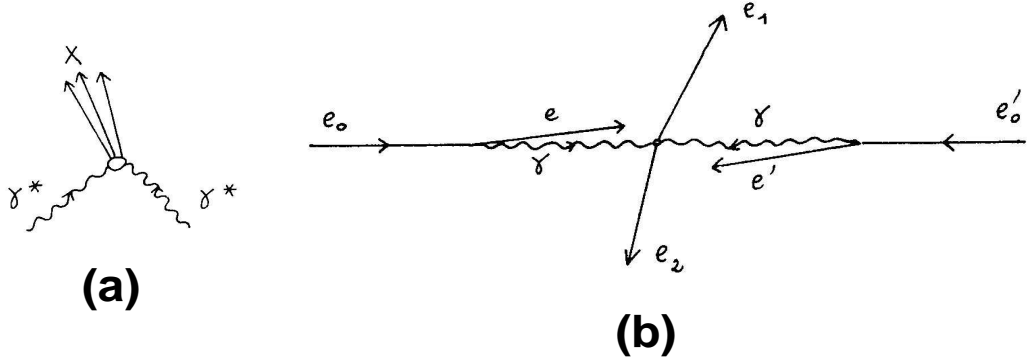}
\vskip 0.25cm

\caption{(a) vertex \`a deux particules virtuelles entrantes ; (b) exemple } \label{vertex34ab}
\end{figure}

\subsection{Troisi\`eme cas (figs \ref{vertex31}, \ref{vertex34ab})}

\vv \nin Consid\'erons maintenant le cas o\`u les deux particules entrantes 1 et 2 du vertex de la figure \ref{vertex31} sont virtuelles et du genre espace. La troisi\`eme ligne, sortante, dont l'impulsion  du genre temps est la somme des impulsions des particules 1 et 2, repr\'esente soit un \'etat quantique r\'eel quelconque $X$ comme dans la figure \ref{vertex34ab} (a),  type de vertex que l'on trouve notamment dans le processus d\'ecrit par le diagramme de la figure \ref{vertex34ab}, soit une particule virtuelle (du genre temps). Nous poserons ici

\beq p^2_1 = - t <0~,~~p^2_2 = -t'~,~~p = p_1 + p_2~,~~p^2=s >0 \enq

\vv \nin En fait, ce troisi\`eme cas diff\`ere peu du premier quant au principe du couplage. Nous proc\`ederons encore par analogie avec les deux cas pr\'ec\'edents.

\vv \nin \ding{182} Pour l'impulsion totale $p$ :

\beq [\,p\,]_h = R(n_3(p) \rightarrow \hat{q}_{12})\,[\,p\,] \enq

\vv \nin o\`u $[p]$ est une t\'etrade arbitraire associ\'ee \`a $[p]$. Le vecteur d'h\'elicit\'e $q_{12}$ repr\'esente ici encore l'impulsion relative des particules 1 et 2 :

\vvv

\beq q_{12} = - q_{21} = p_1 - \di{{p_1 \cdot p}\over s}~p = - p_2 + \di{{p_2 \cdot p}\over s}~p 
\label{q123} \enq

\vv \nin On a ici

$$ q^2_{12} = - \di{\Lambda\over s}~,~~{\rm avec}~~\Lambda= \Lambda(s, -t, -t')~,~~\hat{q}^2_{12} = -1 $$

\vv \nin \ding{183} Pour la particule 1 du genre espace :

\beq [\,p_1\,]_h = [\,\hat{p} \rightarrow h_1\,]\,[\,p\,]_h \enq

\vv \nin $h_1$ \'etant l'impulsion relative unitaire de $p$ et $-p_2$ :

\beq h_1 = h(p_1,p) = e_t(p_1) = \di{{2 \sqrt{t}}\over \sqrt{\Lambda}}~\left(p + \di{{p_1 \cdot p}\over t}~p_1 \right)~,~~h^2_1 = 1~,~~h^0_1 > 0 \label{h123} \enq

\vv \nin Il est facile de v\'erifier que l'on a

\beq  [\,\hat{p} \rightarrow h_1\,]~\hat{q}_{12} = \hat{p}_1 \enq

\vv \nin \ding{184} Pour la particule 2 du genre espace :

\vv \nin Soit $h_2$ l'impulsion relative unitaire de $p$ et $-p_1$, vecteur d'h\'elicit\'e de $p_2$ :

\beq h_2 = h(p_2,p) = e_t(p_2) = \di{{2 \sqrt{t'}}\over \sqrt{\Lambda}}~\left(p + \di{{p_2 \cdot p}\over t'}~p_2 \right)~,~~h^2_2 = 1~,~~h^0_2 > 0 \enq

\vv \nin dans la transformation de Lorentz pure amenant $\hat{p}$ sur ce vecteur, $\hat{q}_{12}$ est transform\'e en $- \hat{p}_2$ : $[\hat{p} \rightarrow h_2]~\hat{q}_{12} = - \hat{p}_2$. Donc, ici \'egalement, il sera n\'ecessaire de renverser au d\'epart l'axe $n_3$ par la rotation $Y$ introduite pr\'ec\'edemment, de fa\c{c}on \`a obtenir $n_3(p_2) = \hat{p}_2$. La t\'etrade associ\'ee \`a la particule 2 sera donc d\'efinie par

\beq [\,p_2\,]_h =  [\,\hat{p} \rightarrow h_2\,]\,[\,p\,]_h~Y \enq

\vv \nin Il est clair que ces d\'efinitions conduisent bien \`a un couplage d'h\'elicit\'e entre les trois particules. On obtient ici encore

$$ e^{(\pm)}(p_1) = e^{(\pm)}(p) = e^{(\mp)}(p_2)~,~~R_{A1} = R_A(p) = R^{-1}_{A2} = [Ap]_h\,A\,[\,p\,]_h $$

\vv \nin Du fait que l'on ne dispose plus d'espace d'\'etats de particules
r\'eelles pour les particules 1 et 2, ce couplage doit \^etre envisag\'e dans
l'espace produit tensoriel des deux espaces de fonctions d'onde d\'ecrivant
les \'etats de spin respectifs de l'une ou l'autre de ces deux particules virtuelles.

\subsection{Extension aux \'etats \`a $N \geq 3$ particules du genre temps\protect\footnote{Cette extension est pr\'esent\'ee sous sa forme la plus g\'en\'erale dans : M. Kummer,  J. Math. Phys. 7, 997 (1966) ; J. Werle, ``Relativistic Theory of Reactions", North-Holland Pub. Comp., Amsterdam (1966) ; P.
Moussa, Th\`ese d'\'etat, Orsay (1968) ; pour le cas $N=3$, voir aussi : S. M. Berman, M. Jacob, Phys. Rev. 139 B, 1023 (1965).}}

\vv \nin Lorsqu'on dispose de plus de deux particules, et donc de plus de deux impulsions, il devient possible de construire, \`a partir de ces impulsions, des bases compl\`etes d'espace-temps, dont les vecteurs se transforment comme les impulsions, c'est-\`a-dire, des vecteurs poss\`edant la propri\'et\'e de covariance \ref{covar0}\footnote{Remarquons ici qu'il peut \^etre utile de choisir les \'el\'ements de la base en fonction de leurs propri\'et\'es de sym\'etrie vis-\`a-vis du groupe des permutations $S_N$.}.  D\`es lors, les projections des spins des particules sur l'un {\it quelconque} de ces vecteurs seront invariantes. Cette possibilit\'e vient du fait que dans le r\'ef\'erentiel du centre de masse du syst\`eme des $N \geq 3$ particules, on peut former, avec  les 
$N-1$ tri-impulsions ind\'ependantes, un v\'eritable solide rigide \`a trois dimensions, auquel on peut attacher un syst\`eme d'axes pouvant servir de base tri-dimensionnelle, ce qui ne peut \^etre r\'ealis\'e pour un \'etat \`a une seule particule (tri-impulsion nulle dans le r\'ef\'erentiel de la particule au repos) ou un \'etat \`a deux particules (une seule tri-impulsion dans ce cas).   Cela est \'evident pour $N \geq 4$ ($N-1 \geq 3$). Pour $N=3$, on note qu'\`a deux tri-vecteurs ind\'ependants $\Vec{p_1}$ et $\Vec{p_2}$ formant un triangle on peut adjoindre le vecteur $\Vec{p_1} \wedge \Vec{p_2}$ perpendiculaire \`a ce triangle pour former un t\'etra\`edre. 
 
\vv \nin Consid\'erons justement le cas $N=3$.  A titre d'exemple, d\'efinissons la t\'etrade d'h\'elicit\'e de l'impulsion totale $P = p_1 + p_2 + p_3$ de la mani\`ere suivante : 

$$ T = \hat{P} = \di{P\over{\sqrt{s}}}~,~~{\rm avec}~~ s = P^2~,~~ Y_\mu = N_y~\varepsilon_{\mu \nu \rho \sigma}~p^\nu_1 p^\rho_2 p^\sigma_3~,~~Z = N_z \left\{ p_3 -\di{{P\cdot p_3}\over s} P \right\}  $$
\beq X_\mu = \varepsilon_{\mu \nu \rho \sigma} T^\nu Y^\rho Z^\sigma \label{choix3} \enq

\vv \nin o\`u l'on note que dans l'\'echange $1 \leftrightarrow 2$, $T$ et $Z$ sont sym\'etriques tandis que $Y$ et $X$ sont antisym\'etriques\footnote{En fait, sous une permutation quelconque de $S_3$, $T$ est compl\`etement sym\'etrique et $Y$ compl\`etement antisym\'etrique.}. On v\'erifie sans peine que ces vecteurs sont bien orthogonaux deux \`a deux. Les facteurs $N_y$ et $N_z$ assurent la normalisation de $Y$ et de $Z$, respectivement ($Y^2 = Z^2 = -1$). On les \'evalue le plus simplement en se pla\c{c}ant dans le r\'ef\'erentiel du centre de masse des trois impulsions o\`u 

$$ Y = (0, \Vec{\,Y\,})~,~~\Vec{\,Y\,} = N_y \left[ \Vec{p_1} \wedge \Vec{p_2} \right]_{\rm CM} =  N_y \left[\Vec{p_2} \wedge \Vec{p_3} \right]_{\rm CM}  = N_y \left[ \Vec{p_3} \wedge \Vec{p_1} \right]_{\rm CM}~,~~$$
$$ Z = (0, \Vec{\,Z\,})~,~~\Vec{\,Z\,} = N_z (\Vec{p_3})_{CM} ~~~(\Vec{p_3}\, = \,-\Vec{p_1} - \Vec{p_2} )_{CM}$$

\vv \nin d'o\`u l'on d\'eduit imm\'ediatement (\`a un signe pr\`es) 

$$ N^{-1}_y = \left\{p_1 p_2 |\sin \Theta_{12} | \right\}^{\rm CM} = \left\{p_2 p_3 |\sin \Theta_{23} | \right\}^{\rm CM} = \left\{p_3 p_1 |\sin \Theta_{31} | \right\}^{\rm CM}~,~~{\rm et}~~N^{-1}_z = p^{CM}_3 $$ 

\nin o\`u, dans ledit r\'ef\'erentiel, $\Theta_{i j}$ est l'angle entre les tri-impulsions des particules $i$ et $j$ et les $p^{CM}_i$ sont les modules de ces tri-impulsions : 

$$ p^{CM}_i = \di{\Lambda^{1/2}(s, m^2_i, W^2_{k \ell})\over {2 \sqrt{s}}} ~,~~ i\neq k \neq \ell $$ 

\nin $W^2_{k \ell} = (p_k + p_\ell)^2$ \'etant la carr\'e de la masse invariante du sous-syst\`eme $( k, \ell)$. Tenant compte de la relation 

$$ \varepsilon_{\sigma \mu \nu \rho}\, \varepsilon^{\sigma \alpha \beta \gamma} = - \left\{ \delta^\alpha_\mu \delta^\beta_\nu \delta^\gamma_\rho + \delta^\alpha_\nu \delta^\beta_\rho \delta^\gamma_\mu + \delta^\alpha_\rho \delta^\beta_\mu \delta^\gamma_\nu - \delta^\alpha_\mu \delta^\beta_\rho \delta^\gamma_\nu - \delta^\alpha_\rho \delta^\beta_\nu \delta^\gamma_\mu - \delta^\alpha_\nu \delta^\beta_\mu \delta^\gamma_\rho \right\} $$

\vv \nin on trouve le vecteur $X$ sous la forme

$$ X = \di{{N_y N_z}\over{\sqrt{s}}}\, \left\{ [ s\, p_2\cdot p_3 - P\cdot p_2\, P\cdot p_3 ] \left( p_1 - \di{{P \cdot p_1}\over s} P \right)  \right.  \hskip 3cm$$
$$ \hskip 3cm \left.  - [ s\, p_1\cdot p_3  -P\cdot p_1 \,P\cdot p_3 ] \left( p_2 - \di{{P \cdot p_2}\over s} P  \right) \right\}  
$$

\vv \nin ou encore 

$$ X = \di{{N_y N_z}\over{\sqrt{s}}}\, \left\{ [ \, p_2\cdot p_3 \, P\cdot p_3 - m^2_3 \,P\cdot p_2 ] \left( p_1 - \di{{p_3 \cdot p_1}\over m^2_3} p_3 \right)  \right.  \hskip 3cm$$
$$ \hskip 3cm \left.  - [ \, p_1\cdot p_3\,P\cdot p_3 - m^2_3\, P\cdot p_1 ] \left( p_2 - \di{{p_3 \cdot p_2}\over m^2_3} p_3 \right) \right\}  
$$

\vv \nin Dans le r\'ef\'erentiel du centre de masse global, ce vecteur s'\'ecrit 

$$ X = (0, \Vec{\,X\,})~,~~{\rm avec}~~ \Vec{\,X\,} = \left. \di{{\Vec{p_3} \wedge \left\{\Vec{p_1} \wedge \Vec{p_2}\right\}}\over{\left|\Vec{p_3} \wedge \left\{\Vec{p_1} \wedge \Vec{p_2}\right\}\right|}}\right|_{CM}$$

\vv \nin A partir de cette t\'etrade $[P]_h$, les t\'etrades d'h\'elicit\'e des particules seront ensuite d\'efinies comme 

$$ [\,p_i\,]_h = [\,\hat{P} \rightarrow \hat{p}_i \,] \, [\,P\,]_h  $$

\vv \nin ce qui conduit \`a l'\'egalit\'e des rotations de Wigner. On obtient ainsi les t\'etrades suivantes\footnote{Le v\'erifier.}. 

\vskip 0.5 cm 

\vv \nin \ding{192} {\bf Particule 1}

$$ T_1 = \hat{p}_1 = \di{{p_1}\over m_1}~,~~Y_1 = Y$$ 
$$ Z_1 = N_z \left\{p_3 - \di{{p_3 \cdot p_1}\over m^2_1} p_1  + \di{1\over{1+ \xi_1}} \left\{\di{{p_1 \cdot p_3}\over{m_1 \sqrt{s}}} +\di{{P\cdot p_3}\over s} \right\}\left[ -P + \di{{P\cdot p_1}\over m^2_1} p_1 \right]   ~\right\} ~$$
$$ X_1 = \di{{N_y N_z}\over{\sqrt{s}}} \left\{  [ s\, p_2\cdot p_3 - P\cdot p_2 P\cdot p_3 ] \di{m_1\over{\sqrt{s}}} \left( -P + \di{{P \cdot p_1}\over m^2_1} p_1 \right)      \right.$$
$$ \left.+ \left[P \cdot p_1 \, P \cdot p_3 - s\, p_1 \cdot p_3 \right] \left[ p_2 - \di{{p_2 \cdot p_1}\over m^2_1} p_1 + \di{1 \over{1 + \xi_1}} \left( \di{{p_1 \cdot p_2}\over{m_1 \sqrt{s}}} + \di{{P\cdot p_2}\over s} \right) \left(-P + \di{{P\cdot p_1}\over m^2_1} p_1 \right) \right] ~\right\} $$
$$ {\rm avec}~~~\xi_1 = \cosh \chi_1 = \di{{P \cdot p_1}\over{m_1 \sqrt{s}}} $$

\vv \nin \ding{193} {\bf Particule 2}

$$ T_2 = \hat{p}_2 = \di{{p_2}\over m_2}~,~~Y_2 = Y$$ 
$$ Z_2 = N_z \left\{p_3 - \di{{p_3 \cdot p_2}\over m^2_2} p_2  + \di{1\over{1+ \xi_2}} 
\left\{\di{{p_2 \cdot p_3}\over{m_2 \sqrt{s}}} +\di{{P\cdot p_3}\over s} \right\}\left[ -P + \di{{P\cdot p_2}\over m^2_2} p_2 \right]   ~\right\} ~$$
$$ X_2 = - \di{{N_y N_z}\over{\sqrt{s}}} \left\{  [ s\, p_1\cdot p_3 - P\cdot p_1 P\cdot p_3 ] \di{m_2\over{\sqrt{s}}} \left( -P + \di{{P \cdot p_2}\over m^2_2} p_2 \right)      \right.$$
$$ \left.+ \left[P \cdot p_2 \, P \cdot p_3 - s\, p_2 \cdot p_3 \right] \left[ p_1 - \di{{p_1 \cdot p_2}\over m^2_2} p_2 + \di{1 \over{1 + \xi_2}} \left( \di{{p_2 \cdot p_1}\over{m_2 \sqrt{s}}} + \di{{P\cdot p_1}\over s} \right) \left(-P + \di{{P\cdot p_2}\over m^2_2} p_2 \right) \right] ~\right\} $$
$$ {\rm avec}~~~\xi_2 = \cosh \chi_2 = \di{{P \cdot p_2}\over{m_2 \sqrt{s}}} $$

\vv

\vv \nin \ding{194} {\bf Particule 3}

$$ T_3 = \hat{p}_3 = \di{{p_3}\over m_3}~,~~X_3 = X~,~~Y_3 = Y$$ 
$$ Z_3 = N_z \di{m_3 \over \sqrt{s}} \left\{ -P +\di{{P\cdot p_3}\over m^2_3}  p_3 \right\} $$

\vv \nin Comme on le voit, le choix \ref{choix3} privil\'egie en fait la particule 3 par rapport aux deux autres, $Z_3$ prenant la forme d'un vecteur d'h\'elicit\'e, selon la d\'efinition \ref{vecelic} que nous lui  avons donn\'ee. Comme le vecteur $Y$ est perpendiculaire \`a l'hyperplan form\'e par les trois impulsions, un autre choix, plus sym\'etrique bien qu'inusuel, consisterait \`a red\'efinir l'h\'elicit\'e des particules comme projection de leurs spins respectifs sur cet axe commun. On peut \'egalement opter 
pour le couplage d'h\'elicit\'e suivant, sym\'etrique lui aussi, qui pr\'esente l'avantage de s'adapter \`a un nombre quelconque de particules, mais qui ne donne pas l'\'egalit\'e des rotations de Wigner.   

\vv \nin Pla\c{c}ons-nous dans le r\'ef\'erentiel du centre de masse global des $N$ particules, de base  

\beq  \left. \hat{P} \right|_{CM} = \ro{T} = (\sqrt{s}, \Vec{0})~,~~ \ro{X} = (0, \Vec{e_x})~,~~ \ro{Y}= (0, \Vec{e_y})~,~~ \ro{Z}= (0, \Vec{e_z}) \label{base-stand} \enq

\vvv
\begin{figure}[hbt]
\centering
\includegraphics[scale=0.3, width=6cm, height=6cm]{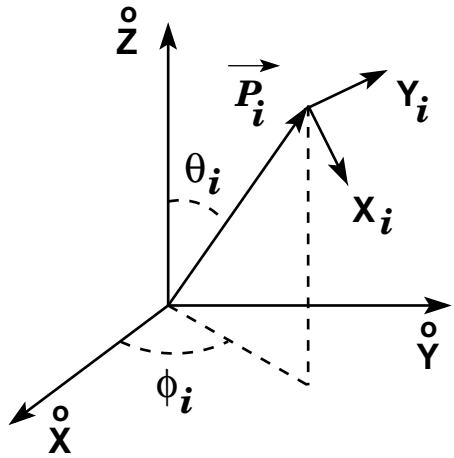}
\vskip 0.25cm

\caption{Param\'etrisation d'une tri-impulsion dans le r\'ef\'erentiel du centre de masse} \label{polar}
\end{figure}

\nin Utilisant la param\'etrisation des coordonn\'ees sph\'eriques(voir figure \ref{polar}), la tri-impulsion de la particule $i$ sera \'ecrite sous la forme

$$ {\Vec{p_i}}_{CM} = \ro{p}_i \, \Vec{e_{ri}} ~,~~{\rm avec} $$
$$ \Vec{e_{ri}} = \cos \theta_i \Vec{e_z} + \sin \theta_i \left( \cos \varphi_i \Vec{e_x} + \sin \varphi_i \Vec{e_y} \right) $$

\vv \nin o\`u $\theta_i$ est l'angle orbital et $\varphi_i$ l'angle azimutal de la particules. Introduisons aussi les vecteurs 

$$ \Vec{e_{\theta i}} = - \sin \theta_i \Vec{e_z} + \cos \theta_i \left( \cos \varphi_i \Vec{e_x} + \sin \varphi_i \Vec{e_y} \right) $$
$$ \Vec{e_{\varphi i}} = - \sin \varphi_i \Vec{e_x} + \cos \varphi_i \Vec{e_y}  $$

\vv \nin et notons 

\beq  \left. X_i \right|_{CM} \equiv (0, \Vec{e_{\theta i}})~,~~\left. Y_i \right|_{CM} \equiv  (0, 
\Vec{e_{\varphi i }})~,~~\left. Z_i \right|_{CM} = (0, \Vec{e_{r i}})  \label{base-standi} \enq 

\vv \nin Les quatre vecteurs $T = \hat{P},~X_i,~Y_i,~Z_i$ forment une base  reli\'ee \`a la base standard \ref{base-stand} par la matrice de rotation

\beq {\cal R}_i = \left( \begin{array}{cccc} 1 & 0 & 0 & 0 \\ 
0 & \cos \theta_i & - \sin \varphi_i & \sin \theta_i \cos \varphi_i \\
0 & \cos \theta_i \sin \varphi_i & \cos \varphi_i & \sin \theta_i \sin \varphi_i \\
0 & - \sin \theta_i & 0 & \cos \theta_i \end{array}             \right) \label{rotri} \enq

\vv \nin On notera que le vecteur $Z_i$ peut \^etre exprim\'e comme 

\beq Z_i = \di{1\over{ \ro{p}_i}} \left( p_i - \di{{P\cdot p_i}\over s} P \right) \enq
$${\rm avec}~~~\ro{p}_i = \di{\Lambda^{1/2}(s, m^2_i, W^2_i)\over{2 \sqrt{s}}}~,~~W^2_i = (P - p_i)^2  $$

\vv \nin et repr\'esente donc pour $P$ un vecteur d'h\'elicit\'e, dans un couplage d'h\'elicit\'e entre le syst\`eme global et la particule $i$. Nous noterons 

\beq [\,P\,]_{h i} = {\cal R}_i \, [\, \ro{P} \,] \enq 

\nin la t\'etrade correspondante, que l'on peut appeler t\'etrade d'h\'elicit\'e de $P$ {\it relativement \`a la particule $i$}. A partir de cette t\'etrade, celle relative \`a $p_i$ sera obtenue par la transformation de Lorentz pure $[ \,\hat{P} \rightarrow \hat{p}_i\,]$ : 

$$ [\, p_i\,]_h = [ \hat{P} \rightarrow \hat{p}_i]\, [\,P\,]_{h i}  ~:  $$
$$ \hat{p}_i = t_i = \cosh \chi_i \,T + \sinh \chi_i \, Z_i ~,~~z_i = \sinh \chi_i \, T + \cosh \chi_i \,Z_i $$ 
$$ x_i = X_i~,~~y_i = Y_i~,~~\cosh \chi_i = \di{{P\cdot p_i}\over{m_i \sqrt{s}}} $$ 
\beq  {\rm et~~donc}~~z_i = \di{{2 m_i }\over{\Lambda^{1/2}(s, m^2_i, W^2_i)} } \left( - P + \di{{p_i \cdot P}\over{m^2_i}} \, p_i \right) \enq

\vv \nin ce qui correspond bien \`a un couplage d'h\'elicit\'e entre les t\'etrades de $P$ et $p_i$. 

\vv \nin Pour terminer, remarquons que ce formalisme peut \^etre \'etendu sans difficult\'e au cas o\`u certaines particules sont virtuelles et du genre espace. 

\vvv

\section{Fonctions d'onde des particules virtuelles}

\vv \nin Mis \`a part le cas sp\'ecial des particules, qui, telles
le photon, ont une masse r\'eelle nulle, cas qui requiert
g\'en\'eralement une \'etude particuli\`ere, la construction
explicite de ces fonctions d'onde, aussi appel\'ees {\it amplitudes spinorielles}\footnote{Voir P. Moussa, R.Stora, loc. cit. ; voir aussi S. Weinberg, Phys. Rev. 133, B1318 (1964).}, ne pose pas de probl\`eme majeur
lorsque l'impulsion $p$ de la particule virtuelle est du genre
temps. Pour les obtenir, il suffit en effet d'effectuer un
prolongement analytique des fonctions d'onde de la particule {\it
sur-couche}, c'est-\`a-dire, lorsque son impulsion (du genre temps
futur) v\'erifie $p^2 = m^2$ o\`u $m$ est sa masse r\'eelle, en
rempla\c{c}ant partout $m$ par $\sqrt{s} = \sqrt{p^2}$, grandeur qui
repr\'esente alors la masse de la particule {\it hors-couche}.
Ainsi, les fonctions d'onde d'un \'electron, resp. d'un positron,
virtuel du genre temps sont simplement les spineurs
$U_{\sigma}([p])$, resp. $V_{\sigma}([p])$, d\'ecrivant les \'etats
de spin d'un \'electron, resp. d'un positron, de masse $\sqrt{s}$.
Ces spineurs v\'erifient les \'equations de Dirac

\beq \left( p \hskip -0.15 cm \slash -\sqrt{s}  \right) U_{\sigma}([\,p\,]) =0~,~~~\left( p \hskip -0.15 cm \slash + \sqrt{s}  \right) V_{\sigma}([p]) =0 \label{eqdirac} \enq

\vv \nin o\`u $ p \hskip -0.15 cm \slash = p_\mu \gamma^\mu$, les $\gamma^\mu,~\mu =0,1,2,3$ \'etant les quatre matrices de Dirac  Elles sont aussi vecteurs propres, avec la valeur propre $\sigma = \pm 1/2$, de la composante suivant $n_3(p)$ de l'op\'erateur de polarisation qui prend ici l'expression

\beq W_\mu(\hat{p}) = \di{i\over 4}~\varepsilon_{\mu \nu \alpha \beta}~\hat{p}^\nu~\gamma^\alpha~\gamma^\beta \enq

\vv \nin Pour une particule de spin $J$ quelconque, les fonctions d'onde de la {\it repr\'esentation spinorielle} correspondant \`a ce spin sont des spineurs \`a $2(2J+1)$ composantes. Ils v\'erifient
une \'equation analogue \`a l'\'equation de Dirac\footnote{Et qui revient \`a \ref{eqdirac} pour $J=1/2$.} :

\begin{eqnarray} & \left[\, \Gamma (\hat{p}) -1 \,\right]~U_{\sigma}([\,p\,]) =0~,~~{\rm avec} & \nonumber \\&~& \nonumber \\
&\Gamma (\hat{p}) = \left( \begin{array}{cc} 0 & {\cal D}^J(\uthp) \\ {\cal D}^J(\,\,\,\tilde{\hat{p}}\,\,\,) & 0 \end{array} \right)~~,~~~{\rm et} ~~~\tilde{\hat{p}} = \uthpm \hat{p}^2 = \uthpm  &
\label{eqn:EDG} \end{eqnarray}

\vv \nin les ${\cal D}^J$ \'etant des matrices carr\'ees (2J+1)$\times$(2J+1).

\vv \nin Les fonctions d'onde des particules virtuelles du genre espace peuvent \^etre construites dans le m\^eme espace de repr\'esentation spinorielle que celui correspondant \`a la particule virtuelle du genre temps. Elles seront alors d\'efinies comme \'etant les vecteurs propres de la composante de l'op\'erateur de polarisation de cette repr\'esentation suivant le vecteur unitaire du genre temps futur $n_0(p)$. Comme nous l'avons d\'ej\`a not\'e, cet op\'erateur et l'op\'erateur $S_3(p)$ pour la particule du genre temps sont les m\^emes, au signe pr\`es, et poss\`edent donc le m\^eme ensemble de valeurs propres. Cependant, selon la valeur $J$ du spin de la repr\'esentation, une difficult\'e se pr\'esente concernant l'\'equation de Dirac g\'en\'eralis\'ee \`a laquelle doivent en principe satisfaire les fonctions d'onde. En effet, l'extension la plus imm\'ediate de \ref{eqn:EDG} est la suivante :

\beq \left[\,\Gamma(\hat{p}) - 1\,\right] \Psi(p) = 0 \label{EDGS} \enq

\vv \nin o\`u cette fois $\hat{p}$ est le vecteur unitaire du genre espace port\'e par $p$ : $\hat{p} = p/\sqrt{t}~,~t= -p^2 >0$. Mais puisque $\Gamma^2(\hat{p}) \Psi(p) = \Gamma(\hat{p}) \Psi(p) = \Psi(p)$ et que $\Gamma^2(\hat{p}) = (-1)^{2J}$, l'\'equation \ref{EDGS} ne peut \^etre valable que si $J$ est entier et ne peut donc \^etre retenue pour les valeurs demi-enti\`eres du spin. Par exemple, dans le cas d'un \'electron, l'\'equation serait

$$ \left( p \hskip -0.15 cm \slash -\sqrt{t}  \right) \Psi(p)=0 $$

\vv \nin d'o\`u l'on d\'eduirait

$$ \left( p \hskip -0.15 cm \slash +\sqrt{t}  \right)~ \left( p \hskip -0.15 cm \slash -\sqrt{t}  \right) \Psi(p)= -2 t \Psi(p) = 0 $$

\vv \nin ce qui conduirait \`a la solution triviale (et inutile) $\Psi(p)=0$.

\vv \nin Pour rem\'edier \`a cette difficult\'e, il peut s'av\'erer tr\`es commode d'introduire, ainsi que l'a fait P. Kessler\footnote{P. Kessler : Nucl. Phys. B \und{15}, 253 (1970).  Voir aussi A. Jaccarini, Canadian Journ. of Phys. \und{51}, 1304 (1973) ; Th\`ese d'Etat, Universit\'e Laval, Qu\'ebec, Canada (1975).}, le nombre $\hat{1}$, dont les propri\'et\'es sont les suivantes :

\beq \hat{1}^\star = \hat{1}~,~~\hat{1}^2 = -1 \enq

\vv \nin L'emploi de ce nombre \'evite celui des nombres complexes et permet d'effectuer simplement certains prolongements analytiques. En particulier, la ``masse" de la particule virtuelle peut \^etre d\'efinie par

\beq \sqrt{p^2} = \hat{1}~\sqrt{t} \enq

\vv \nin et le vecteur unitaire du genre espace $\hat{p}$ peut \^etre transform\'e en un vecteur unitaire du genre temps par la substitution

\beq \hat{p} \rightarrow \di{p\over{\sqrt{p^2}} }= \di{p \over{\hat{1} \sqrt{t}} }= \hat{p}'  \enq

\vv \nin On g\'en\'eralisera d\`es lors l'\'equation \ref{eqn:EDG} au cas ``genre espace- spin demi-entier" par

\beq \left[ \Gamma (\hat{p}') - 1\right] ~\Psi(p) = 0 \enq

\vv \nin Par exemple, pour un \'electron ou un positron, on aura maintenant

$$ \left( p \hskip -0.15 cm \slash -\hat{1} \sqrt{t}  \right) U_{\sigma}([p])=0~,~~\left( p \hskip -0.15 cm \slash +\hat{1} \sqrt{t}  \right) V_{\sigma}([p])=0 $$

\vv \nin Notons \'egalement que si l'on transforme simultan\'ement le vecteur unitaire du genre temps associ\'e \`a $p$ en un vecteur unitaire du genre espace par la substitution

\beq n_0(p) \rightarrow n'_0(p) = \di{{n_0(p)}\over \hat{1}} \enq

\vv \nin la t\'etrade $(\hat{p}', n_1(p), n_2(p), n'_0(p)$ constitue cette fois une base orthonorm\'ee et directe de l'espace-temps, ayant $\hat{p}'$ pour vecteur unitaire du genre temps.

\vv \nin Dans SL(2,C), on peut repr\'esenter la transformation permettant d'\'echanger les r\^oles des vecteurs $n_0$ et $n_3$ d'une t\'etrade de r\'ef\'erence par la matrice

\beq a = \sqrt{\hat{1}}~ \left(\begin{array}{cc} \hat{1} & 0 \\ 0 & 1 \end{array} \right) \enq

\vv \nin qui est telle que

\beq a \tau_0 a = \di{\tau_3 \over {\hat{1}}}~,~~a \tau_3 a = \di{\tau_0 \over {\hat{1}}}~,~~a \tau_i a = \tau_i~~~i=1,2 \enq

\vv \nin D\`es lors, la transformation qui permet d'amener la t\'etrade de r\'ef\'erence sur celle associ\'ee \`a $\hat{p}'$ d\'efinie ci-dessus pourra \^etre repr\'esent\'ee par la matrice

\beq [\,p'\,] = [\,p\,] ~a \enq

\vv \nin $[\,p\,]$ \'etant la t\'etrade de $p$, d\'efinie de la fa\c{c}on habituelle par

\beq [\,p\,] \utnzz [\,p\,]^\dagger = \utn~,~~~n_3(p) = \hat{p}  \enq

\vv \nin L'avantage que l'on retire de ces d\'efinitions est qu'elles permettent de prolonger analytiquement les fonctions d'onde d'une particule du genre temps lorsque celle-ci devient du genre espace. Ceci peut \^etre v\'erifi\'e dans le cas tr\`es simple de l'\'electron, pour lequel les spineurs pour $p^2 <0$ s'obtiennent \`a partir des spineurs pour $p^2 >0$ par la transformation :

\beq \left\{ U_{\sigma} ([p])\right\}_{p^2 >0} \rightarrow \left\{ U_{\sigma} ([p'])\right\}_{p^2 < 0} \enq

\vv \nin ces spineurs \'etant, l'un et l'autre, normalis\'es \`a l'unit\'e.

\section{Premi\`ere application : la m\'ethode d'h\'elicit\'e g\'en\'eralis\'ee}

\vv \nin La m\'ethode dont il sera question ici s'appuie sur le formalisme des couplages d'h\'elicit\'e\footnote{Voir P. Kessler, loc. cit}. Elle permet de clarifier grandement le calcul de certains processus \'el\'ementaires d\'ecrits par des diagrammes de Feynman du type ``arbre". Elle conduit \`a une factorisation de la section efficace relative \`a un tel diagramme  en divers {\it facteurs dynamiques} ou {\it taux de vertex}, dont chacun caract\'erise la dynamique d'un vertex particulier du diagramme. Plus pr\'ecis\'ement, elle fournit le moyen de d\'efinir sans ambigu\"{i}t\'e des facteurs dynamiques, caract\'eristiques, soit de chaque vertex, soit de chaque \'echange de particule virtuelle dans ce diagramme, et d'exprimer directement la section efficace (diff\'erentielle) correspondante comme un produit, en g\'en\'eral de nature tensorielle, de ces divers facteurs. Il en r\'esulte une tr\`es grande clarification des formules, qui se pr\^etent ainsi \`a une meilleure interpr\'etation physique.

\subsection{Processus avec \'echange d'un \'electron du genre espace}

\begin{figure}[hbt]
\centering
\includegraphics[scale=0.3, width=6cm, height=4cm]{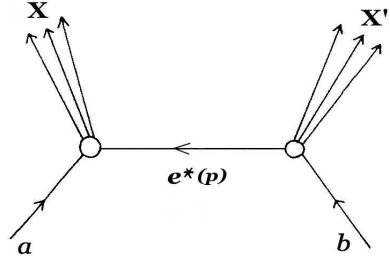}
\vskip 0.25cm

\caption{Diagramme d'un processus avec \'echange d'un \'electron virtuel} \label{ech-elec}
\end{figure}

\nin Le processus envisag\'e est sch\'ematis\'e par le diagramme de la figure \ref{ech-elec}. En mettant \`a part les constantes de couplage et le d\'enominateur du propagateur de l'\'electron \'echang\'e, l'amplitude de transition correspondante s'\'ecrit

\beq T= \Alg{X'}\,\left( p \hskip -0.15 cm \slash +m \right) \,X \enq

\vv \nin $m$ \'etant la masse de l'\'electron sur-couche ; $X$ et $X'$ sont les amplitudes de vertex, de nature spinorielle, se rapportant respectivement au vertex ``de gauche" et au vertex ``de droite" du diagramme ; $\Alg{X}= X^\dagger \gamma_0$ est le spineur adjoint de $X$.

\vvv

\vv \nin D\'ecomposons tout d'abord le propagateur de l'\'electron suivant les fonctions d'onde de cet \'electron virtuel. On obtient :

\beq  p \hskip -0.15 cm \slash +m = \di{\sum_\sigma} \left\{ u\, U_\sigma ([\,p'\,]) \Alg{U}_\sigma([\,p'\,]) + v \,V_\sigma([\,p\,]) \Alg{V}_\sigma([\,p\,]) \right\} ~,~~V = \gamma_5 U \label{decompprop} \enq

\vv \nin o\`u

$$ u = \di{1\over 2}(1 + \di{m\over{m^\star}})~,~~v =  \di{1\over 2}(1 - \di{m\over{m^\star}})~,~~m^\star = \sqrt{p^2} = \hat{1} \sqrt{t} $$

\vv \nin les spineurs \'etant ici normalis\'es selon $\Alg{U}_\sigma ~U_\sigma' = 2 m^\star \delta_{\sigma \sigma'} $

\vv \nin La t\'etrade $[p']$ peut \^etre choisie arbitrairement, puisque la somme figurant dans \ref{decompprop} ne d\'epend pas de ce choix. Aussi, nous la prendrons comme \'etant la t\'etrade d'h\'elicit\'e de $p$, d\'efinie relativement au vertex de gauche. Mais pour pr\'eserver la sym\'etrie entre les deux vertex, nous exprimerons, en une deuxi\`eme \'etape, les spineurs adjoints $\Alg{U}$ et $\Alg{V}$ en fonction des spineurs adjoints $\Alg{U'}$ et $\Alg{V'}$ correspondant \`a la t\'etrade d'h\'elicit\'e de $p$ d\'efinie cette fois relativement au vertex de droite :

\beq \Alg{U}_\lambda = \di{\sum_{\lambda'}} ~r^\star_{\lambda \lambda'}~\Alg{U'}_{\lambda'} \enq

\vv \nin La matrice ${r_{\lambda \lambda'}}$ effectuant ce changement de t\'etrade repr\'esente, en fait, la transformation de Lorentz du petit groupe de $p$, amenant le plan du vertex de gauche sur le plan du vertex de droite (``rotation d'angle imaginaire" autour de $p$). Explicitement, on a ($\lambda$ en indice de ligne, $\lambda'$ en indice de colonne)

\beq
\left\{ r_{\lambda \lambda'}(\theta) \right\} =
\left( \begin{array}{cc} \cos \di{\theta\over 2} & - \sin \di{\theta \over 2} \\
~&~ \\
\sin \di{\theta \over 2} & \cos \di{\theta \over 2}
\end{array}   \right)
 \enq

\vv \nin On obtient ainsi

\beq  p \hskip -0.15 cm \slash +m = \di{\sum_{\lambda, \lambda'}}~r^\star_{\lambda \lambda'}~\left(u~U_\lambda~\Alg{U'}_{\lambda'} + v~V_\lambda~\Alg{V'}_{\lambda'} \right) \enq

\vv \nin Une telle d\'ecomposition du propagateur permet d'exprimer l'amplitude de transition en fonction d'amplitudes d'h\'elicit\'e relatives \`a chacun des deux vertex :

\beq T = \di{\sum_{\lambda, \lambda'}}~r^\star_{\lambda \lambda'}~\left\{ u\, J_\lambda~J'^\star_{\lambda'} + v\, \Alg{J}_\lambda~\Alg{J'}^\star_{\lambda'} \right\} \enq

\vv \nin ces amplitudes \'etant d\'efinies par

\beq J_\lambda = \Alg{X} U_\lambda~,~~~\Alg{J}_\lambda = \Alg{X} V_\lambda \enq

\vv \nin pour le vertex de gauche et par

\beq J'_{\lambda'} = \Alg{X'} U'_{\lambda'}~,~~~\Alg{J'}_{\lambda'} = \Alg{X'} V'_{\lambda'}  \enq

\vv \nin pour le vertex de droite. Supposons que les matrices densit\'es de spin des particules externes du diagramme soient diagonales. Le {\it taux d'interaction} relatif au diagramme est alors donn\'e par

\begin{eqnarray} & {\cal T} = ~\di{\sum_{\lambda, \lambda'}} ~\di{\sum_{\chi, \chi'}} ~\di{\sum_{g,d}} \left\{
u\, J_\lambda ~~J'^\star_{\lambda'} + v\, \Alg{J}_\lambda~\Alg{J'}^\star_{\lambda'} \right\}  r^\star_{\lambda \lambda'}\times r_{\chi \chi'} \left\{
u\, J^\star_\chi ~~J'_{\chi'} + v\, \Alg{J}^\star_\chi~\Alg{J'}_{\chi'} \right\} & \nonumber \\
&= \di{\sum_{\lambda \chi} }~\di{\sum_{\lambda' \chi'} }~r^\star_{\lambda \lambda'} r_{\chi \chi'}~\left\{ u^2 \left\{\di{\sum_g} J_\lambda J^\star_\chi \right\} \left\{\di{\sum_d} J'^\star_{\lambda'} J'_{\chi'} \right\} + v^2  \left\{\di{\sum_g} \Alg{J}_\lambda \Alg{J}^\star_\chi\right\}\left\{\di{\sum_d} \Alg{J'}^\star_{\lambda'} \Alg{J'}_{\chi'}\right\}\right.& \nonumber \\
&\left. + u v \left\{\di{\sum_g} J_\lambda  \Alg{J}^\star_\chi \right\}  \left\{ \di{\sum_d} J'^\star_{\lambda'}
\Alg{J'}_{\chi'} \right\} + u v \left\{ \di{\sum_g} \Alg{J}_\lambda  J^\star_\chi \right\} \left\{ \di{\sum_d}\Alg{ J'}^\star_{\lambda'}
J'_{\chi'}\right\} \right\}& \label{eqn:taux} \end{eqnarray}

\vv \nin o\`u les symboles $\di{\sum_g}$ et $\di{\sum_d}$ signifient que l'on somme sur les indices de spin des particules externes des vertex de gauche et de droite, respectivement.

\vv \nin Dans le diagramme de la figure \ref{ech-elec}, les lignes entrantes $a$ et $b$ repr\'esentent habituellement deux particules allant entrer en collision, d'impulsion et de spin $p_a$, $s_a$ et $p_b$, $s_b$, respectivement. Nous choisirons pour leurs t\'etrades celles qui r\'ealisent le couplage
d'h\'elicit\'e au vertex de gauche et au vertex de droite, respectivement. Quant aux lignes sortantes $X$ et $X'$, elles peuvent a priori repr\'esenter des \'etats finals quelconques r\'esultant de la collision, mais nous supposerons en premier lieu que ces derniers sont des \'etats \`a une particule chacun, d'impulsion et de spin $p_X$, $s_X$ et $p_{X'}$, $s_{X'}$, respectivement, et nous leur attribuerons \'egalement leurs t\'etrades d'h\'elicit\'e respectives au vertex o\`u ils interviennent.
Ainsi, on a, par exemple,

\beq J_\lambda = <X| \Alg{\eta}(0) |a> U_\lambda \equiv
<[\,p_X\,]_h ,\lambda_X |\Alg{\eta}(0) |[\,p_a\,]_h ,\lambda_a > U_\lambda([p]_h) \enq

\vv \nin avec $\Alg{\eta}(0) = \eta^\dagger (0) \gamma_0$ o\`u $\eta(0)$ est un certain op\'erateur de champ, pris au point d'espace-temps $x=0$, qui appartient \`a la m\^eme repr\'esentation du groupe de Lorentz que celle de l'\'electron, c'est-\`a-dire, celle de spin 1/2. Sous une transformation de Lorentz $\Lambda$ repr\'esent\'ee par la matrice (2$\times$2) $A$  de SL(2,C), on a

\begin{eqnarray} & U(A) \,\eta(x)\, U^{-1}(A) = S^{-1}(A) \,\eta(\Lambda x)\,,~~{\rm soit}~~U(A)\, \eta(0) \,U^{-1}(A) = S^{-1}(A) ~\eta(0) & \nonumber \\
& {\rm et} ~~~U(A) \,\Alg{\eta}(0)\, U^{-1}(A) = \Alg{\eta} (0) ~S(A) & \end{eqnarray}

\vv \nin o\`u, dans la repr\'esentation standard des matrices $\gamma$ de Dirac\footnote{C'est-\`a-dire : $\gamma_0 = \left( \begin{array}{cc} \tau_0 & 0 \\ ~&~ \\ 0 & -\tau_0 \end{array} \right),~~\gamma^i = - \gamma_i = \left( \begin{array}{cc} 0 & \tau_i \\ ~&~ \\ - \tau_i & 0 \end{array} \right),~i=1,2,3,~~\gamma_5 = \left( \begin{array}{cc} 0 & \tau_0 \\ ~&~ \\ \tau_0 & 0  \end{array} \right)$},  la matrice
4$\times$4 $S(A)$ a pour expression

\beq S(A) = \di{1\over 2} \left( \begin{array}{cc} A + A^{\dagger - 1} & A - A^{\dagger - 1} \\~\\
A - A^{\dagger - 1} & A + A^{\dagger - 1} \end{array} \right) \enq

\vv \nin Tenant compte de l'unitarit\'e de $U(A)$ ($U^{-1}(A) = U^\dagger(A)$) et de

\newpage

$$S(A) U_\lambda ([\,p\,]) = {\cal D}^{1/2}_{\lambda \lambda'} (R_{A}) U_{\lambda'}([Ap])$$

\vv \nin \'ecrivons

$$ J_\lambda = <X|U^{-1}(A) U(A)\, \Alg{\eta}(0)\, U^{-1}(A) U(A) |a> U_\lambda = $$
$$ \left[ {\cal D}^{s_X }_{\lambda'_X \lambda_X}  (R_{A X})\right]^\star ~{\cal D}^{s_a}_{\lambda'_a \lambda_a} (R_{A a}){\cal D}^{1/2}_{\lambda \lambda'} (R_{A}) <[Ap_X], \lambda'_X| \Alg{\eta}(0) | [Ap_a], \lambda'_a> U_{\lambda'}([Ap]) $$

\vv \nin Or, en couplage d'h\'elicit\'e (dans la voie t),

$$ R_{A X} = R_{A a} = R_A~,~~{\cal D}^J_{\lambda \lambda'}(R_A) = \delta_{\lambda' \lambda} e^{-i \lambda \varphi} $$

\vv \nin d'o\`u il r\'esulte que

\beq J_\lambda \equiv e^{i( \lambda_X - \lambda_a - \lambda) \varphi} ~<[Ap_X]_h, \lambda_X| \Alg{\eta}(0) | [Ap_a]_h, \lambda_a> U_{\lambda}([Ap]_h) \label{invar} \enq

\vv \nin En consid\'erant une rotation dans le bi-plan orthogonal au plan de vertex, $Ap_X = p_X,~Ap_a = p_a, Ap = p$ et l'on trouve alors

\beq J_\lambda \equiv e^{i( \lambda_X - \lambda_a - \lambda) \varphi}~J_\lambda  \enq

\vv \nin ce qui implique la {\it conservation de l'h\'elicit\'e} au vertex consid\'er\'e :

\beq \lambda = \lambda_X - \lambda_a \label{invarhel} \enq

\vv \nin On en d\'eduit par exemple que $\di{\sum_g} J_\lambda  J^\star_\chi
\propto \delta_{\lambda \chi}$ et des propri\'et\'es similaires pour les autres grandeurs tensorielles apparaissant dans \ref{eqn:taux}. Combinant \ref{invar} et \ref{invarhel} on en d\'eduit

\beq J_\lambda(p_X, p_a) = J_\lambda(Ap_X, Ap_a) \label{invaramp} \enq

\vv \nin et que par cons\'equent cette amplitude est invariante relativiste et ne d\'epend donc que d'invariants relativistes du vertex de gauche ($\lambda_X, \lambda_a, t, W, m_a$, etc). Posons ensuite

$$ I_\lambda = \di{\sum_g} |J_\lambda|^2 ~,~~ I'_{\lambda'} = \di{\sum_d} |J'_{\lambda'}|^2~,~~\Alg{I}_\lambda = \di{\sum_g} |\Alg{J}_\lambda|^2~,~~\Alg{I'}_{\lambda'} = \di{\sum_d} |\Alg{J'}_{\lambda'}|^2$$
$$ K_\lambda =\di{\sum_g} J_\lambda  J^\star_\lambda~,~~K'_{\lambda'} =  \di{\sum_d}\Alg{ J'}^\star_{\lambda'} J'_{\lambda'}~,~~R_{\lambda \lambda'} = (r_{\lambda \lambda'})^2 $$

\vv \nin Si l'on suppose de plus que la parit\'e est conserv\'ee \`a chaque vertex, on a les relations\footnote{Voir P. Kessler, loc. cit.}

$$ I_{1/2} = I_{-1/2} = I~,~~I'_{1/2} = I'_{-1/2} = I'~,~~\Alg{I}_{1/2}=\Alg{I}_{-1/2}=\Alg{I}~,~~\Alg{I'}_{1/2}=\Alg{I'}_{-1/2}= \Alg{I'} $$
$$K_{1/2}=-K_{-1/2} = K~,~~K'_{1/2}=-K'_{-1/2} =K' $$

\vv \nin D'o\`u, tous calculs faits,

\beq {\cal T} = 2\left\{ u^2\, I\, I' + v^2\, \Alg{I} \,\Alg{I'} + 2 uv\,\cos \theta\, Re(K K'^\star) \right\} \enq

\vv \nin C'est la formule de factorisation annonc\'ee, o\`u sont clairement s\'epar\'es des ``taux de vertex" tels $I$, $I'$ etc, caract\'eristiques de la dynamique propre \`a chaque vertex, et le facteur  $\cos \theta$ qui, lui, se rattache \`a la propagation de l'\'electron virtuel.

\vv \nin Ce r\'esultat peut \^etre \'etendu dans une certaine mesure au cas o\`u les \'etats finals $X$ et $X'$ comportent plus d'une particule, et dont les \'etats quantiques se construisent dans l'espace produit tensoriel des espaces relatifs aux particules impliqu\'ees. Il suffit en effet de proc\'eder \`a la d\'ecomposition de cet espace en repr\'esentations irr\'eductibles du groupe de Poincar\'e. L'\'etat $|X>$ par exemple sera alors exprim\'e comme une superposition d'\'etats (``ondes partielles") attach\'es chacun \`a une telle repr\'esentation irr\'eductible, avec des nombres quantiques, masse, spin, h\'elicit\'e, parit\'e, etc, bien d\'efinis au regard du groupe de Poincar\'e, comme pour une particule unique, et auquel pourra donc \^etre appliqu\'e le r\'esultat pr\'ec\'edent. On pourra ensuite rattacher les taux de vertex obtenus aux sections efficaces d'\'electro-production virtuelle \`a chaque vertex.

\vv \nin Notons que, l'\'electron \'echang\'e \'etant du genre
espace, le facteur $\cos \theta$ est en fait un cosinus
hyperbolique. Son expression g\'en\'erale se trouve en calculant le
produit scalaire des deux vecteurs d'h\'elicit\'e

\beq q_g = \di{{2 \sqrt{t}}\over{\Lambda^{1/2}(W^2, m^2_a, -t)}}
\left( p_a + \di{{p_a \cdot p}\over t}~p \right),~~q_d = \di{{2
\sqrt{t}}\over{\Lambda^{1/2}({W'}^2, m^2_b, -t)}} \left( p_b +
\di{{p_b \cdot p}\over t}~p \right) \enq

\vv \nin soit

\beq \cos \theta = q_a \cdot q_b = \di{{t(2s -t - W^2 - m^2_a -
{W'}^2 -m^2_b) - (W^2- m^2_a)({W'}^2-m^2_b)}\over{\Lambda^{1/2}(W^2,
m^2_a, -t) \Lambda^{1/2}({W'}^2, m^2_b, -t)} } \label{cosh} \enq
$$ {\rm avec}~~~s = (p_a + p_b)^2,~~m^2_a = p^2_a,~~m^2_b = p^2_b,~~W^2= p^2_X,~~{W'}^2 = p^2_{X'},~~t= -p^2 $$

\vvv

\subsection{Processus avec \'echange d'un photon du genre espace}

\begin{figure}[hbt]
\centering
\includegraphics[scale=0.3, width=6cm, height=4cm]{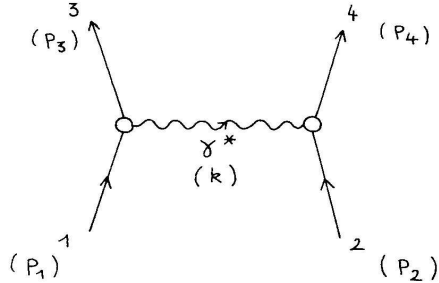}
\vskip 0.25cm

\caption{Diagramme d'un processus avec \'echange d'un photon virtuel} \label{ech-phot}
\end{figure}

\vv \nin Appliquons maintenant le formalisme d'h\'elicit\'e au diagramme de
la figure \ref{ech-phot} qui d\'ecrit la r\'eaction $1+2 \rightarrow 3 + 4$
dans laquelle un photon du genre espace est echang\'e.

\vvv

\vv \nin Dans un premier temps, nous supposerons que les deux lignes sortantes
3 et 4 de l'\'etat final, repr\'esentent, tout comme les lignes entrantes 1 et 2 de l'\'etat initial,
des \'etats \`a une seule particule r\'eelle. Nous poserons

$$ k = p_1 - p_3 = p_4 - p_2,~~k^2 = - t < 0, ~~p^2_1 = m^2,~~p^2_2 = {m'}^2$$
$$p^2_3 = W^2,~~p^2_4 = {W'}^2,~~s=(p_1 + p_2)^2 = (p_3 + p_4)^2 $$

\vv \nin L'\'el\'ement de matrice de transition du diagramme s'\'ecrit

\beq T_{fi} = \di{e^2 \over t}~ J_\mu\, {J'}^\mu \enq

\vv \nin $e$ \'etant la valeur absolue de la charge de l'\'electron
; $J_\mu$ et $J'_\mu$ sont les vecteurs de courant respectifs du
vertex de gauche ($1,3, \gamma^\star)$ et du vertex de droite
$(2,4,\gamma^\star)$ :

\beq J_\mu = <[p_3], \sigma_3| J_\mu(0) |[p_1], \sigma_1>~,~~{J'}_\mu = <[p_4], \sigma_4| J_\mu(0) |[p_2], \sigma_2> \enq

\vv \nin $J_\mu(0)$ \'etant l'op\'erateur courant \'electromagn\'etique au point
d'espace-temps $x=0$. Ici aussi, nous supposerons que les particules initiales ne
sont pas polaris\'ees et qu'on ne cherche pas \`a mesurer les polarisations des
particules finales. Le calcul de la section efficace du processus consid\'er\'e passe par celui de

\beq {\cal T} = \sum~|T_{fi}|^2 = \di{e^4 \over t^2}~I~,~~{\rm avec}~~I =
I_{\mu \nu} ~{I'}^{\mu \nu} \label{defI} \enq

\vv \nin o\`u le symbole $\sum$ signifie que l'on effectue une sommation sur tous les indices de spin de toutes les particules entrantes et sortantes. Les tenseurs

\beq I_{\mu \nu} = \di{\sum_g} ~J_\mu J^\star_\nu~,~~{I'}_{\mu \nu} = \di{\sum_d}~{J'}_\mu {J'}^\star_\nu \label{tenseursI}\enq

\vv \nin relatifs au vertex de gauche et au vertex de droite respectivement sont des sommes sur les indices de spin des particules qui y sont impliqu\'ees. Ces sommes sont \'evidemment ind\'ependantes du choix des t\'etrades choisies pour les particules\footnote{Un changement de t\'etrade laisse invariant un projecteur tel que $\di{\sum_\sigma}~|[p], \sigma><[p], \sigma|$.}. Nous choisirons donc pour celles-ci les t\'etrades d'h\'elicit\'e ad\'equates, lesquelles, comme on le sait d\'ej\`a, apportent une grande simplification.

\vv \nin Compte tenu de l'\'equation de conservation du courant
\'electromagn\'etique, les courants de vertex $J_\mu$ et ${J'}_\mu$
sont orthogonaux \`a l'impulsion $k$ du photon virtuel. Ils peuvent
donc \^etre d\'evelopp\'es suivant une triade quelconque de vecteurs
orthogonaux \`a $k$ et formant avec ce vecteur une base de
l'espace-temps. Pour des raisons \'evidentes de simplicit\'e, nous
choisirons pour ces triades les triades d'h\'elicit\'e appropri\'ees
pour l'un et l'autre vertex. Pour le courant de gauche, on obtient
ainsi

$$ J_\mu = \di{\sum_{\lambda = 0_t, \pm 1} }\eta_\lambda ~e^{(\lambda)}(k)~J_\lambda ,~~{\rm avec}~~\eta_{0_t} = +1, ~\eta_\pm = -1 $$
\beq {\rm et} ~~~J_\lambda = <[\,p_3\,]_h, \lambda_3 | e^{(\lambda)
\star}~J_\mu(0) |[\,p_1\,]_h, \lambda_1> \enq

\vv \nin et de fa\c{c}on similaire,

$$ {J'}_\mu = \di{\sum_{\lambda' = 0_t, \pm 1} }\eta_{\lambda'} ~{e'}^{(\lambda')}(k)~{J'}_{\lambda'} ,~~{\rm avec}~~\eta_{0_t} = +1, ~\eta_\pm = -1 $$
\beq {\rm et} ~~~{J'}_{\lambda'} = <[\,p_4\,]_h, \lambda_4 |
{e'}^{(\lambda') \star}~J_\mu(0) |[\,p_2\,]_h, \lambda_2> \enq

\vv \nin Utilisant la m\^eme technique que celle du paragraphe
pr\'ec\'edent, on trouve facilement la loi de conservation de
l'h\'elicit\'e \`a chacun des vertex :

\beq \lambda = \lambda_1 - \lambda_3~,~~\lambda' = \lambda_2 - \lambda_4 \enq

\vv \nin et que, les amplitudes $J_\lambda$ et ${J'}_{\lambda'}$
sont des invariants relativistes (voir \ref{invarhel},
\ref{invaramp}). Ecrivons ceux-ci sous la forme

$$ J_\lambda = \delta_{\lambda, \lambda_1 -\lambda_3}~J_{\lambda_1 \lambda_3} ~,~~{ J'}_{\lambda'} = \delta_{\lambda', \lambda_2 -\lambda_4}~{J'}_{\lambda_2 \lambda_4}$$

\vv \nin Le tenseur $I_{\mu \nu}$ dans \ref{tenseursI} prend alors la forme

$$ I_{\mu \nu} = \di{\sum_\lambda}~e^{(\lambda)}_\mu (k)\, e^{(\lambda) \star}_\nu(k)\, F_\lambda ~,~~{\rm avec}~~F_\lambda = \di{\sum_{\lambda_1 \lambda_3}}~\delta_{\lambda, \lambda_1 - \lambda_3}\, |J_{\lambda_1 \lambda_3}|^2 $$

\vv \nin La {\it conservation de la parit\'e} dans les interactions \'electromagn\'etiques, combin\'ee avec la conservation de l'h\'elicit\'e, implique la relation (voir Appendice B) $F_{-\lambda} = F_\lambda$. Posons alors

\beq F_{+} = F_{-} = T~,~~F_0 = L \label{parite} \enq

\vv \nin puis utilisons la relation de fermeture des t\'etrades de $k$ :

\beq - \di{\sum_{\lambda = \pm 1}}\, e^{(\lambda)}_\mu(k)\, e^{(\lambda) \star}_\nu(k) = g_{\mu \nu} - \hat{q}_{13 \mu} \hat{q}_{13 \nu} + \di{{k_\mu k_\nu}\over t} \enq

\vv \nin pour obtenir

\beq I_{\mu \nu} = (T+L) \,\hat{q}_{13 \mu} \, \hat{q}_{13 \nu}  -T \left( g_{\mu \nu} + \di{{k_\mu k_\nu}\over t} \right) \label{tensgauche} \enq

\vv \nin Les grandeurs $T$ et $L$ sont les taux de vertex du vertex de gauche. Elles peuvent s'obtenir directement \`a partir de projections du tenseur \ref{tensgauche} :

\beq L = \hat{q}_{13 \mu} \,I^{\mu \nu} \,  \hat{q}_{13 \nu} ~,~~T= e^{(\pm)}_\mu \, I^{\mu \nu} \, e^{(\pm)}_\nu~,~~g_{\mu \nu} I^{\mu \nu} = L - 2 T \enq

\vv \nin Suivant l'interpr\'etation des vecteurs d'h\'elicit\'es de $k$ comme fonctions d'onde du photon virtuel, et compte-tenu des relations pr\'ec\'edentes, ces taux de vertex s'interpr\`etent comme des taux d'interaction\footnote{C'est-\`a-dire, une somme, sur les \'etats de polarisation des particules externes, du carr\'e du module d'une amplitude de transition.} de la r\'eaction virtuelle $\gamma^\star + 1 \rightarrow 3$ avec un photon polaris\'e soit longitudinalement (pour $L$) soit transversalement (pour $T$).

\vv \nin Consid\'erons ensuite le cas o\`u les deux lignes sortantes 3 et 4 figurent des \'etats quantiques comportant un nombre quelconque de particules. Le diagramme \ref{ech-phot} repr\'esente alors, d'une fa\c{c}on g\'en\'erale, un processus \`a \'echange d'un photon du genre espace, avec deux vertex {\it in\'elastiques}.

\vv \nin Ecrivons la section efficace diff\'erentielle qui s'y rapporte sous la forme

\beq d \sigma = \di{1\over{\Lambda^{1/2}(s, m^2, m'^2)}} ~\di{e^4\over t^2}~\di{I\over{n n'}}~ (2 \pi)^4~
\delta^{(4)} (k + p_3 - p_1)~d^4 k~\delta^{(4)} (k+p_2 - p_4) ~d \rho d\rho' \enq

\vv \nin \`a partir de laquelle on obtient la section efficace diff\'erentielle usuelle en effectuant l'int\'egration sur $k$. Dans cette formule, $n$ et $n'$ sont les nombres d'\'etats de spin respectifs des particules entrantes 1 et 2 ; $I$ est d\'efini par \ref{defI} ; $d \rho$ et $d\rho'$ d\'esignent les \'el\'ements d'espace des phases de 3 et 4 respectivement :

$$ d \rho = \di{\prod^N_{i=1}} \,\di{{d^3 q_i}\over{(2 \pi)^3 q^0_i}}$$

\vv \nin En int\'egrant sur l'ensemble des espaces de phases finals, on obtient

$$ d \sigma =  \di{1\over{\Lambda^{1/2}(s, m^2, m'^2)}} ~\di{e^4\over t^2}~\di{{d^4 k}\over{(2 pi)^4}}\, {\cal I} $$

\vv \nin o\`u

$$ {\cal I} = {\cal I}^{\mu \nu}\, {\cal I'}_{\mu \nu} ~~~{\rm avec} $$
\beq {\cal I}_{\mu \nu} = \di{1 \over n} \, \di{\sum} \, \di{\int} \, J_\mu J^\star_\nu \, d{\rm Lips} ~,~~
{\cal I'}_{\mu \nu} = \di{1 \over n'} \, \di{\sum} \, \di{\int} \,{ J'}_\mu {J'}^\star_\nu \, d {\rm Lips'} \label{cal-I} \enq

\vv \nin dans ces expressions, le signe $\sum$ repr\'esente une sommation sur tous les \'etats de spin des particules initiales et finales du vertex consid\'er\'e ; $d$Lips sont les \'el\'ements d'espace de phase incluant les facteurs de conservation de l'\'energie-impulsion\footnote{Lips signifiant {\it Lorentz Invariant Phase Space}.}. Compte tenu de la relation

\beq d^4 k = \di{{d W^2 \,d t\, d {W'}^2\, d \psi}\over{4 \Lambda^{1/2}(s, m^2, m'^2)}} \enq

\vv \nin o\`u $\psi$ est l'angle azimutal du photon \'echang\'e. En int\'egrant sur cet angle, il vient

\beq \di{{d \sigma}\over{dW^2\, dt\,d{W'}^2}} = \di{1\over{{64 \pi^3}}}\,\di{1\over{\Lambda^{1/2}(s, m^2,m'^2)}}\,\di{{e^4 {\cal I}}\over t^2} \label{sec1} \enq

\vv \nin Il est facile de montrer que, l'\'etat final \`a un vertex \'etant indiff\'eremment \`a une ou plusieurs particules, les tenseurs \ref{cal-I} ont une structure similaire \`a celle de \ref{tensgauche}. On a en effet

$$ {\cal I}_{\mu \nu} = \di{1\over n}\, \di{\sum_{\lambda_1}}\, <[p_1]_h, \lambda_1 | J_\mu(0)\,{\cal P}\,J_\nu(0) |[p_1]_h, \lambda_1> $$

\vv \nin o\`u

$${\cal P} = \di{\int}\,\di{\sum_{\sigma_1, \sigma_2,\cdots, \sigma_N}}\, d {\rm Lips} \, |[q_1], \sigma_1 ; [q_2], \sigma_2 ; \cdots ; [q_N], \sigma_N><[q_1], \sigma_1 ; [q_2], \sigma_2 ; \cdots ; [q_N], \sigma_N| $$

\vv \nin est le projecteur sur tous les \'etats possibles du syst\`eme 3. Comme il a  d\'ej\`a \'et\'e sugg\'er\'e au paragraphe pr\'ec\'edent, ce projecteur peut \^etre d\'ecompos\'e en une somme de projecteurs dont chacun correspond \`a une repr\'esentation irr\'eductible du groupe de Poincar\'e, ce qui conduit \`a une d\'ecomposition du tenseur \ref{cal-I} en ondes partielles auxquelles on peut appliquer \`a chacune le traitement valable lorsque le syst\`eme 3 est r\'eduit \`a une seule particule.
Aussi, peut-on \'ecrire

\begin{eqnarray} & n\, {\cal I}_{\mu \nu} =   (T+L) \,\hat{q}_{13 \mu} \, \hat{q}_{13 \nu}  -T \left( g_{\mu \nu} + \di{{k_\mu k_\nu}\over t} \right) & \nonumber \\
& n'\, {\cal I'}_{\mu \nu} =   (T'+L') \,\hat{q}_{24 \mu} \, \hat{q}_{24 \nu}  -T' \left( g_{\mu \nu} + \di{{k_\mu k_\nu}\over t} \right) &\end{eqnarray}

\vv \nin Effectuant le produit scalaire de ces deux tenseurs, on obtient

$$ n\, n'\,{\cal I} = (1 + \cosh^2 \chi) \,T\,T' + \sinh^2 \chi \,(T L' + L T') + \cosh^2 \chi\, L L' $$
\beq {\rm avec}~~~\cosh \chi = \hat{q}_{13} \cdot \hat{q}_{24} \label{facto} \enq

\vv \nin $\chi$ \'etant le param\`etre de la transformation de Lorentz amenant le plan du vertex de gauche $(p_1, p_3,k)$ sur le plan du vertex de droite $(p_2, p_4,k)$, et dont l'expression se d\'eduit de \ref{cosh}.

\vv \nin Envisageons maintenant la r\'eaction de photoproduction virtuelle $1+ \gamma^\star \rightarrow 3$, avec un photon polaris\'e soit transversalement, soit longitudinalement. L'amplitude de transition correspondante s'\'ecrit

$$ - e\, e^{(\lambda) \star}_\mu(k) \, J^\mu $$

\vv \nin o\`u $J_\mu$ est le courant du vertex de gauche et $\lambda = 0_t, \pm 1$. Les matrices de polarisation des particules des syst\`emes 1 et 3 \'etant suppos\'ees diagonales, la section efficace totale correspondant \`a ce processus est

\beq \sigma_\lambda = \di{e^2\over{2(W^2 - m^2)}}\,e^{(\lambda) \star}_\mu(k)\, e^{(\lambda) }_\nu (k) \, {\cal I}^{\mu \nu} \enq

\vv \nin o\`u a \'et\'e introduit un`` facteur de flux" $1/(W^2 - m^2)$ conform\'ement \`a la d\'efinition des sections efficaces virtuelles donn\'ee par L.N. Hand\footnote{Phys. Rev. \und{129}, 1834 (1963).}. Les sections efficaces ``transversale" et ``longitudinale" seront donc d\'efinies par

\beq \sigma_T = \sigma_{+} = \sigma_{-} = \di{e^2\over{2 n (W^2-m^2)}}\, T ~,~~\sigma_L= \sigma_0 =  \di{e^2\over{2 n (W^2-m^2)}}\, L\enq

\vv \nin Les sections efficaces relatives au vertex de droite sont d\'efinies de fa\c{c}on similaire. A l'aide de ces grandeurs, \ref{sec1} prend la forme

$$\di{{d \sigma}\over{dW^2\, dt\,d{W'}^2}} = \di{{(W^2 - m^2)\,(W'^2 - m'^2)}\over{{16 \pi^3 \Lambda^{1/2}(s, m^2, m'^2)}}}\,\di{1\over t^2}\,\times $$
\beq ~~ \label{factoriz} \enq
$$ \left\{ (1 + \cosh^2 \chi) \,\sigma_T\,\sigma'_T + \sinh^2 \chi \,(\sigma_T \sigma'_L +
\sigma_L \sigma'_T) + \cosh^2 \chi\, \sigma_L \sigma'_L  \right\}  $$

\vv \nin Cette formule de factorisation de la section efficace est d'une tr\`es grande transparence physique. Elle est particuli\`erement bien adapt\'ee pour effectuer de fa\c{c}on invariante des approximations de type ``Williams-Weizs\"acker"\footnote{Voir notamment : C. Carimalo, G. Cochard, P. Kessler, J. Parisi, B. Roehner, Phys. Rev. D 10, 1561 (1974) ; C. Carimalo, P. Kessler, J. Parisi, Phys. Rev. D 14, 1819 (1976).}.

\vv \nin Bien entendu, une d\'emarche analogue \`a celle du paragraphe pr\'ec\'edent aurait pu \^etre
utilis\'ee ici pour aboutir \`a \ref{factoriz}, en d\'ecomposant le num\'erateur du propagateur du photon suivant les fonctions d'onde d'h\'elicit\'e de celui-ci :

$$ g_{\mu \nu} = \di{\sum_{m,n}} \, e^{(m)}_\mu(k)~ r^{(1) \star}_{m m'}~ {e'}^{(m') \star}_\nu(k) ~\eta_m \, \eta_{m'} $$

\vv \nin Puisque les courants de vertex sont orthogonaux \`a $k$ il suffit de ne consid\'erer dans cette somme que les termes correspondant aux valeurs $0_t, \pm1$ des h\'elicit\'es ; ${ r^{(1)}_{m m'}}$ est la matrice 3$\times$3 repr\'esentant, pour un spin de valeur 1, la transformation de Lorentz $\hat{q}_{13} \rightarrow \hat{q}_{24}$ :

$$ r^{(1)}_{m m'} = e^{(m)}(k) \cdot {e'}^{(m') \star}(k) $$

\vv \nin Le produit scalaire des deux courants s'\'ecrit donc

$$ J^\mu \, g_{\mu \nu}\, J'^\nu = \di{\sum_{m, m'}}\, \eta_m \,\eta_{m'}\, J_m\, r^{(1) \star}_{m m'}\, J'_{m'} $$

\vv \nin et le taux d'interaction est

$$ I= \di{\sum} \eta_m \, \eta_{m'}\, \eta_n\, \eta_{n'} \, J_m \, J^\star_n \, r^{(1) \star}_{m m'}\, r^{(1)}_{n n'}\, J'^\star_{m'}\, J'_n $$

\vv \nin o\`u la sommation est effectu\'ee sur tous les \'etats de polarisation des particules externes et sur celles du photon. La loi de conservation de l'h\'elicit\'e \`a chaque vertex implique les contraintes $m =n$, $m' = n'$ et conduit \`a l'expression simplifi\'ee de $I$ :

$$ I = \di{\sum_{m, m'}}\, F_m\, F_{m'} |d^{(1)}_{m m'}(\theta)|^2 $$

\vv \nin o\`u les $d^{(1)}_{m m'}(\theta)$ sont les \'el\'ements de la rotation d'angle imaginaire $ \theta = i \chi$ autour du vecteur ``charni\`ere" $k$. Explicitement, avec $m = +1, 0, -1$ et $m' = +1, 0, -1$ en indices de ligne et de colonne respectivement et dans cet ordre des valeurs,

$$ d^{(1)}_{m m'} (\theta)= \left(\begin{array}{ccc} \di{{1 + \cos \theta}\over 2} & - \di{{\sin \theta}\over{\sqrt{2}}} & \di{{1 - \cos \theta}\over 2} \\
 \di{{\sin \theta}\over{\sqrt{2}}} & \cos \theta & - \di{{\sin \theta}\over{\sqrt{2}}} \\
\di{{1 - \cos \theta}\over 2} & \di{{\sin \theta}\over{\sqrt{2}}}  & \di{{1 + \cos \theta}\over 2} \end{array} \right) $$

\vv \nin En effectuant les multiplications matricielles et en tenant compte de la conservation de l'h\'elicit\'e et de la parit\'e \`a chaque vertex, on aboutit \`a la formule de factorisation \ref{facto},
en prenant garde \`a la transformation des fonctions trigonom\'etriques $\cos$ et $\sin$ en fonctions hyperboliques $\cosh$ et $\sinh$.

\begin{figure}[hbt]
\centering
\includegraphics[scale=0.3, width=8cm, height=3cm]{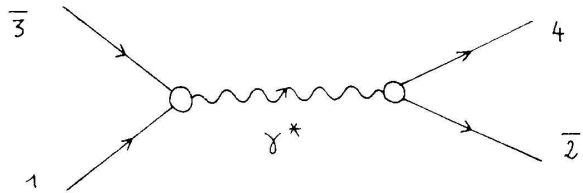}
\vskip 0.25cm

\caption{Processus avec \'echange d'un photon du genre temps} \label{depsg}
\end{figure}

\vv \nin A ce point, il est int\'eressant de noter que, d'apr\`es des principes g\'en\'eraux, le taux d'interaction $I(s,-t)$ (\ref{facto}) est a priori prolongeable analytiquement suivant les deux variables $s$ et $-t$\footnote{Concernant les vertex impliquant des particules hadroniques, c'est sous r\'eserve qu'on puisse r\'ealiser un tel prolongement des facteurs de forme \'eventuellement pr\'esents.}.  Ainsi, en \'echangeant $s$ et $-t$ dans \ref{facto} on obtient le taux d'interaction $I'(s ,-t) = I(-t,s)$ du diagramme de la figure \ref{depsg} d\'ecrivant la ``voie s" de la r\'eaction $1 + \ov{3} \rightarrow \ov{2}+ 4$, $\ov{3}$ et $\ov{2}$ \'etant les anti-particules de 3 et 2 respectivement, tandis que celui de la figure \ref{ech-phot} en repr\'esente la ``voie t". Ainsi,

\vvv

$$ I' = (1+ \cos^2\theta) \,T\, T'+ \sin^2 \theta (T \,L' + L \,T') + \cos^2 \theta\, L\,L' $$

\vv \nin o\`u $\theta$, qui est maintenant un angle r\'eel, est simplement l'angle de diffusion des particules sortantes $\ov{2}$ et $4$ dans le r\'ef\'erentiel du centre de masse de la voie s. On a

$$ \cos [\theta(s, -t) ] = - \hat{q}_{1 \ov{3}} \cdot \hat{q}_{\ov{2} 4} = - \cosh[\theta(-t,s)] $$

\vv \nin $\cosh [\theta(-t,s)] $ \'etant d\'efini par le second membre de \ref{cosh}.

\vv \nin La m\'ethode d'h\'elicit\'e directement appliqu\'ee au diagramme de la figure \ref{depsg} confirme la validit\'e du prolongement analytique. Ce r\'esultat sugg\`ere d'uniformiser le traitement des diagrammes \`a \'echange d'une particule, que celle-ci soit du genre espace ou du genre temps. C'est pr\'ecis\'ement ce point de vue qui fut adopt\'e par P. Kessler (loc. cit.) qui d\'eveloppa \`a cet effet une trigonom\'etrie unifi\'ee de l'espace-temps, utilisant abondamment le nombre $\hat{1}$.

\subsection{Processus avec \'echange de deux photons virtuels du genre espace : exemple de double factorisation}

\begin{figure}[hbt]
\centering
\includegraphics[scale=0.3, width=7cm, height=3cm]{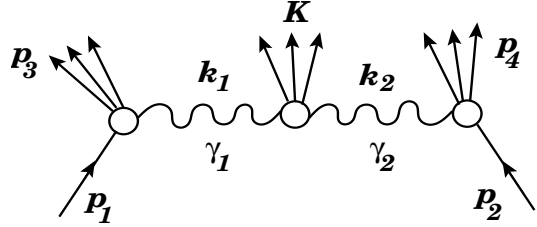}
\vskip 0.25cm

\caption{Processus avec \'echange de deux photons virtuels du genre espace} \label{gam-gam}
\end{figure}

\vv \nin Envisageons ensuite le processus d\'ecrit par le diagramme de la figure \ref{gam-gam}.  Les \'etats initiaux 1 et 2 d'impulsions respectives $p_1$ et $p_2$ du genre temps sont deux particules entrant en collision. Celle-ci s'effectue par un \'echange de deux photons virtuels $\gamma_1$ et $\gamma_2$, d'impulsions respectives $k_1$ et $k_2$, du genre espace. Les \'emissions de ces photons, le premier au vertex de gauche, le second au vertex de droite, sont accompagn\'ees par la production simultan\'ee des \'etats 
finals 3 et 4, d'impulsions du genre temps $p_3$ et $p_4$ respectivement, ces \'etats pouvant \^etre \`a une seule ou plusieurs particules. Les deux photons virtuels entrent ensuite en collision (virtuelle) donnant lieu \`a la production d'un \'etat central $K$ d'impulsion $K$ du genre temps.

\vvv

\vv \nin Les notations sont les suivantes. 

$$ k_1= p_1 - p_3~,~~k^2_1 = - t_1 < 0~;~~k_2 = p_2 - p_4~,~~k^2_2 = - t_2 <0 ~$$
$$p^2_1 = m^2_1~,~~p^2_3 = W^2_3~,~~p^2_2 = m^2_2~,~~p^2_4 = W^2_4~$$
$$P = p_1 + p_2 = p_3 + p_4 + K~,~~s = (p_1 + p_2)^2~,~~K^2 = M^2 $$

\vv \nin D\'efinissons ensuite les diverses t\'etrades d'h\'elicit\'e qui vont intervenir. 
\vv

\vv \nin \leftpointright {\bf  Vertex central}

\vv \nin \ding{192} {\bf  Base attach\'ee \`a $K$}

$$ T = \di{K\over M}~,~~Z = \epsilon^{(0)} (K) = \di{{2 M} \over{\Lambda^{1/2}_c}} \left( k_1 - \di{{K\cdot k_1}\over M^2} K \right) $$ 
$$ {\rm avec}~~\Lambda_c = \Lambda(M^2, -t_1, -t_2) = M^4 + 2 M^2 (t_1 + t_2) + (t_1 - t_2)^2$$

\vv \nin Les impulsions des photons se d\'ecompose relativement \`a ces vecteurs comme : 

$$ k_1 = \omega_1 \,T + k\, Z~,~~k_2 = \omega_2\, T - k\, Z~~~{\rm avec}$$ 
$$ \omega_1 = \di{{M^2 - t_1 +t_2}\over{2 M}}~,~~\omega_2 = \di{{M^2 - t_2 +t_1}\over{2 M}}~,~~k = \di{\Lambda^{1/2}_c \over{2 M}} $$

\vv \nin Le vecteur $Y$ de la base est ensuite choisi comme \'etant :  

$$ Y_\mu = Y_\mu (K) = N\, \varepsilon_{\mu \nu \rho \sigma}\, K^\nu k^\rho_1 q^\sigma $$

\vv \nin o\`u $q$ est l'impulsion de l'une quelconque des particules produite dans l'\'etat final du vertex central ; $N$ est le facteur de normalisation tel que

$$ N^{-1} = M\, k\, \ro{q} \,\sin \theta$$ 

\vv \nin o\`u $\theta$ est, dans le r\'ef\'erentiel du centre de masse de $K$ avec $\Vec{k_1}$ selon l'axe 
des $z$, l'angle d'\'emission de ladite particule par rapport \`a ce dernier axe, et $\ro{q}$ le module de sa tri-impulsion dans ce r\'ef\'erentiel. 

\vv \nin Finalement, le vecteur $X$ de la base se d\'eduit par la formule usuelle $X_\mu = \varepsilon_{\mu \nu \rho \sigma} T^\nu\, Y^\rho\, Z^\sigma$et l'on d\'efinit les vecteurs de 
polarisations circulaires 

$$ \epsilon^{(\pm)} = \epsilon^{(\pm)} (K) = \mp \, \di{1\over{\sqrt{2}}} \left( X \pm i \, Y \right) $$

\vv \nin \ding{193} {\bf Base attach\'ee \`a $k_1$}

$$ T_1 = \epsilon^{(0)}_1 = \epsilon^{(0)} (k_1)  = \di{{2 \sqrt{t_1}}\over{\Lambda^{1/2}_c}} \left( K + \di{{k_1 \cdot K}\over t_1}\, k_1 \right) ~,~~Z_1 = \di{k_1 \over \sqrt{t_1}} $$ 
\beq X_1 = X~,~~Y_1 = Y~,~~{\rm soit}~~\epsilon^{(\pm)}_1 = \epsilon^{(\pm)}  \label{bck1} \enq

\vv \nin \ding{194} {\bf  Base attach\'ee \`a $k_2$}

\beq T_2 = \epsilon^{(0)}_2 = \epsilon^{(0)} (k_2)  = \di{{2 \sqrt{t_2}}\over{\Lambda^{1/2}_c}} \left( K + \di{{k_2 \cdot K}\over t_2}\, k_2 \right) ~,~~Z_2 = \di{k_2 \over \sqrt{t_2}}~,~~\epsilon^{(\pm)}_2 = \epsilon^{(\mp)} \label{bck2} \enq

\vv
\nin \leftpointright {\bf  Vertex de gauche}

\vv \nin On effectue un couplage d'h\'elicit\'e entre $p_1$, $p_3$ et $k_1$, d'o\`u les vecteurs de base suivants. 

$$ {\epsilon '}^{(0)}_1 = {\epsilon '}^{(0)}_1 (k_1) = T'_1 = \di{{2 \sqrt{t_1}} \over{\Lambda^{1/2}_g}} \left( p_1 + \di{{k_1 \cdot p_1}\over t_1} \, k_1 \right) ~,~~Z'_1 = Z_1~$$
$$ {\rm avec}~~~\Lambda_g = \Lambda(W^2_3, m^2_1, -t_1) $$
\beq Y'_{1 \mu} = N_g \, \varepsilon_{\mu \nu \rho \sigma} \, p^\nu_1\, k^\rho_2\, k^\sigma_1~,~~~~
X'_1 =  \varepsilon_{\mu \nu \rho \sigma} \,{T'_1}^\nu\, {Y'_1}^\rho\, {Z'_1}^\sigma \label{bgk1} \enq

\begin{figure}[hbt]
\centering
\includegraphics[scale=0.3, width=8cm, height=6cm]{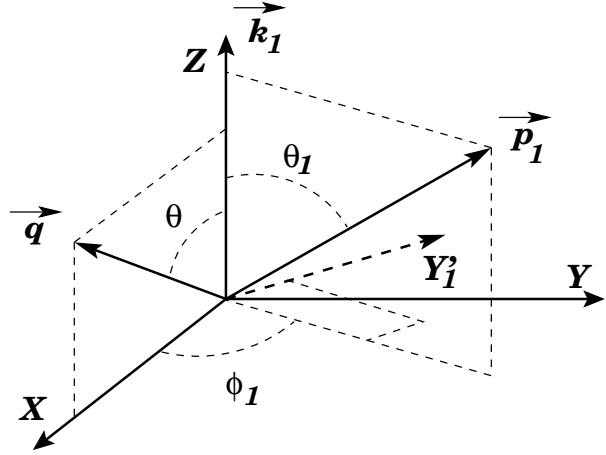}
\vskip 0.25cm

\caption{Disposition des tri-impulsions dans le r\'ef\'erentiel du centre de masse central} \label{cmfc}
\end{figure}

\vv \nin La disposition des tri-impulsions $\Vec{~q~}$, $\Vec{k_1}$ et $\Vec{p_1}$ dans le r\'ef\'erentiel du centre de masse du vertex central est indiqu\'ee \`a la figure \ref{cmfc}. Le facteur de normalisation $N_g$ est tel que 

$$ N^{-1}_g = M\, k\, \ro{q}\, \ro{p}_1\sin \theta_1 $$ 

\nin o\`u, dans ledit r\'ef\'erentiel, $\theta_1$ est l'angle entre $\Vec{p_1}$ et $\Vec{k_1}$, $\ro{p}_1 = |\Vec{p_1}|$ et $\varphi_1$ est l'angle azimutal relatif entre $\Vec{p_1}$ et $\Vec{~q~}$~\footnote{Comme $\Vec{p_3} = \Vec{p_1} - \Vec{k_1}$, on a aussi bien $\ro{p}_3 \sin \theta_3 = \ro{p}_1 \sin \theta$, $\varphi_3 = \varphi_1$.}.  On note que $Y'_1$ s'exprime simplement comme 

\beq Y'_1 = - \sin \varphi_1 \, X + \cos \varphi_1 \, Y \label{yprime1} \enq

\vv \nin Exprimons $T'_1$ dans la base $T_1, X_1, Y_1, Z_1$. Nous poserons 

$$ \cosh \alpha_1 = T'_1 \cdot T_1 $$  

\nin Comme $(T'_1)^2 = 1$ et $T'_1 \cdot Z_1 = 0$, il vient 

$$ (T'_1 \cdot X_1)^2 + (T'_1 \cdot Y_1)^2 = (T'_1 \cdot T_1)^2 - 1 = \sinh^2 \alpha_1 $$ 

\vv \nin Or, $X_1 = X$, $Y_1 = Y$ et $ (X_1~{\rm ou}~Y_1) \cdot T'_1 = \di{{2 \sqrt{t_1}}\over \Lambda^{1/2}_g} \,( X~{\rm ou}~ Y)\cdot p_1 $. D'apr\`es la figure \ref{cmfc}, ceci donne 
\vv
$$ (X~{\rm ou}~Y) \cdot T'_1 = - \di{{2 \sqrt{t_1}}\over \Lambda^{1/2}_g} \,\ro{p}_1 \,\sin \theta_1\, (\cos \varphi_1~{\rm ou}~\sin \varphi_1) $$

\vv \nin Par suite, comme $ \sinh \alpha_1 = \sqrt{(T'_1 \cdot X_1)^2 + (T'_1 \cdot Y_1)^2} $, on obtient 

\beq \sinh \alpha_1 =  \di{{2 \sqrt{t_1}}\over \Lambda^{1/2}_g} \,\ro{p}_1 \,\sin \theta_1 \label{sha1} \enq

\nin et finalement, 

\beq T'_1 = \cosh \alpha_1 \, T_1 + \sinh \alpha_1 \, \left( \cos \varphi_1 \, X + \sin \varphi_1 \, Y \right) \label{tprime1} \enq

\vv \nin Utilisant \ref{bgk1}, \ref{yprime1} et \ref{tprime1}, on en d\'eduit ais\'ement 

\beq X'_1 = \sinh \alpha_1 \, T_1 + \cosh \alpha_1 \, \left( \cos \varphi_1 \, X + \sin \varphi_1\, Y \right) \label{xprime1} \enq 

\vv \nin Ces r\'esultats peuvent \^etre r\'eexprim\'es sous la forme suivante 

$$ T'_1 = \cosh \alpha_1 \, T_1 + \epsilon^{(-)} \, e^{i \varphi_1}\, \di{{\sinh \alpha_1}\over \sqrt{2}} - 
\epsilon^{(+)} \, e^{-i \varphi_1}\, \di{{\sinh \alpha_1}\over \sqrt{2}} $$
\beq~~  \label{basg-basc-k1}  \enq  
$$ {\epsilon'_1}^{(\pm)} = \mp \di{{\sinh \alpha_1}\over \sqrt{2}} \,T_1 +\epsilon^{(-)} \, e^{i \varphi_1}\,\di{1\over 2} \left(1 \mp \cosh \alpha_1 \right) +\epsilon^{(+)} \, e^{-i \varphi_1}\,\di{1\over 2} \left(1 \pm \cosh \alpha_1 \right)$$

\newpage
\nin  \leftpointright {\bf  Vertex de droite}

\vv \nin Proc\'edant de fa\c{c}on similaire pour le vertex de droite tout en choisissant $Y'_2$ comme 

\beq Y'_2 =  - \left(- \sin \varphi_2 \, X + \cos \varphi_2 \,Y  \right) \label{yprime2} \enq 

\nin et donc 

\beq X'_2 = \sinh \alpha_2 \, T_2 + \cosh \alpha_2\, \left( \cos \varphi_2 \, X + \sin \varphi_2 \, Y \right) \label{xprime2} \enq 

\nin on obtient

$$ T'_2 = \cosh \alpha_2 \, T_2 + \epsilon^{(-)} \, e^{i \varphi_2}\, \di{{\sinh \alpha_2}\over \sqrt{2}} - 
\epsilon^{(+)} \, e^{-i \varphi_2}\, \di{{\sinh \alpha_2}\over \sqrt{2}} $$
\beq~~  \label{basg-basc-k2}  \enq  
$$ {\epsilon'_2}^{(\pm)} = \mp \di{{\sinh \alpha_2}\over \sqrt{2}} \,T_2 -\epsilon^{(-)} \, e^{i \varphi_2}\,\di{1\over 2} \left(1 \pm \cosh \alpha_2 \right) -\epsilon^{(+)} \, e^{-i \varphi_2}\,\di{1\over 2} \left(1 \mp \cosh \alpha_2 \right)$$

\vv \nin Dans ces formules, $\theta_2$, $\varphi_2$ et $\ro{p}_2$ sont, respectivement, l'angle orbital,  l'angle azimutal et le module    
de $\Vec{p_2}$ dans le r\'ef\'erentiel du centre de masse central (avec $\Vec{k_1}$ d\'efinissant l'axe des $z$, $\Vec{~q~}$ dans le plan $(x,z)$) et 

\beq \cosh \alpha_2 = T'_2 \cdot T_2~,~~\sinh \alpha_2  =  \di{{2 \sqrt{t_2}}\over \Lambda^{1/2}_d} \,\ro{p}_2 \,\sin \theta_2 \label{cha2-sha2} \enq

\vv \nin \smallpencil L'amplitude de transition de la r\'eaction consid\'er\'ee est de la forme 

\beq T_{fi} = \di{e^4\over {t_1 t_2}}  J^\mu_g\, T^{\mu \nu}\, J^\nu_d \label{ampgamgam} \enq  

\vv \nin o\`u $J^\mu_d$ et $J^\nu_d$ sont les courants \'electromagn\'etiques caract\'eristiques des transitions $ 1 \rightarrow 3 + \gamma_1$ et $2 \rightarrow 4 + \gamma_2$ respectivement et $T^{\mu \nu}$ l'amplitude de la r\'eaction $\gamma_1 + \gamma_2 \rightarrow K$. En supposant les particules externes non polaris\'ees,  le taux d'interaction correspondant est proportionnel \`a 

$$ {\cal I} = {\cal L}_{\mu \rho} \, F^{\mu \rho ; \nu \sigma}\, {\cal R}_{\nu \sigma} $$ 
\beq{\rm avec}~~~~{\cal L}^{\mu \rho}  =  \di{\sum_g}\, J^\mu_g \, J^{\rho \star}_g ~,~~F^{\mu \rho ; \nu \sigma} = \di{\sum_c}\, T^{\mu \nu}\, T^{\rho \sigma \star} ~,~~{\cal R}^{\nu \sigma} = \di{\sum_d}\, J^\nu_d \, J^{\sigma \star}_d \label{lestenseurs} \enq
 
\vv \nin o\`u les diverses sommations portent sur tous les \'etats de spin des particules externes respectivement impliqu\'ees. Dans ${\cal I}$, incorporons maintenant les relations de fermeture 

$$ \di{\sum_m} ~(-1)^m \, \epsilon^{(m)}_{1 \mu} \,\epsilon^{(m) \star}_{1 \rho} = g_{\mu \rho} +\di{{k_{1 \mu} k_{1 \rho}}\over t_1} ~,~~~  \di{\sum_n} ~(-1)^n \, \epsilon^{(n)}_{2 \nu} \,\epsilon^{(n) \star}_{2 \sigma} = g_{\nu \sigma} +\di{{k_{2 \nu} k_{2 \sigma}}\over t_2}$$

\nin dans lesquelles $m$ et $n$, qui prennent les valeurs $0_t, + 1, -1$, sont les indices d'h\'elicit\'e respectifs de $\gamma_1$ et $\gamma_2$,  relatifs \`a leurs couplage d'h\'elicit\'e {\it au vertex central}. Tenant compte du fait que, gr\^ace \`a la conservation des courants \'electromagn\'etiques, les tenseurs de \ref{lestenseurs} sont orthogonaux, suivant leurs indices, soit \`a  $k_1$ (vertex de gauche), soit \`a $k_2$ (vertex de droite), soit \`a $k_1$ et $k_2$ (vertex central), il vient 

$$ {\cal I} = \di{\sum_{m, \ov{m}, n, \ov{n}}} \, {\cal L}_{m \ov{m}}\, F_{m, \ov{m} \,;\, n, \ov{n}} \, {\cal R}_{n \ov{n}} ~,~~~{\rm avec}~~~~ F_{m, \ov{m} \,;\, n, \ov{n}} = \epsilon^{(m)}_{1 \mu} \,\epsilon^{(\ov{m}) \star}_{1 \rho} \,
F^{\mu \rho ; \nu \sigma}\, \epsilon^{(n)}_{2 \nu} \,\epsilon^{(\ov{n}) \star}_{2 \sigma} $$
\beq  {\rm et}~~~{\cal L}_{m \ov{m}} = (-1)^{m + \ov{m}} \, \epsilon^{(m) \star}_{1 \mu} \,\epsilon^{(\ov{m})}_{1 \rho} \,{\cal L}^{\mu \rho} ~,~~~{\cal R}_{n \ov{n}} = (-1)^{n + \ov{n}} \, \epsilon^{(n) \star}_{2 \nu} \,\epsilon^{(\ov{n})}_{2 \sigma} \,{\cal R}^{\nu \sigma} \label{formule1} \enq

\vv \nin Supposons maintenant que chacun des \'etats finals 3 et 4 ou bien soit un \'etat \`a une seule particule, ou bien corresponde \`a une onde partielle d'un \'etat plus complexe, 
selon le proc\'ed\'e indiqu\'e au paragraphe pr\'ec\'edent. Les tenseurs des vertex de gauche (${\cal L}$) et de droite (${\cal R}$) sont alors de la forme  

$$ {\cal L}_{\mu \rho} = ({\cal T}_1 + {\cal L}_1) \, T'_{1 \mu}\, T'_{1 \rho} - {\cal T}_1 \left( g_{\mu \rho} + \di{{k_{1 \mu} k_{1 \rho}}\over t_1} \right),~ {\cal R}_{\nu \sigma} = ({\cal T}_2 + {\cal L}_2) \, T'_{2 \nu}\, T'_{2 \sigma} - {\cal T}_2 \left( g_{\nu \sigma} + \di{{k_{2 \nu} k_{2 \sigma}}\over t_2} \right)$$

\vv \nin On obtient ainsi 

$$ {\cal L}_{ - -} = {\cal L}_{+ +} = \di{1\over 2} \left\{ {\cal T} _1(1+ \cosh^2\alpha_1) + {\cal L}_1 \sinh^2\alpha_1 \right\} $$
$$ {\cal L}_{0 +} = - {\cal L}_{- 0} = - \di{1\over \sqrt{2}} \left( {\cal T}_1 + {\cal L}_1 \right) \cosh \alpha_1 \, \sinh \alpha_1\, e^{i \varphi_1} $$
$$ {\cal L}_{0 -} = - {\cal L}_{+ 0} = \di{1\over \sqrt{2}} \left( {\cal T}_1 + {\cal L}_1 \right) \cosh \alpha_1 \, \sinh \alpha_1\, e^{- i \varphi_1}  $$
$$ {\cal L}_{+ -} = - \di{1\over 2} \left( {\cal T}_1 + {\cal L}_1 \right) \,\sinh^2 \alpha_1 \, e^{-2 i \varphi_1} ~,~~{\cal L}_{- +} = - \di{1\over 2} \left( {\cal T}_1 + {\cal L}_1 \right) \,\sinh^2 \alpha_1 \, e^{2 i \varphi_1} $$
$$ {\cal L}_{0 0} = {\cal T}_1\, \sinh^2 \alpha_1 + {\cal L}_1 \, \cosh^2 \alpha_1 $$

\vv \nin et 

$$ {\cal R}_{ - -} = {\cal R}_{+ +} = \di{1\over 2} \left\{ {\cal T} _2 (1+ \cosh^2\alpha_2) + {\cal L}_2 \sinh^2\alpha_2 \right\} $$
$$ {\cal R}_{0 +} = - {\cal R}_{- 0} =  \di{1\over \sqrt{2}} \left( {\cal T}_2 + {\cal L}_2 \right) \cosh \alpha_2 \, \sinh \alpha_2\, e^{- i \varphi_2} $$
$$ {\cal R}_{0 -} = - {\cal R}_{+ 0} = - \di{1\over \sqrt{2}} \left( {\cal T}_2 + {\cal L}_2 \right) \cosh \alpha_2 \, \sinh \alpha_2\, e^{ i \varphi_2}  $$
$$ {\cal R}_{+ -} =  -\di{1\over 2} \left( {\cal T}_2 + {\cal L}_2 \right) \,\sinh^2 \alpha_2 \, e^{2 i \varphi_2} ~,~~{\cal R}_{- +} = - \di{1\over 2} \left( {\cal T}_2 + {\cal L}_2 \right) \,\sinh^2 \alpha_2 \, e^{- 2 i \varphi_2} $$
$$ {\cal R}_{0 0} = {\cal T}_2\, \sinh^2 \alpha_2 + {\cal L}_2 \, \cosh^2 \alpha_2 $$

\vvv \nin La somme exprimant ${\cal I}$  dans \ref{formule1} contient a priori $3^4 = 81$ termes. Cependant, tous ne sont pas ind\'ependants. En effet, la conservation de la parit\'e combin\'ee avec celle de l'h\'elicit\'e conduit aux relations 

\beq F_{ -m -\ov{m}\, ; \,-n -\ov{n}} = (-1)^{m+\ov{m}+n + \ov{n}}\, F_{m \ov{m}\, ; \,n \ov{n}} 
\enq 

\vv \nin En outre, on a de fa\c{c}on \'evidente la propri\'et\'e d'hermiticit\'e 

\beq  F_{m \ov{m}\, ; \,n \ov{n}}  = \left\{ F_{\ov{m} m \, ; \,\ov{n} n} \right\}^\star \enq 

\vv \nin On peut alors r\'eorganiser ladite somme pour aboutir \`a une somme de 13 termes dont chacun pr\'esente une d\'ependance particuli\`ere vis-\`a-vis des angles azimutaux $\varphi_1$ et $\varphi_2$. Avant d'\'ecrire la formule correspondante, introduisons les notations suivantes. 

$$ \ell_{m \ov{m}} = |{\cal L}_{m \ov{m}}|~,~~\ell_0 = \di{{\ell_{00}}\over{2 \,\ell_{++}}}~,~~\ell_1 = \di{{\ell_{0+}}\over{\ell_{++}}}~,~~\ell_2 = \di{{\ell_{+-}}\over{\ell_{++}}}$$ 
$$ r_{n \ov{n}} = |{\cal R}_{n \ov{n}}|~,~~r_0 = \di{{r_{00}}\over{2 \,r_{++}}}~,~~r_1 = \di{{r_{0+}}\over{r_{++}}}~,~~r_2 = \di{{r_{+-}}\over{r_{++}}}$$ 
\beq {\cal J} = \di{{\cal I}\over{2 \,{\cal L}_{++} {\cal R}_{++}}}~,~~\varphi = \varphi_2 - \varphi_1~,~~\varphi_a = - \varphi_1 \label{lesfi} \enq

\vv \nin Dans le r\'ef\'erentiel du centre de masse central, $\varphi$ et $\varphi_a$ repr\'esentent aussi bien les angles azimutaux respectifs de la particule 4 et de la particule choisie dans le syst\`eme $K$, par rapport \`a celui de la particule 3. Une relation  cin\'ematique entre l'angle $\varphi$ et son analogue $\Phi$ dans le r\'ef\'erentiel du centre de masse global est donn\'ee dans l'appendice C.

\vv \nin Avec ces notations, on a\footnote{On peut trouver la formule d'h\'elicit\'e qui suit, ou des formes similaires, dans : C. E. Carlson, W. K. Tung, Phys. Rev. D4, 2873 (1971) ; R. W. Brown, I. J. Muzinich, Phys. Rev. D4, 1496 (1971) ; C. J. Brown, D. H. Lyth, Nucl. Phys. B53, 323 (1973) ; V. M. Budnev, I. F. Ginzburg, G. V. Medelin, V. G. Serbo, Phys. Rep. 15C, 181 (1975) ; C. Carimalo, P. Kessler, J. Parisi, Phys. Rev. D20, 1057 (1979) ; N. Arteaga, C. Carimalo, P. Kessler, S. Ong, O. Panella, Phys. Rev. 52, 4920 (1995).}   

$$ {\cal J} = F_1 - 2 \ell_1\, F_2 \cos \varphi_a  - 2 \ell_2 \, F_3 \cos 2 \varphi_a  + 2 r_1 \, F_4 \cos(\varphi - \varphi_a) - 2 r_2 \, F_5  \cos 2(\varphi - \varphi_a) $$
$$-2 \ell_1 r_1 \, F_6 \cos \varphi  + \ell_2 r_2 \, F_7 \cos 2 \varphi  - 2 \ell_1 r_1 \, F_8 \cos(2 \varphi_a - \varphi) +\ell_2 r_2 \,F_9 \cos2(2\varphi_a -\varphi) $$
\beq - 2 \ell_2 r_1\, F_{10} \cos(\varphi + \varphi_a) + 2 \ell_1 r_2 \,F_{11} \cos(2 \varphi - \varphi_a) - 2 \ell_2 r_1\, F_{12} \cos(3 \varphi_a - \varphi)  \label{formhelic} \enq
$$+ 2 \ell_1 r_2\, F_{13} \cos(3 \varphi_a - 2\varphi) $$

\vv \nin o\`u les grandeurs $F_i$ ($i = 1,..., 13$) sont des combinaisons lin\'eaires d'\'el\'ements du tenseur d'h\'elicit\'e associ\'e au processus central $\gamma_1 + \gamma_2 \rightarrow K$ et sont donn\'ees ci-apr\`es : 

$$ F_1 = F_{++\, ;\, ++} + F_{++\, ;\,--} + 2 \ell_0 \, F_{00 \,;\, ++} + 2 r_0\, F_{++\,; \,00} + 2 \ell_0 r_0\, F_{00\, ;\, 00} $$
$$ F_2 = Re\left\{ F_{+0 \,;\; ++} - F_{0-\, ;\, ++} + 2 r_0 F_{0+\, ;\, 00} \right\} ~,~~F_3 = Re\left\{ F_{+-\, ;\, ++}\right\} + r_0 F_{+-\, ;\, 00} $$
\beq F_4 = Re\left\{ F_{++ \,;\, +0} - F_{++\, ;\, 0-} + 2 \ell_0 F_{00\, ;\, +0} \right\}~,~~F_5 = Re\left\{F_{++\, ;\, +-}\right\} + \ell_0 F_{00\, ;\, +-} \label{lesF} \enq
$$ F_6 = Re\left\{F_{+0\, ;\, +0} - F_{+0\, ;\, 0-}  \right\}~,~~F_7 = F_{+-\, ;\, +-}~,~~F_8 = Re\left\{F_{+0\, ;\, 0+} - F_{+0\, ;\, - 0}\right\}$$
$$ F_9 = F_{+-\, ;\, -+}~,~~F_{10} = Re\left\{F_{+-\, ;\, +0} \right\}~,~~F_{11} = Re\left\{F_{+0\, ;\, +-}  \right\}$$
$$F_{12} = Re\left\{F_{+-\, ;\, 0+} \right\}~,~~F_{13}= Re\left\{F_{0+\, ;\, +-} \right\}$$

\vvv \nin La formule \ref{formhelic} est {\it exacte}, dans le sens o\`u aucune approximation n'a \'et\'e introduite pour l'\'etablir. Ici encore, le proc\'ed\'e utilis\'e, la m\'ethode d'h\'elicit\'e, s\'epare clairement les grandeurs caract\'eristiques des diff\'erents vertex du diagramme, et par l\`a m\^eme permet une meilleure interpr\'etation physique des diff\'erents termes intervenant dans ${\cal I}$, contrairement \`a ce que pourrait donner une \'evaluation directe, non r\'efl\'echie, du produit tensoriel dans \ref{lestenseurs}.  

\vvv \nin A ce propos, au regard des facteurs $1/t_1$ et $1/t_2$ dans \ref{ampgamgam}, ceux-ci provenant des propagateurs des deux photons virtuels $\gamma_1$ et $\gamma_2$ respectivement, on doit s'attendre \`a ce que, de toutes les configurations cin\'ematiques accessibles dans la r\'eaction globale $1 + 2 \rightarrow 3 + K + 4$, celle pour laquelle les transferts $t_1$ et $t_2$ prennent leurs plus petites valeurs donne les plus forts taux de r\'eaction et soit donc globalement dominante dans cette r\'eaction.  A condition que dans ce domaine $\sqrt{t_1}$ et $\sqrt{t_2}$ soient tr\`es petits vis-\`a-vis des diverses masses invariantes impliqu\'ees, les deux photons peuvent alors \^etre consid\'er\'es comme {\it quasi-r\'eels}, et la r\'eaction s'interpr\`ete alors comme un processus {\it quasi-r\'eel} o\`u les particules initiales 1 et 2 font office de simples {\it g\'en\'erateurs de photons quasi-r\'eels}, photons que l'on fait ensuite entrer en collision pour produire l'\'etat $K$. Cette possibilit\'e de disposer virtuellement de collisions photon-photon, alors qu'il est tr\`es difficile de les r\'ealiser avec des photons strictement r\'eels,  a \'et\'e largement \'etudi\'ee, et l'est encore, tant th\'eoriquement qu'exp\'erimentalement\footnote{Les travaux th\'eoriques pr\'ecurseurs sur les  interactions photon-photon sont d\^us \`a divers auteurs : F. Low, Phys. Rev. 120 (1960) 582 ; F. Calogero, C. Zemach, Phys. Rev. 120 (1960) 1860 ; P. Kessler et collaborateurs : C.R. Acad. Sc., Paris 269B, 113 (1969) ; Phys. Rev. D3, 1569 (1971) ; Phys. Rev. D4, 2927 (1971) ;  S. Brodsky, T. Kinoshita, H. Terazawa, Phys. Rev. Lett. 25, 972 (1970) ; Phys. Rev. D4, 1532 (1971) ; V. E. Balakin, V. M. Budnev, I. F. Ginzburg, Zh. Exp. Teor. Fiz. Pis'ma 11, 559 (1970) - JEPT Lett. 11, 388 (1970).}. Sur ces sujets, nous renvoyons le lecteur \`a la litt\'erature sp\'ecialis\'ee\footnote{Sur un projet futur de collisionneur \`a photons, voir : ``TESLA, The Superconducting Electron Positron Linear Collider with an Integrated X-Ray Laser Laboratory", Technical Design Report, Part VI : Appendices, Chapter 1: Photon Collider at TESLA, DESY-2001-011, ECFA-2001-209 March TESLA-2001-23, TESLA-FEL-2001-05, Mars 2001 ; 
disponible via : 
http:$//$tesla.desy.de/new\_pages/TDR\_CD/PartVI/chapter1.pdf.}. Bien entendu, dans ledit domaine des {\it petits transferts}, on est amen\'e a effectuer des approximations appropri\'ees de ${\cal I}$ que l'utilisation de la formule d'h\'elicit\'e \ref{formhelic} 
permet de bien contr\^oler. De ces approximations \'emerge une formule tr\`es simple, ``lumineuse" pourrait-on dire, exprimant la section efficace diff\'erentielle par une {\it double factorisation}, \`a la {\it  Williams-Weisz\"acker} : 

\beq d \sigma_{1+2 \rightarrow 3+K+4} \approx {\cal P}_1(x_1) dx_1~d \sigma_c~ {\cal P}_2(x_2) d x_2 \enq 

\vv \nin o\`u $d \sigma_c$ est la section efficace diff\'erentielle de la r\'eaction $\gamma_1 + \gamma_2 \rightarrow K$ avec des photons, {\it r\'eels} cette fois\footnote{C'est-\`a-dire, dans laquelle on pose $t_1 = t_2 =0$.}, et ${\cal P}_1(x_1)$ et ${\cal P}_2(x_2)$ sont les {\it spectres} des photons quasi-r\'eels issus respectivement du vertex de gauche et du vertex de droite et transportant la proportion $x_1$ ou $x_2$ de l'\'energie de la {\it particule 
m\`ere} 1 ou 2, respectivement. Ces spectres caract\'erisent \`a la fois le processus de production du photon au vertex consid\'er\'e au moyen des taux de vertex ${\cal T}$ et ${\cal L}$, et sa propagation depuis le plan dudit vertex \`a celui du vertex central par l'interm\'ediaire des fonctions hyperboliques de $\alpha_1$ ou de  $\alpha_2$, selon le cas\footnote{Un premier historique du d\'eveloppement de l'approximation du spectre de photons \'equivalent, dite aussi approximation de Williams-Weizs\"acker et dont on peut faire remonter l'origine dans deux articles de N. Bohr (Phil. Mag. 5, p.10 (1913) ; 30, p.581 (1915)), fut pr\'esent\'e par P. Kessler au 1er colloque international sur les collisions photon-photon qui se tint au Coll\`ege de France \`a Paris en 1973, et dont les comptes-rendus furent publi\'es dans le suppl\'ement au Journal de Physique : Tome 35, Fasc. 3 C2 (1974) ; voir p.97.}.

\newpage
\section{Appendice A : transformation de Lorentz pure $T \rightarrow T'$}

\subsection{Formule g\'en\'erale}

\vv \nin Par d\'efinition, la transformation de Lorentz pure
$\Lambda_{T \rightarrow T'} $ appliqu\'ee \`a la base d'espace-temps
$(T,X,Y,Z)$ transforme celle-ci en la base $(T', X', Y', Z')$ telle
que

\beq T' = \cosh \chi \, T + \sinh \chi\, Z~,~~Z' = \cosh \chi \, Z + \sinh \chi\, T~,~~X' = X~,~~Y' = Y \enq

\vv \nin $\chi$ \'etant la {\it rapidit\'e} de cette transformation. La m\'ethode la plus simple pour obtenir ses \'el\'ements de matrice  est sans doute la suivante. Pour simplifier l'\'ecriture, notons-les $\Lambda_{\mu \nu}$ et posons $c = \cosh \chi = T \cdot T'$, $s = \sinh \chi$. En utilisant la relation de fermeture 

$$ g_{\rho \nu} = T_\rho T_\nu - Z_\rho Z_\nu - X_\rho X_\nu - Y_\rho Y_\nu $$

\nin \'ecrivons\footnote{On notera ici que ce proc\'ed\'e  permet d'exprimer les \'el\'ements de matrice d'une transformation de Lorentz {\it quelconque} comme $\Lambda_{\mu \nu} = {T'}_\mu T_\nu - {Z'}_\mu Z_\nu - {X'}_\mu X_\nu - {Y'}_\mu Y_\nu$.} 

$$ \Lambda_{\mu \nu} = \Lambda^\nu_\rho \, g_{\rho \nu} = \Lambda^\rho_\mu \left( T_\rho T_\nu - Z_\rho Z_\nu - X_\rho X_\nu - Y_\rho Y_\nu \right) =  {T'}_\mu T_\nu - {Z'}_\mu Z_\nu - X_\mu X_\nu - Y_\mu Y_\nu $$
$$ = {T'}_\mu T_\nu - {Z'}_\mu Z_\nu + g_{\mu \nu} -T_\mu T_\nu + Z_\mu Z_\nu  = g_{\mu \nu } + (T' - T)_\mu T_\nu - (Z' -Z)_\mu Z_\nu $$

\vv \nin En remarquant que 

$$ Z = \di{1\over s} \left( T' - c\, T \right) ~~{\rm et}~~Z' = \di{1\over s} \left(- T + c \, T' \right)~,~~{\rm donc}~~Z'-Z = s \, \di{{T'+T}\over{c +1}}  $$

\nin il vient 

$$ \Lambda_{\mu \nu} = g_{\mu \nu} + (T' - T)_\mu T_\nu - \di{{(T'+T)_\mu}\over{c +1}}  (T' - c T)_\nu $$  

\vv \nin En d\'eveloppant et en regroupant des termes, on arrive ais\'ement au r\'esultat : 

\beq \Lambda_{\mu \nu} = g_{\mu \nu} - \di{{T_\mu}\over{1+c}} \, (T_\nu + {T'}_\nu) + \di{{{T'}_\mu}\over{1+c}} \,\left\{ (2 c + 1) T_\nu - {T'}_\nu \right\} \label{lor-pure0} \enq

\vv \nin Cette formule permet d'obtenir l'expression du transform\'e d'un vecteur $W$, connaissant les projections de ce dernier sur $T$ et $T'$  :  

\beq \Lambda_{T \rightarrow T'} (W) = W' = W - T \, \di{{W\cdot T + W\cdot T'}\over{T \cdot T' + 1}} + T' \, \di{{ W\cdot T (2 T \cdot T' + 1) - W\cdot T'}\over{T\cdot T' + 1}} \label{lor-pure} \enq

\vv \nin A l'aide de \ref{lor-pure0}, on v\'erifie facilement la
propri\'et\'e de covariance d'une transformation de Lorentz pure. En
effet, pour une transformation de Lorentz quelconque $\Lambda$, on
trouve

\beq \Lambda\, \Lambda_{T \rightarrow T' } \,\Lambda^{-1} = \Lambda_{\Lambda T \rightarrow \Lambda T'} \label{covar-pure} \enq

\vv \nin On note que pour un vecteur $W$ orthogonal \`a $T$, on a

\beq \Lambda_{T \rightarrow T'} (W) = W - \di{{ W\cdot T'}\over{T \cdot T' + 1}} (T + T') \enq

\vv \nin et que si $W$ est aussi orthogonal \`a $T'$, alors il est
conserv\'e dans la transformation, ce qui, bien entendu, provient  du fait que la transformation laisse invariant le bi-plan
orthogonal au bi-plan form\'e par les deux vecteurs $T$ et $T'$.

\subsection{Cas du vertex de la figure \ref{vertex31}}

\vv \nin Appliquons la formule \ref{lor-pure} pour montrer que dans la transformationde Lorentz pure 
$\Lambda_{\hat{p} \rightarrow \hat{p}_1}$, o\`u $p=p_1 +p_2$, l'impulsion relative unitaire $\hat{q}_{12}$ de $p_1$ et $p_2$ est chang\'ee en vecteur d'h\'elicit\'e $h(p_1,p)$ de $p_1$.  Comme 
$p \cdot q_{12} = 0$, on a 

$$\Lambda_{\hat{p} \rightarrow \hat{p}_1}(q_{12}) = q' = q_{12} - \di{{p_1 \cdot q_{12}}\over{\di{{p_1 \cdot p}\over{\sqrt{s}}} + m_1}} \left(\di{p_1\over m_1} + \di{p\over \sqrt{s}} \right) $$

\vv \nin soit, en explicitant $q_{12}$ 

$$q' = \di{m_1\over \sqrt{s}} \left( -p + \di{{p_1 \cdot p}\over{m^2_1}} p_1 \right) $$ 

\vv \nin D'o\`u 

\beq \Lambda_{\hat{p} \rightarrow \hat{p}_1}(\hat{q}_{12})  = \di{{2 \sqrt{s}}\over{\Lambda^{1/2}(s, m^2_1, m^2_2)}} q' =  \di{{2 m_1}\over{\Lambda^{1/2}(s, m^2_1, m^2_2)}}  \left( -p + \di{{p_1 \cdot p}\over{m^2_1}} p_1 \right) = h(p_1,p) \enq

\vv \nin qui est le r\'esultat annonc\'e.

\subsection{Cas du vertex de la figure \ref{vertex32}}

\vv \nin Ici, l'impulsion relative est donn\'ee par \ref{relcas2}. Nous voulons montrer que dans la transformation $\Lambda_{\hat{q}_{12} \rightarrow \hat{p}_1}$, $\hat{p}$ devient le vecteur d'h\'elicit\'e $h(p_1, p)$ de $p_1$. Notons tout d'abord les relations utiles 

$$ q^2_{12} = p_1 \cdot q_{12} = m^2_1 +\di{(p\cdot p_1)^2\over t} = \di{{\Lambda(m^2_1, m^2_2, -t)}\over{4 t}},~~$$

\vv \nin Compte tenu de $p\cdot q_{12} = 0$, on a 

$$ \Lambda_{\hat{q}_{12} \rightarrow \hat{p}_1}(p) = p' = p - \di{{p\cdot \hat{p}_1}\over{1+ \hat{p}_1 \cdot \hat{q}_{12}}} \left( \hat{p}_1 + \hat{q}_{12} \right) = p - \di{{p\cdot p_1}\over{m_1+ \sqrt{q^2_{12}}} }\left( \di{{p_1}\over{m_1}} + \di{1\over{\sqrt{q^2_{12}}}} ( p_1 + \di{{p_1 \cdot p}\over t}\,p \right)$$
$$ = p \,\left(1 - \di{{(p_1 \cdot p)^2}\over{ t \sqrt{q^2_{12}} ( m_1 + \sqrt{q^2_{12}}) }} \right) -p_1\, \di{{p_1 \cdot p}\over{m_1 \sqrt{q^2_{12}}}} = \di{{m_1}\over{\sqrt{q^2_{12}}}} \left(p - \di{{p_1 \cdot p}\over{m^2_1}} \,p_1\right)$$

\vv \nin et donc 

\beq \Lambda_{\hat{q}_{12} \rightarrow \hat{p}_1}(\hat{q}) = \di{{2 m_1}\over{\Lambda^{1/2}(m^2_1, m^2_2, -t)} }\, \left(p - \di{{p_1 \cdot p}\over{m^2_1}} \,p_1 \right) = h(p_1,p) \enq

\vv \nin ce qui \'etablit le r\'esultat.

\subsection{Cas du vertex \ref{vertex31} avec deux lignes entrantes virtuelles du genre espace}

\vv \nin Il reste \`a consid\'erer le cas o\`u au vertex \ref{vertex31} les impulsions entrantes $p_1$ et $p_2$ sont du genre espace ($p^2_1 = -t < 0~,~p^2_2 = - t' <0$), tandis que leur somme $p= p_1 + p_2$ est du genre temps futur ($p^2 = s>0$). L'impulsion relative $q_{12}$ et $h_1$ \'etant d\'efinis par \ref{q123} et \ref{h123} respectivement. On a 

$$ h_1 \cdot p = \di{\Lambda^{1/2} \over{2 \sqrt{t}}}~,~~h_1 \cdot q_{12} = - \di{{p_1 \cdot p}\over s} \di{\Lambda^{1/2} \over{2 \sqrt{t}}}~,~~\di{{(p \cdot p_1)^2}\over t} = \di{\Lambda \over{4t}} - s~,~~{\rm avec}~~ \Lambda = \Lambda(s,-t,-t') $$

\vv \nin Calculons 

$$ q' = \Lambda_{\hat{p} \rightarrow h_1} (q_{12}) = q_{12} - \di{{q_{12} \cdot h_1}\over{1+ \hat{p} \cdot h_1}} \left( h_1 + \hat{p} \right) = p_1 - \di{{p\cdot p_1}\over s} p + \di{{p\cdot p_1}\over s} \di{\Lambda^{1/2}\over{2 \sqrt{t}}} \di{1\over{1 + \di{\Lambda^{1/2} \over{2 \sqrt{st}}}}} \times $$
$$ \left\{ \di{p\over \sqrt{s}} + \di{{2 \sqrt{t}}\over \Lambda^{1/2}} \left( p + \di{{p \cdot p_1}\over t} p_1 \right)\right\}  =p_1 \left\{ 1 + \di{(p\cdot p_1)^2 \over{t s}} \di{1\over{1 + \di{\Lambda^{1/2} \over{2 \sqrt{st}}}}}  \right\} = p_1 \, \di{\Lambda^{1/2} \over{2 \sqrt{st}}}$$

\vv \nin On a donc bien le r\'esultat 

\beq \Lambda_{\hat{p} \rightarrow h_1} (\hat{q}_{12}) = \hat{p}_1 \enq

\newpage

\section{Appendice B : conservation de la parit\'e \`a un vertex \'electromagn\'etique}

\vv \nin L'op\'eration de parit\'e donne lieu \`a la loi de transformation suivante des \'etats :  

\beq U(\pi) |[\,p_i\,]_h, \lambda_1 > = \eta_i \, {\cal D}^{s_i}_{\lambda'_i \lambda_i}(R Y) |[\,\und{p_i}\, ]_h, \lambda'_i > ~,~~{\rm avec}~~ \und{p_i} = (p^0_i , - \Vec{p_i}) \enq
$$ {\rm et}~~~R\,Y = [\,\und{p_i}\,]^\dagger_h~[\,p_i\,] = [\,\und{k}\,]^\dagger_h~[k]_h $$

\vv \nin o\`u $R$ est une rotation d'angle $\varphi$ autour de $n_3$ ; $\eta_i$ est la parit\'e intrins\`eque de la particule (i) ; $Y$ est la rotation d'angle $\pi$ autour de $n_2$. On note la relation :

$$  {\cal D}^{s}_{\lambda' \lambda}(R Y) = (-1)^{s + \lambda} \, e^{i \lambda \varphi}\, \delta_{\lambda' , - \lambda} $$

\vv \nin Par conservation de la parit\'e dans les interactions \'electromagn\'etiques, on a 

\beq ~^\pi \hskip -0.1cmJ_\mu (0) = U(\pi) J_\mu(0) U^{-1}(\pi)~,~~{\rm avec}~~~~^\pi \hskip -0.1 cmJ_0 (0) = J_0(0)~,~~~
 \Vec{^\pi \hskip -0.1 cm J} (0) = - \Vec{J}(0) \enq

\vv \nin Ecrivons : 

$$J_\lambda =  < [p_3]_h , \lambda_3 | e^{(\lambda) \star}_\mu (k) J^\mu(0) |[p_1]_h, \lambda_1 > = < [\,\und{p_3}\,]_h , -\lambda_3| e^{(\lambda) \star}_\mu (k)  ~^\pi \hskip -0.1 cm J^\mu(0) |[\,\und{p_1}\,]_h, -\lambda_1>\eta $$
$$~~{\rm avec}~~\eta = (-1)^{s_1 + s_3 + \lambda_1 + \lambda_3} \eta_1 \eta_3 e^{i(\lambda_3 - \lambda_1) \varphi} $$

\vv \nin Revenons ensuite, par application de $[\,\und{k}\,]^{-1}_h$, au ``r\'ef\'erentiel de Breit" du vertex de gauche, dans lequel on a 

$$\Vec{q_{13}} = \Vec{0}~,~~ \hat{k} = n_3~,~~e^{(\pm)}(k) = e^{(\pm)}~,~~p_1 = \tilde{p}_1~,~~p_3 = \tilde{p}_3$$

\vv \nin En tenant compte de la relation 

$$U(\pi)\,U(A)\,U^{-1}(\pi) = U(A^{\dagger -1} )$$

\vv \nin il vient 

$$J_\lambda = \eta\, <[\tilde{p}_3,]_h, -\lambda_3| \left([\und{k}]^\dagger_h e^{(\lambda \star}_\mu(k) \right)\, ^\pi \hskip -0.1cm J^\mu (0) | [\tilde{p}_1]_h, -\lambda_1> $$

\vv \nin Or, on a 

$$ \left([\und{k}]^\dagger_h e^{(\lambda \star}_\mu(k) \right) = R Y e^{(\lambda)} = e^{i \varphi \lambda} e^{(-\lambda)} ~~~ {\rm et}~~~ e^{(\lambda) \star}_\mu  J^\mu(0) = (-1)^\lambda e^{(-\lambda)}_\mu J^\mu(0) $$  

\vv \nin On obtient donc 

$$ J_{\lambda_1 \lambda_3} = (-1)^{s_1 + s_3 + \lambda_1 + \lambda_3 + \lambda} \eta_1 \eta_3      J_{- \lambda_1, -\lambda_3} $$ 

\vv \nin et par suite 

\beq \left| J_{\lambda_1 \lambda_3} \right| = \left| J_{-\lambda_1, -\lambda_3} \right| \enq 

\vv \nin d'o\`u r\'esulte la relation $F_{+} = F_{-}$ dans \ref{parite}. 

\newpage
\section{Appendice C : Relation entre l'angle $\varphi$ (\ref{lesfi}) et son homologue $\Phi$ dans le r\'ef\'erentiel du centre de masse global}

\vv \nin D\'efinissons la base associ\'ee au r\'ef\'erentiel du centre de masse global (c'est-\`a-dire, celui du syst\`eme des deux particules 1 et 2) de la fa\c{c}on suivante : 

$$ T_0 = \di{P \over \sqrt{s}} ~,~~Z_0 = \di{{2 \sqrt{s}}\over{\Lambda^{1/2}_0}} \left( p_1 - \di{{P\cdot p_1}\over s} \, P \right)~,~~{\rm avec}~~\Lambda_0 = \Lambda(s, m^2_1, m^2_2)~,$$ 
$$ Y_{0 \mu} = - N_0 \, \varepsilon_{\mu \nu \rho \sigma}\, P^\nu \, p^\rho_3\, p^\sigma_1 ~,~~{\rm avec}~~ N_0 = \di{2\over{\Lambda^{1/2}_0 \,p^\circ_3 \sin \theta^\circ_3}}~,~~X_{0 \mu} =  \varepsilon_{\mu \nu \rho \sigma}\, T^\nu_0 \, Y^\rho_0\, Z^\sigma_0 $$

\vv \nin Dans ce r\'ef\'erentiel, $\Vec{p_1} = - \Vec{p_2}$ est selon l'axe des z, $\Vec{p_3}$ , de module $p^\circ_3$, est dans le plan $(x,z)$ et a pour angle orbital $\theta^\circ_3$. L'axe des y correspondant est donc perpendiculaire au plan des deux tri-vecteurs $\Vec{p_1}$ et $\Vec{p_3}$. Explicitement, les composantes de $Y_0$ dans ce r\'ef\'erentiel sont 

$$ Y_0 = \left(0, \di{{\Vec{p_1} \wedge \Vec{p_3}}\over{ |\Vec{p_1} \wedge \Vec{p_3}|}} \right) = (0,0,1,0)$$

\vv \nin D\'efinissons ensuite la base du r\'ef\'erentiel du centre de masse central comme 

$$ T_c = \di{K\over M}~,~~Z_c = \di{2 \over \Lambda^{1/2}_c} \left (k_1 - \di{{K\cdot k_1}\over M^2}\, K \right) ~,~~Y_{c \mu} = - N_c \, \varepsilon_{\mu \nu \rho \sigma}\,K^\nu p^\rho_1 k^\sigma_1 ~,~~{\rm avec}~~N_c = \di{2 \over{\Lambda^{1/2}_c p^c_1 \sin \theta^c_1}}$$
$$ X_{c \mu} = \varepsilon_{\mu \nu \rho \sigma}\,T^\nu_c Y^\rho_c Z^\sigma_c $$

\vv \nin Dans ce second r\'ef\'erentiel, $\Vec{k_1} = - \Vec{k_2}$ est selon l'axe des z ; les deux tri-vecteurs $\Vec{p_1}$ et $\Vec{p_3}$ sont dans le plan $(x,z)$. Leurs modules et angles orbitaux sont $p^c_1$, $\theta^c_1$ et $p^c_3$, $\theta^c_3$, respectivement, et l'on a 
bien s\^ur (composantes selon l'axe des x)

$$ p^c_1 \sin \theta^c_1 = p^c_3 \sin \theta^c_3 $$  

\vv \nin Ici, l'axe des y correspondant \`a ce r\'ef\'erentiel est perpendiculaire au plan form\'e par les deux tri-vecteurs $\Vec{p_1}$ et $\Vec{p_3}$.

\vv \nin Dans le premier r\'ef\'erentiel, le vecteur $p_4$ a pour composantes  

$$ (E^\circ_4, \,p^\circ_4 \sin \theta^\circ_4 \cos \Phi, \,p^\circ_4 \sin \theta^\circ_4 \sin \Phi, \, p^\circ_4 \cos \theta^\circ_4)_\circ $$ 

\vv \nin  tandis que ses composantes dans le second r\'ef\'erentiel sont\footnote{A noter que $\varphi = \varphi_2 - \varphi_1$ est aussi bien l'angle azimutal relatif entre 4 et 1.} : 

$$ (E^c_4, \, p^c_4 \sin \theta^c_4 \cos \varphi, \, p^c_4 \sin \theta^c_4 \sin \phi,\, p^c_4 \cos \theta^c_4)_c  $$

\vv \nin Ecrivons alors\footnote{Ceci \'equivaut au calcul d'un 4-volume.} 

$$ Y^c \cdot p_4 = - p^c_4 \sin \theta^c_4 \sin \varphi =  - N_c \, \varepsilon_{\mu \nu \rho \sigma}\,K^\nu p^\rho_1 k^\sigma_1 \, p^\mu_4$$

\vv \nin Mais comme $k_1 = p_1 - p_3$ et $K = P - p_3 -p_4$, on peut faire les substitutions

\newpage 
$$\varepsilon_{\mu \nu \rho \sigma}\,K^\nu p^\rho_1 k^\sigma_1 p^\mu_4 \equiv - \varepsilon_{\mu \nu \rho \sigma}\,K^\nu p^\rho_1 p^\sigma_3 p^\mu_4 \equiv \varepsilon_{\mu \nu \rho \sigma}\,P^\nu p^\rho_3 p^\sigma_1 p^\mu_4 = - N^{-1}_0 \, Y_0 \cdot p_4 = N^{-1}_0 \, p^\circ_4 \sin \theta^\circ_4 \sin \Phi $$ 

\nin D'o\`u la relation 

\beq \sin \varphi = \sin \Phi \, \di{\Lambda^{1/2}_0 \over \Lambda^{1/2}_c} \,\kappa ~,~~{\rm avec}~~ \kappa = \di{{p^\circ_3 \sin \theta^\circ_3 \, p^\circ_4 \sin \theta^\circ_4}\over{p^c_3 \sin \theta^c_3 \, p^c_4 \sin \theta^c_4}} = \di{{p^\circ_3 \sin \theta^\circ_3 \, p^\circ_4 \sin \theta^\circ_4}\over{p^c_1 \sin \theta^c_1 \, p^c_2 \sin \theta^c_2}}\label{lefi4} \enq 

\vv \nin entre $\varphi$ et $\Phi$ qui para\^it donc un peu compliqu\'ee dans le cas g\'en\'eral. Elle se simplifie \'enorm\'ement dans la configuration o\`u les angles d'\'emission respectifs $\theta^\circ_3$ et $\theta^\circ_4$ des particules 3 et 4 sont tr\`es petits, o\`u les particules externes 1, 2, 3 et 4 sont ultra-relativistes (\'energie $\gg$ masse),  et o\`u les photons $\gamma_1$ et $\gamma_2$ sont alors quasi-r\'eels. Dans ce cas, \`a des termes du second ordre pr\`es suivant ces faibles angles d'\'emission et les faibles rapports masse/\'energie, on peut faire les approximations suivantes. 

\vv \nin Tout d'abord, on a $E^\circ_1 \approx \sqrt{s}/2$, $E^\circ_2 \approx \sqrt{s}/2$, $\Lambda^{1/2}_0 \approx s$, $\Lambda^{1/2}_c \approx M^2$, puis 

$$ M^2 = (P - p_3 - p_4)^2 = s - 2 \sqrt{s} ( E^\circ_3 + E^\circ_4) + 2 p_3 \cdot p_4 \approx 
s  - 2 \sqrt{s} ( E^\circ_3 + E^\circ_4) + 4 E^\circ_3 E^\circ_4  $$
$$ = 4 (E^\circ_1 - E^\circ_3)(E^\circ_2 - E^\circ_4) = 4 \omega^\circ_1 \omega^\circ_2 $$
\vvv

\begin{figure}[hbt]
\centering
\includegraphics[scale=0.3, width=7cm, height=4cm]{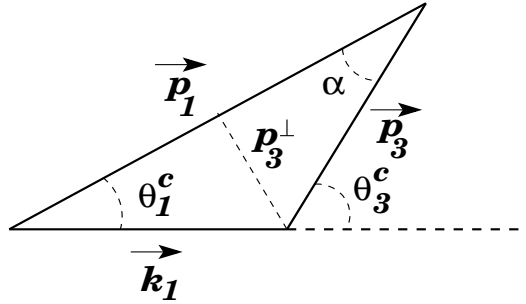}
\vskip 0.25cm

\caption{Tri-impulsions dans le r\'ef\'erentiel du centre de masse central} \label{lespt}
\end{figure}

\vv \nin o\`u $\omega^\circ_1$ et $\omega^\circ_2$ sont les \'energies respectives des photons $\gamma_1$ et $\gamma_2$ dans le r\'ef\'erentiel du centre de masse global. En 
outre, d'apr\`es la figure \ref{lespt} repr\'esentant les tri-impulsions dans le r\'ef\'erentiel du centre de masse central, on a 

$$ k^c_1 \, \sin \theta^c_1 = p^c_3 \, \sin \alpha = p^\perp_3 $$ 

\vv \nin $p^\perp_3$ \'etant la projection de la tri-impulsion de la particule 3 perpendiculairement \`a la direction de propagation de la particule 1. A des termes du second ordre pr\`es suivant les angles $\theta$, cette impulsion transversale se conserve dans le passage d'un r\'ef\'erentiel \`a l'autre :  $p^\perp_3 \approx p^\circ_3 \sin \theta^\circ_3 (\approx E^\circ_3 \theta^\circ_3)$. On a de m\^eme 

$$ k^c_2 \, \sin \theta^c_2 = p^c_4 \, \sin \alpha' = p^\perp_4 \approx p^\circ_4 \sin \theta^\circ_4 (\approx E^\circ_4 \theta^\circ_4)$$ 

\vv \nin D'o\`u 

$$ \kappa \approx \di{{k^c_1 \sin \theta^c_1 \, k^c_2 \sin \theta^c_2}\over{p^c_1 \sin \theta^c_1 \, p^c_2 \sin \theta^c_2}} \approx \di{M^2 \over{4 E^c_1 E^c_2}} $$

\vv \nin Or, 

$$ K \cdot p_1 = M E^c_1 = p_1 \cdot (P- p_3 -p_4) \approx p_1 \cdot (P-p_4) \approx 2 E^\circ_1 (E^\circ_2 - E^\circ_4) = 2 E^\circ_1 \, \omega^\circ_2 $$
$$ {\rm soit}~~E^c_1 \approx E^\circ_1 \sqrt{\omega^\circ_2\over \omega^\circ_1}$$

\vv \nin Sym\'etriquement, on a 

$$E^c_1 \approx E^\circ_1 \sqrt{\omega^\circ_1\over \omega^\circ_2}~,~~ {\rm et~donc}~~E^c_1 E^c_2 \approx E^\circ_1 E^\circ_2 \approx s/4 $$

\vv \nin On aboutit ainsi \`a  

$$ \di{\Lambda^{1/2}_0 \over \Lambda^{1/2}_c} \,\kappa \approx \di{s\over M^2} \, \kappa \approx 1 $$

\vv \nin Dans le contexte de cette approximation, la relation entre $\varphi$ et $\Phi$ se simplifie donc de fa\c{c}on spectaculaire :  

$$ \sin \varphi \approx \sin \Phi~,~~{\rm soit}~~ \varphi \approx \Phi$$

\vv \nin La relation {\it exacte} entre $\varphi_a$ et son \'equivalent $\Phi_a$ dans le r\'ef\'erentiel du centre de masse global est d\'ej\`a moins ``simple" que \ref{lefi4}. Pourtant, dans la configuration pr\'ec\'edente des petits angles d'\'emission, on montre que l'on a aussi, au m\^eme ordre d'approximation : 

$$ \varphi_a \approx \Phi_a$$

\vv \nin r\'esultat d'ailleurs pr\'evisible puisque, comme il a \'et\'e sugg\'er\'e plus haut, \`a l'extr\^eme limite des angles d'\'emission nuls des particules 3 et 4, le passage d'un r\'ef\'erentiel \`a l'autre s'effectue suivant un m\^eme axe et qu'alors toutes les composantes des vecteurs qui sont transversales \`a cet axe restent inchang\'ees.
\vv \nin Pour l'\'etude g\'en\'erale des corr\'elations azimutales dans les ``processus photon-photon" sur la base de la formule g\'en\'erale \ref{formhelic}, le lecteur pourra se reporter \`a l'article de N. Arteaga et co-auteurs d\'ej\`a cit\'e\footnote{N. Arteaga, C. Carimalo, P. Kessler, S. Ong, O. Panella, Phys. Rev. 52, 4920 (1995).}.  

\newpage


\setcounter{page}{45}

\setcounter{chapter}{1}

\setcounter{equation}{0}

\renewcommand{\theequation}{\mbox{2.}\arabic{equation}}

\newcommand{\basic}{\stackrel{\circ}}
\newcommand{\uw}{\uparrow}
\newcommand{\dw}{\downarrow}

\newcommand{\Gi}{\Gamma}
\newcommand{\gs}{\gamma}

\newcommand{\uup}{U^\uparrow}
\newcommand{\ud}{U^\downarrow}
\newcommand{\vup}{V^\uparrow}
\newcommand{\vd}{V^\downarrow}
\newcommand{\wup}{W^\uparrow}
\newcommand{\wdo}{W^\downarrow}

\newcommand{\buup}{\ov{U}^\uparrow}
\newcommand{\bud}{\ov{U}^\downarrow}
\newcommand{\bvup}{\ov{V}^\uparrow}
\newcommand{\bvd}{\ov{V}^\downarrow}
\newcommand{\bwup}{\ov{W}^\uparrow}
\newcommand{\bwd}{\ov{W}^\downarrow} 

\newcommand{\wht}{\widehat}

\newcommand{\bsig}{\bar{\sigma}}

\chapter{Notions et outils de base}

\nin Deux types de particules sont pr\'epond\'erants dans l'univers des particules \'el\'ementaires. Le premier regroupe les particules dites {\it vectorielles} auxquelles ont assigne g\'en\'eralement un r\^ole de m\'ediation dans les interactions. Les unes, comme le photon ou les gluons ont une masse nulle, les autres comme les bosons $W^{\pm}$ et $Z^0$ sont massives. On les d\'ecrit au moyen de la repr\'esentation 4-vectorielle du groupe de Lorentz, sur laquelle repose en fait la d\'efinition de ce groupe. Le second type est celui des particules que l'on d\'ecrit g\'en\'eralement au moyen de {\it spineurs de Dirac} et auquel est attach\'ee la repr\'esentation de spin 1/2 du groupe $SL(2,C)$, groupe de rev\^etement universel du groupe de Lorentz. Certaines de ces particules, les leptons (tels l'\'electron, le muon ou les neutrinos) interviennent dans les interactions \'electro-faibles et sont directement observables ; d'autres, comme les quarks, n'ont jamais \'et\'e observ\'ees \`a l'\'etat libre mais sont consid\'er\'ees comme les structures incontournables \`a partir desquelles sont construites les particules observables hadroniques, et qui d\'ecrivent les interactions de ces derni\`eres, r\'egies au  premier chef par la Chromodynamique Quantique.     

\vv \nin Le but de ce chapitre est de fournir le maximum de propri\'et\'es de ces deux repr\'esentations\footnote{Les notations et formules concernant les groupes de Lorentz, de Poincar\'e et leurs repr\'esentations sont celles des chapitres 4, 5 et 6 du cours ``Introduction \`a la Th\'eorie Lagrangienne", ci-apr\`es nomm\'e ITL (http://physique-univ.fr/physique-th\'eorique/lagrangien.html), et dans le chapitre 1 de ce cours.}. On trouvera \'egalement \`a la fin du chapitre certaines notions compl\'ementaires qu'il nous para\^it utile de conna\^itre.       


\section{La repr\'esentation 4-vectorielle du groupe de Lorentz}

\vv \nin Dans ce qui suit, il est fait un usage intensif des propri\'et\'es du tenseur de Levi-Civita quadri-dimensionnel. La section (\ref{tenseur-LC}) donne la d\'efinition g\'en\'erale et quelques propri\'et\'es des symboles de Levi-Civita d'ordres quelconques.  

\subsection{G\'en\'erateurs de spin 4-vectoriels}

\vv \nin Bien que certaines notions la concernant aient d\'ej\`a \'et\'e abord\'ees au chapitre 1, nous rappelons ici certaines d\'efinitions essentielles de la repr\'esentation 4-vectorielle. Dans cette repr\'esentation, qui utilise des matrices $4\times4$, les g\'en\'erateurs $J^{\mu \nu}$ sont d\'efinis par

\beq (J_{\mu \nu})_{\alpha \beta} = i ( g_{\mu \alpha} \, g_{\nu \beta} - g_{\nu \alpha}\,
g_{\mu \beta} ) \label{Jquadriv} \enq

\vv \nin Le lecteur v\'erifiera que ces op\'erateurs v\'erifient bien les relations de commutations des g\'en\'erateurs du groupe ${\cal L}^\uparrow_{+}$ : 

\beq  \left[~J_{\mu \nu}\, , \,
J_{\rho \sigma} ~\right]  = i \left( g_{\mu \sigma} \, J_{\nu \rho}
- g_{\mu \rho}\, J_{\nu \sigma} - g_{\nu \sigma}\, J_{\mu \rho} +
g_{\nu \rho}\, J_{\mu \sigma} \right) \label{commutJ} \enq

\vv \nin Une transformation de Lorentz du groupe ${\cal L}^\uparrow_{+}$ s'\'ecrit toujours sous la forme $\Lambda = e^G$ o\`u $G = i\, \omega_{\mu \nu} \, J^{\mu \nu}\,/2$, 
$\omega_{\mu \nu}$ \'etant un tenseur antisym\'etrique. D'apr\`es (\ref{Jquadriv}), 
on a simplement $G_{\alpha \beta} = - \omega_{\alpha \beta}$. Si $\Lambda$ appartient au petit groupe d'un 4-vecteur $\eta$, on a $\Lambda(\eta) = \eta$, soit $G(\eta)=0$ et donc $\omega^{\alpha \beta}\, \eta_\beta = 0 $. Il s'ensuit que le tenseur antisym\'etrique $\omega_{\mu \nu}$ est plan et doit \^etre de la forme

\beq \omega_{\mu \nu} = \epsilon_{\mu \nu \rho \sigma}\, p^\rho \, \eta^\sigma \enq

\vv \nin Dans cette expression, le 4-vecteur
$p$ peut toujours \^etre remplac\'e par un 4-vecteur de la forme $\xi = p + a \eta$. Si $\eta$
n'est pas du genre lumi\`ere, le scalaire $a$ peut \^etre d\'etermin\'e de telle sorte que $\xi$
soit orthogonal \`a $\eta$. Dans ce cas, on a

\beq \xi^\alpha =  \frac{1}{2 \eta^2} \, \epsilon^{\alpha \mu \nu \beta}\, \omega_{\mu \nu} \, \eta_\beta \enq

\vv \nin En toute circonstance, on peut \'ecrire

\beq \omega_{\mu \nu} = \epsilon_{\mu \nu \rho \sigma}\, \xi^\rho \, \eta^\sigma \enq

\vv \nin et le g\'en\'erateur $G$ de la transformation est
donc de la forme

\beq G = i\, \xi^\mu \, W_\mu(\eta) \enq

\nin o\`u

\beq W_\mu(\eta) = \frac{1}{2} \, \epsilon_{\mu \nu \rho \sigma}\, \eta^\nu \, J^{\rho \sigma},~~~\left[\,W_\mu(\eta)\,\right]_{\alpha \beta} = i\, \epsilon_{\mu \nu \alpha \beta}\,\eta^\nu \enq

\vv \nin est l'op\'erateur de Pauli-Lubanski associ\'e au 4-vecteur $\eta$.

\vv

\vv \nin \ding{192} {\bf Cas o\`u $\eta$ est du genre temps : $\eta^2 > 0$}

\vv \nin Posons $M = \sqrt{\eta^2}$ et d\'efinissons le 4-vecteur unitaire du genre temps
$t = \pm \eta/M$, le signe \'etant choisi de telle sorte que $t$ pointe vers le futur. Associons \`a
ce 4-vecteur une triade $x, y, z$ de 4-vecteurs du genre espace et orthogonaux entre eux se trouvant
dans l'hyperplan orthogonal \`a $t$, pour former avec $t$ une base d'orientation directe
${\cal B}(t)$. Ils v\'erifient les relations suivantes qui g\'en\'eralisent celles relatives aux bases de l'espace \`a trois dimensions. 

\vv
\beq \begin{array}{c}
 t_\mu\, t_\nu - x_\mu\, x_\nu - y_\mu\, y_\nu - z_\mu\, z_\nu = g_{\mu \nu} ~~~~~({\sf relation~ de ~fermeture})  \\ [0.5cm]
 \epsilon_{\mu \nu \rho \sigma}\,t^\mu\, x^\nu\, y^\rho\, z^\sigma = 1 ~~~~~({\sf 4\hskip-0.07cm -\hskip-0.07cm volume})
\end{array}
\label{relb-1} \enq

\vv
\beq
\begin{array}{c}
t_\mu = \epsilon_{\mu \nu \rho \sigma} \, x^\nu \,
y^\rho\, z^\sigma~,~~~ x_\mu = \epsilon_{\mu \nu \rho
\sigma}\, t^\nu \,y^\rho\, z^\sigma  \\ [0.5cm] 
~\label{baseeps} 
y_\mu = - \epsilon_{\mu \nu \rho \sigma}\, t^\nu \,
x^\rho\, z^\sigma~,~~~ z_\mu = \epsilon_{\mu \nu \rho
\sigma}\, t^\nu \, x^\rho\, y^\sigma 
 \end{array}
\label{relb-2} \enq

\vv

\beq \begin{array}{c}
t_\mu \,x_\nu - t_\nu \,x_\mu = - \epsilon_{\mu \nu
\rho \sigma}\,y^\rho\, z^\sigma~~~~t_\mu \,y_\nu - t_\nu \,y_\mu \,= \, \epsilon_{\mu
\nu \rho \sigma}\,x^\rho\, z^\sigma \\ [0.5cm] 
t_\mu \,z_\nu - t_\nu \,z_\mu =  -\epsilon_{\mu \nu
\rho \sigma}\,x^\rho\, y^\sigma  \\ [0.5cm]  
x_\mu \,y_\nu - x_\nu \,y_\mu \, = \, \epsilon_{\mu
\nu \rho \sigma}\,t^\rho\, z^\sigma ~~~~y_\mu \,z_\nu - y_\nu \,z_\mu \, = \, \epsilon_{\mu
\nu \rho \sigma}\,t^\rho\, x^\sigma ~ \\ [0.5cm] 
z_\mu \,x_\nu - z_\nu \,x_\mu \, =   \epsilon_{\mu
\nu \rho \sigma}\,t^\rho\, y^\sigma ~ 
\end{array}
\label{relb-3} \enq

\vvv 

\beq \begin{array}{c}
 \epsilon_{\mu \nu \rho \sigma}\, t^\mu = - x_\nu\,y_\rho\, z_\sigma - x_\rho\, y_\sigma\, z_\nu - x_\sigma\,y_\nu\, z_\rho + x_\nu\,y_\sigma\, z_\rho + x_\sigma\,y_\rho\, z_\nu + x_\rho\, y_\nu\, z_\sigma  \\ [0.5cm]
= -\,\di{\sum_{k \ell m}}~\epsilon_{k \ell m}\, n_{k \,\mu}\, n_{\ell \,\nu}\, n_{m \,\rho} \\ [0.5cm]
 \epsilon_{\mu \nu \rho \sigma}\, x^\mu = - y_\nu\,z_\rho\, t_\sigma - y_\rho\, z_\sigma\, t_\nu - y_\sigma\,y_\nu\, t_\rho + y_\nu\,z_\sigma\, t_\rho + y_\sigma\,z_\rho\, t_\nu + y_\rho\, z_\nu\, t_\sigma \\ [0.5cm]
 \epsilon_{\mu \nu \rho \sigma}\, y^\mu =  z_\nu\,t_\rho\, x_\sigma + z_\rho\, t_\sigma\, x_\nu + z_\sigma\,t_\nu\, x_\rho - z_\nu\,t_\sigma\, x_\rho - z_\sigma\,t_\rho\, x_\nu - z_\rho\, t_\nu\, x_\sigma \\ [0.5cm]
\epsilon_{\mu \nu \rho \sigma}\, z^\mu = - t_\nu\,x_\rho\, y_\sigma - t_\rho\, x_\sigma\, y_\nu - t_\sigma\,x_\nu\, y_\rho + t_\nu\,x_\sigma\, y_\rho + t_\sigma\,x_\rho\, y_\nu + t_\rho\, x_\nu\, y_\sigma 
\end{array}
\label{relb-4} \enq

\vv \nin o\`u, dans la premi\`ere formule, $n_1 = x, n_2 = y, n_3=z$. Le 4-vecteur $\xi$ envisag\'e plus haut, orthogonal \`a $t$, se d\'eveloppe suivant les trois 4-vecteurs $x$, $y$ et $z$ :

$$ \xi =  \xi^1 \,x  + \xi^2\, y + \xi^3 \, z $$

\vv \nin et le g\'en\'erateur des transformations du petit groupe de $\eta$ s'\'ecrit

\beq
\begin{array}{c}  G = \mp\,i\, M \left( \xi^1\, S_1(t) + \xi^2\, S_2(t) + \xi^3\, S_3(t) \right) 
 \\[0.4cm] 
 {\rm avec}~~~S_1(t) = - x \cdot W(t)~,~~S_2(t) = - y \cdot W(t) ~,~~S_3(t) = - z \cdot W(t) 
\end{array}
\enq

\vv \nin Les trois op\'erateurs $S_1(t)$, $S_2(t)$ et $S_3(t)$ sont les g\'en\'erateurs
infinit\'esimaux du petit groupe associ\'e \`a $t$. On v\'erifie ais\'ement qu'ils constituent une alg\`ebre de Lie
isomorphe \`a su(2)~\footnote{Ce sont alors des op\'erateurs de spin, pas n\'ecessairement hermitiques.}. Ce n'est pas \'etonnant, car, le 4-vecteur $\eta$ \'etant du genre temps,
les op\'erations du petit groupe sont des rotations dans l'hyperplan orthogonal \`a $\eta$. Ledit petit groupe est donc isomorphe \`a $SO(3)$\footnote{Dont le groupe de rev\^etement $SU(2)$ a \'et\'e \'etudi\'e dans ITL, chap. 4.}. Soit $[\,t\,]$ une t\'etrade associ\'ee \`a $t$. En param\'etrisant ce 4-vecteur de la fa\c{c}on suivante  

$$ t = \left(\cosh \chi, \sinh \chi \, \sin \theta\, \cos \varphi, \sinh \chi\, \sin \theta\, \sin \varphi, \sinh \chi\, \cos \theta \right) $$

\vv \nin cette t\'etrade pourra \^etre choisie comme\footnote{Le v\'erifier, et
v\'erifier aussi que le 4-vecteur $z$ du genre espace donn\'e par cette t\'etrade est \\
$z = (\sinh \chi, \,\cosh \chi \,\sin \theta \,\cos \varphi,\, \cosh \chi \,\sin \theta\,\sin \varphi,\,
\cosh \chi\,\cos \theta)$.} 

\beq [\, t\,] = \left( \begin{array}{cc}  e^{-i \varphi/2}& 0 \\
0 &   e^{ i \varphi/2} \end{array} \right)~
\left( \begin{array}{cc} ~ \cos \frac{\theta}{2}&
 - \sin \frac{\theta}{2}  \\
\sin \frac{\theta}{2} &  \cos \frac{\theta}{2}
\end{array} \right) ~\left( \begin{array}{cc}  e^{\chi/2}& 0 \\
0 &   e^{ - \chi/2} \end{array} \right) \label{tetrade-t} \enq

\vv \nin Toute autre t\'etrade $[\, t\,]^\prime$ que l'on pourrait associer au m\^eme 4-vecteur
$t$ diff\`ere de (\ref{tetrade-t}) par une matrice $[\,t\,]^{-1} ~[\, t\,]^\prime$ faisant partie du petit
groupe de $\basic{t}$\,, c'est-\`a-dire, par une matrice de rotation $R$, dont on peut exprimer la forme g\'en\'erale au moyen des angles d'Euler $(\alpha, \beta, \gamma)$ : 

\beq R(\alpha, \beta, \gamma) = \left( \begin{array}{cc} e^{-i \alpha/2} & 0 \\ 0 & e^{i \alpha/2} \end{array} \right) \, \left( \begin{array}{cc} \cos \frac{\beta}{2} & - \sin \frac{\beta}{2} \\ \sin \frac{\beta}{2} & \cos \frac{\beta}{2} \end{array} \right)\,\left( \begin{array}{cc} e^{-i \gamma/2} & 0 \\ 0 & e^{i \gamma/2} \end{array} \right)  \label{REuler} \enq

\vv \nin Nous avons vu qu'une transformation de Lorentz associ\'ee \`a une matrice $A$ de $SL(2,C)$ induit une rotation de Wigner $[\,A t\,]^{-1} A\,[\,t\,]$. Celle-ci est tout aussi bien d\'ecrite par la forme g\'en\'erale (\ref{REuler}).  

\vv \nin De l'expression  

\beq \left\{S_3(t)\right\}_{\alpha \beta}  = i\, \epsilon_{\alpha \beta \mu \nu}\, t^\mu\, z^\nu = i\, \left\{\, x_\alpha \, y_\beta - x_\beta\, y_\alpha \, \right\}\enq

\vv \nin et des formules relatives \`a la base ${\cal B}(t)$, on tire 

\beq S_3(t)[t] = 0,~~~S_3(t)[z] =0,~~~S_3(t)[x] = i\, y,~~~S_3(t)[y] = -i\, x   \enq

\vv \nin Dans l'hyperplan orthogonal \`a $t$, les 4-vecteurs 

\beq e^{(0)} = z,~~~e^{(\pm)} = \mp \di{1\over \sqrt{2}}\left( x \pm i\, y \right)  \label{E0PM} \enq

\vv \nin not\'es $e^{(\lambda)}$ avec $\lambda = 0, \pm 1$, sont vecteurs propres de $S_3(t)$ avec $\lambda$ pour valeur propre respective :

\beq [S_3(t)]_{\alpha \beta}\,e^{(\pm) \beta} = i\, \epsilon_{\alpha \beta \mu \nu}\,t^\mu\, z^\nu \,e^{(\pm) \beta} = \pm \,e^{(\pm)}_\alpha   \label{S3VPM} \enq

\vv \nin Notant que 

\beq g_{\beta \gamma} = t_\beta\, t_\gamma - z_\beta\, z_\gamma - e^{(+)}_\beta\,e^{(+) \star}_\gamma - e^{(-)}_\beta\,e^{(-) \star}_\gamma = t_\beta\, t_\gamma - z_\beta\, z_\gamma + e^{(+)}_\beta\,e^{(-)}_\gamma + e^{(-)}_\beta\,e^{(+)}_\gamma \enq

\vv \nin et calculant  

\beq [S_3(t)]_{\alpha \beta}\, g^{\beta \gamma} = e^{(+)}_\alpha\,e^{(-)\gamma} - e^{(-)}_\alpha\,e^{(+)\gamma} \enq

\vv \nin on d\'eduit les projecteurs

\beq 
e^{(\pm)}_\mu\, e^{(\mp)}_\nu = \di{1\over 2} \left[ \, g_{\mu \nu} - t_\mu\, t_\nu + z_\mu\, z_\nu \pm i\, \epsilon_{\mu \nu \rho \sigma}\,t^\rho\, z^\sigma \, \right] 
\label{projPM} \enq

\vv \nin Les op\'erateurs $S_{+} = \di{1\over \sqrt{2}} \left[ \, S_1 + i S_2 \, \right]$ et $S_{-} = \di{1\over \sqrt{2}} \left[ \, S_1 - i S_2 \, \right]$, qui font respectivement ``monter" et ``descendre" le spin, ont pour \'el\'ements de matrice non nuls les expressions  

\beq [S_{\pm}]_{\alpha \beta} = \mp\, i\, \epsilon_{\alpha \beta \mu \nu}\,t^\mu \, e^{(\pm) \nu} = z_\alpha\, e^{(\pm)}_\beta - z_\beta\, e^{(\pm)}_\alpha  \label{OPPM} \enq

\vv \nin Consid\'erons les tenseurs d\'efinis par : 

\beq \begin{array}{c} 
T_{0\,\mu \nu} = g_{\mu \nu} - t_\mu\, t_\nu + z_\mu\, z_\nu = -\, e^{(+)}_\mu \,e^{(+) \star}_\nu -\, e^{(-)}_\mu\,e^{(-) \star}_\nu  \\ [0.4cm]
T_{1\, \mu \nu} = -\, e^{(+)}_\mu \,e^{(-) \star}_\nu -\, e^{(-)}_\mu\,e^{(+) \star}_\nu  =  e^{(+)}_\mu \,e^{(+)}_\nu +\, e^{(-)}_\mu\,e^{(-)}_\nu   \\ [0.4cm]
T_{2\, \mu \nu} = i\,\left[\, e^{(+)}_\mu \,e^{(-) \star}_\nu -\, e^{(-)}_\mu\,e^{(+) \star}_\nu \, \right] = i\,\left[\, e^{(-)}_\mu \,e^{(-)}_\nu -\, e^{(+)}_\mu\,e^{(+)}_\nu \, \right]  \\ [0.4cm]
T_{3\, \mu \nu} = - \,e^{(+)}_\mu \,e^{(+) \star}_\nu + e^{(-)}_\mu\,e^{(-) \star}_\nu  
\equiv S_3(t) \end{array}
\label{4-pauli-1} \enq

\vv \nin Le tenseur $T_0$ est en fait le projecteur dans le 2-plan $(x,y)$. Quant aux tenseurs  $T_k$ ($k=1,2,3$), on peut les envisager comme des op\'erateurs agissant dans ce bi-plan, lequel, complexifi\'e, est clairement  isomorphe \`a un espace complexe \`a deux dimensions. Ils sont hermitiques car $T^\star_{k\, \mu \nu} = T_{k\, \nu \mu}$ et v\'erifient  

\beq T_k \, T_\ell = \delta_{k \ell}\, T_0 + i\, \epsilon_{k \ell m}\, T_m ,~~~[\, T_k, T_\ell\,] = 2i \,\epsilon_{k \ell m}\, T_m  \label{4-pauli-2} \enq  

\vv \nin Ils engendrent donc une alg\`ebre isomorphe \`a celle des matrices de Pauli, ce qui n'est pas \'etonnant, compte tenu de l'isomorphisme mentionn\'e plus haut. 

\vv \nin Pour \^etre complet, \'ecrivons \'egalement les relations suivantes :

\beq 
\begin{array}{c} 
\epsilon_{\mu \nu \rho \sigma}\, z^\mu\, e^{(\pm) \nu} = \pm\,i\, \left\{\,t_\rho\, e^{(\pm)}_\sigma - t_\sigma\, e^{(\pm)}_\rho \, \right\} \\ [0.4cm]
\epsilon_{\mu \nu \rho \sigma}\,(t + \epsilon\, z)^\mu\, e^{(\pm) \nu} = \pm\,i\,\epsilon\, \left\{\,(t+ \epsilon\, z)_\rho\, e^{(\pm)}_\sigma - (t+ \epsilon z)_\sigma\, e^{(\pm)}_\rho \, \right\} ~~~{\sf avec}~~~\epsilon = \pm 1
\end{array}
\enq

\vvv
\vv \nin \ding{193} {\bf Cas o\`u $\eta$ est du genre espace : $\eta^2 < 0$}

\vv \nin Nous poserons ici $M = \sqrt{- \eta^2}$, puis $z = \eta/M$. Le 4-vecteur $z$ est du genre
espace et unitaire. Nous lui associerons une triade de 4-vecteurs comprenant un 4-vecteur
du genre temps futur et unitaire $t$ et deux vecteurs du genre espace et unitaires $x$ et $y$
de sorte \`a former une base orthonorm\'ee d'espace-temps ${\cal B}(z) : t,\, x, \, y, \, z$. Le 4-vecteur
$\xi$, orthogonal \`a $z$, appartient \`a l'hyperplan engendr\'e par $t$, $x$ et $y$ :

$$ \xi = \xi^0 \, t + \xi^1 \, x + \xi^2 \, y $$

\nin et par suite

\vskip -0.25 cm
\begin{eqnarray} & G = i\, M \left( \xi^0\, W^0(z) - \xi^1\, W^1(z) - \xi^2\, W^2(z) \right) &
\nonumber \\
& {\rm avec}~~~W^0(z) = t \cdot W(z)~,~~W^1(t) = - x \cdot W(z) ~,~~W^2(z) = - y \cdot W(z) &
\end{eqnarray}

\nin Les op\'erateurs $W^0(z)$, $W^1(z)$ et $W^2(z)$, g\'en\'erateurs infinit\'esimaux du petit
groupe de $z$, satisfont aux relations de commutation\footnote{Montrer que
$W^0(z) = x_\mu y_\nu J^{\mu \nu}$, $W^1(z)
= t_\mu y_\nu J^{\mu \nu}$, $W^2(z) = - t_\mu x_\nu J^{\mu \nu}$ et se servir des relations
(\ref{commutJ}).}

\beq [\, W^1 , W^2 \,] ~=~-i\, W^0 ~~,~~[\,W^0\, , W^1 \,] ~=~i\,W^2~~,~~ [\,W^0\, , W^2 \,]~=~- i\,W^1 \enq

\vv \nin qui d\'efinissent l'alg\`ebre de Lie du groupe $L(2,1)$, groupe de Lorentz pour un espace-temps ne comportant que deux dimensions spatiales. Le petit groupe associ\'e \`a $z$ est donc isomorphe \`a $L(2,1)$. Ce r\'esultat \'etait pr\'evisible puisque les transformations
du petit groupe de $z$ agissent dans le 3-espace de Minkowsky ayant une dimension temporelle et deux dimensions spatiales. L'op\'erateur $W^0$ engendre des rotations dans le 2-plan
$(x, y)$, tandis que $W^1$ et $W^2$ engendrent des transformations de Lorentz, dans les 2-plans
$(t, y)$ et $(t, x)$, respectivement. On montre que la forme g\'en\'erale des matrices $M_s$ du petit groupe du 4-vecteur
de r\'ef\'erence (du genre espace) $\basic{z}~ = (0, 0, 0, 1)$ est 

\beq \begin{array}{c}
M_s = \left( \begin{array}{cc}  \cosh \frac{\xi}{2} \, e^{-i (\theta_1 + \theta_2)/2}&
- \sinh \frac{\xi}{2} \,
e^{-i( \theta_1 - \theta_2)/2)} \\
- \sinh \frac{\xi}{2} \,
e^{i( \theta_1 - \theta_2)/2)}&  ~~\cosh \frac{\xi}{2} \, e^{ i (\theta_1+\theta_2)/2}
\end{array} \right)  \label{Ms} ~~~{\rm ou~~encore} \\ [0.7cm]
M_s =  \left( \begin{array}{cc}  e^{-i \theta_1/2}& 0 \\
0 &   e^{ i \theta_1/2} \end{array} \right)~
\left( \begin{array}{cc} ~~ \cosh \frac{\xi}{2}&
~~ \sinh \frac{\xi}{2}  \\
\sinh \frac{\xi}{2} &  \cosh \frac{\xi}{2}~
\end{array} \right) ~\left( \begin{array}{cc}  e^{-i \theta_2/2}& 0 \\
0 &   e^{ i \theta_2/2} \end{array} \right) \end{array} \enq

\vv \nin $\xi$, $\theta_1$ et $\theta_2$ \'etant des r\'eels quelconque. Soit $[\, z\,]$ une t\'etrade associ\'ee au 4-vecteur du genre espace $z$. Si
ce dernier est param\'etris\'e comme

$$z = (\sinh \chi, ~\cosh \chi ~\sin \theta ~\cos \varphi, ~\cosh \chi ~\sin \theta~
\sin \varphi,~ \cosh \chi ~\cos \theta)$$

\vv \nin on peut encore choisir (\ref{tetrade-t}) pour cette t\'etrade, et toute autre t\'etrade $[\, z\,]^\prime$ que l'on pourrait associer au m\^eme 4-vecteur
$z$ en diff\`ere par une matrice $[\,z\,]^{-1} ~[\, z\,]^\prime$ faisant partie du petit
groupe de $\basic{z}$ et qui est donc de la forme (\ref{Ms}). De m\^eme, la transformation de Wigner
$[\, A z \,]^{-1}~A~[\, z\,]$, o\`u $A$ est une matrice quelconque de $SL(2,C)$, est une matrice de la forme (\ref{Ms}) et ne repr\'esente plus une rotation.

\vv

\vv \nin \ding{194} {\bf Cas o\`u $\eta$ est du genre lumi\`ere : $\eta^2 =0 $}

\vv \nin Il est toujours possible de trouver une base ${\cal B}(t)$ ayant un vecteur $z$
convenablement orient\'e, de telle sorte que $\eta$ prenne la forme
$\eta = \kappa (t+z)$, avec $\kappa = t \cdot \eta$. Relativement \`a cette base, les composantes
de l'op\'erateur de Pauli-Lubanski $W(t+z)$ sont\footnote{Comme $(t+z)\cdot W(t+z) = 0$, on a $W^0 = W^3$.}

\beq \begin{array}{c} 
 W^1 = - x \cdot W = - (t+z)_\mu \, y_\nu\, J^{\mu \nu}~~,~~
W^2 = - y \cdot W = (t+z)_\mu \, x_\nu\, J^{\mu \nu}   \\[0.5cm] 
 W^0 = t \cdot W = W^3 = x_\mu \, y_\nu\, J^{\mu \nu}  
\end{array}
\enq 

\nin Elles v\'erifient les relations de commutation

\beq [\, W^1 \, ,\, W^2 \, ] = 0~~,~~[ \, W^0 \, , \, W^1 \, ] = i \, W^2~~,~~
[ \, W^0 \, , \, W^2 \, ] = - i \, W^1 \enq

\vv \nin Comme mentionn\'e pr\'ec\'edemment, cette alg\`ebre est celle du groupe $P(2)$
des d\'eplacements dans un 2-plan euclidien, l'op\'erateur $W^1$ jouant le r\^ole de $-i \partial_x$,
g\'en\'erateur de translation suivant l'axe des $x$ de ce plan, $W^2$ celui de $-i \partial_y$,
g\'en\'erateur de translation suivant l'axe des $y$, et $W^0$ celui de
$-i ( x \partial_y - y \partial_x)$, g\'en\'erateur de rotations dans ledit plan. L'op\'erateur $W^0$ est encore le g\'en\'erateur infinit\'esimal des rotations
dans le 2-plan physique $(x,y)$. Compte tenu de (\ref{Jquadriv}) et des relations

\beq \begin{array}{c} 
\epsilon_{\mu \nu \rho \sigma}~x^\mu~(t+z)^\nu = (t+z)_\rho ~y_\sigma - y_\rho~(t+z)_\sigma \\ [0.5cm] 
\epsilon_{\mu \nu \rho \sigma}~y^\mu~(t+z)^\nu = x_\rho~(t+z)_\sigma - (t+z)_\rho ~x_\sigma 
\end{array}
\label{relb-5} \enq

\nin on a

\vskip -0.5 cm

\beq (W^1)_{\alpha \beta} = -i \left[(t+z)_\alpha ~y_\beta - y_\alpha~(t+z)_\beta \right]~~,~~
(W^2)_{\alpha \beta} = i \left[(t+z)_\alpha ~x_\beta - x_\alpha~(t+z)_\beta \right] \enq

\nin d'o\`u l'on d\'eduit l'action de l'op\'erateur $T =  e^{i(a\,W^1 + b\,W^2)}$ sur la base
${\cal B}(t)$ :

\begin{eqnarray}
T(t) = t + b x - a y + \di{{a^2 + b^2}\over 2}~(t+z)~~, &
T(z) = z - b x + a y - \di{{a^2 + b^2}\over 2}~(t+z)\nonumber \\
T(x) = x + b(t+z)~~~~~~~~,    &  T(y) = y - a (t+z) \end{eqnarray}

\vv \nin Les 4-vecteurs $u$ appartenant \`a l'hyperplan orthogonal \`a $\eta$ sont n\'ecessairement
de la forme $u = u^1 x + u^2 y + u^0 (t+z)$, ($u^0 = u^3$). Pour ceux-ci, on a

\beq T(u) = u + (t+z)( b u^1 - a u^2) \enq

\vv \nin et l'effet sur ces 4-vecteurs du sous-groupe ab\'elien engendr\'e par $W^1$ et $W^2$ est
de les translater parall\`element \`a $\eta$~\footnote{Et notamment $e^{(-)} = \di{{x - i y}\over{\sqrt{2}}}
\rightarrow e^{(-)} + \di{{b+i a}\over \sqrt{2}}~(t+z)$.}. C'est pourquoi on l'appelle
{\em groupe de jauge} de $\eta$.

\vv \nin Les matrices $M_\ell$ de $SL(2,C)$ appartenant
au petit groupe du 4-vecteur du genre lumi\`ere de r\'ef\'erence $\basic{\eta}~ = \,(1,0,0,1)$ ont pour forme g\'en\'erale
 
\beq  M_\ell = \left( \begin{array}{cc} e^{- i \psi/2} & \beta \\ 0 & e^{ i \psi/2}
\end{array} \right) = \left( \begin{array}{cc} 1 & \zeta \\ 0 & 1
\end{array} \right)\left( \begin{array}{cc} e^{- i \psi/2} & 0 \\ 0 & e^{ i \psi/2}
\end{array} \right) =  M_\ell (\zeta,  e^{ i \psi/2}) \label{ML}\enq
$$  = \left( \begin{array}{cc} e^{- i \psi/2} & 0 \\ 0 & e^{ i \psi/2}
\end{array} \right)~\left( \begin{array}{cc} 1 & \zeta^\prime \\ 0 & 1
\end{array} \right),~~ {\rm avec} ~~\zeta^\prime = \zeta~e^{- i \psi}$$

\vv \nin o\`u $\zeta = \beta e^{- i \psi/2}$ est un nombre complexe quelconque. Elles ont pour loi de composition 

\beq  M_\ell (\zeta,  e^{ i \psi/2})~ M_\ell (\zeta^\prime,  e^{ i \psi^\prime/2}) =
M_\ell (\zeta + \zeta^\prime e^{ i \psi},
e^{ i [\psi+ \psi^\prime]/2}) \label{compo-Ml} \enq

\vv \nin qui est bien celle des d\'eplacements dans le plan, assimil\'e au plan complexe (addition
du complexe $\zeta$, repr\'esentant une translation, suivie d'une multiplication par $e^{ i \alpha}$,
repr\'esentant une rotation). Notons $\basic{\ell \,}~ = \kappa \, (1,0,0,1)$ ($\kappa$ \'etant r\'eel) un 4-vecteur de
r\'ef\'erence du genre lumi\`ere,
et $[\, \ell\,]$ une t\'etrade permettant de passer de  $\basic{\ell \,}$
\`a un autre 4-vecteur du genre lumi\`ere $\ell$. Si l'on utilise la param\'etrisation

$$ \ell = \kappa (1,~ \sin \theta~\cos \varphi,~ \sin \theta~\sin \varphi,~ \cos \theta)$$

\vv \nin un exemple de telle t\'etrade est donn\'e par

\beq [\, \ell\,] = \sqrt{\kappa} ~\left( \begin{array}{cc}  e^{-i \varphi/2}& 0 \\
0 &   e^{ i \varphi/2} \end{array} \right)~
\left( \begin{array}{cc} ~ \cos \frac{\theta}{2}&
 - \sin \frac{\theta}{2}  \\
\sin \frac{\theta}{2} &  \cos \frac{\theta}{2}
\end{array} \right) \label{tetl}  \enq

\vv \nin conduisant \`a la base d'espace-temps associ\'ee :

$$ t(\ell) = (1,0,0,0)~,~~
x(\ell) = (0, \cos \theta~\cos \varphi, \cos \theta ~\sin \varphi, - \sin \theta)$$
\vskip -0.5 cm
$$ y(\ell)=( 0, - \sin \varphi, \cos \varphi, 0)~,~~z(\ell)= ( 0,
\sin \theta~\cos \varphi, \sin \theta~\sin \varphi, \cos \theta)$$

\vv \nin Toute autre t\'etrade $[\, \ell \,]^\prime$ associ\'ee au m\^eme 4-vecteur $\ell$ diff\`ere
de (\ref{tetl}) d'une matrice $[\, \ell \,]^{-1} [\, \ell \,]^\prime$ qui, appartenant au petit
groupe de
$\basic{\ell~}$, est n\'ecessairement de la forme (\ref{ML}). Ici aussi, une transformation de Wigner
$[\, A \ell\,]^{-1}~A~[\, \ell\,]$ n'est pas une rotation, mais une matrice $M_\ell$ du petit groupe de
$\basic{\ell~}$, donc de la forme (\ref{ML}). Sous l'action de cette op\'eration, les 4-vecteurs
$e^{(\pm)}([\,\basic{\ell~} ]) = \mp (0, 1, \pm i,0)$ deviennent

\beq E^{(\lambda)}([\,\basic{\ell~} ]) = e^{ i \lambda \psi}~e^{(\lambda)}([\,\basic{\ell~} ])
+ c_\lambda ~\basic{\ell~}~~~~(\lambda = \pm 1) \enq

\vv \nin avec $c_{+} = \frac{\zeta^\star}{\sqrt{2}\kappa}$ et $
c_{-} = -\frac{\zeta}{\sqrt{2}\kappa}$. Il s'ensuit que dans la transformation de
Lorentz repr\'esent\'ee par $A$, le 4-vecteur

\beq e^{(\lambda)} ([\, \ell \,]) = \Lambda([\, \ell \,]) ~e^{(\lambda)}([\,\basic{\ell~} ]) \enq

\vv \nin qui est vecteur propre de l'op\'erateur $W^0$ avec la valeur propre $\lambda$ a pour \'equivalent dans la t\'etrade $[\,A\ell\,]$ 

\beq \begin{array}{c}
e^{(\lambda)}([\,A\ell\,]) = \Lambda([\,A\ell\,])\,e^{(\lambda)}([\,\basic{\ell~} ]) = \Lambda(A\,[\,\ell\,]\,M^{-1}) \,e^{(\lambda)}([\,\basic{\ell~} ])  \\ [0.3cm] 
= \Lambda(A\,[\,\ell\,]) \,\left\{ e^{ i \lambda \psi}~e^{(\lambda)}([\,\basic{\ell~} ])
+ c_\lambda ~\basic{\ell~} \right\} 
 = e^{i \lambda \psi}\,\Lambda(A) ~e^{(\lambda)}([\,\ell\,])  + c_\lambda\,(A\ell) 
\end{array} \label{jauge1} \enq 

\vv \nin Envisageons le cas d'un photon dont l'\'etat est d\'ecrit par une onde plane. La direction de propagation de l'onde et l'\'energie qu'elle transporte sont d\'efinies par le 4-vecteur
\'energie-quantit\'e de mouvement $\ell$ du photon, du genre lumi\`ere. 
L'\'etat de polarisation de l'onde se d\'ecrit quant \`a lui au moyen de deux 4-vecteurs $e^{(+)}$ et
$e^{(-)}$, tels que $e^{(\pm)} \cdot \ell = 0$. Ces 4-vecteurs, dits {\em de polarisation}, repr\'esentent des ondes polaris\'ees
circulairement, a droite et \`a gauche, respectivement\footnote{ITL, \S 4.7.1.}. D'apr\`es (\ref{jauge1}), dans une transformation de Lorentz, ils subissent une translation parall\`element \`a $L = \Lambda(A)(\ell)$. Or, une propri\'et\'e fondamentale des \'equations de Maxwell est qu'elles sont {\em invariantes de jauge}. Au final, cela signifie que les pr\'evisions mesurables de la th\'eorie sont insensibles au remplacement d'un vecteur de polarisation $e^{(\lambda)} (\ell)$ par $e^{(\lambda)} (\ell) + c(\ell)~\ell$, o\`u $c(\ell)$ est un scalaire quelconque. Il s'ensuit que, dans ce cas, le terme $c_\lambda ~L$ est sans effet et peut \^etre tout simplement ignor\'e. Tout se passe alors comme si, dans une transformation de Lorentz, les vecteurs de polarisation subissaient un simple changement de phase $e^{ i \lambda \psi}$. La cons\'equence importante est que l'h\'elicit\'e $\lambda$ du photon peut \^etre consid\'er\'ee comme un v\'eritable invariant relativiste.

\section{Les spineurs de Dirac \protect \footnote{Le lecteur trouvera dans le chapitre 7 de ITL qui leur est consacr\'e toutes les d\'efinitions utiles concernant les spineurs et les matrices de Dirac.}} 

\subsection{Expressions g\'en\'erales}

\vv \nin Les {\em spineurs de Dirac} sont les {\em bi-spineurs} de la repr\'esentation 
$D(\frac{1}{2},0) \oplus D(0,\frac{1}{2})$ de $SL(2,C)$. Dans la suite, nous ne consid\`ererons que le cas o\`u la particule consid\'er\'ee, de spin 1/2, a une masse non nulle. Le cas \'eventuel de la masse nulle sera envisag\'e comme limite du pr\'ec\'edent en faisant tendre $m$ vers z\'ero, lorsque cela est possible\footnote{Le cas des spineurs de Dirac associ\'es \`a une particule de masse nulle est consid\'er\'e dans ITL, \S\,7.6.}. En ``repr\'esentation-$p$", et dans la repr\'esentation que nous appelons ``repr\'esentation initiale", un spineur de Dirac associ\'e \`a un \'etat $|\, \Phi>$ s'exprime comme :  

\beq \Phi(p) = \left( \begin{array}{c} \phi(p) \\ \hat{\phi}(p) \end{array} \right)  \label{spineurp-initial} \enq

\vv \nin au moyen des amplitudes spinorielles 

\beq \phi_\sigma(p) =~ < p, \sigma \, |\, \Phi >~~~{\rm et}~~~~\hat{\phi}_\sigma (p) = ~< \hat{p}, \sigma \, |\, \Phi > \enq

\vv \nin o\`u l'indice de spin $\sigma$ prend les valeurs $+1/2$ et $-1/2$. Il v\'erifie l'\'equation de Dirac : 

\beq \Gi(p)\, \Phi(p) = m\,\Phi(p)  \label{dirac-1} \enq

\vv \nin o\`u $\Gi(p)$ est la matrice $4\times4$ donn\'ee par\footnote{Rappelons que ${\cal D}^{s}(A) \equiv A$ pour $s=1/2$.}$^{\rm ,}${\hskip -0.05 cm ~}\footnote{$p_0 = \sqrt{m^2 + \Vec{\,p\,}^2}$}

\beq  \begin{array}{c} 
\Gi(p) = \left( \begin{array}{cc} 0_2 &  \,p \hskip -0.48 cm \begin{array}[m]{c} ~ \\ \stackrel{~~~}{\sim}
\end{array} \hskip -0.14 cm \\~&~\\ \widetilde{p} & 0_2 \end{array} \right) ~~~~~{\sf avec} \\ [1cm]
 \,p \hskip -0.48 cm \begin{array}[m]{c} ~ \\ \stackrel{~~~}{\sim}
\end{array} \hskip -0.14 cm =\, p_0 ~+ \Vec{\,p\,} \hskip-0.1cm\cdot \hskip-0.1cm \Vec{\tau\,},~~~~\widetilde{p} ~= \, p_0 ~- \Vec{\,p\,} \hskip-0.1cm\cdot \hskip-0.1cm \Vec{\tau\,} 
\end{array}
\label{matriceGp} \enq

\vv \nin et o\`u $0_2$ est la matrice nulle $2\times2$. Sous une transformation $(a, A)$ de $\bar{\cal P}^\uparrow_{+}$, le spineur (\ref{spineurp-initial}) devient 

\beq ~^{(a, A)} \Phi (p)  ~= ~\Phi^{\prime}(p)~= ~e^{i a \cdot p}\,S(A)\, \Phi(A^{-1} p)  ,~~~{\rm avec}~~~S(A) =  \left( \begin{array}{cc} A & 0_2 \\~&~\\  0_2 & A^{\dagger -1} \end{array} \right)  \label{trans-spi} \enq

\vv \nin tandis que la matrice $\Gi(p)$ v\'erifie 

\beq \Gi(p) = S(A)\,  \Gi(A^{-1}p) \,S(A)^{-1} \label{cov-gam} \enq

\vv \nin ce qui fait que l'\'equation de Dirac (\ref{dirac-1}) garde exactement la m\^eme forme dans tout r\'ef\'erentiel galil\'een. Si l'on pose 

\beq \Gi^0 = \Gi_0 =  \left( \begin{array}{cc} 0_2 & \tau_0 \\~&~\\  \tau_0 & 0_2 \end{array} \right),~~~\Gi^k = -\Gi_k = \left( \begin{array}{cc} 0_2 & -\tau_k \\~&~\\  \tau_k & 0_2 \end{array} \right) \label{mat-dirac-i} \enq

\vv \nin o\`u $\tau_0$ est la matrice unit\'e $2\times 2$ et $\tau_k$ ($k=1,2,3$) sont les matrices de Pauli, la matrice (\ref{matriceGp}) peut \^etre exprim\'ee sous forme de produit scalaire :

$$ \Gi(p) = p_\mu\, \Gi^\mu = p^\mu\, \Gi_\mu = p_0 \, \Gi^0 - \di{\sum_k} \, p^k \, \Gi^k $$
  
\nin En posant $q = A^{-1} p $ (ou en rempla\c{c}ant $A$ par $A^{-1}$), l'\'equation (\ref{cov-gam}) devient  

\beq S(A) \, \Gi(q) \, S(A)^{-1} = \Gi(Aq) \enq

\vv \nin d'o\`u l'on d\'eduit ($\left[ A q \,\right]^\nu = \Lambda^\nu_{\, \cdot \, \mu}\, q^\mu$,  $\left[A q\,\right]_\nu = \left[\Lambda^{-1}\right]^\mu_{\, \cdot \, \nu}\, q_\mu$ avec $\Lambda \equiv \Lambda(A)$) :

\beq \fbox{\fbox{\rule[-0.4cm]{0cm}{1cm}~$S(A)\, \Gi_\mu\, S(A)^{-1} = \Lambda^\nu_{\, \cdot \, \mu}\, \Gi_\nu ~~~{\rm et}~~~
 S(A)\, \Gi^\mu\, S(A)^{-1} = \left[\Lambda^{-1}\right]^\mu_{\, \cdot \, \nu}\, \Gi^\nu $~}}         \label{transgamma} \enq

\vv \nin Sous la transformation d\'efinie par $\Gi^{\prime \mu} = S(A)^{-1}\, \Gi^\mu\, S(A)$, les matrices $\Gi^\mu$ se transforment donc comme les composantes contravariantes d'un 4-vecteur. Les matrices (\ref{mat-dirac-i}) v\'erifient les relations\footnote{ITL, section 6.12.} 

$$ (\Gi^0)^2 = \Gi^2_0 = 1 ,~~(\Gi^k)^2 = \Gi^2_k =- 1 $$ 
\beq \Gi^0\, \Gi^k = - \Gi^k\, \Gi^0,~~~\Gi^k\, \Gi^\ell = - \Gi^\ell \, \Gi^k~~(k \neq \ell) \enq

\vv \nin que l'on r\'esume par la relation fondamentale de d\'efinition des matrices de Dirac : 

\beq \fbox{\fbox{\rule[-0.4cm]{0cm}{1cm}~$\Gi^\mu \, \Gi^\nu + \Gi^\nu \, \Gi^\mu = 2\, g^{\mu \nu},~~{\rm ou}~~~\Gi_\mu \, \Gi_\nu + \Gi_\nu \, \Gi_\mu = 2\, g_{\mu \nu} $~}}   \label{def-matd} \enq

\vv \nin Les matrices (\ref{mat-dirac-i}) constituent ce que nous appelons la {\em repr\'esentation initiale} des matrices de Dirac, qui diff\`ere de celle de Weyl par le signe des $\Gi^k$. S'il existe bien une infinit\'e de quadruplets de matrices $4 \times 4$ v\'erifiant  la relation g\'en\'erale (\ref{def-matd}), on montre cependant que deux quadruplets possibles $\{\gamma^\mu\}$ et $\{\gamma^{\prime \mu}\}$ sont n\'ecessairement reli\'es au moyen d'une matrice $4\times 4$ inversible $U$ via la relation\footnote{C'est le {\em th\'eor\`eme fondamental} de W. Pauli, voir {\em Annales de l'I.H.P.} tome 6, $n^\circ$ 2 (1936), p. 109.} :  

\beq \fbox{\rule[-0.4cm]{0cm}{1cm}~$\gamma^{\prime \mu} = U\, \gamma^\mu \, U^{-1} $~} \enq

\vv \nin Toutes les repr\'esentations des matrices de Dirac sont {\em \'equivalentes} et toute propri\'et\'e de ces matrices d\'emontr\'ee dans une repr\'esentation particuli\`ere est applicable \`a toute autre repr\'esentation. La {\em repr\'esentation de Dirac} est la {\em repr\'esentation standard} couramment utilis\'ee des matrices de Dirac. Elle est d\'efinie par   

$$ \gamma^0 = \gamma_0 =  \left( \begin{array}{cc} \tau_0 & 0_2 \\~&~\\  0_2 & -\tau_0 \end{array} \right) = \Gi_5\,,~~\gamma^k = -\gamma_k = \left( \begin{array}{cc} 0_2 & \tau_k \\~&~\\  -\tau_k & 0_2 \end{array} \right) = - \Gi^k $$
\beq \gamma_5 = \left( \begin{array}{cc} 0_2 & \tau_0 \\~&~\\  \tau_0 & 0_2 \end{array} \right) = \Gi^0 \label{mat-dirac-s} \enq

\vv \nin et s'obtient \`a partir de la repr\'esentation initiale (\ref{mat-dirac-i}) par la matrice inversible

\beq {\cal U} = \di{1 \over \sqrt{2}} \,  \left( \begin{array}{cc} \tau_0 & \tau_0 \\~&~\\  \tau_0 & - \tau_0 \end{array} \right)  = \di{1\over \sqrt{2}}\,\left( \Gi_0 + \Gi_5 \right)  = \di{1\over \sqrt{2}}\,\left( \gamma_0+ \gamma_5 \right) = {\cal U}^{-1} \label{matU} \enq

\vv \nin Notons ici qu'en vertu de l'\'equation (6.263), pour toute base d'espace-temps $t, x, y, z$, on a   

\beq \Gi_5 = - i \,\Gi(t)\, \Gi(x)\, \Gi(y)\, \Gi(z) \enq

\vv \nin et que par cons\'equent 

\beq \fbox{\fbox{\rule[-0.4cm]{0cm}{1cm}~$\gamma_5 =  {\cal U} \, \Gi_5 \, {\cal U}^{-1} = - i\, \gamma(t)\, \gamma(x)\, \gamma(y)\, \gamma(z) $~}}  \label{fonda5} \enq
 
\vv \nin o\`u, pour un 4-vecteur $v$ quelconque\footnote{Bien qu'en Physique des Particules on utilise couramment la notation ``slash" de Feynman, $\slashed{v} = v^\mu\, \gs_\mu$, nous utiliserons ici la notation $\gs(v)$, pour la clart\'e du texte.}, 

\beq \gamma(v) = v^\mu\, \gamma_\mu = v_\mu\, \gamma^\mu \enq

\vv \nin On a notamment, pour la base d'espace-temps standard,  

\beq \Gi_5 = i\, \Gi^0\, \Gi^1\, \Gi^2\, \Gi^3~~~~{\rm et}~~~~\gamma_5 = i\, \gamma^0\, \gamma^1\, \gamma^2\, \gamma^3 \label{lesg5} \enq

\vv \nin D'apr\`es (6.264), les matrices $\Gi_5$ et $\gamma_5$ anti-commutent, respectivement, avec les matrices $\Gi^\mu$ et $\gamma^\mu$ :

\beq \Gi_5\, \Gi^\mu = - \Gi^\mu\, \Gi_5,~~~~\gamma_5\, \gamma^\mu = - \gamma^\mu\, \gamma_5 \enq

\vv \nin D'apr\`es (\ref{def-matd}), les matrices $\Gi^\mu$ avec des indices diff\'erents anti-commutent entre elles, ce qui fait que le produit $\Gi^\mu \, \Gi^\nu \, \Gi^\rho \, \Gi^\sigma$ avec les quatre indices $\mu, \nu, \rho, \sigma$ tous diff\'erents est compl\`etement antisym\'etrique suivant ces indices. En introduisant le tenseur compl\`etement antisym\'etrique de Levi-Civita $\epsilon_{\mu \nu \rho \sigma}$, on peut donc \'ecrire 
($\epsilon_{0123} =1$)

$$ \Gi^0 \, \Gi^1 \, \Gi^2\, \Gi^3 = \epsilon_{0123}\, \Gi^0 \, \Gi^1 \, \Gi^2\, \Gi^3 = \di{1\over {4!}} \,\epsilon_{\mu \nu \rho \sigma} \,\Gi^\mu\, \Gi^\nu \, \Gi^\rho \, \Gi^\sigma   $$

\vv \nin Ceci permet de repr\'esenter les matrices $\Gi_5$ et $\gamma_5$ sous la forme de produits contract\'es de tenseurs   

\beq \fbox{\fbox{\rule[-0.4cm]{0cm}{1cm}~$\Gi_5 = \di{i \over{4!}} \, \epsilon_{\mu \nu \rho \sigma} \,\Gi^\mu \, \Gi^\nu \, \Gi^\rho\, \Gi^\sigma~~~{\rm et}~~~\gamma_5 =  \di{i \over{4!}} \, \epsilon_{\mu \nu \rho \sigma} \,\gamma^\mu \, \gamma^\nu \, \gamma^\rho \, \gamma^\sigma $~}}  \label{gs5} \enq

\vv \nin qui se comportent comme des scalaires sous les transformations de $SL(2,C)$. En utilisant (\ref{transgamma}) et en tenant compte de $\det \Lambda =1$, on a en effet  

$$ S(A)^{-1}\, \Gi_5\, S(A) = \di{i \over{4!}}\, \epsilon_{\mu \nu \rho \sigma} \, \Lambda^\mu_{\, \cdot \,\mu'} \,\Lambda^\nu_{\, \cdot \, \nu'} \,\Lambda^\rho_{\, \cdot \,\rho'} \,\Lambda^\sigma_{\, \cdot \,\sigma'} \, \Gi^{\mu'}\, \Gi^{\nu'}\, \Gi^{\rho'}\, \Gi^{\sigma'}$$ 
$$ = \di{i \over{4!}}\, (\det \Lambda) \, \epsilon_{\mu' \nu' \rho' \sigma'} \, \Gi^{\mu'}\, \Gi^{\nu'}\, \Gi^{\rho'}\, \Gi^{\sigma'} = \di{i \over{4!}}\, \epsilon_{\mu' \nu' \rho' \sigma'} \, \Gi^{\mu'}\, \Gi^{\nu'}\, \Gi^{\rho'}\, \Gi^{\sigma'} = \Gi_5$$

\vv \nin Dans la repr\'esentation standard, les spineurs de Dirac de la repr\'esentation-$p$ prennent donc la forme

\beq \Phi_{\rm st}(p) =\left( \begin{array}{c} \phi(p) + \hat{\phi}(p)\\ \phi(p) -\hat{\phi}(p) \end{array} \right) \enq 

\vv \nin \`a un facteur de normalisation pr\`es et $S(A)$ est remplac\'e par   

\beq  L(A) = {\cal U}\, S(A)\, {\cal U}^{-1} = \di{1\over 2}\,\left( \begin{array}{cc} A + A^{\dagger -1} & A - A^{\dagger -1}  \\~&~\\ A - A^{\dagger -1}   & A+ A^{\dagger -1} \end{array} \right)   \label{LA} \enq

\vv \nin Dans la repr\'esentation standard, les spineurs de type U (\'energie positive) et les spineurs de type V (\'energie n\'egative) attach\'es \`a l'\'etat $|\, [\,p\,], \sigma >$ sont donn\'es par\footnote{En mettant de c\^ot\'e un facteur de normalisation $(2 \pi)^3\,2\,p_0\,\delta\left(\Vec{\,p\,} - \Vec{p'}\right)$ o\`u $p_0 = \sqrt{m^2 + \Vec{\,p\,}^2}$.} 

\vv
\beq \fbox{\rule[-1.65cm]{0cm}{3.5cm}~$
\begin{array}{c}
U_\sigma ([\,p\,]) ~=~ \sqrt{\di{m\over 2}}\,\left( \begin{array}{c} \left\{[\,p\,]  + [\,p\,]^{\dagger -1}\right\}_{\cdot \, \sigma}  \\~\\ \left\{[\,p\,]  -[\,p\,]^{\dagger -1}\right\}_{ \cdot \,\sigma} \end{array} \right) \\[0.4cm]~\\
V_\sigma ([\,p\,])~=~ \gs_5\, U_\sigma([\,p\,])~=~\sqrt{\di{m\over 2}}\,\left( \begin{array}{c} \left\{[\,p\,] - [\,p\,]^{\dagger -1}\right\}_{\cdot \, \sigma}  \\~\\ \,\left\{[\,p\,] + [\,p\,]^{\dagger -1}\right\}_{ \cdot \,\sigma} \end{array} 
 \right)  
\end{array}
$~} \label{spinorUVst}  \enq

\vv \nin Ils v\'erifient les relations 

$$ \di{\sum_\sigma}\, U_\sigma \, \ov{U}_\sigma ~=~  m + \gs(p) \, , ~~
\di{\sum_\sigma}\, V_\sigma \, \ov{V}_\sigma ~=~ \gs(p) - m\, , ~~\gs(p)\, U = m\, U\, , ~~\gs(p)\, V = - m\, V $$
\beq \di{\sum_\sigma}~\left[\, U_\sigma \, \ov{U}_\sigma ~-~V_\sigma \, \ov{V}_\sigma \,\right]~=~ 2\, m,~~~~{\sf avec}~~~~\ov{U} = U^\dagger\,\gs_0,~~\ov{V} = V^\dagger\,\gs_0  \label{relspinors}\enq 

\nin et, conform\'ement \`a l'usage, leur normalisation est telle que 

\beq \ov{U}_\sigma \, U_{\sigma'} = 2\, m\, \delta_{\sigma \, \sigma'}\, , ~~~\ov{V}_\sigma \, V_{\sigma'} = - 2\, m\, \delta_{\sigma \, \sigma'}\, , ~~~\ov{U}_{\sigma} \, V_{\sigma'} = \ov{V}_{\sigma}\, U_{\sigma'} =0   \label{normalis} \enq

\vv \nin Les quatre spineurs $U$ et $V$ ainsi d\'efinis constituent une base selon laquelle on peut d\'evelopper tout vecteur unicolonne de $C^4$.

\vv \nin $\bullet$ Lorsque la t\'etrade $[\,p\,]$ correspond \`a un boost le long de $\Vec{\,p\,}$, elle prend la forme de la matrice hermitique ${\cal H}$ donn\'ee par le tableau : 

\beq {\cal H}  =\di{1\over{ \sqrt{ 2 m( E+m)}}}~\times~ \begin{array}{|l|c|c|}  \hline  & &\\ & 1/2 & - 1/2 \\& & \\ \hline & & \\
~~1/2~ & E + m + p_z & p_x - i\, p_y\\ & &\\ \hline & & \\
-1/2~& p_x + i\, p_y& E + m - p_z  \\& & \\ \hline
\end{array} \enq

\vv \nin avec $p_x = p^1, p_y = p^2, p_z = p^3$, $E = p_0$. On obtient alors

\beq \fbox{\rule[-0.8cm]{0cm}{1.8cm}~$\uup ({\cal H}) ~=~ \di{1\over{\sqrt{E+m}}}\,\left( \begin{array}{c} E+m  \\ 0 \\ p_z  \\ p_x + i p_y  \end{array} \right),~~~~\ud ({\cal H}) ~=~ \di{1\over{\sqrt{E+m}}}\,\left( \begin{array}{c} 0 \\ E+m \\p_x - i p_y\\ -p_z \end{array} \right) $~} \label{Uboost} \enq

\beq \fbox{\rule[-0.8cm]{0cm}{1.8cm}~$\vup ({\cal H}) ~=~ \di{1\over{\sqrt{E+m}}}\,\left( \begin{array}{c} p_z \\ p_x + i p_y \\ E+m \\ 0 \end{array} \right),~~~~\vd ({\cal H}) ~=~ \di{1\over{\sqrt{E+m}}}\,\left( \begin{array}{c} p_x - i p_y \\ -p_z \\ 0\\ E+m  \end{array} \right) $~} \label{Vboost} \enq

\vv \nin o\`u les notations $\uw$ (``up") et $\dw$ (``down") se r\'ef\`erent \`a un indice de spin $\sigma$ \'egal \`a $+ 1/2$ et $-1/2$, respectivement. 

  
\vvv
\subsection{Op\'erateurs de spin et projecteurs} \label{projo}

\vv \nin Dans la repr\'esentation standard des spineurs de Dirac, les g\'en\'erateurs de $SL(2,C)$ sont les matrices $\sigma_{\mu \nu} = \di{i\over 4} \left[ \gs_\mu, \gs_\nu \right]$. Posant $t = p/m$,  l'op\'erateur de Pauli-Lubanski associ\'e \`a la 4-impulsion $p$ s'\'ecrit donc 

$$ W_\mu = \di{1 \over 2}\, \epsilon_{\mu \alpha \beta \gamma}\, t^\alpha \, \sigma^{\beta \gamma} =  i \,t^\nu \, \sigma_{\mu \nu}\, \gs_5~~~{\rm soit}  $$
\beq W_\mu = -\di{1 \over 4} \, [\, \gs_\mu\, , \gs_\nu\, ]\, t^\nu\, \gs_5 \label{polarW} \enq

\vv \nin Notons $x$, $y$ et $z$ les trois 4-vecteurs du genre espace associ\'e \`a $t$ dans la t\'etrade $[\,p\,]$ et formant avec $t$ une base d'espace-temps orthonorm\'ee et d'orientation directe. Les g\'en\'erateurs du petit groupe de $p$ (op\'erateurs de spin engendrant un groupe de rotations) sont repr\'esent\'es par les matrices\footnote{A l'aide de (\ref{fonda5}), montrer  que les matrices $S_x$, $S_y$ et $S_z$ v\'erifient bien les relations de commutation de l'alg\`ebre de Lie de $SU(2)$.} 

$$ S_x = - x \cdot W =  \di{1 \over 2}\, \gs(x)\, \gs(t)\, \gs_5,~~S_y = - y \cdot W =  \di{1 \over 2}\, \gs(y)\, \gs(t)\, \gs_5,$$
\beq S_z = - z \cdot W =  \di{1 \over 2}\, \gs(z)\, \gs(t)\, \gs_5 \label{composantesSi} \enq

\vv \nin et, posant $U_\sigma \equiv U_\sigma([\,p\,])$, $V_\sigma \equiv V_\sigma([\,p\,])$,  $S_{\pm} = S_x \pm i S_y$, on a

$$ S_z \, U_\sigma = \sigma\, U_\sigma,~~S_z\, V_\sigma = \sigma\, V_\sigma$$ 
\beq S_+\,\uup = 0\, , ~~S_+ \, \ud = \uup\, , ~~S_+ \, \vup = \vd\, ~~S_+ \, \vd = 0 \label{relaspin} \enq
$$S_{-}\,\uup = \ud\, , ~~S_{-} \, \ud = 0 \, , ~~S_{-}\, \vup = \vd\, ~~S_{-}\, \vd = 0$$

\vv \nin Ces formules conduisent aussi aux suivantes :  

$$ \gs(z)\, U^{\uw, \dw} = \pm\, V^{\uw, \dw},~~ \gs(z)\, V^{\uw, \dw} = \mp\, U^{\uw, \dw} $$
\beq  \gs(x + i y)\, \uup =0, ~~\gs(x+i y)\, \ud = 2\, \vup   \label{gugv} \enq
$$ \gs(x-iy)\, \ud = 0 ,~~\gs(x-i y)\, \uup = 2\, \vd     $$

\vv \nin En appliquant $\gs(z)$ ou $\gs(x \pm i y)$ soit sur les projecteurs soit sur la relation de fermeture de (\ref{relspinors}), on en d\'eduit alors : 

$$ 2\,m\, \gs(z) = \vup\, \buup - \vd\, \bud +\uup\, \bvup - \ud\, \bvd $$
$$ 2\, \gs(z) \, \gs(p)= \vup\, \buup - \vd\, \bud - \uup\, \bvup + \ud\, \bvd $$
\beq  m\, \gs(x+i y) = \vup\, \bud + \uup\, \bvd,~~2\,\gs(x+i y)\, \gs(p) = \vup\, \bud - \uup\, \bvd \enq 
$$ m\, \gs(x-i y) = \vd\, \buup + \ud\, \bvup,~~2\,\gs(x-i y)\, \gs(p) = \vd\, \buup - \ud\, \bvup $$

\vv \nin Montrons maintenant comment l'utilisation conjointe des relations (\ref{relaspin}) et (\ref{relspinors}) permet d'exprimer simplement certains projecteurs au moyen des matrices $\gs$. En additionnant 

$$ S_{+} \, \uup\, \buup = 0\, , ~~{\rm et}~~~S_{+}\, \ud\, \bud = \uup\, \bud\,,~~{\rm on~~obtient} $$

$$ \uup\, \bud = S_{+} \, ( \uup\, \buup +  \ud\, \bud ) = S_{+} \left[\, \gs(p)+m\, \right] $$

\vv \nin soit, puisque $\gs(t) \left[ \,\gs(p) + m \,\right] =  \left[\, \gs(p) + m \, \right] $, 

\beq \fbox{\fbox{\rule[-0.4cm]{0cm}{1cm}~$\uup\, \bud = \di{1\over 2}\, \gs_5\, \gs(x+iy)\, \left[\, \gs(p) + m \,\right]  $~}} \label{updo} \enq

\vv \nin Suivant un proc\'ed\'e similaire, ou bien en op\'erant la conjugaison hermitique de  (\ref{updo}), on obtient  

\beq \fbox{\fbox{\rule[-0.4cm]{0cm}{1cm}~$\ud \, \buup = \di{1\over 2}\, \gs_5 \,\gs(x-iy)\, \left[\, \gs(p) + m \,\right]  $~}} \label{doup} \enq

\vv \nin Comme 

$$ 2\, S_z\, \left[\, \uup\, \buup + \ud\, \bud \, \right] =  \uup\, \buup - \ud\, \bud \, , ~{\rm et}~~~ \uup\, \buup + \ud\, \bud = \gs(p) + m $$

\vv \nin on a 

$$ 2\, U^{\uw , \dw} \, \ov{U}^{\uw , \dw} = \left[\, 1 \pm 2\, S_z\, \right]\, \left[\, \gs(p) + m \, \right] ~~~{\rm soit~~au~~final}$$

\beq \fbox{\fbox{\rule[-0.4cm]{0cm}{1cm}~$U^{\uw , \dw} \, \ov{U}^{\uw , \dw} =\di{1\over 2}\, \left[\, 1 \pm \gs_5 \, \gs(z)\, \right]\, \left[\, \gs(p) + m \, \right] $~}}  \label{uudd} \enq


\subsection{Spineurs propres de $S_z = \Vec{\, \sigma\,}\hskip -0.1cm\cdot\hskip -0.1cm \Vec{\! p \,}\hskip -0.2cm/|\hskip-0.1cm\Vec{\,p\,}\hskip-0.1cm|$} 

\vv \nin Le lecteur v\'erifiera\footnote{Par exemple, en calculant ${\cal H}\, \tau_3\, {\cal H}$.} que les spineurs (\ref{Uboost}) sont vecteurs propres de la projection du vecteur polarisation (\ref{polarW}) selon le 4-vecteur $z$ (orthogonal \`a $p$) ayant pour composantes  

\beq z_0 = \di{p_z \over m},~~ z_x = \di{{p_z p_x}\over{m ( E+m)}}, ~~z_y = \di{{p_z p_y}\over{m ( E+m)}},~~z_z = 1 + \di{{p^2_z}\over{m ( E+m)}}   \enq

\vv \nin et dont la partie spatiale n'est pas colin\'eaire \`a $\Vec{\,p\,}$. Pour obtenir une composante de spin $S_z$ correspondant \`a une projection du spin selon $\Vec{\,p\,}$, il faut pr\'ealablement effectuer une rotation amenant l'axe des $z$ selon la direction de ce 3-vecteur. Notant respectivement $\theta$ et $\varphi$ l'angle orbital et l'angle azimutal de $\Vec{\,p\,}$, la matrice $2\times 2$ repr\'esentant cette rotation s'\'ecrit 

$$ {\cal R}(\theta, \varphi) = \left( \begin{array}{cc} \cos \di{\theta\over 2} & - \sin \di{\theta \over 2}\, e^{- i \varphi} \\
\sin \di{\theta \over 2}\, e^{i \varphi} & \cos \di{\theta \over 2} \end{array} \right) = {\cal R}_z(\varphi) \, {\cal R}_y (\theta)\, {\cal R}^{-1}_z (\varphi) ~~~{\rm avec} $$
\beq {\cal R}_y(\theta) = \left( \begin{array}{cc} \cos \di{\theta\over 2} & - \sin \di{\theta \over 2} \\
\sin \di{\theta \over 2} & \cos \di{\theta \over 2} \end{array} \right),~~~ {\cal R}_z(\varphi) = \left( \begin{array}{cc} e^{i \varphi/2} & 0\\
0 & e^{ - i \varphi/2} \end{array} \right)  \label{rota0} \enq

\vv \nin Il est facile de v\'erifier que l'on a bien 

$$ {\cal R}(\theta, \varphi) \, \tau_3\, {\cal R}^{-1} (\theta, \varphi) = \Vec{\,u\,} \hskip -0.1cm\cdot \hskip -0.1cm\Vec{\, \tau\,} ~~~(\Vec{\,u\,} = \Vec{\,p\,}\hskip -0.2 cm/p,~~~ p = \sqrt{E^2 - m^2})$$

\vv \nin La t\'etrade correspondante s'\'ecrit 

\beq [\,p\,] = {\cal H}\, {\cal R}(\theta, \varphi) \enq

\vv \nin et conduit \`a 

$$ \,z \hskip -0.45 cm \begin{array}[m]{c} ~ \\ \stackrel{~~~}{\sim}
\end{array}\hskip -0.2cm (p)  = [\,p\,]\,\tau_3\,[\,p\,] = {\cal H}\,\Vec{\,u\,} \hskip -0.1cm\cdot \hskip -0.1cm\Vec{\, \tau\,}  \, {\cal H}  = \Vec{\,u\,} \hskip -0.1cm\cdot \hskip -0.1cm\Vec{\, \tau\,} {\cal H}^2 = \di{1\over m}\, \Vec{\,u\,} \hskip -0.1cm\cdot \hskip -0.1cm\Vec{\, \tau\,}  \left( E \,+\hskip -0.1cm \Vec{\,p\,} \hskip -0.1cm\cdot \hskip -0.1cm\Vec{\, \tau\,}  \right)~~~{\rm soit}$$
\beq    \,z \hskip -0.45 cm \begin{array}[m]{c} ~ \\ \stackrel{~~~}{\sim}
\end{array}\hskip -0.2cm (p)  = \di{1\over m}\, \left( p + E\,  \Vec{\,u\,} \hskip -0.1cm\cdot \hskip -0.1cm\Vec{\, \tau}  \right) \enq

\vv \nin Le 4-vecteur $z(p)$ ainsi obtenu a pour composantes 

\beq z_0 = \di{p\over m}\,,~~\Vec{\,z\,} = \di{E\over m}\, \Vec{\,u\,} \enq 

\vv \nin et sa partie spatiale est colin\'eaire \`a $\Vec{\,p\,}$. On a alors 

$$ S_z = \di{1\over{2 m^2}}\, \gs_5\,\left( p \,\gs_0 - E \Vec{\,u\,} \hskip -0.1cm\cdot \hskip -0.1cm\Vec{\,\gs}\, \right)\,\left( E\,\gs_0\, -\hskip -0.1cm \Vec{\,p\,} \hskip -0.1cm\cdot \hskip -0.1cm\Vec{\,\gs\,}  \right) = \di{1\over{2}}\gs_5\, \gs_0\Vec{\,u\,} \hskip -0.1cm\cdot \hskip -0.1cm\Vec{\,\gs}~~~~{\rm soit}$$
\beq S_z = \di{1\over 2} \left(\begin{array}{cc} \Vec{\,u\,} \hskip -0.1cm\cdot \hskip -0.1cm\Vec{\,\tau} & 0_2 \\ 0_2 & \Vec{\,u\,} \hskip -0.1cm\cdot \hskip -0.1cm\Vec{\,\tau} \end{array} \right) \equiv \di{1\over 2}\,\Vec{\,p\,} \hskip -0.1cm\cdot \hskip -0.1cm\Vec{\,\tau}\hskip -0.15cm/p \label{Szp} \enq

\vv \nin Ce choix de t\'etrade correspond donc bien \`a la projection de spin suivant la direction du 3-vecteur $\Vec{\,p\,}$. D'apr\`es la formule g\'en\'erale (\ref{spinorUVst}), et compte tenu de 

$$  \Vec{\,u\,} \hskip -0.1cm\cdot \hskip -0.1cm\Vec{\, \tau\,}   {\cal R}(\theta, \varphi) \, = {\cal R}(\theta, \varphi)  \, \tau_3\,~~~\tau_3 \, \chi^{\uw, \dw}_0 = \pm  \chi^{\uw, \dw}_0,~~~\chi^\uw_0 = \left(\begin{array}{c} 1\\0\end{array}\right),~~~\chi^\dw_0 = \left(\begin{array}{c} 0\\1\end{array}\right)         $$
$$ [\,p\,] + [\,p\,]^{\dagger -1} = \left({\cal H} + {\cal H}^{-1} \right)\, {\cal R}(\theta, \varphi) = \sqrt{\di{2\over m}}\, \sqrt{E+m} \, {\cal R}(\theta, \varphi),$$
$$ [\,p\,] - [\,p\,]^{\dagger -1} = \left({\cal H} + {\cal H}^{-1} \right)\, {\cal R}(\theta, \varphi) = \sqrt{\di{2\over m}}\, \sqrt{E+m} ~ \Delta \, {\cal R}(\theta, \varphi)\, \tau_3,~~~{\rm avec}~~~\Delta= \sqrt{\di{{E-m}\over{E+m}}}$$

\vv \nin les spineurs propres de type U associ\'es sont donn\'es par 

\beq \fbox{\rule[-0.8cm]{0cm}{1.8cm}~$U_\sigma = \sqrt{E+m}\,\left( \begin{array}{c} \left\{{\cal R}(\theta, \varphi)\right\}_{\cdot \, \sigma}  \\~\\ 2\, \sigma\, \Delta\,\left\{{\cal R}(\theta, \varphi)\right\}_{ \cdot \,\sigma} \end{array} \right)  $~} \enq

\vv \nin Explicitement, 

\beq \fbox{\rule[-0.8cm]{0cm}{1.8cm}~$\uup = \sqrt{E+m}\,\left( \begin{array}{c} \cos \di{\theta \over 2}  \\ \sin \di{\theta \over 2}\, e^{i \varphi} \\ \Delta\, \cos \di{\theta \over 2}  \\ \Delta\, \sin \di{\theta \over 2}\, e^{i \varphi}  \end{array} \right),~~~~\ud = \sqrt{E+m}\,\left( \begin{array}{c}  -\sin \di{\theta \over 2}\, e^{-i \varphi} \\ \cos \di{\theta \over 2} \\  \Delta\, \sin \di{\theta \over 2}\, e^{-i \varphi}  \\ - \Delta\, \cos \di{\theta \over 2} \end{array} \right) $~} \label{Uhel} \enq

\vv \nin Ces spineurs admettent des limites finies lorsque $m \rightarrow 0$ qui repr\'esentent les spineurs d'\'energie positive associ\'es \`a des particules de masse nulle : 

\beq \fbox{\rule[-0.8cm]{0cm}{1.8cm}~$\uup_0 = \sqrt{E}\,\left( \begin{array}{c} \cos \di{\theta \over 2}  \\ \sin \di{\theta \over 2}\, e^{i \varphi} \\  \cos \di{\theta \over 2}  \\ \sin \di{\theta \over 2}\, e^{i \varphi}  \end{array} \right),~~~~\ud_0 = \sqrt{E}\,\left( \begin{array}{c}  -\sin \di{\theta \over 2}\, e^{-i \varphi} \\ \cos \di{\theta \over 2} \\   \sin \di{\theta \over 2}\, e^{-i \varphi}  \\ -  \cos \di{\theta \over 2} \end{array} \right) $~} \label{Um0} \enq

\vv \nin Ils correspondent, le premier, $\uup_0$, \`a une particule (de masse nulle) d'h\'elicit\'e $+ \di{1\over 2}$, le second, $\ud_0$, \`a une particule (elle aussi de masse nulle) d'h\'elicit\'e $- \di{1\over 2}$, laquelle particule peut \'eventuellement \^etre l'anti-particule de la premi\`ere. Dans le cas des masses nulles, l'h\'elicit\'e $h$ est un invariant relativiste caract\'erisant la particule consid\'er\'ee et peut \^etre repr\'esent\'ee\footnote{ITL, Eq. 7.180.} par l'op\'erateur $\di{1\over 2} \gs_5$.  On v\'erifie ici que 

\beq \gs_5\, \uup_0 = +\,\uup_0\,,~~~ \gs_5\, \ud_0= -\,\ud_0 \enq

\vvv


\subsection{Expressions ``covariantes" des t\'etrades}

\vv \nin Une transformation $\Lambda = \Lambda(A)$ de ${\cal L}^\uparrow_{+}$ peut \^etre regard\'ee comme celle transformant une base d'espace-temps $T, X, Y, Z$ en la base $t, x, y, z$. On d\'emontre\footnote{ITL, Eqs. 5.204, 7.23.} que la matrice $A$ de $SL(2,C)$ qui lui correspond (au signe pr\`es) peut \^etre exprim\'ee sous la forme 

\beq \fbox{\rule[-0.55cm]{0cm}{1.3cm}~$A = \di{1\over{2~({\rm Tr} \,A)^\star}} \left( ~ t
\hskip -0.42cm \begin{array}[m]{c}
~ \\ \stackrel{~~~}{\sim} \end{array}\hskip -0.1cm \widetilde{T} -
x \hskip -0.45cm \begin{array}[m]{c}
~ \\ \stackrel{~~~}{\sim} \end{array}\hskip -0.1cm\widetilde{X} -
y \hskip -0.45cm \begin{array}[m]{c}
~ \\ \stackrel{~~~}{\sim} \end{array}\hskip -0.1cm\widetilde{Y} -
z \hskip -0.45cm \begin{array}[m]{c}
~ \\ \stackrel{~~~}{\sim} \end{array}\hskip -0.1cm\widetilde{Z}~ \right)
$ } \label{matA2} \enq

\vv \nin avec $|{\rm Tr}\,A |^2 = {\rm Tr}\, \Lambda$, et que la matrice $A^{\dagger -1}$ est donn\'ee par :  

\beq \fbox{\rule[-0.55cm]{0cm}{1.3cm}~$A^{\dagger -1} = \di{1\over{2~{\rm Tr} \,A}} \left(\,\widetilde{T}~\, t
\hskip -0.42cm \begin{array}[m]{c}
~ \\ \stackrel{~~~}{\sim} \end{array}\hskip -0.2cm -
\widetilde{X} ~x\hskip -0.46cm \begin{array}[m]{c}
~ \\ \stackrel{~~~}{\sim} \end{array}  \hskip -0.2cm -
\widetilde{Y} ~y \hskip -0.46cm \begin{array}[m]{c}
~ \\ \stackrel{~~~}{\sim} \end{array} \hskip -0.2cm -
\widetilde{Z} ~z \hskip -0.46cm \begin{array}[m]{c}
~ \\ \stackrel{~~~}{\sim} \end{array} \hskip -0.1cm \right)
$ } \label{matA3} \enq

\vv \nin En posant ${\rm Tr}\, A  = \rho\, e^{i \theta}$ avec $\rho > 0$ et en utilisant (\ref{LA}), on trouve alors l'expressions suivante pour la repr\'esentation $L(A)$ de $\Lambda(A)$ dans l'espace des spineurs de Dirac\footnote{Voir aussi ITL, \S 7.3.5.} :

\vskip -0.2cm
\beq \fbox{\fbox{\rule[-0.9cm]{0cm}{2cm}~$ \begin{array}{c} L(A) = \di{1 \over{2 \,\rho}} \left[ \,\cos \theta + i \sin \theta \,\gs_5 \,\right] \, \left[\, \gs(t)\, \gs(T) - \gs(x)\, \gs(X) - \gs(y) \, \gs(Y) - \gs(z)\, \gs(Z) \,\right]  \\ [0.3cm]
=  \di{1 \over{2 \,\rho}} \left[ \cos \theta + i \sin \theta \,\gs_5 \right] \, \gs^\mu\, \gs^\nu\, \Lambda_{\mu \nu} 
\end{array}$~}} \label{LA1} \enq

\vv \nin Le terme en $\gs_5$ de cette expression n'est absent que si ${\rm Tr}\, A$ est r\'eel.  Comme

 \beq 2 \,({\rm Tr} \,A)^\star = \pm \, \sqrt{4 + ({\rm Tr} \Lambda)^2 - {\rm Tr} (\Lambda^2) +
i\,\epsilon^{\mu \nu \rho \sigma}\, \Lambda_{\mu \nu}\,
\Lambda_{\rho \sigma}} \enq

\vv \nin ceci est r\'ealis\'e pour toute transformation $\Lambda(A)$ v\'erifiant $\epsilon^{\mu \nu \rho \sigma}\, \Lambda_{\mu \nu}\, \Lambda_{\rho \sigma} = 0$, c'est-\`a-dire, si ladite transformation est {\em plane}. C'est le cas pour une transformation de Lorentz pure qui laisse invariants les vecteur d'un 2-plan du genre espace, ou encore celui d'une rotation laissant invariants les vecteurs d'un 2-plan du genre hyperbolique. Dans ce cas, la matrice (\ref{LA1}) s'\'ecrit simplement

\beq \fbox{\fbox{\rule[-0.6cm]{0cm}{1.4cm}~$  L(A) = \di{1 \over{2 \,{\rm Tr}\,A}}\, \left[\, \gs(t)\, \gs(T) - \gs(x)\, \gs(X) - \gs(y) \, \gs(Y) - \gs(z)\, \gs(Z) \,\right]  $~}}  \label{LA2} \enq

\vv \nin D'apr\`es (\ref{LA}), on a alors ${\rm Tr}\,L(A) = 2\,{\rm Tr}\, A$ (car ${\rm Tr}\, A^{\dagger -1} = \left({\rm Tr}\, A^{-1}\right)^\star = \left({\rm Tr}\,A \right)^\star = {\rm Tr}\,A$) et par suite\footnote{Pour les formules de traces de produits de matrices $\gs$, voir ITL, Section 7.3.}\hskip -0.12 cm ~$^{,}$\hskip -0.05cm~\footnote{Dans la suite, nous faisons le choix ${\rm Tr}\,A > 0$.} 

\beq  \left({\rm Tr}\, A\right)^2 =  t \cdot T - x\cdot X - y \cdot Y - z \cdot Z  \enq

\vv \nin \ding{172} Consid\'erons le cas d'une \und{rotation dans le plan $(X,Y)$} pour laquelle $t \equiv T$, $z \equiv Z$ et 

$$ x = \cos \varphi \,X + \sin \varphi\, Y\,,~~~y = - \sin \varphi\, X + \cos \varphi\, Y$$

\vv \nin Il vient 

$$ \gs(x)\,\gs(X) = -\cos \varphi + \sin \varphi\, \gs(Y)\,\gs(X) = - \gs(y)\, \gs(Y)\, ,~~\gs(t)\, \gs(T) = -\gs(z)\, \gs(Z) = 1  $$
$$ {\rm Tr}\, A = \sqrt{2\,(1 - x \cdot X)} = 2 \cos \di{\varphi\over 2} $$

\vv \nin d'o\`u

\beq \fbox{\fbox{\rule[-0.8cm]{0cm}{1.8cm}~$ \begin{array}{c}  L(A) = R_Z(\varphi) = \di{{1 - \gs(x)\,\gs(X)}\over{\sqrt{2\,(1- x\cdot X)}}}  = \cos \di{\varphi \over 2} + \sin \di{\varphi \over 2}\, \gs(X)\,\gs(Y) \\ [0.3cm]
=  \cos \di{\varphi \over 2} -2 i  \sin \di{\varphi \over 2}\, S_Z(T) 
\end{array}$~}} \label{RZ} \enq

\vv \nin  Dans la derni\`ere expression, on a fait appara\^itre le g\'en\'erateur $S_Z(T) = \di{i\over 2}\,\gs(X)\,\gs(Y)$ de la rotation\footnote{Compte tenu de la relation (\ref{fonda5}) $\gs_5 = i\, \gs(X)\,\gs(Y)\,\gs(Z)\,\gs(T)$ et de (\ref{composantesSi}), on a $\gs(X)\,\gs(Y) =- i\, \gs(Z)\,\gs(T)\,\gs_5$.}. Par celle-ci, un spineur $U_\sigma([T])$ de la t\'etrade $[T]$ associ\'ee \`a $T$ est transform\'e en un spineur $U_\sigma([\,t\,])$ de la t\'etrade $[\,t\,] = R_Z(\varphi) [T]$ associ\'ee \`a $t$, tel que 

\beq U_\sigma([\,t\,]) = R_Z(\varphi) \, U_\sigma([T]) = \left(  \cos \di{\varphi \over 2} -2 i  \sigma \sin \di{\varphi \over 2}\right)\,U_\sigma([T]) = e^{-i \sigma \varphi}\,U_\sigma([T])  \enq

\vv \nin \ding{173} Envisageons ensuite une \und{rotation dans le plan $(X,Z)$} pour laquelle $t \equiv T$, $y \equiv Y$ et

$$ z = \cos \varphi \,Z + \sin \varphi\, X\,,~~~x = - \sin \varphi\, Z + \cos \varphi\, X$$

\vv \nin On a cette fois

\beq L(A) = R_Y(\varphi) = \di{{1 - \gs(z)\,\gs(Z)}\over{\sqrt{2\,(1- z\cdot Z)}}}  = \cos \di{\varphi \over 2} + \sin \di{\varphi \over 2}\, \gs(Z)\,\gs(X) \label{RY}\enq
$$ =  \cos \di{\varphi \over 2} -2 i  \sin \di{\varphi \over 2}\, S_Y(T) $$

\vv \nin Comme\footnote{Voir ITL, Eq. 4.35.}

\beq S_Y(T) \, U_\sigma([T]) =  i \sigma \,U_{- \sigma}([T]) \enq

\vv \nin il vient 

\beq U_\sigma([\,t\,]) = R_Y(\varphi) =  \cos \di{\varphi \over 2}\,U_\sigma([T])  + 2 \sigma   \sin \di{\varphi \over 2} \,U_{-\sigma}([T]) \enq

\vv \nin En particulier, pour $\varphi = \pi$, 

\beq \fbox{\fbox{\rule[-0.5cm]{0cm}{1.2cm}~$U_\sigma([\,t\,]) = 2 \sigma  \,U_{-\sigma}([T]) = (-1)^{\frac{1}{2} - \sigma}  \,U_{-\sigma}([T])  $~}}  \label{RPIY} \enq

\vv \nin o\`u ici $[\,t\,] = R_Y(\pi) \,[T]$. 

\newpage

\vv \nin \ding{174} Dans le cas d'une \und{transformation de Lorentz pure}, pour laquelle $x \equiv X$ et $y \equiv Y$ et  

$$ t = \cosh \alpha \,T + \sinh \alpha\, Z\, ,~~~z= \sinh \alpha \,T + \cosh \alpha \, Z $$

\vv \nin on a 

$$ \gs(t)\,\gs(T) = \cosh \alpha + \sinh \alpha\, \gs(Z)\,\gs(T) = - \gs(z)\, \gs(Z)\, ,~~\gs(x)\, \gs(X) = \gs(y)\, \gs(Y) = -1  $$
$$ {\rm Tr}\, A = \sqrt{2\,(1 + t \cdot T)} = 2 \cosh \di{\alpha\over 2} $$

\vv \nin et par suite,  

\beq \fbox{\fbox{\rule[-0.7cm]{0cm}{1.6cm}~$L(A) = S_{T \rightarrow t} = \di{{1 + \gs(t)\,\gs(T)}\over{\sqrt{2\,(1+ t\cdot T)}}}  = \cosh \di{\alpha \over 2} + \sinh \di{\alpha \over 2}\, \gs(Z)\,\gs(T)  $~}} \label{PZ} \enq

\vv \nin  Montrons que, au signe pr\`es, la matrice $\di{i\over 2}\,\gs(Z)\,\gs(T)$ repr\'esente le g\'en\'erateur de cette transformation. Cette derni\`ere laissant $X$ invariant doit appartenir au petit groupe de ce vecteur. Or, l'op\'erateur de Pauli-Lubanski associ\'e \`a $X$ est 

$$ W_\mu(X) = \di{1\over 2} \, \epsilon_{\mu \alpha \beta \gs}\, X^\alpha\,\sigma^{\beta \gs} = - \di{1\over 4} [\,\gs_\mu\,, \gs_\nu\,] \,X^\nu\, \gs_5$$

\vv \nin et parmi les trois composantes de cet op\'erateur, celle qui pr\'eserve le vecteur $Y$ est 

$$ N_Y(X) = - Y\cdot W(X) = -\di{1\over 2}\, \gs(X)\,\gs(Y)\,\gs_5 = \di{i\over 2} \, \gs(Z)\,\gs(T) $$

\vv \nin On peut aussi envisager la transformation comme un \'el\'ement du petit groupe de $Y$. Dans ce cas, on trouve pour g\'en\'erateur $N_X(Y) = - N_Y(X)$. 

\vv \nin Notant que $\gs(Z)\, \gs(T) = 2 S_Z(T) \, \gs_5$, la matrice (\ref{PZ}) transforme le spineur $U_\sigma([T])$ en un spineur $U_\sigma([\,t\,])$ associ\'e \`a la t\'etrade $[\,t\,] = [\,T \rightarrow t\,] [T]$ et tel que 

\beq \fbox{\rule[-0.7cm]{0cm}{1.6cm}~$ U_\sigma([\,t\,]) = S_{T \rightarrow t} \,U_\sigma([T]) = \cosh \di{\alpha\over 2} \, U_\sigma([T]) + 2 \sigma \sinh \di{\alpha \over 2} \, V_\sigma([T]) $~} \label{TUPU} \enq


\section{Produits tensoriels de spineurs}

\vv \nin Dans leurs mod\'elisations de la structure en quarks des hadrons, certains auteurs ont utilis\'e des amplitudes du type $T_{\alpha \beta}$ et $T_{\alpha \beta} U_\delta$, o\`u  $T$ est un tenseur du deuxi\`eme ordre suivant des indices de composantes d'un spineur de Dirac, construit \`a partir de produits de matrices de Dirac \footnote{C. H. Llewellyn Smith, Ann. Phys. (N.Y.) 53 (1969) 327 ; V.L. Chernyak, A.R. Zhitnitsky, Phys. Rep. 112 (1984), 173-318 ; C. Carimalo, ``On the spinor structure of the Proton wave function", J.Math.Phys. 34, (1993), 4930-4963 ; LPC-92-24 App. B. ; G. Eichmann, Dissertation, Universit\'e de Graz (2009), arXiv:0909.0703 [hep-ph].} . Ces formes tensorielles doivent \^etre consid\'er\'ees comme composantes d'\'el\'ements des espaces $E\otimes E$ et $E\otimes E\otimes E$ respectivement, $E$ \'etant l'espace vectoriel \`a quatre dimensions des spineurs de Dirac. L'objet de cette section est de montrer, \`a l'instar de ce qui a \'et\'e fait au paragraphe \ref{projo} pour les projecteurs de spineurs, comment on peut exprimer ces formes tensorielles au moyen des produits tensoriels de spineurs constituant des bases de ces espaces.

\vv \nin Pour ce faire, nous ferons appel aux formules suivantes\footnote{ITL, \S  7.4.3.} . Introduisons les matrices 

\beq  {\cal C} = i\, \gs^2\,\gs_0 = - \left( \begin{array}{cc} 0_2 & C \\ C & 0_2 \end{array} \right) ~~{\rm et}~~~\Omega_c = - \gs_5\,{\cal C} = - {\cal C}\, \gs_5 = \left( \begin{array}{cc} C & 0_2 \\ 0_2 & C \end{array} \right) \enq

\vv \nin La matrice $\Omega_c$ a pour vertu de transformer les matrices $\gs$ en leurs transpos\'ees :  

\beq \Omega_c\, \gs_\mu\, \Omega_c^{-1} = \,^t\gs_\mu,~~~\Omega\, \gs_5\, \Omega_c^{-1} = \,^t\gs_5 = \gs_5 \enq

\vv \nin et est telle que 

\beq \Omega_c^{-1} = -\Omega_c,~~~^t\Omega_c = - \Omega_c,~~~\Omega_c^\star = \Omega_c,~~~\Omega_c\, \gs_0 = \gs_0\, \Omega_c \enq

\vv \nin Des relations

$$ \uup = {\cal C}\,\,^t\bvd =\,^t\hskip -0.1cm\left(\bvd\,^t{\cal C}\right) = -\,^t\hskip-0.1cm \left(\,\bvd {\cal C}\right) ,~~~\ud = \,^t\hskip-0.1cm \left(\,\bvup {\cal C}\right) $$
\beq \vup =  \,^t\hskip-0.1cm \left(\,\bud {\cal C}\right) ,~~~\vd = - \,^t\hskip-0.1cm \left(\,\buup {\cal C}\right) \enq

\vv \nin o\`u les spineurs consid\'er\'es ici sont attach\'es \`a la t\'etrade $[\,p\,]$ d'un 4-vecteur $p$ du genre temps pointant vers le futur, on d\'eduit 

\beq \uup = -\,^t\hskip-0.1cm \left(\,\bud \Omega_c \right) ,~~~\ud =  \,^t\hskip-0.1cm \left(\,\buup \Omega_c\right),~~~ \vup =   \,^t\hskip-0.1cm \left(\,\bvd \Omega_c\right) ,~~~\vd = - \,^t\hskip-0.1cm \left(\,\bvup \Omega_c\right)  \label{tensor1} \enq
$${\rm soit~~encore}~~U_\alpha^\uw = - \left(\,\bud \Omega_c \right)_\alpha ,~U_\alpha^\dw = \left(\,\buup \Omega_c\right)_\alpha,~V_\alpha^\uw = \left(\,\bvd \Omega_c\right)_\alpha ,~V_\alpha^\dw = -\left(\,\bvup \Omega_c\right)_\alpha$$

\subsection{Produits tensoriels \`a 2 spineurs}

\vv \nin Posant $ \wup = \vd,~\wdo = - \vup $, consid\'erons les trois matrices 

$$ \Psi^{(+)} = \uup\,\bwup  =  \uup\, \bvd = - \uup\,\bud\,\gs_5= - \di{1\over{\sqrt{2}}}\,\left[\, m + \gs(p)\,\right]\,\gs(e^{(+)}) $$
$$ \Psi^{(-)} = \ud\,\bwd  = - \ud\, \bvup =  \ud\,\buup\,\gs_5= - \di{1\over{\sqrt{2}}}\,\left[\, m + \gs(p)\,\right]\,\gs(e^{(-)}) $$
$$ \Psi^{(0)} =\di{1\over{\sqrt{2}}} \left[ \, \uup\,\bwd + \ud\, \bwup \, \right] = - \di{1\over{\sqrt{2}}} \left[ \, \uup\,\bvup - \ud\, \bvd \, \right] = \di{1\over{\sqrt{2}}} \left[ \, \uup\,\buup - \ud\, \bud \, \right]\, \gs_5 $$
\beq =  \di{1\over{\sqrt{2}}} \left[ \, m + \gs(p)\, \right]\, \gs(z) \enq

\vv \nin o\`u $e^{(\pm)} = \mp \di{1\over{\sqrt{2}}} ( x \pm i y)$. Il est facile de montrer qu'elles forment une repr\'esentation de $\bar{\cal L}^\uparrow_{+}$ de spin 1. En effet, par transformation de Lorentz, chacune d'elle devient 

\beq {\cal L}(A) [\Psi(p)] = L(A)\, \Psi(A^{-1} p)\, L(A)^{-1} \label{transfo-mat1}\enq

\vv \nin ce qui implique notamment que l'action sur ces matrices du repr\'esentant de la composante de spin suivant $z$ est donn\'ee par\footnote{Notons ici que la d\'efinition des composantes de spin via l'op\'erateur de Pauli-Lubanski a pour vertu d'exclure de cette d\'efinition toute partie ``orbitale" d'un moment cin\'etique, conf\'erant ainsi au spin une valeur ``intrins\`eque". Ainsi, si l'on cherche \`a passer de  (\ref{transfo-mat1}) \`a (\ref{spin-mat1}) en consid\'erant une transformation de Lorentz infinit\'esimale, on ne doit pas tenir compte du terme impliquant les d\'eriv\'ees partielles de $\Psi$ par rapport aux composantes de $p$, terme dont l'apparition est usuellement attribu\'ee \`a un moment cin\'etique ``orbital".} 

\beq {\cal S}_z [\Psi(p) ] = [\, S_z\, , \Psi (p) \,] \label{spin-mat1} \enq 
 
\vv \nin Comme $S_z\, U_\sigma = \sigma\, U_\sigma$ et que 

\beq \bwup\, S_z = - \di{1\over 2}\, \bwup,~~~\bwd\, S_z = + \di{1\over 2}\, \bwd \enq

\vv \nin on trouve ais\'ement que 

\beq   {\cal S}_z [\Psi^{(+)}(p)] = + \Psi^{(+)}(p),~~~ {\cal S}_z [\Psi^{(0)}(p)] =0,~~~ {\cal S}_z [\Psi^{(-)}(p)] = - \Psi^{(-)}(p)\enq 

\vv \nin Les trois matrices en question peuvent ainsi servir de base pour d\'ecrire un syst\`eme particule-antiparticule de spin 1. Quant \`a la matrice  

\beq \di{1\over{\sqrt{2}}}\, \left[\, \uup \, \bwd - \ud\, \bwup\, \right] = \di{1\over{\sqrt{2}}}\,\left[\, m + \gs(p)\, \right] \gs_5 \label{mat-spin0} \enq

\vv \nin qui commute avec tous les op\'erateurs de spin $S_k(p)$, elle peut repr\'esenter, du point de vue du contenu en spin, un syst\`eme quark-antiquark de spin 0.

\vv \nin De (\ref{relspinors}) et (\ref{tensor1}), on tire les formules suivantes\footnote{Pour la clart\'e des formules, nous ommettons le symbole $\otimes$ des produits tensoriels.}.  

$$ \left(1 \times \Omega_c \right)_{\alpha \beta}  = \di{1\over{2m}} \left[ \left(\, \uup \, \buup + \ud \, \bud  -  \vup \, \bvup - \vd \, \bvd \,\right)\Omega_c\right]_{\alpha \beta}~~~{\rm soit}$$
\beq \left[\,\Omega_c\,\right]_{\alpha \beta} =\di{1\over{2m}}\, \left[\,\uup \, \ud - \ud \, \uup  +  \vup \, \vd - \vd \, \vup \, \right]_{\alpha \beta}  \enq 

\vv \nin Puis, successivement,  

$$ \left[\,\gs(p)\,\Omega_c\,\right]_{\alpha \beta} = \di{1\over 2}\left[ \uup \, \ud - \ud \, \uup  -  \vup \, \vd + \vd\, \vup \right]_{\alpha \beta} $$ 
$$\left[\,\gs_5\,\Omega_c\,\right]_{\alpha \beta} = \di{1\over 2}\left[ \vup \, \ud - \vd \, \uup  + \uup \, \vd - \ud\, \vup \right]_{\alpha \beta}  $$
\beq \left[\,\gs_5\,\gs(p)\,\Omega_c\,\right]_{\alpha \beta} = \di{1\over 2}\left[ \vup \, \ud - \vd \, \uup  -  \uup \, \vd + \ud\, \vup \right]_{\alpha \beta}  \enq
$$ \left[ \left(\, m + \gs(p)\, \right)\Omega_c\right]_{\alpha \beta} = \left[\, \uup\, \ud - \ud\, \uup \, \right]_{\alpha \beta} $$

\vv \nin Utilisant les formules\footnote{Que l'on obtient \`a partir des formules du paragraphe  \ref{projo} en calculant des traces, par exemple,  $\ov{U}^{\uw, \dw} \gs_\mu\, U^{\uw, \dw} = {\rm Tr}\,U^{\uw, \dw} \, \ov{U}^{\uw, \dw} \gs_\mu$.}

$$ \ov{U}^{\uw, \dw} \gs_\mu\, U^{\uw, \dw} = 2\, p_\mu,~~ \ov{U}^{\uw, \dw} \gs_\mu\, U^{\dw, \uw} = 0,~~\ov{U}^{\uw, \dw}\, \gs_\mu\, V^{\uw, \dw} = \pm 2\,m\, z_\mu,$$
\beq \ov{U}^{\uw, \dw}\, \gs_\mu\, V^{\dw, \uw} = \mp\, m \,\sqrt{2} \,e^{(\mp)}_\mu \enq 

\vv \nin on d\'eduit $(t_\mu = p_\mu/m)$

$$ 2\, m\, \gs_\mu = t_\mu \left\{ \uup\,\buup +  \ud\,\bud + \vup\,\bvup +  \vd\,\bvd \right\} + $$ 
\beq - z_\mu\, \left\{  \uup\,\bvup -  \ud\,\bvd + \vup\,\buup - \vd\,\bud \right\}  +\label{devel-gmu}  \enq
$$ + \sqrt{2}\, e^{(+)}_\mu\, \left\{\vd\,\buup +\ud\, \bvup \,\right\}  - \sqrt{2}\, e^{(-)}_\mu\, \left\{\uup \,\bvd +\vup\, \bud \,\right\} $$

$$ 2\, m\, \left(\,\gs_\mu \, \Omega_c\,\right)_{\alpha \beta} = t_\mu \left\{ \uup\,\ud - \ud\,\uup - \vup\,\vd +  \vd\,\vup \right\}_{\alpha \beta} + $$ 
\beq - z_\mu\, \left\{  \vup\,\ud + \vd\,\uup - \uup\,\vd - \ud\,\vup \right\}_{\alpha \beta}  + \enq
$$ + \sqrt{2}\, e^{(+)}_\mu\, \left\{\vd\,\ud -\ud\, \vd \,\right\}_{\alpha \beta}  - \sqrt{2}\, e^{(-)}_\mu\, \left\{\uup \,\vup -\vup\, \uup \,\right\}_{\alpha \beta} $$

\beq \gs_\mu\, \uup = t_\mu\, \uup - z_\mu\, \vup + \sqrt{2}\,e^{(+)}_\mu\, \vd \enq

$$ m\, \left\{ \, \gs(e^{(+)})\, \Omega_c\, \right\}_{\alpha \beta} = -\di{1\over{\sqrt{2}}}\, \left\{ \uup\, \vup - \vup \uup \right\}_{\alpha \beta} $$
\beq m\, \left\{ \, \gs(e^{(-)})\, \Omega_c\, \right\}_{\alpha \beta} = -\di{1\over{\sqrt{2}}}\, \left\{ \ud\, \vd - \vd \ud \right\}_{\alpha \beta} \enq

\vv \nin Le d\'eveloppement (\ref{devel-gmu}) permet d'obtenir le commutateur :

$$ m\, \left[\, \gs_\mu , \gs_\nu\, \right] = - \left\{ t_\mu\, z_\nu - z_\mu\, z_\nu \right\} \left\{ \uup\, \bvup - \ud\, \bvd - \vup\, \buup + \vd\,\bud \right\} + $$
$$ - \sqrt{2}\,\left\{ t_\mu\, e^{(+)}_\nu - t_\nu\, e^{(+)}_\mu\right\} \left\{ \vd\, \buup - \ud\, \bvup \right\} + $$
$$ +  \sqrt{2}\,\left\{ t_\mu\, e^{(-)}_\nu - t_\nu\, e^{(-)}_\mu\right\} \left\{ \vup\, \bud - \uup\, \bvd \right\} $$
\beq - \sqrt{2}\,\left\{ z_\mu\, e^{(+)}_\nu - z_\nu\, e^{(+)}_\mu\right\} \left\{ \ud\, \buup - \vd\, \bvup \right\} + \enq
$$ +  \sqrt{2}\,\left\{ z_\mu\, e^{(-)}_\nu - z_\nu\, e^{(-)}_\mu\right\} \left\{ \vup\, \bvd - \uup\, \bud \right\} + $$
$$ - \left\{ e^{(+)}_\mu\, e^{(-)}_\nu - e^{(+)}_\nu\, e^{(-)}_\mu \right\}\left\{ \uup\, \buup + \vd\, \bvd - \ud\, \bud - \vup\, \bvup \right\} $$

\vv \nin d'o\`u 

$$ \di{1 \over 2} [\,\gs_\mu \, , \, \gs_\nu\, ]\, \uup = \left\{ t_\mu z_\nu - t_\nu z_\mu \right\} \vup - 
\left\{ e^{(+)}_\mu e^{(-)}_\nu  - e^{(+)}_\nu e^{(-)}_\mu \right\}  \uup + $$
\beq  - \sqrt{2} \, \left\{ t_\mu e^{(+)}_\nu  - t_\nu e^{(+)}_\mu \right\}  \vd - \sqrt{2} \, \left\{ z_\mu e^{(+)}_\nu  - z_\nu e^{(+)}_\mu \right\} \ud \enq

\vv \nin et 

$$ [ \,\gs_\mu \, , \, \gs(p) \,] = z_\mu \left\{ \uup \bvup - \ud \bvd - \vup \buup + \vd \bud \right\} + $$
\beq + \sqrt{2} \, e^{(+)}_\mu \left\{ \vd \buup - \ud \bvup \right\} - \sqrt{2}\, e^{(-)}_\mu \left\{ \vup \bud - \uup \bvd \right\} \enq

\vv \nin Toutes ces formules indiquent que, comme il se doit, les \'el\'ements d'une quelconque matrice $4\times4$, peuvent \^etre exprim\'es au moyen des tenseurs-spineurs de rang deux : $UU$, $UV$, $VU$ et $VV$.  


\subsection{Produits tensoriels \`a 3 spineurs \protect \footnote{V. Bargmann, E.P. Wigner, Proc. Nat. Acad. Sc. (USA) 34, 211 (1948) ; W. Rarita, J. Schwinger, Phys. Rev. 60, 61 (1941) ; M. D. Nykerk, ``Quantizing spin 3/2 fields", rapport NIKHEF-95-002 (Jan. 1995) ; I. Lovas, K. Sailer, W. Greiner, ``Generalized Rarita-Schwinger equations", Heavy Ion Physics 8 (1998) 237-245. } }

\vv \nin A partir des formules pr\'ec\'edentes, il est facile d'\'etablir les suivantes concernant des tenseurs-spineurs de rang trois. 

$$ 2 m \left\{\,\gs_\mu \,\Omega_c \,\right\}_{\alpha \beta} \left\{ \gs_\mu \uup \right\}_\delta = \left\{ \uup \ud - \ud \uup - \vup \vd + \vd \vup \right\}_{\alpha \beta} \, \uup_\delta $$
\beq - \left\{ \vup \ud - \ud \vup - \uup \vd + \vd \uup \right\}_{\alpha \beta}\, \vup_\delta + \enq
$$ + 2 \left\{ \vup \uup - \uup \vup \right\}_{\alpha \beta} \, \vd_\delta $$

\vv 

$$\left\{ \bsig_{\mu \nu}\, p^\nu \,\Omega_c \right\}_{\alpha \beta} \left\{ \gs^\mu \, \uup \right\}_\delta= \left\{ \uup \vup + \vup \uup \right\}_{\alpha \beta}\, \vd_\delta + $$
\beq - \di{1\over 2} \left\{ \uup \vd + \ud \vup + \vup \ud + \vd \uup \right\}_{\alpha \beta} \, \vup_\delta \enq

$$\left\{ \gs_5 \,\bsig_{\mu \nu}\, p^\nu\, \Omega_c \right\}_{\alpha \beta} \left\{\gs_5  \gs^\mu \, \uup \right\}_\delta= \left\{ \vup \vup + \uup \uup \right\}_{\alpha \beta}\, \ud_\delta + $$
\beq - \di{1\over 2} \left\{ \uup \ud + \ud \uup + \vup \vd + \vd \vup \right\}_{\alpha \beta} \, \uup_\delta \enq

$$\left\{\, \gs_\mu\, \Omega_c \right\}_{\alpha \beta} \left\{ \bsig_{\mu \nu} p^\nu  \, \uup \right\}_\delta= \left\{ \vup \uup - \uup \vup \right\}_{\alpha \beta}\, \vd_\delta + $$
\beq + \di{1\over 2} \left\{ \uup \vd + \ud \vup - \vup \ud - \vd \uup \right\}_{\alpha \beta} \, \vup_\delta \enq

$$\left(\gs_5  \gs_\mu \Omega_c \right)_{\alpha \beta} \left\{ \gs_5 \bsig_{\mu \nu} p^\nu  \, \uup \right\}_\delta= \left\{ \uup \uup - \vup \vup \right\}_{\alpha \beta}\, \ud_\delta + $$
\beq + \di{1\over 2} \left\{ \vup \vd + \vd \vup - \uup \ud - \ud \uup \right\}_{\alpha \beta} \, \uup_\delta \enq

$$ - m \left\{ \bsig_{\mu \nu} \,\Omega_c \right\}_{\alpha \beta} \left\{ \bsig^{\mu \nu}  \uup\right\}_\delta = \left\{ \uup \ud + \ud \uup + \vup \vd + \vd \vup \right\}_{\alpha \beta}\, \uup_\delta + $$ 
\beq + \left\{ \uup \vd + \ud \vup + \vup \ud + \vd \uup \right\}_{\alpha \beta} \, \vup_\delta + \enq
$$ -2 \left\{ \uup \uup + \vup \vup \right\}_{\alpha \beta} \, \ud_\delta - 2 \left\{\vup \uup + \uup \vup \right\}_{\alpha \beta}\, \vd_\delta $$

\vv \nin o\`u l'on a pos\'e $\bsig_{\mu \nu} = [\gs_\mu\, , \, \gs_\nu]/2$. On peut ainsi exprimer \`a l'aide des tenseurs de rang trois construits \`a partir des spineurs $U$ et $V$ toute forme du type $\left(M\, \Omega_c\right)_{\alpha \beta}\,(M' U)_\delta$ o\`u $M$ et $M'$ sont des matrices de la base des 16 matrices de Dirac $1,~ \gs_5, ~\gs_\mu,~ \gs_\mu \, \gs_5, ~\sigma_{\mu \nu} = \di{i \over 4}\, \left[ \gs_\mu, \gs_\nu \right]$. Il est \'evident que le nombre de telles formes \'etant limit\'e \`a 64, elles ne sont pas toutes ind\'ependantes. On trouve par exemple la relation : 

$$ \left\{ \gs_5\, \bsig_{\mu \nu} \,p^\mu \,\Omega_c \right\}_{\alpha \beta} \left\{ \gs_5 \gs^\nu \uup \right\}_\delta + \left\{\,\bsig_{\mu \nu}\, p^\mu\, \Omega_c\,\right\}_{\alpha \beta} \left\{\gs^\nu \uup \right\}_\delta $$
\beq = - \di{m \over 2} \left\{ \bsig_{\mu \nu} \,\Omega_c\, \right\}_{\alpha \beta} \left\{ \bsig^{\mu \nu} \uup \right\}_\delta \enq

\vv \nin Inversement, tout tenseur de rang trois peut \^etre exprim\'e au moyen de telles formes. Par exemple, le tenseur compl\`etement sym\'etrique 

$$ {\cal S} =  \di{1 \over \sqrt{18}} \left\{ \left(\vup \vd + \vd \vup \right) \uup + \left(\uup \vd + \vd \uup \right) \vup + \right. $$
\beq \left. + \left( \vup \uup + \uup \vup \right) \vd - 2 \ud \vup \vup  -2 \vup \ud \vup - 2 \vup \vup \ud\right\} \enq

\vv \nin et le tenseur compl\`etement antisym\'etrique 

 $$ {\cal A} =  \di{1 \over \sqrt{6}} \left\{ \left(\uup \vd - \vd \uup \right) \vup - \left(\vup \vd - \vd \vup \right) \uup + \right. $$
\beq \left. + \left( \vup \uup - \uup \vup \right) \vd \right\} \enq

\vv \nin s'expriment aussi comme 

$$ {\cal S}_{\alpha \beta \delta} = \di{2 \over \sqrt{3}} \left\{ \left(\bsig_{\mu \nu} \,p^\nu\, \Omega_c \right)_{\alpha \beta} \left( \gs^\mu \uup \right)_\delta + \left(\gs^\mu\, \Omega_c\, \gs_5 \right)_{\alpha \beta} \left( \gs_5 \,\bsig_{\mu \nu} \,p^\nu \uup \right)_\delta \right\} $$
\beq {\cal A}_{\alpha \beta \delta} = \di{m \over \sqrt{6}} \left\{ \left( \gs^\mu \Omega_c \right)_{\alpha \beta} \left( \gs_\mu \uup \right)_\delta - (\Omega_c)_{\alpha \beta} \,\uup_\delta + \left(\gs_5 \Omega_c \right)_{\alpha \beta} \left( \gs_5 \uup \right)_\delta \right\} \enq


\newpage

\section{Compl\'ement I : Amplitudes spinorielles de spin 1 et 4-potentiel \protect \footnote{ P. Moussa, R. Stora, loc.cit, p285.}}

\vv \nin Il nous para\^it opportun de pr\'eciser ici le lien entre la description des \'etats d'une particule de spin 1 au moyen d'amplitudes spinorielles et celle, plus courante, \`a l'aide d'un 4-vecteur, appel\'e 4-potentiel en Electromagn\'etisme, s'agissant dans ce cas  du photon. 

\vv \nin \leftpointright~Pour commencer, nous supposerons que la particule consid\'er\'ee a une masse $m$ non nulle. Du point de vue du groupe de Poincar\'e, les \'etats de cette particule appartiennent \`a une repr\'esentation irr\'eductible $[\,m, 1, \eta\,]$ de ce groupe ($\eta$ est la parit\'e de la particule). L'\'etat correspondant \`a une 4-impulsion donn\'ee $p$ et \`a une valeur propre $\sigma$ de la composante de spin suivant un 4-vecteur $z(p)$ associ\'e \`a $p$ dans une t\'etrade $[\,p\,]$, est not\'e $|\, [\,p\,], \sigma >$, $\sigma$ pouvant prendre les valeurs $-1$, $0$ ou $1$ (voir Eq. 1.31 et suivantes). L'ensemble des vecteurs $|\, [\,p\,], \sigma >$ pour diff\'erentes valeurs de $p$ et de $\sigma$ formant une base de l'espace des \'etats de la particule\footnote{Pour simplifier, nous passons ici sous silence l'existence possible d'autres nombres quantiques d\'efinissant compl\`etement la particule.}, un \'etat quelconque $|\, \phi >$ de celle-ci est compl\`etement d\'etermin\'e par la donn\'ee des amplitudes ind\'ependantes : 

\beq \phi_\sigma ([\,p\,]) ~=~ < \sigma, [\,p\,]\,| \, \phi > \label{amplit-1} \enq   

\vv \nin Comme signal\'e au paragraphe 1.3.1, les \'etats $|\, [\,p\,], \sigma >$ d\'ependent du choix de la t\'etrade $[\,p\,]$. Le changement de t\'etrade $[\,p\,] \rightarrow [\,p\,]^\prime$ modifie les amplitudes (\ref{amplit-1}) de la fa\c{c}on suivante\footnote{$R = [\,p\,]^{\prime \,\dagger}\,[\,p\,]^{\dagger -1}$ \'etant une matrice de rotation du petit groupe de $p$, on a $R = R^{\dagger -1} = [\,p\,]^{\prime \,-1}\,[\,p\,] $.} :  

\beq \phi_{\sigma'}([\,p\,]^\prime) = {\cal D}^1_{\sigma' \sigma}([\,p\,]^{\prime \,\dagger}\,[\,p\,]^{\dagger -1})\, \phi_\sigma([\,p\,]) = {\cal D}^1_{\sigma' \sigma}([\,p\,]^{\prime\, -1}[\,p\,])\, \phi_\sigma([\,p\,]) \label{chgt-tetr} \enq

\vv \nin la sommation sur l'indice $\sigma$ \'etant implicite. On montre que les amplitudes 

\beq \varphi_\sigma(p) = {\cal D}^1_{\sigma \sigma'}([\,p\,])\, \phi_{\sigma'}([\,p\,])~~~{\rm et}~~~\wht{\varphi}_\sigma(p) = {\cal D}^1_{\sigma \sigma'}([\,p\,]^{\dagger -1}) \, \phi_{\sigma'}([\,p\,]) \label{ampl-spin} \enq

\vv \nin (avec sommation sur $\sigma'$) sont ind\'ependantes du choix de t\'etrade : ce sont les amplitudes spinorielles attach\'ees \`a l'\'etat $|\, \phi>$. Les deux types d'amplitudes ne peuvent \^etre ind\'ependants car, la valeur de $p$ \'etant fix\'ee, il ne doit y avoir que $2 s+1 = 3$ amplitudes ind\'ependantes. De fait, les amplitudes spinorielles sont li\'ees entre elles par un syst\`eme d'\'equations analogue \`a l'\'equation de Dirac :  

\vskip -0.2cm
\beq \begin{array}{c} 
 \varphi_\sigma(p) = {\cal D}^1_{\sigma \sigma'} ( \,t \hskip -0.4 cm \begin{array}[m]{c} ~ \\ \stackrel{~~~}{\sim}
\end{array} \hskip -0.2cm)\, \wht{\varphi}_{\sigma'}(p)~~~{\rm et}~~~~\wht{\varphi}_{\sigma}(p) = {\cal D}^1_{\sigma \sigma'}(\,\widetilde{t}\,) \,\varphi_{\sigma'}(p)  \\
 {\rm avec}~~~t=p/m,~~~\,t \hskip -0.4 cm \begin{array}[m]{c} ~ \\ \stackrel{~~~}{\sim}
\end{array} \hskip -0.2cm=~[\,p\,]\,[\,p\,]^\dagger = t_0\, + \hskip -0.1cm\Vec{\,t\,}\cdot \Vec{\, \tau\,} ,~~~ \widetilde{t} \,= \,t \hskip -0.4 cm \begin{array}[m]{c} ~ \\ \stackrel{~~~}{\sim}
\end{array} \hskip-0.25cm^{-1} =  t_0\,\, - \hskip -0.1cm\Vec{\,t\,}\cdot \Vec{\, \tau\,} \end{array} \label{dirac-s1} \enq

\nin Par une transformation $(a, A)$ du groupe de Poincar\'e restreint, on a  
\vskip -0.2cm
\beq \begin{array}{c} 
 \,^{(a,A)} \phi_\sigma([\,p\,])~=~e^{i a \cdot p}~{\cal D}^1_{\sigma \sigma'} ([\,p\,]^{-1} A\,[\,A^{-1}p\,]) \,\phi_{\sigma'} ([\,A^{-1} p\,]) \\ ~\\
\,^{(a,A)}\varphi_{\sigma}(p) = e^{i a \cdot p}~{\cal D}^1_{\sigma \sigma'}(A)\, \varphi_{\sigma'}(A^{-1} p) ,~~~^{(a,A)}\wht{\varphi}_\sigma(p) = e^{i a \cdot p}~ {\cal D}^1_{\sigma \sigma'}(A^{\dagger -1})\, \wht{\varphi}_{\sigma'}(A^{-1} p) \label{trans1} \end{array} \enq

\nin Introduisant les vecteurs unicolonnes 

\beq \psi([\,p\,])  = \left( \begin{array}{c} \phi_1([\,p\,])  \\ \phi_0([\,p\,])  \\ \phi_{-1}([\,p\,])  \end{array} \right) ,~~~\varphi(p) = \left( \begin{array}{c} \varphi_1(p) \\ \varphi_0(p) \\ \varphi_{-1} (p)\end{array} \right),~~~\wht{\varphi}(p) = \left( \begin{array}{c} \wht{\varphi}_1(p) \\ \wht{\varphi}_0(p) \\ \wht{\varphi}_{-1}(p) \end{array} \right)\enq

\vv \nin les formules (\ref{chgt-tetr}),  (\ref{ampl-spin}), (\ref{dirac-s1}) et (\ref{trans1}) peuvent \^etre r\'ecrites sous forme matricielle : 

\beq
\begin{array}{c} 
\psi([\,p\,]^\prime) = {\cal D}^1([\,p\,]^{\prime\, -1}[\,p\,])\, \psi([\,p\,]) \\[0.4cm]
\varphi(p) = {\cal D}^1 ([\,p\,])\, \psi([\,p\,]),~~~ \wht{\varphi}(p) = {\cal D}^1 ([\,p\,]^{\dagger -1})\, \psi([\,p\,])  \\ [0.2cm]
\varphi(p) = {\cal D}^1( \,t \hskip -0.4 cm \begin{array}[m]{c} ~ \\ \stackrel{~~~}{\sim}
\end{array} \hskip -0.2cm)\, \wht{\varphi}(p),~~~\wht{\varphi}(p) = {\cal D}^1(\,\widetilde{t}\,) \,\varphi(p) \\[0.4cm]
 \,^{(a,A)} \psi([\,p\,])~=~e^{i a \cdot p}~{\cal D}^1 ([\,p\,]^{-1} A\,[\,A^{-1}p\,]) \,\psi([\,A^{-1} p\,]) \\ [0.4cm]
 \,^{(a,A)}\varphi(p) =e^{ia \cdot p}~ {\cal D}^1(A)\, \varphi (A^{-1} p) ,~~~^{(a,A)}\wht{\varphi} (p) = e^{i a \cdot p}~ {\cal D}^1 (A^{\dagger -1})\, \wht{\varphi} (A^{-1} p) \end{array}   \enq

\vv \nin Donnons ici explicitement l'expression de la matrice ${\cal D}^1(M)$ o\`u $M$ est une matrice de $SL(2,C)$ :

\beq M = \left( \begin{array}{cc} \alpha & \beta \\ \gamma & \delta \end{array} \right) ~\longrightarrow~  {\cal D}^1(M) = \left( \begin{array}{ccc} \alpha^2 & \sqrt{2}\, \alpha \, \beta & \beta^2 \\ 
\sqrt{2}\, \alpha \, \gamma & \alpha \, \delta + \beta\, \gamma & \sqrt{2}\, \beta\, \delta \\
\gamma^2 & \sqrt{2}\, \gamma\, \delta & \delta^2 \end{array} \right) \enq

\vv \nin Il est connu que la relation matricielle $\psi' = {\cal D}^1(M)\, \psi $,  o\`u $\psi$ et $\psi'$ sont des vecteurs \`a trois composantes et $M$ une matrice $2\times 2$ inversible, peut \^etre transcrite en termes de matrices $2\times2$. Repr\'esentons en effet $\psi$ et $\psi'$ par les matrices 
\vv
\beq \begin{array}{c}
 {\bf \Psi} = \left( \begin{array}{cc} \psi_0 & - \sqrt{2}\, \psi_1 \\ \sqrt{2}\, \psi_{-1} &  - \psi_0 \end{array} \right),~{\bf \Psi}' = \left( \begin{array}{cc} \psi'_0 & - \sqrt{2}\, \psi'_1 \\ \sqrt{2}\, \psi'_{-1} &  - \psi'_0 \end{array} \right)~;~~~{\rm on~a~alors} \\[0.7cm]
 {\bf \Psi}' = M\,{\bf \Psi}\,M^{-1} \end{array} 
\enq

\vv \nin En repr\'esentant les vecteurs $\psi([\,p\,])$, $\phi(p)$ et $\wht{\varphi}(p)$ par les matrices  
\vskip 0.1cm
\beq \begin{array}{c}
 {\bf \Psi} = \left( \begin{array}{cc} \phi_0 & - \sqrt{2}\, \phi_1 \\ \sqrt{2}\, \phi_{-1} &  - \phi_0 \end{array} \right)\, , \\ [0.7cm]
 {\bf \Phi} = \left( \begin{array}{cc} \varphi_0 & - \sqrt{2}\, \varphi_1 \\ \sqrt{2}\, \varphi_{-1} &  - \varphi_0 \end{array} \right),~~\wht{{\bf \Phi}} = \left( \begin{array}{cc} \wht{\varphi}_0 & - \sqrt{2}\, \wht{\varphi}_1 \\ \sqrt{2}\, \wht{\varphi}_{-1} &  - \wht{\varphi}_0 \end{array} \right),  \end{array}   ~\label{mat22} \enq

\vv \nin respectivement, les \'equations (\ref{chgt-tetr}),  (\ref{ampl-spin}), (\ref{dirac-s1}) et (\ref{trans1}) prennent la forme\footnote{A la quatri\`eme ligne, on tient compte du fait que $R_W = [\,p\,]^{-1} A\,[\,A^{-1}p\,]$ est une matrice de rotation : $R^{-1}_W = R^{\dagger}_W = [\,A^{-1}p\,]^\dagger \,A^\dagger\,[\,p\,]^{\dagger -1}$.} : 

\vskip -0.2cm
\beq \begin{array}{c}
{\bf \Psi}([\,p\,]^\prime) = [\,p\,]^{\prime -1} [\,p\,]\,{\bf \Psi}([\,p\,])\, [\,p\,]^\dagger \, [\,p\,]^{\prime\, \dagger -1} \\ 
[0.35cm]
 {\bf \Phi}(p) = [\,p\,]\,  {\bf \Psi}([\,p\,])~[\,p\,]^{-1},~~~~\wht{\bf \Phi}(p) =  [\,p\,]^{\dagger -1}\,  {\bf \Psi}([\,p\,])~[\,p\,]^{\dagger} \\
[0.15cm]
 {\bf \Phi}(p) =   \,t \hskip -0.4 cm \begin{array}[m]{c} ~ \\ \stackrel{~~~}{\sim}
\end{array} \hskip -0.12cm\wht{\bf \Phi}(p)~\widetilde{t}~~~~{\rm ou}~~~~ {\bf \Phi}(p) ~ \,t \hskip -0.4 cm \begin{array}[m]{c} ~ \\ \stackrel{~~~}{\sim}
\end{array}  \hskip -0.15cm=  \,t \hskip -0.4 cm \begin{array}[m]{c} ~ \\ \stackrel{~~~}{\sim}
\end{array} \hskip -0.12cm\wht{\bf \Phi}(p)  \\
[0.4cm]
 ~^{(a,A)} {\bf \Psi}([\,p\,]) = e^{ia \cdot p}~[\,p\,]^{-1} A\,[\,A^{-1}p\,]~ {\bf \Psi}([\,A^{-1} p\,])\, [\,A^{-1}p\,]^\dagger\,A^\dagger\,[\,p\,]^{\dagger -1} \\ 
[0.3cm]
\,^{(a,A)} {\bf \Phi}(p) = e^{ia \cdot p}~A~ {\bf \Phi}(A^{-1} p)\, A^{-1},~~~ ^{(a,A)} \wht{{\bf \Phi}}(p) = e^{ia \cdot p}~A^{\dagger -1}~ \wht{{\bf \Phi}}(A^{-1} p)\, A^\dagger \end{array}
\label{relat-mat1} \enq

\vv \nin Les matrices (\ref{mat22}) peuvent aussi s'\'ecrire sous forme ``cart\'esienne" ; on a par exemple 

\beq {\bf \Psi} = \,V \hskip -0.52 cm \begin{array}[m]{c} ~ \\ \stackrel{~~~}{\sim}
\end{array}\hskip -0.2cm = V^0 ~+ \Vec{\,V\,}\hskip -0.1cm\cdot\hskip -0.1cm \Vec{\,\tau\,} ~~~{\rm avec}\enq 
$$ V^0 = 0,~~V^1 = V_x = 
\di{{\psi_{-1} - \psi_1}\over \sqrt{2}},~~V^2 = V_y =  - i \, \di{{\psi_{-1} + \psi_1}\over \sqrt{2}},~~V^3 = V_z = \psi_0 $$

\vv \nin La premi\`ere des relations (\ref{relat-mat1}) montre que la matrice 

\beq \fbox{\fbox{\rule[-0.4cm]{0cm}{1cm}~$~{\cal A} \hskip -0.5 cm \begin{array}[m]{c} ~ \\ \stackrel{~~~}{\sim}
\end{array}\hskip -0.2cm = {\cal A}^0 ~+ \Vec{\,{\cal A}\,}\hskip -0.1cm\cdot\hskip -0.1cm \Vec{\,\tau\,} = [\,p\,]\, {\bf \Psi}([\,p\,])\, [\,p\,]^\dagger ~ $~}} \label{4-pot11}  \enq

\vv \nin est manifestement ind\'ependante du choix de la t\'etrade $[\,p\,]$, tandis que la quatri\`eme de ces relations donne sa loi de transformation :

\beq \fbox{\fbox{\rule[-0.4cm]{0cm}{1cm}~$\,^{(a,A)}{\cal A} \hskip -0.5 cm \begin{array}[m]{c} ~ \\ \stackrel{~~~}{\sim}
\end{array}\hskip -0.2cm(p) =  e^{i a \cdot p}~A~ {\cal A} \hskip -0.5 cm \begin{array}[m]{c} ~ \\ \stackrel{~~~}{\sim}
\end{array}\hskip -0.2cm(A^{-1} p)\,A^\dagger  ~ $~}}   \enq

\vv \nin de laquelle on d\'eduit que les quatre grandeurs ${\cal A}^0$, ${\cal A}^k\, (k=1,2,3)$ se transforment bien comme les composantes contravariantes d'un champ de 4-vecteurs ${\cal A}(p)$. Notons 

\beq e^{(0)}([\,p\,]) \equiv z,~~~e^{(\pm)}([\,p\,]) \equiv \mp \di{1\over \sqrt{2}}\,(x \pm i y)  \enq 

\vv \nin la base de tri-vecteurs associ\'es \`a $t=p/m$ dans la t\'etrade $[\,p\,]$. Comme
\vskip -0.25cm

\beq \begin{array}{c}
 {\bf \Psi}= \phi_0\, \tau_3 - \sqrt{2}~ \phi_1\, \di{1\over 2}\left[\, \tau_1 + i \tau_2\, \right] + \sqrt{2}\, \di{1\over 2}\,\phi_{-1}\, \left[\, \tau_1 - i \tau_2\, \right] \\
 {\rm et~que}~~~~~[\,p\,]\, \tau_3\,[\,p\,]^\dagger = e^{(0)} \hskip -0.87 cm \begin{array}[m]{c} ~ \\ \stackrel{~~~}{\sim}
\end{array} ~\, ([\,p\,]) ,~~~
\mp \,[\,p\,]\, \di{1\over \sqrt{2}}\,\left[\tau_1 \pm i \tau_2\,\right]\,[\,p\,]^\dagger =~ e^{(\pm)} \hskip -0.87 cm \begin{array}[m]{c} ~ \\ \stackrel{~~~}{\sim}
\end{array}~\, ([\,p\,]), 
\end{array} \enq

\nin les composantes covariantes du 4-vecteur ${\cal A}$ s'expriment donc sous la forme

\beq \fbox{\fbox{\rule[-0.4cm]{0cm}{1cm}~$ ~{\cal A}_\mu(p) = \di{\sum_{\lambda = \pm1, 0}} ~e^{(\lambda)}_\mu([\,p\,])~\phi_\lambda ([\,p\,])~ $~}}  \label{4-pot12} \enq

\vv \nin montrant, d'une part, que ce 4-vecteur contient toutes les informations caract\'erisant l'\'etat $|\, \phi>$ et, d'autre part, qu'il est {\em orthogonal \`a $p$} :

\beq \fbox{\fbox{\rule[-0.4cm]{0cm}{1cm}~$ ~ p_\mu\, {\cal A}^\mu(p) = 0 ~ $~}}   \enq

\vv \nin C'est le 4-vecteur ``4-potentiel" recherch\'e dont l'utilisation pour d\'ecrire les \'etats de spin de la particule de spin 1 est \'equivalente \`a celle fournie par les amplitudes spinorielles. A partir de l\`a, on peut d\'efinir les tenseurs antisym\'etriques usuels, duaux l'un vis-\`a-vis de l'autre et orthogonaux \`a $p$\footnote{A noter que le tenseur $F_{\mu \nu}(p) = -i \left[\, p_\mu\, {\cal A}_\nu - p_\nu\, {\cal A}_\mu \right]$ \'equivaut au tenseur $F_{\mu \nu} = \partial_\mu\,{\cal A}_\nu - \partial_\nu\, {\cal A}_\mu$ dans le cas d'une onde plane ${\cal A}_\mu(x) = e^{-i p\cdot x}\,{\cal A}_\mu(p)$.}  :  

\beq F_{\mu \nu} = -i \left( p_\mu\, {\cal A}_\nu - p_\nu\, {\cal A}_\mu \right),~~~G_{\mu \nu} = \di{1\over 2}\, \epsilon_{\mu \nu \rho \sigma}\, F^{\rho \sigma} = -  i\, \epsilon_{\mu \nu \alpha \beta}\, p^\alpha \, {\cal A}^\beta \label{def-FG} \enq 

\vv \nin et de plus {\em invariants de jauge}, c'est-\`a-dire, insensibles \`a la {\em transformation de jauge} consistant \`a effectuer le changement : 

\beq \fbox{\rule[-0.4cm]{0cm}{1cm}~$ ~{\cal A}_\mu \longrightarrow  {\cal A}_\mu + c\, p_\mu~ $~}   \enq

\vv \nin $c$ \'etant un nombre quelconque. Dans le cas o\`u ${\cal A}$ est bien orthogonal \`a $p$, on a, inversement,  

\beq {\cal A}_\mu = -\di{i \over{2 m^2}}\, \epsilon_{\mu \nu \alpha \beta}\, p^\nu\, G^{\alpha \beta} \enq

\vv \nin et dans ce cas, on a aussi les relations

\beq {\cal A} \hskip -0.5 cm \begin{array}[m]{c} ~ \\ \stackrel{~~~}{\sim}
\end{array}\hskip-0.15cm(p) =\, {\bf \Phi}(p) ~ \,t \hskip -0.41 cm \begin{array}[m]{c} ~ \\ \stackrel{~~~}{\sim}
\end{array} \hskip -0.2cm = \,t \hskip -0.4 cm \begin{array}[m]{c} ~ \\ \stackrel{~~~}{\sim}
\end{array} \hskip -0.12cm\wht{\bf \Phi}(p)  \label{matA1} \enq

\vv \nin Pour achever de montrer l'\'equivalence entre la description des \'etats de la particule au moyen de bi-spineurs construits au moyen d'amplitudes spinorielles, et celle utilisant un 4-vecteur, consid\'erons le produit scalaire hermitique des tenseurs ${F_a}_{\mu \nu}(p)$ et ${F_b}_{\mu \nu}(p)$ respectivement associ\'es aux \'etats $|\,a>$ et $|\,b>$ de la particule :

$$ {F^\star_a}_{\mu \nu} \, {F_b}^{\mu \nu} = - G^\star_{a \,\mu \nu}\,G^{\mu \nu}_b = 2\,m^2\, {\cal A}^\star_a \cdot {\cal A}_b = - 2 \,m^2\, \di{\sum_\lambda}\, {\phi}^\star_{a \lambda} \, \phi_{b \lambda}  $$

\nin Dans la repr\'esentation bi-spinorielle des \'etats de la particule\footnote{ ITL, Chap. 6.}, les bi-spineurs \`a 6 composantes 

$$ \Phi_a = \di{1\over{\sqrt{2}}} \left( \begin{array}{c} \phi_a \\ \wht{\phi}_a \end{array} \right)~~~{\rm et}~~~~\Phi_a = \di{1\over \sqrt{2}} \left( \begin{array}{c} \phi_b \\ \wht{\phi}_b \end{array} \right) $$

\vv \nin ont pour produit scalaire invariant\footnote{$\Gamma_0 = \left(\begin{array}{cc} 0 & 1_3 \\ 1_3 & 0 \end{array} \right)$ o\`u $1_3$ est la matrice $3\times 3$ unit\'e.} 

$$ \ov{\Phi}_a\, \Phi_b = \Phi^\dagger_a\, \Gamma_0\, \Phi_b =\di{1\over 2} \left\{ \phi^\dagger_a\, \wht{\phi}_b + \wht{\phi}^{\,\dagger}_a\, \phi_b\right\} =  \di{\sum_\lambda}\, {\phi}^\star_{a \lambda} \, \phi_{b \lambda} $$

\nin On a donc :

\beq \fbox{\rule[-0.4cm]{0cm}{1cm}~$~\di{1\over{2 m^2}}~{F^\star_a}_{\mu \nu} \, {F_b}^{\mu \nu} =  \, {\cal A}^\star_a \cdot {\cal A}_b = -\, \ov{\Phi}_a\, \Phi_b ~ $~}  \label{equival-prod-scal} \enq

\vv \nin Il reste enfin \`a consid\'erer le cas o\`u l'\'etat de la particule est lui-m\^eme un \'etat $|\, [\,p\,], \sigma >$, pour lequel les amplitudes spinorielles contiennent une distribution de Dirac, puisqu'alors  

$$ \phi_{\sigma'}([\,p'\,]) = ~< \sigma', [\,p'\,]\,|\, [\,p\,], \sigma > ~=\, (2 \pi)^2\, 2\, p_0\, \delta(\Vec{\,p'\,} - \Vec{\,p\,})\, \delta_{\sigma, \sigma'}  $$

\vv \nin Dans ce cas, en extrayant la distribution de Dirac et d'autres facteurs, on voit que le 4-vecteur (\ref{4-pot12}) correspondant est simplement \'egal au 4-vecteur de polarisation $e^{(\sigma)}([\,p\,])$.   

\vvv
\vvv
\vv \nin \leftpointright~Passons ensuite au cas d'une particule vectorielle de masse nulle, comme le photon. Du point de vue du groupe de Poincar\'e restreint, les \'etats d'une telle particule n'appartiennent qu'\`a une seule des deux repr\'esentations irr\'eductibles, donc {\em ind\'ependantes}, $[\,m=0, \lambda = +1 \,]$ et $[\, 
m=0, \lambda = -1\,]$ o\`u $\lambda$, l'\'equivalent d'une projection de spin et appel\'e ici h\'elicit\'e de la particule, ne prend pour chacune qu'une seule valeur, invariante relativiste. Cependant, l'op\'eration de parit\'e changeant le signe de $\lambda$, une repr\'esentation irr\'eductible du groupe de Poincar\'e complet (comprenant les sym\'etries parit\'e, renversement du sens du temps et r\'eflexion totale) doit \^etre constitu\'ee par leur somme directe. C'est pourquoi le photon, pour lequel l'op\'eration de parit\'e est une sym\'etrie,  poss\`ede deux \'etats d'h\'elicit\'es $\lambda=-1$ et $\lambda = +1$ oppos\'ees. Dans ce qui suit, nous noterons $\ell$ la 4-impulsion de la particule, avec ici $\ell^2 =0$. Pour un \'etat quelconque $|\, \phi>$, il n'existe donc que deux amplitudes (\ref{amplit-1}) : 

\beq \phi_1 ([\,\ell\,]) = ~< +1, [\, \ell\,]\,|\,\phi>~~{\rm et}~~~\phi_{-1} ([\,\ell\,]) = ~< -1, [\, \ell\,]\,|\,\phi> \label{ampl-0} \enq

\vv \nin dont la d\'ependance vis-\`a-vis du choix de la t\'etrade $[\,\ell\,]$ est r\'ev\'el\'ee par 

\beq   \phi_\lambda ([\,\ell\,]^\prime) = {\cal D}^1_{\lambda \lambda}([\, \ell\,]^{\prime -1}\,[\,\ell\,])\, \phi_\lambda ([\, \ell\,]) \label{chgt-tet-0} \enq

\vv \nin Les amplitudes spinorielles, ind\'ependantes de ce choix, sont de deux types et d\'efinies par 

\beq \varphi^{+}_\sigma (\ell) = {\cal D}^1_{\sigma, +1}([\,\ell\,])\, \phi_1([\,\ell\,]) ~~{\rm et}~~~\varphi^{-}_\sigma (\ell) = {\cal D}^1_{\sigma, -1}([\,\ell\,]^{\dagger -1})\, \phi_{-1}([\,\ell\,]) \label{ampl0-1} \enq

\vv \nin avec $\sigma = -1,\,0$ ou $+1$. Elles v\'erifient les \'equations 

\beq {\cal D}^1(~\ell \hskip -0.42 cm \begin{array}[m]{c} ~ \\ \stackrel{~~~}{\sim}
\end{array}\hskip -0.15cm ) \,\varphi^+(\ell) = 0\, , ~~~ {\cal D}^1 (\,\widetilde{\ell}~) \, \varphi^{-}(\ell) = 0 \label{dirac-s0} \enq

\nin Toutes ces amplitudes ont pour lois de transformation  

\beq \begin{array}{c}
 \,^{(a,A)}\phi_\lambda([\, \ell\,])~=~e^{i a \cdot \ell}~{\cal D}^1_{\lambda \lambda}([\,\ell\,]^{-1} A\,[\,A^{-1}\ell\,])\, \phi_\lambda([\, A^{-1}\ell\,]) \\[0.4cm]
\,^{(a, A)}\varphi^+_\sigma(\ell) =  e^{i a \cdot \ell}~{\cal D}^1_{\sigma \sigma'}(A)\, \varphi^+_{\sigma'}(A^{-1} \ell)\,,~~ \,^{(a, A)}\varphi^{-}_\sigma(\ell) =  e^{i a \cdot \ell}~{\cal D}^1_{\sigma \sigma'}(A^{\dagger -1})\, \varphi^{-}_{\sigma'}(A^{-1} \ell) \end{array} 
\label{transfo-0} \enq

\vv \nin (avec sommation sur $\sigma'$). 

\vv \nin Rappelons ici que le petit groupe d'un 4-vecteur du genre lumi\`ere $\ell$ n'est plus $SU(2)$ mais un groupe isomorphe au groupe des d\'eplacements dans un 2-plan euclidien, $P(2)$. Agissant sur les 4-vecteurs, il comprend un sous-groupe ab\'elien de rotations dans un 2-plan $(x,y)$ orthogonal \`a $\ell$ et un autre sous-groupe ab\'elien translatant parall\`element \`a $\ell$ les 4-vecteurs de ce plan (groupe de jauge de $\ell$). Les matrices de $SL(2,C)$ appartenant au petit groupe du 4-vecteur isotrope de r\'ef\'erence $\basic{\ell} ~=~ \kappa (1,0,0,1)$ sont de la forme indiqu\'ee par la formule (\ref{ML}) :

$$ M(\zeta, \theta) = \left( \begin{array}{cc} e^{-i \theta/2} & \zeta\, e^{i \theta/2} \\ 0 & e^{i \theta/2} \end{array} \right) \label{mat-pgr-01} $$

\vv \nin $\zeta$ \'etant un nombre complexe quelconque. Puisque $[\, \ell\,]^{\prime -1}\,[\,\ell\,]$ appartient \`a ce petit groupe, on a notamment ${\cal D}^j_{m m}(M(\zeta, \theta)) = e^{-i m \theta}$. En se reportant \`a (\ref{chgt-tet-0}), on voit qu'un changement de t\'etrade provoque un simple changement de phase des amplitudes (\ref{ampl-0}).

\vv \nin A l'\'evidence, l'\'equivalent pour masse nulle du 4-vecteur (\ref{4-pot12}) doit \^etre le suivant :  

\beq \fbox{\fbox{\rule[-0.4cm]{0cm}{1cm}~$~{\cal A}_\mu(\ell) = ~e^{(+)}_\mu([\,\ell\,])~\phi_1 ([\,\ell\,]) +  ~e^{(-)}_\mu([\,\ell\,])~\phi_{-1} ([\,\ell\,])~ $~}}  \label{4-pot-01} \enq

\vv \nin qui est manifestement orthogonal \`a $\ell$. Examinons tout d'abord sa d\'ependance vis-\`a-vis du choix de t\'etrade. D\'efinissons 

\beq {\bf \Psi}([\,\ell\,]) = \left( \begin{array}{cc} 0 & - \sqrt{2}\, \phi_1([\,\ell\,]) \\
\sqrt{2}\, \phi_{-1}([\,\ell\,]) & 0 \end{array} \right) \enq

\vv \nin Compte tenu de ce que 

\beq 
\begin{array}{c}
e \hskip -0.45 cm \begin{array}[m]{c} ~ \\ \stackrel{~~~}{\sim}
\end{array}\hskip -0.2cm^{(+)}([\,\ell\,]) = - \di{1\over \sqrt{2}}\, [\,\ell\,]\,\left(\tau_1 + i \tau_2 \right) \,[\,\ell\,]^\dagger = - \sqrt{2}\, [\,\ell\,]\, \left( \begin{array}{cc} 0 & 1 \\ 0 & 0 \end{array} \right) \, [\, \ell\,]^\dagger  \\ [0.4cm]
e \hskip -0.45 cm \begin{array}[m]{c} ~ \\ \stackrel{~~~}{\sim}
\end{array}\hskip -0.2cm^{(-)}([\,\ell\,]) =  \di{1\over \sqrt{2}}\, [\,\ell\,]\,\left(\tau_1 - i \tau_2 \right) \,[\,\ell\,]^\dagger = \sqrt{2}\, [\,\ell\,]\, \left( \begin{array}{cc} 0 & 0 \\ 1 & 0 \end{array} \right) \, [\, \ell\,]^\dagger 
\end{array}
\enq

\vv \nin la transcription de (\ref{4-pot-01}) en matrices $2 \times 2$ est  

\beq  {\cal A} \hskip -0.5 cm \begin{array}[m]{c} ~ \\ \stackrel{~~~}{\sim}
\end{array}\hskip -0.2cm(\ell) = [\,\ell\,]\, {\bf \Psi}([\,\ell\,])\, [\,\ell\,]^\dagger       \enq

\vv \nin Posons alors 

\beq M = [\,\ell\,]^{\prime\, -1}\,[\,\ell\,] ~=~\left( \begin{array}{cc} \alpha & \beta \\ 0 & \delta \end{array} \right) ~~~{\rm avec}~~~\alpha = e^{-i \theta/2} = \delta^\star,~~~\beta = \zeta\, \delta \label{mat-chgt-tet-0} \enq

\vv \nin et calculons : 

$$ M\, {\bf \Psi}([\,\ell\,])\, M^\dagger = \sqrt{2}\,\left( - \alpha^2\, \zeta^\star\, \phi_1 +\zeta \,\delta^2 \, \phi_{-1} \right)\, \left( \begin{array}{cc} 1 & 0 \\ 0 & 0 \end{array} \right)  + \left( \begin{array}{cc} 0 & - \alpha^2 \, \sqrt{2}\, \phi_1 \\ \delta^2 \, \sqrt{2}\, \phi_{-1} & 0 \end{array} \right) $$

\vv \nin Or, d'une part, d'apr\`es (\ref{chgt-tet-0}), on a $\phi_1([\,\ell\,]^\prime) = \alpha^2\, \phi_1([\,\ell\,])$, $\phi_{-1}([\,\ell\,]^\prime) = \delta^2\, \phi_{-1}([\,\ell\,])$ ; d'autre part\footnote{La t\'etrade $[\,\ell\,]$ est telle que $[\, \ell\,]\,[\,\ell\,]^\dagger~ = ~ t \hskip -0.4 cm \begin{array}[m]{c} ~ \\ \stackrel{~~~}{\sim}
\end{array}$\hskip -0.2cm, $[\, \ell\,]\,\tau_3\,[\,\ell\,]^\dagger~ = ~ z \hskip -0.4 cm \begin{array}[m]{c} ~ \\ \stackrel{~~~}{\sim}
\end{array}$\hskip -0.2cm, $\ell = \kappa ( t + z)$, $\kappa = \ell \cdot t$.}, 

$$  \left( \begin{array}{cc} 1 & 0 \\ 0 & 0 \end{array} \right) = \di{1\over{2 \kappa}} ~
\basic{\ell} \hskip -0.52 cm \begin{array}[m]{c} ~ \\ \stackrel{~~~}{\sim}
\end{array}\hskip -0.2cm $$

\vv \nin En posant $\xi =(\zeta^\star\, \phi_1([\,\ell\,]^\prime) - \zeta\,\phi_{-1}([\,\ell\,]^\prime) /(2 \kappa)$, on obtient ainsi

$$ M\, {\bf \Psi}([\,\ell\,])\, M^\dagger = - \xi\, \basic{\ell} \hskip -0.52 cm \begin{array}[m]{c} ~ \\ \stackrel{~~~}{\sim}
\end{array} \hskip-0.1cm +\, {\bf \Psi}([\,\ell\,]^\prime) $$

\vv \nin et finalement 

\beq \fbox{\rule[-0.4cm]{0cm}{1cm}~$~ {\cal A} \hskip -0.5 cm \begin{array}[m]{c} ~ \\ \stackrel{~~~}{\sim}
\end{array}\hskip -0.2cm([\,\ell\,]^\prime)~=~ {\cal A} \hskip -0.5 cm \begin{array}[m]{c} ~ \\ \stackrel{~~~}{\sim}
\end{array}\hskip -0.2cm([\,\ell\,]) \, +\, \xi~\ell\hskip -0.42 cm \begin{array}[m]{c} ~ \\ \stackrel{~~~}{\sim}
\end{array} $}  \enq

\vv \nin Ainsi, le changement de t\'etrade provoque une translation du vecteur ${\cal A}$ parall\`element \`a $\ell$. Ce r\'esultat n'est pas surprenant, compte tenu des propri\'et\'es du petit groupe de $\ell$ rappel\'ees plus haut. On peut d\'efinir un 4-potentiel ind\'ependant de la t\'etrade en lui imposant de rester dans le 2-plan engendr\'e par les 4-vecteurs $e^{(+)}$ et $e^{(-)}$. Comme il est d\'ej\`a orthogonal \`a $\ell$, cela \'equivaut \`a lui imposer la condition subsidiaire 

\beq \fbox{\rule[-0.4cm]{0cm}{1cm}~$\, {\cal A} \cdot t = 0 ~~$} \label{subsid} \enq

\vv \nin Les changements de t\'etrade reviennent alors \`a de simples rotations dans le 2-plan orthogonal \`a $\ell$ et $t$ (qui est aussi celui orthogonal \`a $t$ et $z$), ce qui \'equivaut aussi \`a poser $\zeta = 0$ dans la matrice correspondante (\ref{mat-pgr-01}). Dans cette transformation, le 4-vecteur $e^{(\lambda)}$ subit un changement de phase oppos\'e \`a celui de l'amplitude $\phi_\lambda$ \`a laquelle il est associ\'e, laissant ainsi le 4-vecteur ${\cal A}$ inchang\'e. 

\vv \nin La loi de transformation (\ref{transfo-0}) fait intervenir la matrice $N = [\,\ell\,]^{-1}\, A\,[\,A^{-1}\ell\,]$ qui elle aussi appartient au petit groupe de $\basic{\ell}$. En cons\'equence, la transformation correspondante du 4-vecteur (\ref{4-pot-01}) provoque encore une translation parall\`element \`a $\ell$. Le lecteur v\'erifiera que l'on a 

\beq \begin{array}{c} 

\,^{(a,A)}{\cal A} \hskip -0.5 cm \begin{array}[m]{c} ~ \\ \stackrel{~~~}{\sim}
\end{array}\hskip -0.2cm([\,\ell\,]) ~=~e^{i a \cdot \ell}~A~ {\cal A} \hskip -0.5 cm \begin{array}[m]{c} ~ \\ \stackrel{~~~}{\sim}
\end{array}\hskip -0.2cm([A^{-1}\ell\,]) \,A^\dagger + \,\eta~ \ell \hskip -0.42 cm \begin{array}[m]{c} ~ \\ \stackrel{~~~}{\sim}
\end{array}\hskip -0.2cm \\ [0.5cm]
 e^{(+)}_\mu([\, A \ell\,]) = \Lambda^\nu_{\, \mu}(A) \left[\,e^{i \theta}\,e^{(+)}_\nu([\,\ell\,]) + \zeta^\star\, \ell_\nu \right], \\ [0.5cm]  
e^{(-)}_\mu([\, A \ell\,]) = \Lambda^\nu_{\, \mu}(A) \left[\,e^{-i \theta}\,e^{(-)}_\nu([\,\ell\,]) + \zeta\, \ell_\nu \right] 
\end{array} \enq

\vv \nin Ici encore, si l'on se restreint aux 4-vecteurs ${\cal A}$ v\'erifiant la condition subsidiaire (\ref{subsid}), on trouve pour ceux-ci la loi de transformation usuelle des 4-vecteurs :

\beq \fbox{\fbox{\rule[-0.4cm]{0cm}{1cm}~$ \,^{(a,A)}{\cal A} \hskip -0.5 cm \begin{array}[m]{c} ~ \\ \stackrel{~~~}{\sim}
\end{array}\hskip -0.2cm([\,\ell\,]) ~=~e^{i a \cdot \ell}~A\, {\cal A} \hskip -0.5 cm \begin{array}[m]{c} ~ \\ \stackrel{~~~}{\sim}
\end{array}\hskip -0.2cm([A^{-1}\ell\,]) \,A^\dagger  \,$~}}  \enq

\vv \nin Dans le cas de la masse nulle, on introduit aussi des tenseurs antisym\'etriques et invariants de jauge similaires \`a ceux d\'efinis en (\ref{def-FG}) o\`u $\ell$ prend la place de $p$. Cependant, lorsqu'on veut r\'eexprimer de fa\c{c}on unique le 4-vecteur ${\cal A}$ qui y appara\^it en fonction du tenseur $G_{\mu \nu}$, il ne suffit plus ici d'imposer que ${\cal A}$ soit orthogonal \`a $\ell$. En effet, comme $\ell$ est isotrope, la transformation de jauge ${\cal A}^\prime = {\cal A} + \xi\, \ell$ conduit \`a un nouveau 4-vecteur ${\cal A}^\prime$ admissible, puisque $\ell \cdot {\cal A}^\prime = \ell \cdot {\cal A} = 0$. Une condition suppl\'ementaire doit donc \^etre introduite. Si l'on adopte (\ref{subsid}),  on obtient alors\footnote{$F_{\mu \nu} = -\di{1\over 2}\,\epsilon_{\mu \nu \rho \sigma}\, G^{\rho \sigma}$.} 

\beq {\cal A}_\alpha = \di{i \over {2 \kappa}}~\epsilon_{\alpha \beta \mu \nu}~t^\beta\,G^{\mu \nu} \enq

\vv \nin La transcription des formules (\ref{ampl0-1}) et (\ref{dirac-s0}) s'effectue comme suit. Posons 

\beq 
\begin{array}{c} 

[\,\ell\,] = \left( \begin{array}{cc} a & b \\ c & d \end{array} \right), \\[0.4cm]
K_{+} = \di{1\over 2}\, \left( 1 + \tau_3 \right) = \left( \begin{array}{cc} 1 & 0 \\ 0 & 0 \end{array} \right),~~~K_{-} =  \di{1\over 2}\, \left( 1 - \tau_3 \right) = \left( \begin{array}{cc} 0 & 0 \\ 0 & 1 \end{array} \right) 
\end{array} \enq

\vv \nin D'apr\`es (\ref{ampl0-1}), on a 

$$ \varphi^{+}_1 = a^2\, \phi_1,~~\varphi^{+}_0 = \sqrt{2}\, a\,c\, \phi_1,~~\varphi^{+}_{-1} = c^2\, \phi_1 $$
$$ \varphi^{-}_1 = [c^\star]^2\, \phi_{-1},~~\varphi^{-}_0 = - \sqrt{2}\, a^\star\,c^\star\, \phi_{-1},~~\varphi^{+}_{-1} = [a^\star]^2\, \phi_{-1} $$

\vv \nin ce qui se traduit par les matrices

\beq \begin{array}{c} 
{\bf \Phi}^{+}(\ell) = \left( \begin{array}{cc} \varphi^{+}_0 & - \sqrt{2}\, \varphi^{+}_1 \\ \sqrt{2}\, \varphi^{+}_{-1} & - \varphi^{+}_0 \end{array} \right) = [\,\ell\,]\,K_{+}\, {\bf \Psi}\, K_{-}\, [\,\ell\,]^{-1}  \\ [0.5cm]
{\bf \Phi}^{-}(\ell) = \left( \begin{array}{cc} \varphi^{-}_0 & - \sqrt{2}\, \varphi^{-}_1 \\ \sqrt{2}\, \varphi^{-}_{-1} & - \varphi^{-}_0 \end{array} \right) =  [\,\ell\,]^{\dagger -1} \,K_{-}\, {\bf \Psi}\, K_{+}\, [\,\ell\,]^\dagger 
\end{array} \label{mat-FIPM} \enq

\vv \nin Posant $M = [\,\ell\,]^{\prime \, -1}\, [\,\ell\,]$, on trouve 

$$ M\,K_{+}\,{\bf \Psi}([\,\ell\,])\,K_{-}\, M^{-1} = K_{+}\,{\bf \Psi}([\,\ell\,]^\prime)\, K_{-} ,~~~M^{\dagger\,-1}\,K_{-}\,{\bf \Psi}([\,\ell\,])\,K_{+}\, M^\dagger = K_{-}\,{\bf \Psi}([\,\ell\,]^\prime)\, K_{+}$$ 

\vv \nin d'o\`u l'on d\'eduit que les deux matrices ${\bf \Phi}^{+}(\ell)$ et ${\bf \Phi}^{-}(\ell)$ sont bien ind\'ependantes du choix de la t\'etrade associ\'ee \`a $\ell$. Par une transformation de Poincar\'e $(a,A)$, elles deviennent\footnote{Le v\'erifier.}

\beq \begin{array}{c}
\,^{(a,A)}{\bf \Phi}^{+} (\ell) = e^{i a \cdot \ell}~A\,{\bf \Phi}^{+} (A^{-1} \ell)\,A^{-1} \\ [0.3cm]
\,^{(a,A)}{\bf \Phi}^{-} (\ell) = e^{i a \cdot \ell}~A^{\dagger\,-1} \,{\bf \Phi}^{-}(A^{-1} \ell)\,A^\dagger 
\end{array} \enq

\vv \nin En outre, comme 

$$ [\,\ell\,]\,K_{+}\,[\,\ell\,]^\dagger = \di{1\over {2 \kappa}} ~\ell \hskip -0.43 cm \begin{array}[m]{c} ~ \\ \stackrel{~~~}{\sim}
\end{array}~~{\rm soit}~~~[\,\ell\,]^{-1} ~\ell \hskip -0.43 cm \begin{array}[m]{c} ~ \\ \stackrel{~~~}{\sim}
\end{array} \hskip-0.2cm= 2 \kappa [\,\ell\,]\,K_{+} ~~{\rm ou}~~~\ell \hskip -0.43 cm \begin{array}[m]{c} ~ \\ \stackrel{~~~}{\sim}
\end{array} \hskip-0.1cm [\,\ell\,]^{\dagger\,-1} =2 \kappa\,[\,\ell\,]\,K_{+}  $$
$${\rm et}~~~K_{+}\,K_{-} = K_{-} K_{+} = 0$$

\nin lesdites matrices satisfont les \'equations  

\beq \fbox{\rule[-0.4cm]{0cm}{1cm}~$ {\bf \Phi}^{+}(\ell)~~\ell \hskip -0.43 cm \begin{array}[m]{c} ~ \\ \stackrel{~~~}{\sim}
\end{array} \hskip -0.2cm = ~\ell \hskip -0.43 cm \begin{array}[m]{c} ~ \\ \stackrel{~~~}{\sim}
\end{array} {\bf \Phi}^{-}(\ell) = 0  $~}  \enq

\vv \nin traduisant en termes de matrices les \'equations de Dirac (\ref{dirac-s0}) ;  de plus, comme 

$$ [\,\ell\,]\, K_{+} = \di{1\over {2 \kappa}}\, ~\ell \hskip -0.43 cm \begin{array}[m]{c} ~ \\ \stackrel{~~~}{\sim}
\end{array} \hskip-0.1cm [\,\ell\,]^{\dagger\,-1} ~~{\rm ou}~~~K_{+}\,[\ell\,]^\dagger = \di{1\over{2\kappa}}\, [\,\ell\,]^{-1} ~\ell \hskip -0.43 cm \begin{array}[m]{c} ~ \\ \stackrel{~~~}{\sim}
\end{array} ~~{\rm et~que}~~~ \widetilde{\ell} ~\ell \hskip -0.43 cm \begin{array}[m]{c} ~ \\ \stackrel{~~~}{\sim}
\end{array} \hskip-0.2cm=\,\ell \hskip -0.43 cm \begin{array}[m]{c} ~ \\ \stackrel{~~~}{\sim}
\end{array} \hskip-0.1cm\widetilde{\ell}\, = \,\ell^2 = \,0$$

\vv \nin elles v\'erifient aussi

\beq \fbox{\rule[-0.4cm]{0cm}{1cm}~$ ~~\widetilde{\ell}~{\bf \Phi}^{+}(\ell)~= ~{\bf \Phi}^{-}(\ell)~\widetilde{\ell}  = 0 ~~$~}  \enq

\vskip 0.7cm
\vv \nin \leftpointright~Rappelons enfin que dans les deux cas, masse non nulle et masse nulle, pour un observateur donn\'e auquel est attach\'e une base d'espace-temps $(n_0, n_1, n_2, n_3)$,
le tenseur $F_{\mu \nu}$ est d\'ecomposable en une partie ``\'electrique" et une partie ``magn\'etique", \'egales respectivement \`a $n_0 \wedge E$ et $^\star(n_0 \wedge B)$, les 4-vecteurs ``\'electrique" $E$ et ``magn\'etique" $B$ \'etant donn\'es par\footnote{ITL, \S 3.3.2.} 

\beq E_\mu = - F_{\mu \nu}\, n^\nu_0,~~~B_\mu = \di{1\over 2}\,\epsilon_{\mu \nu \rho \sigma}\,n^\nu_0 \,F^{\rho \sigma} = G_{\mu \nu}\, n^\nu_0 \enq

\subsection{Lien avec la repr\'esentation $D(\frac{1}{2}, \frac{1}{2})$ du groupe $SL(2,C)$  \protect \footnote{ITL, \S 5.5.1}}

\vv \nin Du point de vue du groupe des rotations, la repr\'esentation $D(\frac{1}{2}, \frac{1}{2})$ de $SL(2,C)$ est d\'ecomposable en une
repr\'esentation scalaire (spin 0) et une repr\'esentation vectorielle
(spin 1), l'ensemble formant un 4-vecteur. Pour cette raison, elle est couramment consid\'er\'ee  comme la repr\'esentation 4-vectorielle par excellence de $SL(2,C)$. On peut l'obtenir en
effectuant le produit tensoriel de la repr\'esentation fondamentale $D(\frac{1}{2},0)$ de $SL(2,C)$ avec sa contragr\'ediente $D(0, \frac{1}{2})$, ou encore, avec sa repr\'esentation conjugu\'ee, \'equivalente \`a la contragr\'ediente. Consid\'erons donc les
produits tensoriels

\beq \Psi^{a \, \dot{b}} ~=~u^a\, v^{\dot{b}} \enq

\vv \nin des composantes contravariantes d'un spineur $u$ d'ordre 1 de la repr\'esentation fondamentale (\`a 2 composantes avec indices non point\'es) 
et d'un spineur $v^\star$ d'ordre 1 de la repr\'esentation conjugu\'ee (lui aussi \`a 2 composantes, mais avec indices point\'es), d\'efinissant les quatre
composantes d'un tenseur mixte $\Psi$ d'ordre 2 ayant un indice non point\'e et un indice point\'e. Si on l'\'ecrit sous forme matricielle :

\beq \Psi \equiv \left( \begin{array}{cc} u^1 \, v^{\dot{1}} & u^1\, v^{\dot{2}} \\ u^2\, v^{\dot{1}} &
u^2 \, v^{\dot{2}} \end{array} \right) \enq

\vv \nin sa loi de transformation par une matrice $A$ de $SL(2,C)$ est

\beq ~^A \Psi = A\, \Psi \, A^\dagger \enq

\vv \nin Cette matrice $\Psi$ peut \^etre mise en relation
biunivoque avec un 4-vecteur $V$ (de composantes a priori complexes) au moyen de la formule

\vv \nin

$$ \Psi~ =~V \hskip -0.53cm \begin{array}[m]{c} ~
\\ \stackrel{~~~}{\sim}\end{array} \hskip -0.1 cm =~ V^0~ + \Vec{~V ~} \cdot \Vec{~\tau~}  ~~~{\rm avec} $$ \vskip -0.57cm
\beq \begin{array}{cc}
V^0 = \frac{1}{2}\,{\rm Tr}\, \Psi ~=~\frac{1}{2} (u^1 \, v^{\dot{1}} + u^2 \, v^{\dot{2}})~~, & V^3 =
\di{\frac{1}{2}}\,{\rm Tr}\, \Psi \, \tau_3~=~\frac{1}{2} (u^1 \, v^{\dot{1}} - u^2 \, v^{\dot{2}})
 \\~\\
V^1 = \di{\frac{1}{2}}\,{\rm Tr}\, \Psi \, \tau_1~=~\frac{1}{2} (u^1 \, v^{\dot{2}} +
u^2 \, v^{\dot{1}})~~, & V^2 = \di{\frac{1}{2}}\,{\rm Tr}\, \Psi \, \tau_2~=~\frac{i}{2}
(u^1 \, v^{\dot{2}} - u^2 \, v^{\dot{1}}) 
\end{array} \label{quavec} \enq

\vv \nin et ce 4-vecteur $V$ a pour propri\'et\'e d'\^etre du genre lumi\`ere. On v\'erifie que sa loi de transformation est bien celle d'un 4-vecteur, puisque

$$ ~^AV^\mu = \frac{1}{2}\,{\rm Tr}\,\tau_\mu\, A\, \Psi\, A^\dagger ~=~V^\nu \,
\frac{1}{2}\,{\rm Tr}\,\tau_\mu\, A\,
\tau_\nu \, A^\dagger~\equiv~ \Lambda^\mu_\nu \, V^\nu $$

\vv \nin o\`u les grandeurs $\Lambda^\mu_\nu$ sont les \'el\'ements de matrice de la transformation de Lorentz $\Lambda$ associ\'ee \`a $A$.

\newpage

\section{Compl\'ement II : Symboles de Levi-Civita \protect \footnote{G. Ricci-Curbastro et T. Levi-Civita, `` M\'ethodes de calcul diff\'erentiel absolu et leurs applications", 
Mathematische Annalen, Springer Verlag, vol. 54, no 1-2, mars 1900, p. 125-201.} \label{tenseur-LC} }

\vv \nin Le {\em symbole de Levi-Civita} d'ordre $N$, not\'e $\epsilon_{r_1 r_2 \cdots r_N}$, encore appel\'e {\em symbole indicateur de volume de Kronecker}, ou encore {\em tenseur dualiseur}, est donn\'e par le d\'eterminant 

\beq \epsilon_{r_1 r_2 \cdots r_N}~ = ~\begin{array}{|ccccc|} 
\delta_{r_1, 1} & \delta_{r_1, 2} & \cdots & \delta_{r_1, N-1} & \delta_{r_1, N}  \\[0.2cm]
\delta_{r_2, 1} & \delta_{r_2, 2} & \cdots & \delta_{r_2, N-1} & \delta_{r_2, N} \\ [0.1cm]
. & . & \cdots& . & . \\
. & . & \cdots & . & . \\
\delta_{r_{N-1}, 1} & \delta_{r_{N-1}, 2} & \cdots & \delta_{r_{N-1}, N-1} & \delta_{r_{N-1}, N}  \\[0.2cm]
\delta_{r_N, 1} & \delta_{r_N, 2} & \cdots & \delta_{r_N, N-1} & \delta_{r_N, N} 
\end{array} \label{LCN} \enq

\vv \nin $\delta_{a,b}$ \'etant le symbole de Kronecker. Ledit symbole est compl\`etement antisym\'etrique suivant ses $N$ indices $r_1, \cdots, r_N$, chacun courant de $1$ \`a $N$,  et tel que   

\beq \epsilon_{1 2 \cdots N} = 1 \enq 

\vv \nin Il peut aussi s'\'ecrire comme 

\beq \epsilon_{r_1 r_2 \cdots r_N}~ =~\di{\prod_{1 \leq i \leq j \leq N}}\, {\rm sgn}(r_j - r_i) ~=~\sigma(P)  \enq

\vv \nin o\`u ${\rm sgn}(a-b)$ le signe de la diff\'erence $a-b$, pris \'egal \`a $0$ si $a=b$ et $\sigma(P)$ la signature de la permutation $[1,2,\cdots, N] \longrightarrow [r_1,r_2,\cdots r_N]$ (soit $(-1)^p$ o\`u $p$ est la parit\'e de la permutation). Par permutation de ses colonnes, le d\'eterminant (\ref{LCN}) se voit multipli\'e par la signature de cette permutation, et il est donc \'evident que l'on peut \'ecrire 
\vv
\beq  \begin{array}{|ccccc|} 
\delta_{r_1, s_1} & \delta_{r_1, s_2} & \cdots & \delta_{r_1, s_{N-1}} & \delta_{r_1, s_N}  \\[0.2cm]
\delta_{r_2, s_1} & \delta_{r_2, s_2} & \cdots & \delta_{r_2, s_{N-1}} & \delta_{r_2, s_N} \\ [0.1cm]
. & . & \cdots & . & . \\
. & . & \cdots & . & . \\
\delta_{r_{N-1}, s_1} & \delta_{r_{N-1}, s_2} & \cdots & \delta_{r_{N-1}, s_{N-1}} & \delta_{r_{N-1}, s_N}  \\[0.2cm]
\delta_{r_N, s_1} & \delta_{r_N, s_2} & \cdots & \delta_{r_N, s_{N-1}} & \delta_{r_N, s_N} 
\end{array} ~= ~ \epsilon_{r_1 r_2 \cdots r_N}~\epsilon_{s_1 s_2 \cdots s_N}~  \label{LC-0}\enq
\vv

\vv \nin Le symbole de Levi-Civita intervient notamment dans l'expression du d\'eterminant d'une matrice $A$ ($N \times N$). On a les formules suivantes, utilisant la convention de sommation d'Einstein avec des indices r\'ep\'et\'es courant chacun de $1$ \`a $N$.

\beq \begin{array}{c} 
 {\rm det} A =\epsilon_{r_1 r_2 \cdots r_N}~A^{r_1}_1\, A^{r_2}_2 \cdots A^{r_N}_N \\ [0.3cm]
  {\rm det} A = \di{1\over{N!}}~\epsilon_{r_1 r_2 \cdots r_N}~\epsilon_{s_1 s_2 \cdots s_N}~A^{r_1}_{s_1}\, A^{r_2}_{s_2} \cdots A^{r_N}_{s_N}  \\ [0.5cm]
  {\rm det} A ~\epsilon_{s_1 s_2 \cdots s_N}~=~ \epsilon_{r_1 r_2 \cdots r_N}~~A^{r_1}_{s_1}\, A^{r_2}_{s_2} \cdots A^{r_N}_{s_N} \end{array}
\label{LC-1} \enq

\vv \nin Du point de vue des transformations du groupe $GL(N,C)$ agissant sur un espace vectoriel complexe $E_N$ de dimension $N$, le symbole de Levi-Civita repr\'esente un tenseur d'ordre $N$. La derni\`ere formule dans (\ref{LC-1}) montre que sous l'effet d'une transformation de ce groupe repr\'esent\'ee par la matrice $A$, ce tenseur est simplement multipli\'e par le d\'eterminant de la matrice et reste m\^eme invariant si ${\rm det}\,A =1$, c'est-\`a-dire, s'il s'agit d'une transformation du sous-groupe $SL(N,C)$. En fait, il est facile de montrer qu'\`a un facteur pr\`es, le tenseur de Levi-Civita est le seul tenseur d'ordre $N$ compl\`etement antisym\'etrique\footnote{Voir H. Bacry, loc. cit. Chap.\,4.}.  
 
\vv \nin D\'eveloppons le d\'eterminant (\ref{LC-0}) suivant les \'el\'ements de sa premi\`ere colonne, tout en exprimant le r\'esultat \`a l'aide des symboles de Levi-Civita d'ordre $N-1$ :  

$$ \epsilon_{r_1 r_2 \cdots r_N}~\epsilon_{s_1 s_2 \cdots s_N} = \left[~\delta_{r_1 s_1}~\epsilon_{r_2 r_3 \cdots r_N} - \delta_{r_2 s_1}\, \epsilon_{r_1 r_3 \cdots r_N} + \cdots +(-1)^{N-1}\,\delta_{r_N s_1}\,  \epsilon_{r_1 r_2 \cdots r_{N-1}} \right]\, \epsilon_{s_2 s_3 \cdots s_N}$$

\vv \nin Choisissons les indices $s_k$ tous diff\'erents, auquel cas $\epsilon_{s_1 s_2 \cdots s_N}$ et $\epsilon_{s_2 s_3 \cdots s_N}$ sont diff\'erents de z\'ero, mais prenons tous les indices $r_k$ dans une suite de $N-1$ valeurs diff\'erentes. Le symbole d'ordre $N$ $\epsilon_{r_1 r_2 \cdots r_N}$ est alors nul car deux de ses indices sont certainement identiques. On en d\'eduit la relation suivante entre symboles de Levi-Civita d'ordre $N-1$ :  

\vv
\beq \fbox{\rule[-0.4cm]{0cm}{1cm}~$~ \delta_{r_1 s_1}~\epsilon_{r_2 r_3 \cdots r_N} - \delta_{r_2 s_1}\, \epsilon_{r_1 r_3 \cdots r_N} + \cdots +(-1)^{N-1}\, \delta_{r_N s_1} \,\epsilon_{r_1 r_2 \cdots r_{N-1}} =0~$~}  \label{LC-2} \enq

\vv
\vv \nin Voici quelques autres formules g\'en\'erales :

\beq \di{\sum^N_{r_1 r_2 \cdots r_N = 1}} ~\epsilon_{r_1 r_2 \cdots r_N}\, \epsilon_{r_1 r_2 \cdots r_N}~=~N\,! \enq

\beq \di{\sum^N_{r_1 r_2 \cdots r_k = 1}} ~\epsilon_{r_1 r_2 \cdots r_k r_{k+1} \cdots r_N}\, \epsilon_{r_1 r_2 \cdots r_k s_{k+1} \cdots s_N}~=~k\,!\,\left(N -k \right) \,! ~\delta^{\,s_{k+1} \cdots s_N}_{\,r_{k+1} \cdots r_N}\enq

\vv \nin o\`u $\delta^{\,j_1 \cdots j_q}_{\,i_1 \cdots i_q}$ est un symbole de Kronecker g\'en\'eralis\'e d'ordre $q$ donn\'e par\footnote{Ici, on ne fait aucune distinction entre les symboles $\delta_{a b}$ et $\delta^b_a$.} 

\beq 
\delta^{\,j_1 \cdots j_q}_{\,i_1 \cdots i_q} = \di{\sum_{P}} ~ \sigma(P)~\delta^{j_{P{(1)}}}_{i_1}~ \delta^{j_{P{(2)}}}_{i_2}\cdots \delta^{j_{P{(q)}}}_{i_q}~=~  \begin{array}{|ccccc|} 
\delta^{j_1}_{i_1} & \delta^{j_1}_{i_2} & \cdots & \delta^{j_1}_{i_{q-1}} & \delta^{j_1}_{i_q}  \\[0.2cm]
\delta^{j_2}_{i_1} & \delta^{j_2}_{i_2} & \cdots & \delta^{j_2}_{i_{q-1}} & \delta^{j_2}_{i_q} \\ [0.1cm]
. & . & \cdots & . & . \\
. & . & \cdots & . & . \\
\delta^{j_{q-1}}_{i_1} & \delta^{j_{q-1}}_{i_2} & \cdots & \delta^{j_{q-1}}_{i_{q-1}} & \delta^{j_{q-1}}_{i_q}  \\[0.2cm]
\delta^{j_q}_{i_1} & \delta^{j_q}_{i_2} & \cdots & \delta^{j_q}_{i_{q-1}} & \delta^{j_q}_{i_q} 
\end{array} \enq

\vv \nin $P$ \'etant une permutation de $q$ \'el\'ements et $\sigma(P)$ sa signature. Ci-apr\`es, nous consid\'erons plus particuli\`erement les cas $N=3$ et $N=4$.

\vv \nin \ding{192} {\bf Cas de 3 dimensions euclidiennes}

\vv \nin Comme on sait, le symbole de Levi-Civita d'ordre 3 peut s'exprimer simplement \`a partir d'un produit mixte. Dans une base orthonorm\'ee $\left\{\Vec{e_a}, \Vec{e_b},\Vec{e_c} \right\}$, on a en effet 

\beq \epsilon_{a b c} =  \Vec{e_a} \cdot \left( \Vec{e_b} \wedge\Vec{e_c} \right) =\, \pm 1  \enq

\vv \nin Cette forme met clairement en \'evidence l'invariance du tenseur $\{\epsilon_{a b c}\}$ vis-\`a-vis du groupe des rotations $SO(3)$ : c'est une grandeur {\em scalaire} vis-\`a-vis de ce groupe. Comme on sait, le qualificatif de {\em pseudo-scalaire} qui lui est attribu\'e vient de son comportement dans l'op\'eration de parit\'e. Sous celle-ci, un vecteur ordinaire se voit changer de sens\footnote{Et pour cette raison est aussi appel\'e {\em vecteur vrai} ou encore {\em vecteur polaire}.}. Les vecteurs de base \'etant suppos\'es se comporter de cette fa\c{c}on, leur produit mixte $\epsilon_{a b c}$ change de signe dans l'op\'eration, alors qu'un ``vrai scalaire" (tel le produit scalaire de deux vrais vecteurs) ne change pas de signe. 

\vv \nin La m\'etrique euclidienne utilis\'ee dans ce cas permet d'identifier les composantes contravariantes aux composantes covariantes : $\epsilon^{a b c} = \epsilon_{abc}$. Dans les formules suivantes on utilise la convention de sommation d'Einstein. 

\beq \begin{array}{c}
 \epsilon_{a b c}\, \epsilon_{\alpha \beta \gamma} = \delta_{a \alpha}\, \delta_{b \beta}\, \delta_{c \gamma} + \delta_{a \beta}\, \delta_{b \gamma}\, \delta_{c \alpha} + \delta_{a \gamma}\, \delta_{b \alpha}\, \delta_{c \beta} \\ [0.25cm] 
-\delta_{a \alpha}\, \delta_{b \gamma}\, \delta_{c \beta} - \delta_{a \gamma}\, \delta_{b \beta}\, \delta_{c \alpha} - \delta_{a \beta}\, \delta_{b \alpha}\, \delta_{c \gamma} \\ [0.25cm]
\epsilon_{a b s}\,\epsilon^{c \alpha s} + \epsilon_{b c  s}\,\epsilon^{a \alpha s} +\epsilon_{c a  s}\,\epsilon^{b \alpha s} =0 \\ [0.25cm]
\delta_{ra}\, \epsilon_{bcd} - \delta_{rb}\,\epsilon_{acd} + \delta_{rc}\, \epsilon_{abd} - \delta_{rd}\, \epsilon_{abc} = 0 \\ [0.25cm]
\epsilon_{a b c} \, \epsilon^{a \beta \gamma} = \delta^\beta_b\, \delta^\gamma_c - \delta^\gamma_b\, \delta^\beta_c ,~~~\epsilon_{a b c} \, \epsilon^{a b \gamma} = 2\, \delta^\gamma_c,~~~\epsilon_{a b c} \, \epsilon^{a b c} = 6
\end{array}
\enq 

\vv \nin \ding{193} {\bf Cas des 4 dimensions d'espace-temps} 

\vv \nin On sait que dans ce cas, la m\'etrique pseudo-euclidienne de signature $(+\,-\,-\,-)$ est utilis\'ee pour passer des composantes contravariantes de tenseurs \`a leurs composantes covariantes (ex. : $T_{\mu \nu} = g_{\mu \alpha}\, g_{\nu \beta}\,T^{\alpha \beta}$) et l'on a

\beq \fbox{\rule[-0.4cm]{0cm}{1cm}~$\epsilon^{\mu \nu \rho \sigma} = -\, \epsilon_{\mu \nu \rho \sigma},~~~\epsilon_{0123} = 1 = -\, \epsilon^{0123} $~}  \enq

\vv \nin Des propri\'et\'es g\'en\'erales mentionn\'ees plus haut, on d\'eduit les formules suivantes  

\beq \begin{array}{c}

\epsilon_{\mu \nu \rho \sigma}\,\epsilon^{\alpha \beta \gamma
\sigma} = - \left[\, \delta^\alpha_\mu \, \delta^\beta_\nu \,
\delta^\gamma_\rho + \delta^\alpha_\nu \, \delta^\beta_\rho \,
\delta^\gamma_\mu + \delta^\alpha_\rho \, \delta^\beta_\mu \,
\delta^\gamma_\nu - \delta^\alpha_\mu \, \delta^\beta_\rho \,
\delta^\gamma_\nu - \delta^\alpha_\nu \, \delta^\beta_\mu \,
\delta^\gamma_\rho - \delta^\alpha_\rho \, \delta^\beta_\nu \,
\delta^\gamma_\mu \,\right] \\ [0.4cm] 
\epsilon_{\mu \nu \rho \sigma}\,\epsilon^{\alpha
\beta \rho \sigma} = - 2 \left[\,\delta^\alpha_\mu \,
\delta^\beta_\nu -\delta^\alpha_\nu \, \delta^\beta_\mu \, \right] \\ [0.4cm] 
\epsilon_{\mu \nu \rho \sigma}\,\epsilon^{\alpha \nu \rho \sigma} =
- 6 \,\delta^\alpha_\mu~,~~~\epsilon_{\mu \nu \rho
\sigma}\,\epsilon^{\mu \nu \rho \sigma} = - 24 \end{array} 
\label{epsilon-4d}
\enq

\beq \fbox{\rule[-0.4cm]{0cm}{1cm}~$\epsilon_{\nu \rho \sigma \lambda}\, g_{\mu \omega} -\epsilon_{\mu \rho \sigma \lambda }\, g_{\nu \omega} +\epsilon_{ \mu \nu \sigma \lambda}\, g_{\rho \omega}  -\epsilon_{\mu \nu \rho \lambda}\, g_{\sigma \omega}  + \epsilon_{\mu \nu \rho \sigma}\, g_{\lambda \omega}   =0 $~}  \label{eps-4d-2} \enq

\vv \nin La derni\`ere relation (\ref{eps-4d-2}), directement d\'eduite de (\ref{LC-2}), se retrouve notamment dans le calcul de la trace d'un produit de matrices de Dirac\footnote{ITL, Eq. 7.56.}.

\newpage

\section{Compl\'ement III : D\'eveloppement multipolaire d'une matrice densit\'e de spin \protect \footnote{Ugo Fano : ``Description of States in Quantum Mechanics by Density Matrix and Operators Techniques", Rev. Mod. Phys. 29 (1957) 74. }} 

\vvv
\subsection{Pr\'eliminaires}

\vv \nin \ding{192} Tout espace vectoriel complexe ${\cal E}$ de dimension $N$ peut \^etre envisag\'e  comme un espace de repr\'esentation irr\'eductible $D^j$ du groupe des rotations, ou plus pr\'ecis\'ement du groupe $SU(2)$, correspondant \`a une valeur du spin \'egale \`a $j=(N-1)/2$. Dans la base canonique de ${\cal E}$, la matrice correspondant \`a la composante $J_3$ du spin suivant un axe de r\'ef\'erence $z$ n'est pas difficile \`a contruire : il s'agira de la matrice diagonale dont les \'el\'ements sont, dans l'ordre, $-j, - j +1, -j +2, \cdots, j-1, j$. Chaque \'el\'ement de la base canonique sera ainsi consid\'er\'e comme un vecteur propre de $J_3$, ayant pour valeur propre $m$ avec $-j \leq m \leq j$, et pour cette raison, sera not\'e $|\, j, m>$. La construction des matrices $J_1$ et $J_2$ correspondant aux composantes du spin suivant les deux autres axes  $x$ et $y$, formant avec $z$ un rep\`ere cart\'esien, ne pose pas plus de difficult\'e, car on connait parfaitement les actions des matrices $J_{\pm} = J_1 \pm i \, J_2$ sur les vecteurs propres de $J_3$ :   

\beq J_{\pm}\, |\, j, m>~=\,\sqrt{j(j+1) - m(m \pm 1)}~ |\, j, m \pm 1> \enq

\vv \nin On obtient ainsi 3 matrices $J_1$, $J_2$ et $J_3$, hermitiques, de traces nulles, qui satisfont clairement les relations usuelles de l'alg\`ebre de Lie su(2). On a notamment 

\beq [\, J_3\, , \, J_\pm \, ] ~=~\pm\, J_{\pm},~~~\Vec{\,J\,}^2=~J^2_1 + J^2_2 + J^2_3  = j(j+1) \enq 

\vv \nin Par l'op\'eration $J'_k = U\, J_k\,U^\dagger $, une matrice unitaire quelconque $U$ transforme $J_1$, $ J_2$ et $ J_3$ en $J'_1$, $J'_2$ et $J'_3$ respectivement. Ces nouvelles matrices sont \'egalement hermitiques, de trace nulle, poss\`edent vis-\`a-vis de su(2) les m\^emes propri\'et\'es que les pr\'ec\'edentes, mais $J'_3$ peut ne plus \^etre diagonal dans la base canonique.  R\'eciproquement, deux triplets $J_1, J_2, J_3$ et $J'_1, J'_2, J'_3$ v\'erifiant les relations de commutation de su(2) et $J^2_1 +J^2_2 +J^2_3 = j(j+1)$ sont n\'ecessairement reli\'es par une matrice unitaire\footnote{Etablir cette r\'eciproque. A noter aussi que pour $SU(2)$, toute repr\'esentation est \'equivalente \`a sa contragr\'ediente ainsi qu'\`a sa conjugu\'ee.}. 

\vv \nin \ding{193} Suivant la base des $|\, j, m>$, une matrice $N \times N$ quelconque $Q$ peut de fa\c{c}on \'evidente \^etre d\'evelopp\'ee comme suit : 

\beq Q = \di{\sum^{m = + j}_{m = -j}~\sum^{m' = +j}_{m' = -j}} ~|\, j, m> Q_{m m'} <m', j\, | ~~~{\rm avec}~~~Q_{m m'} = ~<m, j\,|\,Q\,|\,j, m'> \label{develQ} \enq 

\vv \nin Soit alors $U(R)$ la matrice unitaire repr\'esentant dans ${\cal E}$ la rotation $R$ d'angle $\theta$ effectu\'ee autour d'un axe de vecteur unitaire $\Vec{\,n\,}$ dans l'espace tri-dimensionnel. Elle s'\'ecrit 

$$  U(R) = \exp\left[{-i \,\theta \Vec{\,n\,}\hskip-0.1cm\cdot\hskip-0.1cm \Vec{\,J\,}}\right]  $$ 

\vv \nin et son action sur sur les vecteurs $|\, j, m>$ ainsi que sur les matrices $N \times N$ est donn\'ee par\footnote{L'expression de ${\cal D}^j(R)$ peut \^etre trouv\'ee dans ITL, chapitre 4, Eq. 4.66.} 

$$ U(R)\, |\, j, m>\,~ =~\di{\sum^{m' = +j}_{m'=-j}}~{\cal D}^j_{m' m}(R)\, |\, j, m'> ~~~{\rm et} $$
\beq Q' = U(R)\,Q\,U^\dagger(R) = \di{\sum_{m_1, m'_1}\sum_{m_2, m'_2}}\,{\cal D}^j_{m_1 m'_1}(R)\,\left[ {\cal D}^j_{m_2 m'_2}(R)\right]^\star\, Q_{m_1 m_2}\, |\, j, m'_1>\,<m'_2, j\, |\enq
$$ {\rm soit}~~~~Q'_{m'_1 m'_2} = ~ <m'_1, j\,|\,Q'\,|\,j, m'_2> ~= \di{\sum_{m_1}\sum_{m_2}}~{\cal D}^j_{m_1 m'_1}(R)\,\left[ {\cal D}^j_{m_2 m'_2}(R)\right]^\star\, Q_{m_1 m_2}$$

\vv \nin La derni\`ere relation met en \'evidence le fait que l'espace ${\cal E}^\star$ des matrices complexes $N \times N$ peut lui m\^eme \^etre envisag\'e comme un espace de repr\'esentation de $SU(2)$, construit comme produit tensoriel de la repr\'esentation $D^j$ et de sa repr\'esentation conjugu\'ee $D^{j \star}$ (ou de sa contragr\'ediente, \'equivalente \`a cette derni\`ere\footnote{ITL, \S 4.3.3.}). Un tel produit tensoriel est r\'eductible par rapport \`a $SU(2)$ selon le sch\'ema bien connu

\beq \fbox{\rule[-0.4cm]{0cm}{1cm}~$\begin{array}{rcl}
&{\scriptstyle L=2 j}& \\[-1.2mm]
 D^j \, \otimes\, D^{j \star} =& {\bf{ \oplus}}& D^L\\[-1.2mm]
&{\scriptstyle L=0}& 
\end{array}  $~}  \label{decompExE} \enq

\vv \nin On est ainsi conduit \`a envisager une matrice $N\times N$ quelconque $Q$ comme une somme de matrices $Q^L$, $L$ courant de 0 \`a $2j$, $Q^L$ \'etant associ\'ee \`a la repr\'esentation irr\'eductible $D^L$ de $SU(2)$ : 

\beq Q = \di{\sum^{L= 2j}_{L=0}}\, Q^L \enq 

\vv \nin somme auquel on donne le nom de {\em d\'eveloppement multipolaire} de la matrice $Q$. 

 \vv \nin \ding{194} Une matrice complexe quelconque $N\times N$ est d\'efinie par $N^2$ nombres complexes a priori ind\'ependants. Si la trace de la matrice est nulle, seuls $N^2 -1$ de ces nombres restent a priori ind\'ependants.  De cette constatation on d\'eduit qu'une base de l'espace ${\cal E}^\star$ des matrices complexes $N\times N$ peut \^etre construite au moyen de $N^2 -1$ matrices ind\'ependantes, ayant chacune une trace nulle, auxquelles on adjoint la matrice identit\'e. Dans la suite, nous montrons comment on peut construire une telle base de matrices tout en se conformant \`a la d\'ecomposition (\ref{decompExE}). 

\subsection{D\'eveloppement multipolaire d'une matrice $N \times N$}

\vv \nin  \ding{192} D'un point de vue technique, la d\'ecomposition (\ref{decompExE}) peut \^etre men\'ee en utilisant la repr\'esenta-tion adjointe des matrices de ${\cal E}^\star$. Celle-ci, agissant sur ${\cal E}^\star$, est d\'efinie par l'application ($Q$ est une matrice $N \times N$ quelconque) 

\beq  A \longrightarrow  \Delta_A : \Delta_A(Q) ~=~[\, A,\,Q\,] \enq

\vv \nin laquelle constitue un homomorphisme d'alg\`ebre de Lie lorsque l'ensemble des matrices $A$ envisag\'ees forment elles-m\^emes une alg\`ebre de Lie. Ainsi, en utilisant l'identit\'e de Jacobi 

$$ [\,A\, , [\,B, C\,]\,] +  [\,B\, , [\,C, A\,]\,] +  [\,C\, , [\,A, B\,]\,]  = 0 $$

\vv \nin on v\'erifie ais\'ement que les op\'erateurs adjoints des $J_k$, not\'es ${\cal J}_k$ ($k=1,2,3)$, satisfont les relations de commutation de su(2) : 

$$  [\,{\cal J}_a , {\cal J}_b\,](Q) = [\,J_a,[\,J_b, Q\,]\,] - [\,J_b,[\,J_a, Q\,]\,] = [\,J_a,[\,J_b, Q\,]\,] + [\,J_b,[\,Q, J_a\,]\,] $$
\beq=[\,[\,J_a,\, J_b\,], Q\,] = i\, \epsilon_{abc}\,[\,J_c, Q\,] =i\, \epsilon_{a b c}\,{\cal J}_c(Q) \enq

\vv \nin Une matrice $Q^L_M$ relevant de la repr\'esentation $D^L$ doit v\'erifier les \'equations 

$$ \Vec{\cal J}^{\,2}\hskip-0.15cm(Q^L_M) = \di{\sum^3_{a=1}} ~[\,J_a\,[\,J_a, Q^L_M\,]\,] = L(L+1)\, Q^L_M $$
\beq {\cal J}_3(Q^L_M) = [\,J_3, Q^L_M\,] = M\, Q^L_M~~~~{\rm avec}~~~ - L \leq M \leq L \label{EQVP} \enq

\vv \nin En choisissant des combinaisons lin\'eaires ad\'equates de matrices dans $D^L$, on peut toujours trouver des matrices $Q^L_M$ v\'erifiant de plus les formules standard 

\beq [\, J_{\pm},\, Q^L_M\,] = \sqrt{L(L+1) -M(M\pm 1)} ~ Q^L_{M\pm1} \label{OPTI} \enq 

\vv \nin Selon la nomenclature consacr\'ee, l'ensemble des matrice $Q^L_M$ avec $-L\leq M \leq L$ constitue un {\em op\'erateur tensoriel irr\'eductible} \footnote{Voir par exemple A. Messiah : ``M\'ecanique Quantique", Dunod, Paris (1964), Tome 2, Chap. XIII, \S 31, 32.}. En raison du th\'eor\`eme de Wigner-Eckart, l'\'el\'ement de matrice $<m', j\,|\,Q^L_M\,|\, j, m>$ est proportionnel au coefficient de Clebsch-Gordan 
$<j, - m \, ; j, m'\,|\, L, M>$, lui-m\^eme proportionnel \`a $\delta_{M, m'-m}$.

\vv
\vv \nin \ding{193} Dans l'espace ${\cal E}^\star$, on d\'efinit couramment le produit scalaire hermitien de deux matrices $A$ et $B$ par\footnote{Montrer que ${\rm Tr} \,F\,G = 0$ si $F$ s'\'ecrit sous la forme $F = [\, G, H\,]$.}  
\vskip-0.25cm

\beq \fbox{\rule[-0.4cm]{0cm}{1cm}~$(A, B) = {\rm Tr}\,A^\dagger\, B $~}  \label{prodscalmat}  \enq

\nin Il est facile de montrer que les op\'erateurs $\Vec{\cal J}^{\,2}$ et ${\cal J}_3$ sont hermitiques par rapport \`a ce produit scalaire. On sait qu'alors deux matrices ``vecteurs propres" $Q^{L_1}_{M_1}$ et $Q^{L_2}_{M_2}$ correspondant \`a des valeurs propres diff\'erentes de ces op\'erateurs sont n\'ecessairement orthogonales selon ledit produit scalaire. Par une normalisation convenable, on peut faire en sorte que l'on ait\footnote{Auquel cas, $<m', j\,|\,Q^L_M\,|\, j, m>~ = A_L\, <j, m' \,|\, j, m \,; L, M>$ o\`u $A_L$, ind\'ependant de $m$ et $M$, a pour module $\sqrt{\di{{2 L +1}\over{2j+1}}}$.} 

\beq \fbox{\rule[-0.4cm]{0cm}{1cm}~${\rm Tr} \left[Q^{L_1}_{M_1}\right]^\dagger Q^{L_2}_{M_2} = \delta_{L_1 L_2}\,\delta_{M_1 M_2}  $~} \label{normaQ}  \enq

\vv \nin Chacune des repr\'esentations $D^L$ n'appara\^it qu'une seule fois dans la d\'ecomposition  (\ref{decompExE}). Aussi, pour une base $\left\{Q^L_M \right\}$ donn\'ee, chaque vecteur propre $Q^L_M$ est {\em unique}, \`a un facteur de phase pr\`es\footnote{On v\'erifie que $\di{\sum^{2j}_{L=0}}~(2L+1) = (2j+1)^2$.}. Or, par conjugaison hermitique, on a 

$$ [\,J_3, Q^L_M\,]^\dagger = - [\,J_3, \left[ Q^L_M \right]^\dagger\,] = M\, \left[ Q^L_M \right]^\dagger $$

\vv \nin d'o\`u il ressort que $\left[ Q^L_M \right]^\dagger$ est vecteur propre de ${\cal J}_3$ avec la valeur propre $-M$ et, d'apr\`es ce qui pr\'ec\`ede, ne peut donc diff\'erer de $Q^L_{-M}$ que par un facteur de phase. On peut adapter celui-ci de telle sorte \`a avoir, conform\'ement \`a la convention standard\footnote{On a alors $A^\star_L = A_L$.}, 

\beq \fbox{\rule[-0.4cm]{0cm}{1cm}~$\left[ Q^L_M \right]^\dagger = (-1)^M\, Q^L_{-M}  $~}  \label{conjQ} \enq

\vv \nin Enfin, la repr\'esentation $D^0$ correspond manifestement \`a la matrice identit\'e. L'application de (\ref{normaQ}) o\`u l'une des deux matrices est proportionnelle \`a l'identit\'e montre que pour $L \geq 1$, les matrices $Q^L_M$ ont donc une trace nulle. Nous d\'efinirons $Q^0_0 = \di{1\over \sqrt{N}}$.

\vv \nin \ding{194} Le but recherch\'e est atteint : l'ensemble des matrices $Q^L_M$ v\'erifiant (\ref{EQVP}), (\ref{OPTI}), (\ref{normaQ}) et (\ref{conjQ}) constitue une base de l'espace ${\cal E}^\star$, orthonorm\'ee selon le produit scalaire (\ref{prodscalmat}), et telle que tous ses \'el\'ements correspondant \`a $L \geq 1$ sont de trace nulle. Une matrice $N \times N$ quelconque $A$ admet ainsi le d\'eveloppement multipolaire\footnote{Ainsi nomm\'e car il s'apparente au d\'eveloppement d'un champ scalaire d\'ependant des coordonn\'ees sph\'eriques $r, \theta, \varphi$, selon les harmoniques sph\'eriques $Y^\ell_m(\theta, \varphi)$, les coefficients $Q^\ell_m$ intervenant dans ce d\'eveloppement \'etant qualifi\'es de {\em multip\^oles}.} :

\beq \fbox{\fbox{\rule[-0.4cm]{0cm}{1cm}~$A = \di{\sum^{L=2 j}_{L=0} \sum^{M= +L}_{M= -L}} ~A^L_M \, Q^L_M ~~~{\rm avec}~~~A^L_M = {\rm Tr} \left[ Q^L_M \right]^\dagger \hskip -0.1cm A  $~}}  \label{devel-multip} \enq

\vv \nin La norme de la matrice $A$, telle que d\'efinie par le produit scalaire (\ref{prodscalmat}) , prend alors la forme 

\beq ||A|| = \sqrt{\,\di{\sum_{L,M}}~|A^L_M|^2 } \label{normA} \enq

\vvv

\subsection{Matrice densit\'e de spin \protect \footnote{Voir, par exemple,  J. Werle, loc. cit.,  Chap. III \S 26, Chap IV, \S 32.}}

\vv \nin \ding{192} Admettons que l'espace ${\cal E}$ soit effectivement celui des vecteurs d'\'etats de spin d'une particule massive de spin $j$. On sait que quel que soit l'\'etat physique dans lequel se pr\'esente la particule, \'etat pur ou \'etat de m\'elange, en tout cas ici, \'etat de spin, celui-ci peut toujours \^etre d\'ecrit au moyen d'une matrice densit\'e, ici matrice densit\'e de spin, usuellement not\'ee $\rho$. Cet op\'erateur, hermitique et de  trace est \'egale \`a 1 permet de calculer la moyenne d'une observable ou d'un op\'erateur quelconque $A$ dans ledit \'etat physique, au moyen de la trace  

\beq <\,A\,> ~=~{\rm Tr} \,\rho A \enq

\vv \nin L'op\'erateur densit\'e de spin n'est pas n\'ecessairement diagonal dans la base canonique consid\'er\'ee pr\'ec\'edemment. Cependant, \'etant hermitique, il peut \^etre diagonalis\'e en prenant pour base de ${\cal E}$ celle de ses vecteurs propres $|\, \varphi_a >$ : 

\beq \rho ~=~\di{\sum_a}~|\, \varphi_a>\,p_a\,<\varphi_a\,|  \enq

\vv \nin Dans cette formule, $p_a$ est la valeur propre (d\'eg\'en\'er\'ee ou non) de $\rho$ correspondant au vecteur propre $|\, \varphi_a >$. Les valeurs propres $p_a$ sont non n\'egatives et satisfont 

\beq   \di{\sum_a}~p_a = 1     \enq

\nin La valeur propre $p_a$ s'interpr\`ete comme la probabilit\'e de trouver l'\'etat d\'ecrit par $|\,\varphi_a>$ dans l'\'etat physique d'\'ecrit par $\rho$. Lorsque la particule se trouve dans un \'etat pur $|\, \varphi>$ norm\'e \`a l'unit\'e, la matrice densit\'e correspondante est le projecteur  

\beq \rho ~=~|\, \varphi ><\varphi\,| \label{etatpur} \enq 

\vv \nin G\'en\'eralement, on a, de fa\c{c}on \'evidente

\beq \fbox{\rule[-0.4cm]{0cm}{1cm}~$~ {\rm Tr}\, \rho^2 \leq 1 ~ $~}  \label{crit1}  \enq

\vv \nin L'\'egalit\'e dans (\ref{crit1}) n'est en fait r\'ealis\'ee que {\em si et seulement si} l'op\'erateur densit\'e est associ\'e \`a un \'etat pur, c'est-\`a-dire, est de la forme (\ref{etatpur})\footnote{D\'emontrer cette assertion.}. L'\'egalit\'e ${\rm Tr}\, \rho^2 =1$ repr\'esente donc un crit\`ere pour reconna\^itre si la particule se trouve ou non dans un \'etat pur. 

\vv
\vv \nin \ding{193} Comme toute matrice $N \times N$, la matrice densit\'e de spin admet le d\'eveloppement multipolaire :

\beq \fbox{\fbox{\rule[-0.4cm]{0cm}{1cm}~$\rho = \di{\sum^{L=2 j}_{L=0} \sum^{M= +L}_{M= -L}} ~\rho^L_M \, Q^L_M ~~~{\rm avec}~~~\rho^L_M = {\rm Tr} \left[ Q^L_M \right]^\dagger \hskip -0.1cm \rho  $~}}  \label{dvl-mult-ro} \enq

\vv \nin La matrice $\rho$ \'etant hermitique, on a 

\beq \left[\,\rho^L_M\,\right]^\star = (-1)^M\,\rho^L_{-M} \enq 

\vv \nin Explicitons dans ce d\'eveloppement les termes correspondant \`a $L=0$ et $L=1$. Pour $L=0$, il n'y a qu'un seul terme, pour lequel $Q^0_0 = 1/\sqrt{N}$, et 

$$ \rho^0_0 = {\rm Tr}\, Q^0_0\, \rho = \di{1\over \sqrt{N}}\, {\rm Tr}\, \rho = \di{1\over \sqrt{N}} ,~~~{\rm donc}~~~\rho^0_0\, Q^0_0 = \di{1\over N} $$

\vv \nin Si tous les \'etats de spin de la particule sont {\em \'equiprobables}, auquel cas la particule est dite {\em non polaris\'ee}, seul ce terme est pr\'esent et l'on a donc 

\beq \rho_{\rm np} = \di{1\over N} = \di{1 \over{2j+1}} \enq

\vv \nin A l'aide du d\'eveloppement (\ref{dvl-mult-ro}), l'in\'egalit\'e g\'en\'erale (\ref{crit1}) prend la forme  

\beq \di{1\over N}~ +~\di{\sum^{L=2j}_{L=1}\,\sum^{M= +L}_{M=-L} } \,|\,\rho^L_M\,|^2 ~\leq ~1~~~{\rm soit}~~~N\,\di{\sum^{L=2j}_{L=1}\,\sum^{M= +L}_{M=-L} } \,|\,\rho^L_M\,|^2~\leq~N-1 = 2j  \label{inegal1} \enq 

\vv \nin On est alors amen\'e \`a d\'efinir un {\em degr\'e de polarisation} par 

\beq \zeta ~=~\sqrt{  \di{N \over{2j}}~ \di{\sum^{L=2j}_{L=1}\,\sum^{M= +L}_{M=-L} } \,|\,\rho^L_M\,|^2  }\enq

\vv \nin Ce param\`etre est compris entre 0 et 1. Le cas $\zeta =1$ correspond \`a un \'etat pur, tandis que si $\zeta =0$, la particule est compl\`etement non polaris\'ee. Si $0 < \zeta < 1$, la particule est dite {\em partiellement polaris\'ee}.

\vv \nin Pour $L=1$, on doit avoir $Q^1_1 \propto J_1 + i J_2$ et $Q^1_0 \propto J_3$. On trouve les facteurs de normalisation en calculant 

$$ {\rm Tr}\, J^2_1~ =~ {\rm Tr}\, J^2_2~ =~ {\rm Tr}\,J^2_3 ~= ~\di{1\over 3}\, {\rm Tr} \Vec{\,J\,}^2 ~=~ \di{1\over 3}\, N\,j(j+1) $$

\vv \nin On trouve ainsi 

$$ Q^1_1 = -\, \sqrt{3\over{2 N j (j+1)}}\, \left( J_1 + i J_2 \right) ,~~Q^1_{-1} = - \left[\, Q^1_1\,\right]^\dagger = \, \sqrt{3\over{2 N j (j+1)}}\, \left( J_1 - i J_2 \right),$$ 
\beq Q^1_0 = \, \sqrt{3\over{N j (j+1)}}\, J_3\enq

\vv \nin D'o\`u 

\beq \rho^1_1 = -\, \sqrt{3\over{2 N j (j+1)}}\, < J_1 + i J_2 >,~~~\rho^1_0 =  \sqrt{3\over{N j (j+1)}}\,< J_3> \enq 

\vv \nin Il vient alors 

\beq \di{\sum^{M=+1}_{M=-1}} \, \rho^1_M\, Q^1_M = \di{3\over{N j(j+1)}}\, <\Vec{\,J\,}> \cdot \hskip -0.1cm \Vec{\,J\,}  \label{ro1} \enq

\vv \nin D\'efinissant le {\em vecteur de polarisation} de la particule par

\beq \fbox{\rule[-0.45cm]{0cm}{1.3cm}~$ \Vec{\, \xi\,}  = \sqrt{\di{3 \over {2 j^2 (j + 1)}}}\, < \Vec{\,J\,}>  $~}  \label{VP-1}\enq

\vv \nin on obtient finalement 

\beq \di{\sum^{M=+1}_{M=-1}} \, \rho^1_M\, Q^1_M = \di{1\over N}\,\sqrt{\di{6\over {j+1}}}\, \Vec{\,\xi\,} \cdot \Vec{\,J\,} ~~~~{\rm et}~~~~~\di{ N\over{2j}} \, \di{\sum^{M=+1}_{M=-1}} \, |\rho^1_M\,|^2 = |\hskip -0.12cm\Vec{\,\xi\,}\hskip-0.12cm|^2 \enq
~~~~
\vv \nin et la relation (\ref{inegal1}) induit la contrainte 

\beq \fbox{\rule[-0.4cm]{0cm}{1cm}~$~~ |\hskip -0.12cm\Vec{\,\xi\,}\hskip-0.12cm|^2 \leq 1  ~~$~}  \enq

\vv 
\vv \nin \ding{194}~On sait que l'op\'eration de parit\'e (ou r\'eflexion d'espace) {\em commute} avec les rotations. L'espace ${\cal E}$ \'etant suppos\'e \^etre un espace de repr\'esentation irr\'eductible du groupe des rotations, l'application du lemme de Schur\footnote{ I. Schur : ``Untersuchungen \"uber die Darstellung der endlichen Gruppen durch gebrochenen linearen Substitutionen", J. Reine. Angew. Math., vol. 132, 1907, p. 85-137 ; voir aussi : H. Bacry, loc. cit., p. 59.} montre alors que ladite op\'eration est repr\'esent\'ee dans cet espace par un multiple de la matrice unit\'e, et sa repr\'esentation {\em unitaire} est un simple facteur de phase. Il en r\'esulte que tous les tenseurs irr\'eductibles du d\'eveloppement (\ref{dvl-mult-ro}) sont tous {\em pairs} vis-\`a-vis de l'op\'eration de parit\'e. Le tenseur de rang 1 notamment, qui repr\'esente un vecteur est en fait un {\em pseudo-vecteur}. Le vecteur de polarisation est donc un pseudo-vecteur.

\vv
\vv \nin \ding{195}~Dans le cas o\`u $j =1/2$, le d\'eveloppement  (\ref{dvl-mult-ro}) ne comporte que les deux premiers termes correspondant \`a $L=0$ et $L=1$. On a ici $\Vec{\,J\,} = \Vec{\, \tau\,}\hskip-0.15cm/2$ o\`u $\Vec{\,\tau\,}$ repr\'esente l'ensemble des matrices de Pauli. La formule g\'en\'erale (\ref{VP-1}) donne ici $\Vec{\, \xi\,} \,=~ <\Vec{\,\tau\,}>$ ~et la matrice densit\'e correspondante s'\'ecrit 

\beq \fbox{\rule[-0.4cm]{0cm}{1cm}~$~ \rho = \di{1\over 2}\, \left\{ 1~ + \Vec{\,\xi\,}\hskip-0.1cm \cdot \hskip -0.1cm \Vec{\,\tau\,} \right\} ~$~}  \label{dens-un-demi} \enq

\vv \nin Elle est donc compl\`etement d\'etermin\'ee par le vecteur de polarisation. 

\vv
\vv \nin \ding{196}~Il est clair qu'en effectuant les produits de divers ordres des op\'erateurs $J_1$, $J_2$ et $J_3$, lesquels se transforment par rotation comme les composantes d'un 3-vecteur, on peut construire des composantes d'op\'erateurs tensoriels irr\'eductibles correspondant chacun \`a une valeur donn\'ee de $L$. Ainsi, pour $L=2$ et $j \geq 1$, on obtient\footnote{D\'emontrer que ${\rm Tr} \, J^4_3 = \di{{N j(j+1)}\over 15}\,[\,3 j(j+1) -1\,]$.}  

\beq 
\begin{array}{c}
Q^2_2 = \di{1\over D}\,J^2_{+}\,,~~~Q^2_1 = -\di{1\over D}\,\left(J_{+}\,J_3 + J_3\,J_{+}\, \right),~~~Q^2_0 = \sqrt{\di{2\over 3}} \,\di{1\over  D} \left[\,3\, J^2_3 - j(j+1) \,\right], \\ [0.4cm] 
 {\sf avec}~~~~~D = \sqrt{ \di{{2 N j(j+1) [\, 4j(j+1) -3\,]}\over 15}} \\ [0.4cm] 
Q^2_{-1} = - Q^{2 \dagger}_1,~~~Q^2_{-2} = Q^{2 \dagger}_2 
\end{array} 
\enq 

\vv \nin En tenant compte du fait que $j(j+1) \equiv J^2_1 + J^2_1 + J^2_3$, le terme du d\'eveloppement de $\rho$ correspondant \`a $L=2$ s'\'ecrit sous la forme 

\vskip-0.3cm
\beq \begin{array}{c}
\rho^{(2)} = \di{\sum^3_{i, j =1}}~a_{i j}\, J_i\, J_j  ~~~~~{\sf avec} \\ [0.5cm] 
a_{11} = \di{1\over D}\, \left[\,\rho^2_2 +\rho^{2 \star}_2 - \sqrt{\di{2\over 3}}\, \rho^2_0\, \right],~~~ a_{22} = -\,\di{1\over D}\, \left[\,\rho^2_2 +\rho^{2 \star}_2 + \sqrt{\di{2\over 3}}\, \rho^2_0\, \right] \\ [0.4cm]
a_{33} = \di{2\over D}\,\rho^2_0\, \sqrt{\di{2\over 3}},~~~a_{12} = a_{21} =  \di{i\over D}\,\left[\, \rho^2_2 - \rho^{2 \star}_2\,\right],~~~a_{13} = a_{31} = -\,\di{1\over D}\,\left[\, \rho^2_1 + \rho^{2 \star}_1 \, \right] \\ [0.4cm] 
a_{23} = a_{32} = -\, \di{i\over D}\, \left[\, \rho^2_1 - \rho^{2 \star}_1 \, \right]  
\end{array}
\label{ro2} \enq 

\vv \nin On constate que les coefficients $a_{ij}$ sont tous r\'eels et sym\'etriques suivant leurs indices $i$ et $j$. En outre, 
\vskip-0.2cm
\beq \di{\sum^3_{i=1}}~a_{ii} = 0 \enq

\vv \nin On peut tout aussi bien consid\'erer l'espace ${\cal E}$ comme celui de la   repr\'esentation irr\'eductible ${\cal D}(j,0)$ de $SL(2,C)$, pour laquelle les g\'en\'erateurs de ce groupe sont repr\'esent\'es par les matrices $J_{k \ell} = \epsilon_{k \ell m}\, J_m$ et $J^{0 k} = i\,J_k$. Dans ce contexte, une base de spin de vecteurs $|\, j, m>$ est n\'ecessairement d\'efinie par rapport \`a une ``t\'etrade" attach\'ee \`a un 4-vecteur $t$ donn\'e, suppos\'e ici du genre temps, pointant vers le futur, et unitaire. Cette t\'etrade est constitu\'ee en adjoignant \`a $t$ trois 4-vecteurs $n_1, n_2, n_3$ du genre espace, formant avec lui une base d'espace-temps orthonorm\'ee et d'orientation directe, laquelle s'obtient \`a partir d'une base de r\'ef\'erence au moyen d'une transformation de Lorentz repr\'esent\'ee par une matrice de $SL(2,C)$ not\'ee $[\,t\,]$, appel\'ee aussi ``t\'etrade". On est ainsi amen\'e \`a poser  

\beq J_k \equiv - \, n_k \cdot W ~~(k=1,2,3)~~~{\sf avec}  ~~~W_\mu = \di{1\over 2}\, \epsilon_{\mu \nu \alpha \beta}\, t^\nu \, J^{\alpha \beta}  \enq

\nin les matrices $J^{\alpha \beta}$ \'etant les repr\'esentants des g\'en\'erateurs de $SL(2,C)$, d\'efinis comme indiqu\'e ci-dessus, et $W_\mu$ l'op\'erateur de polarisation appropri\'e. 
L'expression de $\rho^{(2)}$ peut alors \^etre r\'ecrite sous la forme d'un produit scalaire\footnote{Ici, nous ne consid\`erons pas le cas plus g\'en\'eral o\`u dans (\ref{develQ}) le ket $|\, j, m>$ correspondrait au 4-vecteur $t$ tandis que le bra $<m',j\,|$ correspondrait \`a un autre 4-vecteur $t'$ et donc \`a une autre t\'etrade $[\, t'\,]$ ; voir P. Moussa, R. Stora, loc. cit. p.297.}

\vskip -0.4cm
\beq \rho^{(2)} = \varphi_{\mu \nu}\, W^\mu\, W^\nu ~~~~{\sf avec}~~~~\varphi_{\mu \nu} = \di{\sum^3_{i, j=1}}\, a_{i j}\, n_{i \mu}\, n_{j \nu} \enq

\nin L'op\'erateur densit\'e se comportant comme une grandeur scalaire sous les transformations de Lorentz : 

\vskip-0.25cm
\beq U(a, A)\,\rho([\,t\,])\,U^\dagger(a,A) = \rho([\,At\,]) \enq

\nin les $\varphi_{\mu \nu}$ sont les composantes d'un tenseur de rang 2. De par la sym\'etrie des $a_{ij}$ et les relations d'orthogonalit\'e $t \cdot n_i = 0$, $n_i \cdot n_j = - \delta_{ij}$, ce tenseur est :

\vv \nin $\bullet$ {\em r\'eel}, {\em sym\'etrique} suivant ses deux indices $\mu$ et $\nu$ ; 

\vv \nin $\bullet$ et v\'erifie ~~~$t^\mu\, \varphi_{\mu \nu} =0$ ;~~~$\varphi^\mu_\mu ~= ~\di{\sum_i}\,a_{ii} = 0 $.

\vv \nin On notera que le terme de rang 1 (\ref{ro1}) peut s'\'ecrire  sous une forme similaire :

\beq \rho^{(1)} = \varphi_\mu \, W^\mu~~~~{\sf avec}~~~~t^\mu\, \varphi_\mu\,=0  \enq 

\nin L. Michel a montr\'e que l'on peut g\'en\'eraliser ce qui pr\'ec\`ede \`a tous les termes du d\'eveloppement de la matrice densit\'e et \'ecrire celle-ci comme\footnote{L . Michel : ``Covariant description of Polarization", Nuovo Cimento, suppl. 14, (1959), p.95.}$^{,}$ \footnote{Prendre garde au fait que les composantes $W_\mu$ ne commutent pas : $[\, W_\mu , W_\nu\,] = i\, \epsilon_{\mu \nu \rho \sigma}\,t^\rho\, W^\sigma$.}

\vskip-0.4cm
\beq \rho = \di{1\over {2j+1}} \, \left\{\, 1 + \varphi_\mu\, W^\mu + \varphi_{\mu \nu}\, W^\mu\, W^\nu + \cdots + \varphi_{\mu_1  \mu_2  \cdots \mu_{2j}}\, W^{\mu_1}\, W^{\mu_2} \cdots W^{\mu_{2j}} \, \right\} \enq

\nin Les $2j$ tenseurs $\varphi_{\mu_1 \cdots \mu_q}$ ($ 1\leq q \leq 2j$) intervenant dans cette somme sont tous r\'eels, sym\'etriques, orthogonaux \`a $t$ et tels que $\varphi^\mu_{\mu \mu_3 \cdots \mu_q} = 0$ pour $q \geq 2$.

\vv 
\vv \nin \ding{198}~Ainsi que nous l'avons rappel\'e dans une note pr\'ec\'edente, toute repr\'esentation irr\'eductible $D$ de $SU(2)$ est \'equivalente \`a sa contragr\'ediente $^tD^{-1}$, elle-m\^eme \'equivalente \`a la repr\'esentation conjugu\'ee $D^\star$. En effet, d'une part, dans la repr\'esentation standard introduite au d\'ebut de cette section, on a $J^\star_1 = J_1,\, J^\star_2 = - J_2,\, J^\star_3 = J_3$ ; d'autre part, il est facile de montrer que  l'op\'erateur unitaire $U = e^{i \pi J_2}$ est tel que $U\, J_1 \, U^{-1} = - J_1,~U\,J_2 \, U^{-1} =  J_2,~U\,J_3 \, U^{-1} = - J_3$, soit $U\, J_k\, U^{-1} = - J^\star_k = -\,^t\hskip -0.06cmJ_k$. Ce r\'esultat nous permet d'\'ecrire : 

\vskip-0.4cm
\beq  \begin{array}{c}
{\rm Tr}~J_{k_1} \cdots J_{k_q} \equiv  {\rm Tr}~J^\prime_{k_1} \cdots J^\prime_{k_q} 
= (-1)^q ~{\rm Tr}~J^\star_{k_1} \cdots J^\star_{k_q} = (-1)^q~\left[\, {\rm Tr}~J_{k_1} \cdots J_{k_q} \,\right]^\star \\ [0.3cm]
= (-1)^q~ {\rm Tr}~^t\hskip-0.05cm J_{k_1} \cdots\,^t \hskip-0.03cm J_{k_q} = (-1)^q~{\rm Tr} ~J_{k_q} \cdots J_{k_1} 
\end{array}
\enq 

\nin On en d\'eduit notamment que la trace d'un produit d'un nombre {\em  pair} de matrices $J_k$ est {\em r\'eelle} ; c'est un nombre {\em imaginaire pur} si ce produit comprend un nombre {\em impair} de matrices $J_k$ ; soit encore, la trace ${\rm Tr}~W_{\mu_1} \cdots W_{\mu_q}$ est r\'eelle si $q$ est pair, imaginaire pure si $q$ est impair. 

\vv \nin \ding{197}~Nous laissons au lecteur le soin de d\'emontrer les relations suivantes\footnote{ Pour calculer la trace d'un produit de $q$ matrices $J_k$, on peut proc\'eder de deux fa\c{c}ons. La premi\`ere consiste \`a consid\'erer la trace de $q+1$ de ces matrices, d'utiliser leurs relations de commutation pour obtenir des traces de produits de $q$ matrices, puis de contracter le r\'esultat par un symbole de Levi-Civita appropri\'e. Cependant, comme cette m\'ethode introduit in\'evitablement une dissym\'etrie, le r\'esultat final ne devient pr\'esentable qu'au prix d'une op\'eration de sym\'etrisation qui alourdit notablement le calcul lorsque le nombre $q$ de matrices est grand. La seconde est plus m\'ethodique, plus sym\'etrique aussi,  mais devient \'egalement compliqu\'ee \`a mesure que $q$ devient grand. Elle consiste \`a consid\'erer la trace d'un produit de $q$ rotations $R_k = e^{-i \theta_k \Vec{n_k} \cdot \Vec{J}}$\hskip-0.1cm. Ce produit est aussi une rotation $R= e^{-i \psi_q \Vec{N_q} \cdot \Vec{J}}$ dont la trace est donn\'ee par 

$$ F(\psi^2_q) = \di{{\sin (\psi_q( j+ \frac{1}{2}))}\over{\sin(\psi_q/2)}} = (2j+1)\,j (j+1) \left\{ 1 - \di{{\psi^2_q}\over{3!}} + \di{{\psi^4_q}\over{5!}} \left[j(j+1) - \di{1\over3}\right] + \cdots       \right\}$$  

\vv \nin Pour des petits angles, on a $R_k = 1 - i \theta_k \Vec{n_k}\cdot\hskip-0.1cm \Vec{J}$.   On voit ainsi que la trace de $R$ contient le terme $(-i)^q\, \theta_1 \cdots \theta_q\, {\rm Tr} \,\Vec{n_1}\cdot \hskip-0.1cm\Vec{J}\cdots \Vec{n_q}\cdot \hskip-0.1cm\Vec{J}$. L'\'etape suivante consiste \`a extraire du d\'eveloppement de $F(\psi^2_q)$ le terme proportionnel au produit $\theta_1 \cdots \theta_q$ et d'obtenir la trace recherch\'ee par identification. D'une part, ce calcul implique celui de $ z_q = \sin(\psi_q/2) = \sqrt{1 - \cos^2(\psi_q/2)}$ que l'on peut mener \`a bien au moyen des formules de r\'ecurrence 

$$ \cos \di{{\psi_{k+1}}\over 2} = \cos \di{\theta_{k+1} \over 2} \cos\di{\psi_k \over 2} - \sin\di{\theta_{k+1} \over 2}\,\sin\di{\psi_k \over 2} \Vec{n}\hskip-0.2cm_{k+1} \cdot \Vec{N_k} $$
$$ \sin\di{\psi_{k+1}\over 2} \Vec{N}\hskip-0.2cm_{k+1} = \cos \di{\theta_k \over 2} \sin\di{\psi_k \over 2} \Vec{N_k} + \cos\di{\psi_k \over 2} \sin \di{\theta_{k+1}\over 2} \Vec{n}\hskip-0.2cm_{k+1} + \sin\di{\psi_k \over 2} \sin \di{\theta_{k+1} \over 2} \Vec{N_k} \wedge \Vec{n}\hskip-0.2cm_{k+1}$$

\vv \nin D'autre part, il doit tenir compte du d\'eveloppement 

$$ \di{\psi_q \over 2} = z_q + \di{{1.z^3_q} \over{2.3}} + \di{{1.3.z^5_q}\over{2.4.5}} + \di{{1.3.5.z^7_q}\over{2.4.6.7}} + \cdots$$}, o\`u $K= \di{1\over 3} \,N\,j(j+1)$. 
\beq \begin{array}{c}
{\rm Tr}\, J_k\, J_\ell = K~\delta_{k \ell},~~~{\rm Tr}\, W_\mu\, W_\nu = K \left(\,t_\mu\,t_\nu - g_{\mu \nu} \, \right) \\[0.4cm]
{\rm Tr}\, J_k\, J_\ell \,J_m= i\,\di{K\over 2}~\epsilon_{k \ell m},~~~{\rm Tr}\, W_\mu\, W_\nu\, W_\rho = i\, \di{K\over 2}~\epsilon_{\mu \nu \rho \sigma}\,t^\sigma \\ [0.5cm]

\di{\sum_r}~J_r\,J_k\,J_r = \left[\,j(j+1) -1 \right] J_k \\[0.3cm]
\di{\sum_r}~J_r\, J_k\, J_\ell\, J_r = \left[\,j(j+1)-2\,\right] J_k\,J_\ell -J_\ell\,J_k + \delta_{k \ell}\,j(j+1)\\ [0.4cm]
\di{\sum_r}~J_r\, J_k\, J_\ell\, J_m\, J_r = \left[\, j(j+1)-3\,\right] J_k\, J_\ell\, J_m - J_\ell\,J_k\,J_m -J_m\,J_\ell\,J_k  -J_k\,J_m\,J_\ell \\[0.3cm]
+j(j+1)\left[\,\delta_{k \ell}\, J_m + \delta_{km}\, J_\ell +\delta_{\ell m}\, J_k \, \right] -\delta_{km} J_\ell\\[0.4cm]

{\rm Tr}\, J_k \, J_\ell \, J_m\, J_n = \di{K \over {10}}\,\left\{ 2 \, j\,(j+1) \left[\, \delta^k_m\, \delta^\ell_n + \delta^k_\ell\, \delta^m_n + \delta^k_n\, \delta^\ell_m \, \right] + \delta^k_n\, \delta^\ell_m + \delta^k_\ell\, \delta^m_n - 4\, \delta^k_m\, \delta^\ell_n  \right\}  \\[0.4cm]
{\rm Tr}\, J_k \, J_\ell \, J_m\, J_n\,J_p = i\,\di{K \over {10}}\, \left\{\, j\,(j+1) \left[\, \delta_{k \ell}\, \epsilon_{mnp} + \delta_{km}\, \epsilon_{\ell np} + \delta_{k n}\,\epsilon_{\ell mp}  + \delta_{k p}\, \epsilon_{\ell mn} \right. \right.\\ [0.4cm]
\left. \left.+ \delta_{\ell m}\, \epsilon_{nkp} +\, \delta_{\ell n}\, \epsilon_{mpk}  + \delta_{\ell p}\, \epsilon_{mnk} + \delta_{mn}\, \epsilon_{pk \ell} + \delta_{mp}\, \epsilon_{n k \ell} + \delta_{np}\, \epsilon_{k \ell m}\,  \right]\right. \\ [0.4cm] 
\left.  - \delta_{k m}\,\epsilon_{\ell np} - \delta_{kn}\, \epsilon_{\ell mp} - \delta_{\ell n}\, \epsilon_{kmp} - \delta_{\ell p}\, \epsilon_{kmn} - \delta_{mp}\,\epsilon_{k \ell n}\,\right\} 
\\[0.3cm]

{\rm Tr}\left[\Vec{\,n\,}\hskip-0.1cm \cdot \hskip-0.1cm \Vec{\,J\,}\right]^{2p+ 1} \hskip-0.1cm=\, 0  \\[0.3cm]
{\rm Tr}\left[\Vec{\,n\,}\hskip-0.1cm \cdot \hskip-0.1cm \Vec{\,J\,}\right]^{2p} = \di{{(2p)!}\over{p!}} \left\{\di{{d^p }\over{ d \chi^p}} \left[ \di{{\sinh (\sqrt{\chi} (j+ \frac{1}{2}))}\over{\sinh( \sqrt{\chi}/2)}}\right] \right\}_{\chi =0}
\end{array} \enq

\vv
\newpage

\section{Compl\'ement IV : Matrice densit\'e de spin du photon et son expression covariante}

\vv \nin On sait qu'un photon (particule de masse nulle) de 4-impulsion donn\'ee ne peut se trouver que dans deux \'etats d'h\'elicit\'e oppos\'ees, $+1$ ou $-1$. Envisageons l'espace vectoriel complexe de dimension 2 engendr\'e par ces deux \'etats que nous noterons $|\,+>$ et $|\,->$, respectivement. L'analogie avec la description des \'etats de spin d'une particule de spin 1/2 appara\^it clairement et l'on peut utiliser ici le m\^eme formalisme bi-dimensionnel avec les matrices de Pauli pour construire la matrice densit\'e de spin du photon. Il est ainsi \'evident que dans ce sch\'ema cette matrice doit prendre la forme (\ref{dens-un-demi}) :  

\beq~ \rho = \di{1\over 2}\, \left\{ 1~ + \Vec{\,\xi\,}\hskip-0.1cm \cdot \hskip -0.1cm \Vec{\,\tau\,} \right\} ~~~~{\sf avec}~~~~\Vec{\,\xi\,} = {\rm Tr}~\rho \hskip -0.1cm\Vec{\,\tau\,} \enq

\vv \nin Consid\'erons le cas d'un photon d\'ecrit par un \'etat pur $|\,c> ~= c_{+}\,|\,+> +~ c_{-}\,|\,->$ normalis\'e \`a 1 ($|c_{+}|^2 + |c_{-}|^2 = 1$). On a 

\beq  \begin{array}{c} 
\rho = |\,c><c\,| = \left(\begin{array}{cc} |c_{+}|^2 & c_{+}\, c^\star_{-} \\ c_{-}\,c^\star_{+} & |c_{-}|^2 \end{array} \right) ~~~~~{\sf et}  \\ [0.45cm] 
\xi_1 =~ <\tau_1>~=~2\,\Re [c_{-}\,c^\star_{+}],~~~\xi_2 =~<\tau_2>~= 2\, \Im[c_{-}\,c^\star_{+}]  \\ [0.4cm] 
\xi_3 =~<\tau_3>~=~|c_{+}|^2 - |c_{-}|^2,~~~~\Vec{\,\xi\,}^2 = \xi^2_1 + \xi^2_2 + \xi^2_3 = 1 
\end{array}
\enq

\vv \nin et l'on v\'erifie que ${\rm Tr}\, \rho^2 = \di{1\over 2}\left[~1\,+\hskip-0.1cm \Vec{\,\xi\,}^2 \,\right] = 1$. Les composantes $\xi_k$ du vecteur de polarisation sont les {\em param\`etres de Stokes} qui caract\'erisent l'\'etat pur consid\'er\'e\footnote{Voir ITL, section 4.7.}. 

\vv \nin Il est facile d'\'etablir l'expression covariante de la matrice densit\'e du photon en tenant compte de l'isomorphisme entre les matrices $2\times2$ de Pauli et les tenseurs d\'efinis en (\ref{4-pauli-1}). Dans cette description covariante, la matrice densit\'e est alors un tenseur de rang 2 dont les \'el\'ements sont 

 \beq \rho_{\mu \nu} = \di{1\over 2}\, \left\{ T_{0 \mu \nu} + \di{\sum_k}~\xi_k\,T_{k \mu \nu} \right\} \enq 

\vv \nin Les expressions de la matrice densit\'e donn\'ees ci-dessus restent valables lorsqu'on a affaire \`a un m\'elange statistique d'\'etats, mais la norme du vecteur $\Vec{\,\xi\,}$ est dans ce cas inf\'erieure \`a 1.


\setcounter{page}{89}

\setcounter{chapter}{2}

\setcounter{equation}{0}

\renewcommand{\theequation}{\mbox{3.}\arabic{equation}}

\newcommand{\kmi}{\chi^{(-)}_{12}} 
\newcommand{\kpi}{\chi^{(+)}_{12}} 
\newcommand{\kmf}{\chi^{(-)}_{34}} 
\newcommand{\kpf}{\chi^{(+)}_{34}} 

\newcommand{\kmg}{\chi^{(-)}_{31}} 
\newcommand{\kpg}{\chi^{(+)}_{31}} 
\newcommand{\kmd}{\chi^{(-)}_{42}} 
\newcommand{\kpd}{\chi^{(+)}_{42}} 

\newcommand{\cTi}{{\cal T}_{12}}
\newcommand{\cZi}{{\cal Z}_{12}}
\newcommand{\cTf}{{\cal T}_{34}}
\newcommand{\cZf}{{\cal Z}_{34}}

\newcommand{\dzp}{\zeta^{(+)}} 
\newcommand{\dzm}{\zeta^{(-)}}

\newcommand{\xpg}{\xi^{(+)}_{13}} 
\newcommand{\xmg}{\xi^{(-)}_{13}} 
\newcommand{\xpd}{\xi^{(+)}_{24}} 
\newcommand{\xmd}{\xi^{(-)}_{24}} 

\newcommand{\cT}{{\cal T}}
\newcommand{\cZ}{{\cal Z}}

\newcommand{\cTg}{{\cal T}_{13}}
\newcommand{\cZg}{{\cal Z}_{13}}
\newcommand{\cTd}{{\cal T}_{24}}
\newcommand{\cZd}{{\cal Z}_{24}}

\newcommand{\cU}{{\cal U}}
\newcommand{\cV}{{\cal V}}

\newcommand{\Ld}{\Lambda^{\frac{1}{2}}} 
\newcommand{\Ldg}{\Lambda^{\frac{1}{2}}_g}
\newcommand{\Ldd}{\Lambda^{\frac{1}{2}}_d}

\newcommand{\vp}{\varphi}


\chapter{Spineurs de Dirac en couplages d'h\'elicit\'e}

\section{Introduction} 

\nin En Physique des particules, on a souvent \`a consid\'erer des r\'eactions \'el\'ementaires faisant intervenir, \`a l'\'etat initial ou \`a l'\'etat final, des particules de spin 1/2, leptons ou quarks. Consid\'erons l'exemple simple de l'effet Compton en Electrodynamique Quantique : 

\beq e + \gs \rightarrow e + \gs \label{compton} \enq 

\vv \nin $e$ repr\'esentant l'\'electron, $\gs$ le photon. A part des facteurs ici inessentiels, l'amplitude de transition de ce processus prend la forme  

\beq T_{\sigma', \lambda' ;\, \sigma, \lambda} = \epsilon^{\star \mu}_{\lambda'} (k')\ov{U}_{\sigma'}(p')\, {\cal M}_{\mu \nu} \,U_{\sigma}(p) \epsilon^\nu_{\lambda} (k) \label{ampl} \enq

\vv \nin o\`u ${\cal M}_{\mu \nu}$ est un tenseur que l'on \'ecrit suivant les r\`egles bien connues de Feynman et qui se pr\'esente, en toute g\'en\'eralit\'e, comme une somme de produits de matrices $\gs$ de Dirac, la pr\'esence de celles-ci venant d'une part du couplage vectoriel \'electron-photon en $\gs_\delta$, et d'autre part de facteurs de propagation de l'\'electron, du type $m + \gs(q)$, $m$ \'etant la masse de l'\'electron. 

\vv \nin Si l'on s'int\'eresse uniquement \`a la section efficace totale de cette r\'eaction, on doit tout d'abord calculer le taux d'interaction :   

\beq {\cal T} = \di{\sum_{\lambda' \lambda}}\,\di{\sum_{\sigma, \sigma'}}\,\left| T_{\sigma' \sigma ;\, \lambda' \lambda} \right|^2 \label{taux} \enq

\nin Ceci est usuellement men\'e \`a terme sans qu'il soit n\'ecessaire de d\'efinir les \'etats de spin des particules entrantes et sortantes, en transformant la somme dans (\ref{taux}) de fa\c{c}on \`a faire appara\^itre des projecteurs    

$$ {\cal T} = \di{\sum_{\lambda'}}\, \epsilon^{\star \mu}_{\lambda'} (k')\,\epsilon^\rho_{\lambda'}  (k')\,\di{\sum_{\lambda}}\, \epsilon^\nu_{\lambda} (k)\,\epsilon^{\star \delta}_{\lambda}  (k)\, \times $$
$$ \times {\rm Tr} \left\{\di{\sum_{\sigma'} } \,U_{\sigma'}(p')\, \ov{U}_{\sigma'}(p') \,{\cal M}_{\mu \nu}\, \di{\sum_\sigma}\,U_\sigma(p) \ov{U}_\sigma(p) \,\ov{{\cal M}}_{\rho \delta } \right\}$$

\nin  o\`u $\ov{\cal M}_{\rho \delta} = \gs_0\,{\cal M}^\dagger_{\rho \delta} \,\gs_0$, puis en  effectuant les substitutions\footnote{En tenant compte de l'invariance de jauge : $k^{\prime \mu}\,\bar{U}_{\sigma'}(p')\, {\cal M}_{\mu \nu} \,U_{\sigma}(p) =0$,  $ k^\nu\,\bar{U}_{\sigma'}(p')\, {\cal M}_{\mu \nu} \,U_{\sigma}(p) =0$.} 

$$ \di{\sum_{\lambda'}}\, \epsilon^{\star \mu}_{\lambda'} (k')\,\epsilon^\rho_{\lambda'}  (k') \rightarrow - g_{\mu \rho}\,,~~~\di{\sum_{\lambda}}\, \epsilon^\nu_{\lambda} (k)\,\epsilon^{\star \delta}_{\lambda}  (k) \rightarrow - g_{\nu \delta} $$
$$ \di{\sum_{\sigma'} } \,U_{\sigma'}(p')\, \ov{U}_{\sigma'}(p')  = m + \gs(p') \,,~~~ \di{\sum_\sigma}\,U_\sigma(p) \ov{U}_\sigma(p) = m + \gs(p) $$

\nin de sorte que ${\cal T}$ prenne la forme d'une trace 

\beq {\cal T} = {\rm Tr}\, \left[\,m + \gs(p') \,\right] {\cal M}^{\mu \nu} \left[\,m + \gs(p)\,\right] \ov{\cal M}_{\mu \nu} \enq

\vv \nin dont le calcul est une affaire de ``$\gs$-gymnastique"\footnote{ITL, section 7.3.}. Si le nombre maximum de matrices $\gs$ impliqu\'ees n'est pas trop \'elev\'e, ce calcul se fait ais\'ement \`a la main. Pour l'effet Compton consid\'er\'e \`a l'ordre le plus bas suivant la constante \'electromagn\'etique\footnote{Ou constante de structure fine.} $\alpha = 1/137$, il est de trois, ce qui ne pose aucune difficult\'e. Par contre, pour des processus plus compliqu\'es, ce nombre peut \^etre \'elev\'e et dans ce cas on effectue g\'en\'eralement le calcul sur ordinateur, \`a l'aide d'un logiciel de calcul symbolique (par exemple, ``Mathematica"). Certes, ce moyen permet un gain ind\'eniable en rapidit\'e d'ex\'ecution du calcul tout en pr\'evenant les erreurs. Cependant,  l'exp\'erience montre que ce gain est g\'en\'eralement contrecarr\'e par la grande longueur et le manque de transparence du r\'esultat ainsi obtenu, qui devient d\`es lors herm\'etique \`a l'interpr\'etation physique. En effet, l'exp\'erience montre aussi que tout calcul analytique comporte des astuces permettant des simplifications et des regroupements de termes, ce que le logiciel de calcul\footnote{Ou plut\^ot, son concepteur !} ignore g\'en\'eralement. C'est d'ailleurs la difficult\'e d'exploiter le r\'esultat analytique d\'elivr\'e par ledit logiciel et la facilit\'e d'un traitement informatique global, qui pourraient en partie expliquer le fait que dans les revues sp\'ecialis\'ees, nombre d'articles traitant de tels processus ne pr\'esentent quasiment plus de formules analytiques, mais seulement des r\'esultats num\'eriques finals, sous forme de courbes ou de jolies repr\'esentations 3D tr\`es en vogue...  

\vv \nin Le calcul de chacune des amplitudes telles que (\ref{ampl}), s'av\`ere n\'ecessaire lorsqu'on recherche des effets li\'es aux \'etats de spin des particules impliqu\'ees dans une r\'eaction. D'une part, on accomplit ainsi  une analyse th\'eorique fine de celle-ci, ce qui est d\'ej\`a en soi tr\`es satisfaisant, et d'autre part, par comparaison avec les mesures exp\'erimentales, on peut v\'erifier plus pr\'ecis\'ement ces pr\'evisions th\'eoriques  et \'eventuellement, selon la r\'eaction consid\'er\'ee, de d\'eceler plus clairement des d\'eviations par rapport \`a ces pr\'evisions, qui seraient r\'ev\'elatrices de nouveaux ph\'enom\`enes, comme par exemple des couplages jusqu'alors inconnus entre particules. Outre qu'il soit incontournable pour ce type d'\'etude, un calcul pas \`a pas du processus, amplitude par amplitude, appara\^it plus m\'ethodique techniquement parlant, tout en se pr\^etant mieux \`a l'interpr\'etation physique\footnote{Certaines amplitudes peuvent se r\'ev\'eler plus importantes que d'autres dans le processus \'etudi\'e.}, ces deux aspects se compl\'etant dans une telle d\'ecomposition du calcul, ce qui permet aussi d'\'eviter des erreurs\footnote{Ajoutons que m\^eme dans cette d\'emarche, il est toujours possible de s'aider astucieusement d'un ordinateur !}. 

\vv \nin On a souvent consid\'er\'e les effets de spin comme n'apportant que des complications de calcul. Sans doute ce jugement provient-il, au moins en partie, du fait que l'on peut se trouver perplexe devant les nombreux choix possibles pour d\'efinir les \'etats de spin des particules, ne sachant lequel pourrait \^etre le plus judicieux. Or, comme nous l'avons esquiss\'e au chapitre 1, et comme nous l'illustrerons dans la suite, l'utilisation des couplages d'h\'elicit\'e se r\'ev\`ele un  proc\'ed\'e efficace et clarificateur pour le calcul d'amplitudes envisag\'e. C'est pourquoi l'objet de ce chapitre est de d\'efinir les couplages d'h\'elicit\'e impliquant des spineurs de Dirac et d'\'etablir des formules utiles s'y rapportant.     

\begin{figure}[hbt] 
\centering
\includegraphics[scale=0.3, width=3.5cm, height=3cm]{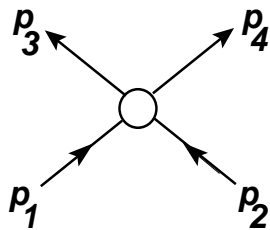} 
\vskip 0.25cm  
\caption{Sch\'ema d'une r\'eaction $1 + 2 \rightarrow 3 + 4$} \label{fig:proc4}
\end{figure}

\vv \nin Consid\'erons un processus $1+2 \rightarrow 3+4$ (deux particules $1$ et $2$ entrant en collision donnant deux particules $3$ et $4$ \`a l'\'etat final), sch\'ematis\'e \`a la figure (\ref{fig:proc4}). Les impulsions et les masses respectives des particules $1$, $2$, $3$ et $4$ seront not\'ees $(p_1, m_1)$, $(p_2,m_2)$, $(p_3, m_3)$ et $(p_4, m_4)$. Dans un premier temps, les masses sont  suppos\'ees non nulles et toutes diff\'erentes. 

\vvv

\begin{figure}[hbt]
\centering
\includegraphics[scale=0.3, width=5cm, height=3cm]{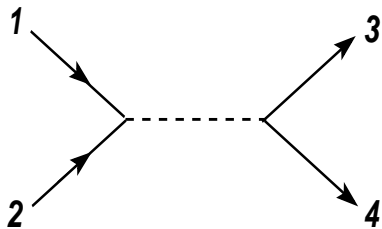}
\vskip 0.25cm

\caption{R\'eaction avec \'echange de particule dans la voie $s$} \label{fig:echs}
\end{figure}

\nin Le couplage le plus couramment utilis\'e est celui de la voie $s$, entre 1 et 2 d'une part, et 3 et 4 d'autre part. C'est surement le plus judicieux pour des processus tels que celui repr\'esent\'e par le diagramme de la figure (\ref{fig:echs}), dans lequel une particule (virtuelle) est \'echang\'ee dans la voie $s$, ou lorsqu'une r\'esonance est form\'ee dans cette voie.   

\vvv

\begin{figure}[hbt]
\centering
\includegraphics[scale=0.3, width=9cm, height=3.5cm]{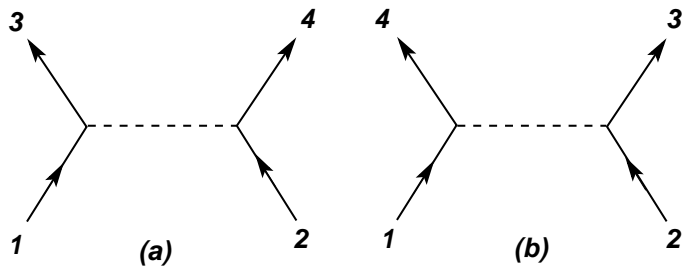}
\vskip 0.25cm

\caption{(a) : R\'eaction avec \'echange de particule dans la voie $t$ ;  \\\hskip -1.08cm 
(b) : diagramme d'\'echange $3 \leftrightarrow 4$. } \label{fig:echt}
\end{figure}

\vv \nin Par contre, pour des processus r\'epondant au sch\'ema de la figure (\ref{fig:echt}) (a),  o\`u une particule (virtuelle) est \'echang\'ee dans la voie $t$, le couplage dans la voie $t$ s'impose, entre 1 et 3 d'une part, et 2 et 4 d'autre part.  

\vvv

\vv \nin Cependant, l'affaire se complique si les particules impliqu\'ees sont toutes de la m\^eme esp\`ece, car en plus du diagramme (a), on doit consid\'erer le diagramme (b) d\'eduit du premier par l'\'echange entre les particules 3 et 4. Comme les \'etats de spin de chaque particule doivent \^etre d\'efinis une fois pour toutes pour le processus global, le couplage dans la voie $t$ utilis\'e pour le diagramme (a) n'est plus adapt\'e au diagramme (b). Dans ce cas, il vaut mieux utiliser le couplage dans la \mbox{voie $s$.} 

\vv \nin Pour des raisons pratiques \'evidentes, les r\'eactions (collisions) r\'ealis\'ees en laboratoire avec des acc\'el\'erateurs se font entre \und{deux} particules entrantes uniquement. Si ces particules sont composites (hadrons constitu\'es de quarks), les ph\'enom\`enes sont g\'en\'eralement interpr\'et\'es en termes de collisions entre deux des constituants \'el\'ementaires de ces particules (quark ou gluon). Par contre, l'\'etat final peut contenir plus de deux particules. Il est alors pr\'ef\'erable d'introduire un autre type de couplage d'h\'elicit\'e dans cet \'etat final\footnote{Voir Chapitre 1, \S 1.3.4, et C. Carimalo, ``Jet-like QED Processes : On General Properties of Impact Factors" hep-ph arXiv:1401.4407.}.

\vv \nin Faisons aussi la remarque suivante. Envisageons l'amplitude de transition d'un processus relevant du mod\`ele standard \'electro-faible ou de la chromodynamique quantique, et impliquant des leptons ou des quarks externes. Il est connu qu'en effectuant des anti-commutations appropri\'ees de matrices $\gs$ et en utilisant les \'equations de Dirac relatives aux spineurs   
externes, on peut faire dispara\^itre tous les facteurs de masse des num\'erateurs des propagateurs internes de spineurs. On obtient alors une amplitude de transition o\`u les matrices $4\times 4$ ``sandwich\'ees" sont des produits de matrices $\gs$ o\`u celles-ci interviennent \und{en nombre impair}. Fondamentalement, ceci est une cons\'equence de la nature du couplage 
$\gs_\mu$ ou $\gs_\mu \gs_5$ entre les leptons ou les quarks, et les particules vectorielles ($\gs, W, Z$, gluon). A titre d'illustration, consid\'erons  l'effet  
Compton (\ref{compton}), au plus bas ordre en $\alpha$. Notons $p_1$ et $p_2$ les impulsions respectives de l'\'electron et du photon initiaux, $p_3$ et $p_4$ les impulsions respectives de l'\'electron et du photon finals. L'amplitude de transition s'\'ecrit 

$$ T_{\sigma_2, \lambda_2 ;\, \sigma_1 \lambda_1} = 4 \pi \alpha\, \epsilon^{\mu\, \star}_{\lambda_4} (p_4)\,\ov{U}_{\sigma_3}(p_3)\, {\cal M}_{\mu \nu} \,U_{\sigma_1}(p_1)\, \epsilon^\nu_{\lambda_2} (p_2) ~~~{\rm avec}$$
$$ {\cal M}_{\mu \nu} = \gs_\mu\, \di{{m + \gs(p_1 + p_2)}\over{m^2 - (p_1 + p_2)^2}} \,\gs_\nu + \gs_\nu\, \di{{m + \gs(p_1 - p_4)}\over{m^2 -(p_1-p_4)^2 }} \,\gs_\mu $$

\vv \nin o\`u $\epsilon_{\lambda_2}(p_2)$ et  $\epsilon_{\lambda_4}(p_4)$ sont les vecteurs de polarisation respectifs du photon initial et du photon final. Comme 

$$ \left[\, m+ \gs(p_1) \,\right]\,\gs_\delta \, U_{\sigma_1}(p_1) = \left\{ \gs_\delta \left[\, m - \gs(p_1) \, \right] + 2 p_{1 \delta} \right\} \,U_{\sigma_1}(p_1) \equiv 2 p_{1 \delta}\,U_{\sigma_1}(p_1) $$

\vv \nin on voit que ${\cal M}_{\mu \nu}$ peut \^etre remplac\'e par 

$${\cal M}'_{\mu \nu} = \di{{\gs_\mu \gs(p_2) \gs_\nu + 2 p_{1 \nu} \gs_\mu}\over{m^2 - (p_1 + p_2)^2}} +\di{{ 2 p_{1 \mu} \gs_\nu - \gs_\nu \gs(p_4) \gs_\mu}\over{m^2 - (p_1 - p_4)^2}}  $$

\vv \nin qui s'\'ecrit bien sous la forme attendue. Projet\'ee sur la base des 16 matrices de Dirac\footnote{Voir ITL, section 7.2.}, une telle expression n'a de composantes que sur les matrices $\gs_\delta$ et $\gs_\delta\, \gs_5$ : 

\beq {\cal M}'_{\mu \nu} = A_{\mu \nu \delta}\, \gs^\delta + B_{\mu \nu \delta}\, \gs^\delta\, \gs_5 \label{decomp} \enq

\vv \nin Par cons\'equent, pour ces processus, il suffirait de savoir calculer uniquement des expressions telles que $\ov{U}_{\sigma_3}(p_3)\, \gs_\delta\,\,U_{\sigma_1}(p_1)$ et $\ov{U}_{\sigma_3}(p_3)\, \gs_\delta\,\gs_5\,\,U_{\sigma_1}(p_1)$. Cependant, pour des processus plus compliqu\'es que l'effet Compton, la d\'ecomposition (\ref{decomp}) peut s'av\'erer difficile \`a \'etablir et prendre finalement plus de temps et de place qu'un calcul direct gardant ${\cal M}'_{\mu \nu}$ en l'\'etat\footnote{En consid\'erant aussi qu'il existe toujours des astuces de calcul.}. 
Par ailleurs, il est int\'eressant de conna\^itre les expressions $\ov{U}_{\sigma_3}(p_3)\, \Gamma\,\,U_{\sigma_1}(p_1)$ o\`u $\Gamma$ est l'une des 16 matrices de Dirac, dans la perspective de mod\`eles th\'eoriques ``au-del\`a du mod\`ele standard", utilisant des couplages d'interaction  
diff\'erents des couplages vectoriels. Pour ces raisons, nous \'etablirons aussi des formules  
pour tous les \'el\'ements de matrice $\ov{\Psi}_{\sigma_3}(p_3)\, \Gamma\,\,\Phi_{\sigma_1}(p_1)$, avec $\Psi$, $\Phi$ pris comme des spineurs de type $U$ ou de type $V$\! ($= \gs_5\,U$). 

\vv \nin Les formules de base utilis\'ees sont les suivantes. Au paragraphe 2.2.4, nous avons montr\'e que dans la transformation de Lorentz pure effectuant le passage de la base $(T,\,X,\,Y,\,Z)$ \`a la base $(t,\,x=X,\,y=Y,\,z)$, le spineur $U_\sigma([\,T\,])$ attach\'e \`a la premi\`ere base devient un spineur $U_\sigma([\,t\,])$ attach\'e \`a la seconde d\'efinie par\footnote{Rappelons que $[\,T \rightarrow t\,]$ est la matrice (au signe pr\`es) de $SL(2,C)$ qui repr\'esente ladite transformation de Lorentz pure.} $[\,t\,] = [\,T \rightarrow t\,]\,[\,T\,]$, et l'on a 

\beq U_\sigma([\,t\,]) = S_{T \rightarrow t}\,U_\sigma([\,T\,])~~~{\rm avec}~~~S_{T \rightarrow t} = \di{{1 + \gs(t)\, \gs(T) }\over{\sqrt{2(1 + t \cdot T)}}}  \label{spt} \enq 

\vv \nin Les relations 

$$1 + t\cdot T = 1 + \cosh \chi = 2 \cosh^2 \di{\chi \over 2}$$ 
$$\left[\,1 + \gs(t)\, \gs(T) \,\right]\,U_\sigma([\,T\,]) = \left[\,1 + \gs(t) \,\right]\,U_\sigma([\,T\,]) ~~{\rm car}~~\gs(T)\, U_\sigma([\,T\,])= U_\sigma([\,T\,]) ~~{\rm et}$$
$$\left[\,1 + \gs(t) \,\right]\,U_\sigma([\,T\,]) = \left[\, 1+ \cosh \chi \,\gs(T) + \sinh \chi\, \gs(Z)\, \right] \,\,U_\sigma([\,T\,]) $$
$$ = 2 \cosh \di{\chi \over 2} \left[\, \cosh \di{\chi \over 2} + \sinh \di{\chi \over 2} \, \gs(Z) \, \right]\,U_\sigma([\,T\,])$$

\nin permettent de r\'ecrire (\ref{spt}) comme 

\beq \fbox{\rule[-0.5cm]{0cm}{1.2cm}~$ U_\sigma([\,t\,]) = \left[\, \cosh \di{\chi \over 2} + \sinh \di{\chi \over 2} \, \gs(Z) \, \right]\,U_\sigma([\,T\,])$~}  \label{F1} \enq

\vv \nin ou encore, en utilisant (2.72), 

\beq \fbox{\fbox{\rule[-0.5cm]{0cm}{1.2cm}~$ U_\sigma([\,t\,]) = \cosh \di{\chi \over 2}\, U_\sigma([\,T\,]) + 2 \sigma \,\sinh \di{\chi \over 2} \,V_\sigma([\,T\,])  $~}} \label{F2} \enq

\vv \nin De la m\^eme mani\`ere, dans la transformation de Lorentz pure $(T,\,X,\,Y,\,Z) \rightarrow (t',\,x'=X,\,y'=Y,\,z')$, on a  
\vskip -0.2cm
\beq U_\sigma([\,t'\,]) = S_{T \rightarrow t'}\,U_\sigma([\,T\,])~~~{\rm avec}~~~S_{T \rightarrow t'} = \di{{1 + \gs(t')\, \gs(T) }\over{\sqrt{2(1 + t'\cdot T)}}}~~{\rm et}~~ [\,t'\,] = [\,T \rightarrow t'\,]\,[\,T\,] \label{sptp} \enq 

\nin La t\'etrade $[\,T\,]$ peut servir de r\'ef\'erence pour d\'efinir les spineurs attach\'es aux vecteurs $t$ et $t'$, au moyen des formules (\ref{spt}) et (\ref{sptp}) ou (\ref{F2}). Cependant, on doit prendre garde au fait que les spineurs qui y figurent doivent avoir la m\^eme normalisation, par exemple $\ov{U}\,U =2$. La particularit\'e des deux transformations pr\'ec\'edentes est qu'elles agissent dans le m\^eme 2-plan $(T, Z)$. Elles peuvent \^etre envisag\'ees comme des rotations d'angles complexes autour de l'axe $Y$, conservant l'axe $X$. Supposons que l'on ait     
  
$$ t = \cosh \chi \,T + \sinh \chi\, Z\, ,~~~z= \sinh \chi \,T + \cosh \chi \, Z,  $$
$$ t' = \cosh \chi' \,T - \sinh \chi'\, Z\, ,~~~z'= - \sinh \chi' \,T + \cosh \chi' \, Z $$

\vv \nin La transformation $(t,\,X,\,Y,\,z) \rightarrow (t',\,X,\,Y,\,z')$ est aussi une transformation de Lorentz pure agissant dans le 2-plan $(T,Z)$, et l'on a\footnote{Il est facile de v\'erifier que  $S_{T \rightarrow t'} \,S^{-1}_{T \rightarrow t} = \di{{1 + \gs(t)\, \gs(T) }\over{\sqrt{2(1 + t' \cdot T)}}} \di{{1 + \gs(T)\, \gs(t) }\over{\sqrt{2(1 + t \cdot T)}}} =  \di{{1 + \gs(t')\, \gs(t) }\over{\sqrt{2(1 + t \cdot t')}}} $.} 

\vskip -0.2cm

$$ t' = \cosh 2 \dzp \,t - \sinh 2\dzp\, z\, ,~~z'= - \sinh 2\dzp \,t + \cosh 2\dzp\, z,~~~{\rm avec}~~~\dzp = \di{{\chi + \chi'}\over 2}  $$
\beq U_\sigma([\,t'\,]) = S_{t \rightarrow t'}\,U_\sigma([\,t\,])~~~{\rm avec}~~~S_{t \rightarrow t'} = \di{{1 + \gs(t')\, \gs(t) }\over{\sqrt{2(1 + t \cdot t')}}} ~~{\rm et}~~[\,t'\,] = [\,t \rightarrow t'\,]\,[\,t\,]  \label{ututp} \enq 
$$ U_\sigma([\,t'\,]) = \cosh \dzp\, U_\sigma([\,t\,]) - 2 \sigma \,\sinh \dzp \,V_\sigma([\,t\,]) $$

\vv \nin Introduisons les 4-vecteurs
\vskip -0.2cm
\beq  \cT = \cosh \dzm  \,T + \sinh \dzm \, Z,~~ \cZ = \cosh \dzm  \,Z + \sinh \dzm\,T, ~~{\rm o\grave{u}}~~~~\dzm = \di{{\chi - \chi'}\over 2}, \enq

\vv \nin le premier du genre temps pointant vers le futur, le second du genre espace, orthogonal au premier. On v\'erifie ais\'ement que 

$$ \cT = \cosh \dzp  \,t - \sinh \dzp \, z = \cosh \dzp  \,t'  + \sinh \dzp \, z' $$
$$\cZ = \cosh \dzp \,z - \sinh \dzp\,t = \cosh \dzp  \,z' + \sinh \dzp\,t' $$
\beq {\rm o\grave{u}}~~~~\dzp = \di{{\chi + \chi'}\over 2} \enq

\vv \nin L'ensemble $(\cT, X, Y, \cZ)$ forme \'egalement une base orthonorm\'ee d'orientation directe et l'on a 

\beq \fbox{\fbox{\rule[-0.5cm]{0cm}{1.2cm}~$ \gs(\cT)\, \gs(\cZ) = \gs(T)\,\gs(Z) = \gs(t)\, \gs(z) = \gs(t')\, \gs(z') = 2 S_Z \gs_5$~}}  \label{ega-heli} \enq

\vv \nin On peut alors r\'ecrire la relation entre spineurs dans (\ref{ututp}) comme 

$$ U_\sigma([\,t'\,]) = \left[ \hskip -0.1cm \di{{}\over{}} \cosh \dzp - \sinh \dzp \,\gs(z) \, \gs(t) \,\right]  U_\sigma([\,t\,]) $$
$$= \left[ \hskip -0.1cm \di{{}\over{}} \cosh \dzp\, \gs(t)  - \sinh \dzp \,\gs(z) \, \right] \gs(t) \,U_\sigma([\,t\,]) ~~~{\rm soit} $$

\beq \fbox{\fbox{\rule[-0.5cm]{0cm}{1.2cm}~$ U_\sigma([\,t'\,]) = \gs(\cT)\,\gs(t)\, U_\sigma([\,t\,]) \equiv \gs(\cT)\,U_\sigma([\,t\,])  $~}}  \label{F3}\enq

\vv \nin Une telle relation (\ref{F3}) sera syst\'ematiquement utilis\'ee pour relier les spineurs de Dirac de deux particules en couplage d'h\'elicit\'e. Dans celle-ci, les spineurs peuvent avoir tous deux la normalisation usuelle $\ov{U}\,U = 2\,m$. Supposons que $T$, $t$ et $t'$ soient tels que   

\beq T = \kappa \left(t + t' \right) \label{gtttp} \enq  

\vv \nin o\`u $\kappa$ est un r\'eel positif. On a dans ce cas $t \cdot T = t^\prime \cdot T = \kappa\, ( 1 + t \cdot t^\prime)$, d'o\`u $\chi = \chi'$\footnote{$\sqrt{2(1 + t\cdot t')} = \kappa^{-1}$ car $T^2 =1$.}, puis $\cT \equiv T$, $\cZ \equiv Z$, ce qui simplifie encore (\ref{ututp}). La relation (\ref{gtttp}) est r\'ealis\'ee lorsqu'on consid\`ere deux particules \und{de m\^eme masse} $m$, dont on d\'efinit les \'etats de spin par un couplage d'h\'elicit\'e, dans la voie $s$ ou dans la voie $t$ (voir chapitre 1). En effet, dans le premier cas, la t\'etrade de r\'ef\'erence est celle attach\'ee \`a l'impulsion totale $P = p_1 + p_2$ o\`u $p_1$ et $p_2$ sont les impulsions respectives des deux particules : notant $P^2 = s$, $T = P/\sqrt{s}$, $t_1 = p_1/m$, $t_2=p_2/m$, on a $T = \di{m\over \sqrt{s}} (t_1 + t_2)$. Dans le second cas, la t\'etrade de r\'ef\'erence a pour vecteur unitaire du genre temps (voir Eqs. 1.57, 1.58) 

\beq T =  \di{{p_1 + p_2}\over{\sqrt{t + 4 m^2}}} = \di{m\over{\sqrt{t + 4 m^2}}} (t_1+t_2) \enq

\nin o\`u ici $t = -(p_1 -p_2)^2 > 0$. 

\vv \nin Faisons une derni\`ere remarque sur l'interpr\'etation des h\'elicit\'es des particules. Supposant qu'elles soient de spin 1/2, l'op\'erateur d'h\'elicit\'e de la premi\`ere est 

$$ S_{z_1}(t_1) = \di{1\over 2} \gs_5\, \gs(z_1)\, \gs(t_1) $$ 

\nin Appliquant (\ref{ega-heli}), on a 

\beq  S_{z_1}(t_1)  = S_Z(T) \label{hel1} \enq

\vv \nin Cette \'egalit\'e est \'egalement valable pour la seconde particule dans le cas d'un couplage  dans la voie $t$. A noter aussi qu'elle n'est vraie que pour les composantes de spin suivant les axes ``$z$" respectifs des particules. Dans le cas du couplage dans la voie $s$, on sait qu'une rotation d'angle $\pi$ autour de $Y$ doit \^etre introduite pour d\'efinir la t\'etrade de la seconde particule. Son op\'erateur d'h\'elicit\'e est alors l'oppos\'e de (\ref{hel1}).

\vv \nin Dans le r\'ef\'erentiel propre de $T$, la 3-impulsion de la particule 1 est selon le troisi\`eme axe, et (\ref{hel1}) devient 

 \beq S_{z_1}(t_1) \equiv \di{1\over 2} \, \tau_3 \enq 

\vv \nin autrement dit, dans ce r\'ef\'erentiel, l'h\'elicit\'e de ladite particule repr\'esente la composante de son spin suivant sa 3-impulsion\footnote{L'h\'elicit\'e d'une particule est couramment d\'efinie comme la composante de son spin suivant sa 3-impulsion, quel que soit le r\'ef\'erentiel consid\'er\'e. Mais comme nous l'avons vu, sauf pour une particule sans masse, il ne s'agit pas d'une grandeur invariante relativiste.}.  S'il s'agit d'un couplage dans la voie $s$, le r\'ef\'erentiel propre de $T$ est celui du centre de masse des deux particules. Dans celui-ci, la 3-impulsion de la seconde particule est aussi selon l'axe 3, mais en sens oppos\'e. Du fait du renversement de signe mentionn\'e plus haut, son h\'elicit\'e y repr\'esente aussi la composante de son spin suivant sa 3-impulsion. Par contre, s'il s'agit d'un couplage dans la voie $t$, le r\'ef\'erentiel propre de $T$ est le r\'ef\'erentiel de Breit dans lequel la 3-impulsion de la seconde particule est dans le sens oppos\'e de l'axe 3. L'h\'elicit\'e de cette particule y repr\'esente alors l'oppos\'ee de la composante de son spin suivant sa 3-impulsion.   

\vv \nin Dans la suite, sauf indication contraire, les spineurs de Dirac ont tous la m\^eme normalisation : 

\vvv
\beq \fbox{\rule[-0.5cm]{0cm}{1.2cm}~$~~ \ov{U}\,U = -\, \ov{V}\, V = 2~~ $~}  \label{N2}\enq

\vvv \vvv
\section{Le couplage d'h\'elicit\'e dans la voie $s$}

\subsection{D\'efinition des t\'etrades} 

\vv \nin Consid\'erons le processus de la figure (\ref{fig:proc4}) o\`u nous supposons dans un premier temps  que les quatre particules sont toutes de spin 1/2. Nous utiliserons les notations suivantes.

$$ P = p_1 + p_2 = p_3 + p_4\,,~~s = P^2 ~\left(~\geq {\rm Max}\left\{ (m_1 +m_2)^2, (m_3 + m_4)^2\right\}~ \right)$$
$$ t_i = \di{{p_i}\over m_i}\,,~~T= \di{P\over \sqrt{s}} \,,~~ \cosh \chi_i = t_i \cdot T $$
$$t= -(p_1 -p_3)^2 = - (p_2 -p_4)^2\,,~~u = -(p_1 -p_4)^2 = - (p_2 - p_3)^2 $$

\vv \nin Dans le r\'ef\'erentiel du centre de masse (o\`u $\Vec{\,P\,} = \Vec{\,0\,}$), les \'energies et modules de 3-impulsion de chaque particule sont 

$$ E_1 = \di{{s + m^2_1 - m^2_2}\over{ 2 \,\sqrt{s}}} = m_1 \, \cosh \chi_1\,,~~ E_2 = \di{{s + m^2_2 - m^2_1}\over{ 2 \,\sqrt{s}}} = m_2 \, \cosh \chi_2$$
$$ \left|\Vec{p_1}\right| =\left|\Vec{p_2}\right| = m_1 \sinh \chi_1 = m_2 \sinh \chi_2 = \di{\Lambda_{12}^{{\frac{1}{2}}}\over{2\, \sqrt{s}}} ~~{\rm avec}~~\Lambda_{12} = \Lambda(s, m^2_1,m^2_2) $$
$$ \left|\Vec{p_3}\right| =\left|\Vec{p_4}\right| = m_3 \sinh \chi_3 = m_4 \sinh \chi_4 = \di{\Lambda_{34}^{{\frac{1}{2}}}\over{2\, \sqrt{s}}} ~~{\rm avec}~~\Lambda_{34} = \Lambda(s, m^2_3,m^2_4) $$

\nin De par la conservation de la 4-impulsion totale, trois seulement des impulsions sont ind\'ependantes. Pour constituer une base d'espace-temps, on peut par exemple utiliser les trois impulsions $P$, $p_1$ et $p_3$. Le dernier vecteur devant compl\'eter la base sera choisi perpendiculaire \`a l'hyperplan de ces trois vecteurs. C'est donc un vecteur du genre espace que l'on prendra comme axe $Y$,  commun \`a toutes les t\'etrades. On montre que le vecteur unitaire correspondant s'\'ecrit 

\beq Y_\mu = A\, \epsilon_{\mu \nu \rho \delta} P^\nu\,p^\nu_1\, p^\delta_3~~~~{\rm avec}~~~A = \di{{4 \sqrt{s}}\over{\Lambda_{12}^{\frac{1}{2}}\, \Lambda_{34}^{\frac{1}{2}}\,\sin \theta}} \label{VY} \enq  

\vv \nin o\`u, dans le r\'ef\'erentiel du centre de masse, $\theta$ est l'angle entre $\Vec{p_3}$ et $\Vec{p_1}$ (angle de diffusion).

\vv \nin En suivant ce qui a \'et\'e dit au chapitre 1, les t\'etrades associ\'ees \`a chacune des particules dans ce  couplage d'h\'elicit\'e sont les suivantes. 

\vv \nin \ding{172} \und{\bf Pour la particule 1} 

$$ t_1 = \cosh \chi_1 \,T + \sinh \chi_1 \,Z\,,~~z_1 = \sinh \chi_1\,T + \cosh \chi_1\, Z $$
$$ x_1 = X\,,~~~y_1 = Y\, ,~~~\epsilon^{(\pm)}_1 = E^{(\pm)} = \mp \di{1\over \sqrt{2}} \left[ X \pm i Y\right] $$

\nin Ce choix correspond bien \`a un couplage d'h\'elicit\'e entre la t\'etrade $(T, X, Y, Z)$ associ\'ee \`a $T$ et celle, $(t_1,x_1,y_1,z_1)$, associ\'ee \`a $t_1$ car 

$$ Z \equiv \di{{2 \,\sqrt{s}}\over{\Lambda_{12}^{\frac{1}{2}}}} \left( \, p_1 - \di{{(p_1\cdot P)}\over s} P \right)~~~{\rm et}~~~ z_1 \equiv \di{{2 \,m_1}\over{\Lambda_{12}^{\frac{1}{2}}}} \left( \, - P + \di{{(p_1\cdot P)}\over m^2_1} p_1 \right)$$

\vv \nin On passe de la t\'etrade $\left\{ [\,T\,] : (T, X, Y, Z)\right\}$ \`a la t\'etrade $\left\{ [\,t_1\,]_s : (t_1,x_1,y_1,z_1)\right\}$ par une transformation de Lorentz pure agissant dans le 2-plan $(T,Z)$ ($[\,t_1\,]_s = [\,T \rightarrow t_1\,]\,[\,T\,]$). Avec ce choix, la 3-impulsion de la particule 1 dans le r\'ef\'erentiel $(T, X, Y, Z)$ est selon l'axe des z. 

\vv \nin \ding{173} \und{\bf Pour la particule 2} 

$$ t_2 = \cosh \chi_2 \,T - \sinh \chi_2 \,Z\,,~~z_2 =  \sinh \chi_2\,T - \cosh \chi_2\, Z $$
$$ x_2 = - X\,,~~~y_2 = Y\, ,~~~\epsilon^{(\pm)}_2 = E^{(\mp)} = \pm \di{1\over \sqrt{2}} \left[ X \mp i Y\right]  $$

\nin On a ici 

$$ Z \equiv \di{{2 \,\sqrt{s}}\over{\Lambda_{12}^{\frac{1}{2}}}} \left( \,- p_2 + \di{{(p_2\cdot P)}\over s} P \right)~~~{\rm et}~~~ z_2 \equiv \di{{2 \,m_2}\over{\Lambda_{12}^{\frac{1}{2}}}} \left( \, - P + \di{{(p_2\cdot P)}\over m^2_2} p_2 \right)$$

\vv \nin On passe de la t\'etrade $[\,T\,]$ \`a la t\'etrade $\left\{ [\,t_2\,]_s : (t_2,x_2,y_2,z_2)\right\}$ en effectuant d'abord une rotation de $\pi$ autour de $Y$ dans le r\'ef\'erentiel li\'e \`a $T$ ($R_Y(\pi)$), puis une transformation de Lorentz pure agissant dans le 2-plan $(T,Z)$ ($[\,t_2\,]_s = [\,T \rightarrow t_2\,]\,R_Y(\pi)\,[\,T\,]$). Dans le r\'ef\'erentiel $(T, X, Y, Z)$ la 3-impulsion de la particule 2 est en sens oppos\'e \`a celui de l'axe des z. 

\vv
\vv \nin \ding{174} \und{\bf Pour les particules 3 et 4} 

\vv \nin On effectue tout d'abord une rotation d'angle $\theta$ autour de $Y$, $R_Y(\theta)$,  transformant $X$ et $Z$ en 

$$ X' = \cos \theta \, X - \sin \, \theta Z\,,~~~Z' = \cos \, \theta \,Z + \sin\, \theta\, X $$

\vv \nin Une nouvelle t\'etrade ($[\,T\,]' = R_Y(\theta)\,[\,T\,]$ est ainsi associ\'ee \`a $T$ et la t\'etrade associ\'ee \`a la particule 3 est alors  

$$ t_3 = \cosh \chi_3\, T + \sinh \chi_3\, Z'\,,~~~z_3 = \sinh \chi_3\,T + \cosh \chi_3\, Z' $$
$$ x_3 = X'\,,~~~y_3 = Y\,,~~~\epsilon^{(\pm)}_3 = \mp \di{1\over \sqrt{2}} \left[X' \pm i Y\right] $$
$$ {\rm soit}~~~\epsilon^{(\pm)}_3 = \pm \di{1\over \sqrt{2}} \sin \theta\,Z + \di{1\over 2} \left[ 1 + \cos \theta \, \right] E^{(\pm)}  + \di{1\over 2} \left[ 1 - \cos \theta \, \right] E^{(\mp)}   $$

\nin On a donc $[\,t_3\,]_s = [\,T \rightarrow t_3\,]'\,[\,T\,]'$, o\`u $[\,T \rightarrow t_3\,]'$ est une transformation de Lorentz pure agissant dans le 2-plan $(T, Z')$. Enfin, la t\'etrade associ\'ee \`a la particule 4 est $[\,t_4\,]_s = [\,T \rightarrow t_4\,]'\,R_Y(\pi)\,[\,T\,]'$, soit

$$t_4 = \cosh \chi_4\, T - \sinh \chi_4\, Z'\,,~~~ z_4 =  \sinh \chi_4\,T - \cosh \chi_4\, Z' $$
$$ x_4 = - X'\,,~~~y_4 = Y\,,~~~\epsilon^{(\pm)}_4 = \epsilon^{(\mp)}_3 = \pm \di{1\over \sqrt{2}} \left[X' \mp i Y\right]  $$

\nin Notons $U_\sigma$ et $V_\sigma = \gs_5\, U_\sigma$ les spineurs de Dirac associ\'es \`a la t\'etrade de r\'ef\'erence $[\,T\,]$. Dans l'espace des spineurs, les rotations $R_Y(\pi)$ et $R_Y(\theta)$ sont repr\'esent\'ees par les matrices 

$$ R_Y(\pi) = \gs(Z)\, \gs(X) = - 2 i\, S_Y \,,~~~R_Y(\theta) = \cos \di{\theta \over 2} + \sin \di{\theta \over 2} \,\gs(Z)\,\gs(X) $$

\vv \nin et les transformations de Lorentz pures pr\'ec\'edentes par 

$$ [\, T \rightarrow t_1\,] \rightarrow S_{T \rightarrow t_1} = \cosh \di{\chi_1 \over 2} + \sinh \di{\chi_1 \over 2}\, \gs(Z) \, \gs(T) ~~~\left( \, \gs(Z)\, \gs(T) = 2 \gs_5\, S_Z\, \right)$$
$$ [\, T \rightarrow t_2\,] \rightarrow S_{T \rightarrow t_2} = \cosh \di{\chi_2 \over 2} - \sinh \di{\chi_2 \over 2}\, \gs(Z) \, \gs(T) $$
$$ [\, T \rightarrow t_3\,]' \rightarrow S'_{T \rightarrow t_3} = \cosh \di{\chi_3 \over 2} + \sinh \di{\chi_3 \over 2}\, \gs(Z') \, \gs(T) ~~~\left( \, \gs(Z')\, \gs(T) = 2 \gs_5\, S_{Z'}\, \right)$$
$$ [\, T \rightarrow t_4\,]' \rightarrow S'_{T \rightarrow t_4} = \cosh \di{\chi_4 \over 2} - \sinh \di{\chi_4 \over 2}\, \gs(Z') \, \gs(T) $$

\vv \nin Compte tenu de la relation $S_Y \, U_\sigma =  i \sigma\, U_{-\sigma}$, les spineurs associ\'es \`a la t\'etrade $[\,T\,]'$ sont donn\'es par 

\beq  U'_\sigma = R_Y(\theta)\,U_\sigma = \cos \di{\theta \over 2} \,U_\sigma + 2 \sigma \sin \di{\theta \over 2} \,U_{-\sigma} \,,~~V'_\sigma = \cos \di{\theta \over 2}\, V_\sigma + 2 \sigma \sin \di{\theta \over 2} \,V_{-\sigma} \label{uprime} \enq 

\vv \nin Appliquant sur $U_\sigma$ les transformations appropri\'ees, on obtient facilement les spineurs de Dirac associ\'es dans ce couplage aux particules 1, 2, 3 et 4, respectivement. Ainsi :   

\vv \nin \ding{172} \und{\bf Particule 1} 

\beq {U_1}_{ \sigma} = S_{T \rightarrow t_1}\, U_\sigma = \cosh \di{\chi_1 \over 2}\,U_\sigma + 2\, \sigma\,\sinh \di{\chi_1 \over 2}\,V_\sigma \label{us1} \enq

\vv \nin \ding{173} \und{\bf Particule 2} 

\beq {U_2}_{ \sigma} = S_{T \rightarrow t_2}\,R_Y(\pi)\, U_\sigma = 2 \sigma \left[ \cosh \di{\chi_2 \over 2}\,U_{-\sigma} + 2\, \sigma\,\sinh \di{\chi_2 \over 2}\,V_{-\sigma} \right] \label{us2} \enq

\vv \nin \ding{174} \und{\bf Particule 3} 

\beq {U_3}_{ \sigma} = S'_{T \rightarrow t_3}\, R_Y(\theta)\,U_\sigma = \cosh \di{\chi_3 \over 2}\,U'_\sigma + 2\, \sigma\,\sinh \di{\chi_3 \over 2}\,V'_\sigma \label{us3} \enq

\vv \nin \ding{175} \und{\bf Particule 4} 

\beq {U_4}_{ \sigma} = S'_{T \rightarrow t_4}\,R_Y(\theta)\,R_Y(\pi)\, U_\sigma = 2 \sigma \left[ \cosh \di{\chi_4 \over 2}\,U'_{-\sigma} + 2\, \sigma\,\sinh \di{\chi_4 \over 2}\,V'_{-\sigma} \right] \label{us4} \enq

\vv \nin Voyons comment ces spineurs peuvent \^etre reli\'es les uns aux autres. Posons 

\beq \kpi = \di{{\chi_1 + \chi_2}\over 2}\, ,~~~ \kmi = \di{{\chi_1 - \chi_2}\over 2}\, ,~~~\kpf = \di{{\chi_3 + \chi_4}\over 2}\, ,~~~ \kmf = \di{{\chi_3 - \chi_4}\over 2} \enq

\vv \nin Combinant (\ref{us1}) et (\ref{us2}), on obtient

\beq {U_2}_{ \sigma} = 2 \sigma \, \left[ \cosh \kpi \,{U_1}_{ -\sigma} + 2\, \sigma\,\sinh \kpi \,{V_1}_{ -\sigma} \right] \label{r12} \enq  

\vv \nin De m\^eme, \`a l'\'etat final, on a 

\beq {U_4}_{ \sigma} = 2 \sigma \, \left[ \cosh \kpf \,{U_3}_{ -\sigma} + 2\, \sigma\,\sinh \kpf \,{V_3}_{ -\sigma} \right]\enq  

\nin Enfin, en posant $\kmg= \di{{\chi_3 - \chi_1}\over 2} $~~et~ $\kmd = \di{{\chi_4 - \chi_2}\over 2}$, on trouve 

$$ {U_3}_{ \sigma} = R_Y(\theta)\, \left[\, \cosh \kmg \, {U_1}_{ \sigma} + 2 \sigma\, \sinh \kmg \,{V_1}_{ \sigma} \right] $$
\beq {U_4}_{ \sigma} = R_Y(\theta)\, \left[\, \cosh \kmd\, {U_2}_{ \sigma} + 2 \sigma\, \sinh \kmd \,{V_2}_{ \sigma} \right] \enq

\vv \nin La relation (\ref{r12}) peut \^etre r\'ecrite sous forme condens\'ee en observant que $\gs(t_1)\,U_1 = U_1$ et que $- 2 \sigma\, {V_1}_{- \sigma} = \gs(z_1)\, {U_1}_{-\sigma}$. On obtient en effet 

$$ {U_2}_{ \sigma}  =  2 \sigma \,\gs(\cTi) \,{U_1}_{ -\sigma}~~~{\rm avec} $$ 
\beq \cTi= t_1\, \cosh \kpi - z_1\, \sinh \kpi = T\,\cosh \kmi + Z\,\sinh \kmi  
\label{rn12} \enq  

\vv \nin Pour l'\'etat final, on obtient de fa\c{c}on similaire 

$$ {U_4}_{ \sigma} = 2 \sigma \,\gs(\cTf) \,{U_3}_{ -\sigma}~~~{\rm avec}$$
\beq \cTf= t_3\, \cosh \kpf - z_3\, \sinh \kpf = T\,\cosh \kmf + Z'\,\sinh \kmf  
\label{rn34} \enq  

\vv \nin A l'aide des deux vecteurs $\cTi$ et $\cTf$ on obtient donc des formules du type (\ref{F3}). Ce sont  des vecteurs du genre temps et unitaires, et les t\'etrades 

$$ \cTi\,,~~\cZi =  Z\,\cosh \kmi + T\,\sinh \kmi\,,~~X\,,~~\, Y $$  
\beq \cTf\,,~~\cZf =  Z'\,\cosh \kmf + T\,\sinh \kmf\,,~~X'\,,~~\, Y \enq  

\vv \nin qui leur sont respectivement associ\'ees se d\'eduisent, la premi\`ere de la t\'etrade $(T,X,Y,Z)$ par la transformation de Lorentz pure de param\`etre $\kmi$ dans le 2-plan $(T,Z)$ ; la seconde, de la t\'etrade $(T,X',Y,Z')$ par la transformation de Lorentz pure de param\`etre $\kmf$ dans le 2-plan $(T,Z')$. 

\vv \nin On voit alors que la t\'etrade $(t_1, X,Y,z_1)$ se d\'eduit de la t\'etrade $(\cTi, X,Y,\cZi)$ par la transformation de Lorentz pure $L_{12}$ de param\`etre $\kpi$ dans le 2-plan $(T,Z)$, tandis que la t\'etrade $(t_2, -X, Y, z_2)$ est obtenue \`a partir de cette t\'etrade en effectuant d'abord une rotation de $\pi$ autour de $Y$ et en appliquant ensuite $L_{12}$. On a ainsi 

$$ t_1 = \cosh \kpi \,\cTi + \sinh \kpi\, \cZi\,,~~~  t_2 = \cosh \kpi \,\cTi - \sinh \kpi\, \cZi\,, ~~~{\rm donc} $$
\beq t_1 + t_2 = 2\,\cosh \kpi \,\cTi \,,~~~t_1 - t_2 = 2\, \sinh \kpi \,\cZi \enq

\vv \nin Notons que lorsque $m_1 = m_2$, soit $\kmi =0$, (\ref{rn12}) devient  similaire \`a  (\ref{F3}) : 

\beq  {U_2}_{ \sigma} = 2 \sigma \,\gs(T) \,{U_1}_{ -\sigma} \enq

\nin et que pour $\kmg = 0$, soit, pour le cas qui nous int\'eressera, $m_1 = m_3$ \und{et} $m_2 = m_4$, la relation entre $U_3$ et $U_1$ se r\'eduit \`a 

\beq {U_3}_{ \sigma} = R_Y(\theta)\, {U_1}_{ \sigma}  \label{egm31} \enq 

\vv
\subsection{Projecteurs $U_1\,\ov{U}_2$ du couplage en voie $s$}

\vvv \nin Rappelons tout d'abord certaines formules 

\vvv
\beq  \fbox{\rule[-0.7cm]{0cm}{1.5cm}~$ \begin{array}{c} ~\\
U_\sigma\, \ov{U}_{\sigma'} = \delta_{\sigma',\sigma} \di{1\over 2} \left[ 1 + (2\sigma) \gs_5\, \gs(z) \right] \left[ 1+ \gs(t)\right] \\
+ \delta_{\sigma', - \sigma} \di{1\over 2} \gs_5 \left(x + i (2\sigma) y \right)\left[1+\gs(t)\right] \\~\\
\hline\\
\ov{U}_{\sigma'}\, U_\sigma = 2 \delta_{\sigma', \sigma},~~~~\ov{U}_{\sigma'}\,\gs_5\, U_\sigma =0,~~~~\ov{U}_{\sigma'}\,\gs_\mu\, U_\sigma = 2 \delta_{\sigma', \sigma}\,t_\mu, \\~\\

~~~~\ov{U}_{\sigma'}\,\gs_\mu\,\gs_5\, U_\sigma = 2 (2\sigma) \delta_{\sigma', \sigma}\,z_\mu + 2 \delta_{\sigma', -\sigma} \left[x+i(2\sigma) y\right]_\mu \\~\\
\hline \\
\ov{U}_{\sigma'}\,\sigma_{\mu \nu}\, U_\sigma = (2\sigma) \delta_{\sigma', \sigma}\,\left[ x_\mu y_\nu - x_\nu y_\mu \right] \\
+ i (2 \sigma) \delta_{\sigma',-\sigma} \left\{  z_\mu \left[x+ i (2\sigma) y\right]_\nu - z_\nu \left[x+ i (2\sigma) y\right]_\mu \right\} \\~\\
\hline \\
\ov{U}_{\sigma'}\,\sigma_{\mu \nu}\,\gs_5\, U_\sigma = i (2\sigma) \delta_{\sigma', \sigma}\,\left[ t_\mu z_\nu - t_\nu z_\mu \right] \\
+ i \delta_{\sigma',-\sigma} \left\{  t_\mu \left[x+ i (2\sigma) y\right]_\nu - t_\nu \left[x+ i (2\sigma) y\right]_\mu \right\} \\~\\
\hline \\
\gs_5\, \gs(x + i \epsilon y) = \epsilon\, \gs(x + i \epsilon y) \gs(z) \gs(t)~~~(\epsilon = \pm 1) \\~\\

 \gs(z_1) \gs(t_1) = \gs(Z) \gs(T) = \gs(\cZi)\,\gs(\cTi) \\~
\end{array} $} \label{projo}\enq

\vv\nin et rappelons aussi\footnote{ITL, Chap. 7, Eq. 7.80.} que $\gs^\dagger_\mu = \gs_0 \, \gs_\mu\,\gs_0$. 

\newpage
\nin \ding{172} \und{\bf Projecteur $\uup_1\,\ov{\uup_2}$}

\vv \nin D'apr\`es (\ref{rn12}), on a 
\vskip -0.3cm
$$ \uup_1\,\ov{\uup_2} = \uup_1 \, \ov{\ud_1} \, \gs(\cTi) = \di{1\over 2}\, \gs_5\, \gs(X + i Y) \, \left[1 + \gs(t_1) \right] \,\gs(\cTi) = \di{1\over 2}\, \left[1 + \gs(t_1) \right] \, \gs_5\, \gs(X + i Y) \, \gs(\cTi) $$
$$= - \di{1\over 2}\, \left[1 + \gs(t_1) \right] \, \gs(\cZi)\, \gs(X + i Y) = \di{1\over 2}\, \gs(X+iY)\,\left[1 - \gs(t_1) \right] \, \gs(\cZi) $$

\nin Or, ~~$\left[1 - \gs(t_1) \right]  \,\gs(\cZi)  = \gs(\cZi)   - \cosh \kmi \gs(\cTi)\,\gs(\cZi)  + \sinh \kmi  =  \gs(\cZi)\,\left[\,1+\gs(t_2)\,\right],~~{\rm d'o\grave{u}}  $

\beq \fbox{\fbox{\rule[-0.7cm]{0cm}{1.5cm}~$ \uup_1\,\ov{\uup_2} = \di{1\over 2} \left[ \,1 + \gs(t_1)\,\right] \gs(X+iY)\,\gs(\cZi) 
= \di{1\over 2}\,\gs(X+iY) \,\gs(\cZi)\,\left[\,1 + \gs(t_2)\, \right]$~}}  \label{p12uu}\enq

\vv \nin \ding{173} \und{\bf Projecteur $\ud_1\,\ov{\ud_2}$}

\vv \nin Le m\^eme type de calcul conduit \`a 
\vskip -0.2cm
\beq \fbox{\fbox{\rule[-0.7cm]{0cm}{1.5cm}~$ \ud_1\,\ov{\ud_2} = - \,\di{1\over 2} \left[ \,1 + \gs(t_1)\,\right]  \gs(X-i Y) \,\gs(\cZi)
= -\di{1\over 2}\,\gs(X-iY) \,\gs(\cZi)\,\left[\,1 + \gs(t_2)\, \right]$~}}  \label{p12dd}\enq

\vv \nin \ding{174} \und{\bf Projecteur $\ud_1\,\ov{\uup_2}$}

\vv \nin On a \hskip 1cm $\ud_1\,\ov{\uup_2} = \ud_1\,\ov{\ud_1} \, \gs(\cTi) = \di{1\over 2} \left[ 1 - \gs_5\,\gs(z_1)\,\gs(t_1)\, \right] \left[ 1+ \gs(t_1)\,\right] \, \gs(\cTi) $ 
$$= \di{1\over 2} \left[ 1+ \gs(t_1)\,\right] \left[ 1 - \gs_5\,\gs(z_1)\,\gs(t_1)\, \right] \gs(\cTi) $$

\nin Mais \hskip 0.7cm $\left[ 1 - \gs_5\,\gs(z_1)\,\gs(t_1)\, \right] \gs(\cTi) =  \left[ 1 - \gs_5\,\gs(\cZi)\,\gs(\cTi)\, \right] \gs(\cTi) = \gs(\cTi) - \gs_5 \gs(\cZi) ~~~{\rm et}$ 
$$ [\,1 + \gs(t_1)\,]\,\gs(\cTi) = \gs(\cTi)\,\left[\, 1 + \gs(t_2)\,\right],~~~  [\,1 - \gs(t_1)\,]\,\gs(\cZi) = \gs(\cZi)\,\left[\, 1 + \gs(t_2)\,\right], ~~{\rm d'o\grave{u}}$$

\beq \fbox{\fbox{\rule[-0.7cm]{0cm}{1.5cm}~$\ud_1\,\ov{\uup_2} = \di{1\over 2} \left[ \, 1 + \gs(t_1)\, \right]\,\left[ \,\gs(\cTi) - \gs_5 \gs(\cZi)\,\right] =\di{1\over 2} \,\left[ \,\gs(\cTi) - \gs_5 \gs(\cZi)\,\right] \,\left[ \, 1 + \gs(t_2)\, \right] $~}} \label{p12du} \enq

\vv \nin \ding{175} \und{\bf Projecteur $\uup_1\,\ov{\ud_2}$}

\vv \nin On trouve de la m\^eme mani\`ere 

\beq \fbox{\fbox{\rule[-0.9cm]{0cm}{2cm}~$\begin{array}{c} \uup_1\,\ov{\ud_2} = - \di{1\over 2} \left[ \, 1 + \gs(t_1)\, \right]\,\left[ \,\gs(\cTi) + \gs_5 \gs(\cZi)\,\right] \\~\\=- \di{1\over 2} \,\left[ \,\gs(\cTi) + \gs_5 \gs(\cZi)\,\right] \,\left[ \, 1 + \gs(t_2)\, \right] \end{array} $~}} \label{p12ud} \enq

\vvv
\subsection{Formes bilin\'eaires $\ov{U_2}_{\sigma'}\,\Gamma\,{U_1}_{\sigma}$ et $\ov{V_2}_{\sigma'}\,\Gamma\,{U_1}_{\sigma}$ du couplage en voie $s$}

\vv \nin Celles-ci se calculent ais\'ement en les exprimant comme des traces et en utilisant les expressions pr\'ec\'edentes des projecteurs :  $\ov{U_2}_{ \sigma'}\,\Gamma\,{U_1}_{ \sigma} = {\rm Tr}\, \Gamma \,U_{1 \sigma}\ov{U_2}_{\sigma'}$, $\ov{V_2}_{\sigma'}\,\Gamma\,{U_1}_{ \sigma} = - {\rm Tr}\,\gs_5\, \Gamma \,{U_1}_{ \sigma}\ov{U_2}_{\sigma'} $. Rappelons que les traces des matrices $\gs$ ainsi que celle de $\gs_5$ sont nulles, que ${\rm Tr}\, \gs_\mu\, \gs_\nu\, \gs_5 =0$, et que la trace du produit d'un nombre impair de matrices $\gs$ est nul\footnote{Pour la ``$\gs$-gymnastique", voir ITL, section 7.3.}. On tient compte \'egalement des relations d'orthogonalit\'e entre les vecteurs de la t\'etrade $(\cTi,X,Y,\cZi)$.

\vskip 1cm
\begin{center} 
\fbox{\fbox{\rule[-2cm]{0cm}{4cm}
\begin{minipage}{0.75\textwidth}
\vvv
\beq
\fbox{\rule[-0.2cm]{0cm}{0.6cm}~$  \ov{U_2}_{ \sigma'}\,U_{1 \sigma}  $~}  
\label{u2unitu1}\enq 
$$\ov{\uup_2} \, \uup_1 = \ov{\ud_2} \, \ud_1 =  \ov{\vup_2} \,\gs_5\, \uup_1 = \ov{\vd_2} \,\gs_5\, \ud_1 = 0$$ 
$$\ov{\uup_2} \, \ud_1 = - \ov{\ud_2} \, \uup_1  = - \ov{\vup_2} \,\gs_5 \,\ud_1 =  \ov{\vd_2} \, \gs_5\,\uup_1 = 2\, \cosh \kpi $$ 
---------------------------------------------------------------------------------------------
\beq 
\fbox{\rule[-0.2cm]{0cm}{0.6cm}~$  \ov{U_2}_{ \sigma'}\,\gs_5\,U_{1 \sigma}  $~}  
\label{u2g5u1} \enq 
$$\ov{\uup_2} \, \gs_5\,\uup_1 = \ov{\ud_2} \, \gs_5\,\ud_1 =  \ov{\vup_2} \,\uup_1 = \ov{\vd_2} \, \ud_1 = 0 $$
$$\ov{\uup_2} \, \gs_5\,\ud_1 =  \ov{\ud_2} \,\gs_5\, \uup_1  = - \ov{\vup_2} \,  \ud_1 =  -\ov{\vd_2} \, \uup_1 = - 2\, \sinh \kpi  $$
---------------------------------------------------------------------------------------------
\beq 
\fbox{\rule[-0.2cm]{0cm}{0.6cm}~$ \ov{U_2}_{ \sigma'}\,\gs_\mu\,U_{1 \sigma} $~} 
\label{u2gmuu1} 
\enq
$$\ov{\uup_2} \, \gs_\mu\,\uup_1 = \ov{\vup_2} \,\gs_\mu\,\gs_5\,\uup_1= 2\,\left[X+ iY\right]_\mu\,\sinh \kpi $$
$$\ov{\ud_2} \, \gs_\mu\,\ud_1  = \ov{\vd_2} \,\gs_\mu\,\gs_5\, \ud_1 =   2\,\left[X- iY\right]_\mu\,\sinh \kpi $$
$$\ov{\uup_2} \, \gs_\mu\,\ud_1 = -\ov{\ud_2} \,\gs_\mu\, \uup_1  =  \ov{\vup_2} \,\gs_\mu\,\gs_5  \ud_1 = - \ov{\vd_2} \,\gs_\mu\,\gs_5\,  \uup_1 =  2\,{\cTi}_\mu $$
---------------------------------------------------------------------------------------------
\beq
\fbox{\rule[-0.2cm]{0cm}{0.6cm}~$ \ov{U_2}_{\sigma'}\,\gs_\mu\,\gs_5\, {U_1}_{ \sigma}  $~} 
\label{u2gmug5u1}
\enq
$$\ov{\uup_2} \, \gs_\mu\,\gs_5\, \uup_1 = \ov{\vup_2} \,\gs_\mu\,\uup_1= 2\,\left[X+ iY\right]_\mu\,\cosh \kpi $$  
$$\ov{\ud_2} \, \gs_\mu\,\gs_5\,\ud_1  = \ov{\vd_2} \,\gs_\mu\, \ud_1 =  - 2\,\left[X- iY\right]_\mu\,\cosh \kpi  $$
$$\ov{\uup_2} \, \gs_\mu\,\gs_5\,\ud_1 = \ov{\ud_2} \,\gs_\mu\,\gs_5\, \uup_1  =  \ov{\vup_2} \,\gs_\mu\, \ud_1 =  \ov{\vd_2} \,\gs_\mu\, \uup_1 =  - 2\,{\cZi}_\mu $$
\vskip 0.2cm
\end{minipage}}}
\end{center}

\newpage

\begin{center} 
\fbox{\fbox{\rule[-2cm]{0cm}{4cm}
\begin{minipage}{0.75\textwidth}
\vvv
\beq
\fbox{\rule[-0.2cm]{0cm}{0.6cm}~$ \ov{U_2}_{ \sigma'}\,\sigma_{\mu \nu} \, {U_1}_{ \sigma} $~} 
\label{u2sigmunuu1}
\enq
$$\ov{\uup_2} \, \sigma_{\mu \nu}\, \uup_1 = - \ov{\vup_2} \,\sigma_{\mu \nu}\,\gs_5\,\uup_1 =  i\, \left\{\, {\cZi}_\mu\, (X+i Y)_\nu\, - {\cZi}_\nu\, (X+i Y)_\mu \,\right\} $$
$$\ov{\ud_2} \, \sigma_{\mu \nu}\, \ud_1 = - \ov{\vd_2} \,\sigma_{\mu \nu}\,\gs_5\,\ud_1  =  i\, \left\{\,  {\cZi}_\mu\, (X-i Y)_\nu\, - {\cZi}_\nu\, (X-i Y)_\mu \, \right\} $$
$$\ov{\uup_2} \, \sigma_{\mu \nu}\,\ud_1 = - \ov{\vup_2} \,\gs_\mu\, \ud_1  = i\,\sinh \kpi\,\left[ \,{\cTi}_\mu\, {\cZi}_\nu - {\cTi}_\nu\, {\cZi}_\mu\, \right]  $$
$$- \cosh \kpi\, \left[\,X_\mu\, Y_\nu - X_\nu\,Y_\mu\, \right] $$
$$\ov{\ud_2} \, \sigma_{\mu \nu}\,\uup_1 = - \ov{\vd_2} \,\gs_\mu\, \uup_1 = - i\,\sinh \kpi \,\left[ \,{\cTi}_\mu\, {\cZi}_\nu - {\cTi}_\nu\, {\cZi}_\mu\, \right] $$
$$- \cosh \kpi\, \left[\,X_\mu\, Y_\nu - X_\nu\,Y_\mu\, \right]  $$
---------------------------------------------------------------------------------------------
\beq
\fbox{\rule[-0.2cm]{0cm}{0.6cm}~$ \ov{U_2}_{ \sigma'}\,\sigma_{\mu \nu} \, \gs_5\, {U_1}_{ \sigma} $~} 
\label{u2sigmunug5u1}
\enq
$$\ov{\uup_2} \, \sigma_{\mu \nu}\,\gs_5\, \uup_1 = - \ov{\vup_2} \,\sigma_{\mu \nu}\,\uup_1 =  i\, \left\{\, {\cTi}_\mu\, (X+i Y)_\nu\, - {\cTi}_\nu\, (X+i Y)_\mu \,\right\} $$
$$\ov{\ud_2} \, \sigma_{\mu \nu}\, \gs_5\,\ud_1 = - \ov{\vd_2} \,\sigma_{\mu \nu}\,\ud_1  = - i\, \left\{\,  {\cTi}_\mu\, (X-i Y)_\nu\, - {\cTi}_\nu\, (X-i Y)_\mu \, \right\}$$
$$\ov{\uup_2} \, \sigma_{\mu \nu}\,\gs_5\,\ud_1 = - \ov{\vup_2} \,\sigma_{\mu \nu}\, \ud_1  =-  i\,\cosh \kpi\,\left[ \,{\cTi}_\mu\, {\cZi}_\nu - {\cTi}_\nu\, {\cZi}_\mu\, \right]  $$
$$+ \sinh \kpi\, \left[\,X_\mu\, Y_\nu - X_\nu\,Y_\mu\, \right]  $$
$$\ov{\ud_2} \, \sigma_{\mu \nu}\,\gs_5\,\uup_1 = - \ov{\vd_2} \,\sigma_{\mu \nu}\, \uup_1 = - i\,\cosh \kpi \,\left[ \,{\cTi}_\mu\, {\cZi}_\nu - {\cTi}_\nu\, {\cZi}_\mu\, \right] $$
$$- \sinh \kpi\, \left[\,X_\mu\, Y_\nu - X_\nu\,Y_\mu\, \right] $$
\vskip 0.2cm
\end{minipage}}}
\end{center}

\vv \nin Pour obtenir (\ref{u2sigmunuu1}), on a utilis\'e la relation $\gs(\cTi)\,\gs(\cZi)\, \gs_5 = -i \gs(X)\,\gs(Y)$.  
On notera les \'egalit\'es 
\vskip -0.2cm 
$$ {\cTi}_\mu\, {\cZi}_\nu - {\cTi}_\nu\, {\cZi}_\mu = T_\mu\,Z_\nu - T_\nu\,Z_\mu = t_{1\mu} \,z_{1\nu} - t_{1\nu} \,z_{1\mu} = - \epsilon_{\mu \nu \rho \delta}\, X^\rho Y^\delta$$ 
\beq Y_\mu\, X_\nu - Y_\nu\, X_\mu = i \left\{\, E^{(+)}_\mu\, E^{(-)}_\nu - E^{(+)}_\nu\, E^{(-)}_\mu \, \right\} = \epsilon_{\mu \nu \rho \delta} \,T^\rho\, Z^\delta \enq

\beq \fbox{\fbox{\rule[0.2cm]{0cm}{3cm}~$\begin{array}{c} \ov{U_3}_{\sigma_3} \, {U_4}_{\sigma_4} = \ov{U_3}_{\sigma_3} \,\gs_5 \, {V_4}_{\sigma_4}  = 2\, (2 \sigma_4) \,\delta_{\sigma_3, - \sigma_4} \, \cosh \kpf \\~\\

\ov{U_3}_{\sigma_3} \,\gs_5\, {U_4}_{\sigma_4} = \ov{U_3}_{\sigma_3} \, {V_4}_{\sigma_4}  = 2\, \delta_{\sigma_3, - \sigma_4} \, \sinh \kpf \\~\\

\ov{U_3}_{\sigma_3} \,\gs_\mu\, {U_4}_{\sigma_4} = \ov{U_3}_{\sigma_3} \,\gs_\mu\, \gs_5\, {V_4}_{\sigma_4}  = 2\, \delta_{\sigma_3, - \sigma_4} \,(2 \sigma_4)\, {\cTf}_\mu \\~\\
 +\, 2\, \delta_{\sigma_3, \sigma_4}\, \sinh \kpf \, ( X'- 2i \sigma_4 Y )_\mu \\~\\

\ov{U_3}_{\sigma_3} \,\gs_\mu\, \gs_5\, {U_4}_{\sigma_4} = \ov{U_3}_{\sigma_3} \,\gs_\mu\, {V_4}_{\sigma_4}  =-\, 2\, \delta_{\sigma_3, - \sigma_4} \, {\cZf}_\mu  \\~\\
+\, 2\, (2 \sigma_4)\,\delta_{\sigma_3, \sigma_4}\,\cosh \kpf \, ( X'- 2i \sigma_4 Y )_\mu
\\~\\
\end{array}  $~}}  \label{u3v4} \enq

\vv \nin Les formules pour l'\'etat final $(3,4$) sont tout \`a fait similaires aux pr\'ec\'edentes, notamment celles pr\'esent\'ees dans (\ref{u3v4}).

\vvv \vvv
\subsection{Formes bilin\'eaires $\ov{U_3}_{\sigma'}\,\Gamma\,{U_1}_{\sigma}$ et $\ov{V_3}_{\sigma'}\,\Gamma\,{U_1}_{\sigma}$ du couplage en voie $s$}

\vv \nin A partir de (\ref{uprime}) et (\ref{projo}) on d\'eduit les formules  

\vskip -0.1cm
\beq \fbox{\fbox{\rule[-1.6cm]{0cm}{8.5cm}~$\begin{array}{c}   
\ov{U'}_{\sigma'} \, U_\sigma = 2\, \left[\, \cos \di{\theta \over 2}\, \delta_{\sigma' \sigma} - 2 \sigma\, \sin \di{\theta \over 2}\, \delta_{\sigma', - \sigma} \right] \, ,~~~\ov{U'}_{\sigma'} \, \gs_5\,U_\sigma =0 \\~\\

 \ov{U'}_{\sigma'} \,\gs_\mu\, U_\sigma = 2\,T_\mu \left[\,\cos \di{\theta \over 2}\, \delta_{\sigma' \sigma} - 2 \sigma\, \sin \di{\theta \over 2}\, \delta_{\sigma', - \sigma} \right] \\~\\

\ov{U'}_{\sigma'} \,\gs_\mu\,\gs_5\,U_\sigma = 2\,\delta_{\sigma' \sigma} (2 \sigma) \left\llbracket \, Z_\mu \cos \di{\theta \over 2}\, +  \sin \di{\theta \over 2}\, (X + 2 \sigma i Y)_\mu \,\right\rrbracket \\~\\

+ 2\,\delta_{\sigma', -\sigma} \left\llbracket\,  (X + 2 \sigma i Y)_\mu \cos \di{\theta \over 2}\, -  \sin \di{\theta \over 2}\, Z_\mu \,\right\rrbracket \\~\\

\ov{U'}_{\sigma'} \,\sigma_{\mu \nu}\,U_\sigma = \delta_{\sigma' \sigma} \left\llbracket \,(2 \sigma) \cos \di{\theta \over 2}\left( X_\mu Y_\nu - X_\nu Y_\mu \right) +  i\,\sin \di{\theta \over 2}\, 
\left[ \hskip-0.1cm \di{{}\over{}}  Z_\mu (X + 2 i \sigma Y)_\nu \right. \right.   \\~\\

\left.\left. \di{{}\over{}}  - Z_\nu ( X+ 2 i \sigma Y)_\nu \right] \,\right\rrbracket 

+  \delta_{\sigma', -\sigma} \left\llbracket \, (2 \sigma)\,i\, \cos \di{\theta \over 2} \left[\hskip-0.1cm \di{{}\over{}} Z_\mu (X + 2 i \sigma Y)_\nu \right. \right.\\~\\

\left. \left. - Z_\nu ( X+ 2 i \sigma Y)_\nu \right] -  \sin \di{\theta \over 2} (X_\mu Y_\nu - X_\nu Y_\mu)\, \right\rrbracket  \\~\\

\ov{U'}_{\sigma'} \,\sigma_{\mu \nu}\,\gs_5\,U_\sigma = i\,\delta_{\sigma' \sigma} (2\sigma) \left\llbracket \,\cos \di{\theta \over 2}\left( T_\mu Z_\nu - T_\nu Z_\mu \right) +  \sin \di{\theta \over 2}\, 
\left[ \hskip-0.1cm \di{{}\over{}}  T_\mu (X + 2 i \sigma Y)_\nu \right. \right.\\~\\

\left. \left.  \di{{}\over{}}  - T_\nu ( X+ 2 i \sigma Y)_\nu \right] \,\right\rrbracket  

+i\, \delta_{\sigma', -\sigma} \left\llbracket \,  \cos \di{\theta \over 2} \left[\hskip-0.1cm \di{{}\over{}}  T_\mu (X + 2 i \sigma Y)_\nu \right. \right.\\~\\

\left. \left. \di{{}\over{}}  - T_\nu ( X+ 2 i \sigma Y)_\nu \right] -  \sin \di{\theta \over 2} (T_\mu Z_\nu - T_\nu Z_\mu) \,\right\rrbracket \\~

\end{array}  $~}} \enq

\vv\nin Utilisant (\ref{us1}) et (\ref{us3}), on a 

$$\ov{U_3}_{\sigma'}\,\Gamma\,{U_1}_{\sigma} = \cosh \di{\chi_3 \over 2}\, \cosh \di{\chi_1 \over 2} ~\ov{U'}_{\sigma'} \,\Gamma\, U_\sigma - (2 \sigma)(2 \sigma') \sinh \di{\chi_3 \over 2}\, \sinh \di{\chi_1 \over 2}~\ov{U'}_{\sigma'} \,\gs_5\,\Gamma\, \gs_5\,U_\sigma $$
$$ + (2 \sigma) \cosh \di{\chi_3 \over 2}\, \sinh \di{\chi_1 \over 2} ~\ov{U'}_{\sigma'} \,\Gamma\, \gs_5\,U_\sigma -  (2 \sigma') \sinh \di{\chi_3 \over 2}\, \cosh \di{\chi_1 \over 2} ~\ov{U'}_{\sigma'} \,\gs_5\,\Gamma\,U_\sigma $$

\vv \nin Nous distinguerons alors le cas o\`u $\Gamma$ commute avec $\gs_5$, r\'ealis\'e par les matrices $1,~\gs_5,\, \sigma_{\mu \nu},\, \sigma_{\mu \nu}\,\gs_5$, de celui  o\`u $\Gamma$ anticommute avec $\gs_5$, r\'ealis\'e par les matrices $\gs_\mu$ et $\gs_\mu\, \gs_5$. Pour simplifier l'\'ecriture, nous poserons $c_{+}\! =\, \cosh \chi^{(+)}_{31}$, $c_{-} \!=\, \cosh \chi^{(-)}_{31}$, $s_{+} \!=\, \sinh \chi^{(+)}_{31}$, $s_{-} \!=\, \sinh \chi^{(-)}_{31}$. 

\vv
\vv \nin \ding{172} \und{\bf $\Gamma$ commute avec $\gs_5$}

$$ \ov{U_3}_{\sigma'}\,\Gamma\,{U_1}_{\sigma} = \delta_{\sigma', \sigma} \left\{ c_{-} \ov{U'}_{\sigma} \,\Gamma\, U_\sigma  - (2 \sigma) s_{-}\,\ov{U'}_{\sigma} \,\Gamma\,\gs_5\, U_\sigma\right\} $$
\beq +\, \delta_{\sigma', - \sigma} \left\{ c_{+}\,\ov{U'}_{-\sigma} \,\Gamma\, U_\sigma +(2 \sigma) s_{+}\,  \ov{U'}_{-\sigma} \,\Gamma\, \gs_5\,U_\sigma \right\}\enq

\vv \nin \ding{172} \und{\bf $\Gamma$ anticommute avec $\gs_5$}

$$ \ov{U_3}_{\sigma'}\,\Gamma\,{U_1}_{\sigma} = \delta_{\sigma', \sigma} \left\{ c_{+} \ov{U'}_{\sigma} \,\Gamma\, U_\sigma  + (2 \sigma) s_{+}\,\ov{U'}_{\sigma} \,\Gamma\,\gs_5\, U_\sigma\right\} $$
\beq +\, \delta_{\sigma', - \sigma} \left\{ c_{-}\,\ov{U'}_{-\sigma} \,\Gamma\, U_\sigma -(2 \sigma) s_{-}\,  \ov{U'}_{-\sigma} \,\Gamma\, \gs_5\,U_\sigma \right\}\enq

\vv \nin On en d\'eduit les formes bilin\'eaires ci-dessous, pr\'esent\'ees  de fa\c{c}on synth\'etique, comme il sera fait dor\'enavant. 

\vskip 1cm
\begin{center} 
\fbox{\fbox{\rule[-2cm]{0cm}{4cm}
\begin{minipage}{0.87\textwidth}
\vskip 0.4cm
\beq
 \fbox{\rule[-0.2cm]{0cm}{0.6cm}~$ \ov{U_3}_{ \sigma'}\, {U_1}_{ \sigma}  $~} 
\enq
$$\ov{U_3}_{ \sigma'}\, {U_1}_{ \sigma} = 2\,\delta_{\sigma', \sigma} \, c_{-}\, \cos \di{\theta \over 2} - 2 (2 \sigma) \,\delta_{\sigma', - \sigma} \,c_{+}\, \sin \di{\theta \over 2} $$
$$=  - \ov{V_3}_{ \sigma'}\,\gs_5\,{U_1}_{ \sigma} =   - \ov{V_3}_{ \sigma'}\,{V_1}_{ \sigma} $$
------------------------------------------------------------------------------------------------------------
\beq
\fbox{\rule[-0.2cm]{0cm}{0.6cm}~$ \ov{U_3}_{\sigma'}\,\gs_5\,{U_1}_ {\sigma} $~} 
\enq
$$\ov{U_3}_{\sigma'}\,\gs_5\,{U_1}_ {\sigma} = - 2 \left\llbracket \,\delta_{\sigma', \sigma} \,(2 \sigma) \, s_{-}\, \cos \di{\theta \over 2} +  \delta_{\sigma', - \sigma} \,s_{+}\,  \sin \di{\theta \over 2} \, \right\rrbracket $$
$$=  - \ov{V_3}_{ \sigma'}\,{U_1}_{ \sigma} =   \ov{U_3}_{ \sigma'}\,{V_1}_{ \sigma} $$
------------------------------------------------------------------------------------------------------------
\beq
\fbox{\rule[-0.2cm]{0cm}{0.6cm}~$ \ov{U_3}_{ \sigma'}\,\gs_\mu\,U_{1 \sigma} $~} 
\enq
$$\ov{U_3}_{ \sigma'}\,\gs_\mu\,U_{1 \sigma} = 2\, \delta_{\sigma', \sigma} \left \llbracket \,T_\mu\,c_{+} \cos \di{\theta \over 2} + s_{+}\left[ Z_\mu\, \cos \di{\theta \over 2} +(X+ 2i \sigma Y)_\mu\, \sin \di{\theta \over 2} \,\right] \right\rrbracket $$
$$- 2\,(2 \sigma)\, \delta_{\sigma', - \sigma}\, \left\llbracket\, T_\mu\, c_{-} \sin \di{\theta \over 2} + s_{-}\left[ - Z_\mu\, \sin \di{\theta \over 2} +(X+ 2i \sigma Y)_\mu\, \cos \di{\theta \over 2} \,\right] \right\rrbracket $$
$$ = \ov{V_3}_{ \sigma'}\,\gs_\mu\,\gs_5\,U_{1 \sigma} = \ov{V_3}_{ \sigma'}\,\gs_\mu\,V_{1 \sigma} = \ov{U_3}_{ \sigma'}\,\gs_\mu\,\gs_5\,V_{1 \sigma} $$
--------------------------------------------------------------------------------------------------------------
\beq
\fbox{\rule[-0.2cm]{0cm}{0.6cm}~$ \ov{U_3}_{ \sigma'}\,\gs_\mu\,\gs_5\, U_{1 \sigma} $~} 
\enq
$$ \ov{U_3}_{ \sigma'}\,\gs_\mu\,\gs_5\,U_{1 \sigma} = 2\, \delta_{\sigma', \sigma}\, (2 \sigma) \left\llbracket \,  T_\mu\,s_{+} \cos \di{\theta \over 2} + c_{+}\left[ Z_\mu \cos \di{\theta \over 2} +(X+ 2i \sigma Y)_\mu \sin \di{\theta \over 2} \,\right]\, \right\rrbracket  $$
$$ + 2\, \delta_{\sigma', - \sigma}\, \left\llbracket \,  T_\mu\, s_{-} \sin \di{\theta \over 2} + c_{-}\left[ - Z_\mu\, \sin \di{\theta \over 2} +(X+ 2i \sigma Y)_\mu\, \cos \di{\theta \over 2} \,\right] \right\rrbracket $$
$$= \ov{V_3}_{ \sigma'}\,\gs_\mu\,U_{1 \sigma} = \ov{U_3}_{ \sigma'}\,\gs_\mu\,V_{1 \sigma} = \ov{V_3}_{ \sigma'}\,\gs_\mu\,\gs_5\,V_{1 \sigma} $$
\vskip 0.2cm
\end{minipage}}}
\end{center}

\newpage

\begin{center} 
\fbox{\fbox{\rule[-2cm]{0cm}{4cm}
\begin{minipage}{0.88\textwidth}
\vskip 0.4cm
\beq
 \fbox{\rule[-0.2cm]{0cm}{0.6cm}~$ \ov{U_3}_{ \sigma'}\,\sigma_{\mu \nu}\, U_{1 \sigma} $~} 
\enq
$$\ov{U_3}_{ \sigma'}\,\sigma_{\mu \nu}\,U_{1 \sigma} = \delta_{\sigma', \sigma} \left\llbracket\, c_{-} \left\{  (2\sigma) \cos \di{\theta \over 2} (X_\mu Y_\nu - X_\nu Y_\mu)  \right.  \right.  $$
$$ \left.   + i \sin \di{\theta \over 2} \,\left[ \!\di{{}\over{}}Z_\mu (X+ 2i \sigma Y)_\nu- Z_\nu (X+2i \sigma Y)_\mu \,\right] \di{{}\over{}} \right\} $$
$$ \left.- i s_{-} \left\{ \cos \di{\theta \over 2} (T_\mu Z_\nu - T_\nu Z_\mu)  +  \sin \di{\theta \over 2} \left[ \!\di{{}\over{}}T_\mu (X+ 2i \sigma Y)_\nu- T_\nu (X+2i \sigma Y)_\mu \,\right] \right\} \right\rrbracket  $$
$$+ \delta_{\sigma', - \sigma}\, \left\llbracket\, c_{+} \left\{ i  (2\sigma)\cos \di{\theta \over 2} \left[\! \di{{}\over{}} Z_\mu (X+ 2i \sigma Y)_\nu - Z_\nu (X+2i \sigma Y)_\mu \right] \right. \right. $$ 
$$ \left. - \sin \di{\theta \over 2} (X_\mu Y_\nu - X_\nu Y_\mu ) \right\} + (2\sigma)\, i s_{+} \left\{ \cos \di{\theta \over 2}  \left[ \!\di{{}\over{}}T_\mu (X+ 2i \sigma Y)_\nu- T_\nu (X+2i \sigma Y)_\mu \,\right] \right.    $$
$$ \left. \left. -  \sin \di{\theta \over 2} (T_\mu Z_\nu - T_\nu Z_\mu) \right\} \right\rrbracket   $$
$$= -\ov{V_3}_{ \sigma'}\,\sigma_{\mu \nu} \gs_5\,\,U_{1 \sigma} = - \ov{V_3}_{ \sigma'}\,\sigma_{\mu \nu}\,V_{1 \sigma} = \ov{U_3}_{ \sigma'}\,\sigma_{\mu \nu} \,\gs_5\,V_{1 \sigma} $$
--------------------------------------------------------------------------------------------------------------
\beq
\fbox{\rule[-0.2cm]{0cm}{0.6cm}~$ \ov{U_3}_{ \sigma'}\,\sigma_{\mu \nu}\,\gs_5\, U_{1 \sigma}  $~} 
\enq
$$ \ov{U_3}_{ \sigma'}\,\sigma_{\mu \nu}\, \gs_5\,U_{1 \sigma} = \delta_{\sigma', \sigma} \left\llbracket\, (2 \sigma)\, i c_{-} \left\{  \cos \di{\theta \over 2} (T_\mu Z_\nu - X_\nu Y_\mu)  \right.  \right.  $$
$$ \left.   +  \sin \di{\theta \over 2} \,\left[ \!\di{{}\over{}}T_\mu (X+ 2i \sigma Y)_\nu- T_\nu (X+2i \sigma Y)_\mu \,\right] \di{{}\over{}} \right\} - s_{-} \left\{ \cos \di{\theta \over 2} (X_\mu Y_\nu - X_\nu Y_\mu) \right.
$$
$$ \left. \left.+ i (2 \sigma)  \sin \di{\theta \over 2} \left[ \!\di{{}\over{}}Z_\mu (X+ 2i \sigma Y)_\nu- Z_\nu (X+2i \sigma Y)_\mu \,\right] \right\} \right\rrbracket  $$
$$ + i\,\delta_{\sigma', - \sigma}\, \left\llbracket\, c_{+} \left\{ \cos \di{\theta \over 2} \left[\! \di{{}\over{}} T_\mu (X+ 2i \sigma Y)_\nu - T_\nu (X+2i \sigma Y)_\mu \right] \right. \right. $$ 
$$ \left. - \sin \di{\theta \over 2} (T_\mu Z_\nu - T_\nu Z_\mu ) \right\} + s_{+} \left\{  \cos \di{\theta \over 2}  \left[ \!\di{{}\over{}}Z_\mu (X+ 2i \sigma Y)_\nu- Z_\nu (X+2i \sigma Y)_\mu \,\right] \right.  $$
$$ \left. \left. + i (2\sigma)\sin \di{\theta \over 2} (X_\mu Y_\nu - X_\nu Y_\mu) \right\} \right\rrbracket  $$
$$= -\ov{V_3}_{ \sigma'}\,\sigma_{\mu \nu} \,U_{1 \sigma} = - \ov{V_3}_{ \sigma'}\,\sigma_{\mu \nu}\,\gs_5\,V_{1 \sigma} = \ov{U_3}_{ \sigma'}\,\sigma_{\mu \nu} \,V_{1 \sigma} $$
\vskip 0.2cm
\end{minipage}}}
\end{center}

\vv \nin A titre d'exercice, nous proposons au lecteur : 

\vv \nin $\bullet$ de montrer que 

$$ {U_1}_\sigma \ov{U_3}_{\sigma'} = \cos \di{\theta \over 2}\, {U_1}_\sigma\, \ov{U_1}_{\sigma'}\, \left[ \cosh \kmg - (2 \sigma')\, \sinh \kmg\, \gs_5 \right] $$
\beq+ \sin \di{\theta \over 2}\, {U_1}_\sigma\, \ov{U_1}_{-\sigma'}\, \left[ (2 \sigma') \cosh \kpg - \sinh \kpg \gs_5 \right] \enq

\vv \nin et d'utiliser ce projecteur pour retrouver les expressions des formes bilin\'eaires ci-dessus ;    

\newpage
\vv \nin $\bullet$ de montrer que 

$$ {U_2}_\sigma \ov{U_4}_{\sigma'} = \cos \di{\theta \over 2}\, {U_2}_\sigma\, \ov{U_2}_{\sigma'}\, \left[ \cosh \kmd - (2 \sigma')\, \sinh \kmd\, \gs_5 \right] $$
\beq+ \sin \di{\theta \over 2}\, {U_2}_\sigma\, \ov{U_2}_{-\sigma'}\, \left[ (2 \sigma') \cosh \kpd - \sinh \kpd \gs_5 \right] \enq

\vv \nin et d'utiliser ce projecteur pour \'etablir les expressions des formes bilin\'eaires ~$\ov{U_4}_{\sigma'}\, \Gamma\, {U_2}_\sigma$.    

\subsection{Couplage en voie $s$ : cas $m_1=m_2=m$}

\nin Puisqu'alors $\chi^{(-)}_{12} = 0$, on a $ \cTi =T$, $\cZi = Z$, $\chi^{(+)}_{12} = \chi_1 = \chi_2$. Ci-dessous nous donnons les expressions, correspondant \`a ce cas, du projecteur ${U_1}_{\sigma}\, \ov{U_2}_{\sigma'}$ et des formes bilin\'eaires $\ov{\Psi_2}_{\sigma'}\, \Gamma\,{\Psi_1}_\sigma$ ($\Psi = U, V$). On notera qu'ici $\cosh \chi_1 = \di{ \sqrt{s} \over{2m}}$, $\sinh \chi_1 = \di{ \sqrt{s} \over{2m}}\, \beta$ o\`u $\beta = \sqrt{1 - \di{{4 m^2}\over s}} $.

\vvv 
\vv \centerline{\und{\bf Projecteur ${U_1}_{\sigma}\, \ov{U_2}_{\sigma'}$} }
\vskip 0.1cm
\beq \fbox{\fbox{\rule[-0.7cm]{0cm}{3cm}~$\begin{array}{c}  {U_1}_{\sigma}\, \ov{U_2}_{\sigma'}= \delta_{\sigma' , \sigma} \, \di{1\over 2} \left[ 1 + \gs(t_1) \right] \gs(X + 2 i \sigma Y) \gs(Z) \\
- \,\delta_{\sigma' , -\sigma} \di{1\over 2} \left[ 1 + \gs(t_1) \right] \left[ (2 \sigma) \gs(T) + \gs_5 \gs(Z) \right]\\~\\
= \delta_{\sigma' , \sigma} \, \di{1\over 2} \gs(X + 2 i \sigma Y) \gs(Z) \left[ 1 + \gs(t_2) \right] \\
- \,\delta_{\sigma' , -\sigma}  \di{1\over 2} \left[ (2 \sigma) \gs(T) +  \gs_5 \gs(Z) \right] \left[ 1 + \gs(t_2) \right] \\~

\end{array}$~}}  \enq


\begin{center} 
\fbox{\fbox{\rule[-2cm]{0cm}{4cm}
\begin{minipage}{0.7\textwidth}
\vskip 0.4cm
\beq
\fbox{\rule[-0.2cm]{0cm}{0.6cm}~$ \ov{U_2}_{ \sigma'}\, {U_1}_{ \sigma} $~} 
\enq
$$\ov{U_2}_{ \sigma'}\, {U_1}_{ \sigma} = - 2 \cosh \chi_1 (2 \sigma) \,\delta_{\sigma', -\sigma} 
=  - \ov{V_2}_{ \sigma'}\,\gs_5\,{U_1}_{ \sigma} =   - \ov{V_2}_{ \sigma'}\,{V_1}_{ \sigma} $$
---------------------------------------------------------------------------------------
\beq
 \fbox{\rule[-0.2cm]{0cm}{0.6cm}~$ \ov{U_2}_{ \sigma'}\, \gs_5\,{U_1}_{ \sigma} $~} 
\enq
$$\ov{U_2}_{ \sigma'}\,\gs_5\, {U_1}_{ \sigma} = - 2 \sinh \chi_1  \,\delta_{\sigma', -\sigma} 
=  - \ov{V_2}_{ \sigma'}\,{U_1}_{ \sigma} =   - \ov{V_2}_{ \sigma'}\,\gs_5\,{V_1}_{ \sigma}$$
---------------------------------------------------------------------------------------
\beq 
\fbox{\rule[-0.2cm]{0cm}{0.6cm}~$ \ov{U_2}_{ \sigma'}\, \gs_\mu\,{U_1}_{ \sigma} $~} 
\enq
$$ \ov{U_2}_{ \sigma'}\,\gs_\mu\, {U_1}_{ \sigma} = 
2\, \delta_{\sigma', \sigma} \sinh \chi_1 \, (X + 2 i \sigma Y)_\mu -  2 (2 \sigma) \,\delta_{\sigma', -\sigma} \,T_\mu $$ 
$$ =   \ov{V_2}_{ \sigma'}\,\gs_\mu\, \gs_5\,{U_1}_{ \sigma} = \ov{U_2}_{ \sigma'}\,\gs_\mu\, \gs_5\,{V_1}_{ \sigma}  = \ov{V_2}_{ \sigma'}\,\gs_\mu\,{V_1}_{ \sigma}  $$

\vskip 0.2cm
\end{minipage}}}
\end{center}

\begin{center} 
\fbox{\fbox{\rule[-2cm]{0cm}{4cm}
\begin{minipage}{0.7\textwidth}
\vskip 0.4cm
\beq
\fbox{\rule[-0.2cm]{0cm}{0.6cm}~$ \ov{U_2}_{ \sigma'}\, \gs_\mu\,\gs_5\, {U_1}_{ \sigma} $~} 
\enq
$$ \ov{U_2}_{ \sigma'}\,\gs_\mu\, \gs_5\,{U_1}_{ \sigma} = 
2\, (2 \sigma) \,\delta_{\sigma', \sigma} \cosh \chi_1 \, (X + 2 i \sigma Y)_\mu -  2  \,\delta_{\sigma', -\sigma}\, Z_\mu $$
$$ 
=   \ov{V_2}_{ \sigma'}\,\gs_\mu\, {U_1}_{ \sigma} =  \ov{U_2}_{ \sigma'}\,\gs_\mu\, {V_1}_{ \sigma} =  \ov{V_2}_{ \sigma'}\,\gs_\mu\,\gs_5\,{V_1}_{ \sigma} $$
---------------------------------------------------------------------------------------
\beq
\fbox{\rule[-0.2cm]{0cm}{0.6cm}~$ \ov{U_2}_{ \sigma'}\,\sigma_{\mu \nu} \, {U_1}_{ \sigma}  $~}  
\enq
$$ \ov{U_2}_{ \sigma'}\,\sigma_{\mu \nu} \, {U_1}_{ \sigma}  =  i\, \delta_{\sigma', \sigma} \left\llbracket\di{{}\over{}}  Z_\mu\, (X+2 i\sigma  Y)_\nu\, - Z_\nu\, (X+2 i \sigma Y)_\mu \,\right\rrbracket $$
$$ - \delta_{\sigma', - \sigma} \left\llbracket\di{{}\over{}}  i (2 \sigma) \sinh \chi_1 \left[ T_\mu Z_\nu - T_\nu Z_\mu \right]   + \cosh \chi_1 (X_\mu Y_\nu - X_\nu Y_\mu) \,\right\rrbracket $$
$$= -\ov{V_2}_{ \sigma'}\,\sigma_{\mu \nu} \,\gs_5\, {U_1}_{ \sigma} = \ov{U_2}_{ \sigma'}\,\sigma_{\mu \nu} \, \gs_5\,{V_1}_{ \sigma}  = -\ov{V_2}_{ \sigma'}\,\sigma_{\mu \nu} \, {V_1}_{ \sigma} $$
---------------------------------------------------------------------------------------
\beq
\fbox{\rule[-0.2cm]{0cm}{0.6cm}~$ \ov{U_2}_{ \sigma'}\, \sigma_{\mu \nu} \,\gs_5\, {U_1}_{ \sigma} $~} 
\enq
$$\ov{U_2}_{ \sigma'}\,\sigma_{\mu \nu} \, \gs_5\,{U_1}_{ \sigma}  =  i\, \delta_{\sigma', \sigma} (2 \sigma) \left\llbracket  \di{{}\over{}}  T_\mu\, (X+2 i\sigma  Y)_\nu\, - T_\nu\, (X+2 i \sigma Y)_\mu \,\right\rrbracket $$
$$ - \delta_{\sigma', - \sigma} \left\llbracket\di{{}\over{}}  i \cosh \chi_1 \left[ T_\mu Z_\nu - T_\nu Z_\mu \right]   +(2 \sigma) \sinh \chi_1 (X_\mu Y_\nu - X_\nu Y_\mu) \,\right\rrbracket $$
$$ = -\ov{V_2}_{ \sigma'}\,\sigma_{\mu \nu} \, {U_1}_{ \sigma} = \ov{U_2}_{ \sigma'}\,\sigma_{\mu \nu} \,{V_1}_{ \sigma}  = -\ov{V_2}_{ \sigma'}\,\sigma_{\mu \nu} \,\gs_5\, {V_1}_{ \sigma} $$
\vskip 0.2cm
\end{minipage}}}
\end{center}

\vvv\vv
\subsection{Couplage en voie $s$ : cas $m_1= m_3$ et $m_2 = m_4$ ($\kmg = \kmd =0$)}

\vskip 1cm
\begin{center} 
\fbox{\fbox{\rule[-2cm]{0cm}{4cm}
\begin{minipage}{0.7\textwidth}
\vv
\beq
\fbox{\rule[-0.2cm]{0cm}{0.6cm}~$ \ov{U_3}_{ \sigma'}\, {U_1}_{ \sigma} $~} 
 \label{e3u1} 
\enq
$$\ov{U_3}_{ \sigma'}\, {U_1}_{ \sigma} = 2\,\delta_{\sigma', \sigma} \,  \cos \di{\theta \over 2} - 2 (2 \sigma) \,\delta_{\sigma', - \sigma} \,\cosh \chi_1\, \sin \di{\theta \over 2}  $$
$$=  - \ov{V_3}_{ \sigma'}\,\gs_5\,{U_1}_{ \sigma} =   - \ov{V_3}_{ \sigma'}\,{V_1}_{ \sigma} $$
---------------------------------------------------------------------------------------
\beq
\fbox{\rule[-0.2cm]{0cm}{0.6cm}~$ \ov{U_3}_{\sigma'}\,\gs_5\,{U_1}_ {\sigma} $~} 
 \label{e3g51} 
\enq
$$\ov{U_3}_{\sigma'}\,\gs_5\,{U_1}_ {\sigma} = - 2\, \delta_{\sigma', - \sigma} \,\sinh \chi_1\,  \sin \di{\theta \over 2}  
=  - \ov{V_3}_{ \sigma'}\,{U_1}_{ \sigma} =   \ov{U_3}_{ \sigma'}\,{V_1}_{ \sigma} $$
 
\vskip 0.2cm
\end{minipage}}}
\end{center}

\newpage
~\vskip 2cm
\begin{center} 
\fbox{\fbox{\rule[-2cm]{0cm}{4cm}
\begin{minipage}{0.8\textwidth}
\vskip 0.4cm
\beq
\fbox{\rule[-0.2cm]{0cm}{0.6cm}~$ \ov{U_3}_{ \sigma'}\,\gs_\mu\,U_{1 \sigma} $~} 
\label{eg3mu1}
\enq
$$ \ov{U_3}_{ \sigma'}\,\gs_\mu\,U_{1 \sigma} = 2\, \delta_{\sigma', \sigma} \left \llbracket \,T_\mu\,\cosh \chi_1 \cos \di{\theta \over 2}  \right. $$
$$ \left. + \sinh \chi_1\left[ Z_\mu\, \cos \di{\theta \over 2} +(X+ 2i \sigma Y)_\mu\, \sin \di{\theta \over 2} \,\right] \right \rrbracket  
 - 2\,(2 \sigma)\, \delta_{\sigma', - \sigma}\, T_\mu\, \sin \di{\theta \over 2} $$
$$= \ov{V_3}_{ \sigma'}\,\gs_\mu\,\gs_5\,U_{1 \sigma} = \ov{V_3}_{ \sigma'}\,\gs_\mu\,V_{1 \sigma} = \ov{U_3}_{ \sigma'}\,\gs_\mu\,\gs_5\,V_{1 \sigma} $$
---------------------------------------------------------------------------------------------------
\beq
\fbox{\rule[-0.2cm]{0cm}{0.6cm}~$ \ov{U_3}_{ \sigma'}\,\gs_\mu\,\gs_5\, U_{1 \sigma} $~} 
\label{eg3mu51} 
\enq
$$ \ov{U_3}_{ \sigma'}\,\gs_\mu\,\gs_5\,U_{1 \sigma} = 2\, \delta_{\sigma', \sigma}\, (2 \sigma) \left\llbracket \, T_\mu\,\sinh \chi_1  \cos \di{\theta \over 2} 
+ \cosh \chi_1 \left[ Z_\mu \cos \di{\theta \over 2} \right. \right. $$
$$ \left. \left.+(X+ 2i \sigma Y)_\mu \sin \di{\theta \over 2} \,\right] \right\rrbracket  
+ 2\, \delta_{\sigma', - \sigma}\, \left\llbracket  - Z_\mu\, \sin \di{\theta \over 2} +(X+ 2i \sigma Y)_\mu\, \cos \di{\theta \over 2} \,\right \rrbracket $$
$$= \ov{V_3}_{ \sigma'}\,\gs_\mu\,U_{1 \sigma} = \ov{U_3}_{ \sigma'}\,\gs_\mu\,V_{1 \sigma} = \ov{V_3}_{ \sigma'}\,\gs_\mu\,\gs_5\,V_{1 \sigma} $$
---------------------------------------------------------------------------------------------------
\beq
\fbox{\rule[-0.2cm]{0cm}{0.6cm}~$ \ov{U_3}_{ \sigma'}\,\sigma_{\mu \nu}\, U_{1 \sigma} $~} 
\enq
$$ \ov{U_3}_{ \sigma'}\,\sigma_{\mu \nu}\,U_{1 \sigma} = \delta_{\sigma', \sigma} \left \llbracket \,(2\sigma) \cos \di{\theta \over 2} (X_\mu Y_\nu - X_\nu Y_\mu)   
  + i \sin \di{\theta \over 2} \,\left[ \!\di{{}\over{}}Z_\mu (X+ 2i \sigma Y)_\nu \right. \right. $$
$$ \left. \left. - Z_\nu (X+2i \sigma Y)_\mu \di{{ }\over{}} \right] \,\right \rrbracket 
+\, \delta_{\sigma', - \sigma}\, \left \llbracket \,\cosh \chi_1 \left\{ i  (2\sigma)\cos \di{\theta \over 2} \left[\! \di{{}\over{}} Z_\mu (X+ 2i \sigma Y)_\nu \right. \right. \right. $$
$$ \left. \left. - Z_\nu (X+2i \sigma Y)_\mu  \di{{}\over{}}\right] \right. 
\left. - \sin \di{\theta \over 2} (X_\mu Y_\nu - X_\nu Y_\mu ) \right\} $$
$$ + (2\sigma)\, i \sinh \chi_1 \left\{ \cos \di{\theta \over 2}  \left[ \!\di{{}\over{}}T_\mu (X+ 2i \sigma Y)_\nu- T_\nu (X+2i \sigma Y)_\mu \,\right] \right.  $$
$$ \left. \left. -  \sin \di{\theta \over 2} (T_\mu Z_\nu - T_\nu Z_\mu) \right\} \,\right \rrbracket   $$
$$ = -\ov{V_3}_{ \sigma'}\,\sigma_{\mu \nu} \gs_5\,\,U_{1 \sigma} = - \ov{V_3}_{ \sigma'}\,\sigma_{\mu \nu}\,V_{1 \sigma} = \ov{U_3}_{ \sigma'}\,\sigma_{\mu \nu} \,\gs_5\,V_{1 \sigma} $$
\vskip 0.2cm
\end{minipage}}}
\end{center}

\newpage 

\begin{center} 
\fbox{\fbox{\rule[-2cm]{0cm}{4cm}
\begin{minipage}{0.85\textwidth}
\vv
\beq
\fbox{\rule[-0.2cm]{0cm}{0.6cm}~$ \ov{U_3}_{ \sigma'}\,\sigma_{\mu \nu}\,\gs_5\, U_{1 \sigma} $~}  
\enq
$$ \ov{U_3}_{ \sigma'}\,\sigma_{\mu \nu}\, \gs_5\,U_{1 \sigma} = \delta_{\sigma', \sigma} \left \llbracket\, (2 \sigma)\, i \left\{  \cos \di{\theta \over 2} (T_\mu Z_\nu - X_\nu Y_\mu)  + \sin \di{\theta \over 2} \,\left[ \!\di{{}\over{}}T_\mu (X+ 2i \sigma Y)_\nu\right.  \right. \right. $$
$$ \left.  \left. - T_\nu (X+2i \sigma Y)_\mu \di{{}\over{}} \right]  \right\}  $$
$$ + i\,\delta_{\sigma', - \sigma}\, \left \llbracket \, \cosh \chi_1 \left\{ \cos \di{\theta \over 2} \left[\! \di{{}\over{}} T_\mu (X+ 2i \sigma Y)_\nu - T_\nu (X+2i \sigma Y)_\mu \right] \right. \right. $$ 
$$ \left. - \sin \di{\theta \over 2} (T_\mu Z_\nu - T_\nu Z_\mu ) \right\} + \sinh \chi_1 \left\{  \cos \di{\theta \over 2}  \left[ \!\di{{}\over{}}Z_\mu (X+ 2i \sigma Y)_\nu- Z_\nu (X+2i \sigma Y)_\mu \,\right] \right.  $$
$$ \left. \left. + i (2\sigma)\sin \di{\theta \over 2} (X_\mu Y_\nu - X_\nu Y_\mu) \right\} \,\right \rrbracket   $$
$$ = -\ov{V_3}_{ \sigma'}\,\sigma_{\mu \nu} \,U_{1 \sigma} = - \ov{V_3}_{ \sigma'}\,\sigma_{\mu \nu}\,\gs_5\,V_{1 \sigma} = \ov{U_3}_{ \sigma'}\,\sigma_{\mu \nu} \,V_{1 \sigma} $$

\vskip 0.2cm
\end{minipage}}}
\end{center}

\vvv

\section{Le couplage d'h\'elicit\'e dans la voie $t$}

\subsection{D\'efinition des t\'etrades} 

\vv \nin Consid\'erons \`a nouveau le processus de la figure (\ref{fig:proc4}) o\`u nous supposons que les quatre particules sont toutes de spin 1/2. On effectue maintenant des couplages d'h\'elicit\'e entre les particules $1$ et $3$ d'une part, et $2$ et $4$ d'autre part, ces deux groupes de particules ayant en commun le 4-vecteur de ``transfert" du genre espace $q = p_1 - p_3 = -( p_2 - p_4)$ (``quadri-transfert"), pour lequel nous poserons $q^2 = - t < 0$. Dans le couplage de la voie $t$, les t\'etrades associ\'ees aux quatre particules ont en commun le 4-vecteur $Y$ d\'efini en (\ref{VY}) et le 4-vecteur $Q = q/\sqrt{t}$, ce dernier servant \`a d\'efinir un axe des z . Pour la commodit\'e, les associations $(p_1, p_3,q)$ et $(p_2, p_4, q)$ seront respectivement appel\'ees vertex de gauche et vertex de droite.  

\vv \nin Les t\'etrades associ\'ees \`a chacun des vertex dans ce couplage d'h\'elicit\'e sont les suivantes. 

\vv \nin \leftpointright~ \und{\bf Pour le vertex de gauche} 

\vv \nin $\bullet$ La t\'etrade associ\'ee \`a $q$ est :  

$$ T_g = \di{{2 \sqrt{t}}\over \Ldg} \left( p_1 + q \,\di{{ p_1\cdot q}\over t} \right) = \di{{2 \sqrt{t}}\over \Ldg} \left( p_3 + q \,\di{{ p_3\cdot q}\over t} \right),~~Z_g = Q,~~Y,~~{\rm avec}~~\Lambda_g = \Lambda(-t, m^2_1, m^2_3), $$ 
$$ {\rm et}~~X_{g \mu} =\epsilon_{ \mu \nu \rho \sigma} T^\nu_g Y^\rho Z^\sigma_g \,;$$

\vv \nin $\bullet$ celle associ\'ee \`a $p_1$ est\footnote{Bien que les m\^emes lettres soient utilis\'ees pour les d\'efinir, les 4-vecteurs introduits ici ne doivent pas \^etre confondus avec ceux du couplage de la voie $s$ !}  

$$ t_1 = \di{{p_1}\over m_1} = \cosh \xi_1 \,T_g + \sinh \xi_1 \,Z_g\,,~~$$
$$ z_1 =  \di{{2 m_1}\over \Ldg} \left( q - p_1\, \di{{p_1 \cdot q}\over m^2_1} \right) = \sinh \xi_1\,T_g + \cosh \xi_1\, Z_g $$
$${\rm avec}~~~\cosh \xi_1 = \di{ \Ldg \over{2 m_1 \sqrt{t}}},~~\sinh \xi_1 = \di{{t + m^2_3 - m^2_1}\over{2 m_1 \sqrt{t}}}  ~~( -2\,p_1\cdot q = t+m^2_3 - m^2_1)$$ 
$$ x_1 = X_g\,,~~~y_1 = Y\, ,~~~\epsilon^{(\pm)}_1 = E^{(\pm)}_g = \mp \di{1\over \sqrt{2}} \left[ X_g \pm i Y\right]\, ; $$

\vv\nin $\bullet$ et celle associ\'ee \`a $p_3$ est : 

$$ t_3 = \di{{p_3}\over m_3} = \cosh \xi_3 \,T_g - \sinh \xi_3 \,Z_g\,,$$ 
$$ z_3 = \di{{2 m_3}\over \Ldg} \left( q - p_3\, \di{{p_3 \cdot q}\over m^2_3} \right) = - \sinh \xi_3\,T_g + \cosh \xi_3\, Z_g $$
$${\rm avec}~~~\cosh \xi_3 = \di{ \Ldg \over{2 m_3 \sqrt{t}}},~~\sinh \xi_3 = \di{{t + m^2_1 - m^2_3}\over{2 m_3 \sqrt{t}}}~~(2\, p_3 \cdot q = t+ m^2_1 - m^2_3)  $$ 
$$ x_3 = X_g\,,~~~y_3 = Y\, ,~~~\epsilon^{(\pm)}_3 = \epsilon^{(\pm)}_1 = \mp \di{1\over \sqrt{2}} \left[ X_g \pm i Y\right]\, ; $$

\vv \nin Dans le r\'ef\'erentiel $(T_g,X_g,Y, Z_g)$, les 3-vecteurs $\Vec{\,p_1\,}$, $\Vec{\,q\,}$ et $\Vec{\,p_3\,}$ sont tous les trois selon l'axe des z, dans le sens positif pour les deux premiers, dans le sens n\'egatif pour le troisi\`eme. On passe de la t\'etrade $\left\{ [\,T_g\,] : (T_g, X_g, Y,Z_g)\right\}$ aux t\'etrades $\left\{ [\,t_1\,]_t : (t_1, x_1, y_1, z_1)\right\}$ et $\left\{[\,t_3\,]_t : (t_3, x_3, y_3, z_3)\right\}$ au moyen des transformations de Lorentz pures $[ T_g \rightarrow t_1]$ et $[T_g \rightarrow t_3]$ respectivement :

$$ [\,t_1\,]_t = [ T_g \rightarrow t_1]\,[ T_g ],~~~ [\,t_3\,]_t = [ T_g \rightarrow t_3]\,[ T_g ]$$

\vv\nin lesquelles s'effectuent dans le 2-plan $(T_g, Z_g)$ orthogonal au 2-plan $(X_g,Y)$ et dont les repr\'esentations agissant dans l'espace des spineurs de Dirac sont 

\beq  S_{T_g \rightarrow t_1} = \cosh \di{\xi_1\over 2} + \sinh \di{\xi_1 \over 2} \,\gs(Z_g) \, \gs(T_g) ~~{\rm et}~~S_{T_g \rightarrow t_3} = \cosh \di{\xi_3\over 2} - \sinh \di{\xi_3 \over 2} \,\gs(Z_g) \, \gs(T_g) \label{tlg-ct}\enq

\nin On notera que 

\beq \fbox{\rule[-0.5cm]{0cm}{1.2cm}~$\gs(Z_g)\,\gs(T_g)  = \gs(z_1)\,\gs(t_1) = \gs(z_3)\,\gs(t_3) = 2\, S_{z g}\, \gs_5  $~}  \label{sping-ct} \enq

\vv \nin Notons $U_{g \sigma}$ et $V_{g \sigma} = \gs_5\,U_{g \sigma}$ les spineurs de Dirac associ\'es \`a la t\'etrade $[\,T_g\,]$, prise comme r\'ef\'erence pour le vertex de gauche. Ces spineurs sont suppos\'es normalis\'es selon $\ov{U}\,U = 2$. Les spineurs de Dirac correspondant aux t\'etrades $[\,t_1\,]_t$ et $[\,t_2\,]_t$ s'en d\'eduisent en utilisant (\ref{tlg-ct}) et (\ref{sping-ct}) : 

\vv \nin \ding{172} \und{\bf pour la particule 1} 

\beq U_{1 \sigma} =  S_{T_g \rightarrow t_1}\, U_{g \sigma} = \cosh \di{\xi_1\, \over 2} U_{g \sigma} + (2 \sigma) \sinh \di{\xi_1\over 2}\, V_{g \sigma} \label{ct-u1} \enq

\nin \ding{173} \und{\bf pour la particule 3} 

\beq U_{3 \sigma} =  S_{T_g \rightarrow t_3}\, U_{g \sigma} = \cosh \di{\xi_3\, \over 2} U_{g \sigma} - (2 \sigma) \sinh \di{\xi_3\over 2}\, V_{g \sigma} \label{ct-u3} \enq

\vv \nin Les deux t\'etrades $[\,t_1\,]_t$ et $[\,t_3\,]_t$ peuvent \^etre directement reli\'ees par la transformation de Lorentz pure $ [ T_g \rightarrow t_3]\, [ T_g \rightarrow t_1]^{-1}$ s'effectuant dans le m\^eme 2-plan $(T_g, Z)$. Exprimons sa repr\'esentation sur les spineurs de Dirac. Compte tenu de (\ref{sping-ct}), on a 

$$ S_{t_1 \rightarrow t_3} = S_{T_g \rightarrow t_3}\, S^{-1}_{T_g \rightarrow t_1}\,  = \left[ \cosh \di{\xi_3\over 2} - \sinh \di{\xi_3 \over 2} \,\gs(z_1) \, \gs(t_1) \right] \left[\cosh \di{\xi_1\over 2} - \sinh \di{\xi_1 \over 2} \,\gs(z_1) \, \gs(t_1) \right] $$
$$ = \cosh \xpg - \sinh \xpg\, \gs(z_1)\, \gs(t_1) $$

\vv \nin o\`u l'on a pos\'e $\xpg  =\di{{\xi_1 + \xi_3}\over 2}$. Cette expression peut \^etre r\'ecrite comme 

\vv
\beq \fbox{\rule[-0.2cm]{0cm}{1.3cm}~$\begin{array}{c}  ~\\  S_{t_1 \rightarrow t_3} = \gs(\cTg) \, \gs(t_1) ~~~{\rm avec} \\~\\
\cTg = \cosh \xpg\, t_1 - \sinh \xpg\, z_1 = \cosh \xmg\, T_g + \sinh \xmg\, Z_g ~~~{\rm o\grave{u}}  \\~\\
 \xmg = \di{{\xi_1 - \xi_3}\over 2}  \\~
\end{array}  $~}   \enq

\vv\nin Puisque $\gs(t_1)\, U_1 = U_1$, les deux spineurs (\ref{ct-u3}) et (\ref{ct-u1}) sont donc simplement reli\'es par  :    

\beq \fbox{\rule[-0.5cm]{0cm}{1.2cm}~$ U_{3 \sigma} = \gs(\cTg)\, U_{1 \sigma} $~}  \label{rt-13} \enq

\vv \nin \leftpointright~ \und{\bf Pour le vertex de droite} 

\vv \nin Pour le vertex de droite, le 4-transfert est $q' = p_2 - p_4 = - q$. Cependant, nous garderons le m\^eme 4-vecteur $q$ pour d\'efinir la t\'etrade de r\'ef\'erence associ\'ee \`a ce vertex. Bien que cela induise une certaine dissym\'etrie entre les deux vertex, revenant \`a inverser les r\^oles de la particule entrante 2 et de la particule sortante 4, on en tire l'avantage que les t\'etrades de r\'ef\'erence des deux vertex sont alors reli\'ees par une simple transformation de Lorentz dans le 2-plan orthogonal au 2-plan $(Q, Y)$.  Compte tenu de ce choix, les t\'etrades du vertex de droite seront d\'efinies comme suit.

\vv \nin $\bullet$ La t\'etrade associ\'ee \`a $q$ est :  

$$ T_d = \di{{2 \sqrt{t}}\over \Ldd} \left( p_2 + q\,\di{{ p_2\cdot q}\over t} \right) = \di{{2 \sqrt{t}}\over \Ldd} \left( p_4 + q \,\di{{ p_4\cdot q}\over t} \right),~~Z_d= Z_g = Q,~~Y, $$ 
$${\rm avec}~~\Lambda_d = \Lambda(-t, m^2_2, m^2_4), ~~{\rm et}~~X_{d \mu} =\epsilon_{ \mu \nu \rho \sigma} T^\nu_d Y^\rho Z^{\sigma}_d $$

\vv \nin Elle sera prise comme r\'ef\'erence pour le vertex de droite.

\vv \nin $\bullet$ La t\'etrade associ\'ee \`a $p_2$ est

$$ t_2 = \di{{p_2}\over m_2} = \cosh \xi_2 \,T_d - \sinh \xi_2 \,Z_d\,,~~$$
$$ z_2 =  \di{{2 m_2}\over \Ldd} \left( -q + p_2\, \di{{p_2 \cdot q}\over m^2_2} \right) = -\sinh \xi_2\,T_d + \cosh \xi_2\, Z_d $$
$${\rm avec}~~~\cosh \xi_2 = \di{ \Ldd \over{2 m_2 \sqrt{t}}},~~\sinh \xi_2 = \di{{t + m^2_4 - m^2_2}\over{2 m_2 \sqrt{t}}}  ~~( 2\,p_2\cdot q = t+m^2_4 - m^2_2)$$ 
$$ x_2 = X_d\,,~~~y_2 = Y\, ,~~~\epsilon^{(\pm)}_2 = E^{(\pm)}_d = \mp \di{1\over \sqrt{2}} \left[ X_d \pm i Y\right]\, ; $$

\vv\nin $\bullet$ et celle associ\'ee \`a $p_4$ est : 

$$ t_4 = \di{{p_4}\over m_4} = \cosh \xi_4 \,T_d + \sinh \xi_4 \,Z_d\,,$$ 
$$ z_4 = \di{{2 m_4}\over \Ldd} \left( -q + p_4\, \di{{p_4 \cdot q}\over m^2_4} \right) =  \sinh \xi_4\,T_d + \cosh \xi_4\, Z_d $$
$${\rm avec}~~~\cosh \xi_4 = \di{ \Ldd \over{2 m_4 \sqrt{t}}},~~\sinh \xi_4 = \di{{t + m^2_2 - m^2_4}\over{2 m_4 \sqrt{t}}}~~(-2\, p_4 \cdot q = t+ m^2_2 - m^2_4)  $$ 
$$ x_4 = X_d\,,~~~y_4 = Y\, ,~~~\epsilon^{(\pm)}_4 = \epsilon^{(\pm)}_2 = \mp \di{1\over \sqrt{2}} \left[ X_d \pm i Y\right]\, ; $$

\vv \nin Dans le r\'ef\'erentiel $(T_d,X_d,Y, Z_d)$, les 3-vecteurs $\Vec{\,p_4\,}$, $\Vec{\,q\,}$ et $\Vec{\,p_2\,}$ sont tous les trois selon l'axe des z, dans le sens positif pour les deux premiers, dans le sens n\'egatif pour le troisi\`eme. On passe de la t\'etrade $\left\{ [\,T_d\,] : (T_d, X_d, Y,Z_d)\right\}$ aux t\'etrades $\left\{ [\,t_2\,]_t : (t_2, x_2, y_2, z_2)\right\}$ et $\left\{[\,t_4\,]_t : (t_4, x_4, y_4, z_4)\right\}$ au moyen des transformations de Lorentz pures $[ T_d \rightarrow t_2]$ et $[T_d \rightarrow t_4]$ respectivement :

$$ [\,t_2\,]_t = [ T_2 \rightarrow t_2]\,[ T_d ],~~~ [\,t_4\,]_t = [ T_d \rightarrow t_4]\,[ T_d ]$$

\vv\nin lesquelles s'effectuent dans le 2-plan $(T_d, Z_d)$ orthogonal au 2-plan $(X_d,Y)$ et dont les repr\'esentations agissant dans l'espace des spineurs de Dirac sont 

\beq  S_{T_d \rightarrow t_2} = \cosh \di{\xi_2\over 2} - \sinh \di{\xi_2 \over 2} \,\gs(Z_d) \, \gs(T_d) ~~{\rm et}~~S_{T_d \rightarrow t_4} = \cosh \di{\xi_4\over 2} + \sinh \di{\xi_4 \over 2} \,\gs(Zd) \, \gs(T_d) \label{tld-ct}\enq

\nin On notera la relation sym\'etrique de (\ref{sping-ct}) :  

\beq \fbox{\rule[-0.5cm]{0cm}{1.2cm}~$\gs(Z_d)\,\gs(T_d)  = \gs(z_2)\,\gs(t_2) = \gs(z_4)\,\gs(t_4) = 2\, S_{z d}\, \gs_5  $~}  \label{spind-ct} \enq

\vv \nin Notons $U_{d \sigma}$ et $V_{d \sigma} = \gs_5\,U_{d \sigma}$ les spineurs de Dirac associ\'es \`a la t\'etrade $[\,T_d\,]$, prise comme r\'ef\'erence pour le vertex de droite, dont notera bien qu'elle est diff\'erente de la t\'etrade de r\'ef\'erence du vertex de gauche. Ces spineurs sont aussi suppos\'es normalis\'es selon $\ov{U}\,U = 2$. Les spineurs de Dirac correspondant aux t\'etrades $[\,t_2\,]_t$ et $[\,t_4\,]_t$ se d\'eduisent de ces spineurs de r\'ef\'erence en utilisant (\ref{tld-ct}) et (\ref{spind-ct}) : 

\vv \nin \ding{172} \und{\bf pour la particule 2} 

\beq U_{2 \sigma} =  S_{T_d \rightarrow t_2}\, U_{d \sigma} = \cosh \di{\xi_2\, \over 2} U_{d \sigma} - (2 \sigma) \sinh \di{\xi_2\over 2}\, V_{d \sigma} \label{ct-u2} \enq

\nin \ding{173} \und{\bf pour la particule 4} 

\beq U_{4 \sigma} =  S_{T_d \rightarrow t_4}\, U_{d \sigma} = \cosh \di{\xi_4\, \over 2} U_{d \sigma} + (2 \sigma) \sinh \di{\xi_4\over 2}\, V_{d \sigma} \label{ct-u4} \enq

\vv \nin Comme dans le cas du vertex de gauche, les deux t\'etrades $[\,t_2\,]_t$ et $[\,t_4\,]_t$ peuvent aussi \^etre directement reli\'ees par une transformation de Lorentz pure, \`a savoir $ [ T_d \rightarrow t_4]\, [ T_d \rightarrow t_2]^{-1}$, laquelle s'effectue dans 2-plan $(T_d, Z')$. Compte tenu de (\ref{spind-ct}), on a 

$$ S_{t_2 \rightarrow t_4} = S_{T_d \rightarrow t_4}\, S^{-1}_{T_d \rightarrow t_2}\,  = \left[ \cosh \di{\xi_4\over 2} + \sinh \di{\xi_4 \over 2} \,\gs(z_2) \, \gs(t_2) \right] \left[\cosh \di{\xi_2\over 2} + \sinh \di{\xi_2 \over 2} \,\gs(z_2) \, \gs(t_2) \right] $$
$$ = \cosh \xpd + \sinh \xpd\, \gs(z_2)\, \gs(t_2) $$

\vv \nin o\`u l'on a pos\'e $\xpd  =\di{{\xi_2 + \xi_4}\over 2}$. Cette expression peut \^etre r\'ecrite comme 

\vv
\beq \fbox{\rule[-0.2cm]{0cm}{1.3cm}~$\begin{array}{c}  ~\\  S_{t_2 \rightarrow t_4} = \gs(\cTd) \, \gs(t_2) ~~~{\rm avec} \\~\\
\cTd = \cosh \xpd\, t_2 + \sinh \xpd\, z_2 = \cosh \xmd\, T_d - \sinh \xmd\, Z_d ~~~{\rm o\grave{u}}  \\~\\
 \xmd = \di{{\xi_2 - \xi_4}\over 2}  \\~
\end{array}  $~}   \enq

\vv\nin Puisque $\gs(t_2)\, U_2 = U_2$, les deux spineurs (\ref{ct-u2}) et (\ref{ct-u4}) sont simplement reli\'es par  :    

\beq \fbox{\rule[-0.5cm]{0cm}{1.2cm}~$ U_{4 \sigma} = \gs(\cTd)\, U_{2 \sigma} $~}  \label{rt-24} \enq

\subsection{Projecteurs $U_1\,\ov{U}_3$ et $U_2\,\ov{U}_4$ du couplage en voie $t$}

\vv \nin D\'efinissons tout d'abord  

$$ \cZg = \cosh \xmg\, Z + \sinh \xmg\, T_g = \cosh \xpg\, z_1 - \sinh \xpg\, t_1  $$ 
$$ = \cosh \xpg\, z_3 + \sinh \xpg\, t_3 ~~~{\rm et} $$
\beq \cZd = \cosh \xmd\, Z_d - \sinh \xmd\, T_d = \cosh \xpd\, z_2 + \sinh \xpd\, t_2 \enq 
$$ = \cosh \xpd\, z_4 - \sinh \xpd\, t_4 $$ 

\nin On v\'erifie que 

\beq \gs(\cZg)\, \gs(\cTg) = \gs(z_1)\, \gs(t_1) = \gs(z_3)\,\gs(t_3),~~\gs(\cZd)\, \gs(\cTd) = \gs(z_2)\, \gs(t_2)= \gs(z_4)\,\gs(t_4)   \label{spinct} \enq

\vv \nin Rappelons que pour une base quelconque $(T,X,Y,Z)$, on a 

$$ \gs_5\, \gs(X + i (2 \sigma) Y ) = (2 \sigma) \gs(X + i (2 \sigma) Y ) \, \gs(Z)\, \gs(T) $$

\vv \nin Combinant alors (\ref{rt-13}) et 

\beq  {U_1}_\sigma \,\ov{U_1}_{\sigma'} = \delta_{\sigma', \sigma} \di{1\over 2} \left[  1 + (2 \sigma) \gs_5\, \gs(z_1) \right][1+ \gs(t_1)]  + \delta_{\sigma', - \sigma} \di{1\over 2} \gs_5\, \gs(X_g + (2 \sigma) i Y ) \left[ 1+ \gs(t_1) \right] \enq

\vv \nin il n'est pas difficile de d\'eduire les projecteurs $U_1\,\ov{U}_3$ : 

\beq \fbox{\fbox{\rule[-0.5cm]{0cm}{1.2cm}~$\begin{array}{c}   
~\\
 {U_1}_\sigma \,\ov{U_3}_{\sigma'} =  {U_1}_\sigma \,\ov{U_1}_{\sigma'} \, \gs(\cTg) =\delta_{\sigma', \sigma} \di{1\over 2} [1+ \gs(t_1)] \left[ \gs(\cTg) + (2 \sigma) \gs_5\, \gs(\cZg) \right]  \\~\\

+\delta_{\sigma', - \sigma} \, (2 \sigma) \di{1\over 2}  [ 1+ \gs(t_1)]\, \gs(X_g + (2 \sigma) i Y )\,\gs(\cZg) \\~\\

= \gs(\cTg)\, {U_3}_\sigma \,\ov{U_3}_{\sigma'} = \delta_{\sigma', \sigma} \di{1\over 2} \left[ \gs(\cTg) + (2 \sigma) \gs_5\, \gs(\cZg) \right] [1+ \gs(t_3)] \\~\\

+\delta_{\sigma', - \sigma} \, (2 \sigma) \di{1\over 2} \, \gs(X_g + (2 \sigma) i Y )\,\gs(\cZg)  [ 1+ \gs(t_3)] \\~

\end{array}  $~}}  \enq
\vskip 0.1cm
\nin Sym\'etriquement, on a 
\vskip -0.1cm
\beq \fbox{\fbox{\rule[-0.5cm]{0cm}{1.2cm}~$\begin{array}{c}   
~\\
 {U_2}_\sigma \,\ov{U_4}_{\sigma'} =  {U_2}_\sigma \,\ov{U_2}_{\sigma'} \, \gs(\cTd) =\delta_{\sigma', \sigma} \di{1\over 2} [1+ \gs(t_2)] \left[ \gs(\cTd) + (2 \sigma) \gs_5\, \gs(\cZd) \right]  \\~\\

+\delta_{\sigma', - \sigma} \, (2 \sigma) \di{1\over 2}  [ 1+ \gs(t_2)]\, \gs(X_d + (2 \sigma) i Y )\,\gs(\cZd) \\~\\

= \gs(\cTd)\, {U_4}_\sigma \,\ov{U_4}_{\sigma'} = \delta_{\sigma', \sigma} \di{1\over 2} \left[ \gs(\cTd) + (2 \sigma) \gs_5\, \gs(\cZd) \right] [1+ \gs(t_4)] \\~\\

+\delta_{\sigma', - \sigma} \, (2 \sigma) \di{1\over 2} \, \gs(X_d + (2 \sigma) i Y )\,\gs(\cZd)  [ 1+ \gs(t_4)] \\~

\end{array}  $~}}  \enq

\subsection{Formes bilin\'eaires $\ov{U_3}_{\sigma'}\,\Gamma\,{U_1}_{\sigma}$ du couplage en voie $t$}

\nin Nous laissons au lecteur le soin de v\'erifier les formules ci-dessous. 

\vvv \vvv
\nin \leftpointright ~\und{\bf Vertex de gauche} 
\vvv 

\begin{center} 
\fbox{\fbox{\rule[-2cm]{0cm}{4cm}
\begin{minipage}{0.7\textwidth}
\vskip 0.3cm
\beq
\fbox{\rule[-0.2cm]{0cm}{0.6cm}~$ \ov{U_3}_{ \sigma'}\, {U_1}_{ \sigma}  $~} 
\enq
$$ \ov{U_3}_{ \sigma'}\, {U_1}_{ \sigma} = 2\,\delta_{\sigma', \sigma} \, \cosh \xpg $$
$$\ov{U_3}_{ \sigma'}\,\gs_5\,{V_1}_{ \sigma}=  - \ov{V_3}_{ \sigma'}\,\gs_5\,{U_1}_{ \sigma} =   - \ov{V_3}_{ \sigma'}\,{V_1}_{ \sigma} $$
---------------------------------------------------------------------------------------
\beq 
\fbox{\rule[-0.2cm]{0cm}{0.6cm}~$ \ov{U_3}_{\sigma'}\,\gs_5\,{U_1}_ {\sigma} $~} 
\enq
$$\ov{U_3}_{\sigma'}\,\gs_5\,{U_1}_ {\sigma} =  2 \,(2 \sigma)\,\delta_{\sigma', \sigma} \, \sinh \xpg $$
$$ \ov{U_3}_{\sigma'}\,{V_1}_ {\sigma} =  - \ov{V_3}_{ \sigma'}\,{U_1}_{ \sigma} =  - \ov{V_3}_{ \sigma'}\,\gs_5\,{V_1}_{ \sigma} $$
\vskip 0.2cm
\end{minipage}}}
\end{center}

\newpage 
~\vskip 2cm

\begin{center} 
\fbox{\fbox{\rule[-2cm]{0cm}{4cm}
\begin{minipage}{0.8\textwidth}
\vskip 0.3cm
\beq
\fbox{\rule[-0.2cm]{0cm}{0.6cm}~$ \ov{U_3}_{ \sigma'}\,\gs_\mu\,U_{1 \sigma} $~} 
\enq
$$\ov{U_3}_{ \sigma'}\,\gs_\mu\,U_{1 \sigma} = 2\, \delta_{\sigma', \sigma} \, {\cTg}_\mu
+ 2\,(2 \sigma)\, \delta_{\sigma', - \sigma}\,\sinh \xpg (X_g + i(2 \sigma) Y)_\mu  $$
$$= \ov{V_3}_{ \sigma'}\,\gs_\mu\,\gs_5\,U_{1 \sigma} = \ov{V_3}_{ \sigma'}\,\gs_\mu\,V_{1 \sigma} = \ov{U_3}_{ \sigma'}\,\gs_\mu\,\gs_5\,V_{1 \sigma} $$
---------------------------------------------------------------------------------------------------
\beq
\fbox{\rule[-0.2cm]{0cm}{0.6cm}~$ \ov{U_3}_{ \sigma'}\,\gs_\mu\,\gs_5\, U_{1 \sigma} $~} 
\enq
$$ \ov{U_3}_{ \sigma'}\,\gs_\mu\,\gs_5\,U_{1 \sigma} = 2\, \delta_{\sigma', \sigma}\, (2 \sigma) {\cZg}_\mu  
+ 2\, \delta_{\sigma', - \sigma}\, \cosh \xpg\,(X_g+ i (2\sigma) Y)_\mu  $$
$$= \ov{V_3}_{ \sigma'}\,\gs_\mu\,U_{1 \sigma} = \ov{U_3}_{ \sigma'}\,\gs_\mu\,V_{1 \sigma} = \ov{V_3}_{ \sigma'}\,\gs_\mu\,\gs_5\,V_{1 \sigma} $$
---------------------------------------------------------------------------------------------------
\beq
\fbox{\rule[-0.2cm]{0cm}{0.6cm}~$ \ov{U_3}_{ \sigma'}\,\sigma_{\mu \nu}\, U_{1 \sigma} $~} 
\enq
$$ \ov{U_3}_{ \sigma'}\,\sigma_{\mu \nu}\,U_{1 \sigma} = \delta_{\sigma', \sigma} \left\llbracket  \di{{}\over{}}  i \sinh \xpg \left[ {\cTg}_\mu {\cZg}_\nu - {\cTg}_\nu {\cZg}_\mu \right] \right. $$
$$ \left. \di{{}\over{}} + (2 \sigma) \cosh \xpg \left[ {X_g}_\mu Y_\nu - {X_g}_\nu Y_\mu \right] \right\rrbracket  + i(2 \sigma) \delta_{\sigma', - \sigma} \left[ \di{{}\over{}} {\cZg}_\mu (X_g + i(2\sigma) Y)_\nu \right. $$
$$ \left. \di{{}\over{}} - {\cZg}_\nu (X_g + i(2\sigma)Y)_\mu  \right] $$
$$ = -\ov{V_3}_{ \sigma'}\,\sigma_{\mu \nu} \gs_5\,\,U_{1 \sigma} = - \ov{V_3}_{ \sigma'}\,\sigma_{\mu \nu}\,V_{1 \sigma} = \ov{U_3}_{ \sigma'}\,\sigma_{\mu \nu} \,\gs_5\,V_{1 \sigma} = -\ov{V_3}_{ \sigma'}\,\sigma_{\mu \nu} \,\,V_{1 \sigma} $$
---------------------------------------------------------------------------------------------------
\beq
\fbox{\rule[-0.2cm]{0cm}{0.6cm}~$ \ov{U_3}_{ \sigma'}\,\sigma_{\mu \nu}\,\gs_5\, U_{1 \sigma}  $~} 
\enq
$$ \ov{U_3}_{ \sigma'}\,\sigma_{\mu \nu}\, \gs_5\,U_{1 \sigma} = \delta_{\sigma', \sigma} \left\llbracket\di{{}\over{}} i (2 \sigma)  \cosh \xpg ({\cTg}_\mu {\cZg}_\nu - {\cTg}_\nu {\cZg}_\mu)  \right.  $$
$$ \left. \left. +\sinh \xpg (X_\mu Y_\nu - X_\nu Y_\mu) \right\} \right\rrbracket   + i \delta_{\sigma', - \sigma} \left[ \di{{}\over{}} {\cTg}_\mu (X_g + i(2 \sigma) Y)_\nu \right. $$
$$\left. \di{{}\over{}} - {\cTg}_\nu (X_g + i (2 \sigma) Y)_\mu \right]   $$
$$ = -\ov{V_3}_{ \sigma'}\,\sigma_{\mu \nu} \,U_{1 \sigma} = - \ov{V_3}_{ \sigma'}\,\sigma_{\mu \nu}\,\gs_5\,V_{1 \sigma} = \ov{U_3}_{ \sigma'}\,\sigma_{\mu \nu} \,V_{1 \sigma} = -\ov{V_3}_{ \sigma'}\,\sigma_{\mu \nu}\, \gs_5 \,V_{1 \sigma} $$
\vskip 0.2cm
\end{minipage}}}
\end{center}

\beq  {\rm Note~:}~~~~{\cTg}_\mu {\cZg}_\nu - {\cTg}_\nu {\cZg}_\mu = t_{1 \mu}\, z_{1 \nu} - t_{1 \nu}\, z_{1 \mu} =  t_{3 \mu}\, z_{3 \nu} - t_{3 \nu}\, z_{3 \mu} \enq

\newpage
\nin \leftpointright ~\und{\bf Vertex de droite} 

\vskip 1cm

\begin{center} 
\fbox{\fbox{\rule[-2cm]{0cm}{4cm}
\begin{minipage}{0.8\textwidth}
\vskip 0.3cm
\beq
\fbox{\rule[-0.2cm]{0cm}{0.6cm}~$ \ov{U_4}_{ \sigma'}\, {U_2}_{ \sigma}  $~} 
\enq
$$\ov{U_4}_{ \sigma'}\, {U_2}_{ \sigma} = 2\,\delta_{\sigma', \sigma} \, \cosh \xpd $$
$$ \ov{U_4}_{ \sigma'}\,\gs_5\,{V_2}_{ \sigma}=  - \ov{V_4}_{ \sigma'}\,\gs_5\,{U_2}_{ \sigma} =   - \ov{V_4}_{ \sigma'}\,{V_2}_{ \sigma} $$
---------------------------------------------------------------------------------------------------
\beq
\fbox{\rule[-0.2cm]{0cm}{0.6cm}~$ \ov{U_4}_{\sigma'}\,\gs_5\,{U_2}_ {\sigma} $~} 
\enq 
$$\ov{U_4}_{\sigma'}\,\gs_5\,{U_2}_ {\sigma} =  - 2 \,(2 \sigma)\,\delta_{\sigma', \sigma} \, \sinh \xpd $$
$$ \ov{U_4}_{\sigma'}\,{V_2}_ {\sigma} =  - \ov{V_4}_{ \sigma'}\,{U_2}_{ \sigma} =  - \ov{V_4}_{ \sigma'}\,\gs_5\,{V_2}_{ \sigma} $$
---------------------------------------------------------------------------------------------------
\beq
\fbox{\rule[-0.2cm]{0cm}{0.6cm}~$ \ov{U_4}_{ \sigma'}\,\gs_\mu\,U_{2 \sigma} $~} 
\enq
$$\ov{U_4}_{ \sigma'}\,\gs_\mu\,U_{2 \sigma} = 2\, \delta_{\sigma', \sigma} \, {\cTd}_\mu
- 2\,(2 \sigma)\, \delta_{\sigma', - \sigma}\,\sinh \xpd (X_d + i(2 \sigma) Y)_\mu  $$
$$ = \ov{V_4}_{ \sigma'}\,\gs_\mu\,\gs_5\,U_{2 \sigma} = \ov{V_4}_{ \sigma'}\,\gs_\mu\,V_{2 \sigma} = \ov{U_4}_{ \sigma'}\,\gs_\mu\,\gs_5\,V_{2 \sigma} $$
---------------------------------------------------------------------------------------------------
\beq
\fbox{\rule[-0.2cm]{0cm}{0.6cm}~$ \ov{U_4}_{ \sigma'}\,\gs_\mu\,\gs_5\, U_{2 \sigma} $~} 
\enq
$$\ov{U_4}_{ \sigma'}\,\gs_\mu\,\gs_5\,U_{2 \sigma} = 2\, \delta_{\sigma', \sigma}\, (2 \sigma) {\cZd}_\mu  
+ 2\, \delta_{\sigma', - \sigma}\, \cosh \xpd\,(X_d+ i (2\sigma) Y)_\mu  $$
$$ = \ov{V_4}_{ \sigma'}\,\gs_\mu\,U_{2 \sigma} = \ov{U_4}_{ \sigma'}\,\gs_\mu\,V_{2 \sigma} = \ov{V_4}_{ \sigma'}\,\gs_\mu\,\gs_5\,V_{2 \sigma} $$
---------------------------------------------------------------------------------------------------
\beq
\fbox{\rule[-0.2cm]{0cm}{0.6cm}~$ \ov{U_4}_{ \sigma'}\,\sigma_{\mu \nu}\, U_{2 \sigma} $~} 
\enq
$$ \ov{U_4}_{ \sigma'}\,\sigma_{\mu \nu}\,U_{2 \sigma} = \delta_{\sigma', \sigma} \left\llbracket  \di{{}\over{}} - i \sinh \xpd \left[ {\cTd}_\mu {\cZd}_\nu - {\cTd}_\nu {\cZd}_\mu \right] \right. $$
$$ \left. \di{{}\over{}} + (2 \sigma) \cosh \xpd \left[ {X_d}_\mu Y_\nu - {X_d}_\nu Y_\mu \right] \right\rrbracket  + i(2 \sigma) \delta_{\sigma', - \sigma} \left[ \di{{}\over{}} {\cZd}_\mu (X_d + i(2\sigma) Y)_\nu \right. $$
$$\left. \di{{}\over{}} - {\cZd}_\nu (X_d + i(2\sigma)Y)_\mu  \right] $$
$$= -\ov{V_4}_{ \sigma'}\,\sigma_{\mu \nu} \gs_5\,\,U_{2 \sigma} = - \ov{V_4}_{ \sigma'}\,\sigma_{\mu \nu}\,V_{2 \sigma} = \ov{U_4}_{ \sigma'}\,\sigma_{\mu \nu} \,\gs_5\,V_{2 \sigma} = -\ov{V_4}_{ \sigma'}\,\sigma_{\mu \nu} \,\,V_{2 \sigma} $$
\vskip 0.2cm
\end{minipage}}}
\end{center}

\vvv
\beq  {\rm Note~:}~~~~{\cTd}_\mu {\cZd}_\nu - {\cTd}_\nu {\cZd}_\mu = t_{2 \mu}\, z_{2 \nu} - t_{2 \nu}\, z_{2 \mu} =  t_{4 \mu}\, z_{4 \nu} - t_{4 \nu}\, z_{4 \mu} \enq
\newpage

\begin{center} 
\fbox{\fbox{\rule[-2cm]{0cm}{4cm}
\begin{minipage}{0.8\textwidth}
\vskip 0.3cm
\beq
\fbox{\rule[-0.2cm]{0cm}{0.6cm}~$ \ov{U_4}_{ \sigma'}\,\sigma_{\mu \nu}\,\gs_5\, U_{2 \sigma}  $~} 
\enq
$$\ov{U_4}_{ \sigma'}\,\sigma_{\mu \nu}\, \gs_5\,U_{2 \sigma} = \delta_{\sigma', \sigma} \left\llbracket\di{{}\over{}}  i (2 \sigma)  \cosh \xpd ({\cTd}_\mu {\cZd}_\nu - {\cTd}_\nu {\cZd}_\mu)  \right.  $$
$$\left. \left. -\sinh \xpd (X_{d \mu} Y_\nu - X_{d \nu} Y_\mu) \right\} \right\rrbracket   + i \delta_{\sigma', - \sigma} \left[ \di{{}\over{}} {\cTd}_\mu (X_d + i(2 \sigma) Y)_\nu \right. $$
$$ \left. \di{{}\over{}} - {\cTd}_\nu (X_d + i (2 \sigma) Y)_\mu \right]    $$
$$ = -\ov{V_4}_{ \sigma'}\,\sigma_{\mu \nu} \,U_{2 \sigma} = - \ov{V_4}_{ \sigma'}\,\sigma_{\mu \nu}\,\gs_5\,V_{2 \sigma} = \ov{U_4}_{ \sigma'}\,\sigma_{\mu \nu} \,V_{2 \sigma} = -\ov{V_4}_{ \sigma'}\,\sigma_{\mu \nu}\, \gs_5 \,V_{2 \sigma} $$
\vskip 0.2cm
\end{minipage}}}
\end{center}

\vv
\subsection{Couplage en voie $t$ : cas $m_1=m_3=m$ (ou $m_2 = m_4 = m$)}

\vv \nin On a alors $\xpg = \xi_1 = \xi_3$ avec  

\beq \fbox{\rule[-0.5cm]{0cm}{1.2cm}~$\begin{array}{c}  
\cosh \xi_1 = \di{{\sqrt{t+ 4 m^2}}\over {2m}},~~~\sinh \xi_1 = \di{ \sqrt{t}\over{2m}} 
\end{array}  $~} \enq

\vv \nin et $\cTg = T_d$, $\cZg = Z_g$. On a maintenant ${U_3}_\sigma = S^{-1}_{T_g \rightarrow t_1} U_{g \sigma} = S^{-2}_{T_g \rightarrow t_1} {U_1}_\sigma$. Les formes bilin\'eaires du pr\'ec\'edent paragraphe prennent alors les expressions ci-dessous. 

\vvv \vvv
\nin \leftpointright~\und{Vertex de gauche} 
\vvv \vvv

\begin{center} 
\fbox{\fbox{\rule[-2cm]{0cm}{4cm}
\begin{minipage}{0.8\textwidth}
\vskip 0.3cm
\beq
\fbox{\rule[-0.2cm]{0cm}{0.6cm}~$ \ov{U_3}_{ \sigma'}\, {U_1}_{ \sigma}  $~} 
\enq
$$\ov{U_3}_{ \sigma'}\, {U_1}_{ \sigma} = 2\,\delta_{\sigma', \sigma} \, \cosh \xi_1 $$
$$\ov{U_3}_{ \sigma'}\,\gs_5\,{V_1}_{ \sigma}=  - \ov{V_3}_{ \sigma'}\,\gs_5\,{U_1}_{ \sigma} =   - \ov{V_3}_{ \sigma'}\,{V_1}_{ \sigma} $$
---------------------------------------------------------------------------------------------------
\beq
\fbox{\rule[-0.2cm]{0cm}{0.6cm}~$ \ov{U_3}_{\sigma'}\,\gs_5\,{U_1}_ {\sigma} $~}  
\enq
$$ \ov{U_3}_{\sigma'}\,\gs_5\,{U_1}_ {\sigma} =  2 \,\delta_{\sigma', \sigma} \,(2 \sigma)\, \sinh \xi_1 $$
$$ \ov{U_3}_{\sigma'}\,{V_1}_ {\sigma} =  - \ov{V_3}_{ \sigma'}\,{U_1}_{ \sigma} =  - \ov{V_3}_{ \sigma'}\,\gs_5\,{V_1}_{ \sigma} $$
---------------------------------------------------------------------------------------------------
\beq
\fbox{\rule[-0.2cm]{0cm}{0.6cm}~$ \ov{U_3}_{ \sigma'}\,\gs_\mu\,U_{1 \sigma} $~} 
\enq
$$\ov{U_3}_{ \sigma'}\,\gs_\mu\,U_{1 \sigma} = 2\, \delta_{\sigma', \sigma} \, T_{g \mu}
+ 2\,(2 \sigma)\, \delta_{\sigma', - \sigma}\,\sinh \xi_1 (X_g + i(2 \sigma) Y)_\mu  $$
$$= \ov{V_3}_{ \sigma'}\,\gs_\mu\,\gs_5\,U_{1 \sigma} = \ov{V_3}_{ \sigma'}\,\gs_\mu\,V_{1 \sigma} = \ov{U_3}_{ \sigma'}\,\gs_\mu\,\gs_5\,V_{1 \sigma} $$
\vskip 0.2cm
\end{minipage}}}
\end{center}

\newpage
~\vvv
\begin{center} 
\fbox{\fbox{\rule[-2cm]{0cm}{4cm}
\begin{minipage}{0.8\textwidth}
\vskip 0.3cm
\beq 
\fbox{\rule[-0.2cm]{0cm}{0.6cm}~$ \ov{U_3}_{ \sigma'}\,\gs_\mu\,\gs_5\, U_{1 \sigma} $~} 
\enq
$$\ov{U_3}_{ \sigma'}\,\gs_\mu\,\gs_5\,U_{1 \sigma} = 2\, \delta_{\sigma', \sigma}\, (2 \sigma) \,Z_{g \mu}  
+ 2\, \delta_{\sigma', - \sigma}\, \cosh \xi_1\,(X_g+ i (2\sigma) Y)_\mu  $$
$$= \ov{V_3}_{ \sigma'}\,\gs_\mu\,U_{1 \sigma} = \ov{U_3}_{ \sigma'}\,\gs_\mu\,V_{1 \sigma} = \ov{V_3}_{ \sigma'}\,\gs_\mu\,\gs_5\,V_{1 \sigma} $$
---------------------------------------------------------------------------------------------------
\beq
\fbox{\rule[-0.2cm]{0cm}{0.6cm}~$ \ov{U_3}_{ \sigma'}\,\sigma_{\mu \nu}\, U_{1 \sigma} $~}
\enq
$$\ov{U_3}_{ \sigma'}\,\sigma_{\mu \nu}\,U_{1 \sigma} = \delta_{\sigma', \sigma} \left\llbracket  \di{{}\over{}}  i \sinh \xi_1 \left[ T_{g \mu} Z_{g \nu} - T_{g \nu} Z_{g \mu} \right] \right. $$
$$\left. \di{{}\over{}} + (2 \sigma) \cosh \xi_1 \left[ {X_g}_\mu Y_\nu - {X_g}_\nu Y_\mu \right] \right\rrbracket  + i(2 \sigma) \delta_{\sigma', - \sigma} \left[ \di{{}\over{}} Z_{g \mu} (X_g + i(2\sigma) Y)_\nu \right. $$
$$\left. \di{{}\over{}} - Z_{g \nu} (X_g + i(2\sigma)Y)_\mu  \right] $$
$$= -\ov{V_3}_{ \sigma'}\,\sigma_{\mu \nu} \gs_5\,\,U_{1 \sigma} = - \ov{V_3}_{ \sigma'}\,\sigma_{\mu \nu}\,V_{1 \sigma} = \ov{U_3}_{ \sigma'}\,\sigma_{\mu \nu} \,\gs_5\,V_{1 \sigma} = -\ov{V_3}_{ \sigma'}\,\sigma_{\mu \nu} \,\,V_{1 \sigma} $$
---------------------------------------------------------------------------------------------------
\beq
\fbox{\rule[-0.2cm]{0cm}{0.6cm}~$ \ov{U_3}_{ \sigma'}\,\sigma_{\mu \nu}\,\gs_5\, U_{1 \sigma}  $~} 
\enq
$$\ov{U_3}_{ \sigma'}\,\sigma_{\mu \nu}\, \gs_5\,U_{1 \sigma} = \delta_{\sigma', \sigma} \left\llbracket\di{{}\over{}} i (2 \sigma)  \cosh \xi_1 (T_{g \mu} Z_{g\nu} - T_{g \nu} Z_{g \mu})  \right.  $$
$$\left. \left. +\sinh \xi_1 (X_{g \mu} Y_\nu - X_{g \nu} Y_\mu) \right\} \right\rrbracket   + i \delta_{\sigma', - \sigma} \left[ \di{{}\over{}} T_{g \mu} (X_g + i(2 \sigma) Y)_\nu \right. $$
$$\left. \di{{}\over{}} - T_{g \nu} (X_g + i (2 \sigma) Y)_\mu \right]   $$
$$= -\ov{V_3}_{ \sigma'}\,\sigma_{\mu \nu} \,U_{1 \sigma} = - \ov{V_3}_{ \sigma'}\,\sigma_{\mu \nu}\,\gs_5\,V_{1 \sigma} = \ov{U_3}_{ \sigma'}\,\sigma_{\mu \nu} \,V_{1 \sigma} = -\ov{V_3}_{ \sigma'}\,\sigma_{\mu \nu}\, \gs_5 \,V_{1 \sigma} $$
\vskip 0.2cm
\end{minipage}}}
\end{center}

\vvv
\nin \leftpointright~\und{Vertex de droite} 
\vvv
\vvv

\begin{center} 
\fbox{\fbox{\rule[-2cm]{0cm}{4cm}
\begin{minipage}{0.8\textwidth}
\vskip 0.3cm
\beq
\fbox{\rule[-0.2cm]{0cm}{0.6cm}~$ \ov{U_4}_{ \sigma'}\, {U_2}_{ \sigma}  $~} 
\enq
$$ \ov{U_4}_{ \sigma'}\, {U_2}_{ \sigma} = 2\,\delta_{\sigma', \sigma} \, \cosh \xi_1 $$
$$ \ov{U_4}_{ \sigma'}\,\gs_5\,{V_2}_{ \sigma}=  - \ov{V_4}_{ \sigma'}\,\gs_5\,{U_2}_{ \sigma} =   - \ov{V_4}_{ \sigma'}\,{V_2}_{ \sigma} $$
---------------------------------------------------------------------------------------------------
\beq
\fbox{\rule[-0.2cm]{0cm}{0.6cm}~$ \ov{U_4}_{\sigma'}\,\gs_5\,{U_2}_ {\sigma} $~}   
\enq
$$ \ov{U_4}_{\sigma'}\,\gs_5\,{U_2}_ {\sigma} = - 2 \,\delta_{\sigma', \sigma} \,(2 \sigma)\, \sinh \xi_1 $$
$$ \ov{U_4}_{\sigma'}\,{V_2}_ {\sigma} =  - \ov{V_4}_{ \sigma'}\,{U_2}_{ \sigma} =  - \ov{V_4}_{ \sigma'}\,\gs_5\,{V_2}_{ \sigma} $$
\vskip 0.2cm
\end{minipage}}}
\end{center}

\newpage
~\vskip 2.2cm 
\begin{center} 
\fbox{\fbox{\rule[-2cm]{0cm}{4cm}
\begin{minipage}{0.8\textwidth}
\vskip 0.3cm
\beq
\fbox{\rule[-0.2cm]{0cm}{0.6cm}~$ \ov{U_4}_{ \sigma'}\,\gs_\mu\,U_{2 \sigma} $~} 
\enq
$$ \ov{U_4}_{ \sigma'}\,\gs_\mu\,U_{2 \sigma} = 2\, \delta_{\sigma', \sigma} \, T_{d \mu}
- 2\,(2 \sigma)\, \delta_{\sigma', - \sigma}\,\sinh \xi_1 (X_d + i(2 \sigma) Y)_\mu  $$
$$= \ov{V_4}_{ \sigma'}\,\gs_\mu\,\gs_5\,U_{2 \sigma} = \ov{V_4}_{ \sigma'}\,\gs_\mu\,V_{2 \sigma} = \ov{U_4}_{ \sigma'}\,\gs_\mu\,\gs_5\,V_{2 \sigma} $$
---------------------------------------------------------------------------------------------------
\beq 
\fbox{\rule[-0.2cm]{0cm}{0.6cm}~$ \ov{U_4}_{ \sigma'}\,\gs_\mu\,\gs_5\, U_{2 \sigma} $~} 
\enq
$$ \ov{U_4}_{ \sigma'}\,\gs_\mu\,\gs_5\,U_{2 \sigma} = 2\, \delta_{\sigma', \sigma}\, (2 \sigma) \,Z_{d \mu}  
+ 2\, \delta_{\sigma', - \sigma}\, \cosh \xi_1\,(X_d+ i (2\sigma) Y)_\mu  $$
$$= \ov{V_3}_{ \sigma'}\,\gs_\mu\,U_{1 \sigma} = \ov{U_3}_{ \sigma'}\,\gs_\mu\,V_{1 \sigma} = \ov{V_3}_{ \sigma'}\,\gs_\mu\,\gs_5\,V_{1 \sigma}  $$
---------------------------------------------------------------------------------------------------
\beq 
\fbox{\rule[-0.2cm]{0cm}{0.6cm}~$ \ov{U_4}_{ \sigma'}\,\sigma_{\mu \nu}\, U_{2 \sigma} $~} 
\enq 
$$ \ov{U_4}_{ \sigma'}\,\sigma_{\mu \nu}\,U_{2 \sigma} = \delta_{\sigma', \sigma} \left\llbracket  \di{{}\over{}}  -i \sinh \xi_1 \left[ T_{d \mu} Z_{d \nu} - T_{d \nu} Z_{d \mu} \right] \right. $$
$$ \left. \di{{}\over{}} + (2 \sigma) \cosh \xi_1 \left[ {X_d}_\mu Y_\nu - {X_d}_\nu Y_\mu \right] \right\rrbracket  + i(2 \sigma) \delta_{\sigma', - \sigma} \left[ \di{{}\over{}} Z_{d \mu} (X_d + i(2\sigma) Y)_\nu \right. $$
$$ \left. \di{{}\over{}} - Z_{d \nu} (X_d + i(2\sigma)Y)_\mu  \right] $$
$$ = -\ov{V_3}_{ \sigma'}\,\sigma_{\mu \nu} \gs_5\,\,U_{1 \sigma} = - \ov{V_3}_{ \sigma'}\,\sigma_{\mu \nu}\,V_{1 \sigma} = \ov{U_3}_{ \sigma'}\,\sigma_{\mu \nu} \,\gs_5\,V_{1 \sigma} = -\ov{V_3}_{ \sigma'}\,\sigma_{\mu \nu} \,\,V_{1 \sigma} $$
---------------------------------------------------------------------------------------------------
\beq
\fbox{\rule[-0.2cm]{0cm}{0.6cm}~$ \ov{U_4}_{ \sigma'}\,\sigma_{\mu \nu}\,\gs_5\, U_{2 \sigma}  $~} 
\enq
$$ \ov{U_4}_{ \sigma'}\,\sigma_{\mu \nu}\, \gs_5\,U_{2 \sigma} = \delta_{\sigma', \sigma} \left\llbracket\di{{}\over{}} i (2 \sigma)  \cosh \xi_1 (T_{d \mu} Z_{d\nu} - T_{d \nu} Z_{d \mu})  \right.  $$
$$ \left. \left. -\sinh \xi_1 (X_{d \mu} Y_\nu - X_{d \nu} Y_\mu) \right\} \right\rrbracket   + i \delta_{\sigma', - \sigma} \left[ \di{{}\over{}} T_{d \mu} (X_d + i(2 \sigma) Y)_\nu \right. $$
$$\left. \di{{}\over{}} - T_{d \nu} (X_d + i (2 \sigma) Y)_\mu \right]   $$
$$= -\ov{V_4}_{ \sigma'}\,\sigma_{\mu \nu} \,U_{2 \sigma} = - \ov{V_4}_{ \sigma'}\,\sigma_{\mu \nu}\,\gs_5\,V_{2 \sigma} = \ov{U_4}_{ \sigma'}\,\sigma_{\mu \nu} \,V_{2 \sigma} = -\ov{V_4}_{ \sigma'}\,\sigma_{\mu \nu}\, \gs_5 \,V_{2 \sigma} $$
\vskip 0.2cm
\end{minipage}}}
\end{center}

\newpage
\subsection{Relations entre vertex de gauche et vertex de droite}

\vv \nin Ayant les deux vecteurs $Y$ et $Q$ en commun, les t\'etrades de r\'ef\'erence $(T_g, X_g,Y, Q)$ et $(T_d, X_d, Y,Q)$ des vertex de gauche et de droite, respectivement, sont reli\'ees par une tranformation de Lorentz pure agissant dans le 2-plan orthogonal au 2-plan $(Y, Q)$. Soit $\Theta$ son param\`etre, tel que $ T_g \cdot T_d = \cosh \Theta$. Exprimons cette grandeur\footnote{Voir : C. Carimalo et al. ``Nuclei as Generators of Quasireal Photons", Phys.Rev. D10 (1974) 1561 ; C. Carimalo, ``Les noyaux comme g\'en\'erateurs de photons quasi-r\'eels", Th\`ese d'Etat, UPMC, Paris, 1977.}. Compte tenu de $T_d \cdot q = 0$, on a 

$$ T_g \cdot T_d = \di{4 \over{\Ld_{13}\, \Ld_{24}}} \,p_1 \cdot  \left[ t\, p_2 + q\, (p_2 \cdot q) \right] $$
$$= \di{1 \over{\Ld_{13}\, \Ld_{24}}} \left[2  t (s - m^2_1 - m^2_2) -( t+ m^2_3 - m^2_1)(t + m^2_4 - m^2_2) \right]~~~{\rm soit}  $$

\beq \fbox{\rule[-0.65cm]{0cm}{1.5cm}~$\begin{array}{c} \cosh \Theta = \di{1 \over{\Ld_{13}\, \Ld_{24}}} \left[ t( 2 s - t - \Sigma) - (m^2_3 - m^2_1)( m^2_4 - m^2_2) \right] \\  
{\rm o\grave{u}}~~\Sigma = m^2_1 + m^2_2 + m^2_3 + m^2_4  \end{array}
 $~}  \label{tetagd} \enq

\vv \nin On montre que

\beq \fbox{\rule[-1.2cm]{0cm}{3.5cm}~$\begin{array}{c}  \sinh^2 \Theta = \di{{ 4 s\,t\,(t - t_{\rm min} )( t_{\rm max} - t)}\over{\Lambda_{13}\, \Lambda_{24}} }~~~{\rm o\grave{u}}  \\
t_{\rm max} = \di{1\over 2} \left[ s - \Sigma + \di{{(m_2^2 - m^2_1)(m^2_4 - m^2_3)}\over s} + \di{{\Ld_{12}\,\Ld_{34}}\over s}\right] \\
 t_{\rm min} = \di{1\over 2} \left[ s - \Sigma + \di{{(m_2^2 - m^2_1)(m^2_4 - m^2_3)}\over s} - \di{{\Ld_{12}\,\Ld_{34}}\over s}\right] \\~

 \end{array}
 $~}  \enq

\vv \nin les grandeurs $t_{\rm max}$ et $t_{\min}$ sont, respectivement, les valeurs maximum et minimum du transfert $t$, impos\'ees par la cin\'ematique de la r\'eaction $1+2 \rightarrow 3+4$. Utilisant les variables relatives au r\'ef\'erentiel du centre de masse, on a   

$$ t = - m^2_1 - m^2_3 + 2 E_1\, E_3 - 2\, p_i\, p_f \cos \theta ~~{\rm avec}~~p_i = \di{{\Ld_{12}}\over{ 2 \sqrt{s}}},~~p_f = \di{{\Ld_{34}}\over{ 2 \sqrt{s}}},$$ 

\nin  et $t_{\rm max}$ et $t_{\rm min}$ correspondent en fait \`a $\cos \theta = -1$ et $\cos \theta = +1$ respectivement, ce qui donne 

$$ (t - t_{\rm min} )( t_{\rm max} - t) = 4 p^2_i p^2_f \sin^2 \theta $$

\nin d'o\`u cette autre expression : 

\beq \fbox{\rule[-0.43cm]{0cm}{1.cm}~$\sinh \Theta = \di{{ 4 \sqrt{s\,t}\, p_i p_f\, \sin \theta}\over{\Ld_{13}\, \Ld_{24}} } 
 $~}  \enq

\nin Calculons maintenant : 

$$ T_d \cdot X_g = \epsilon_{\mu \nu \rho \pi} T^\mu_d T^\nu_g Y^\rho Q^\pi = \di{{4 t}\over{\Ld_{13} \Ld_{24}}} p^\mu_2 p^\nu_1 \epsilon_{\mu \nu \rho \pi} Y^\rho Q^\pi $$

\nin Or, dans la base $(T, X,Y,Z)$ attach\'ee au r\'ef\'erentiel du centre de masse de la r\'eaction, on a $p_1 = E_1 T + p_i Z$, $p_2 = E_2 - p_i Z$ (avec $E_1 + E_2 = \sqrt{s}$) et donc 

$$p^\mu_2\, p^\nu_1  \,\epsilon_{\mu \nu \rho \pi} Y^\rho = p_i \sqrt{s}\, \epsilon_{\mu \nu \rho \pi} T^\mu Z^\nu Y^\rho = p_i \sqrt{s}\, X_\pi $$ 

\nin et comme $\sqrt{t}\, X\cdot Q =  X \cdot (p_1 - p_3) = - X \cdot p_3 = p_f \sin \theta$, on en d\'eduit 

$$ T_d \cdot X_g = + \sinh \Theta $$ 

\nin En conclusion, 

\beq T_d = \cosh \Theta\, T_g - \sinh \Theta\, X_g  \label{td-tgxg}\enq

\nin Enfin, puisque 

$$ X_{d \mu} = \epsilon_{\mu \nu \rho \pi}\, T^\nu_d Y^\rho Q^\pi = \cosh \Theta \,\epsilon_{\mu \nu \rho \pi}\, T^\nu_g Y^\rho Q^\pi - \sinh \Theta \,\epsilon_{\mu \nu \rho \pi}\, X^\nu_d Y^\rho Q^\pi $$
$$ {\rm et~puisque} ~~~X_{g \mu} = \epsilon_{\mu \nu \rho \pi}\, T^\nu_g Y^\rho Q^\pi,~~~T_{g \mu} = \epsilon_{\mu \nu \rho \pi}\, X^\nu_d Y^\rho Q^\pi  $$

\nin on a aussi 

\beq  X_d = \cosh \Theta\, X_g - \sinh \Theta\, T_g  \label{xd-tgxg} \enq

\vv \nin Les relations (\ref{td-tgxg}) et (\ref{xd-tgxg}) d\'efinissent la transformation de Lorentz entre les deux vertex, laquelle est, pour ce couplage de la voie $t$, l'analogue de la rotation $R_Y(\theta)$ pour le couplage dans la voie $s$. Sa repr\'esentation spinorielle est 

\beq S_{T_d \rightarrow T_g} = \cosh \di{\Theta \over 2} - \sinh \di{\Theta \over 2}\, \gs(X_g) \gs(T_g) 
\label{stgtd} \enq 

\nin Compte tenu des relations g\'en\'erales $\gs(X)\, \gs(T) = 2 S_x \gs_5$ et $S_x U_\sigma = \di{1\over 2} U_{-\sigma} $, le spineur de r\'ef\'erence $U_d$ du vertex de droite se d\'eduit ainsi de celui $U_g$ du vertex de gauche par la relation 

\beq  U_{d \sigma} = S_{T_d \rightarrow T_g}\,U_{g \sigma} = \cosh \di{\Theta \over 2} \,U_{g \sigma} - \sinh \di{\Theta \over 2}\, V_{g - \sigma} \label{udug-1} \enq 

\vv \nin On remarquera que puisque $\gs(Y_g) \gs(Z_g) = \gs(Y_d)\, \gs(Z_d) = \gs(Y)\, \gs(Q)$, c'est la composante suivant l'axe ``x" du spin associ\'e \`a $T_d$ ou \`a $T_g$ qui est conserv\'ee dans le passage d'un vertex \`a l'autre : 

\beq \gs(X_d)\, \gs(T_d) = \gs(X_g)\, \gs(T_g),~~~{\rm donc}~~~S_{d x} = S_{gx} \enq

\vv \nin Notons que l'op\'erateur (\ref{stgtd}) peut \^etre mis sous la forme 

$$ S_{T_d \rightarrow T_g} =  \gs(\cT_{gd}) \gs(T_g) ~~~{\rm o\grave{u}}~~~\cT_{gd} = \cosh \di{\Theta \over 2}\,T_g - \sinh \di{\Theta \over 2}\,X_g $$

\nin de sorte que la relation entre $U_d$ et $U_g$ peut \^etre exprim\'ee comme 

\beq U_{d \sigma} =  \gs(\cT_{gd}) \gs(T_g) \,U_{g \sigma} \equiv \gs(\cT_{gd}) \,\,U_{g \sigma}  \label{udug-2} \enq

\vv \nin A partir de (\ref{udug-1}), ou de (\ref{udug-2}), et de (\ref{projo}), on d\'eduit les formes bilin\'eaires $\ov{U_d}_{\sigma'}\, \Gamma\, {U_g}_\sigma$ : 

~\vskip 1cm
\beq \fbox{\fbox{\rule[-0.7cm]{0cm}{4cm}~$\begin{array}{c}  ~\\
 \ov{U_d}_{ \sigma'}\, {U_g}_{ \sigma} = 2\,\delta_{\sigma', \sigma} \, \cosh \di{\Theta \over 2},~~~ 
\ov{U_d}_{ \sigma'}\,\gs_5\,{U_g}_{ \sigma}=  2\,\delta_{\sigma', -\sigma} \, \sinh \di{\Theta \over 2}  \\~\\
\ov{U_d}_{ \sigma'}\,\gs_\mu\,{U_g}_{ \sigma} = 2\,\delta_{\sigma', \sigma} \left\llbracket \,\cosh \di{\Theta \over 2} \,T_{g \mu} - \sinh \di{\Theta \over 2}  \left[ X_g + i (2\sigma) Y\right]_\mu \,\right\rrbracket \\~\\
- (2 \sigma) 2 \delta_{\sigma', - \sigma}\, Z_{g \mu} \\~\\
\ov{U_d}_{ \sigma'}\,\gs_\mu\,\gs_5\,{U_g}_{ \sigma} = 2\,\delta_{\sigma', \sigma} (2\sigma) \cosh \di{\Theta \over 2} \,Z_{g \mu} + 2 \delta_{\sigma', - \sigma} \left\llbracket \, \cosh \di{\Theta \over 2}  \left[X_g + i (2\sigma) Y \right]_\mu \right.  \\
\left.-  \sinh \di{\Theta \over 2}  T_{g \mu} \,\right\rrbracket \\~\\

 \ov{U_d}_{ \sigma'}\,\sigma_{\mu \nu}\, {U_g}_{ \sigma} =  \delta_{\sigma', \sigma}  \left\llbracket \di{{}\over{}} (2\sigma) \cosh \di{\Theta \over 2}\left[ X_{g \mu} Y_\nu - X_{g \nu} Y_\mu \right]  \right. \\ \left.\di{{}\over{}}+ i \sinh \di{\Theta \over 2} \left\{ \hskip -0.1cm \di{{}\over{}} T_{g \mu} \left[ X_g + i(2\sigma) Y \right]_\nu -T_{g \nu} \left[ X_g + i(2\sigma) Y \right]_\mu \right\} \, \right\rrbracket \\
+ i (2 \sigma) \delta_{\sigma', -\sigma} \left\llbracket \hskip -0.1cm \di{{}\over{}}  \cosh \di{\Theta \over 2} \left\{\hskip -0.1cm\di{{}\over{}}  Z_{g \mu} \left[ X_g + i (2\sigma) Y \right]_\nu -  Z_{g \nu} \left[ X_g + i (2\sigma) Y \right]_\mu \right\}             \right.\\
\left. \di{{}\over{}}~~~ + \sinh \di{\Theta \over 2} \left\{\hskip -0.1cm\di{{}\over{}} T_{g \mu} Z_{g \nu} - T_{g \nu} Z_{g \mu} \right\} \,\right\rrbracket \\~\\

 \ov{U_d}_{ \sigma'}\,\sigma_{\mu \nu}\, \gs_5\, {U_g}_{ \sigma} = i(2 \sigma)  \delta_{\sigma', \sigma} \left\llbracket \hskip -0.1cm\di{{}\over{}} \cosh \di{\Theta \over 2} \left[\hskip -0.1cm  \di{{}\over{}}  T_{g \mu} Z_{g \nu} - T_{g \nu} Z_{g \mu} \right] \right. \\
\left. + \sinh \di{\Theta \over 2} \left\{\hskip -0.1cm\di{{}\over{}}  Z_{g \mu} \left[ X_g + i (2\sigma) Y \right]_\nu -  Z_{g \nu} \left[ X_g + i (2\sigma) Y \right]_\mu \right\}   \right\rrbracket     \\

+ i \delta_{\sigma', -\sigma} \left\llbracket \hskip -0.1cm \di{{}\over{}}  \cosh \di{\Theta \over 2} \left\{\hskip -0.1cm\di{{}\over{}}  T_{g \mu} \left[ X_g + i (2\sigma) Y \right]_\nu -  T_{g \nu} \left[ X_g + i (2\sigma) Y \right]_\mu \right\}        \right.     \\
\left. + (2 \sigma) \sinh \di{\Theta \over 2} \left[\hskip -0.1cm\di{{}\over{}}  X_{g \mu} Y_\nu - X_{g \nu} Y_\mu \right] \,\right\rrbracket \\~

\end{array}  $~}} \enq
 
\vvv
\subsection{Formes bilin\'eaires $\ov{U_2}_{\sigma'}\,\Gamma\,{U_1}_{\sigma}$ du couplage en voie $t$}

\vv \nin Pour exprimer ces formes, nous proc\`ederons comme au paragraphe 3.2.4. Ecrivons : 

$$\ov{U_2}_{\sigma'}\,\Gamma\,{U_1}_{\sigma} = \cosh \di{\xi_2 \over 2}\, \cosh \di{\xi_1 \over 2} ~\ov{U}_{ d\sigma'} \,\Gamma\, U_{g \sigma} + (2 \sigma)(2 \sigma') \sinh \di{\xi_2 \over 2}\, \sinh \di{\xi_1 \over 2}~\ov{U}_{d \sigma'} \,\gs_5\,\Gamma\, \gs_5\,U_{g \sigma} $$
$$ + (2 \sigma) \cosh \di{\xi_2 \over 2}\, \sinh \di{\xi_1 \over 2} ~\ov{U}_{d \sigma'} \,\Gamma\, \gs_5\,U_{g \sigma} +  (2 \sigma') \sinh \di{\xi_2 \over 2}\, \cosh \di{\xi_1 \over 2} ~\ov{U}_{d\sigma'} \,\gs_5\,\Gamma\,U_{g\sigma} $$

\vv \nin et distinguons le cas o\`u $\Gamma$ commute avec $\gs_5$ de celui  o\`u $\Gamma$ anticommute avec $\gs_5$. Pour simplifier l'\'ecriture, nous poserons ici encore $c_{+}\! =\, \cosh \xi^{(+)}_{12}$, $c_{-} \!=\, \cosh \xi^{(-)}_{12}$, $s_{+} \!=\, \sinh \xi^{(+)}_{12}$, $s_{-} \!=\, \sinh \xi^{(-)}_{12}$ avec les d\'efinitions 

\beq \xi^{(+)}_{12} = \di{{\xi_1 + \xi_2}\over 2},~~~\xi^{(-)}_{12} = \di{{\xi_1 - \xi_2}\over 2}  \enq

\newpage
\vv \nin \ding{172} \und{\bf $\Gamma$ commute avec $\gs_5$}

$$ \ov{U_2}_{\sigma'}\,\Gamma\,{U_1}_{\sigma} = \delta_{\sigma', \sigma} \left\{ c_{+} \ov{U}_{d \sigma} \,\Gamma\, U_{g \sigma}  + (2 \sigma) s_{+}\,\ov{U}_{d \sigma} \,\Gamma\,\gs_5\, U_{g \sigma} \right\} $$
\beq +\, \delta_{\sigma', - \sigma} \left\{ c_{-}\,\ov{U}_{d -\sigma} \,\Gamma\, U_{g \sigma} +(2 \sigma) s_{-}\,  \ov{U}_{d -\sigma} \,\Gamma\, \gs_5\,U_{g \sigma} \right\}\enq

\vv \nin \ding{173} \und{\bf $\Gamma$ anticommute avec $\gs_5$}

$$ \ov{U_2}_{\sigma'}\,\Gamma\,{U_1}_{\sigma} = \delta_{\sigma', \sigma} \left\{ c_{-} \ov{U}_{d \sigma} \,\Gamma\, U_{g \sigma}  + (2 \sigma) s_{-}\,\ov{U}_{d \sigma} \,\Gamma\,\gs_5\, U_{g \sigma} \right\} $$
\beq +\, \delta_{\sigma', - \sigma} \left\{ c_{+}\,\ov{U}_{d -\sigma} \,\Gamma\, U_{g \sigma} + (2 \sigma) s_{+}\,  \ov{U}_{d -\sigma} \,\Gamma\, \gs_5\,U_{g \sigma} \right\}\enq

\vvv
\vv \nin D'o\`u les formes bilin\'eaires cherch\'ees : 


\vskip 1cm
\begin{center} 
\fbox{\fbox{\rule[-2cm]{0cm}{4cm}
\begin{minipage}{0.75\textwidth}
\vvv
\beq
\fbox{\rule[-0.2cm]{0cm}{0.6cm}~$  \ov{U_2}_{ \sigma'}\,U_{1 \sigma}  $~}  
\label{u2unitu1}\enq 
$$ \ov{U_2}_{ \sigma'}\,U_{1 \sigma}  = 2\, \delta_{\sigma', \sigma} \,c_{+} \cosh \di{\Theta \over 2} + 2\,(2 \sigma) \,\delta_{\sigma', - \sigma} \, s_{-} \sinh \di{\Theta \over 2}  $$ 
---------------------------------------------------------------------------------------------
\beq 
\fbox{\rule[-0.2cm]{0cm}{0.6cm}~$  \ov{U_2}_{ \sigma'}\,\gs_5\,U_{1 \sigma}  $~}  
\label{u2g5u1} \enq 
$$  \ov{U_2}_{ \sigma'}\,\gs_5\,U_{1 \sigma}  = 2\,\delta_{\sigma', \sigma} (2 \sigma) s_{+} \cosh \di{\Theta \over 2}  + 2\,\delta_{\sigma', - \sigma} \, c_{-} \sinh \di{\Theta \over 2} $$
---------------------------------------------------------------------------------------------
\beq 
\fbox{\rule[-0.2cm]{0cm}{0.6cm}~$ \ov{U_2}_{ \sigma'}\,\gs_\mu\,U_{1 \sigma} $~} 
\label{u2gmuu1} 
\enq
$$ \ov{U_2}_{ \sigma'}\,\gs_\mu\,U_{1 \sigma} = 2\,\delta_{\sigma', \sigma}  \left\llbracket \,c_{-} \left\{\cosh \di{\Theta \over 2} \,T_{g \mu} - \sinh \di{\Theta \over 2}  \left[ X_g + i (2\sigma) Y\right]_\mu \,\right\} \right.$$
$$\left.  +  s_{-} \cosh \di{\Theta \over 2} \,Z_{g \mu}     \right\rrbracket + 2\,(2 \sigma)  \delta_{\sigma', - \sigma} \left\llbracket \,  s_{+}  \left\{ \cosh \di{\Theta \over 2}  \left[X_g + i (2\sigma) Y \right]_\mu \right. \right.  $$
$$\left. \left.  - \sinh \di{\Theta \over 2} T_{g \mu} \right\}  - c_{+} \, Z_{g \mu} \, \right\rrbracket $$ 
---------------------------------------------------------------------------------------------
\beq
\fbox{\rule[-0.2cm]{0cm}{0.6cm}~$ \ov{U_2}_{\sigma'}\,\gs_\mu\,\gs_5\, {U_1}_{ \sigma}  $~} 
\label{u2gmug5u1}
\enq
$$\ov{U_2}_{\sigma'}\,\gs_\mu\,\gs_5\, {U_1}_{ \sigma}  = 2\, (2 \sigma) \,\delta_{\sigma', \sigma}  \left\llbracket \, c_{-}  \cosh \di{\Theta \over 2} \,Z_{g \mu}  +  s_{-} \left\{ \,\cosh \di{\Theta \over 2} \,T_{g \mu} \right. \right.$$
$$\left. \left. - \sinh \di{\Theta \over 2}  \left[ X_g + i (2\sigma) Y\right]_\mu \right\}  \right\rrbracket +2\, \delta_{\sigma', - \sigma} \left\llbracket \, c_{+}  \left\{ \cosh \di{\Theta \over 2}  \left[X_g + i (2\sigma) Y \right]_\mu  \right.   \right.  $$
$$ \left. \left. -  \sinh \di{\Theta \over 2}  T_{g \mu} \right\} - s_{+} \, Z_{g \mu} \,\right\rrbracket $$ 
\vskip 0.2cm
\end{minipage}}}
\end{center}

\newpage

~\vskip 1.5cm
\begin{center} 
\fbox{\fbox{\rule[-2cm]{0cm}{4cm}
\begin{minipage}{0.85\textwidth}
\vvv
\beq
\fbox{\rule[-0.2cm]{0cm}{0.6cm}~$  \ov{U_2}_{ \sigma'}\,\sigma_{\mu \nu}\,U_{1 \sigma}  $~}  
\label{u2unitu1}\enq 
$$ \ov{U_2}_{ \sigma'}\,\sigma_{\mu \nu}\,U_{1 \sigma}  =  \delta_{\sigma', \sigma} \left\llbracket   \, c_{+} \left\{ (2\sigma) \cosh \di{\Theta \over 2}\left[ X_{g \mu} Y_\nu - X_{g \nu} Y_\mu \right]  \right. \right.$$
$$\left. \left.\di{{}\over{}}+ i \sinh \di{\Theta \over 2} \left[ \hskip -0.1cm \di{{}\over{}} T_{g \mu} \left[ X_g + i(2\sigma) Y \right]_\nu -T_{g \nu} \left[ X_g + i(2\sigma) Y \right]_\mu \right]   \right\}  \right.$$
$$  + i s_{+} \left\{ \cosh \di{\Theta \over 2} \left[\hskip -0.1cm  \di{{}\over{}}  T_{g \mu} Z_{g \nu} - T_{g \nu} Z_{g \mu} \right] 
 + \sinh \di{\Theta \over 2} \left[\hskip -0.1cm\di{{}\over{}}  Z_{g \mu} \left[ X_g + i (2\sigma) Y \right]_\nu \right. \right.$$
$$\hskip -0.3cm  \di{{ }\over{}} \left.\left.\left. -  Z_{g \nu} \left[ X_g + i (2\sigma) Y \right]_\mu \right]  \,\right\} \right\rrbracket  + i (2\sigma) \delta_{\sigma', -\sigma} \left\llbracket  \, c_{-} \left\{  \cosh \di{\Theta \over 2} \left\{\hskip -0.1cm\di{{}\over{}}  Z_{g \mu} \left[ X_g + i (2\sigma) Y \right]_\nu  \right. \right.\right.$$
$$ \left.\left.-  Z_{g \nu} \left[ X_g + i (2\sigma) Y \right]_\mu   + \sinh \di{\Theta \over 2} \left[\hskip -0.1cm\di{{}\over{}}  T_{g \mu} Z_{g \nu} - T_{g \nu} Z_{gc\mu} \right] \,\right\}  \right. $$ 
$$  +  s_{-} \left\{ (2 \sigma) \cosh \di{\Theta \over 2} \left[\hskip -0.1cm\di{{}\over{}}  T_{g \mu} \left[ X_g + i (2\sigma) Y \right]_\nu -  T_{g \nu} \left[ X_g + i (2\sigma) Y \right]_\mu \right]  \right.$$ 
$$ \left.\left. + \sinh \di{\Theta \over 2} \left[\hskip -0.1cm\di{{}\over{}}  X_{g \mu} Y_\nu - X_{g \nu} Y_\mu \right] \right\}\,\right\rrbracket    $$ 
---------------------------------------------------------------------------------------------------------
\beq
\fbox{\rule[-0.2cm]{0cm}{0.6cm}~$ \ov{U_2}_{\sigma'}\,\sigma_{\mu \nu}\,\gs_5\, {U_1}_{ \sigma}  $~} 
\label{u2gmug5u1}
\enq
$$\ov{U_2}_{\sigma'}\,\sigma_{\mu \nu}\,\gs_5\, {U_1}_{ \sigma}  =  (2 \sigma) \delta_{\sigma', \sigma} \left\llbracket\, i c_{+} \left\{ \cosh \di{\Theta\over 2} \left[ T_{g \mu} Z_{g \nu} - T_{g \nu} Z_{gc\mu} \right]          \right. \right.$$ 
$$\left. +\sinh \di{\Theta \over 2} \left[ \hskip -0.1cm\di{{}\over{}}  Z_{g \mu} \left[ X_g + i (2\sigma) Y \right]_\nu -  Z_{g \nu} \left[ X_g + i (2\sigma) Y \right]_\mu \right] \right\} $$
$$ + s_{+} \left\{ (2\sigma) \cosh \di{\Theta \over 2}\left[ X_{g \mu} Y_\nu - X_{g \nu} Y_\mu \right]  
+ i \sinh \di{\Theta \over 2} \left[ \hskip -0.1cm \di{{}\over{}} T_{g \mu} \left[ X_g + i(2\sigma) Y \right]_\nu \right.\right. $$
$$ \left. \left. \left.\di{{}\over{}}  -T_{g \nu} \left[ X_g + i(2\sigma) Y \right]_\mu \right] \right\} \right\rrbracket + i  \delta_{\sigma', \sigma} \left\llbracket \, c_{-} \left\{ \cosh \di{\Theta \over 2} \left[\hskip -0.1cm\di{{}\over{}}  T_{g \mu} \left[ X_g + i (2\sigma) Y \right]_\nu \right. \right. \right. $$ 
$$\left.  \left. - T_{g \nu} \left[ X_g + i (2\sigma) Y \right]_\mu  \right]   + (2 \sigma) \sinh \di{\Theta \over 2} \left[\hskip -0.1cm\di{{}\over{}}  X_{g \mu} Y_\nu - X_{g \nu} Y_\mu \right] \right\} $$
$$ + s_{-} \left\{ \cosh \di{\Theta \over 2} \left[ \hskip -0.1cm\di{{}\over{}}  Z_{g \mu} \left[ X_g + i (2\sigma) Y \right]_\nu -  Z_{g \nu} \left[ X_g + i (2\sigma) Y \right]_\mu \right] \right.$$
$$\left. \left. + \sinh \di{\Theta \over 2} \left[\hskip -0.1cm\di{{}\over{}} T_{g \mu} Z_{g \nu} - T_{g \nu} Z_{g \mu} \right] \right\} \right\rrbracket$$
\vskip 0.2cm
\end{minipage}}}
\end{center}

\newpage 
\section{Cas o\`u toutes les masses sont \'egales, puis nulles}

\vv \nin Consid\'erons maintenant le cas o\`u les quatre particules de spin 1/2 qui participent \`a la r\'eaction $1+2 \rightarrow 3+4$ ont la m\^eme masse $m$, masse que nous ferons tendre vers z\'ero par la suite. Tout d'abord, pr\'ecisons encore les notations qui seront utilis\'ees. Dans le couplage en voie $s$, la base de r\'ef\'erence de l'\'etat initial est 

$$ T = \di{{p_1 + p_2}\over \sqrt{s}},~~Z_i =   \di{{p_1 - p_2}\over{\beta\, \sqrt{s}}}, ~~{\rm avec}~~s = (p_1 + p_2)^2,~~\beta = \sqrt{1 - \di{{4m^2}\over s}} $$  
\beq Y_\mu \,\propto \,\epsilon_{\mu \nu \rho \omega} \,p^\nu_1\, p^\nu_2 \,p^\omega_3,~~X_{i \mu} = \epsilon_{\mu \nu \rho \omega} T^\nu Y^\rho Z^\omega_i  \label{Bei} \enq  

\vv \nin Le vecteur $Y$ \'etant en fait d\'efini comme en (\ref{VY}). Dans ce m\^eme couplage, la base de r\'ef\'erence de l'\'etat final est 

$$ T = \di{{p_3 + p_4}\over \sqrt{s}},~~Z_f =   \di{{p_3 - p_4}\over{\beta\, \sqrt{s}}},~~Y,~~X_f = \epsilon_{\mu \nu \rho \omega} T^\nu Y^\rho Z^\omega_f ,~~~{\rm et~l'\hskip-0.04cmon~a} $$
\beq Z_f = Z_i \cos \theta + \sin \theta  X_i, ~~ X_f = - Z_i \sin \theta +  \cos \theta  X_i  \label{Bef} \enq  

\vv \nin Cette base (\ref{Bef}) s'obtient \`a partir de la premi\`ere (\ref{Bei}) par une rotation d'angle $\theta$ autour de l'axe $Y$. Rappelons que 

$$ t_1 = T \cosh \chi + Z_i \sinh \chi,~~z^{(s)}_1 = T \sinh \chi + Z_i \cosh \chi, $$
$$t_2 = T \cosh \chi - Z_i \sinh \chi,~~z^{(s)}_2 =  T \sinh \chi - Z_i \cosh \chi,$$
\beq t_3 = T \cosh \chi + Z_f \sinh \chi,~~z^{(s)}_3 = T \sinh \chi + Z_f \cosh \chi, \enq
$$t_4 = T \cosh \chi - Z_f \sinh \chi, ~~z^{(s)}_4 = T \sinh \chi - Z_f \cosh \chi, $$
$$ {\rm o\grave{u}}~~~\cosh \chi = \di{\sqrt{s} \over{2m }},~~\sinh \chi = \beta \,\di{\sqrt{s} \over{2m }}$$

\vv \nin Dans le couplage en voie $t$, la base de r\'ef\'erence du vertex de gauche est 

$$ T_g = \di{{p_1 + p_3}\over \sqrt{t + 4 m^2}},~~Z_g = \di{{p_1 - p_3}\over \sqrt{t}},~~Y,~~X_{g \mu} = \epsilon_{\mu \nu \rho \omega} T^\nu_g Y^\rho Z^\omega_g , ~~{\rm avec}$$
\beq  t = - (p_1-p_3)^2 = s \beta^2 \sin^2 \di{\theta \over 2} \label{Beg} \enq

\vv \nin Celle du vertex de droite est 

\beq  T_d = \di{{p_2 + p_4}\over \sqrt{t+ 4 m^2}},~~Z_d = \di{{p_4 - p_2 }\over \sqrt{t}} = Z_d,~~Y,~~X_{d \mu} = \epsilon_{\mu \nu \rho \omega} T^\nu_d Y^\rho Z^\omega_d 
\label{Bed} \enq

\vv \nin Et l'on \'ecrit 

$$ t_1 = T_g \cosh \xi + Z_g \sinh \xi,~~z^{(t)}_1 = T_g \sinh \xi + Z_g \cosh \xi, $$
$$ t_3 =  T_g \cosh \xi - Z_g \sinh \xi,~~z^{(t)}_3 = - T_g \sinh \xi + Z_g \cosh \xi, $$ 
\beq  t_2 = T_d \cosh \xi - Z_d \sinh \xi,~~z^{(t)}_2 = -T_d \sinh \xi + Z_d \cosh \xi, \enq
$$t_4 =  T_d \cosh \xi + Z_d \sinh \xi, ~~z^{(t)}_4 = T_d \sinh \xi + Z_d \cosh \xi, $$
$$ {\rm o\grave{u}}~~~\cosh \xi = \di{\sqrt{t+4m^2} \over{2m }},~~\sinh \chi = \,\di{\sqrt{t} \over{2m }}$$

\vv \nin Bien entendu, nous excluons ici les situations o\`u $\theta =0$ ou bien $\theta = \pi$. Sous cette condition, on trouve les d\'ecompositions 

$$ T_g = T \cosh \Upsilon + \sinh \Upsilon \left[ Z_i \cos \di{\theta \over 2} + X_i \sin \di{\theta \over 2} \right],$$
$$ X_g = T \sinh \Upsilon +  \cosh \Upsilon \left[ Z_i \cos \di{\theta \over 2} + X_i \sin \di{\theta \over 2} \right],$$
$$ T_d = T \cosh \Upsilon - \sinh \Upsilon \left[ Z_i \cos \di{\theta \over 2} + X_i \sin \di{\theta \over 2} \right],$$
$$ X_d = -T \sinh \Upsilon +  \cosh \Upsilon \left[ Z_i \cos \di{\theta \over 2} + X_i \sin \di{\theta \over 2} \right],$$
\beq Z_g = Z_d = Z_i \sin \di{\theta \over 2} - X_i \cos \di{\theta \over 2}, \enq
$$ {\rm avec}~~~\cosh \Upsilon = \sqrt{\di{s\over{t+ 4 m^2}}},~~~\sinh \Upsilon = \beta \cos \di{\theta \over 2}  \sqrt{\di{s\over{t+ 4 m^2}}} $$

\vv \nin desquelles il ressort que la t\'etrade $\left\{ [\,T_g\,] : T_g, X_g, Y, Z_g\right\} $ se d\'eduit de la t\'etrade $\left\{ [\,T\,]_i : T, X_i, Y, Z_i\right\}$ par une transformation de Lorentz pure de rapidit\'e $\Upsilon$ dans le 2-plan $(T,X)$, suivie d'une rotation d'angle $\psi = \di{{\theta - \pi}\over 2}$ autour de l'axe $Y$. On remarque que

$$ \cosh \Upsilon = \di{{\cosh \chi}\over{\cosh \xi}} ~~~{\rm que}$$  
\beq \sqrt{\di{{t+4m^2}\over s}} \,X_g = T \beta \cos \di{\theta\over 2} + Z_i \cos \di{\theta\over 2} + X_i \sin \di{\theta\over 2} \label{XG} \enq

\vv \nin et que $\Upsilon = \Theta/2$ o\`u $\Theta$ est le param\`etre d\'efini en (\ref{tetagd}), o\`u les masses sont  prises toutes \'egales \`a $m$.  

\vv \nin Les t\'etrades associ\'ees, par exemple \`a la particule 1, dans les deux types de couplage, en voie $s$ et en voie $t$, ont en commun les deux vecteurs $t_1$ et $Y$. On doit donc passer de l'une \`a l'autre par une rotation autour de l'axe $Y$, dans le 2-plan orthogonal au 2-plan $(t_1, Y)$. Les formules ci-apr\`es permettent d'obtenir, {\it \`a $2 \pi$ pr\`es}, l'angle de cette rotation. On a   

$$ z^{(t)}_1 \cdot X_i = - \sinh \xi\, \sinh \Upsilon \sin \di{\theta \over 2} = \di{{2m}\over \sqrt{s} }\cosh \Upsilon \cos \di{\theta \over 2} $$ 
$$z^{(t)}_1 \cdot z^{(s)}_1 = \sinh \chi \sinh \xi \cosh \Upsilon - \cos \di{\theta \over 2} \cosh \chi \sinh \xi - \sin \di{\theta \over 2} \cosh \chi \cosh \xi = - \sin \di{\theta \over 2} \cosh \Upsilon, ~~{\rm et}$$
$$ X_g \cdot X_i = - \sin \di{\theta \over 2} \cosh \Upsilon,~~X_g \cdot z^{(s)}_1 = - \di{{2m}\over \sqrt{s} }\cosh \Upsilon \cos \di{\theta \over 2} $$ 

\nin On peut donc \'ecrire ($ x^{(s)}_1 = X_i$)

$$ z^{(t)}_1 = z^{(s)}_1 \cos \psi_1 + x^{(s)}_1  \sin \psi_1,~~x^{(t)}_1 = X_g = -z^{(s)}_1 \sin \psi_1 + x^{(s)}_1  \cos \psi_1,~~~{\rm avec}$$
\beq  \cos \psi_1 = \sin \di{\theta \over 2} \cosh \Upsilon,~~\sin \psi_1 =  -\di{{2m}\over \sqrt{s} }\cosh \Upsilon \cos \di{\theta \over 2} \enq 

\nin On trouve de m\^eme 

$$ z^{(t)}_3 = - z^{(s)}_3 \cos \psi_1 + x^{(s)}_3 \sin \psi_1,~~x^{(t)}_3 = X_g = - z^{(s)}_3 \sin \psi_1 - x^{(s)}_3 \cos \psi_1,$$
$$ {\rm soit}~~\psi_3 = \pi - \psi_1, $$

\nin puis ($x^{(s)}_2 = -X_i, ~x^{(s)}_4 = - X_f$)

$$ z^{(t)}_2 = - z^{(s)}_2 \cos \psi_1 - x^{(s)}_2 \sin \psi_1,~~x^{(t)}_2 = X_d =  z^{(s)}_2 \sin \psi_1 - x^{(s)}_2 \cos \psi_1,$$
$$ {\rm soit}~~\psi_2 = \pi + \psi_1, $$
$$ z^{(t)}_4 =  z^{(s)}_4 \cos \psi_1 - x^{(s)}_4 \sin \psi_1,~~x^{(t)}_4 = X_d =  z^{(s)}_4 \sin \psi_1 + x^{(s)}_4 \cos \psi_1,$$
$$ {\rm soit}~~\psi_4 = 2 \pi - \psi_1, $$

\vv \nin et l'on observe les relations (\`a $2 \pi$ pr\`es)

$$ \psi_2 - \psi_1 = \psi_4 - \psi_3 = \pi,~~\psi_2 + \psi_3 = \psi_1 + \psi_4 = 2 \pi  $$

\vv \nin Lesdites rotations \'etant planes, on peut \'evidemment utiliser la formule (2.91) du chapitre 2 pour exprimer, au signe pr\`es, les matrices qui les repr\'esentent, par exemple, 

$$ L_1 = L([\,t_1\,]_t [\,t_1\,]^{-1}_s) = \di{1 \over{{\rm Tr} \left([\,t_1\,]_t [\,t_1\,]^{-1}_s\right)}} \left[ \gs(t_1) \gs(t_1) - \gs(x^{(s)}_1) \gs(x^{(t)}_1) - \gs(Y) \gs(Y) \right. $$ 
\beq \left.  -\gs(z^{(s)}_1) \gs(z^{(t)}_1)  \right]   = \pm \left[ \cos \di{\psi_1 \over 2} + \sin \di{\psi_1 \over 2}  \gs(z^{(s)}_1) \gs(x^{(s)}_1) \right]  \label{L11} \enq 

\nin D'un autre c\^ot\'e, on a 

$$ [\,t_1\,]_t = \left[T_g \rightarrow t^{(t)}_\xi \right] [\,T_g\,],~~ [\,t_1\,]_s = \left[T \rightarrow t^{(s)}_\chi \right]  [\,T\,], ~~{\rm et}$$
$$ [\,t_1\,]_t [\,t_1\,]^{-1}_s =  \left[T_g \rightarrow t^{(t)}_\xi \right]\, L \,\left[T \rightarrow t^{(s)}_\chi \right]^{-1}~~{\rm avec}~~~L =  [\,T_g\,][\,T\,]^{-1}, $$
$$ t^{(t)}_\xi = T_g \cosh \xi + Z_g \sinh \xi,~~~t^{(s)}_\chi = T \cosh \chi +Z \sinh \chi $$

\vv \nin  On a 

$$ L^{-1} \, \left[T_g \rightarrow t^{(t)}_\xi \right]\,L =    \left[T \rightarrow t^{(s)}_\xi \right],~~{\rm et~donc}~~~[\,t_1\,]_t [\,t_1\,]^{-1}_s = L \left[\, T \rightarrow t^{(s)}_{\xi - \chi}\,\right]$$

\vv \nin Comme expliqu\'e plus haut, la transformation $L$ est le produit d'une transformation Lorentz pure dans le 2-plan $(T,X)$ et d'une rotation autour de l'axe $Y$. En repr\'esentation spinorielle de Dirac, on a donc 

$$L_1 = \left[ \cos \di{\psi \over 2} + \sin \di{\psi \over 2} \gs(Z) \gs(X) \right] \left[ \cosh \di{\Upsilon \over 2} + \sinh \di{\Upsilon \over 2} \gs(X) \gs(T) \right]\times $$
\beq \times \left[ \cosh \di{{\chi - \xi}\over2} - \sinh \di{{\chi - \xi}\over 2} \gs(Z) \gs(T) \right] \label{L12} \enq

\vv \nin Il est instructif, bien que fastidieux, de montrer que l'expression (\ref{L12}) est \'equivalente \`a l'expression (\ref{L11}). Ceci est fait en appendice.

\vv \nin Le fait important ici est que lorsque $m \rightarrow 0$, $\psi_1 \rightarrow 0$, $\psi_2 \rightarrow  \pi$, $\psi_3 \rightarrow  \pi$, $\psi_4 \rightarrow 2 \pi$, {\it \`a $2\pi$ pr\`es}. Par cons\'equent, \`a la limite des masses nulles, on a, {\it aux signes pr\`es}, les \'equivalences

\beq \fbox{\fbox{\rule[-0.5cm]{0cm}{1.2cm}~$  U^{(t)}_{1_\sigma} \simeq U^{(s)}_{1_\sigma},~~U^{(t)}_{2_\sigma} \simeq (2 \sigma) \,U^{(s)}_{2_{-\sigma}},~~U^{(t)}_{3_\sigma} \simeq (2 \sigma) \,U^{(s)}_{3_{-\sigma}},~~U^{(t)}_{4_\sigma} \simeq U^{(s)}_{4_\sigma}   $~}} \label{Uequiv}  \enq

\vv \nin Il en r\'esulte que, pour obtenir les formes bilin\'eaires dans les deux types de couplage lorsque $m=0$, il suffit de consid\'erer uniquement les formules \'etablies dans le couplage en voie $s$\footnote{On notera le changement du signe de l'h\'elicit\'e des particules 2 et 3 d'un couplage \`a l'autre.}. A cette fin, il faut tout d'abord utiliser des spineurs normalis\'es selon $ \ov{U} \, U = 2\, m$ puis poser $m =0$ dans les expressions obtenues pour les projecteurs ou les formes bilin\'eaires. On trouve ainsi les expressions ci-apr\`es. 

\vvv
\beq
\fbox{\fbox{\rule[-2cm]{0cm}{4.2cm} ~$
\begin{array}{c}
~ \\
\ov{U_2}_{ \sigma'}\, {U_1}_{ \sigma} = - \sqrt{s}\, (2 \sigma) \,\delta_{\sigma', -\sigma} 
= (2 \sigma)\,\ov{U_2}_{ \sigma'}\,\gs_5\, {U_1}_{ \sigma} \\ ~\\ \hline \\
\ov{U_2}_{ \sigma'}\,\gs_\mu\, {U_1}_{ \sigma} = 
 \delta_{\sigma', \sigma} \sqrt{s} \, (X + 2 i \sigma Y)_\mu = (2\sigma)\,\ov{U_2}_{ \sigma'}\,\gs_\mu\, \gs_5\,{U_1}_{ \sigma} \\
~ \\ \hline \\
 \ov{U_2}_{ \sigma'}\,\sigma_{\mu \nu} \, {U_1}_{ \sigma}  = 
 - \delta_{\sigma', - \sigma}\,\di{\sqrt{s}\over 2} \, \left\llbracket\di{{}\over{}}  i (2 \sigma) \left[ T_\mu Z_\nu - T_\nu Z_\mu \right]   +  (X_\mu Y_\nu - X_\nu Y_\mu) \,\right\rrbracket \\~\\ 
= (2\sigma)\,\ov{U_2}_{ \sigma'}\,\sigma_{\mu \nu} \, \gs_5\,{U_1}_{ \sigma}  \\~
\end{array}$~}}  \label{0-12} 
\enq
\vvv
\vv \nin Le tableau (\ref{0-12}) indique qu'\`a la limite $m=0$, on a la relation $\ov{U_2}_{\sigma'} \Gamma \gs_5\,{U_1}_\sigma = (2 \sigma) \,\ov{U_2}_{\sigma'} \Gamma \,{U_1}_\sigma$. Ce r\'esultat m\'erite une explication. Consid\'erons une particule de spin 1/2, de 4-impulsion $p$ et de masse $m$. Il est toujours possible de d\'efinir une base de r\'ef\'erence $\left\{T,X,Y,Z\right\}$ telle que la t\'etrade $\left\{ [\,t\,] : t,x,y,z\right\}$ associ\'ee \`a $p$ puisse s'\'ecrire 

$$ t = \di{p \over m} = T \cosh \chi + Z \sinh \chi,~~z = T \sinh \chi + Z \cosh \chi,~~x = X,~~y=Y $$ 
\beq {\rm o\grave{u}}~~~\cosh \chi = \di{E\over m},~~~~\sinh \chi = \di{E\over m}\, \beta,~~{\rm avec}~~\beta = \sqrt{1 - \di{m^2\over E^2}} \enq
$$ {\rm soit}~~~e^\chi = \di{E\over m} (1 + \beta),~~e^{-\chi} = \di{m\over E}\di{1\over{1+\beta}} $$
\vv \nin Les spineurs $U_\sigma$ de ladite particule v\'erifiant 

$$ S_z U_\sigma = \di{1\over 2} \gs_5 \gs(z) \gs(t) U_\sigma = \sigma\, U_\sigma,~~~\gs(t) U_\sigma = U_\sigma, $$

\vv \nin on a 

\beq  \gs(t\pm z) U_\sigma = e^{\pm \chi} \,\gs(T+Z) U_\sigma = \left[1 \pm (2\sigma) \gs_5 \right] U_\sigma \enq

\vv \nin d'o\`u l'on d\'eduit qu'\`a la limite $m=0$ ($e^{-\chi} \rightarrow 0$), les spineurs deviennent vecteurs propres de la chiralit\'e $\gs_5$ : 

\beq \gs_5 \,U_\sigma  \simeq (2 \sigma)\,U_\sigma \label{chiral} \enq

\vvv \nin ce qui se traduit aussi par la relation $V_\sigma \simeq (2 \sigma)\,U_\sigma$, d'o\`u le r\'esultat en question. 


\newpage

\beq
\fbox{\fbox{\rule[-2cm]{0cm}{4.2cm} ~$
\begin{array}{c}
~ \\
\ov{U_3}_{ \sigma'}\, {U_1}_{ \sigma} =-  (2 \sigma) \,\delta_{\sigma', - \sigma} \,\sqrt{t} 
=  (2 \sigma)\,\ov{U_3}_{\sigma'}\,\gs_5\,{U_1}_ {\sigma} 
\\ ~\\ \hline \\
\ov{U_3}_{ \sigma'}\,\gs_\mu\,U_{1 \sigma} = \delta_{\sigma', \sigma} \, \sqrt{t}\, \left\llbracket \,\left[ T_\mu + Z_\mu\right] \cot \di{\theta \over 2} + (X+ 2i \sigma Y)_\mu \di{{}\over{}} \right\rrbracket  
 \\~ \\
 = (2 \sigma)\,\ov{U_3}_{ \sigma'}\,\gs_\mu\,\gs_5\,U_{1 \sigma} \\ ~\\ \hline \\~\\
\ov{U_3}_{ \sigma'}\,\sigma_{\mu \nu}\,U_{1 \sigma} =
 \delta_{\sigma', - \sigma}\, \di{\sqrt{t}\over 2} \,\left\llbracket \, i  (2\sigma)\cot \di{\theta \over 2} \left[ \di{{}\over{}} (T+ Z)_\mu (X+ 2i \sigma Y)_\nu \right. \right. 
\\~\\
\left. \left. \di{{}\over{}} -(T+ Z)_\nu (X+2i \sigma Y)_\mu  \di{{}\over{}}\right]  
- (X_\mu Y_\nu - X_\nu Y_\mu )  - (2\sigma)\, i   (T_\mu Z_\nu - T_\nu Z_\mu) \right\rrbracket 
\\~\\
= (2 \sigma)\,\ov{U_3}_{ \sigma'}\,\sigma_{\mu \nu}\, \gs_5\,U_{1 \sigma}  \\~
\end{array}$~}}  \label{0-13} 
\enq

\vv \nin Dans les deux tableaux (\ref{0-12}) et (\ref{0-13}), la forme bilin\'eaire vectorielle s'exprime uniquement \`a l'aide des vecteurs de polarisations circulaires correspondant au couplage d'h\'elicit\'e qui appara\^it le plus naturel entre les deux particules consid\'er\'ees. Ainsi, pour les particules 1 et 2, coupl\'ees naturellement en voie $s$, lesdits vecteurs de polarisation sont 

$$ \epsilon^{(\lambda)}_{(s)} = -  \lambda\,\di{1\over \sqrt{2}}\, \left[ X + i \lambda Y \right],~~~{\rm et~l'\hskip-0.03cmon~a} $$
\beq \ov{U_2}_{ \sigma'}\,\gs_\mu\, {U_1}_{ \sigma} =-(2 \sigma)\, \delta_{\sigma', \sigma}\, \sqrt{2 s} \,\left[ \epsilon^{(2 \sigma)}_{(s)}\right]_\mu \enq

\vv \nin Pour les particules 1 et 3, coupl\'ees naturellement en voie $t$, les vecteurs de polarisation font intervenir $X_g$. Or, d'apr\`es (\ref{XG}), pour $m=0$ on a  

$$ X_g = X + \cot \di{\theta \over 2}\,(T+Z),~~~{\rm et} $$
\beq    X_g + i (2 \sigma) Y =  \cot \di{\theta \over 2}\,(T+Z) + X + i (2 \sigma) Y = - (2 \sigma) \,\epsilon^{(2 \sigma)}_{(t)},~~~{\rm~ d'o\grave{u}} \label{eps-t} \enq
$$ \ov{U_3}_{ \sigma'}\,\gs_\mu\,U_{1 \sigma} = - (2 \sigma)\, \delta_{\sigma', \sigma}\, \sqrt{2 t} \,\left[ \epsilon^{(2 \sigma)}_{(t)}\right]_\mu $$

\vv \nin Le lecteur v\'erifiera que l'on a un r\'esultat similaire pour les formes bilin\'eaires vectorielles  $\ov{U_4}_{ \sigma'}\,\gs_\mu\,U_{1 \sigma}$ et $\ov{U_3}_{ \sigma'}\,\gs_\mu\,U_{2 \sigma}$, pour lesquelles on a cette fois une r\'eminiscence de couplage ``en voie $u$". Ainsi, 

$$ \ov{U_4}_{ \sigma'}\,\gs_\mu\,U_{1 \sigma} = - (2 \sigma) \, \delta_{\sigma', \sigma}\, \sqrt{2 u} \,\left[ \epsilon^{(2 \sigma)}_{(u)}\right]_\mu ,~~~\ov{U_3}_{ \sigma'}\,\gs_\mu\,U_{2 \sigma} = - (2 \sigma) \, \delta_{\sigma', \sigma}\, \sqrt{2 u} \,\left[ {\epsilon'}^{(2 \sigma)}_{(u)}\right]_\mu, $$
\beq {\rm avec}~~~ - \sqrt{2}\,(2 \sigma)\, \epsilon^{(2 \sigma)}_{(u)} = X_u + i(2 \sigma) Y,~~~X_u = X - \tan \di{\theta \over 2} \,(T+Z),  \label{autrescoups} \enq
$$ - \sqrt{2}\,(2\sigma)\, {\epsilon'}^{(2 \sigma)}_{(u)} = X'_u + i(2 \sigma) Y,~~~X'_u = X + \tan \di{\theta \over 2} \,(T-Z), $$
$$ u = s \cos^2 \di{\theta \over 2} $$

\newpage

\section{Couplage sym\'etrique}

\vv \nin Lorsque le nombre de particules apparaissant dans l'\'etat final d'une r\'eaction est sup\'erieur \`a deux, il est pr\'ef\'erable d'utiliser un couplage d'h\'elicit\'e plus sym\'etrique pour cet \'etat final. Notons ici encore $T,\,X,\,Y,\,Z$ les vecteurs de base d\'efinissant le r\'ef\'erentiel du centre de masse de la r\'eaction. Relativement \`a cette base, l'impulsion $p$ d'une particule de masse 
$m$ participant \`a la r\'eaction s'\'ecrit
\beq
p = T E  + q Z \cos \theta  + q \sin\theta \left( X \cos\vp  +
Y \sin \vp \right)
\enq
avec  $q = \sqrt{E^2 -m^2}$. Associons \`a $T$ la triade de 4-vecteurs du genre espace $X', Y', Z'$
d\'efinis par 

\begin{eqnarray}
&X'= -Z \sin \theta + \cos\theta \left( X \cos\vp + Y \sin \vp \right),~~~
Y'= -X \sin\vp  + Y \cos \vp & \nonumber \\
&Z'= Z \cos \theta + \sin\theta \left(X  \cos\vp +
Y \sin \phi \right)&
\end{eqnarray}

\vv \nin On r\'ealise un couplage d'h\'elicit\'e entre $T$ et $p$ en associant \`a $p$ la base : 

$$
t = \di{p\over m},~~~x= X',~~~y=Y',~~~z= T \sinh\chi  + Z' \cosh\chi,~~~{\rm o\grave{u}} 
$$
\beq \cosh\chi = \di{E\over m},~~\sinh \chi = \di{q\over m} = \beta \,\di{E\over m},~~{\rm avec}~~\beta = \sqrt{1 - \di{m^2 \over E^2}}    \label{ths}  \enq

\vv \nin Ce couplage, du type couplage en voie $s$, est compl\`etement sym\'etrique vis-\`a-vis des particules de l'\'etat final. Le choix du r\'ef\'erentiel du centre de masse comme r\'ef\'erence garantit la propri\'et\'e de covariance qui est requise pour la d\'efinition d'une h\'elicit\'e invariante. Le cas d'un \'etat final \`a 3 particules a \'et\'e abord\'e au chapitre 1. La transformation de Lorentz ${\cal L}$ permettant de passer de ladite base de r\'ef\'erence $(T,X,Y,Z)$ \`a la base d'h\'elicit\'e $(t,x,y,z)$ est le produit de trois transformations : une transformation de Lorentz pure ${\cal H}_Z(\chi)$ de rapidit\'e $\chi$ le long de l'axe $Z$, suivie d'une rotation ${\cal R}_Y(\theta)$ d'angle $\theta$ autour de l'axe $Y$, elle-m\^eme suivie d'une rotation ${\cal R}_Z(\vp)$ d'angle $\vp$ autour de l'axe $Z$. En repr\'esentation spinorielle de Dirac : 

\beq
\begin{array}{c}
{\cal L} = {\cal R}_Z(\vp) {\cal R}_Y(\theta) {\cal H}_Z(\chi),~~{\rm avec}  \\~\\
{\cal R}_Z(\vp) = \cos \di{\vp\over 2} - 2 i \sin \di{\vp\over 2} S_Z = \cos \di{\vp\over 2} + \sin \di{\vp\over 2}  \gs(X) \gs(Y), \\~\\
 {\cal R}_Y(\theta) = \cos \di{\theta\over 2}  - 2 i \sin \di{\theta\over 2}  S_Y =\cos \di{\theta\over 2}  + \sin \di{\theta\over 2}  \gs(Z) \gs(X), \\~\\
 {\cal H}_Z(\chi) = \cosh \di{\chi\over 2}  + 2
\sinh \di{\chi\over 2}  \gs_5 S_Z = \cosh \di{\chi\over 2}  + 
\sinh \di{\chi\over 2}  \gs(Z) \gs(T) 
\end{array}
 \label{tr-sym} \enq

\vv \nin Dans la repr\'esentation spinorielle \`a deux dimensions, ces transformations sont  d\'ecrites par les matrices 

\beq
\begin{array}{c}
{\cal R}^{(2)}_Z(\vp) = \cos \di{\vp\over 2} -i  \sin \di{\vp\over 2} \,Z  \hskip -0.5cm \begin{array}[m]{c} ~
\\ \stackrel{~~~}{\sim}
\end{array} \hskip -0.1cm\widetilde{T}, ~~~~{\cal R}^{(2)}_Y(\theta) = \cos \di{\theta\over 2} - i \sin \di{\theta\over 2}  \,Y  \hskip -0.52cm \begin{array}[m]{c} ~
\\ \stackrel{~~~}{\sim}
\end{array} \hskip -0.1cm\widetilde{T},\\~\\
 {\cal H}^{(2)}_Z(\chi) =  \cosh \di{\chi\over 2} + 
\sinh \di{\chi\over 2}\,Z  \hskip -0.5cm \begin{array}[m]{c} ~
\\ \stackrel{~~~}{\sim}
\end{array} \hskip -0.1cm\widetilde{T} 
\end{array}
\enq

\nin et la trace de leur produit ${\cal L}^{(2)}$ est calcul\'ee comme suit : 

$$ {\rm Tr}\,{\cal L}^{(2)} = 2 a_0 = {\rm Tr} [\,T\,] [\,T\,]^{-1} {\cal R}^{(2)}_Z(\vp){\cal R}^{(2)}_Y(\theta) {\cal H}^{(2)}_Z(\chi) $$
$$ = {\rm Tr}\, \left[ \cos \di{\vp\over 2} -i  \sin \di{\vp\over 2} \tau_3 \right] \left[  \cos \di{\theta\over 2} - i \sin \di{\theta\over 2} \tau_2\right] \left[ \cosh \di{\chi\over 2} + 
\sinh \di{\chi\over 2} \tau_3 \right] $$ 
$$ = 2 \cos \di{\theta \over 2} \left[\cosh \di{\chi\over 2}   \cos \di{\vp\over 2} - i \sinh \di{\chi\over 2} \sin \di{\vp\over 2} \right]$$

\nin On a donc 

$$ a_0 = \cos \di{\theta \over 2} \left[\cosh \di{\chi\over 2}   \cos \di{\vp\over 2} - i \sinh \di{\chi\over 2} \sin \di{\vp\over 2} \right] ~~~{\rm et}  $$
\beq 2 |a_0|^2 = \cos^2 \di{\theta \over 2}  \left[ \,\cosh \chi + \cos \vp\, \right] \enq

\vv \nin La transformation $(T,X,Y,Z) \rightarrow (t,x,y,z)$ n'est pas plane et c'est pourquoi le param\`etre $a_0$ poss\`ede une partie imaginaire. Pour $\chi \neq 0$, cette partie imaginaire est imputable \`a la rotation ${\cal R}_Z(\vp)$\footnote{A ce propos, le lecteur v\'erifiera que la transformation $[\,t_3\,]_s [\,t_1\,]^{-1}_s$ est plane.}. Utilisant la formule (7.98) de I.T.L, la matrice ${\cal L}$ dans (\ref{tr-sym})  peut \^etre r\'ecrite sous la forme\footnote{En explicitant les 4-vecteurs dans (\ref{tr-sym-2}), le lecteur v\'erifiera aussi, en usant abondamment de la relation $\gs_5 = i \gs(X) \gs(Y)\gs(Z)\gs(T)$, que l'expression obtenue co\"incide bien avec le d\'eveloppement du produit (\ref{tr-sym}).}

$${\cal L} = \di{{ \cosh \di{\chi \over 2}   \cos \di{\vp\over 2} - i \sinh \di{\chi\over 2} \sin \di{\vp\over 2}\, \gs_5}\over {2 \,\cos {\di{\theta \over 2} \,\left[ \cosh \chi + \cos \vp \right] }}}\, \times $$
\beq \times \left[ \hskip-0.08cm \di{{}\over{}} \gs(t)\gs(T) - \gs(x) \gs(X) - \gs(y) \gs(Y) - \gs(z) \gs(Z) \right] 
\label{tr-sym-2} \enq

\vv \nin Notons $U_\sigma$ les spineurs de Dirac associ\'es \`a la t\'etrade d'h\'elicit\'e $[\,t\,]$ de $p$ et ayant la normalisation $\ov{U}_{\sigma'}\,U_\sigma = 2\, \delta_{\sigma', \sigma}$. On les obtient en appliquant la matrice ${\cal L}$ aux spineurs de Dirac associ\'es au r\'ef\'erentiel de r\'ef\'erence (t\'etrade $[\,T\,]$), not\'es ${U_0}_\sigma$ et  normalis\'es de la m\^eme fa\c{c}on : $ U_\sigma = {\cal L}\, {U_0}_\sigma $. On peut aussi les relier plus simplement aux spineurs de Dirac $U'_\sigma$ associ\'es \`a la t\'etrade d'h\'elicit\'e $\left\{ [\,T\,]' : T,X',Y', Z'\right\}$, d\'efinis par  

$$ U'_\sigma =  {\cal R}_Z(\vp)\, {\cal R}_Y(\theta) \,{U_0}_\sigma $$
\beq = \cos \di{\theta \over 2} e^{-i \sigma \vp}\, {U_0}_\sigma +(2 \sigma) \sin \di{\theta \over 2} e^{i \sigma \vp} \,{U_0}_{- \sigma}, ~~~{\rm et~l'\hskip-0.05cmon~a}    \label{spineur-cousym}  \enq
$$ U_\sigma = \cosh \di{\chi \over 2}\,U'_\sigma + (2 \sigma) \sinh \di{\chi \over 2}\,V'_{\sigma} $$

\vvv

\vv \nin Si la particule consid\'er\'ee est de spin 1/2, ses spineurs sont d\'efinis comme ci-dessus. 
S'il s'agit d'un photon r\'eel ou d'un gluon, sa 4-impulsion $k$ est du genre lumi\`ere 
et de la forme $k= E(T+Z')$. Ses vecteurs de polarisations circulaires sont 

\beq
\epsilon^{(\lambda)} =  - \di{1\over{\sqrt 2}}\left(\lambda x + i y \right)
\enq

\nin $x$ et $y$ \'etant d\'efinis comme dans (\ref{ths}). Notons ici la relation 

\beq
x =  X \cos\vp + Y \sin \vp -\left(Z+Z'\right) \tan\di{\theta\over 2}
\enq

\vv \nin qui permet de r\'ecrire ces vecteurs comme 

$$ \epsilon^{(\lambda)} = E^{(\lambda)}\exp(- i \lambda \vp)  +\di{\lambda \over
\sqrt{2}} \tan \di{\theta\over 2}\left[(T+Z') -(T-Z)\right] ~~~{\rm o\grave{u}} $$
\beq E^{(\lambda)} = -\di{1\over\sqrt{2}} \left[ \lambda X +i Y \right] \enq

\vv \nin Si l'on consid\`ere des amplitudes issues de la th\'eorie \'electro-faible ou de la Chromodynamique Quantique, celles-ci \'etant invariantes de jauge vis-\`a-vis du photon ou du gluon, on peut ignorer le terme $\propto\, (T+Z')\, \propto \,k$ dans l'expression ci-dessus et red\'efinir les vecteurs de polarisations circulaires par 
 
$$
\epsilon^{(\lambda)} =\exp(- i \lambda \vp)\left[ E^{(\lambda)} + \xi (T-Z) \right],~~~{\rm o\grave{u}} 
$$
\beq
\xi = - \di{\lambda\over
\sqrt{2}} \tan \di{\theta\over 2} \exp( i \lambda \vp)
\enq

\vv \nin Il est facile de v\'erifier que ces nouveaux vecteurs sont encore orthogonaux \`a la fois \`a $k$ et \`a $T - Z$.

\vv \nin Pour les applications dans le domaine dit ``ultra-relativiste", celui des tr\`es hautes \'energies pour lesquelles $E \gg m$, il peut \^etre utile de pouvoir distinguer nettement, parmi les composantes de spineurs,  celles qui sont les plus importantes dans ce domaine (les ``grandes composantes") de celles pouvant y \^etre n\'eglig\'ees (les ``petites composantes"). Comme nous allons le voir, l'utilisation conjointe de la param\'etrisation (\ref{spineur-cousym}) et de {\it projecteurs de chiralit\'e} permet d'effectuer une telle distinction, o\`u la rapidit\'e $\chi$, d\'efinie en (\ref{ths}), est d\'eterminante ; et ce, tout en \'evitant d'avoir \`a donner une repr\'esentation explicite des spineurs, en donnant de surcro\^it la possibilit\'e de contr\^oler efficacement les approximations effectu\'ees.    

\vv \nin Au spineur (\ref{spineur-cousym}), appliquons le projecteur de chiralit\'e $\di{1\over 2}\left[ 1 + \eta\, \gs_5 \right]$ o\`u $\eta = \pm1$. On obtient  

$$ \di{1\over 2}\left[ 1 + \eta\, \gs_5 \right]\,U_\sigma = \di{1\over 2}\left[ 1 + \eta\, \gs_5 \right]\,\left[ \cosh \di{\chi \over 2} + \eta (2 \sigma)  \sinh \di{\chi \over 2} \right]\, U'_\sigma $$ 

\vv \nin Choisissons $\eta = \pm (2 \sigma)$ et posons $P^{(\pm)}_\sigma =   \di{1\over 2}\left[ 1 \pm (2 \sigma)\, \gs_5 \right]$. Il vient 

$$  P^{(+)}_\sigma\,U_\sigma = e^{\chi/2} \,P^{(+)}_\sigma\,U'_\sigma\,,~~ P^{(-)}_\sigma\,U_\sigma = e^{-\chi/2} \,P^{(-)}_\sigma\,U'_\sigma$$

\vv \nin Pour $E \gg m$, on a clairement $e^{\chi/2} \gg e^{-\chi/2}$, et l'on voit que la d\'ecomposition d'un spineur en spineurs propres de la chiralit\'e, qui revient \`a une d\'ecomposition de Weyl dans la repr\'esentation initiale des spineurs, permet de faire une distinction claire entre composantes, dans leurs comportements \`a tr\`es haute \'energie. Dor\'enavant, nous utiliserons des spineurs normalis\'es selon $\ov{U}\,U = 2 m$, tout en gardant pour les spineurs $U'$ la normalisation $\ov{U'} \,U' =2$. Nous poserons 

$$  U^{(\pm)}_\sigma = P^{(\pm)}_\sigma \, U_\sigma = \sqrt{m} \, e^{\pm \chi/2}\, P^{(\pm)}_\sigma\,U'_\sigma\, , ~~~{\rm et} $$ 
\beq U_\sigma = U^{(+)}_\sigma + U^{(-)}_\sigma \label{weyl} \enq

\vv \nin Manifestement, les spineurs $U^{(\pm)}$ ne d\'ependent de l'\'energie $E$ que par les facteurs $\sqrt{m}\, e^{\pm \chi/2}$. On a 

$$ e^\chi = \cosh \chi + \sinh \chi = \di{E\over m} (1 + \beta),~~ {\rm et}$$
\beq  \sqrt{m}\, e^{\chi/2} = \sqrt{ E(1 + \beta)},~~ \sqrt{m}\, e^{-\chi/2} =  \di{m\over{\sqrt{E(1+\beta)}}} \enq

\vv \nin Du point de vue de leurs d\'ependances vis-\`a-vis de l'\'energie, il y a donc un rapport $m/E$ entre $U^{(-)}$ et $U^{(+)}$, le premier spineur portant ainsi les petites composantes du spineur $U$, tandis que le second en porte les grandes. Il est aussi int\'eressant de noter le fait suivant. On a   

$$  U^{(\pm)}_\sigma = \sqrt{m}\,e^{\pm \chi/2}\, P^{(\pm)}_\sigma\, U'_\sigma = \sqrt{m}\,e^{\pm \chi/2}\,    
{\cal R}_Z(\vp) {\cal R}_Y(\theta) P^{(\pm)}_\sigma\, U_{0 \sigma},~~~{\rm et} $$
$$ P^{(\pm)}_\sigma\, U_{0 \sigma} = \di{1\over 2} \left[ 1\pm  2 \gs_5 S_Z \right] U_{0 \sigma} \equiv \di{1\over 2} \gs(T\pm Z) U_{0 \sigma} \,;~~~{\rm or,}$$

$${\cal R}_Y(\theta) \gs(T \pm Z) = \cos \di{\theta \over 2} \, \gs(T \pm Z) - \sin \di{\theta \over 2}\, \gs(T \mp Z) 2 i S_Y,~~~{\rm de~sorte~que} $$

\beq  U^{(\pm)}_\sigma = \di{1\over 2}\,\sqrt{m}\,e^{\pm \chi/2}\,{\cal R}_Z(\vp) \left\{ \cos \di{\theta \over 2} \gs(T \pm Z) U_{0 \sigma} +(2 \sigma) \sin \di{\theta \over 2} \gs(T \mp Z) \, U_{0 - \sigma} \right\} \label{uplmo} \enq

\vv \nin On voit alors qu'en appliquant $\gs(T \pm Z)$ aux spineurs ``chiraux" $U^{(\pm)}$, on peut s\'electionner d'autres termes dans les spineurs. On a en effet :  
 
$$ \gs(T-Z)\, U^{(+)}_\sigma = \sqrt{E(1+\beta)}\, {\cal R}_Z(\vp) \cos \di{\theta \over 2} \left[ 1 - (2 \sigma) \gs_5 \right]\,U_{0 \sigma} $$ 

$$ \gs(T+Z)\, U^{(+)}_\sigma = (2 \sigma)\,\sqrt{E(1+\beta)}\, {\cal R}_Z(\vp) \sin \di{\theta \over 2} \left[ 1 - (2 \sigma) \gs_5 \right]\,U_{0 -\sigma} $$ 

$$ \gs(T+Z)\, U^{(-)}_\sigma = \di{m \over{\sqrt{E(1+\beta)}}} \, {\cal R}_Z(\vp) \cos \di{\theta \over 2} \left[ 1 + (2 \sigma) \gs_5 \right]\,U_{0 \sigma} $$ 

$$ \gs(T-Z)\, U^{(-)}_\sigma = (2 \sigma)\,\di{m \over{\sqrt{E(1+\beta)}}} \, {\cal R}_Z(\vp) \sin \di{\theta \over 2} \left[ 1 + (2 \sigma) \gs_5 \right]\,U_{0 -\sigma} $$ 

\vv \nin Dans une configuration de haute \'energie et de petit angle de diffusion $\theta$, les projections ci-dessus se classent respectivement en terme d'ordre z\'ero pour la premi\`ere, termes d'ordre 1 pour la seconde et la troisi\`eme\footnote{A ce sujet, voir par exemple, C. Carimalo, hep-ph arXiv:1401.4407}.  

\vv \nin Etant donn\'e deux particules de spin 1/2 \'etiquet\'ees par les indices $(i)$ et $(j)$, consid\'erons les formes bilin\'eaires $F = \ov{U_j}^{(\epsilon_j)}_{\sigma_j}\,\Gamma\, {U_i}^{(\epsilon_i)}_{\sigma_i}$. Comme 

$$ \gs_5\, {U_i}^{(\epsilon_i)}_{\sigma_i} = \epsilon_i (2 \sigma_i) {U_i}^{(\epsilon_i)}_{\sigma_i},~~~{\rm il~ vient}  ~~~\epsilon_i (2 \sigma_i) \,F = \ov{U_j}^{(\epsilon_j)}_{\sigma_j}\,\Gamma\,\gs_5\, {U_i}^{(\epsilon_i)}_{\sigma_i} = - \epsilon\, \epsilon_j (2 \sigma_j) F$$

\vv \nin avec $\epsilon = +1$ ou $\epsilon= -1$, selon que $\Gamma$ commute ou anti-commute avec $\gs_5$, respectivement. La forme bilin\'eaire consid\'er\'ee est donc a priori non nulle si $\epsilon_i (2 \sigma_i) = - \epsilon\, \epsilon_j (2 \sigma_j)$, soit 
\vv
\beq \fbox{\rule[-0.5cm]{0cm}{1.2cm}~$ \sigma_j = - \epsilon \, \epsilon_i\, \epsilon_j \, \sigma_i  $~}  \label{cond1} \enq

\vv \nin Si l'on a envisage un r\'egime ultra-relativiste o\`u $U_\sigma \approx U^{(+)}_\sigma$, on sera plut\^ot enclin \`a  consid\'erer principalement le cas $\epsilon_i = \epsilon_j = 1$.  La relation (\ref{cond1}) indique alors que pour $\Gamma = 1, \gs_5, \sigma_{\mu \nu}$, ($\epsilon = +1$), la forme bilin\'eaire correspondante n'est a priori non nulle que si $\sigma_j = - \sigma_i$ ; tandis que les formes bilin\'eaires correspondant \`a $\Gamma = \gs_\mu, \gs_\mu \gs_5$, ($\epsilon = -1$), ne sont a priori non nulles que si $\sigma_j = \sigma_i$. On retrouve ici le fait que pour les th\'eories du mod\`ele standard, qui reposent sur des couplages vectoriels entre particules, l'h\'elicit\'e se conserve. Dans cette m\^eme perspective, calculons explicitement la forme bilin\'eaire correspondant \`a $\Gamma = \gs_\mu$ ; celle avec $\Gamma = \gs_\mu \gs_5$ s'en d\'eduit facilement, puisque $\gs_5 \,{U_i}^{(+)}_{\sigma} = (2 \sigma)\,{U_i}^{(+)}_{\sigma}$. On a 

$$ U^{(+)}_\sigma = \di{1\over 2} \sqrt{E(1+\beta)}\, \left[1 + (2 \sigma) \gs_5 \right] \left[ e^{-i \sigma \vp} \cos \di{\theta \over 2} \, U_{0 \sigma} + (2 \sigma) e^{i \sigma \vp} \sin \di{\theta \over 2} \, U_{0 -\sigma} \right]~~~{\rm et} $$

$$ \ov{U_j}^{(+)}_{\sigma'}\,\gs_\mu\, {U_i}^{(+}_{\sigma} = \sqrt{E_j E_i (1+ \beta_j)(1+ \beta_i)}\, G,~~~{\rm avec}$$
$$ G = \di{1\over 4} \left[  e^{i \sigma' \vp_j} \cos \di{\theta_j \over 2} \, \ov{U_0}_{\sigma'} +  (2 \sigma') e^{- i \sigma' \vp_j} \sin \di{\theta_j \over 2} \, \ov{U_0}_{ -\sigma'} \right] \left[1 -(2 \sigma') \gs_5 \right] \gs_\mu \left[ 1+ (2 \sigma) \gs_5 \right] \times $$
$$ \times \left[ e^{-i \sigma \vp_i} \cos \di{\theta_i \over 2} \, U_{0 \sigma} + (2 \sigma) e^{i \sigma \vp_i} \sin \di{\theta_i \over 2} \, U_{0 -\sigma} \right]$$ 

\vv \nin Comme $\left[1 -(2 \sigma') \gs_5 \right] \gs_\mu \left[ 1+ (2 \sigma) \gs_5 \right] = \gs_\mu \left[1+ (2 \sigma') \gs_5 \right] \left[ 1+ (2 \sigma) \gs_5 \right] = 2 \delta_{\sigma, \sigma'} \gs_\mu \left[ 1 + (2 \sigma) \gs_5 \right]$, il vient $ G = \delta_{\sigma', \sigma}\, G'$, o\`u 
 
$$ G' = \di{1\over 2} \left[  e^{i \sigma \vp_j} \cos \di{\theta_j \over 2} \, \ov{U_0}_{\sigma} +  (2 \sigma) e^{- i \sigma \vp_j} \sin \di{\theta_j \over 2} \, \ov{U_0}_{ -\sigma} \right] \,\gs_\mu \left[ 1 + (2 \sigma) \gs_5 \right] \, \times $$
$$ \times \, \left[ e^{-i \sigma \vp_i} \cos \di{\theta_i \over 2} \, U_{0 \sigma} + (2 \sigma) e^{i \sigma \vp_i} \sin \di{\theta_i \over 2} \, U_{0 -\sigma} \right] $$

\vv \nin Rappelons que 

$$ \ov{U_0}_{\sigma'} \, \gs_\mu\, U_{0 \sigma} = 2 \,\delta_{\sigma', \sigma}\, T_\mu ,~~~\ov{U_0}_{\sigma'} \, \gs_\mu\,\gs_5\, U_{0 \sigma} = 2 (2 \sigma) \,\delta_{\sigma', \sigma} \,Z_\mu + 2 \delta_{\sigma', - \sigma} \left[X+i(2 \sigma)Y \right]_\mu,~~{\rm donc} $$
$$ \ov{U_0}_{\sigma'} \, \gs_\mu \left[ 1 + \eta\,(2 \sigma) \gs_5 \right] \, U_{0 \sigma} =2\, \delta_{\sigma', \sigma} \left[ T +\eta \, Z \right]_\mu + 2(2 \sigma)\, \eta\,\delta_{\sigma', - \sigma} \left[X + i (2 \sigma) Y \right]_\mu ~~~(\eta = \pm1) $$

\vv \nin Pour $\Gamma = \gs_\mu$, on en d\'eduit 

\beq \fbox{\rule[-0.5cm]{0cm}{1.2cm}~$ \begin{array}{c} 
~\\
 G'_\mu =  e^{i \sigma( \vp_j - \vp_i) } \cos \di{\theta_j \over 2}  \cos \di{\theta_i\over 2} \left[ \,T + Z\, \right]_\mu +  e^{- i \sigma( \vp_j - \vp_i) } \sin \di{\theta_j \over 2}  \sin \di{\theta_i\over 2} \left[ \,T - Z\, \right]_\mu \\~\\
+  e^{- i \sigma( \vp_j + \vp_i) } \sin \di{\theta_j \over 2}  \cos \di{\theta_i\over 2}  \left[X + i (2 \sigma) Y \right]_\mu  
+ e^{i \sigma( \vp_j + \vp_i) } \cos \di{\theta_j \over 2}  \sin \di{\theta_i\over 2}  \left[X - i (2 \sigma) Y \right]_\mu \\~
\end{array}
 $~}  \label{Gprime} \enq

\vv \nin Il est facile de v\'erifier que 

$$ \left[G'\right]^2 = 0,~~~G' \cdot \left[\,G'\right]^\star = - \left[1 - \cos \theta_{ji}\right],~~{\rm o\grave{u}}  $$
\beq  \cos \theta_{ji} = \cos \theta_j \cos \theta_i + \sin \theta_j \sin \theta_i \cos(\vp_j- \vp_i) 
\label{NormGp} \enq

\vv \nin Le vecteur $G'$, isotrope et tel que $G' \cdot \left[\,G'\right]^\star < 0$, s'apparente \`a un vecteur de polarisation circulaire. Ceci nous am\`ene \`a poser  

\vvv
\beq \fbox{\fbox{\rule[-0.5cm]{0cm}{1.2cm}~$\begin{array}{c}
~\\
\epsilon^{(2 \sigma)}_{ji} = - (2 \sigma) \di{G' \over{ \sqrt{2} \left|\sin \di{\theta_{ji} \over 2} \right|}},~~~{\rm de~sorte~que} \\~\\
 \ov{U_j}^{(+)}_{\sigma'}\,\gs_\mu\, {U_i}^{(+)}_{\sigma} =(2 \sigma) \, \ov{U_j}^{(+)}_{\sigma'}\,\gs_\mu\, {V_i}^{(+)}_{\sigma} \\~\\
 = - (2 \sigma)\,\delta_{\sigma', \sigma} \sqrt{\di{(1+\beta_j)(1+\beta_i)}\over 4}\, \sqrt{ 2 Q_{ji}} \,\left[\epsilon^{(2 \sigma)}_{ji}\right]_\mu~~~{\rm o\grave{u}}  \\~\\
 Q_{ji} = 2 E_j E_i (1 - \cos \theta_{ji} ) \\~
\end{array}  $~}}  \label{fbsym}\enq

\newpage

\nin L'expression de la forme bilin\'eaire pr\'esent\'ee dans (\ref{fbsym}) est tout \`a fait similaire \`a celles dans (\ref{eps-t}) et (\ref{autrescoups}), lorsque les masses des particules peuvent \^etre n\'eglig\'ees. En effet, dans ces conditions, la grandeur $Q_{ji}$ peut tout aussi bien \^etre interpr\'et\'ee comme le carr\'e $M^2_{ji} = (p_j +p_i)^2$ de la masse invariante du syst\`eme des deux particules $(j)$ et $(i)$ si celles-ci sont toutes deux dans l'\'etat final\footnote{Auquel cas, la particule $(i)$ est une anti-particule, plut\^ot repr\'esent\'ee par le spineur ${V_i}^{(+)}_\sigma = \gs_5\, {U_i}^{(+)}_\sigma$.}, que comme le transfert $\Delta_{ji} = -(p_i-p_j)^2$, si la particule $(j)$ est dans l'\'etat final tandis que la particule $(i)$ est dans l'\'etat initial ; de plus,  $(1+\beta_j)(1+\beta_i)/4 \simeq 1$. Le vecteur  
$\epsilon^{(2 \sigma)}_{ji}$ doit donc repr\'esenter un vecteur de polarisation circulaire d'un couplage en voie $s$ ou en voie $t$. Pour mettre ce fait en \'evidence, prenons $m_j = m_i =0$ et d\'efinissons 

\beq T_{ji} = \di{{p_j + p_i}\over \sqrt{ Q_{ji}}},~~~X_{ji} = -(2 \sigma) \sqrt{2} \,\,\Re\left\{\epsilon^{(2 \sigma)}_{ji}\right\},~~~Y_{ji} = - \sqrt{2} \,\,\Im \left\{\epsilon^{(2 \sigma)}_{ji}\right\}  \label{baseji} \enq

\vv \nin C'est un excellent exercice formateur que de d\'emontrer la relation 

\beq \fbox{\rule[-0.5cm]{0cm}{2cm}~$ \begin{array}{c} 
~\\
\epsilon_{\mu \nu \rho \delta}\,X^\rho_{ji}\, Y^\delta_{ji} = - \left[T_{ji}\right]_\mu \left[Z_{ji}\right]_\nu + \left[T_{ji}\right]_\nu \left[Z_{ji}\right]_\mu,~~~{\rm o\grave{u} } \\~\\
 Z_{ji} = \di{{p_i - p_j}\over \sqrt{Q_{ji}}}  \\~
\end{array}
 $~}  \enq

\vv \nin qui montre que les quatre vecteurs $T_{ji}, X_{ji}, Y_{ji}, Z_{ji}$ forment une base orthonorm\'ee et d'orientation directe, et que cette base est associ\'ee \`a un couplage d'h\'elicit\'e entre les deux particules concern\'ees, en voie $s$ ou en voie $t$ selon la position des particules dans la r\'eaction\footnote{A noter qu'\`a la limite des masses nulles, on a $\gs(p_i)\, U^{(+)}_i \hskip -0.1cm = \gs(p_j)\,U^{(+)}_j \hskip -0.1cm = 0$ et par cons\'equent $p_i \cdot \epsilon^{(2 \sigma)}_{ji} \hskip -0.1cm = p_j \cdot \epsilon^{(2 \sigma)}_{ji} =0$, soit encore, $p_i \cdot X_{ji} = p_j \cdot X_{ji} = p_i \cdot Y_{ji} = p_j \cdot Y_{ji} =0$, conform\'ement \`a un couplage d'h\'elicit\'e.}. Ceci g\'en\'eralise les observations faites \`a la fin du pr\'ec\'edent paragraphe.

\vv \nin Examinons maintenant le projecteur $ P^{(+)}_{ij}(\sigma, \sigma')  =  {U_i}^{(+)}_{\sigma} \, \ov{U_j}^{(+)}_{\sigma'}$ (avec $m \neq 0$). Comme toute matrice $4 \times 4$, il peut \^etre d\'ecompos\'e sur la base des 16 matrices de Dirac, dont les coefficients sont les formes bilin\'eaires consid\'er\'ees plus haut. 

$$ P^{(+)}_{ij}(\sigma, \sigma')  = S + P\, \gs_5+  V^\mu\, \gs_\mu + A^\mu\, \gs_\mu\, \gs_5 + T^{\mu \nu}\, \sigma_{\mu \nu} , ~~{\rm avec}$$  
$$ {\rm Tr}\, \left[ \Gamma_\alpha \, P^{(+)}_{ij}(\sigma, \sigma') \right] =  \ov{U_j}^{(+)}_{\sigma'} \, \Gamma_\alpha\,  {U_i}^{(+)}_{\sigma} $$

\vv \nin Il est avantageux de d\'ecomposer pr\'ealablement le projecteur comme suit 

$$ P^{(+)}_{ij}(\sigma, \sigma') = \delta_{\sigma', \sigma} \,P^{(+)}_{ij}(\sigma, \sigma)  + \delta_{\sigma', - \sigma}\, P^{(+)}_{ij}(\sigma, -\sigma) $$   

\vv \nin D'apr\`es ce qui pr\'ec\`ede, la partie $P^{(+)}_{ij}(\sigma, \sigma) $ du projecteur n'a de composantes que sur les matrices $\gs_\mu$ et $\gs_\mu \gs_5$, tandis que l'autre partie $P^{(+)}_{ij}(\sigma, -\sigma) $ se d\'ecompose uniquement sur les matrices $1, \gs_5$ et $\sigma_{\mu \nu}$. On a 

$$ P^{(+)}_{ij}(\sigma, \sigma) = \di{1\over 4} \left[ \gs_\mu \, \ov{U_j}^{(+)}_{\sigma} \, \gs^\mu\,{U_i}^{(+)}_{\sigma}  - \gs_\mu \, \gs_5 \,\ov{U_j}^{(+)}_{\sigma} \, \gs^\mu\,\gs_5\, {U_i}^{(+)}_{\sigma} \right] $$
\beq  = \di{1\over 4} \, \gs_\mu\, \left[ 1 - (2 \sigma) \gs_5 \right]\, \ov{U_j}^{(+)}_{\sigma} \, \gs^\mu\,{U_i}^{(+)}_{\sigma} \enq

\vv \nin o\`u la forme $\ov{U_j}^{(+)}_{\sigma} \, \gs^\mu\,{U_i}^{(+)}_{\sigma} $ peut \^etre exprim\'ee selon (\ref{fbsym}). Calculons ensuite 

$$ \ov{U_j}^{(+)}_{-\sigma} \,{U_i}^{(+)}_{\sigma} = (2 \sigma)\, \ov{U_j}^{(+)}_{-\sigma} \,\gs_5\,{U_i}^{(+)}_{\sigma}= \,\sqrt{E_j E_i (1+\beta_j)(1+\beta_i)}\, F^\sigma_{ji} $$

$$ {\rm o\grave{u}} ~~~F^\sigma_{ji} = \di{1\over 2} \left[  e^{-i \sigma \vp_j} \cos \di{\theta_j \over 2} \, \ov{U_0}_{-\sigma} -  (2 \sigma) e^{ i \sigma \vp_j} \sin \di{\theta_j \over 2} \, \ov{U_0}_{\sigma} \right]  \left[ 1+ (2 \sigma) \gs_5 \right] \times $$
$$ \times \left[ e^{-i \sigma \vp_i} \cos \di{\theta_i \over 2} \, U_{0 \sigma} + (2 \sigma) e^{i \sigma \vp_i} \sin \di{\theta_i \over 2} \, U_{0 -\sigma} \right],~~{\rm soit} $$

\beq \fbox{\rule[-0.5cm]{0cm}{1.2cm}~$ F^\sigma_{ji} = (2 \sigma) \left[  e^{-i \sigma (\vp_j - \vp_i)} \, \cos \di{\theta_j \over 2}  \sin \di{\theta_i \over 2} - e^{i \sigma (\vp_j - \vp_i)} \, \sin \di{\theta_j \over 2}  \cos \di{\theta_i \over 2}   \right]  $~}  \enq

\vv \nin et l'on a  

\beq \fbox{\rule[-0.5cm]{0cm}{1.2cm}~$\left|\, F^\sigma_{ji} \,\right|^2 = \sin^2 \di{\theta_{ji} \over 2}   $~}  \enq

\vv \nin Puis 

$$ \ov{U_j}^{(+)}_{-\sigma} \,\sigma_{\mu \nu} \,{U_i}^{(+)}_{\sigma} =  \,\sqrt{E_j E_i (1+\beta_j)(1+\beta_i)}\, \left[ T^\sigma_{ji} \right]_{\mu \nu}$$
$$ {\rm o\grave{u}} ~~~\left[ T^\sigma_{ji} \right]_{\mu \nu}= \di{1\over 2} \left[  e^{-i \sigma \vp_j} \cos \di{\theta_j \over 2} \, \ov{U_0}_{-\sigma} -  (2 \sigma) e^{ i \sigma \vp_j} \sin \di{\theta_j \over 2} \, \ov{U_0}_{\sigma} \right]  \left[ 1+ (2 \sigma) \gs_5 \right] \, \sigma_{\mu \nu} \,\times $$
$$ \times \left[ e^{-i \sigma \vp_i} \cos \di{\theta_i \over 2} \, U_{0 \sigma} + (2 \sigma) e^{i \sigma \vp_i} \sin \di{\theta_i \over 2} \, U_{0 -\sigma} \right],~~{\rm soit} $$ 

$$ \left[ T^\sigma_{ji} \right]_{\mu \nu}= \di{1 \over 2} i(2 \sigma)\left\llbracket \,e^{-i \sigma (\vp_j + \vp_i)} \cos \di{\theta_j \over 2} \cos \di{\theta_i \over 2} \left[\di{{}\over{}}  (T+Z)_\mu (X + i(2 \sigma)Y)_\nu \right. \right. $$ 
$$\left. \di{{}\over{}}  - (T+Z)_\mu (X + i(2 \sigma)Y)_\nu \right] + \left[ e^{i \sigma (\vp_j - \vp_i)} \sin \di{\theta_j \over 2} \cos \di{\theta_i \over 2} + e^{-i \sigma (\vp_j - \vp_i)} \cos \di{\theta_j \over 2} \sin \di{\theta_i \over 2} \right] \times$$
$$ \times \left[ \di{{}\over{}} T_\mu Z_\nu - T_\nu Z_\mu - i (2 \sigma) \left[X_\mu Y_\nu-X_\nu Y_\mu \right] \, \right]  - e^{i \sigma (\vp_j + \vp_i)} \sin \di{\theta_j \over 2} \sin \di{\theta_i \over 2}\, \times $$
$$ \left. \di{{}\over{}} \times \left[\di{{}\over{}} (T-Z)_\mu (X-i(2\sigma)Y)_\nu -  (T-Z)_\nu (X-i(2\sigma)Y)_\mu \,\right] \,\right\rrbracket $$

\vv \nin et finalement 

\beq  P^{(+)}_{ij}(\sigma, -\sigma) = \sqrt{E_j E_i (1+\beta_j)(1+\beta_i)}\,\left\llbracket\, \di{1\over 4} \left[1 + (2 \sigma) \gs_5 \right]  F^\sigma_{ji} + \di{1\over 2} \sigma_{\mu \nu} T^{\mu \nu}_{ji} \,\right\rrbracket  \enq

\vv \nin Dans un premier temps, supposons non nulles, et toutes deux \'egales \`a $m$ pour simplifier, les masses des particules $(i)$ et $(j)$. Couplons les deux particules en h\'elicit\'e, avec pour base de r\'ef\'erence 

\beq  T_{ji} = \di{{p_j + p_i}\over \sqrt{s_{ji}}},~~~Z_{ji} = \di{{p_j - p_i}\over \sqrt{s_{ji}}},~~X_{ji},~~Y_{ji}, ~~{\rm avec}~~s_{ji}= 2 m^2 + 2 p_j \cdot p_i \label{baseji-nn} \enq
 
\vv \nin et o\`u les vecteurs $X_{ji}$ et $Y_{ji}$ ne co\"incident avec ceux d\'efinis en (\ref{baseji}) que lorsque $m=0$. Soit $[\,t_i\,]'$ la t\'etrade associ\'ee \`a $t_i$ dans ce couplage, similaire \`a celles \'etudi\'ees dans le couplage en voie $t$. Comme cette t\'etrade et la t\'etrade $[\,t_i\,]$ d\'efinie dans le couplage sym\'etrique ont le vecteur $t_i$ en commun, on passe de l'une \`a l'autre par une rotation. Prenons ensuite $m=0$. Dans ce cas, la t\'etrade de r\'ef\'erence (\ref{baseji-nn}) devient exactement celle d\'efinie dans (\ref{baseji}). Nous invitons alors le lecteur \`a \'etablir\footnote{Et, en l'occurrence, \`a v\'erifier.}  et \`a interpr\'eter\footnote{Revoir le paragraphe 2.1.1, \ding{174}.} les relations 

\vv
\beq \fbox{\fbox{\rule[-0.5cm]{0cm}{1.2cm}~$ \begin{array}{c} 
~\\
 X_{ji} + i (2 \sigma) Y_{ji} = e^{i \Psi_s} \left[ X_i + i (2 \sigma) Y_i \right] +  e^{i \Psi_c}  (T+ Z_i)\left|\cot \di{\theta_{ji}\over 2}\right| \\~\\
 = - e^{- i \Psi_s} \left[ X_j - i (2 \sigma) Y_j \right] +e^{i \Psi_c}   (T+ Z_j) \left|\cot \di{\theta_{ji}\over 2}\right|, ~~{\rm o\grave{u}}  \\~\\
 e^{i \Psi_s} = \di{1 \over {\left|\sin \di{\theta_{ji} \over 2}\right|}}\left[ \sin \di{\theta_j \over 2} \cos \di{\theta_i \over 2} e^{-i \sigma(\vp_j - \vp_i)}-  \cos \di{\theta_j \over 2} \sin \di{\theta_i \over 2} e^{i \sigma(\vp_j - \vp_i)} \right], \\~\\
e^{i \Psi_c} = \di{1 \over {\left|\cos \di{\theta_{ji} \over 2}\right|}}\left[ \cos \di{\theta_j \over 2} \cos \di{\theta_i \over 2} e^{i \sigma(\vp_j - \vp_i)} +  \sin \di{\theta_j \over 2} \sin \di{\theta_i \over 2} e^{-i \sigma(\vp_j - \vp_i)} \right]
\\~
\end{array}
 $~}}  \enq

\vv
\vv \nin Nous invitons \'egalement le lecteur \`a v\'erifier que dans ce couplage d'h\'elicit\'e sym\'etrique, l'h\'elicit\'e d'une particule, consid\'er\'ee dans le r\'ef\'erentiel du centre de masse de la r\'eaction, y est repr\'esent\'ee par la projection de son spin sur le vecteur unitaire de sa 3-impulsion. 

\vv \nin Pour terminer, notons que le choix des axes spatiaux $X, Y, Z$ du r\'ef\'erentiel du centre de masse est g\'en\'eralement dict\'e par des consid\'erations pratiques. Par exemple, l'axe de collision des deux particules de l'\'etat initial est pris de fa\c{c}on naturelle comme axe $Z$. Cependant, comme nous l'avons d\'ej\`a remarqu\'e, il peut \^etre judicieux de choisir les vecteurs spatiaux de la base de r\'ef\'erence de l'\'etat final en fonction de leurs propri\'et\'es au regard du groupe des permutations des particules composant cet \'etat. La nouvelle base ainsi choisie se d\'eduit alors de celle de l'\'etat initial par une simple rotation.

\newpage
\section{Appendice : \'equivalence entre (\ref{L12}) et (\ref{L11}) } 

\vv \nin Le d\'eveloppement du produit de matrices (\ref{L12}), donne une expression du type

$$ L_1 = a_1 + a_2 \gs(X)\gs(T) + a_3 \gs(Z)\gs(T) + a_4 \gs(X) \gs(Z) $$ 

\vv \nin Il est inutile d'expliciter ici les coefficients $a_i$, car cette expression doit \^etre transform\'ee en exprimant les vecteurs $T$ et $Z$ en fonction de $t_1$ et $z^{(s)}_1$, de mani\`ere \`a faire appara\^itre le g\'en\'erateur $\gs(z^{(s)}_1) \gs(X)$ de la rotation attendue. Cette op\'eration conduit \`a  

$$ L_1 = A_1 + A_2 \,\gs(z^{(s)}_1) \gs(X) + A_3 \,\gs(X) \gs(t_1) + A_4 \,\gs(z^{(s)}_1) \gs(t_1), ~~~{\rm o\grave{u}} $$ 

$$ A_1 = \cos \di{\psi \over 2} \cosh \di{\Upsilon \over 2} \cosh \di{{\chi - \xi}\over 2} + \sin \di{\psi \over 2} \sinh \di{\Upsilon \over 2} \sinh \di{{\chi - \xi}\over 2} $$

$$A_2 = \cos \di{\psi \over 2} \sinh \di{\Upsilon \over 2} \sinh \di{{\chi + \xi}\over 2} + \sin \di{\psi \over 2} \cosh \di{\Upsilon \over 2} \cosh \di{{\chi + \xi}\over 2} $$

$$A_3 = \cos \di{\psi \over 2} \sinh \di{\Upsilon \over 2} \cosh \di{{\chi + \xi}\over 2} + \sin \di{\psi \over 2} \cosh \di{\Upsilon \over 2} \sinh \di{{\chi + \xi}\over 2} $$

$$- A_4 = \cos \di{\psi \over 2} \cosh \di{\Upsilon \over 2} \sinh \di{{\chi - \xi}\over 2} + \sin \di{\psi \over 2} \sinh \di{\Upsilon \over 2} \cosh \di{{\chi - \xi}\over 2} $$

\vv \nin Il reste maintenant \`a montrer que cette expression peut \^etre mise sous la forme (\ref{L11}), et notamment \`a montrer en premier lieu que $A_3 = A_4 =0$. Pour la commodit\'e d'\'ecriture, nous poserons ici $c = \cos \di{\theta \over 2}$, $s = \sin \di{\theta \over 2}$, $ r = \sqrt{1 - \beta^2 c^2} $. Une premi\`ere \'etape consiste \`a exprimer des fonctions hyperboliques et trigonom\'etriques, comme il est fait ci-apr\`es.   

$$ \hskip -0.5cm (1) ~~~1+ \cosh (\chi -\xi) = 1+ \cosh \chi \cosh \xi - \sinh \chi \sinh \xi = \di{1\over{1 - \beta^2}} \left[ r +1 - \beta^2 (s + 1)  \right] $$
$$ =\di{1\over{1 - \beta^2}} \di{{1-r}\over{1-r}}  \left[ r +1 - \beta^2 (s + 1)  \right] = \di{\beta^2\over{1 - \beta^2}} \,(1+s)\, \di{{r-s}\over{1-r}} $$
$$= \di{{(1+s)(1+r)}\over{r+s}}  = 2 \cosh^2 \left(\di{{\chi -\xi}\over 2} \right)$$

$$\hskip -0.5 cm (2)~~~  \cosh (\chi -\xi) -1 = \di{1\over{1 - \beta^2}} \left[ r -1 + \beta^2 ( 1- s)  \right] = \di{1\over{1 - \beta^2}} \di{{1+r}\over{1+r}}  \left[ r -1 + \beta^2 (1-s )  \right] $$
$$ =  \di{\beta^2\over{1 - \beta^2}} \,(1-s)\, \di{{r-s}\over{1+r}} = \di{{(1-s)(1-r)}\over{r+s}}  = 2 \sinh^2 \left(\di{{\chi -\xi}\over 2} \right)$$ 

\vv \nin Le passage de $ \chi - \xi$ \`a $\chi+\xi$ \'equivaut \`a changer $s$ en $-s$. Donc 

$$ \hskip -1.7cm (3) \hskip 0.7cm 2 \cosh^2 \left(\di{{\chi +\xi}\over 2} \right) = \di{{(1-s)(1+r)}\over{r-s}},~~~ 2 \sinh^2 \left(\di{{\chi +\xi}\over 2} \right)= 
\di{{(1+s)(1-r)}\over{r-s}}  $$

$$ \hskip -4.3cm (4) \hskip 3cm 2 \cosh^2 \di{\Upsilon \over 2} = \di{{1+r}\over r},~~~ 2 \sinh^2 \di{\Upsilon \over 2} = \di{{1-r}\over r} $$
 
$$ \hskip -0.3cm (5)~~~\cos \di{\psi\over 2} = \di{1\over \sqrt{2}} \left[ \cos \di{\theta \over 4} +  \sin \di{\theta \over 4} \right] = \sqrt{\di{{1+s}\over 2}},~~~\sin \di{\psi\over 2} = \di{1\over \sqrt{2}} \left[ \sin \di{\theta \over 4} -  \cos \di{\theta \over 4} \right] = -\sqrt{\di{{1-s}\over 2}} $$

\newpage

\nin Utilisant ces expressions, on trouve  

$$ A_3 = \di{1\over{2 \sqrt{2 \,r\,(r -s)} }} \left[ \hskip -0.05cm \di{{}\over{}} \sqrt{1+s} \sqrt{1-r} \sqrt{(1-s)(1+r)} - \sqrt{1-s} \sqrt{1+r} \sqrt{(1+s)(1-r)}\right] =0 $$

$$ - A_4 = \di{1\over{2 \sqrt{2 \,r\,(r +s)}}} \left[ \hskip -0.05cm \di{{}\over{}} \sqrt{1+s} \sqrt{1+r} \sqrt{(1-s)(1-r)} - \sqrt{1-s} \sqrt{1-r} \sqrt{(1+s)(1+r)} \right] = 0 $$

\vv \nin Puis 

$$ A_1 = \di{1\over{2 \sqrt{2 \,r\,(r +s)}}} \left[ \sqrt{1+s} \sqrt{1+r} \sqrt{(1+s)(1+r)} - \sqrt{1-s} \sqrt{1-r} \sqrt{(1-s)(1-r)} \right] $$
$$ =  \di{1\over{2 \sqrt{2 \,r\,(r +s)}}} \times 2(r+s) = \di{1\over \sqrt{2}} \sqrt{1 + \di{s\over r}} =  \cos \di{\psi_1 \over 2} , ~~~~(\cos \psi_1 = \di{s\over r} >0 )$$

\nin et 

$$ A_2 = \di{1\over{2 \sqrt{2 \,r\,(r -s)}}} \left[ \sqrt{1+s} \sqrt{1-r} \sqrt{(1+s)(1-r)} -  \sqrt{1-s} \sqrt{1+r} \sqrt{(1-s)(1+r)} \right]$$
$$ = \di{1\over{2 \sqrt{2 \,r\,(r -s)}}} \times 2(s-r) = - \di{1\over \sqrt{2}} \sqrt{1 - \di{s\over r}} =  \sin \di{\psi_1 \over 2}~~~(\sin \psi_1 <0).  $$

\vv \nin En conclusion, le produit matriciel (\ref{L12}) a bien la forme (\ref{L11}) attendue. 

\newpage

\setcounter{page}{143}

\setcounter{chapter}{3}

\setcounter{equation}{0}

\renewcommand{\theequation}{\mbox{4.}\arabic{equation}}

\chapter{Applications}
\section{Amplitudes d'h\'elicit\'e de la diffusion Compton, avec couplage d'h\'elicit\'e dans la voie $s$}

\vv \nin L'exemple typique d'une r\'eaction o\`u, de fa\c{c}on naturelle, on couple en h\'elicit\'e une particule de spin 1/2, lepton ou quark, et une particule vectorielle de masse nulle\footnote{D\'enomm\'ee ci-dessus ``photon r\'eel" pour la simplicit\'e.}, photon pour un lepton, ou photon ou gluon   
pour un quark, ces particules \'etant bien s\^ur toutes deux entrantes ou sortantes, est celui de l'effet Compton $e^- + \gs \rightarrow e^- + \gs$ vers lequel nous revenons maintenant. Le calcul des amplitudes d'h\'elicit\'e de ce processus servira de test des formules \'etablies au chapitre 3 et fournira l'occasion de pr\'esenter certaines astuces de calcul. Pour ne pas perturber le lecteur, nous garderons les m\^emes notations que celles de la figure (3.1). 

\vvv
\begin{figure}[hbt]
\centering
\includegraphics[scale=0.3, width=9cm, height=3cm]{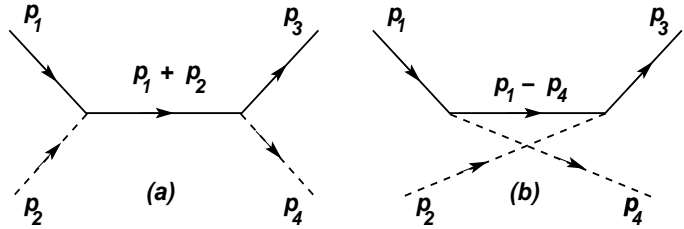}
\vskip 0.25cm

\caption{Diagrammes de Feynman d\'ecrivant l'effet Compton} \label{compton2}
\end{figure}

\vv \nin Au plus bas ordre suivant la constante $\alpha$, l'effet est d\'ecrit par les deux diagrammes de Feynman de la figure (\ref{compton2}), o\`u la ligne en trait plein est une ligne d'\'electron et les lignes en tirets sont des lignes de photons. L'amplitude g\'en\'erique correspondante s'\'ecrit 

$$  {\cal M} = 4 \pi \alpha\, T_c\,,~~~ {\rm avec} $$
\beq T_c = \ov{U_3} \left[ \gs(\epsilon^\star_4) \di{{m + \gs(p_1 + p_2)}\over{m^2 - (p_1 + p_2)^2}} \gs(\epsilon_2) + \gs(\epsilon_2) \di{{m + \gs(p_1 - p_4)}\over{m^2 - (p_1 - p_4)^2}} \gs(\epsilon^\star_4)  \right] U_1 \enq 

\vv \nin o\`u : $U_1$ et $U_3$ sont les spineurs de Dirac respectifs des \'electrons entrant et sortant, $p_1$ et $p_3$ leurs 4-impulsions respectives, $m$ leur masse ; $\epsilon_2$ et $\epsilon_4$ sont les vecteurs de polarisation respectifs des photons entrant et sortant, $p_2$ et $p_4$ leurs 4-impulsions respectives. Nous avons ici $p^2_1 = p^2_3 = m^2$, $p^2_2 = p^2_4 = 0$. Nous poserons $s = (p_1 + p_2)^2 = (p_3 + p_4)^2$, $u = - (p_1 - p_4)^2 = - (p_2 - p_3)^2$. En couplage d'h\'elicit\'e entre 1 et 2 d'une part, 3 et 4 d'autre part, on a 

$$ p_1 \cdot \epsilon_2 = p_2 \cdot \epsilon_2 = 0\, ,~~~p_3 \cdot \epsilon_4 = p_4 \cdot \epsilon_4 =0$$ 

\vv \nin En tenant compte de l'\'equation de Dirac $\left[ m - \gs(p_1) \right] U_1 = 0$ et apr\`es quelques anti-commutations de matrices $\gs$, on obtient  

$$ T_c =  \ov{U_3} \left\llbracket  - \di{{ \gs(\epsilon^\star_4) \gs(p_2) \gs(\epsilon_2) }\over{s - m^2}} + \di{1 \over{u+ m^2}} \left\{ \hskip -0.1cm\di{{}\over{}} \gs(\epsilon^\star_4) \gs(p_4) \gs(\epsilon_2) + 2\, p_1 \cdot \epsilon^\star_4 \,\gs(\epsilon_2)   \right. \right.$$
$$\left. \left.  - 2 \,p_4 \cdot \epsilon_2\, \gs(\epsilon^\star_4) + 2 \,\epsilon_2 \cdot \epsilon^\star_4\, \gs(p_4) \di{{}\over{}} \hskip -0.1cm \right\} \right\rrbracket U_1$$

\vv \nin On voit alors que le calcul n\'ecessite de conna\^itre des expressions telles que $\gs(\epsilon_2) U_1$ et $\gs(p_2) U_1$. Notons $T, X, Y, Z$ la base associ\'ee au r\'ef\'erentiel du centre de masse. En couplage d'h\'elicit\'e dans la voie $s$, on choisit $Z$ de telle sorte que

$$ p_1 = E\,T + p\, Z\, ,~~~p_2 = p\,(T-Z) \,,~~~ \epsilon^{(\lambda_1)}_1 = E^{(\lambda_1)} = - \di{1\over{\sqrt{2}}} \left( \lambda_1 X + i Y\right) \,,~~~\epsilon^{(\lambda_2)}_2 = \,\epsilon^{(-\lambda_2)}_1 $$

\vv \nin et, comme pr\'econis\'e pr\'ec\'edemment, l'axe $Y$ est orthogonal aux 4-impulsions de toutes les particules. Pour l'\'etat final, on a 

$$ p_3 = E T + p Z'\,,~~~p_4 = p\,(T - Z')\,,~~~\epsilon^{(\lambda_3)}_3 = - \di{1\over{\sqrt{2}}} \left( \lambda_3 X' + i Y\right)\,,~~~\epsilon^{(\lambda_4)}_4 = \,\epsilon^{(-\lambda_4)}_3 $$
$${\rm avec}~~~~~~Z' = Z \cos \theta  + X \sin \theta\,,~~~X' = X \cos \theta - Z \sin \theta $$

\vv \nin $\theta$ \'etant, dans ledit r\'ef\'erentiel, l'angle de diffusion de l'\'electron \'emergeant, par rapport \`a la direction de propagation $Z$ de l'\'electron initial ; dans ce m\^eme r\'ef\'erentiel, les deux \'electrons (initial et final) ont la m\^eme \'energie $E$ et le m\^eme module $p$ de quantit\'e de mouvement : 

$$ E = \di{{s + m^2}\over{2 \,\sqrt{s}}}\,,~~~p = \di{{s-m^2}\over{2\, \sqrt{s}}} $$ 

\vv \nin tandis que les deux photons (initial et final) ont la m\^eme \'energie, \'egale \`a $p$ et \'egale aux modules de leurs quantit\'es de mouvement.  

\vv \nin \ding{45} ~ On a 

$$\gs(p_2) \gs(p_1) = p\, (E+p) \left[ 1 - \gs(Z)\,\gs(T) \right] = p \sqrt{s}\, \left[ 1 - \gs(z_1) \gs(t_1) \right] $$

\vv \nin o\`u l'on a tenu compte du fait qu'en couplage d'h\'elicit\'e, $\gs(Z) \gs(T) = \gs(z_1) \gs(t_1)$. Or, d'une part, $\gs(z_1)\gs(t_1) = 2 \gs_5\, S_{z 1}$ et, d'autre part, $\gs(p_1) U_1 = m \,U_1$. Appliquant $\gs(p_2) \gs(p_1)$ \`a ${U_1}_{\sigma_1}$, on en d\'eduit

\beq \fbox{\rule[-0.5cm]{0cm}{1.2cm}~$\gs(p_2)\, {U_1}_{\sigma_1} = \di{{p \sqrt{s}}\over m}  \left[ 1 - (2 \sigma_1) \gs_5 \right] {U_1}_{\sigma_1}  $~} \label{gp2u1} \enq

\vv \nin \ding{45} ~Rappelons que 

$$ S_{x 1} {U_1}_{\sigma_1} = \di{1\over 2} \gs_5\,\gs(X)\, \gs(t_1) {U_1}_{\sigma_1} = \di{1\over 2} \gs_5\,\gs(X)\, {U_1}_{\sigma_1} = \di{1\over 2}  {U_1}_{-\sigma_1} $$
$$ S_{y 1} {U_1}_{\sigma_1} = \di{1\over 2} \gs_5\,\gs(Y)\, \gs(t_1) {U_1}_{\sigma_1} = \di{1\over 2} \gs_5\,\gs(Y)\, {U_1}_{\sigma_1} = i \sigma_1  {U_1}_{-\sigma_1} $$

\vv \nin d'o\`u l'on tire ais\'ement que 

\beq \gs(\epsilon^{(\lambda)}_1)\, {U_1}_{\sigma_1} = \delta_{\lambda, - 2 \sigma_1}\, \sqrt{2}\, (2 \sigma_1)\,    \gs_5\,{U_1}_{-\sigma_1} \label{geps1u1} \enq

\vv \nin soit encore 

\beq \fbox{\rule[-0.5cm]{0cm}{1.2cm}~$\gs(\epsilon^{(\lambda_2)}_2)\, {U_1}_{\sigma_1} = \delta_{\lambda_2,  2 \sigma_1}\, \sqrt{2}\, (2 \sigma_1)\,  \gs_5 \, {U_1}_{-\sigma_1}  $~} \label{geps2u1} \enq

\vv \nin \ding{45}~ Combinant (\ref{gp2u1}) et (\ref{geps2u1}), il vient 

\beq \fbox{\rule[-0.5cm]{0cm}{1.2cm}~$\gs(p_2)\, \gs(\epsilon^{(\lambda_2)}_2)\, {U_1}_{\sigma_1} = - \delta_{\lambda_2,  2 \sigma_1}\, \sqrt{2}\,(2 \sigma_1) \di{{p \sqrt{s}}\over m}  \left[ \gs_5 + (2 \sigma_1)  \right] {U_1}_{-\sigma_1}  $~} \label{gp2u1} \enq

\vv \nin Les m\^emes types de relations valent \'egalement pour l'\'etat final : 

\beq \fbox{\rule[-0.6cm]{0cm}{2.7cm}~$ \begin{array}{c} \gs(p_4)\, {U_3}_{\sigma_3} = \di{{p \sqrt{s}}\over m}  \left[ 1 - (2 \sigma_3) \gs_5 \right] {U_3}_{\sigma_3} \\~\\  
\gs(\epsilon^{(\lambda_4)}_4)\, {U_3}_{\sigma_3} = \delta_{\lambda_4,  2 \sigma_4}\, \sqrt{2}\, (2 \sigma_3)\,  \gs_5 \, {U_3}_{-\sigma_3} \\~\\
\gs(p_4)\, \gs(\epsilon^{(\lambda_4)}_4)\, {U_3}_{\sigma_3} = - \delta_{\lambda_4,  2 \sigma_3}\, \sqrt{2}\,(2 \sigma_3) \di{{p \sqrt{s}}\over m}  \left[ \gs_5 + (2 \sigma_3)  \right] {U_3}_{-\sigma_3} \\~\\
\end{array} 
$~}  \enq

\vvv
\vv \nin \ding{45}~ Calculons alors\footnote{On prendra garde au fait que la matrice $\gs_5$ anti-commute avec 
la matrice $\gs_0$ intervenant dans la d\'efinition de $\ov{U}$.} 

$$ A = \ov{U_3}_{\sigma_3}\, \gs(\epsilon^{(\lambda_4) \star}_4)\,\gs(p_2) \,\gs(\epsilon^{(\lambda_2)}) \, {U_1}_{\sigma_1}  = 2\, \delta_{\lambda_2,  2 \sigma_1}\, \delta_{\lambda_4,  2 \sigma_3}\, \lambda_2\,\lambda_4 \di{{p \sqrt{s}}\over m}\, \ov{U_3}_{- \sigma_3} \left[\, 1 + (2 \sigma_1)\, \gs_5\, \right] {U_1}_{-\sigma_1} $$

\vv \nin Les spineurs \'etant ici normalis\'es selon $\ov{U} \,U = 2\,m$, les formules (3.67) et (3.68) donnent  

$$ \ov{U_3}_{\sigma_3}\,{U_1}_{\sigma_1} = 2\,m\, \left\{ \delta_{\sigma_3, \sigma_1}\, \cos \di{\theta \over 2} - (2 \sigma_1)\, \delta_{\sigma_3, - \sigma_1} \cosh \chi_1\, \sin \di{\theta \over 2}  \right\} $$
$$  \ov{U_3}_{\sigma_3}\,\gs_5\,{U_1}_{\sigma_1} = - 2\,m\, \delta_{\sigma_3, - \sigma_1}\, \sinh \chi_1\, \sin \di{\theta \over 2} $$
$${\rm avec}~~~~\cosh \chi_1 = \di{E\over m}\,,~~~\sinh \chi_1 = \di{p\over m} $$ 

\vv \nin Les appliquant \`a $A$, on trouve ($|\lambda_2| = |\lambda_4| =1$)

\beq \begin{array}{c} A = 2\,(s - m^2)\, \lambda_2\,\lambda_4\, \delta_{\lambda_2,  2 \sigma_1}\, \delta_{\lambda_4,  2 \sigma_3}\, \left\{ \delta_{\sigma_3, \sigma_1} \cos \di{\theta\over 2} \right. \\
\left. + (2 \sigma_1)\,\delta_{\sigma_3, - \sigma_1}\, \left(\cosh \chi_1 - \sinh \chi_1 \right) \sin \di{\theta \over 2}  \,\right\}  = 2\,(s - m^2)\,\delta_{\lambda_2,  2 \sigma_1}\, \delta_{\lambda_4,  2 \sigma_3}\, \left\{ \delta_{\sigma_3, \sigma_1} \cos \di{\theta\over 2}\right. \\
\left. - (2 \sigma_1)\,\delta_{\sigma_3, - \sigma_1}\, \left(\cosh \chi_1 - \sinh \chi_1 \right) \sin \di{\theta \over 2}  \,\right\} 
 \end{array} \enq

\vv \nin On trouve de m\^eme que  

\beq  B= \ov{U_3}_{\sigma_3}\, \gs(\epsilon^{(\lambda_4) \star}_4) \gs(p_4) \gs(\epsilon^{(\lambda_2)}_2) \,{U_1}_{\sigma_1} \equiv A \enq

\vv \nin Remarquons que cette \'egalit\'e n'est pas fortuite. D'une part, les masses v\'erifiant les \'egalit\'es $m_1 = m_3 =m$ et $m_2 = m_4\,(=0)$, la t\'etrade associ\'ee \`a $p_3$ se d\'eduit de celle associ\'ee \`a $p_1$ par une simple rotation $R_Y(\theta)$ d'angle $\theta$ autour de $Y$ (Eq. 3.35), et l'on a

$$ {U_3}_{\sigma_3} = R_Y(\theta)\, {U_1}_{\sigma_3}\, ,~~~R_Y(\theta)\,\gs(p_1)\,R_Y(\theta)^{-1} = \gs(p_3) \, ,~~~R_Y(\theta)\,\gs(p_2)\,R_Y(\theta)^{-1} = \gs(p_4) \, ,~~~ $$
$$R_Y(\theta)\,\gs(\epsilon^{\lambda_2)}_2 )\,R_Y(\theta)^{-1} = \gs(\epsilon^{(\lambda_2)}_4) $$

\vv \nin On peut donc \'ecrire\footnote{On rappelle que $\gs^\dagger_\mu = \gs_0\, \gs_\mu\,\gs_0$, voir ITL, \S 7.3.4, Eq. 7.80.}

$$ A(\sigma_1, \sigma_3 ; \lambda_2, \lambda_4) = \ov{U_1}_{\sigma_3} R_Y(\theta)^{-1} \gs(\epsilon^{(\lambda_4) \star}_4)\,R_Y(\theta)\, R_Y(\theta)^{-1} \,\gs(p_2) \,R_Y(\theta)\,\times $$
$$ R_Y(\theta) \,\gs(\epsilon^{(\lambda_2)}_2) \,R(\theta)^{-1}\, {U_3}_{\sigma_1} =  \ov{U_1}_{\sigma_3} \gs(\epsilon^{(\lambda_4) \star}_2)\,\gs(p_4)\,\gs(\epsilon^{(\lambda_2)}_4) \, {U_3}_{\sigma_1} $$
$$ = \left[ \ov{U_3}_{\sigma_1} \gs(\epsilon^{(\lambda_2) \star}_4) \gs(p_4) \gs(\epsilon^{(\lambda_4)}_2) \,{U_1}_{\sigma_3} \right]^\star = B^\star(\sigma_1, \sigma_3 ; \lambda_4, \lambda_2) $$
$${\rm soit}~~~~~~~B(\sigma_1, \sigma_3 ; \lambda_2, \lambda_4) = A^\star(\sigma_3, \sigma_1; \lambda_4, \lambda_2) $$

\vv \nin D'autre part\footnote{Voir ITL, \S 7.3.4 et \S 7.4.3.}, en prenant le complexe conjugu\'e de $A$ et en utilisant la matrice $U_c = i \gs^2\, \gs_5$ telle que 

$$ U_c\, U_\sigma = -(2 \sigma)\, U^\star_{-\sigma} ~~~{\rm et}~~~U_c\, \gs_\mu\,U^{-1}_c = \gs^\star_\mu$$ 

\vv \nin on obtient, compte tenu de $\epsilon^{(\lambda \star} = - \epsilon^{(-\lambda)}$,  

$$A^\star(\sigma_1, \sigma_3 ; \lambda_2, \lambda_4) = {}^t{U_3}_{\sigma_3}\,\gs_0\, \gs^\star(\epsilon^{(\lambda_4)}_4)\,\gs^\star(p_2) \,\gs^\star(\epsilon^{(\lambda_2) \star}) \, {U^\star_1}_{\sigma_1} $$
$$= (2 \sigma_3)\,(2 \sigma_1) \,\ov{U_3}_{-\sigma_3}\, \gs(\epsilon^{(-\lambda_4) \star}_4)\,\gs(p_2) \,\gs(\epsilon^{(-\lambda_2)}) \, {U_1}_{-\sigma_1} =(2\sigma_1)(2 \sigma_3)\, A(-\sigma_1, - \sigma_3 ; -\lambda_2, -\lambda_4) $$

\vv \nin On en d\'eduit 

$$ B(\sigma_1, \sigma_3; \lambda_2, \lambda_4) = (2 \sigma_1)(2 \sigma_3)\,A(- \sigma_3, -\sigma_1; -\lambda_4, -\lambda_2) $$

\vv \nin d'o\`u l'\'egalit\'e $B=A$, compte tenu de la forme de $A$. 

\vv \nin \ding{45}~ Puis

$$ 2\, p_1\cdot \epsilon^\star_4 = \sqrt{2}\, \lambda_4\,p\, \sin \theta\,,~~~2\, p_4\cdot \epsilon_2 = \sqrt{2} \,\lambda_2\, p\,\sin \theta \,,~~~\epsilon_2 \cdot \epsilon^\star_4 = - \di{1\over 2} \left( 1 + \lambda_2 \lambda_4 \cos \theta \right) $$
$$ \ov{U_3} \, \gs(\epsilon_2)\,U_1 = \delta_{\lambda_2, 2 \sigma_1}\, \sqrt{2}\, \left(- 2m\, \delta_{\sigma_3, \sigma_1}\, \sinh \chi_1\, \sin \di{\theta \over 2} \,\right) $$
$$ \ov{U_3} \, \gs(\epsilon^\star_4)\,U_1 = \delta_{\lambda_4, 2 \sigma_3}\, \sqrt{2}\, \left( 2m\, \delta_{\sigma_3, \sigma_1}\, \sinh \chi_1\, \sin \di{\theta \over 2} \,\right) $$
$$ \ov{U_3} \, \gs(p_4)\,U_1 = (s-m^2)\,\left\{ \delta_{\sigma_3, \sigma_1}\, \cos \di{\theta \over 2} - (2 \sigma_1) \delta_{\sigma_3, - \sigma_1}\, (\cosh \chi_1 -\sinh \chi_1 )\, \sin \di{\theta \over 2} \right\} $$

\vv \nin \ding{45} ~ Collectant tous ces r\'esultats, il vient 

$$T_c = T_c(\sigma_1, \sigma_3 ; \lambda_2, \lambda_4) = 2 \,\lambda_2\, \lambda_4\, \delta_{\lambda_2, 2 \sigma_1}\,\delta_{\lambda_4, 2 \sigma_3}\, \left[ -1 + \di{{s-m^2}\over{u+m^2}} \right] \left[ \delta_{\sigma_3, \sigma_1} \cos \di{\theta\over 2} \right.$$ 
$$ \left. + (2 \sigma_1) \delta_{\sigma_3, - \sigma_1} \di{m \over \sqrt{s}} \sin \di{\theta\over 2} \right] 
- \di{{(s-m^2)^2}\over{s(u+m^2)}} \sin \theta \sin \di{\theta \over 2} \lambda_2 \lambda_4 \left[ \delta_{\lambda_2, 2 \sigma_1} + \delta_{\lambda_4, 2 \sigma_3} \right] \delta_{\sigma_3, \sigma_1} $$
$$- (1+ \lambda_2 \lambda_4 \cos \theta) \di{{s-m^2}\over{u+m^2}} \left[ \delta_{\sigma_3, \sigma_1} \cos \di{\theta\over 2} - (2 \sigma_1) \delta_{\sigma_3, - \sigma_1} \di{m \over \sqrt{s}} \sin \di{\theta\over 2} \right] $$

\vv \nin soit, en rempla\c{c}ant le produit $\lambda_2 \lambda_4$ par 1 lorsqu'il est en facteur de $\delta_{\lambda_2, 2 \sigma_1}\,\delta_{\lambda_4, 2 \sigma_3}\,\delta_{\sigma_3, \sigma_1}$ et par -1 lorsqu'il est en facteur de $\delta_{\lambda_2, 2 \sigma_1}\,\delta_{\lambda_4, 2 \sigma_3}\,\delta_{\sigma_3, -\sigma_1}$

$$T_c = \delta_{\sigma_3, \sigma_1} \left\llbracket  \,2 \delta_{\lambda_2, 2 \sigma_1} \delta_{\lambda_4, 2 \sigma_3} \cos \di{\theta \over 2} \, \left[-1 + \di{{s-m^2}\over{u+m^2}} \right]  \right.$$
$$\left. - \di{{(s-m^2)^2}\over{s(u+m^2)}} \sin \theta \sin \di{\theta \over 2} \lambda_2 \lambda_4 \left[ \delta_{\lambda_2, 2 \sigma_1} + \delta_{\lambda_4, 2 \sigma_3} \right] - (1+ \lambda_2 \lambda_4 \cos \theta) \di{{s-m^2}\over{u+m^2}} \, \cos \di{\theta\over 2} \,\right\rrbracket $$
$$ + \delta_{\sigma_3, - \sigma_1} (2 \sigma_1) \di{m \over \sqrt{s}} \, \sin \di{\theta \over 2} \left\llbracket       (1 + \lambda_2 \lambda_4 \cos \theta) \di{{s-m^2}\over{u+m^2}} -2 \delta_{\lambda_2, 2\sigma_1}\delta_{\lambda_4, 2 \sigma_3} \left[-1 + \di{{s-m^2}\over{u+m^2}} \,\right] \right\rrbracket $$

\vv \nin Simplifions encore cette expression en explicitant $u$ en fonction de $\theta$ : 

$$ u +m^2 = 2 p_1\cdot p_4 =  2\, p\, \left[ E + p \cos \theta \right] = (s - m^2) \left[1 - \beta' \sin^2 \di{\theta \over 2} \right] ~~~{\rm o\grave{u}} ~~~\beta' = \di{{s-m^2}\over s} $$ 

\vv \nin On prendra garde \`a ne pas confondre $\beta'$ avec $\beta = \di{p\over E} = 
\di{{s-m^2}\over{s+m^2}}$. Tous calculs effectu\'es, on obtient : 

\beq \fbox{\fbox{\rule[-0.9cm]{0cm}{3.5cm}~$ \begin{array}{c}  T_c(\sigma_1, \sigma_3; \lambda_2, \lambda_4) = \delta_{\sigma_3, \sigma_1}\, \di{{\cos \di{\theta\over 2}}\over{ 1- \beta' \sin^2 \di{\theta \over 2}}} \left\llbracket \,2 \beta' \sin^2\di{\theta \over 2} \,\delta_{\lambda_2, 2 \sigma_1} \delta_{\lambda_4, 2 \sigma_3}  \right. \\
 \left.   - (1 + \lambda_2 \lambda_4 \cos \theta ) - 2 \beta'\,\sin^2\di{\theta \over 2} \,  \lambda_2 \lambda_4\,\left[ \delta_{\lambda_2, 2 \sigma_1} + \delta_{\lambda_4, 2 \sigma_3} \right] \,  \right\rrbracket   \\
+ (2 \sigma_1)\,\delta_{\sigma_3, -\sigma_1}\, \di{m \over \sqrt{s}}\,\di{{\sin \di{\theta\over 2}}\over{ 1- \beta' \sin^2 \di{\theta \over 2}}} \left\llbracket \di{{}\over{}}  
(1 + \lambda_2 \lambda_4 \cos \theta )\right. 
 \left.  \di{{}\over{}} - 2 \beta' \sin^2\di{\theta \over 2} \, \delta_{\lambda_2, 2 \sigma_1} \delta_{\lambda_4, 2 \sigma_3}    \,  \right\rrbracket \\~\\ \end{array}
 $~}}  \label{AmpCompt-1} \enq

\vv \nin On remarque imm\'ediatement que les amplitudes avec changement d'h\'elicit\'e de l'\'electron sont proportionnelles \`a $m/\sqrt{s}$. Il s'ensuit qu'\`a tr\`es haute \'energie, soit plus pr\'ecis\'ement pour $\sqrt{s} \gg m$, ces amplitudes ont une tr\`es faible contribution : {\it \`a tr\`es haute \'energie, l'h\'elicit\'e de l'\'electron se conserve.} En fait, il s'agit l\`a d'un r\'esultat g\'en\'eral pour les processus de l'Electrodynamique Quantique ou ceux de la Chromodynamique Quantique, pour lesquels les masses peuvent \^etre n\'eglig\'ees. En effet, dans ces conditions, les amplitudes font intervenir des produits de nombres impairs de matrices $\gs$, se d\'eveloppant uniquement sur les matrices $\gs_\mu$ et $\gs_\mu \, \gs_5$. Les spineurs \'etant normalis\'es selon $\ov{U} \,U = 2\,m$, les formules (3.69) et (3.70) montrent qu'en faisant tendre $m$ vers z\'ero, seuls subsistent des termes conservant l'h\'elicit\'e : 

\beq \ov{U_3}_{\sigma'}\, \gs_\mu\, {U_1}_{\sigma} \approx (2 \sigma)\, \ov{U_3}_{ \sigma'}\,\gs_\mu\,\gs_5\,U_{1 \sigma} \approx \sqrt{s} \, \delta_{\sigma', \sigma} \left[ \, (T+ Z)_\mu\, \cos \di{\theta \over 2} +(X +  i (2 \sigma) Y )_\mu \sin \di{\theta \over 2} \right] \enq

\vv \nin Les tableaux ci-apr\`es donnent les expressions explicites des 16 amplitudes d'h\'elicit\'e de l'effet Compton, tir\'ees de (\ref{AmpCompt-1}), dans le cas $m \neq 0$ et dans le cas $m=0$.  On remarque aussi que  les amplitudes avec changement d'h\'elicit\'e du photon sont d\'efavoris\'ees \`a haute \'energie par au moins un facteur $m^2/s$ : {\it \`a tr\`es haute \'energie, l'h\'elicit\'e du photon est donc aussi conserv\'ee}. Cela implique que dans ce domaine et pour ce processus, l'h\'elicit\'e {\it totale} est conserv\'ee\footnote{A noter que les h\'elicit\'es de l'\'etat initial et de l'\'etat final sont respectivement $\sigma_1 - \lambda_2$ et $\sigma_3 - \lambda_4$.}.

$$\fbox{\fbox{\rule[-1cm]{0cm}{5.5cm}~$\begin{array}{c}  

T_c(\uw,\uw\, ;+,+ ) = T_c(\dw,\dw\, ;-,- ) = - \di{{ 2\cos \di{\theta\over 2}}\over{ 1- \beta' \sin^2 \di{\theta \over 2}}} \left[ 1 - \di{m^2 \over s} \sin^2 \di{\theta \over 2}\, \right] \\

T_c(\uw,\uw\, ;-,- ) = T_c(\dw,\dw\, ;+,+ ) =  - \di{{ 2\cos^3 \di{\theta\over 2}}\over{ 1- \beta' \sin^2 \di{\theta \over 2}}} \\

T_c(\uw,\uw\, ;+,- ) = T_c(\uw,\uw\, ;-,+ ) = T_c(\dw,\dw\, ;+,- ) = T_c(\dw,\dw\, ;-,+ ) = - \di{m^2 \over s}\, \di{{ 2\cos \di{\theta\over 2}\sin^2 \di{\theta\over 2}}\over{ 1- \beta' \sin^2 \di{\theta \over 2}}} \\

 T_c(\uw,\dw\, ;+,+ ) = T_c(\uw,\dw\, ;-,- ) = - T_c(\dw,\uw\, ;+,+ ) = -T_c(\dw,\uw\, ;-,- ) =\di{m \over \sqrt{s}}\, \di{{ 2\cos^2 \di{\theta\over 2}\sin \di{\theta\over 2}}\over{ 1- \beta' \sin^2 \di{\theta \over 2}}} \\

T_c(\uw,\dw\, ;+,- ) = T_c(\uw,\dw\, ;-,+ ) =  -T_c(\dw,\uw\, ;+,- ) = -T_c(\dw,\uw\, ;-,+ ) = \di{m^3 \over s^{3/2}}\, \di{{ 2\sin^3 \di{\theta\over 2}}\over{ 1- \beta' \sin^2 \di{\theta \over 2}}} \\~\\

\end{array}
  $~}}  $$

\vv \vv
\centerline{\bf Tableau I - Amplitudes d'h\'elicit\'e de la diffusion Compton}

\vskip 0.75cm 

$$\fbox{\fbox{\rule[-0.4cm]{0cm}{2cm}~$\begin{array}{c}  

T_c(\uw,\uw\, ;+,+ ) = T_c(\dw,\dw\, ;-,- ) = - \di{ 2\over{ \cos \di{\theta \over 2}}} \\

T_c(\uw,\uw\, ;-,- ) = T_c(\dw,\dw\, ;+,+ ) =  - 2\cos \di{\theta\over 2} \\~\\

\end{array}
  $~}}  $$

\vv \vv
\centerline{\bf Tableau II - Amplitudes d'h\'elicit\'e non nulles de la diffusion Compton pour $m=0$}

\newpage
\section{Amplitudes d'h\'elicit\'e de la diffusion Compton avec couplage dans la voie $t$}

\vv \nin Il est int\'eressant de comparer le calcul pr\'ec\'edent avec celui o\`u l'on effectue un couplage d'h\'elicit\'e de voie $t$ entre les particules, bien que ce dernier couplage soit plut\^ot inhabituel pour ce type de processus. Bien entendu, les amplitudes obtenues avec le couplage en voie $t$ ne sont pas directement comparables \`a celles du couplage en voie $s$, car les h\'elicit\'es des particules \'etant d\'efinies par rapport \`a d'autres axes sont diff\'erentes, sauf peut-\^etre pour ce qui concerne les photons. Pr\'ecisons ce dernier point. Comme au chapitre 3, appelons ``vertex de gauche" l'association (1,3) des deux \'electrons et ``vertex de droite" celle, (2,4), des deux photons. On a 

\nin $~~~ X_{d \mu} = \epsilon_{\mu \nu \rho \omega} \,T^\nu_d Y^\rho Z^\omega_d = \di{2\over \Ldd} \epsilon_{\mu \nu \rho \omega}\, p^\nu_2\, Y^\rho p^\omega_4$,~~~avec ici~~~~$p_2 = p ( T-Z),~~p_4 = p (T- Z')$,~~et

\nin $~~~\Ldd = t = 2 p^2 (1 - \cos \theta)$,~~d'o\`u~~ $ X_{d \mu} = \di{1\over{1 - \cos \theta}} \,\epsilon_{\mu \nu \rho \omega} \left[ - T^\nu Y^\rho Z^{\prime \omega} - Z^\nu Y^\rho T^\omega + Z^\nu Y^\rho Z^{\prime \omega} \right] $

\nin $= \di{1\over{1 - \cos \theta}} \left[ -X'_\mu + X_\mu - T_\mu \sin \theta \right]$, ~~soit~~~$ X_d = X - (T-Z) \cot \di{\theta \over 2} = - X' - (T-Z') \cot \di{\theta \over 2} $

\vvv \nin Il s'ensuit que 

\beq \epsilon^{(\lambda)}_d = \epsilon^{(\lambda)} + \di{\lambda \over \sqrt{2}} (T-Z) \cot \di{\theta \over 2} =\epsilon^{\prime (-\lambda)} + \di{\lambda \over \sqrt{2}} (T-Z') \cot \di{\theta \over 2}\enq

\vv \nin On voit ainsi que le vecteur de polarisation d'un photon en voie $t$ ne diff\`ere de celui en voie $s$ que par un terme proportionnel \`a la 4-impulsion de ce photon. Or, ce terme ne contribue pas, du fait de l'invariance de jauge de l'amplitude tensorielle g\'en\'erique ($p^\mu_2  \,T_{\mu \nu}= 0$ ou $T_{\mu \nu} \,p^\nu_4 =0$). Les h\'elicit\'es $\lambda'_2$ et $\lambda'_4$ des photons dans la voie $t$ sont ainsi reli\'ees \`a celles, $\lambda_2$ et $\lambda_4$ de la voie $s$ par $\lambda'_2 = - \lambda_2$, $\lambda'_4 = \lambda_4$.    

\vv  \nin Un avantage apparent du couplage en voie $t$ est que les photons y ont les m\^emes vecteurs de polarisation, ce qui peut apporter des simplifications. Cependant, une difficult\'e appara\^it concernant le cas limite $\theta =0$. En effet, les photons \'etant sans masse, on a $t = 2p^2 (1 - \cos \theta)$, soit $t = t_{\rm min} =0$ pour $\theta=0$. Pour cette valeur du transfert, la transformation de Lorentz permettant de passer du vertex de gauche au vertex de droite perd toute signification. Pour $t \neq 0$, le param\`etre $\Theta$ de cette transformation est tel que 

\beq  \cosh \Theta = \di{{2(s-m^2) -t}\over{\sqrt{t(t+4m^2)}}},~~~\sinh \Theta = 2\,\sqrt{ \di{{ \left[(s-m^2)^2 - st \right]}\over{t(t+4m^2)}}}  \label{hyperb-1} \enq

\vv \nin et, selon ces formules, tend vers l'infini lorsqu'on fait tendre $t$ tend vers z\'ero.  Or, comme le montre le calcul avec le couplage en voie $s$, aucune divergence ne doit appara\^itre dans les amplitudes pour $\theta =0$, m\^eme si $m=0$. On veillera donc \`a ce que les fonctions hyperboliques (\ref{hyperb-1}), qui sont utilis\'ees naturellement dans le couplage en voie $t$, n'induisent aucune divergence des amplitudes lorsque $t$ tend vers z\'ero, attention qui peut en l'occurrence constituer un fil conducteur du calcul : comme $\sqrt{t}\, \cosh \Theta$ et $\sqrt{t}\, \sinh \Theta \rightarrow (s-m^2)/m$, ces fonctions devront  appara\^itre au moins multipli\'ees par un facteur $\sqrt{t}$, et \'eventuellement par un facteur $m$ suppl\'ementaire, car, comme le montre le tableau II, les amplitudes doivent \'egalement rester finies pour $m \rightarrow 0$, tant que $\theta \neq \pi$. Ecrivons l'amplitude g\'en\'erique sous la forme

$$ T_c = \ov{U_3} \left\llbracket \di{1\over{m^2 - s }} \left[ \hskip -0.1cm \di{{}\over{}} (2 p_1 \cdot \epsilon_2) \gs(\epsilon^\star_4) +  \gs(\epsilon^\star_4) \gs(p_2)  \gs(\epsilon_2) \right] \right. $$
\beq \left. + \di{1\over{u+m^2}} \left[  \hskip -0.1cm \di{{}\over{}} (2 p_3 \cdot \epsilon_2) \gs(\epsilon^\star_4)  - \gs(\epsilon_2) \gs(p_2) \gs(\epsilon^\star_4)  \right] \right\rrbracket U_1 
\label{Tct1} \enq 

\nin avec cette fois 

$$ \epsilon_2 = \epsilon^{(\lambda_2)}_d = - \di{1\over \sqrt{2}} \left[ \lambda_2 X_d + i Y \right],~~\epsilon_4 = \epsilon^{(\lambda_4)}_d = - \di{1\over \sqrt{2}} \left[ \lambda_4 X_d + i Y \right],~~~(\lambda_{2, 4} = \pm1) $$
$$ T_d = T_g \cosh \Theta  - X_g \sinh \Theta ,~~X_d = X_g \cosh \Theta - T_g \sinh \Theta $$
$$\ov{U_3}_{ \sigma_3}\,\gs_\mu\,U_{1 \sigma_1} = 2\, m\,\delta_{\sigma_3, \sigma_1} \, T_{g \mu}
+ 2\,m\,(2 \sigma_1)\, \delta_{\sigma_3, - \sigma_1}\,\sinh \xi (X_g + i(2 \sigma_1) Y)_\mu  $$
$$\ov{U_3}_{ \sigma_3}\,\gs_\mu\,\gs_5\,U_{1 \sigma_1} = 2\,m\, \delta_{\sigma_3, \sigma_1}\, (2 \sigma_1) \,Z_{g \mu}  
+ 2\,m\, \delta_{\sigma_3, - \sigma_1}\, \cosh \xi\,(X_g+ i (2\sigma_1) Y)_\mu  $$
$$ \cosh \xi = \di{{\sqrt{t+ 4 m^2}}\over {2m}},~~~\sinh \xi = \di{ \sqrt{t}\over{2m}} $$
$$ T_d = \di{{p_2+ p_4}\over \sqrt{t}},~~Z_d = \di{{p_4- p_2}\over \sqrt{t}},~~p_2 = \di{\sqrt{t}\over 2} \left[T_d - Z_d\right] = p \left[T-Z\right], $$
$$ p_4 =\di{\sqrt{t}\over 2} \left[T_d + Z_d\right]  = p \left[T - Z'\right] $$
\vv
$$ \hskip -0.4cm{\rm On ~a} \hskip 0.7cm\gs(\epsilon^\star_4) \gs(p_2) \gs(\epsilon_2) =- \gs(p_2) \gs(\epsilon^\star_4)  \gs(\epsilon_2) = - \gs(p_2) \di{1\over 2}\left[ - 1 - \lambda_2 \lambda_4 - i (\lambda_2 + \lambda_4) \gs(X_d) \gs(Y) \right]  $$
$$ = \delta_{\lambda_2, \lambda_4} \gs(p_2) \left[ 1 + i \lambda_2 \gs(X_d) \gs(Y) \right]$$
$$  = \delta_{\lambda_2, \lambda_4}\gs(p_2) \left[ 1- \lambda_2 \gs_5 \right]~~~{\rm car}~~~ \gs(t-z) \gs_5 = i \gs(t-z) \gs(x) \gs(y) $$

$$\hskip -4.45cm {\rm De~m\hat{e}me}~\hskip 2.6cm\gs(\epsilon_2) \gs(p_2) \gs(\epsilon^\star_4) = \delta_{\lambda_2, \lambda_4}\gs(p_2) \left[ 1+ \lambda_2 \gs_5 \right]$$
\vv 
$$\hskip -4.5cm {\rm Puis} \hskip 4.cm p_1 \cdot \epsilon_2 = p_3 \cdot \epsilon_2 = m\,\di{\lambda_2 \over \sqrt{2}} \cosh \xi \sinh \Theta $$

\vv \nin Dans un premier temps, on obtient donc 

$$ T_c = m\,\sqrt{2} \,\lambda_2  \cosh \xi \sinh \Theta \,\left[\di{1\over{u+m^2}} - \di{1\over{s-m^2}} \right] \ov{U_3}\, \gs(\epsilon^\star_4)\, U_1 $$
\beq - \delta_{\lambda_2, \lambda_4} \left\llbracket  \di{1 \over{s-m^2}}  \ov{U_3}\, \gs(p_2) \left[ 1- \lambda_2 \gs_5 \right] U_1 + \di{1 \over{u+m^2}}  \ov{U_3}\, \gs(p_2) \left[ 1+ \lambda_2 \gs_5 \right] U_1 \right\rrbracket \label{Tct2} \enq

\vv \nin Comme 

$$ T_g \cdot \epsilon^\star_4 = - \di{\lambda_4 \over \sqrt{2}} \,T_g \cdot X_d  =  \di{\lambda_4 \over \sqrt{2}} \sinh \Theta $$ 
$$ (X_g + i(2 \sigma_1) Y) \cdot \epsilon^\star_4 = - \di{1\over \sqrt{2}} \left[ \lambda_4 X_d \cdot X_g - (2 \sigma_1) \right] = \di{1\over \sqrt{2}} \left[ \lambda_4 \cosh \Theta + (2 \sigma_1)\right] ,~~~{\rm il~vient}$$

$$ \ov{U_3}\, \gs(\epsilon^\star_4)\, U_1 = 2\,m\, \delta_{\sigma_3, \sigma_1} \, T_g \cdot \epsilon^\star_4 
+ 2\,m\, (2 \sigma_1)\, \delta_{\sigma_3, - \sigma_1}\,\sinh \xi (X_g + i(2 \sigma_1) Y) \cdot \epsilon^\star_4  $$
$$= m \sqrt{2} \, \delta_{\sigma_3, \sigma_1} \,\lambda_4\, \sinh \Theta + m \sqrt{2} \, \delta_{\sigma_3, - \sigma_1}\,\sinh \xi \left[ 1 + \lambda_4 (2 \sigma_1) \cosh \Theta \right] $$

\nin Puis 

$$\ov{U_3}_{ \sigma_3}\,\gs(p_2)\,U_{1 \sigma_1} = 2\, m\,\delta_{\sigma_3, \sigma_1} \, T_g \cdot p_2 
+ 2\,m\,(2 \sigma_1)\, \delta_{\sigma_3, - \sigma_1}\,\sinh \xi (X_g + i(2 \sigma_1) Y)\cdot p_2   $$
$$ = m\,\sqrt{t}\,\delta_{\sigma_3, \sigma_1} \,\cosh \Theta + m\,\sqrt{t}\,\delta_{\sigma_3, -\sigma_1} \,(2 \sigma_1)\,\sinh \xi \, \sinh \Theta$$

$$\ov{U_3}_{ \sigma_3}\,\gs(p_2)\,\gs_5\,U_{1 \sigma_1} =2\, m\, \delta_{\sigma_3, \sigma_1}\, (2 \sigma_1) \,Z_g \cdot p_2   
+ 2\,m\, \delta_{\sigma_3, - \sigma_1}\, \cosh \xi\,(X_g+ i (2\sigma_1) Y)\cdot p_2   $$
$$= m\, \sqrt{t}\, \delta_{\sigma_3, \sigma_1}\, (2 \sigma_1)   
+ m\, \sqrt{t}\,\delta_{\sigma_3, - \sigma_1}\, \cosh \xi\,\sinh \Theta   $$

$$\ov{U_3}_{ \sigma_3}\,\gs(p_2)\,\left[ 1 \pm \lambda_2 \gs_5\right]\,U_{1 \sigma_1} = m\, \sqrt{t} \left\llbracket \hskip -0.05cm \di{{}\over{}}  \delta_{\sigma_3, \sigma_1} \left[ \cosh \Theta \pm \lambda_2 (2 \sigma_1) \right]    \right. $$
$$ \left. \di{{}\over{}} + \,\delta_{\sigma_3, - \sigma_1} \sinh \Theta \left[ (2 \sigma_1) \sinh \xi \pm \lambda_2 \cosh \xi \right]\right\rrbracket$$

\vv \nin L'amplitude (\ref{Tct2}) peut \^etre r\'ecrite sous la forme 

$$ T_c = \delta_{\sigma_3, \sigma_1}\,T_{+} + \delta_{\sigma_3, - \sigma_1}\, T_{-} ~~{\rm avec} $$
$$ T_{+} = 2 \,m^2 \lambda_2 \lambda_4 \cosh \xi \sinh^2 \Theta \left[ \di{1\over{u+m^2}} - \di{1\over{s-m^2}} \right]  - \delta_{\lambda_2, \lambda_4} m \sqrt{t} \cosh \Theta \left[ \di{1\over{s-m^2}} + \di{1\over{u+m^2}} \right] $$
\beq - \delta_{\lambda_2, \lambda_4} m \sqrt{t}\,\lambda_2 (2 \sigma_1) \left[ \di{1\over{u+m^2}} - \di{1\over{s-m^2}} \right], \enq
$$ T_{-} = 2\,m^2 \lambda_2 \cosh \xi \sinh \xi \, \sinh \Theta\,\left[1 + \lambda_4 (2 \sigma_1) \cosh \Theta \right] \left[ \di{1\over{u+m^2}} - \di{1\over{s-m^2}} \right] $$
$$- \delta_{\lambda_2, \lambda_4} m \sqrt{t}\,\sinh \Theta \,\left[ \di{{(2\sigma_1) \sinh \xi - \lambda_2 \cosh \xi }\over{s-m^2}}   +  \di{{(2\sigma_1) \sinh \xi + \lambda_2 \cosh \xi }\over{u+ m^2}} \right] $$

\vv \nin De la relation $s = t + u + 2 m^2$, on tire 

$$s-m^2 -(u+m^2) =t,$$
$$ s-m^2 + u+m^2 = s+u = 2(s-m^2) - t  = \sqrt{t(t+4m^2)} \,\cosh \Theta $$

\vv \nin Explicitons alors $T_{+}$ et $T_{-}$ : 

$$ T_{+} = \di{{m t}\over{ (s-m^2)(u+m^2)}} \left[ \hskip -0.05cm \di{{}\over{}}  \sqrt{t+4m^2}\left[ \lambda_2 \lambda_4  \sinh^2 \Theta   - \,\delta_{\lambda_2, \lambda_4}  \cosh^2 \Theta \right] \right. $$
$$\left.\di{{}\over{}}  -  \,\delta_{\lambda_2, \lambda_4}  \sqrt{t} \,\lambda_2 (2\sigma_1) \,\right]$$

$$ \hskip -3.3cm {\rm Or,}~~~~\lambda_2 \lambda_4  \sinh^2 \Theta   - \delta_{\lambda_2, \lambda_4}  \cosh^2 \Theta = - \delta_{\lambda_2, \lambda_4} +\left[ \lambda_2 \lambda_4 - \delta_{\lambda_2, \lambda_4}  \right] \sinh^2 \Theta$$
$$  = - \delta_{\lambda_2, \lambda_4} - \delta_{\lambda_2, - \lambda_4} \sinh^2 \Theta,$$ 
$$\hskip 1.6cm{\rm car}~~~\lambda_2 \lambda_4 - \delta_{\lambda_2, \lambda_4} =-\delta_{\lambda_2, - \lambda_4} , ~~{\rm d'o\grave{u} } $$

$$ T_{+} =- \di{m \over{ (s-m^2)(u+m^2)}} \left[ \hskip -0.05cm \di{{}\over{}} t\, \delta_{\lambda_2, \lambda_4} \left[ \sqrt{t+ 4m^2}  + \lambda_2 (2 \sigma_1) \sqrt{t} \right] \right. $$
$$\left. +4\, \delta_{\lambda_2, - \lambda_4}  \di{{(s-m^2)^2 - st}\over{\sqrt{t+4 m^2}}} \right]$$

\nin Puis, 

$$ T_{-} = \di{{t \sinh \Theta}\over{2 (s-m^2)(u+m^2)}} \left[  \hskip -0.05cm \di{{}\over{}}  \lambda_2 \,\sqrt{t(t+4m^2)} \left[ 1 - \delta_{\lambda_2, \lambda_4}+   \lambda_4 (2\sigma_1) \cosh \Theta \right]  \right. $$
$$ \left.\di{{}\over{}}  - \delta_{\lambda_2, \lambda_4} (2\sigma_1) \,(s+u) \right]$$

$$ = \di{{t \sinh \Theta}\over{2 (s-m^2)(u+m^2)}} \left[  \hskip -0.05cm \di{{}\over{}}  \lambda_2\, \delta_{\lambda_2, -\lambda_4}  \,\sqrt{t(t+4m^2)}  +(2 \sigma_1) (s+u) \left[ \lambda_2 \lambda_4  -\delta_{\lambda_2, \lambda_4}\,  \right] \right] $$

$$ = \delta_{\lambda_2, - \lambda_4} \di{{t \sinh \Theta}\over{2 (s-m^2)(u+m^2)}} \left[  \hskip -0.05cm \di{{}\over{}}  \lambda_2\, \sqrt{t(t+4m^2)} 
-(2 \sigma_1) (s+u) \right] $$

\vv \nin On obtient finalement 

 \beq \fbox{\fbox{\rule[-0.2cm]{0cm}{2.5cm}~$\begin{array}{c} ~\\
T_c = - \delta_{\sigma_3, \sigma_1} \di{m \over{ (s-m^2)(u+m^2)}} \left[ \hskip -0.05cm \di{{}\over{}} t\, \delta_{\lambda_2, \lambda_4} \left[ \sqrt{t+ 4m^2}  + \lambda_2 (2 \sigma_1) \sqrt{t} \right] \right. \\~\\ 
 \left. + 4\, \delta_{\lambda_2, - \lambda_4}\, \di{{(s-m^2)^2 - st}\over{\sqrt{t+4 m^2}}} \right] \\~\\
+\, \delta_{\sigma_3, - \sigma_1} \delta_{\lambda_2, - \lambda_4} \,\di{{t \sinh \Theta}\over{2 (s-m^2)(u+m^2)}} \left[  \hskip -0.05cm \di{{}\over{}}  \lambda_2  \,\sqrt{t(t+4m^2)} 
-(2 \sigma_1) (s+u) \right] \\~
\end{array} $~}}
\enq

\vv \nin On remarque que dans ce sch\'ema, ce sont les amplitudes sans changement d'h\'elicit\'e qui sont d\'efavoris\'ees dans le domaine cin\'ematique o\`u la masse $m$ peut \^etre prise \'egale \`a z\'ero. Bien s\^ur, cela provient encore de la nature vectorielle du couplage \'electromagn\'etique puisqu'alors

$$\ov{U_3}_{ \sigma_3}\,\gs_\mu\,U_{1 \sigma_1} \approx (2 \sigma_1) \,\ov{U_3}_{ \sigma_3}\,\gs_\mu\,\gs_5\,U_{1 \sigma_1} \approx  \sqrt{t} \,(2 \sigma_1)\, \delta_{\sigma_3, - \sigma_1}\, (X_g + i(2 \sigma_1) Y)_\mu  $$

\vv \nin De m\^eme, on observe dans ce domaine un changement de signe de l'h\'elicit\'e du photon. Ces r\'esultats ne sont pas en contradiction avec ceux du couplage en voie $s$, en raison des changements de signe des h\'elicit\'es, expliqu\'es ailleurs, lorsqu'on passe du couplage en voie $s$ \`a celui en voie $t$. 

\vv \nin Lorsque $m \rightarrow 0$ tandis que $t$ reste fini, on a $t \approx s\, \sin^2 \di{\theta\over 2}$, $u \approx s\, \cos^2 \di{\theta\over 2}$, $\sinh \Theta \approx 2 \di{\sqrt{su}\over t}$, et 

$$ T_c \approx \, \delta_{\sigma_3, - \sigma_1} \delta_{\lambda_2, - \lambda_4} \,\di{1 \over{ \sqrt{su}}} \left[ \lambda_2 t - (2\sigma_1) (s+u)    \right] $$
$$ \approx - 2 (2 \sigma_1) \delta_{\sigma_3, - \sigma_1} \delta_{\lambda_2, - \lambda_4} \left[ \delta_{\lambda_2, 2 \sigma_1} \sqrt{u\over s} + \delta_{\lambda_2, - 2 \sigma_1} \sqrt{s\over u} \right]$$
$$ = - 2 (2 \sigma_1) \delta_{\sigma_3, - \sigma_1} \delta_{\lambda_2, - \lambda_4} \left[ \delta_{\lambda_2, 2 \sigma_1} \cos \di{\theta \over 2}  + \delta_{\lambda_2, - 2 \sigma_1} \di{1\over{\cos \di{\theta\over 2}}}  \right]$$

\vv \nin On trouve alors des amplitudes similaires \`a celles du tableau II.

\newpage
\section{Amplitudes d'h\'elicit\'e de $\gs + \gs \rightarrow e^{-} + e^{+} $}

\begin{figure}[hbt]
\centering
\includegraphics[scale=0.3, width=9cm, height=3cm]{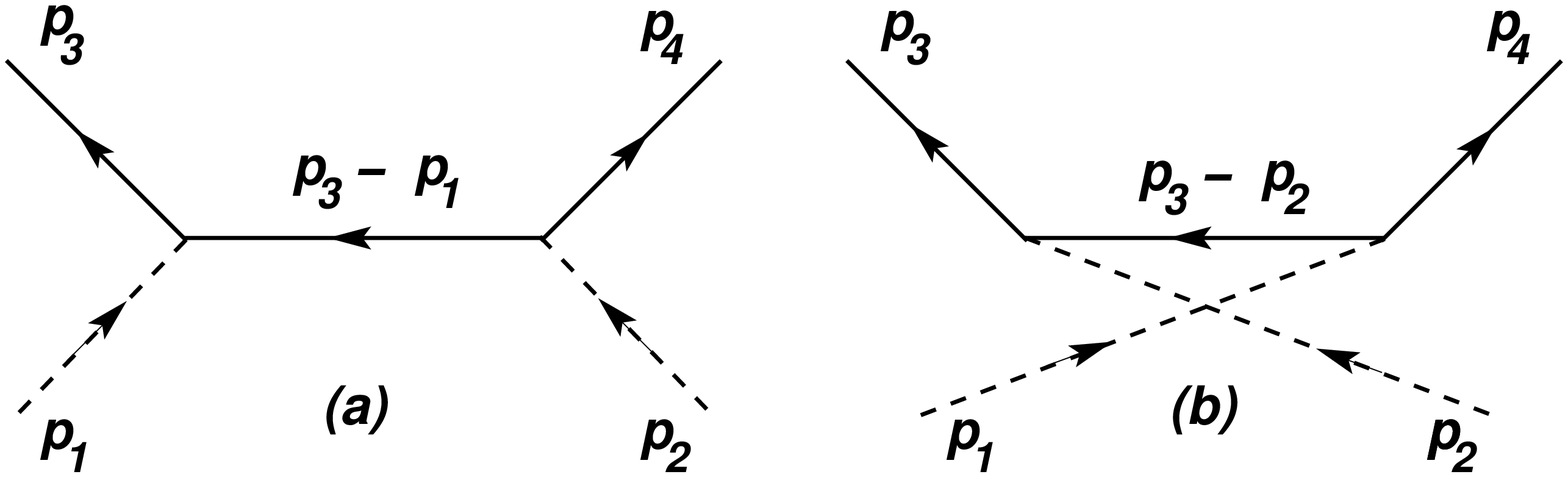}
\vskip 0.25cm

\caption{Diagrammes de Feynman pour $\gs + \gs \rightarrow e^- + e^+$} \label{creat}
\end{figure}

\vv \nin La production d'une paire particule-antiparticule $e^{-}\,e^{+}$ par collision de deux photons r\'eels est  l'un des deux processus ``crois\'es" de la diffusion Compton, le dernier \'etant la production d'une paire de photons par annihilation d'une paire $e^-\, e^+$. Ce processus donne un exemple de couplage d'h\'elicit\'e en voie $s$ d'un syst\`eme particule-antiparticule de spin 1/2 d'une part, et de deux photons (ici r\'eels) d'autre part. Les diagrammes de Feynman d\'ecrivant ce processus au plus bas ordre en $\alpha$ sont repr\'esent\'es \`a la figure (\ref{creat}). Apr\`es extraction de la constante de couplage $4 \pi \alpha$, l'amplitude g\'en\'erique correspondante s'\'ecrit  

\beq T = \ov{U_3} \left[ \gs(\epsilon_1) \di{{m + \gs(p_3 - p_1)}\over{m^2 - (p_1 - p_3)^2}} \gs(\epsilon_2) +  \gs(\epsilon_2) \di{{m + \gs(p_3 - p_2)}\over{m^2 - (p_2 - p_3)^2}} \gs(\epsilon_1)\right]  V_4  \enq

\vvv

\vv \nin Les notations sont les suivantes. L'\'electron final est repr\'esent\'e par le spineur ${U_3}_{\sigma_3}$ d'h\'elicit\'e $\sigma_3$ et $p_3 = E \,T + p\, Z'$ est sa 4-impulsion ; le positron final est repr\'esent\'e par le spineur ${V_4}_{\sigma_4} = \gs_5\, {U_4}_{\sigma_4}$ d'h\'elicit\'e $\sigma_4$ et $p_4 = E \,T - p\, Z'$ est sa 4-impulsion ; un premier photon initial a pour vecteur de polarisation $\epsilon^{(\lambda_1)}_1 = - \di{1\over \sqrt{2}} \left( \lambda_1 X + i Y \right)$ et a pour 4-impulsion $p_1 = E (T+Z)$ ;  le second photon initial a pour vecteur de polarisation $\epsilon^{(\lambda_2)}_2 = \epsilon^{(-\lambda_2)}_1 = - \di{1\over \sqrt{2}} \left(- \lambda_2 X + i Y \right)$ et a pour 4-impulsion $p_2 = E (T-Z)$. On pose $s = (p_1 +p_2)^2 = (p_3+p_4)^2$, $ t = -(p_1 -p_3)^2 = - (p_2 -p_4)^2$, $u = -(p_2 -p_3)^2 = -(p_1-p_4)^2$, et l'on a ici $E = \di{\sqrt{s}\over 2}$, $p = \beta E$ avec $\beta = \sqrt{1 - \di{{4 m^2}\over s}}$, $m$ \'etant la masse de l'\'electron (ou du positron !). 

\vv \nin Utilisant l'\'equation de Dirac pour l'\'electron, l'amplitude devient 

$$ T = \ov{U_3} \left\{ \di{1\over{m^2 +t}} \left[ \hskip -0.1cm\di{{}\over{}} 2\, p_3 \cdot \epsilon_1\, \gs(\epsilon_2) - \gs(\epsilon_1) \gs(p_1) \gs(\epsilon_2) \right] \right. $$
$$ \left.+  \di{1\over{m^2 +u}} \left[ \hskip -0.1cm\di{{}\over{}} 2\, p_3 \cdot \epsilon_2\, \gs(\epsilon_1) - \gs(\epsilon_2) \gs(p_2) \gs(\epsilon_1) \right] \right\}  V_4  $$

\vv \nin \ding{45}~ On a 
$$ - \gs(\epsilon_1) \gs(p_1) \gs(\epsilon_2) =  \gs(p_1) \gs(\epsilon_1)\gs(\epsilon_2)\,,~~~ \gs(\epsilon_1)\gs(\epsilon_2) = \di{1\over 2} \left[ 1 + \lambda_1 \lambda_2 + i(\lambda_1 + \lambda_2) \gs(X) \gs(Y) \right] $$
$$ = \delta_{\lambda_1, \lambda_2} \left[ 1 + i \lambda_1 \gs(X) \gs(Y) \right] ~~~{\rm car}~~~\lambda_1 \lambda_2 = \pm 1$$

\vv \nin Or, de $\gs_5 = i\, \gs(X)\,\gs(Y)\,\gs(Z)\,\gs(T)$, on tire  

$$ \gs(X)\, \gs(Y)\, \gs(T + \epsilon Z) =  i\, \epsilon\,\gs(T + \epsilon Z)\, \gs_5\,,~~~{\rm avec}~~~\epsilon = \pm 1\,,~~~{\rm d'o\grave{u}}$$
\beq \gs(p_1)\,\gs(X)\,\gs(Y) = i\,\gs(p_1)\, \gs_5\,,~~~\gs(p_2)\,\gs(X)\,\gs(Y) = -i\, \gs(p_2)\,\gs_5 \enq

\vv \nin Il vient ainsi 

$$ - \gs(\epsilon_1) \gs(p_1) \gs(\epsilon_2) = \delta_{\lambda_1, \lambda_2} \gs(p_1) \left[ 1 - \lambda_1 \gs_5 \right] ~~~{\rm et} $$
\beq - \gs(\epsilon_2) \gs(p_2) \gs(\epsilon_1) = \delta_{\lambda_1, \lambda_2} \gs(p_2) \left[ 1 - \lambda_1 \gs_5 \right] \enq

\vv\vv \nin \ding{45}~ Exprimons ci-dessous quelques produits scalaires : 

$$ {\rm comme}~~~p_3 = E \left( T + \beta\, Z' \right),~~~{\rm il~ vient} $$
$$ p_3 \cdot \epsilon_1 = E \,\beta\, Z' \cdot \epsilon_1 =  - \di{{E \beta \lambda_1} \over \sqrt{2}}\, Z' \cdot X =\di{{E \beta \lambda_1} \over \sqrt{2}}\,  \sin \theta\,,~~~ p_3 \cdot \epsilon_2 = - \di{{E \beta \lambda_2} \over \sqrt{2}}\,  \sin \theta\,;$$ 
$$ {\rm puis} $$
$$ p_1 \cdot \left[X' - i (2 \sigma_3 )Y \right] = p_1 \cdot X' = E \sin \theta = -p_2 \cdot  \left[ X' - i (2 \sigma_3)Y \right]\,,~~~p_1 \cdot Z' = - E \cos \theta $$

$$ \epsilon_1 \cdot \left[ X' - i (2 \sigma_3) Y \right] = \di{1\over \sqrt{2}} \left[ \lambda_1 \cos \theta + (2 \sigma_3) \right] \,,~~~\epsilon_2 \cdot \left[ X' - i (2 \sigma_3) Y \right] = \di{1\over \sqrt{2}} \left[ -\lambda_2 \cos \theta + (2 \sigma_3) \right] $$

$$ \epsilon_1 \cdot Z' = \di{\lambda_1 \over \sqrt{2}} \sin \theta\,,~~~ \epsilon_2 \cdot Z' = -\di{\lambda_2 \over \sqrt{2}} \sin \theta $$

\vv
\vv\nin \ding{45}~ Adaptant (\ref{u3v4}) \`a l'\'etat final $e^- e^+$, on a les formules 

 $$ \ov{U_3}_{ \sigma_3}\,\gs_\mu\, {V_4}_{ \sigma_4} = 
2\, m \left[\hskip -0.1cm \di{{}\over{}}  (2 \sigma_3) \,\delta_{\sigma_4, \sigma_3} \cosh \chi \, (X' - 2 i \sigma_3 Y)_\mu -  \delta_{\sigma_4, -\sigma_3}\, Z'_\mu \right] $$
\beq   \ov{U_3}_{ \sigma_3}\,\gs_\mu\, \gs_5\,{V_4}_{ \sigma_4} = 
2\,m \left[\hskip-0.1cm \di{{}\over{}} \delta_{\sigma_4, \sigma_3} \sinh \chi \, (X' - 2 i \sigma_3 Y)_\mu -   (2 \sigma_3) \,\delta_{\sigma_4, -\sigma_3} \,T_\mu \right] \enq

\vv \nin avec $ 2 m \cosh \chi = \sqrt{s}$, $2 m \sinh \chi = \beta \sqrt{s}$. On en d\'eduit  

$$ \ov{U_3}_{ \sigma_3}\,\gs(\epsilon_1)\, {V_4}_{ \sigma_4} = (2 \sigma_3) \sqrt{s} \,\delta_{\sigma_4, \sigma_3} \,\di{1\over \sqrt{2}} \left[ \lambda_1 \cos \theta + (2 \sigma_3) \right]  -  (2m)\,\delta_{\sigma_4, -\sigma_3}\,\di{\lambda_1 \over \sqrt{2}} \sin \theta
$$
$$ \ov{U_3}_{ \sigma_3}\,\gs(\epsilon_2)\, {V_4}_{ \sigma_4} = (2 \sigma_3) \sqrt{s} \,\delta_{\sigma_4, \sigma_3} \,\di{1\over \sqrt{2}} \left[ -\lambda_2 \cos \theta + (2 \sigma_3) \right]  +  (2m)\,\delta_{\sigma_4, -\sigma_3}\,\di{\lambda_2 \over \sqrt{2}} \sin \theta
$$

$$ \ov{U_3}_{ \sigma_3}\,\gs(p_1)\, {V_4}_{ \sigma_4} = - \,\ov{U_3}_{ \sigma_3}\,\gs(p_2)\, {V_4}_{ \sigma_4} =  
  (2 \sigma_3) \,\delta_{\sigma_4, \sigma_3} \di{s\over 2} \sin \theta +  \delta_{\sigma_4, -\sigma_3}\, m\,\sqrt{s} \cos \theta $$

$$ \ov{U_3}_{ \sigma_3}\,\gs(p_1)\,\gs_5\, {V_4}_{ \sigma_4}  = \beta\, \di{s\over 2} \,\delta_{\sigma_4, \sigma_3} \, \sin \theta -  (2 \sigma_3) \,\delta_{\sigma_4, -\sigma_3}\,m\,\sqrt{s}  $$

$$ \ov{U_3}_{ \sigma_3}\,\gs(p_2)\,\gs_5\, {V_4}_{ \sigma_4}  = - \,\beta\, \di{s\over 2} \,\delta_{\sigma_4, \sigma_3} \, \sin \theta -  (2 \sigma_3) \, \delta_{\sigma_4, -\sigma_3}\,m\,\sqrt{s}  $$

\nin puis 

$$ -\,  \ov{U_3} \, \gs(\epsilon_1) \gs(p_1) \gs(\epsilon_2)\, {V_4}_{ \sigma_4}  = \delta_{\lambda_1, \lambda_2} \left\llbracket  \delta_{\sigma_4, \sigma_3} \di{s\over 2} \sin \theta \left[ (2 \sigma_3) - \lambda_1 \beta \right]       +  \delta_{\sigma_4, -\sigma_3} m \sqrt{s} \left[ \cos \theta + (2 \sigma_3) \lambda_1 \right] \right\rrbracket$$

$$ -\,  \ov{U_3} \, \gs(\epsilon_2) \gs(p_2) \gs(\epsilon_1)\, {V_4}_{ \sigma_4}  = -\,\delta_{\lambda_1, \lambda_2} \left\llbracket   \,\delta_{\sigma_4, \sigma_3} \di{s\over 2} \sin \theta \left[ (2 \sigma_3) - \lambda_1 \beta \right]       +  \delta_{\sigma_4, -\sigma_3} m \sqrt{s} \left[ \cos \theta - (2 \sigma_3) \lambda_1 \right] \right\rrbracket$$

\vv \nin \ding{45}~ Collectant tous ces r\'esultats, et explicitant ~$t+m^2 = \di{s \over 2} \left[ 1 - \beta \cos \theta \right]$, $u+m^2 = \di{s \over 2} \left[ 1 + \beta \cos \theta \right]$, on trouve

\newpage 
~\vskip -0.7cm
\beq \fbox{\fbox{\rule[-0.35cm]{0cm}{4cm}~$\begin{array}{c}  

T = T(\lambda_1, \lambda_2 ; \sigma_3, \sigma_4)  = \di{1\over{1 - \beta \cos \theta}} \left\llbracket 
\di{{}\over{}} \beta \lambda_1 \sin \theta \left\{\hskip -0.1cm \di{{}\over{}} \left[1 - (2 \sigma_3)  \lambda_2 \cos \theta \right] \delta_{\sigma_4, \sigma_3}  
\right. \right. \\~\\

 \left. + \di{{2 m}\over \sqrt{s}} \lambda_2 \sin \theta \,\delta_{\sigma_4, - \sigma_3} \right\}  
 \left. + \delta_{\lambda_1, \lambda_2} \left\{\hskip-0.1cm \di{{}\over{}} \delta_{\sigma_4, \sigma_3} \sin \theta \left[ (2 \sigma_3) - \lambda_1 \beta \right]  \right. \right. \\~\\

 \left. \left. +\di{{2m}\over \sqrt{s}}\, \delta_{\sigma_4, - \sigma_3} \left[ \cos \theta + (2 \sigma_3) \lambda_1 \right] \right\} \,\right\rrbracket \\~\\

- \,\di{1\over{1 + \beta \cos \theta}} \left\llbracket \,\beta \lambda_2 \sin \theta \left\{\hskip-0.1cm \di{{}\over{}}  \left[ 1+ (2 \sigma_3)  \lambda_1 \cos \theta \right] \delta_{\sigma_4, \sigma_3} - \di{{2 m}\over \sqrt{s}} \lambda_1 \sin \theta \,\delta_{\sigma_4, - \sigma_3} \right\} \right. \\~\\

\left. + \delta_{\lambda_1, \lambda_2} \left\{ \delta_{\sigma_4, \sigma_3} \sin \theta \left[ (2 \sigma_3) - \lambda_1 \beta \right] +\di{{2m}\over \sqrt{s}}\, \delta_{\sigma_4, - \sigma_3} \left[ \cos \theta - (2 \sigma_3) \lambda_1 \right] \right\} \,\right\rrbracket \\~\\

\end{array} \label{ampcrea-1} 
  $~}}  \enq

\vv \nin Les tableaux ci-apr\`es donnent les expressions explicites des 16 amplitudes d'h\'elicit\'e du processus, tir\'ees de (\ref{ampcrea-1}), dans le cas $m \neq 0$ et dans le cas $m=0$.

$$\fbox{\fbox{\rule[-0.85cm]{0cm}{5cm}~$\begin{array}{c}  

T(+,+\, ; \uw,\uw) = T(+,+\, ; \dw,\dw) =T(-,-\, ; \uw,\uw) =T(-,-\, ; \dw,\dw) = 0\\~\\

T(+,- \,;\uw,\uw) = - T(-,+\, ; \dw, \dw ) =   \di{{ 2 \beta \sin \theta (1 + \cos \theta) }\over{ 1- \beta^2 \cos^2 \theta}}  \\~\\

T(+,- \,;\dw,\dw) = - T(-,+\, ; \uw, \uw ) =   \di{{ 2 \beta \sin \theta (1 - \cos \theta) }\over{ 1- \beta^2 \cos^2 \theta}}  \\~\\

T(+,+ \,;\uw,\dw) =  T(-,-\, ; \dw, \uw ) = \di{{4 m \beta} \over \sqrt{s}}   \di{{ 1+ \beta }\over{ 1- \beta^2 \cos^2 \theta}}  \\~\\

T(-,- \,;\uw,\dw) =  T(+,+\, ; \dw, \uw ) = -\,\di{{4 m \beta} \over \sqrt{s}}   \di{{ 1- \beta }\over{ 1- \beta^2 \cos^2 \theta}}  \\~\\

T(+,- \,;\uw,\dw) =  T(-,+\, ; \uw, \dw ) = T(+,- \,;\dw,\uw) =  T(-,+\, ; \dw, \uw ) = -\,\di{{4 m \beta} \over \sqrt{s}}   \di{{ \sin^2 \theta }\over{ 1- \beta^2 \cos^2 \theta}}  \\~\\

\end{array}
  $~}}  $$

\vvv
\centerline{\bf Tableau III - Amplitudes d'h\'elicit\'e de $\gs + \gs \rightarrow e^- + e^+$}

$$\fbox{\fbox{\rule[-0.4cm]{0cm}{2cm}~$\begin{array}{c}  

T(+,- \,;\uw,\uw) = - T(-,+\, ; \dw, \dw ) =   2 \cot \di{\theta \over 2}  \\~\\

T(+,- \,;\dw,\dw) = - T(-,+\, ; \uw, \uw ) =   2 \tan \di{\theta \over 2} \\~\\

\end{array}
  $~}}  $$

\vv 
\centerline{\bf Tableau IV - Amplitudes d'h\'elicit\'e non nulles de $\gs + \gs \rightarrow e^- + e^+$ pour $m=0$}

\newpage 

\vv \nin Ici aussi, on observe que certaines amplitudes s'annulent pour $m=0$ ( en fait, pour $\sqrt{s} \gg m$) : celles pour lesquelles les h\'elicit\'es de l'\'electron et du positron sont {\it oppos\'ees}, ph\'enom\`ene similaire \`a celui observ\'e pour la diffusion Compton, et qui r\'esulte, comme d\'ej\`a indiqu\'e, de la nature vectorielle de l'interaction \'electromagn\'etique.   

\section{Amplitudes d'h\'elicit\'e de $\gs^\star + \gs^\star \rightarrow e^{-} +\, e^{+} $ \label{secggee} }

\vv\nin Le processus envisag\'e ici diff\`ere de celui du paragraphe pr\'ec\'edent en ce que les deux photons entrant en collision sont maintenant suppos\'es {\it virtuels}, cette appellation signifiant que ces photons sont en fait des interm\'ediaires d\'ecrivant une interaction \'electromagn\'etique entre deux vertex d'un diagramme de Feynman, comme par exemple ceux des diagrammes des figures (1.4) et (1.5) du chapitre 1, et que de ce fait, ils ne sont pas observables. Nous supposerons qu'ils sont du genre espace et poserons $p^2_1 = - t_1 <0$, $p^2_2 = - t_2 <0$. Leur t\'etrades d'h\'elicit\'e respectives dans un couplage en voie $s$ sont d\'efinies comme au paragraphe 1.5.3 : 

\vv \nin \ding{172} \und{\bf photon 1}  

$$ \epsilon^{(0)}_1 = \di{{2 \sqrt{t_1}}\over{\Lambda^{\frac{1}{2}}}} \left( P + p_1 \di{{P\cdot p_1}\over t_1} \right) = \di{{2 \sqrt{t_1}}\over{\Lambda^{\frac{1}{2}}}} \left( p_2 + p_1 \di{{p_2\cdot p_1}\over t_1} \right)  = T_1 $$
$$ {\rm avec}~~~P= p_1 + p_2\,,~~P^2=s\,,~~~\Lambda = \Lambda(s, -t_1,-t_2) $$
$$ z_1 = \di{{p_1}\over \sqrt{t_1}} \,,~~~\epsilon^{(\pm)}_1 = \mp \di{1\over \sqrt{2}} \left( X \pm i Y \right) = \epsilon^{(\pm)} $$

\nin \ding{173} \und{\bf photon 2}  

$$ \epsilon^{(0)}_2 = \di{{2 \sqrt{t_2}}\over{\Lambda^{\frac{1}{2}}}} \left( P + p_2 \di{{P\cdot p_2}\over t_2} \right) = T_2\,,~~~z_2 = \di{{p_2}\over \sqrt{t_2}} \,,~~~\epsilon^{(\pm)}_2 = \epsilon^{(\mp)}_1$$

\vv \nin Notons que  

$$ T_1 = \cosh \eta_1 \,T + \sinh \eta_1\,Z\,,~~~z_1 = \cosh\, \eta_1\,Z + \sinh \eta_1\, T $$ 
$$ T_2 = \cosh \eta_2 \,T - \sinh \eta_2\,Z\,,~~~z_2 = -\cosh\, \eta_2\,Z + \sinh \eta_2\, T $$ 
$$ T_1 = \cosh \eta\, T_2 - \sinh \eta\, z_2 \,,~~~z_1 = - \cosh \eta\, z_2 + \sinh \eta\, T_2 $$
$$ T_2 = \cosh \eta\, T_1 - \sinh \eta\, z_1 \,,~~~z_2 = - \cosh \eta\, z_1 + \sinh \eta\, T_1 $$

\nin avec 

$$ \cosh \eta_1 = \di{ \Ld \over{ 2 \sqrt{s t_1}}} \,,~~\sinh \eta_1 = \di{{s + t_2 -t_1}\over{2 \sqrt{s t_1}}}\,,~~ \cosh \eta_2 = \di{ \Ld \over{ 2 \sqrt{s t_2}}} \,,~~\sinh \eta_2 = \di{{s + t_1 -t_2}\over{2 \sqrt{s t_2}}} $$
$$ \eta = \eta_1 + \eta_2\,,~~\cosh \eta =\di{{s + t_1 + t_2}\over{2 \sqrt{t_1 t_2}}} \,,~~\sinh \eta = \di{ \Ld \over{ 2 \sqrt{ t_1 t_2 }}} $$

\nin et que, inversement, 

$$ T = \cosh \eta_1\, T_1 - \sinh \eta_1 \, z_1 = \cosh \eta_2\, T_2 - \sinh \eta_2\, z_2 $$
$$ Z = \cosh \eta_1\, z_1 - \sinh \eta_1 \, T_1 = - \cosh \eta_2\, z_2 + \sinh \eta_2\, z_2 $$

\vv \nin Les photons virtuels disposent chacun de trois \'etats de polarisation : $\epsilon^{(0)}_1,\, \epsilon^{(\pm)}_1$ pour le premier, $\epsilon^{(0)}_2,\, \epsilon^{(\pm)}_2$ pour le second, correspondant aux indices d'h\'elicit\'e $0$ et $\pm 1$, ce qui conduit \`a d\'eterminer 36 amplitudes d'h\'elicit\'e pour ce processus. Nous \'ecrirons son amplitude g\'en\'erique sous la forme  

$$ T = \ov{U_3} \left\{ \di{1\over a} \left[ \hskip -0.1cm\di{{}\over{}} 2\, p_3 \cdot \epsilon_1\, \gs(\epsilon_2) - \gs(\epsilon_1) \gs(p_1) \gs(\epsilon_2) \right] \right.  \left.+  \di{1\over b} \left[ \hskip -0.1cm\di{{}\over{}} 2\, p_3 \cdot \epsilon_2\, \gs(\epsilon_1) - \gs(\epsilon_2) \gs(p_2) \gs(\epsilon_1) \right] \right\}  V_4  $$
$$ {\rm avec}~~~ a = m^2 - (p_3-p_1)^2 = m^2 + t\,,~~~~b = m^2 -(p_3 - p_2)^2 = m^2 + u $$

\vv \nin Exprimons ici les d\'enominateurs $a$ et $b$ en fonction de $\theta$ :  

$$ a = t_1 + 2\, p_1\cdot p_3= t_1 + \sqrt{s t_1}  \left( \sinh \eta_1 - \beta\, \cosh \eta_1\, \cos \theta \right) = t_1 + \di{1\over 2} \left( s+ t_2 - t_1 \right) - \di{1\over 2} \beta \Ld \cos \theta ~~{\rm soit} $$
$$ a = \sqrt{t_1 t_2} \left( \cosh \eta - \beta \sinh \eta \cos \theta \right) ~~{\rm et~de~m\hat{e}me}~~b = \sqrt{t_1 t_2} \left( \cosh \eta + \beta \sinh \eta \cos \theta \right)$$

\vvv
\vv \nin Nous d\'eterminerons s\'epar\'ement les amplitudes $T(0,0\, ; \sigma_3, \sigma_4)$, $T(0, \lambda_2 \,; \sigma_3, \sigma_4)$, $T(\lambda_1, 0\, ; \sigma_3, \sigma_4)$ et $T(\lambda_1, \lambda_2 \,; \sigma_3, \sigma_4)$, avec ici $\lambda_1, \lambda_2 = \pm 1$. 

\vvv
\vv \nin $\bullet$ \und{\bf Amplitudes $T(0,0\, ;\sigma_3, \sigma_4)$}

\vv \nin Pour ces amplitudes, $\epsilon_1 = T_1$, $\epsilon_2 = T_2$. On a 
$$ \gs(\epsilon^{(0)}_1) \,\gs(p_1) \, \gs(\epsilon^{(0)}_2) = \sqrt{t_1}\, \gs(T_1)\, \gs(z_1)\, \gs(T_2) ~~{\rm et~comme}~~~\gs(T_1)\, \gs(z_1) = - \gs(T_2)\, \gs(z_2),~~{\rm il~vient}$$
$$ \gs(\epsilon^{(0)}_1) \,\gs(p_1) \, \gs(\epsilon^{(0)}_2) = \sqrt{t_1}\,\gs(T_1) \,\gs(z_1) \, \gs(T_2)  = - \sqrt{t_1}\,\gs(T_2) \,\gs(z_2) \, \gs(T_2)  =\sqrt{t_1}\, \gs(z_2) =  \sqrt{\di{{t_1}\over{t_2}}}\, \gs(p_2) $$

$$ {\rm De~m\hat{e}me}, ~~ \gs(\epsilon^{(0)}_2) \,\gs(p_2) \, \gs(\epsilon^{(0)}_1) =  \sqrt{t_2}\, \gs(z_1) =  \sqrt{\di{{t_2}\over{t_1}}}\, \gs(p_1) $$

\vv \nin Il est judicieux d'utiliser ici l'\'equation $ P^\mu\,   \ov{U_3}\,\gs_\mu\, V_4 =0$, qui r\'esulte de la conservation d'un courant et qui permet d'\'ecrire  

$$   \ov{U_3}\,\gs(p_2) \, V_4 = - \,\ov{U_3}\,\gs(p_1) \, V_4\,,~~ \ov{U_3}\,\gs(T_1) \, V_4 = \di{{\sinh \eta_1}\over{\cosh \eta_1}} \,\ov{U_3}\,\gs(z_1) \, V_4 = \di{{\sinh \eta_1}\over{\sqrt{t_1}\,\cosh \eta_1}} \,\ov{U_3}\,\gs(p_1) \, V_4 $$
$$ \ov{U_3}\,\gs(T_2) \, V_4 = \di{{\sinh \eta_2}\over{\cosh \eta_2}} \,\ov{U_3}\,\gs(z_2) \, V_4 = \di{{\sinh \eta_2}\over{\sqrt{t_2}\,\cosh \eta_2}} \,\ov{U_3}\,\gs(p_2) \, V_4 = -\,\di{{\sinh \eta_2}\over{\sqrt{t_2}\,\cosh \eta_2}} \,\ov{U_3}\,\gs(p_1) \, V_4 $$

\vv \nin On obtient ainsi 

$$T(0,0\, ;\sigma_3, \sigma_4) = \ov{U_3}\,\gs(p_1) \, V_4 \times A~~~{\rm avec} $$
$$ A = \di{1\over a} \left\{ - 2 p_3 \cdot T_1\,\di{{\sinh \eta_2}\over{\sqrt{t_2}\,\cosh \eta_2}}  + \sqrt{t_1 \over t_2} \right\} + \di{1\over b} \left\{ 2 p_3\cdot T_2\, \di{{\sinh \eta_1}\over{\sqrt{t_1}\,\cosh \eta_1}}  - \sqrt{t_2 \over t_1} \right\} $$ 

\vv \nin Comme $2 p_3\cdot T_1 = \sqrt{s} \left( \cosh \eta_1 - \beta \sinh \eta_1 \cos \theta \right)$, $2 p_3\cdot T_2 = \sqrt{s} \left( \cosh \eta_2 + \beta \sinh \eta_2 \cos \theta \right)$, $\sqrt{t_1} \cosh \eta_1 = \sqrt{t_2} \cosh \eta_2 = \sqrt{\di{ {t_1 t_2}\over s} }\sinh \eta$, les coefficients respectifs $A_1$ et $B_1$ de $1/a$ et $1/b$ s'\'ecrivent  

$$ A_1 = - \sqrt{s} \di{{\sinh \eta_2}\over{\sqrt{t_2}\,\cosh \eta_2}} \left( \cosh \eta_1 - \beta \sinh \eta_1 \cos \theta \right) + \sqrt{ t_1 \over t_2} =  \di{{s\beta\, \sinh \eta_2 \,\sinh \eta_1 \,\cos \theta } \over{\sqrt{t_1 t_2} \sinh \eta} } $$
$$ + \sqrt{ t_1 \over t_2 } - \di{1\over{ 2 \sqrt{t_1 t_2}}} ( s + t_1 - t_2 ),~~~{\rm soit}$$ 

$$ A_1 =  \di{{s\beta\, \sinh \eta_2 \,\sinh \eta_1 \,\cos \theta } \over{\sqrt{t_1 t_2} \sinh \eta} } -  \di{1\over{ 2 \sqrt{t_1 t_2}}} ( s - t_1 - t_2 )\,;$$

$$ B_1 =  \sqrt{s} \di{{\sinh \eta_1}\over{\sqrt{t_1}\,\cosh \eta_1}} \left( \cosh \eta_2 + \beta \sinh \eta_2 \cos \theta \right) - \sqrt{ t_2 \over t_1}$$
$$ = \di{{s\beta\, \sinh \eta_1 \,\sinh \eta_2 \,\cos \theta } \over{\sqrt{t_1 t_2} \sinh \eta} } +  \di{1\over{ 2 \sqrt{t_1 t_2}}} ( s - t_1 - t_2 ) $$

\vv \nin Ainsi, 

$$ a\,b\,A =  \di{{s\beta\, \sinh \eta_1 \,\sinh \eta_2 \,\cos \theta } \over{\sqrt{t_1 t_2} \sinh \eta} } (a + b) +  \di{1\over{ 2 \sqrt{t_1 t_2}}} ( s - t_1 - t_2 ) (a - b) = \beta \cos \theta \left\{ \di{{2 \, s\,\sinh \eta_1 \,\sinh \eta_2 \,\cosh \eta } \over{\sinh \eta} } \right.$$
$$ \left. \di{{}\over{}} - \sinh \eta \,(s - t_1 -t_2) \right\} = \di{{\beta \cos \theta}\over{ 2\,\sqrt{t_1 t_2} \,\Ld}}  \left[\hskip-0.1 cm \di{{}\over{}}  (s + t_2 - t_1)(s + t_1 - t_2)(s + t_1 + t_2) - ( s-t_1 -t_2) \Lambda \right],~~~{\rm soit} $$

$$ A = \di{{ 4 s\, \sqrt{t_1 t_2} \,\beta \cos \theta}\over{ a\,b\, \,\Ld}} $$

\vv \nin Calculons ensuite 

$$ \ov{U_3}\,\gs(p_1) \, V_4 = 2\, m \left[\hskip -0.1cm \di{{}\over{}}  (2 \sigma_3) \,\delta_{\sigma_4, \sigma_3} \cosh \chi \,\, p_1 \cdot X'  -  \delta_{\sigma_4, -\sigma_3}\,p_1 \cdot  Z' \right] $$
$$ = \sqrt{t_1}\,\cosh \eta_1 \left[ (2 \sigma_3) \,\delta_{\sigma_4, \sigma_3} \sqrt{s} \,\, \sin \theta  + 2\,m\,\delta_{\sigma_4, -\sigma_3}\,\cos \theta \right] $$
$$ = \di{\Ld \over 2} \left[ (2 \sigma_3) \,\delta_{\sigma_4, \sigma_3} \, \sin \theta  + \di{{2\,m}\over \sqrt{s}}\,\delta_{\sigma_4, -\sigma_3}\,\cos \theta \right] $$

\vv \nin Les amplitudes cherch\'ees ont donc pour expression g\'en\'erale 

\beq \fbox{\rule[-0.6cm]{0cm}{1.5cm}~$T(0,0\, ;\sigma_3, \sigma_4) =  \di{{ 2 s\, \sqrt{t_1 t_2} \,\beta \cos \theta}\over{ a\,b}} \left[ (2 \sigma_3) \,\delta_{\sigma_4, \sigma_3} \, \sin \theta  + \di{{2\,m}\over \sqrt{s}}\,\delta_{\sigma_4, -\sigma_3}\,\cos \theta \right]   $~}  \enq

\vv \nin soit, explicitement,

\beq \fbox{\fbox{\rule[-0.5cm]{0cm}{1.2cm}~$ \begin{array}{c}
~\\T(0,0\, ;\uw, \uw) = -T(0,0\, ;\dw, \dw) =  \di{{ 2 s\, \sqrt{t_1 t_2} \,\beta \cos \theta \sin \theta }\over{ a\,b}}    \\~\\
T(0,0\, ;\uw, \dw) = T(0,0\, ;\dw, \uw) =  \di{{ 4 m \sqrt{s \,t_1 t_2} \,\,\beta \cos^2 \theta}\over{ a\,b}}  \\~\\
\end{array}   $~}}  \enq

\vvv
\vv \nin $\bullet$ \und{\bf Amplitudes $T(0,\lambda_2\, ;\sigma_3, \sigma_4)$}

\vv \nin Pour calculer ces amplitudes, il appara\^it plus avantageux d'exprimer leur forme g\'en\'erique comme 

$$T = \ov{U_3} \left\{ \di{1\over a} \left[ \hskip -0.1cm\di{{}\over{}} 2\, p_3 \cdot \epsilon_1\, \gs(\epsilon_2) - \gs(\epsilon_1) \gs(p_1) \gs(\epsilon_2) \right] \right.  \left.+  \di{1\over b} \left[ \hskip -0.1cm\di{{}\over{}} - 2\, p_4 \cdot \epsilon_1\, \gs(\epsilon_2) + \gs(\epsilon_2) \gs(p_1) \gs(\epsilon_1) \right] \right\}  V_4 $$

\vv \nin En effet, on a maintenant ($\lambda_2 = \pm 1)$

$$\gs(\epsilon_1) \gs(p_1) \gs(\epsilon_2) = \sqrt{t_1}\, \gs(T_1)\,\gs(z_1)\,\gs(\epsilon^{(-\lambda_2)}_1) =  \lambda_2\,\sqrt{t_1}\,\gs(\epsilon^{(-\lambda_2)}_1) \, \gs_5$$
$$ 2 p_3 \cdot T_1 = \sqrt{s} \left( \cosh \eta_1 - \beta \sinh \eta_1 \cos \theta \right)\,,~~2 p_4 \cdot T_1 = \sqrt{s} \left( \cosh \eta_1 + \beta \sinh \eta_1 \cos \theta \right) $$

\nin de sorte que 

$$ a\,b\, T = \ov{U_3} \, \gs(\epsilon_2)\, V_4 \left\{ \hskip -0.1cm \di{{}\over{}} \sqrt{s}\, \cosh \eta_1 (b -a) - \sinh \eta_1 \sqrt{s} \beta \cos \theta (a+ b) \right\} $$
$$ + \ov{U_3} \, \gs(\epsilon_2)\, \gs_5\,V_4 \left\{ \hskip -0.1cm \di{{}\over{}} \lambda_2 \sqrt{t_1}\, (a + b) \right\} = 2 \sqrt{s\,t_1\,t_2} \, \beta \cos \theta  \,\sinh \eta_2~\ov{U_3} \, \gs(\epsilon_2)\, V_4 $$
$$ + 2 \lambda_2 \,t_1\, \sqrt{t_2}\, \cosh \eta ~\ov{U_3} \, \gs(\epsilon_2)\,\gs_5\, V_4, ~~~{\rm soit} $$

$$ T = \di{ \sqrt{t_1} \over{a\,b}} \left\{ \hskip -0.1cm \di{{}\over{}}  ( s+t_1 - t_2) \beta \cos \theta~\ov{U_3} \, \gs(\epsilon_2)\, V_4 +\lambda_2\, (s + t_1 + t_2) ~\ov{U_3} \, \gs(\epsilon_2)\, \gs_5\,V_4 \right\}$$

\nin Or, 

$$\ov{U_3} \, \gs(\epsilon_2)\, V_4 = \delta_{\sigma_3, \sigma_4}\, \sqrt{\di{s \over 2}} \left[ \hskip -0.1cm\di{{}\over{}} 1 - \lambda_2 (2 \sigma_3) \cos \theta \right] + \di{{2\,m}\over \sqrt{2}} \lambda_2 \, \sin \theta \, \delta_{\sigma_3, - \sigma_4} $$ 
$$\ov{U_3} \, \gs(\epsilon_2)\,\gs_5\, V_4 =  \delta_{\sigma_3, \sigma_4}\,(2 \sigma_3) \,\beta\, \sqrt{\di{s \over 2}} \left[ \hskip -0.1cm\di{{}\over{}} 1 - \lambda_2 (2 \sigma_3) \cos \theta \right] $$

\nin d'o\`u l'amplitude 

\beq \fbox{\rule[-0.1cm]{0cm}{1.5cm}~$ \begin{array}{c} ~\\ T(0,\lambda_2\, ;\sigma_3, \sigma_4) = \sqrt{ \di{{s\,t_1} \over 2}} \, \di{ \beta \over{a\,b}} \left\{ \hskip -0.1cm\di{{}\over{}}   \delta_{\sigma_3, \sigma_4} \left[ \hskip -0.1cm\di{{}\over{}} 1 - \lambda_2 (2 \sigma_3) \cos \theta \right]  \left[ \hskip -0.1cm\di{{}\over{}} (s + t_1 - t_2) \cos \theta \right. \right.  \\~\\
\left. \left. \hskip -0.1cm\di{{}\over{}}  + (2 \sigma_3) \lambda_2 (s + t_1 + t_2) \right] +\delta_{\sigma_3, - \sigma_4} \di{{2\, m}\over \sqrt{s}} \lambda_2 \,(s+t_1 -t_2)\,\sin \theta \right\} \\~ 
\end{array} $~}  \enq

\vv \nin Explicitement, 

\beq \fbox{\fbox{\rule[-0.5cm]{0cm}{1.2cm}~$ \begin{array}{c}
~\\T(0,+\, ;\uw, \uw) = T(0,-\, ;\dw, \dw) = \sqrt{ \di{{s\,t_1} \over 2}} \, \di{ \beta \over{a\,b}}  \left[ \hskip -0.1cm\di{{}\over{}} 1 -  \cos \theta \right] \left[ \hskip -0.1cm\di{{}\over{}} (s + t_1 - t_2) \cos \theta \right. \\~\\
\left.\hskip -0.1cm\di{{}\over{}}  +  (s + t_1 + t_2) \right]  \\~\\
T(0,+\, ;\dw, \dw) = T(0,-\, ;\uw, \uw) =   \sqrt{ \di{{s\,t_1} \over 2}} \, \di{ \beta \over{a\,b}}  \left[ \hskip -0.1cm\di{{}\over{}} 1 +  \cos \theta \right] \left[ \hskip -0.1cm\di{{}\over{}} (s + t_1 - t_2) \cos \theta \right. \\~\\
\left.\hskip -0.1cm\di{{}\over{}}  -  (s + t_1 + t_2) \right]  \\~\\
T(0,+\,; \uw, \dw) = T(0,+\,; \dw, \uw) = - T(0,-\,; \uw, \dw) =  - T(0,-\,; \dw, \uw)  \\~\\
= \di{{m \beta\, \sqrt{2\, t_1}} \over{a\,b}}\, (s + t_1 - t_2)\, \sin \theta \cos \theta \\~
\end{array}   $~}}  \enq

\newpage
\vv \nin $\bullet$ \und{\bf Amplitudes $T(\lambda_1,0\, ;\sigma_3, \sigma_4)$}

\vv \nin Sym\'etriquement au cas pr\'ec\'edent, l'amplitude g\'en\'erique sera ici pr\'esent\'ee sous la forme 

\vskip -0.2cm 
$$T = \ov{U_3} \left\{ \di{1\over a} \left[ \hskip -0.1cm\di{{}\over{}} -2\, p_4 \cdot \epsilon_2\, \gs(\epsilon_1) + \gs(\epsilon_1) \gs(p_2) \gs(\epsilon_2) \right] \right.  \left.+  \di{1\over b} \left[ \hskip -0.1cm\di{{}\over{}}  2\, p_3 \cdot \epsilon_2\, \gs(\epsilon_1) - \gs(\epsilon_2) \gs(p_2) \gs(\epsilon_1) \right] \right]  V_4 $$ 

\nin o\`u~ $\epsilon_2 = T_2$, $\epsilon_1 = \epsilon^{(\lambda_1)}_1,~\lambda_1 = \pm 1$. Ensuite, on utilise les identit\'es $\gs(\epsilon_1)\, \gs(p_2)\, \gs(T_2) = \sqrt{t_2} \,\lambda_1\, \gs(\epsilon_1) \, \gs_5$, \vskip 0.1cm
\nin $ 2 p_3\cdot T_2 = \sqrt{s} \left( \cosh \eta_2 - \beta \sinh \eta_2 \cos \theta \right),~~  2 p_4\cdot T_2 = \sqrt{s} \left( \cosh \eta_2 + \beta \sinh \eta_2 \cos \theta \right)$, pour obtenir

$$ a\,b\,T = \sqrt{t_2} \,\lambda_1 \,\ov{U_3} \, \gs(\epsilon_1) \,\gs_5\,V_4\,(a+b) +\sqrt{s}~ \ov{U_3} \, \gs(\epsilon_1) \,V_4 \left[\di{{}\over{}}  \cosh \eta_2 \,(a-b) + \beta\, \sinh \eta_2\, \cos \theta (a+b) \right]$$
$$ = \sqrt{t_2}\, \left[\di{{}\over{}} \lambda_1 ( s+ t_1 + t_2)\, \ov{U_3} \, \gs(\epsilon_1) \,\gs_5\,V_4 - (s +t_2 - t_1) \, \beta\, \cos \theta\, \ov{U_3} \, \gs(\epsilon_1) \,V_4\right] $$

\vv \nin Les remplacements  

$$\ov{U_3} \, \gs(\epsilon_1)\, V_4 = \delta_{\sigma_3, \sigma_4}\, \sqrt{\di{s \over 2}} \left[ \hskip -0.1cm\di{{}\over{}} 1 + \lambda_1 (2 \sigma_3) \cos \theta \right] - \di{{2\,m}\over \sqrt{2}} \lambda_1 \, \sin \theta \, \delta_{\sigma_3, - \sigma_4} $$ 
$$\ov{U_3} \, \gs(\epsilon_1)\,\gs_5\, V_4 =  \delta_{\sigma_3, \sigma_4}\,(2 \sigma_3) \,\beta\, \sqrt{\di{s \over 2}} \left[ \hskip -0.1cm\di{{}\over{}} 1 + \lambda_1 (2 \sigma_3) \cos \theta \right] $$

\nin conduisent enfin \`a l'expression 

\beq \fbox{\rule[-0.1cm]{0cm}{1.5cm}~$ \begin{array}{c} ~\\ T(\lambda_1, 0\, ;\sigma_3, \sigma_4) = \sqrt{ \di{{s\,t_2} \over 2}} \, \di{ \beta \over{a\,b}} \left\{ \hskip -0.1cm\di{{}\over{}}   \delta_{\sigma_3, \sigma_4} \left[ \hskip -0.1cm\di{{}\over{}} 1 + \lambda_1 (2 \sigma_3) \cos \theta \right]  \left[ \hskip -0.1cm\di{{}\over{}} - (s + t_2 - t_1) \cos \theta \right. \right.  \\~\\
\left. \left. \hskip -0.1cm\di{{}\over{}}  + (2 \sigma_3) \lambda_1 (s + t_1 + t_2) \right] +\delta_{\sigma_3, - \sigma_4} \di{{2\, m}\over \sqrt{s}} \lambda_2 \,(s+t_2 -t_1)\,\sin \theta \right\} \\~ 
\end{array} $~}  \enq

\vv\nin et explicitement aux amplitudes 

\beq \fbox{\fbox{\rule[-0.5cm]{0cm}{1.2cm}~$ \begin{array}{c}
~\\T(+,0\, ;\uw, \uw) = T(-,0\, ;\dw, \dw) = \sqrt{ \di{{s\,t_2} \over 2}} \, \di{ \beta \over{a\,b}}  \left[ \hskip -0.1cm\di{{}\over{}} 1 + \cos \theta \right] \left[ \hskip -0.1cm\di{{}\over{}} - (s + t_2 - t_1) \cos \theta \right. \\~\\
\left.\hskip -0.1cm\di{{}\over{}}  +  (s + t_1 + t_2) \right]  \\~\\
T(+,0\, ;\dw, \dw) = T(-,0\, ;\uw, \uw) =  - \sqrt{ \di{{s\,t_2} \over 2}} \, \di{ \beta \over{a\,b}}  \left[ \hskip -0.1cm\di{{}\over{}} 1 -  \cos \theta \right] \left[ \hskip -0.1cm\di{{}\over{}} (s + t_2 - t_1) \cos \theta \right. \\~\\
\left.\hskip -0.1cm\di{{}\over{}}  +  (s + t_1 + t_2) \right]  \\~\\
T(+,0\,; \uw, \dw) = T(+,0\,; \dw, \uw) = - T(-,0\,; \uw, \dw) =  - T(-,0\,; \dw, \uw)  \\~\\
= \di{{m \beta\, \sqrt{2\, t_2}} \over{a\,b}}\, (s + t_2 - t_1)\, \sin \theta \cos \theta \\~
\end{array}   $~}}  \enq

\newpage
\vv \nin $\bullet$ \und{\bf Amplitudes $T(\lambda_1,\lambda_2\, ;\sigma_3, \sigma_4)$, ($\lambda_1, \lambda_2 = \pm 1$)}

\vv \nin Ecrivons l'amplitude g\'en\'erique sous la forme 

$$T = \ov{U_3} \left\{ \di{1\over a} \left[ \hskip -0.1cm\di{{}\over{}} 2\, p_3 \cdot \epsilon_2\, \gs(\epsilon_2) - \gs(\epsilon_1) \gs(p_1) \gs(\epsilon_2) \right] \right.  \left.+  \di{1\over b} \left[ \hskip -0.1cm\di{{}\over{}}  -2\, p_4 \cdot \epsilon_1\, \gs(\epsilon_2) + \gs(\epsilon_2) \gs(p_1) \gs(\epsilon_1) \right] \right]  V_4 $$ 

\vv \nin avec $\epsilon_1 = \di{1\over \sqrt{2}} \left( - \lambda_1 X - i Y \right)$, $\epsilon_2 = \di{1\over \sqrt{2}} \left(  \lambda_2 X - i Y \right)$. On a $ 2 \,p_3 \cdot \epsilon_1 = - 2\, p_4\cdot \epsilon_1 = \lambda_1\, \sqrt{\di{s \over 2}}\,\beta\, \sin \theta $, et de $\gs(\epsilon_1)\,\gs(\epsilon_2) = \delta_{\lambda_1, \lambda_2} \left[  1 - \lambda_1 \gs(T_1)\, \gs(z_1)\, \gs_5 \right]$, $\gs(\epsilon_2)\,\gs(\epsilon_1) = \delta_{\lambda_1, \lambda_2} \left[  1 +\lambda_1 \gs(T_1)\, \gs(z_1)\, \gs_5 \right]$, on tire 

$$ \gs(\epsilon_1)\, \gs(p_1)\, \gs(\epsilon_2) = \sqrt{t_1}\,\delta_{\lambda_1, \lambda_2} \left[\, \gs(z_1) - \lambda_1\, \gs(T_1)\, \gs_5 \,\right]$$ 
$$\gs(\epsilon_2)\, \gs(p_1)\, \gs(\epsilon_1) = \sqrt{t_1}\,\delta_{\lambda_1, \lambda_2} \left[\, \gs(z_1) + \lambda_1\, \gs(T_1)\, \gs_5 \,\right]$$

\vv \nin D'o\`u 

$$ a\,b\, T = (a+b)\, \left\{ \beta\, \lambda_1\, \sqrt{\di{s \over 2}} \sin \theta~ \ov{U_3} \,\gs(\epsilon_2)\, V_4 - \delta_{\lambda_1, \lambda_2}\, \sqrt{t_1} ~\ov{U_3} \,\gs(T_1)\,\gs_5\, V_4 \right\} $$
$$+ \delta_{\lambda_1, \lambda_2}\, \sqrt{t_1}\,(b-a)~\ov{U_3} \,\gs(z_1)\, V_4 =  (s+t_1 +t_2)\, \left\{ \beta\, \lambda_1\, \sqrt{\di{s \over 2}} \sin \theta~ \ov{U_3} \,\gs(\epsilon_2)\, V_4 - \delta_{\lambda_1, \lambda_2}\, \sqrt{t_1} ~\ov{U_3} \,\gs(T_1)\,\gs_5\, V_4 \right\} $$
$$+ \delta_{\lambda_1, \lambda_2}\, \sqrt{t_1}\,\beta\,\cos \theta\,\Ld~\ov{U_3} \,\gs(z_1)\, V_4 $$

\vv \nin Exprimons les formes bilin\'eaires : 

$$ \ov{U_3} \,\gs(\epsilon_2)\, V_4 = \di{1\over \sqrt{2}} \left[ \hskip -0.1cm \di{{}\over{}} \delta_{\sigma_3, \sigma_4}\, \sqrt{s} \, \left[ 1 - \lambda_2 (2 \sigma_3) \cos \theta \right] + 2\, m\, \lambda_2\, \delta_{\sigma_3, - \sigma_4} \sin \theta \right] $$
$$ \ov{U_3} \,\gs(T_1)\,\gs_5\, V_4 = \delta_{\sigma_3, \sigma_4}\, \beta\,\sqrt{s}\, \sin \theta\, \sinh \eta_1 -2\,m\,(2 \sigma_3)\,\delta_{\sigma_3, - \sigma_4}\, \cosh \eta_1\,\cos \theta $$
$$ = \di{1\over{2 \sqrt{s t_1}}} \left[\hskip -0.1cm \di{{}\over{}}  (s + t_2 - t_1) \delta_{\sigma_3, \sigma_4}\, \beta\,\sqrt{s}\, \sin \theta -2\,m\,(2 \sigma_3)\,\delta_{\sigma_3, - \sigma_4}\, \Ld \cos \theta \right] $$
$$ \ov{U_3} \,\gs(z_1)\, V_4 = \cosh \eta_1 \left[\hskip -0.1cm \di{{}\over{}} (2 \sigma_3)\,\sqrt{s}\, \delta_{\sigma_3, \sigma_4}\,\sin \theta + 2\,m\, \delta_{\sigma_3, - \sigma_4}\,\cos \theta \right] $$
$$ = \di{\Ld\over{2 \sqrt{s t_1}}}  \left[\hskip -0.1cm \di{{}\over{}} (2 \sigma_3)\,\sqrt{s}\, \delta_{\sigma_3, \sigma_4}\,\sin \theta + 2\,m\, \delta_{\sigma_3, - \sigma_4}\,\cos \theta \right] $$

\vv \nin Posant $Z = s+t_1+t_2$ et $X = s+t_2-t_1$, on en d\'eduit l'amplitude g\'en\'erique

\beq \fbox{\fbox{\rule[-0.2cm]{0cm}{1.5cm}~$ \begin{array}{c} ~\\ T(\lambda_1, \lambda_2\,; \sigma_3, \sigma_4) = \delta_{\sigma_3, \sigma_4}\,\di{{\beta \sin \theta}\over{2\,a\,b}}  \left\{\hskip -0.1cm \di{{}\over{}} \lambda_1\,s\,Z \left[ 1 -\lambda_2(2 \sigma_3) \cos \theta \right] \right.  \\~\\
\left.\di{{}\over{}}  - \delta_{\lambda_1, \lambda_2}\,\lambda_1\, X\,Z + \delta_{\lambda_1, \lambda_2}\,(2 \sigma_3)\, \Lambda\, \cos \theta \right\} \\~\\
+ \delta_{\sigma_3, -\sigma_4}\,\di{m \over {a b \,\sqrt{s}}} \left\{\hskip -0.1cm \di{{}\over{}} \lambda_1\, \lambda_2\, s \beta Z \sin^2 \theta + \lambda_1 (2 \sigma_3) \, \delta_{\lambda_1, \lambda_2}\, Z \Ld \right.\\~\\
 \left.\di{{}\over{}}  + \delta_{\lambda_1, \lambda_2}\,\beta\, \Lambda \,\cos^2 \theta \right\} \\~
\end{array} $~}}  \enq

\newpage 

\nin et explicitement les 16 amplitudes suivantes :  

\beq \fbox{\fbox{\rule[-0.1cm]{0cm}{2cm}~$ \begin{array}{c}
~\\T(+,+\, ;\uw, \uw) = -T(-,-\, ;\dw, \dw) =\di{{\beta \sin \theta}\over{2\,a\,b}} \left\{ \hskip -0.1cm\di{{}\over{}}  s Z (1 - \cos \theta) + \Lambda\, \cos \theta - X\,Z   \right\} \\~\\
T(+,+\, ;\dw, \dw) = - T(-,-\, ;\uw, \uw) =\di{{\beta \sin \theta}\over{2\,a\,b}} \left\{ \hskip -0.1cm\di{{}\over{}}  s Z (1 + \cos \theta) - \Lambda\, \cos \theta - X\,Z   \right\} \\~\\

T(+,+\, ;\uw, \dw) = T(-,-\, ;\dw, \uw) =  \di{m \over{a\,b\, \sqrt{s}}} \left\{\hskip -0.1cm \di{{}\over{}} s \beta Z \sin^2 \theta + \beta \Lambda \cos^2 \theta + Z \Ld  \right\} \\~\\
T(+,+\, ;\dw, \uw) = T(-,-\, ;\uw, \dw) =  \di{m \over{a\,b\, \sqrt{s}}} \left\{\hskip -0.1cm \di{{}\over{}} s \beta Z \sin^2 \theta + \beta \Lambda \cos^2 \theta - Z \Ld  \right\} \\~\\
T(+,-\,; \uw, \uw) = -T(-,+\,; \dw, \dw) =  \di{{s Z \beta \sin \theta}\over{2\,a\,b}} (1 + \cos \theta)  \\~\\
T(+,-\,; \dw, \dw) = -T(-,+\,; \uw, \uw) =  \di{{s Z \beta \sin \theta}\over{2\,a\,b}} (1 - \cos \theta)  \\~\\
T(+,-\,; \uw, \dw) = T(+,-\,; \dw, \uw) = T(-,+\,; \uw, \dw) = T(-,+\,; \dw, \uw) \\~\\
= - \di{{m\, \beta \, Z\,\sqrt{s}\,\sin^2 \theta}\over{a\,b}} \\~ 

\end{array}   $~}}  \enq

\vv
\section{Amplitudes d'h\'elicit\'e de $\gs^\star +\, \gs^\star \rightarrow \pi^{-} +\, \pi^{+} $}

\vv \nin A titre de comparaison avec les amplitudes d'h\'elicit\'e du paragraphe pr\'ec\'edent, il nous para\^it int\'eressant de calculer celles correspondant \`a la production, par collision de deux photons virtuels, d'une paire de particules sans spin, telles que les pions $\pi^-$ et $\pi^+$. Les pions \'etant des particules hadroniques, leur interaction \'electromagn\'etique est certainement plus compliqu\'ee que celle obtenue en appliquant na\"ivement le principe de couplage minimum au lagrangien libre d'un pion charg\'e. Cependant, dans notre optique de comparaison, il nous suffira de consid\'erer l'amplitude de Born dudit processus, d\'eduite de ce principe. Utilisant les m\^emes notations, elle s'\'ecrit 

$$ A(\lambda_1, \lambda_2) = 4 \pi \alpha\, \epsilon^\mu_1\, \epsilon^\nu_2\, T_{\mu \nu}~~~{\rm avec} $$ 

$$ T_{\mu \nu} = - 2 g_{\mu \nu} + \di{1\over a}\, (2 p_3 - p_1)_\mu (2 p_4 - p_2)_\nu + \di{1\over b} (2p_4 - p_1)_\mu (2p_3 -p_2)_\nu $$

\vv \nin $\bullet$ \und{\bf Amplitude $T(0,0)$}

\vv \nin En tenant compte du fait que $p^\mu_1\, T_{\mu \nu} = T_{\mu \nu}\, p^\nu_2 = 0$, on obtient 

$$T(0,0) = \di{{4 \sqrt{t_1 t_2}}\over \Ld}\, P^\mu  P^\nu T_{\mu \nu} = \di{{4 \sqrt{t_1 t_2}}\over \Ld}\,\left[\hskip -0.1cm  \di{{}\over{}} - 2s + 4(P\cdot p_3)(P \cdot p_4) \left( \di{1\over a} + \di{1 \over b} \right) \right] ~~~{\rm soit} $$

\beq \fbox{\fbox{\rule[-0.5cm]{0cm}{1.2cm}~$ \begin{array}{c}
T(0,0) =  \di{{ 4 s\, \sqrt{t_1 t_2} }\over{ \Ld}} \left[\, \di{{s Z}\over{a\,b}} - 2 \,\right]    
\end{array}   $~}}  \enq

\newpage
\nin $\bullet$ \und{\bf Amplitudes $T(\lambda_1,0)$, ($\lambda_1 = \pm 1$)}

$$ T(\lambda_1, 0) = \di{{2 \sqrt{t_2}}\over \Ld} \left[\di{4\over a}\,(\epsilon_1 \cdot p_3)(P\cdot p_4)+ \di{4\over b}\,(\epsilon_1 \cdot p_4)(P\cdot p_3) \right]= \di{{8 \sqrt{t_2}}\over \Ld} (\epsilon_1 \cdot p_3)(P\cdot p_4)\left[ \di{1\over a} - \di{1\over b} \right]$$
$$ = \di{{2 s \sqrt{t_2}}\over \Ld} \left[  \lambda_1 \sqrt{\di{s\over 2}} \beta \sin \theta \right] \left[ \di{1\over a} - \di{1\over b} \right] ~~~{\rm soit} $$

\beq \fbox{\fbox{\rule[-0.5cm]{0cm}{1.2cm}~$ \begin{array}{c}
T(\lambda_1,0) = \lambda_1\di{{s \sqrt{2 s t_2}\, \beta^2 \sin \theta \cos \theta} \over{a\, b}}       
\end{array}   $~}}  \enq

\vv \nin $\bullet$ \und{\bf Amplitudes $T(0, \lambda_2)$, ($\lambda_2 = \pm 1$, $\epsilon_2 = \epsilon^{(-\lambda_2)}_1$)}

$$ T(0, \lambda_2) = \di{{2 \sqrt{t_1}}\over \Ld} \left[\di{4\over a}\,(P\cdot p_3)(\epsilon_2 \cdot p_4)+ \di{4\over b}\,(P\cdot p_4) (\epsilon_2 \cdot p_3)\right]= \di{{8 \sqrt{t_1}}\over \Ld} (P\cdot p_3)(\epsilon_2 \cdot p_4)\left[ \di{1\over a} - \di{1\over b} \right]$$
$$ = \di{{2 s \sqrt{t_1}}\over \Ld} \left[  \lambda_2 \sqrt{\di{s\over 2}} \beta \sin \theta \right] \left[ \di{1\over a} - \di{1\over b} \right] ~~~{\rm soit} $$

\beq \fbox{\fbox{\rule[-0.5cm]{0cm}{1.2cm}~$ \begin{array}{c}
T(0,\lambda_2) = \lambda_2\di{{s \sqrt{2 s t_1}\, \beta^2 \sin \theta \cos \theta} \over{a\, b}}       
\end{array}   $~}}  \enq

\vv \nin $\bullet$ \und{\bf Amplitudes $T(\lambda_1, \lambda_2)$, ($\lambda_1, \lambda_2 = \pm 1$)}

$$ T(\lambda_1, \lambda_2) = -2 \delta_{\lambda_2, \lambda_1} + \di{{4(\epsilon_1 \cdot p_3)(\epsilon_2 \cdot p_4)}\over a} +  \di{{4(\epsilon_1 \cdot p_4)(\epsilon_2 \cdot p_3)}\over b}  =  -2 \delta_{\lambda_2, \lambda_1}  $$
$$ + \lambda_1\, \lambda_2\, s \,\beta^2 \sin^2 \theta \left[ \di{1\over a} + \di{1\over b} \right]~~~~{\rm soit} $$

\beq \fbox{\fbox{\rule[-0.5cm]{0cm}{1.2cm}~$ \begin{array}{c}
T(\lambda_1 ,\lambda_2) = -2 \, \delta_{\lambda_2, \lambda_1} + \lambda_1\, \lambda_2\, \di{{s Z\,\beta^2 \sin^2 \theta}\over{a\,b\,}}  
\end{array}   $~}}  \enq

\newpage
\section{Amplitudes d'h\'elicit\'e de $e^- +\, e^- \rightarrow e^{-} +\, e^{-} $ \label{secmoller}}

\vvv
\begin{figure}[hbt]
\centering
\includegraphics[scale=0.3, width=9cm, height=3.5cm]{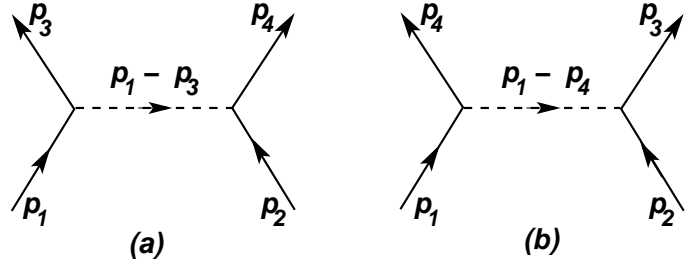}
\vskip 0.25cm

\caption{(a) : Diagrammes de Feynman de la diffusion M\o ller. } \label{moller}
\end{figure}

\nin Dans ce processus, appel\'e  diffusion M\o ller\footnote{C. M\o ller, Ann. d. Physik 14, 568 (1932).}, les particules initiales et finales sont toutes identiques et donc de m\^eme masse $m$. Les notations sont celles de la figure (3.1). Au plus bas ordre en $\alpha$, elle est repr\'esent\'ee par les deux diagrammes de Feynman de la figure (\ref{moller}) et son amplitude, divis\'ee par $4 \pi \alpha$ s'\'ecrit\footnote{Le signe ``-" entre les deux termes de l'amplitude provient du fait que le passage de l'un \`a l'autre implique l'\'echange de deux fermions.}

\beq T_m = \di{1 \over t} \,\ov{U_3} \,\gs_\mu\, U_1\, \ov{U_4}\, \gs^\mu \,U_2 - \di{1\over u}\, \ov{U_4}\, \gs_\mu\, U_1\, \ov{U_3}\, \gs^\mu\, U_2 \enq
$$ t = - (p_1 - p_3)^2\, ,~~~u = -(p_1 - p_4)^2 $$

\vv \nin Les facteurs $X_1 = \ov{U_3}_{\sigma_3} \,\gs_\mu\, {U_1}_{\sigma_1}\, \ov{U_4}_{\sigma_4}\, \gs^\mu \,{U_2}_{\sigma_2}$ et $X_2 = \ov{U_4}_{\sigma_4} \,\gs_\mu\, {U_1}_{\sigma_1}\, \ov{U_3}_{\sigma_3}\, \gs^\mu \,{U_2}_{\sigma_2}$ peuvent \^etre calcul\'es facilement en utilisant des formules du type  (3.69), avec des spineurs normalis\'es selon $\ov{U} U = 2 m$. En effet, prenant

$$\ov{U_3}_{ \sigma_3}\,\gs_\mu\,U_{1 \sigma_1} = 2\,m\, \delta_{\sigma_3, \sigma_1} \left \llbracket \,T_\mu\,\cosh \chi \cos \di{\theta \over 2}  \right.$$
$$\left. + \sinh \chi\left[ Z_\mu\, \cos \di{\theta \over 2} +(X+ 2i \sigma_1 Y)_\mu\, \sin \di{\theta \over 2} \,\right] \right \rrbracket  
 - 2\,m\,(2 \sigma_1)\, \delta_{\sigma_3, - \sigma_1}\, T_\mu\, \sin \di{\theta \over 2} $$

$$\ov{U_4}_{ \sigma_4}\,\gs_\mu\,U_{2 \sigma_2} = 2\, m\,\delta_{\sigma_4, \sigma_2} \left \llbracket \,T_\mu\,\cosh \chi \cos \di{\theta \over 2}  \right.$$
$$\left. + \sinh \chi\left[ -Z_\mu\, \cos \di{\theta \over 2} +(-X+ 2i \sigma_2 Y)_\mu\, \sin \di{\theta \over 2} \,\right] \right \rrbracket  
 - 2\,(2 \sigma_2)\,m\, \delta_{\sigma_4, - \sigma_2}\, T_\mu\, \sin \di{\theta \over 2} $$

$$\ov{U_3}_{ \sigma_3}\,\gs_\mu\,U_{2 \sigma_2} = 2\, m\,\delta_{\sigma_3, \sigma_2} \left \llbracket \,T_\mu\,\cosh \chi \sin \di{\theta \over 2}  \right.$$
$$\left. + \sinh \chi\left[ -Z_\mu\, \sin \di{\theta \over 2} +(X- 2i \sigma_2 Y)_\mu\, \sin \di{\theta \over 2} \,\right] \right \rrbracket  
 + 2\,(2 \sigma_2)\,m\, \delta_{\sigma_3, - \sigma_2}\, T_\mu\, \cos \di{\theta \over 2} $$

$$\ov{U_4}_{ \sigma_4}\,\gs_\mu\,U_{1 \sigma_1} = -\,2\, m\,\delta_{\sigma_4, \sigma_1} \left \llbracket \,T_\mu\,\cosh \chi \sin \di{\theta \over 2}  \right.$$
$$\left. + \sinh \chi\left[ Z_\mu\, \sin \di{\theta \over 2} -(X +2i \sigma_1 Y)_\mu\, \sin \di{\theta \over 2} \,\right] \right \rrbracket  
 - 2\,(2 \sigma_1)\,m\, \delta_{\sigma_4, - \sigma_1}\, T_\mu\, \cos \di{\theta \over 2} $$

\vv \nin et effectuant les produits scalaires appropri\'es tout en tenant compte de $1 + 4 \sigma_1 \sigma_2 = 2\, \delta_{\sigma_2, \sigma_1}$, on trouve 

\beq \fbox{\rule[-0.2cm]{0cm}{3.2cm}~$ \begin{array}{c} 

X_1 = 4\,m^2\, \left[ \cosh \chi\,\cos \di{\theta \over 2} \,\delta_{\sigma_3, \sigma_1}  - (2 \sigma_1) \delta_{\sigma_3, - \sigma_1} \sin \di{\theta \over 2}\right] \left[ \cosh \chi\,\cos \di{\theta \over 2} \,\delta_{\sigma_4, \sigma_2}  \right. \\~\\
\left. - (2 \sigma_2) \delta_{\sigma_4, - \sigma_2} \sin \di{\theta \over 2} \right]   + 4\,m^2\, \delta_{\sigma_3, \sigma_1}\, \delta_{\sigma_4, \sigma_2} \sinh^2 \chi\, \left[ \cos^2 \di{\theta \over 2} + 2 \,\delta_{\sigma_2, \sigma_1} \sin^2 \di{\theta \over 2} \right]  \\~\\

X_2 = - 4\,m^2\, \left[   \cosh \chi \, \sin \di{\theta \over 2}\,\delta_{\sigma_4, \sigma_1} + (2 \sigma_1)\delta_{\sigma_4, - \sigma_1}  \cos \di{\theta \over 2}   \right]   \left[   \cosh \chi \, \sin \di{\theta \over 2}\,\delta_{\sigma_3, \sigma_2} \right. \\~\\
\left.+ (2 \sigma_2) \cos \di{\theta \over 2} \delta_{\sigma_3, \sigma_2}  \right]   - 4\, m^2\, \delta_{\sigma_4, \sigma_1} \delta_{\sigma_3, \sigma_2}\, \sinh^2 \chi\,\left[ \sin^2 \di{\theta \over 2} + 2 \delta_{\sigma_2, \sigma_1} \cos^2 \di{\theta \over 2} \right] \\~
\end{array} 
$~}  
\enq

\vv\nin On en d\'eduit les expressions des amplitudes $T_m(\sigma_1, \sigma_2\,, ; \sigma_3, \sigma_4)$ du tableau V.

\beq \fbox{\fbox{\rule[-0.7cm]{0cm}{6cm}~$ \begin{array}{c} 

T_m( \uw, \uw\, ; \uw, \uw) = T_m( \dw, \dw\, ; \dw, \dw)  =  \di{4 \over \beta^2}\left[ \di{{1+\beta^2}\over \sin^2 \theta} - \di{2 m^2 \over s } \right] \\~\\

 T_m(\uw, \uw\, ; \dw, \dw) = T_m(\dw, \dw\, ; \uw, \uw)  = \di{{8 m^2}\over{s \beta^2}} \\~\\

T_m(\uw, \uw\, ; \uw, \dw) = T_m(\uw, \uw\, ; \dw, \uw) = - T_m(\uw, \dw\, ; \uw, \uw) = -  T_m(\dw, \uw\, ; \uw, \uw) \\~\\
=  - T_m(\dw, \dw\, ; \dw, \uw) =  - T_m(\dw, \dw\, ; \uw, \dw) = T_m(\dw, \uw\, ; \dw, \dw) =  T_m(\uw, \dw\, ; \dw, \dw) \\~\\
= - \di{{4m}\over{\sqrt{s} \,\beta^2}} \di{{\cos \theta}\over{\sin \theta}} \\~\\ 
T_m(\uw, \dw\, ; \dw, \uw) = T_m(\dw, \uw\, ; \uw, \dw) = \di{1\over \beta^2} \left[ \di{{1 + \beta^2}\over{\sin^2 \di{\theta \over 2}}} -2 \right] \\~\\ 

T_m(\uw, \dw\, ; \uw, \dw) = T_m(\dw, \uw\, ; \dw, \uw) = \di{1\over \beta^2} \left[ \di{{1 + \beta^2}\over{\cos^2 \di{\theta \over 2}}} -2 \right] \\~

\end{array} 
$~}}  
\enq

\vv 
\centerline{\bf Tableau V - Amplitudes d'h\'elicit\'e de la diffusion M\o ller}

\vv \nin A titre de v\'erification de ces formules, nous proposons au lecteur de retrouver l'expression bien 
connue\footnote{Voir par exemple, V. Berestetski, E. Lifchitz, L. Pitayevski, ``Electrodynamique Quantique", Cours de Physique Th\'eorique de L. Landau et E. Lifchitz, Tome 4, \S\hskip 0.05cm 81, Ed. Librairie du Globe et Ed. Mir (1989).} de la somme des carr\'es des amplitudes $T_m$ (obtenue par un calcul ordinaire de trace) :

\beq  \di{\sum_{\sigma_i}}\,\, |\,T_m\,|^2 = 16\left(\di{{1 + \beta^2}\over \beta^2} \right)^2 \left\{ \di{4 \over{ \sin^4 \theta}} - \di{3 \over{\sin^2 \theta}} + \left( \di{\beta^2 \over{1 + \beta^2}} \right)^2 \left[ 1 + \di{4 \over{\sin^2 \theta}} \right] \right\} \enq

\section{Amplitudes d'h\'elicit\'e de \,$\gs + e^- \rightarrow e^- + \mu^- + \mu^+$ \label{sectrident}}

\begin{figure}[hbt]
\centering
\includegraphics[scale=0.3, width=13cm, height=5cm]{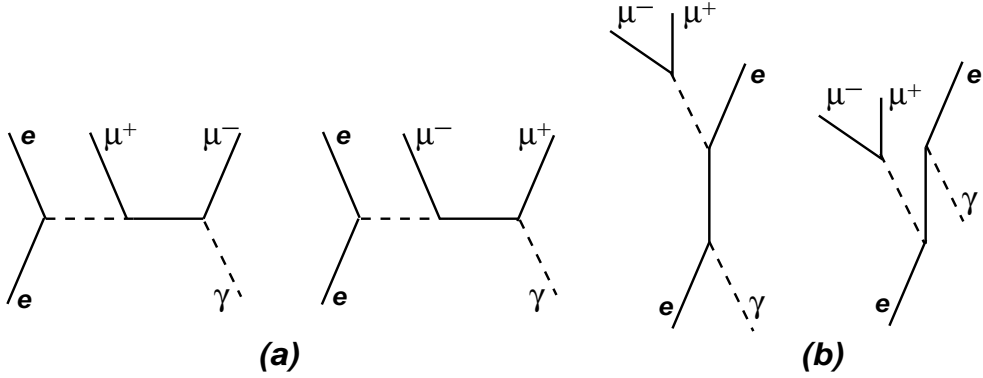}
\vskip 0.25cm

\caption{Diagrammes de Feynman pour $\gs + e^- \rightarrow e^- + \mu^- + \mu^+$, sans \'echange de $Z$, ni production de r\'esonance vectorielle} \label{fig:trident2}
\end{figure}

\vv \nin Ce processus de production de trois leptons (``trident") par collision \'electron-photon, dont une paire muon $\mu^-$ $-$ anti-muon $\mu^+$, sera ici envisag\'e du seul point de vue de l'Electrodynamique Quantique, excluant ainsi un \'echange possible de la particule $Z$ ou la production de particule (hormis un photon) ou de r\'esonance se d\'esint\'egrant ensuite en une paire $\mu^- \mu^+$. Consid\'er\'e \`a l'ordre le plus bas vis-\`a-vis de la constante \'electromagn\'etique $\alpha$, il est d\'ecrit par les diagrammes de Feynman de la figure \ref{fig:trident2}. Cette figure contient deux s\'eries de diagrammes, repr\'esentant chacune une voie possible pour ce processus. La premi\`ere s\'erie (a) se rapporte \`a ``l'\'emission" par l'\'electron d'un photon virtuel du genre espace, conduisant au sous-processus $\gs + \gs^\star \rightarrow \mu^- + \mu^+$. Dans la seconde s\'erie (b) appara\^it, comme sous-processus, une diffusion Compton virtuelle $\gs + e \rightarrow \gs^\star + e$, o\`u le photon virtuel \'emis est cette fois du genre temps, ce photon ``lourd" se ``d\'esint\'egrant" par la suite en une paire $\mu^- \mu^+$. 

\vvv

\vv \nin Dans ce qui suit, les amplitudes d'h\'elicit\'e dudit processus seront calcul\'ees uniquement dans le cas des masses nulles, en utilisant le couplage d'h\'elicit\'e sym\'etrique pour l'\'etat final et le couplage en voie $s$ pour l'\'etat initial. L'amplitude g\'en\'erique est la somme des amplitudes correspondant aux diagrammes (a) et (b) respectivement :   

$$ T_{\rm trident}  = \epsilon^\mu_2 \left[ T^a_\mu + T^b_\mu \right],~~~{\rm avec} $$
$$T^a_\mu  = - \di{{\ov{U}_3 \,\gs^\nu\,U_1}\over{(p_1 - p_3)^2}}\, \ov{U}_4 \,T^a_{\mu \nu}\,  V_5,~~~T^b_\mu =    - \di{{\ov{U}_4 \,\gs^\nu\,V_5}\over{(p_4 + p_5)^2}}\, \ov{U}_3 \,T^b_{\mu \nu}\, U_1,~~~{\rm o\grave{u}}  $$
 \beq T^a_{\mu \nu} = \left[ \di{p_{4 \mu}\over{p_2 \cdot p_4} } - \di{p_{5 \mu}\over{p_2 \cdot p_5} } \right] \gs_\nu - \di{1\over{2 p_2 \cdot p_4}} \gs_\mu \, \gs(p_2)\,\gs_\nu + \di{1\over{2 p_2 \cdot p_5}} \gs_\nu \, \gs(p_2)\,\gs_\mu, \label{TaTb} \enq 
$$ T^b_{\mu \nu} = \left[ \di{p_{3 \mu}\over{p_2 \cdot p_3} } - \di{p_{1 \mu}\over{p_2 \cdot p_1} } \right] \gs_\nu - \di{1\over{2 p_2 \cdot p_3}} \gs_\mu \, \gs(p_2)\,\gs_\nu - \di{1\over{2 p_2 \cdot p_1}} \gs_\nu \, \gs(p_2)\,\gs_\mu$$

\vv \nin La constante de couplage $e^3 = (4 \pi \alpha)^{3/2}$ a \'et\'e extraite. Les indices sont attribu\'es ainsi : indice 1 pour l'\'electron incident, indices 2 et $\mu$ pour le photon r\'eel incident, indice 3 pour l'\'electron \'emergeant, indice 4 pour le muon $\mu^-$, indice 5 pour son anti-particule $\mu^+$.  L'\'electron incident est suppos\'e se propager selon l'axe $Z$ du r\'ef\'erentiel du centre de masse de la r\'eaction, et le photon incident, en sens inverse. Par cons\'equent, on a 

$$ p_1 = \di{\sqrt{s} \over 2} \left[T+Z \right],~~~p_2 =  \di{\sqrt{s} \over 2} \left[T-Z \right],~~~\epsilon^{(\lambda)}_2 = \di{1\over \sqrt{2}} \left[ \lambda X - i Y \right],~~~{\rm o\grave{u}} $$
$$ s = (p_1 + p_2)^2 = (p_3 + p_4 + p_5)^2$$


\vv \nin Nous utiliserons la d\'ecomposition 

$$ \gs_\mu \gs_\alpha \gs_\nu = g_{\mu \alpha}\, \gs_\nu - g_{\mu \nu} \gs_\alpha + g_{\alpha \nu}\gs_\mu - i \epsilon_{\mu \alpha \nu \omega} \gs^\omega \gs_5 $$

\vv \nin pour exprimer les tenseurs (\ref{TaTb}) sous la forme 

$$T^a_{\mu \nu} = A^a_{\mu \nu \alpha} \, \gs^\alpha + i \,B^a_{\mu \nu \alpha} \, \gs^\alpha \gs_5,~~~ T^b_{\mu \nu} = A^b_{\mu \nu \alpha} \, \gs^\alpha + i \,B^b_{\mu \nu \alpha} \, \gs^\alpha \gs_5,~~~{\rm avec}$$

$$A^a_{\mu \nu \alpha} =  \left[ \di{p_{4 \mu}\over{p_2 \cdot p_4} } - \di{p_{5 \mu}\over{p_2 \cdot p_5} } \right] g_{\nu \alpha} + \di{1\over 2} \left[ \di{1\over{ p_2\cdot p_5}} - \di{1\over{ p_2 \cdot p_4}} \right] \left[ p_{2 \mu} g_{\nu \alpha} -g_{\mu \nu} p_{2 \alpha} + p_{2 \nu} g_{\mu \alpha} \right]$$
$$B^a_{\mu \nu \alpha} = \di{1\over 2} \left[ \di{1\over{ p_2\cdot p_5}} + \di{1\over{ p_2 \cdot p_4}} \right]\,\epsilon_{\mu \nu \alpha \omega}\, p^\omega_2 $$

$$A^b_{\mu \nu \alpha} =  \left[ \di{p_{3 \mu}\over{p_2 \cdot p_3} } - \di{p_{1 \mu}\over{p_2 \cdot p_1} } \right] g_{\nu \alpha} - \di{1\over 2} \left[ \di{1\over{ p_2\cdot p_1}} + \di{1\over{ p_2 \cdot p_3}} \right] \left[ p_{2 \mu} g_{\nu \alpha} -g_{\mu \nu} p_{2 \alpha} + p_{2 \nu} g_{\mu \alpha} \right]$$
$$B^b_{\mu \nu \alpha} = \di{1\over 2} \left[ \di{1\over{ p_2\cdot p_3}} - \di{1\over{ p_2 \cdot p_1}} \right]\,\epsilon_{\mu \nu \alpha \omega}\, p^\omega_2 $$

\vv \nin Supposant nulles les masses des leptons, on peut utiliser les formules suivantes\footnote{Voir section 3.5.}  

$$ V_5 = \gs_5 U_5 = (2 \sigma_5) U_5 $$

$$ \ov{U}_\ell\, \gs^\mu\,U_k = \delta_{\sigma_\ell \sigma_k} \, \sqrt{(2 E_\ell)(2E_k)} \, G^\mu_{\ell k}(\sigma_k) ,~~~{\rm avec}$$

$$  G^\mu_{\ell k}(\sigma) =  e^{i \sigma ( \vp_\ell - \vp_k ) } \cos \di{\theta_\ell \over 2}  \cos \di{\theta_k\over 2} \left[ \,T + Z\, \right]^\mu +  e^{- i \sigma( \vp_\ell - \vp_k ) } \sin \di{\theta_\ell \over 2}  \sin \di{\theta_k\over 2} \left[ \,T - Z\, \right]^\mu $$
$$ + \, e^{- i \sigma( \vp_\ell + \vp_k) } \sin \di{\theta_\ell \over 2}  \cos \di{\theta_k\over 2}  \left[X + i (2 \sigma) Y \right]^\mu  
+ e^{i \sigma( \vp_\ell + \vp_k) } \cos \di{\theta_\ell \over 2}  \sin \di{\theta_k\over 2}  \left[X - i (2 \sigma) Y \right]^\mu $$

\vv \nin On notera que 

$$ G_{n m}(\sigma') \cdot G_{\ell k}(\sigma) = 2\, \delta_{\sigma', \sigma}\, H_{n \ell}(\sigma) \, H^\star_{m k}(\sigma) + 2\, \delta_{\sigma', - \sigma}\,H_{m \ell}(\sigma)\,H^\star_{n k}(\sigma), ~~~{\rm o\grave{u}} $$
$$ H_{n \ell}(\sigma) = e^{i \sigma (\vp_n - \vp_\ell)} \, \cos \di{\theta_n \over 2}  \sin \di{\theta_\ell \over 2}  
- e^{- i \sigma (\vp_n - \vp_\ell)} \, \sin \di{\theta_n \over 2}  \cos \di{\theta_\ell \over 2} = - H_{\ell n}(\sigma),~~~{\rm avec} $$
$$\left|H_{n \ell}(\sigma)\right|^2 = \sin^2 \di{\theta_{n \ell} \over 2}  $$

\vv \nin \ding{172}~\und{\bf Amplitudes relatives \`a la s\'erie de diagrammes (a)} 

\vv \nin  Ecrivons 

\beq T^a = \epsilon^\mu_2\,T^a_\mu = \di{1 \over{2 p_1 \cdot p_3} }\, \sqrt{(2E_1)(2E_3)(2E_4)(2E_5) }\,\delta_{\sigma_3, \sigma_1}\, \delta_{\sigma_4, \sigma_5} (2 \sigma_4)\,X_a, 
\label{aTa}  \enq

$$ {\rm avec}~~~~~X_a = \epsilon^\mu_2\, G^\nu_{31}(\sigma_1)\, G^\alpha_{45}(\sigma_4) \left[ A^a_{\mu \nu \alpha} + i\, (2 \sigma_4)\,B^a_{\mu \nu \alpha} \right],    $$  

$${\rm soit}~~~~~ X_a = \left[ \di{{\epsilon_2 \cdot p_4}\over{p_2 \cdot p_4} } - \di{{\epsilon_2 \cdot p_5}\over{p_2 \cdot p_5} } \right] G_{45}(\sigma_4) \cdot G_{31}(\sigma_1)\,\, + $$
$$ +\,\, \di{1\over 2} \left[ \di{1\over{ p_2\cdot p_5}} - \di{1\over{ p_2 \cdot p_4}} \right] \left[  p_2 \cdot G_{31}(\sigma_1) \,\epsilon_2 \cdot G_{45} (\sigma_4) -  \epsilon_2 \cdot G_{31}(\sigma_1)\, p_2 \cdot G_{45}(\sigma_4)   \right] $$
$$ +\,\,(2 \sigma_4)  \di{ i\over 2} \, \epsilon_{\mu \nu \alpha \omega}\, \epsilon^\mu_2 \, p^\omega_2\, G^\nu_{31}(\sigma_1)\,G^\alpha_{45}(\sigma_4) \left[\di{1\over p_2 \cdot p_4} + \di{1 \over p_2 \cdot p_5} \right]$$

\nin Comme 

$$ \epsilon_{\mu \nu \alpha \omega}\, \epsilon^\mu_2 \, p^\omega_2 = - i \lambda \left[ p_{2 \nu} \, \epsilon_{2 \alpha} - p_{2 \alpha}\, \epsilon_{2 \nu} \right] $$

\vv \nin $X_a$ peut \^etre r\'ecrit sous la forme  

$$ X_a =  \left[ \di{{\epsilon_2 \cdot p_4}\over{p_2 \cdot p_4} } - \di{{\epsilon_2 \cdot p_5}\over{p_2 \cdot p_5} } \right] G_{45}(\sigma_4) \cdot G_{31}(\sigma_1)  $$
$$ + \,\left[  p_2 \cdot G_{31}(\sigma_1) \,\epsilon_2 \cdot G_{45} (\sigma_4) -  \epsilon_2 \cdot G_{31}(\sigma_1)\, p_2 \cdot G_{45}(\sigma_4)   \right] \left[ \di{{\delta_{\lambda, 2 \sigma_4} } \over p_2 \cdot p_5}      - \di{{\delta_{\lambda, - 2 \sigma_4} } \over p_2 \cdot p_4} \right] $$

\vv \nin Compte tenu du fait que $\theta_1 =0$, $\vp_1 =0$, $\sigma_3 = \sigma_1$, on a 

$$ p_2 \cdot G_{31}(\sigma_1) = \sqrt{s} \,\cos \di{\theta_3 \over 2}\, e^{i \sigma_1 \vp_3} ,~~~\epsilon_2 \cdot G_{31}(\sigma_1) =- \sqrt{2}\, \lambda\,\delta_{\lambda,  2 \sigma_1}\, \sin \di{\theta_3 \over 2}\,e^{-i \sigma_1 \vp_3} $$

\nin Puis 
$$ p_2 \cdot G_{45}(\sigma_4) = \sqrt{s}\, \cos \di{\theta_4 \over 2}\, \cos \di{\theta_5 \over 2}\,e^{i \sigma_4 (\vp_4 - \vp_5)} ,$$
$$ \epsilon_2 \cdot G_{45}(\sigma_4) = - \sqrt{2}\, \lambda\, \left[ e^{-i \sigma_4( \vp_4 + \vp_5)}\,\sin \di{\theta_4 \over 2}\,\cos \di{\theta_5 \over 2}\,\delta_{\lambda, 2 \sigma_4} + e^{i \sigma_4( \vp_4 + \vp_5)}\,\cos \di{\theta_4 \over 2}\,\sin \di{\theta_5 \over 2}\,\delta_{\lambda,- 2 \sigma_4} \right],$$
$$ \epsilon_2 \cdot p_k = -\di{\lambda \over \sqrt{2}} \,E_k\, \sin \theta_k\, e^{-i \lambda \vp_k},~~~p_2 \cdot p_k = \sqrt{s}\,E_k\, \cos^2 \di{\theta_k \over 2}  $$

\nin Calculons alors 

$$Q_1 =  p_2 \cdot G_{31}(\sigma_1) \,\epsilon_2 \cdot G_{45} (\sigma_4) -  \epsilon_2 \cdot G_{31}(\sigma_1)\, p_2 \cdot G_{45}(\sigma_4)  = - \lambda \, \sqrt{2 s} \,\cos \di{\theta_3 \over 2}\, e^{i \sigma_1 \vp_3} \, \times$$
$$\times \, \left\llbracket \,e^{-i \sigma_4( \vp_4 + \vp_5)}\,\sin \di{\theta_4 \over 2}\,\cos \di{\theta_5 \over 2}\,\delta_{\lambda, 2 \sigma_4} + e^{i \sigma_4( \vp_4 + \vp_5)}\,\cos \di{\theta_4 \over 2}\,\sin \di{\theta_5 \over 2}\,\delta_{\lambda,- 2 \sigma_4}   \right\rrbracket + \lambda \, \sqrt{2 s} \,\times$$
$$\times \, \delta_{\lambda,  2 \sigma_1}\, \sin \di{\theta_3 \over 2}\,e^{-i \sigma_1 \vp_3} \, \cos \di{\theta_4 \over 2}\, \cos \di{\theta_5 \over 2}\,e^{i \sigma_4 (\vp_4 - \vp_5)} $$

\vv \nin Utilisant la d\'ecomposition 

$$ Q_1 = \delta_{\sigma_4, \sigma_1}\,Q_1(\sigma_4 = \sigma_1) + \delta_{\sigma_4, - \sigma_1}\, Q_1(\sigma_4 = - \sigma_1), $$

\nin on trouve 

$$ Q_1 = - \lambda\, \sqrt{2s} \left\llbracket \hskip -0.02cm \di{{}\over {}} \delta_{\sigma_4, \sigma_1} 
\left\{ \cos \di{\theta_3 \over 2}\, e^{i \sigma_1 \vp_3} \, \left[ e^{-i \sigma_1 (\vp_4 + \vp_5)}  \,\sin \di{\theta_4 \over 2}\,\cos \di{\theta_5 \over 2}\,\delta_{\lambda, 2 \sigma_1}   \right.    \right.\right. $$
$$\left. \left.+\,  e^{i \sigma_1( \vp_4 + \vp_5)}\,\cos \di{\theta_4 \over 2}\,\sin \di{\theta_5 \over 2}\,\delta_{\lambda,- 2 \sigma_1} \right]  - \delta_{\lambda,  2 \sigma_1}\, \sin \di{\theta_3 \over 2}\,e^{-i \sigma_1 \vp_3} \, \cos \di{\theta_4 \over 2}\, \cos \di{\theta_5 \over 2}\,e^{i \sigma_1 (\vp_4 - \vp_5)} \right\}$$
$$+ \delta_{\sigma_4, - \sigma_1} \left\{  \cos \di{\theta_3 \over 2}\, e^{i \sigma_1 \vp_3} \, \left[  \,e^{i \sigma_1 ( \vp_4 + \vp_5)}\,\sin \di{\theta_4 \over 2}\,\cos \di{\theta_5 \over 2}\,\delta_{\lambda,  -2 \sigma_1} + e^{- i \sigma_1 ( \vp_4 + \vp_5)}\,\cos \di{\theta_4 \over 2}\,\sin \di{\theta_5 \over 2}\,\delta_{\lambda, 2 \sigma_1}   \right] \right. $$
$$\left. \left.  -  \delta_{\lambda,  2 \sigma_1}\, \sin \di{\theta_3 \over 2}\,e^{-i \sigma_1 \vp_3} \, \cos \di{\theta_4 \over 2}\, \cos \di{\theta_5 \over 2}\,e^{-i \sigma_1 (\vp_4 - \vp_5)} \,\right\} \,\right\rrbracket,~~{\rm soit} $$


$$ Q_1 = - \lambda\, \sqrt{2s} \left\llbracket \hskip -0.02cm \di{{}\over {}} \delta_{\sigma_4, \sigma_1} \left\{ \delta_{\lambda, - 2\sigma_1}\, \cos \di{\theta_3 \over 2}\,\cos \di{\theta_4 \over 2}\,\sin \di{\theta_5 \over 2}\,e^{i \sigma_1 (\vp_3 + \vp_4 + \vp_5)}  \right. \right. $$
$$\left. - \, \delta_{\lambda, 2 \sigma_1}\,\cos \di{\theta_5 \over 2}\, e^{- i \sigma_1 \vp_5} \, H_{43}(\sigma_1) \right\} + \delta_{\sigma_4, -\sigma_1} \left\{  \delta_{\lambda, - 2\sigma_1}\, \cos \di{\theta_3 \over 2}\,\cos \di{\theta_5 \over 2}\,\sin \di{\theta_4 \over 2}\,e^{i \sigma_1 (\vp_3 + \vp_4 + \vp_5)}  \right. $$
$$ \left.  \left.- \, \delta_{\lambda, 2 \sigma_1}\,\cos \di{\theta_4 \over 2}\, e^{-i \sigma_1 \vp_4} \, H_{53}(\sigma_1)  \right\}\,\right\rrbracket  $$

\nin Puis 

$$ Q_1 \left[ \di{{\delta_{\lambda, 2 \sigma_4} } \over p_2 \cdot p_5}      - \di{{\delta_{\lambda, - 2 \sigma_4} } \over p_2 \cdot p_4} \right] = - \lambda\, \sqrt{2s} \left\llbracket \,  \di{1\over p_2 \cdot p_5} \left[  \delta_{\sigma_4, -\sigma_1} \, \delta_{\lambda, - 2\sigma_1}\, \cos \di{\theta_3 \over 2}\,\cos \di{\theta_5 \over 2}\,\sin \di{\theta_4 \over 2}\,e^{i \sigma_1 (\vp_3 + \vp_4 + \vp_5)}   \right.  \right.$$
$$\left. - \,\delta_{\sigma_4, \sigma_1} \delta_{\lambda, 2\sigma_1} \cos \di{\theta_5 \over 2}\, e^{- i \sigma_1 \vp_5} \, H_{43}(\sigma_1) \right] - \di{1\over p_2 \cdot p_4} \left[  \delta_{\sigma_4, \sigma_1}  \delta_{\lambda, - 2\sigma_1}\, \cos \di{\theta_3 \over 2}\,\cos \di{\theta_4 \over 2}\,\sin \di{\theta_5 \over 2}\,e^{i \sigma_1 (\vp_3 + \vp_4 + \vp_5)}   \right. $$
\beq \left.\left. -\,\delta_{\sigma_4, -\sigma_1} \delta_{\lambda, 2 \sigma_1}\,\cos \di{\theta_4 \over 2}\, e^{-i \sigma_1 \vp_4} \, H_{53}(\sigma_1) \,\right] \,\right\rrbracket = Q'_1 \label{Qp1} \enq

\vv \nin Calculons ensuite 

$$Q_2 =  \left[ \di{{\epsilon_2 \cdot p_4}\over{p_2 \cdot p_4} } - \di{{\epsilon_2 \cdot p_5}\over{p_2 \cdot p_5} } \right] G_{45}(\sigma_4) \cdot G_{31}(\sigma_1) = \lambda \,\sqrt{2} \left[ \di{{E_4 \sin \theta_4 e^{-i \lambda \vp_4}}\over p_2 \cdot p_4} - \di{{E_5 \sin \theta_5 e^{-i \lambda \vp_5}}\over p_2 \cdot p_5} \right] \times $$
$$ \times \left[ \delta_{\sigma_4, \sigma_1} \sin \di{\theta_5 \over 2}\, e^{i \sigma_1 \vp_5}\, H_{43}(\sigma_1)                 +  \delta_{\sigma_4, -\sigma_1} \sin \di{\theta_4 \over 2}\, e^{i \sigma_1 \vp_4}\, H_{53}(\sigma_1)                \right]$$

\vv \nin On a 

$$ \di{{E_4 \sin \theta_4 e^{-i \lambda \vp_4}}\over p_2 \cdot p_4} - \di{{E_5 \sin \theta_5 e^{-i \lambda \vp_5}}\over p_2 \cdot p_5} = \di{2 \over \sqrt{s}} \left[ \tan \di{\theta_4 \over 2} \,e^{-i \lambda (\vp_4 - \vp_5)/2} - \tan \di{\theta_5 \over 2} \,e^{i \lambda (\vp_4 - \vp_5)/2}   \right] \times $$
$$ \times e^{-i \lambda (\vp_4 + \vp_5)/2}
 = -\di{2 \over \sqrt{s}} e^{-i \lambda (\vp_4 + \vp_5)/2} \di{{H_{45}(\lambda/2)}\over{ \cos \di{\theta_4 \over 2} \cos \di{\theta_5 \over 2}}},~~~{\rm d'o\grave{u}} $$

$$ Q_2 = - \di{{2 \lambda \sqrt{2}}\over \sqrt{s}}  \,\di{{ e^{-i \lambda (\vp_4 + \vp_5)/2} }\over{\cos \di{\theta_4 \over 2} \cos \di{\theta_5 \over 2}}} \left[ \hskip -0.05cm \di{{}\over{}} \delta_{\lambda, 2 \sigma_1}\, H_{45}(\sigma_1) + \delta_{\lambda, - 2 \sigma_1} \,H^\star_{45}(\sigma_1) \right] \,\times$$
\beq \times \left[ \delta_{\sigma_4, \sigma_1} \sin \di{\theta_5 \over 2}\, e^{i \sigma_1 \vp_5}\, H_{43}(\sigma_1)                 +  \delta_{\sigma_4, -\sigma_1} \sin \di{\theta_4 \over 2}\, e^{i \sigma_1 \vp_4}\, H_{53}(\sigma_1)                \right]  \label{Q2} \enq

\vv \nin Pour simplifier l'\'ecriture, nous poserons $x_i = \di{{2 E_i}\over \sqrt{s}}$, $c_i = \cos \di{ \theta_i \over 2}$, $s_i = \sin \di{\theta_i \over 2}$, pour $i = 3,4,5$. Compte tenu des expressions (\ref{Qp1}) et (\ref{Q2}), l'amplitude $X_a$ 
sera \'ecrite sous la forme 

$$  X_a = \di{{2 \sqrt{2} (2 \sigma_1)}\over \sqrt{s}} X'_a, ~~~{\rm avec} $$
\beq X'_a = \delta_{\sigma_4, \sigma_1}\, \delta_{\lambda, 2 \sigma_1}\,X_1 + \delta_{\sigma_4, -\sigma_1}\, \delta_{\lambda, 2 \sigma_1}\,X_2 +  \delta_{\sigma_4, \sigma_1}\, \delta_{\lambda, -2 \sigma_1}\,X_3  + \delta_{\sigma_4, -\sigma_1}\, \delta_{\lambda, -2 \sigma_1}\,X_4  \label{Xadelta}  \enq

\vv \nin On trouve 

\beq \fbox{\fbox{\rule[-0.4cm]{0cm}{1cm}~$ \begin{array}{c} 
~\\
X_1 = \di{{ e^{-i \sigma_1 \vp_5}}\over{x_5 \,c_5\, c_4}} \,H_{43}(\sigma_1)\left[ c_4 - x_5\,s_5\,e^{i \sigma_1(\vp_5 - \vp_4)}\, H_{45}(\sigma_1) \right]  \\~\\
X_2 = - \di{{e^{-i \sigma_1 \vp_4}} \over{x_4 \,c_4\, c_5}}\,H_{53}(\sigma_1)\left[ c_5 + x_4\,s_4\,e^{i \sigma_1(\vp_4 - \vp_5)}\, H_{45}(\sigma_1) \right]  \\~\\
X_3 = - \di{{ s_5\, e^{i \sigma_1(\vp_3 + \vp_4+ \vp_5)}} \over{x_4 \,c_4\, c_5}} \,\left[ c_3\,c_5 - x_4\,\,e^{i \sigma_1(\vp_5 - \vp_3)}\, H^\star_{45}(\sigma_1) H_{43}(\sigma_1)\right]  \\~\\
X_4 =  \di{{s_4\,e^{i \sigma_1(\vp_3 + \vp_4+ \vp_5)}} \over{x_5 \,c_5\, c_4}} \,\left[ c_3\,c_4 + x_5\,e^{i \sigma_1(\vp_4 - \vp_3)}\, H^\star_{45}(\sigma_1) H_{53}(\sigma_1)\right] \\~
\end{array}
 $~}} \label{Xtrida}  \enq

\vv \nin \ding{173}~\und{\bf Amplitudes relatives \`a la s\'erie de diagrammes (b)} 

\vv \nin A part quelques petites variantes, le calcul des amplitudes des diagrammes (b) est similaire au pr\'ec\'edent. Ecrivant

\beq T^b = \epsilon^\mu_2\,T^b_\mu =- \di{1 \over{2 p_4 \cdot p_5} }\, \sqrt{(2E_1)(2E_3)(2E_4)(2E_5) }\,\delta_{\sigma_3, \sigma_1}\, \delta_{\sigma_4, \sigma_5} (2 \sigma_4)\,X_b, 
\label{aTb}  \enq

$$ {\rm o\grave{u}}~~~~~X_b = \epsilon^\mu_2\, G^\alpha_{31}(\sigma_1)\, G^\nu_{45}(\sigma_4) \left[ A^b_{\mu \nu \alpha} + i\, (2 \sigma_1)\,B^b_{\mu \nu \alpha} \right],    $$  

$${\rm soit}~~~~~ X_b = \di{{\epsilon_2 \cdot p_3}\over{p_2 \cdot p_3} } \, G_{45}(\sigma_4) \cdot G_{31}(\sigma_1)\,\, + $$
$$ +\,\, \di{1\over 2} \left[ \di{1\over{ p_2\cdot p_3}} + \di{1\over{ p_2 \cdot p_1}} \right] \left[  p_2 \cdot G_{31}(\sigma_1) \,\epsilon_2 \cdot G_{45} (\sigma_4) -  \epsilon_2 \cdot G_{31}(\sigma_1)\, p_2 \cdot G_{45}(\sigma_4)   \right] $$
$$ +\,\,(2 \sigma_1)  \di{ i\over 2} \, \epsilon_{\mu \nu \alpha \omega}\, \epsilon^\mu_2 \, p^\omega_2\, G^\alpha_{31}(\sigma_1)\,G^\nu_{45}(\sigma_4) \left[\di{1\over p_2 \cdot p_3} - \di{1 \over p_2 \cdot p_1} \right]$$

\vv \nin ou encore 

$$ X_b =  \di{{\epsilon_2 \cdot p_3}\over{p_2 \cdot p_3} } \, G_{45}(\sigma_4) \cdot G_{31}(\sigma_1)\,\, +$$
$$+\,\,\left[ \di{\delta_{\lambda, 2 \sigma_1} \over{ p_2\cdot p_1}} + \di{\delta_{\lambda, - 2 \sigma_1} \over{ p_2 \cdot p_3}} \right] \left[  p_2 \cdot G_{31}(\sigma_1) \,\epsilon_2 \cdot G_{45} (\sigma_4) -  \epsilon_2 \cdot G_{31}(\sigma_1)\, p_2 \cdot G_{45}(\sigma_4)   \right] $$ 

\vv \nin on obtient $X_b$ sous la forme 
$$ X_b = \di{{2 \sqrt{2} (2 \sigma_1)}\over \sqrt{s}} X'_b, ~~~{\rm avec} $$
\beq  X'_b = \delta_{\sigma_4, \sigma_1}\, \delta_{\lambda, 2 \sigma_1}\,Y_1 + \delta_{\sigma_4, -\sigma_1}\, \delta_{\lambda, 2 \sigma_1}\,Y_2 +  \delta_{\sigma_4, \sigma_1}\, \delta_{\lambda, -2 \sigma_1}\,Y_3  + \delta_{\sigma_4, -\sigma_1}\, \delta_{\lambda, -2 \sigma_1}\,Y_4  \label{Xbdelta}  \enq

\vv \nin et 

\beq \fbox{\fbox{\rule[-0.4cm]{0cm}{1cm}~$ \begin{array}{c} 
~\\
Y_1 = \di{{e^{-i \sigma_1 \vp_5}} \over{c_3}} \,H_{43}(\sigma_1)\left[ c_3\, c_5 + s_3\,s_5\,e^{2 i \sigma_1(\vp_5 - \vp_3)}\,\right]  \\~\\
Y_2 =  \di{{ e^{-i \sigma_1 \vp_4}} \over{c_3}} \,H_{53}(\sigma_1)\left[ c_3\, c_4 + s_3\,s_4\,e^{2 i \sigma_1(\vp_4 - \vp_5)}\,\right]  \\~\\
Y_3 = \di{{s_5\,e^{i \sigma_1(\vp_3 + \vp_4+ \vp_5)}} \over{x_3 \,c_3}} \,\left[ c_4 - x_3\,s_3\,\,e^{i \sigma_1(\vp_3 - \vp_4)}\, H_{43}(\sigma_1) \right]  \\~\\
Y_4 =  \di{{s_4\,e^{i \sigma_1(\vp_3 + \vp_4+ \vp_5)}} \over{x_3 \,c_3}} \,\left[ c_5 - x_3\,s_3\,e^{i \sigma_1(\vp_3 - \vp_5)}\, H_{53}(\sigma_1)\right] \\~
\end{array}
 $~}} \label{Ytrida}  \enq

\vvv
\vv \nin \ding{174}~\und{\bf Formules de cin\'ematique} 

\vv \nin Les variables \'energies et angles des particules finales (de masses nulles) consid\'er\'ees ici sont relatives au r\'ef\'erentiel $(T,X,Y,Z)$ du centre de masse de la r\'eaction. Elles satisfont les relations  

$$ \sqrt{s} = E_3 + E_4 + E_5 $$
\beq  0 = E_3 \cos \theta_3 + E_4 \cos \theta_4 + E_5 \cos \theta_5 \label{cin1} \enq
$$ 0 = E_3 \sin \theta_3 \cos \vp_3 + E_4 \sin \theta_4 \cos \vp_4 + E_5 \sin \theta_5 \cos \vp_5 $$ 
$$ 0 = E_3 \sin \theta_3 \sin \vp_3 + E_4 \sin \theta_4 \sin \vp_4 + E_5 \sin \theta_5 \sin \vp_5 $$ 

\vv \nin qui montrent que les \'energies des particules doivent s'exprimer en fonctions des variables angulaires. Dans ledit r\'ef\'erentiel, les 3-impulsions des particules finales sont dans un m\^eme plan. Notant $\theta_{k \ell}$ l'angle, inf\'erieur \`a $\pi$, entre les 3-impulsions    
des particules $k$ et $\ell$, on a $\theta_{34} + \theta_{35} + \theta_{45} = 2 \pi$, on trouve ($x_k = 2 E_k/\sqrt{s}$)

\beq x_3 = 1 - \cot \di{\theta_{34}\over 2} \cot \di{\theta_{35} \over 2} ,~~x_4 = 1 - \cot \di{\theta_{34}\over 2} \cot \di{\theta_{45} \over 2},~~x_5 = 1 - \cot \di{\theta_{35}\over 2} \cot \di{\theta_{45} \over 2} \enq

\vv \nin Il est utile de rappeler que 

\beq \cos \theta_{k \ell} = \cos \theta_k \cos \theta_\ell + \sin \theta_k \sin \theta_\ell \cos(\vp_k - \vp_\ell) \label{coskl} \enq

\vv \nin Combinant les deux premi\`eres relations de (\ref{cin1}), on trouve aussi 

$$ x_3 c^2_3 + x_4 c^2_4 + x_5 c^2_5 = 1,~~x_3 s^2_3 + x_4 s^2_4 + x_5 s^2_5 = 1,~~{\rm et} $$
\beq x_3 + x_4 + x_5 = 2 \label{cin2} \enq

\vv \nin Les carr\'es des masses invariantes $W^2_{k \ell} = (p_k + p_\ell)^2$ et le transfert $t = - (p_1 - p_3)^2$ s'expriment comme suit

$$ t = s\,x_3\,s^2_3,~~W^2_{34} = s\,x_3\,x_4\, \sin^2 \di{\theta_{34} \over 2} = s(1-x_5), $$
\beq W^2_{35} = s\,x_3\,x_5\, \sin^2 \di{\theta_{35} \over 2} = s(1-x_4),~~~ W^2_{45} = s\,x_4\,x_5\, \sin^2 \di{\theta_{45} \over 2} = s(1-x_3)  \label{cin3} \enq

\vvv
\vv \nin \ding{175}~\und{\bf Taux d'interaction} 

\vv \nin L'amplitude totale du processus s'\'ecrit 

$$ T = T_a + T_b = 2 \sqrt{2 s}\,(2 \sigma_4) (2\sigma_1)\, \delta_{\sigma_3, \sigma_1}\,\delta_{\sigma_5, \sigma_4}\, \sqrt{x_3\,x_4\,x_5} \left[ \di{X'_a \over t} - \di{X'_b \over W^2_{45}} \right]$$

\nin et le taux d'interaction correspondant est 

$$ {\cal I} = \di{\sum_{\sigma_k, \,\lambda}} \left|\,T\,\right|^2 = 16\, s \, \di{\sum^4_{i = 1}} |\,Z_i\,|^2 , ~~~{\rm avec}~~~Z_i = x_3 \,x_4\,x_5 \left[ \di{X_i \over t} - \di{Y_i \over W^2_{45}} \right] $$

\nin En utilisant les formules de cin\'ematique \'etablies pr\'ec\'edemment, on trouve 

\vvv
\beq \fbox{\fbox{\rule[-0.4cm]{0cm}{1cm}~$ \begin{array}{c}~\\
 x_3\,x_4\,x_5\,|\,X_1\,|^2 = \di{1 \over{x_4\, c^2_4\,x_5\,c^2_5}}\, x_3\,s^2_3\,(1-x_5)^2 
\\~\\ x_3\,x_4\,x_5\,|\,X_2\,|^2 = \di{1 \over{x_4\, c^2_4\,x_5\,c^2_5}}\, x_3\,s^2_3\,(1-x_4)^2
\\~\\
x_3\,x_4\,x_5\,|\,X_3\,|^2 = \di{1 \over{x_4\, c^2_4\,x_5\,c^2_5}}\,x_3\,s^2_3\,x^2_5\,s^4_5
\\~\\
x_3\,x_4\,x_5\,|\,X_4\,|^2 = \di{1 \over{x_4\, c^2_4\,x_5\,c^2_5}}\,x_3\,s^2_3\,x^2_4\,s^4_4,
\\~\\
x_3\,x_4\,x_5\,|\,Y_1\,|^2 = \di{1\over{x_3\,c^2_3}} (1-x_3)(1-x_5)^2
\\~\\
x_3\,x_4\,x_5\,|\,Y_2\,|^2 = \di{1\over{x_3\,c^2_3}} (1-x_3)(1-x_4)^2
\\~\\
x_3\,x_4\,x_5\,|\,Y_3\,|^2 = \di{1\over{x_3\,c^2_3}}\, (1-x_3)\,x^2_5\,s^4_5
\\~\\
x_3\,x_4\,x_5\,|\,Y_3\,|^2 = \di{1\over{x_3\,c^2_3}}\, (1-x_3)\,x^2_4\,s^4_4
\\~
\end{array}
 $~}}  \enq

\vv \nin On en d\'eduit les taux d'interaction correspondant aux diagrammes (a) et (b) pris  s\'epar\'ement : 

$$ {\cal I}_a = \di{{16}\over t^2}\times t \left[ (1-x_4)^2 + (1-x_5)^2 + x^2_4\,s^4_4 + x^2_5\,s^4_5 \,\right] \times \di{1\over{x_4 c^2_4\,x_5 c^2_5}}$$
\beq {\cal I}_b = \di{{16}\over W^4_{45}}\times W^2_{45} \left[ (1-x_4)^2 + (1-x_5)^2 + x^2_4\,s^4_4 + x^2_5\,s^4_5 \,\right] \times \di{1\over x_3 c^2_3} \label{tauxab} \enq

\vv \nin Nous ferons ici trois commentaires. D'apr\`es (\ref{tauxab}), le taux ${\cal I}_a$ est proportionnel \`a $1/t$ et non pas \`a $1/t^2$. Ceci pouvait \^etre pr\'evu en consid\'erant la formule de factorisation (1.139) du chapitre 1. En effet, pour un vertex leptonique et dans le cas des masses nulles, on trouve $L=8 m^2 =0$ et $T=2t$, ce qui fait tomber une puissance de $t$ au d\'enominateur de la formule (1.142). Ensuite, on observe une grande similarit\'e entre ${\cal I}_a$ et ${\cal I}_b$. Ceci n'est pas \'etonnant car les deux s\'eries de diagrammes ont la m\^eme structure et l'on passe des premiers aux seconds en effectuant les substitutions (compte tenu des masses nulles, donc \'egales) des 4-vecteurs : 

\beq p_1  \rightarrow - p_5,~~p_2 \rightarrow p_2,~~p_3 \rightarrow p_4,~~p_4 \rightarrow p_3,~~ p_5 \rightarrow - p_1 \enq   

\vv \nin Or, le taux d'interaction des diagrammes (a), exprim\'e en fonction des invariants de la r\'eaction, admet un prolongement analytique, et celui-ci peut \^etre mis en oeuvre pour obtenir ${\cal I}_b$ par ladite substitution. On obtient ainsi 

$$ {\cal I}_a = \di{8 \over{(p_1 \cdot p_3)(p_2 \cdot p_4)(p_2 \cdot p_5)}} \left[ (p_3 \cdot p_5)^2 + (p_3 \cdot p_4)^2 + (p_1 \cdot p_4)^2 + (p_1 \cdot p_5)^2 \right] $$
$$ \longrightarrow \di{8 \over{(p_5 \cdot p_4)(p_2 \cdot p_3)(p_2 \cdot p_1)}} \left[ (p_4 \cdot p_1)^2 + (p_4 \cdot p_3)^2 + (p_5 \cdot p_3)^2 + (p_5 \cdot p_1)^2 \right] $$
$$ = \di{{16}\over W^2_{45}\, x_3 c^2_3}  \left[  x^2_4\,s^4_4  + (1-x_5)^2 + (1-x_4)^2 + x^2_5\,s^4_5 \,\right] = {\cal I}_b$$

\vv \nin On note en passant que ce prolongement montre que ${\cal I}_b $ est proportionnel \`a $1/W^2_{45}$ et non pas \`a $1/W^4_{45}$, ce qui pouvait aussi \^etre pr\'evu au regard d'une formule de factorisation concernant les processus \`a \'echange d'un photon du genre temps.

\vv \nin Enfin, on v\'erifie que l'expression de ${\cal I}_a$ est bien conforme \`a la formule (\ref{XG}) du Compl\'ement I. Dans cette formule, faisons d'abord les changements $1 \rightarrow 2$, tout en posant  $p^2_2 = - t_1 \neq 0$, $3 \rightarrow 4$, $4 \rightarrow 5$, $t_2 = t$. Puis prenons $F_{\mu \rho} = - \left[ g_{\mu \rho} + \di{{p_{2 \mu} p_{2 \rho}} \over t_1} \right]$, $G_ {\nu \sigma} = {\rm Tr} \gs(p_1) \gs_\nu \gs(p_3) \gs_\sigma$ et posons finalement $t_1 = 0$, $m =0$. On obtient 

$$  X_G = - 4\,G \left\{ ab \left( \di{1 \over a^2} + \di{1 \over b^2} \right)  -\di{{2 t W^2_{45}}\over {ab}}  \right\} 
 + \di{{16 t} \over{a b}} \left[ G_{44} + G_{55} \right] $$

\vv \nin Or, $G= -4 t$, $G_{44} = 8 (p_1 \cdot p_4) (p_3 \cdot p_4) = 2 s^2 x_4 s^2_4 (1 - x_5)$, $G_{55} = 8 (p_1 \cdot p_5) (p_3 \cdot p_5) = 2 s^2 x_5 s^2_5 (1 - x_4)$, $a = s x_4 c^2_4$, $b=s x_5 c^2_5$. D'o\`u

$$ \di{1\over t^2} X_G = \di{{16}\over{t\,x_4\, c^2_4 \,x_5\, c^2_5}} \left[x^2_4 c^2_4 + x^2_5 c^4_5 -2 x_3 s^2_3 (1-x_3)  + 2 x_4 s^2_4 (1-x_5) +2  x_5 s^2_5 (1-x_4) \right] $$

\nin Mais

$$ x^2_4 c^2_4 + x^2_5 c^4_5 -2 x_3 s^2_3 (1-x_3)  + 2 x_4 s^2_4 (1-x_5) + 2 x_5 s^2_5 (1-x_4) = x^2_4 s^4_4 -2 x^2_4 c^2_4 + x^2_4 + x^2_5 s^4_5 $$
$$ - 2 x^2_5 s^2_5 + x^2_5 - 2 x_3 s^2_3 (1-x_3) + 2 x_4 s^2_4(1-x_5) + 2 x_5 s^2_5 (1-x_4) $$
$$ = x^2_4 s^4_4 + x^2_5 s^4_5  -2 x_3 s^2_3 (1 - x_3) + 2 x_4 s^2_4 ( 1 - x_5 - x_4) + 2 x_5 s^2_5 ( 1 - x_4 - x_5)  +x^2_4 + x^2_5 $$
$$=x^2_4 s^4_4 + x^2_5 s^4_5 -2(1-x_3) [ x^3 s^2_3 + x_4 s^2_4 + x_5 s^2_5 ] + x^2_4 + x^2_5$$ 
$$ = x^2_4 s^4_4 + x^2_5 s^4_5 - 2(1 - x_3) + x^2_4 + x^2_5 =  x^2_4 s^4_4 + x^2_5 s^4_5 + 1- 2 x_4 +1 - 2 x_5 +x^2_4 + x^2_5 $$
$$ = x^2_4 s^4_4 + x^2_5 s^4_5 +(1-x_4)^2 + (1-x_5)^2 $$

\vv \nin et l'on a bien $X_G/t^2 = {\cal I}_a$. 

\vv \nin Disons quelques mots sur le terme d'interf\'erence ${\cal I}_{ab} = \di{\sum}\,\, T_a\,T^\star_b$ entre les amplitudes des diagrammes (a) et (b). Il appara\^it que ce terme est antisym\'etrique dans l'\'echange des particules 4 ($\mu^-$) et 5 ($\mu^+$), m\^eme en tenant compte de leur masse. La cons\'equence est qu'il donne une contribution nulle \`a la section efficace totale du processus, car l'int\'egration sur les variables desdites particules est sym\'etrique vis-\`a-vis de cet \'echange. Ce fait \'etait pr\'evisible pour la raison suivante. Le taux d'interaction total peut tout aussi bien \^etre calcul\'e comme la somme des carr\'es des amplitudes correspondant chacune \`a un \'etat du syst\`eme particule-antiparticule $(4,5)$ associ\'e \`a un moment orbital $L$ et \`a un spin total $S$ donn\'es. Un tel \'etat, neutre en charge, a une C-parit\'e bien d\'efinie, \'egale \`a $C= (-1)^{L+S}$. Or, dans les diagrammes (a), la paire $(4,5)$ est produite par un syst\`eme de deux photons dont la C-parit\'e totale est $(-1)(-1) = +1$, tandis que dans les diagrammes (b), elle est issue d'un photon virtuel de C-parit\'e \'egale \`a $-1$. La C-parit\'e \'etant une grandeur conserv\'ee, les \'etats $(L,S)$ impliqu\'es respectivement dans les amplitudes (a) et dans les amplitudes (b) sont donc diff\'erents, et l'interf\'erence totale entre les deux s\'eries de diagrammes est donc nulle.


\newpage



\newpage
\section{Amplitudes d'h\'elicit\'e de \,$\gs + \gs \rightarrow W^- + W^+$\protect \footnote{J-F Loiseau, ``Photoproduction de paires de bosons lourds dans le champ \'electromagn\'etique des noyaux", Th\`ese de 3\`eme cycle, UPMC, Paris, juin 1973 ; M. Baillargeon, G. B\'elanger, F. Boudjema, ``Effects of nonstandard trilinear couplings in photon-photon collisions : 1. $\gs + \gs \rightarrow W^- + W^+$", Nucl. Phys. B500 (1997) 224.} }

\vvv
\begin{figure}[hbt]
\centering
\includegraphics[scale=0.3, width=10cm, height=3.5cm]{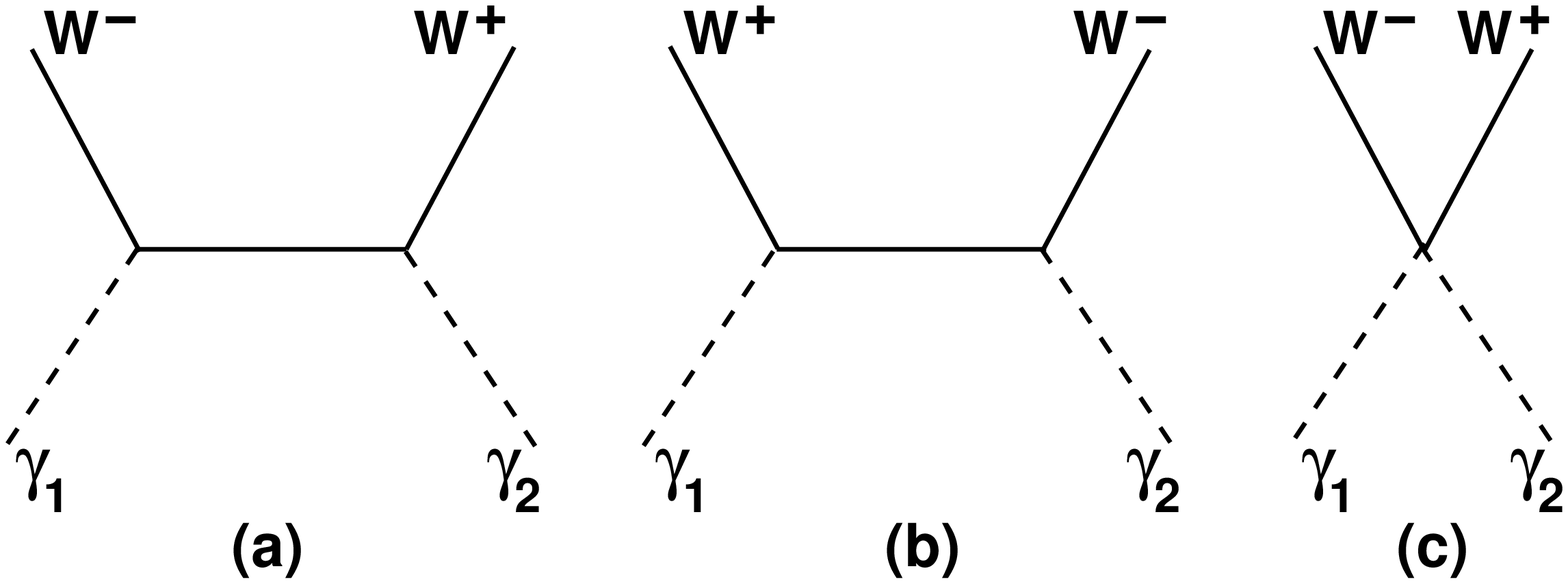}
\vskip 0.25cm

\caption{Diagrammes de Feynman pour $\gs + \gs \rightarrow W^- + W^+$} \label{fig:ggWW}
\end{figure}

\vv \nin A l'ordre le plus bas suivant $\alpha$, l'amplitude g\'en\'erique du processus de production d'une paire $W^-\,W^+$ de bosons vecteurs lourds $W$ par collision photon-photon, d\'ecrite principalement par les diagrammes de la figure (\ref{fig:ggWW}), est donn\'ee par 

$$ {\cal M} = 4 \pi \alpha \,\epsilon^{\star \rho}_3 \epsilon^{\star \sigma}_4\, T_{\rho, \sigma ; \mu, \nu} \,\epsilon^\mu_1 \epsilon^\nu_2,$$ 

\nin Elle fait intervenir le tenseur de rang 4 :  

\beq \begin{array}{c} 
T_{\rho, \sigma ; \mu, \nu} = A_{\rho, \sigma ; \mu, \nu} +B_{\rho, \sigma ; \mu, \nu} +C_{\rho, \sigma ; \mu, \nu}, ~~{\rm o\grave{u}} \\~\\
 A_{\rho, \sigma ; \mu, \nu} = \left\llbracket \hskip -0.1cm \di{{}\over{}}  A_{1\, \mu \rho \omega} + (\kappa -1) \left[ p_{1 \rho} \,g_{\omega \mu} - p_{1 \omega} \,g_{\rho \mu} \right] \right\rrbracket a^{\omega \gs} \times \\~\\
\times\left\llbracket \hskip -0.1cm \di{{}\over{}}  A_{2\, \nu \sigma \gs} + (\kappa -1) \left[ p_{2 \sigma} \,g_{\gs \nu} - p_{2 \gs} \,g_{\sigma \nu} \right]\right\rrbracket, \\~\\
B_{\rho, \sigma ; \mu, \nu} = \left\llbracket \hskip -0.1cm \di{{}\over{}}  B_{1\, \mu \sigma \omega} +(\kappa -1 ) \left[ p_{1 \sigma}\, g_{ \omega \mu} - p_{1 \omega}\, g_{\sigma \mu} \right] \right\rrbracket b^{\omega \gs} \times \\~\\
\times \left\llbracket \hskip -0.1cm \di{{}\over{}}  B_{2\, \nu \rho \gs} +(\kappa -1) \left[ p_{2 \rho}\, g_{\gs \nu} - p_{2 \gs}\, g_{\rho \nu} \right]\right\rrbracket, \\~\\
A_{1\, \mu \rho \omega} =  [ 2 p_3 - p_1 ]_\mu\, g_{\rho \omega} + [p_1 - p_3]_\omega\, g_{\rho \mu} +2 \left[ p_{1 \rho} \,g_{\omega \mu} - p_{1 \omega} \,g_{\rho \mu} \right], \\~\\
A_{2\, \nu \sigma \gs} =  [ 2p_4 - p_2]_\nu\, g_{\sigma \gs} + [p_2 - p_4]_\gs\, g_{\sigma \nu} + 2\left[ p_{2 \sigma} \,g_{\gs \nu} - p_{2 \gs} \,g_{\sigma \nu} \right], \\~\\
 B_{1\, \mu \sigma \omega} = [ 2 p_4 - p_1 ]_\mu \,g_{\sigma \omega} + [p_1 - p_4]_\omega \,g_{\sigma \mu} +2 \left[ p_{1 \sigma}\, g_{ \omega \mu} - p_{1 \omega}\, g_{\sigma \mu} \right], \\~\\
 B_{2\, \nu \rho \gs} = [ 2p_3 - p_2]_\nu\, g_{\rho \gs} + [p_2 - p_3]_\gs\, g_{\rho \nu} + 2 \left[ p_{2 \rho}\, g_{\gs \nu} - p_{2 \gs}\, g_{\rho \nu} \right], \\~\\
C_{\rho, \sigma ; \mu, \nu} = g_{\rho \mu} g_{\sigma \nu} + g_{\rho \nu} g_{\sigma \mu} -2\, g_{\rho \sigma} \, g_{\mu \nu} , \\~\\
 a^{\beta \gs} = \di{1\over{m^2 -p^2}} \left[  g^{\beta \gs} - \di{{ p^\beta p^\gs}\over m^2} \right],~~~p = p_1 - p_3 = p_4 - p_2 ,\\~\\ 
b^{\beta \gs} = \di{1\over{m^2 -p^{ \prime 2}}} \left[  g^{\beta \gs} - \di{{ p^{\prime \beta} p^{\prime \gs}}\over m^2} \right],~~~p^\prime = p_1 - p_4 = p_3 - p_2. \\~ 
\end{array}  \label{agWW}
\enq

\newpage
~\vskip -0.5cm 
\nin La 4-impulsion et la polarisation de chacune des particules sont : 

\vv \nin $\bullet$ $p_1$, $\epsilon_{1 \mu}$ avec l'indice d'h\'elicit\'e $\lambda_1 = \pm 1$ pour le photon 1 ;  
 
\vv \nin $\bullet$ $p_2$, $\epsilon_{2 \nu}$ avec l'indice d'h\'elicit\'e $\lambda_2 = \pm 1$ pour le photon 2 ;  
 
\vv \nin $\bullet$ $p_3$, $\epsilon_{3 \rho}$ avec l'indice d'h\'elicit\'e $\lambda_3 = 0, \pm 1$ pour le boson $W^-$ ;  

\vv \nin $\bullet$ $p_4$, $\epsilon_{4 \sigma}$ avec l'indice d'h\'elicit\'e $\lambda_4 = 0, \pm 1$ pour le boson $W^+$ ;  

\vv \nin Etant massif, le boson vectoriel $W$ dispose d'un troisi\`eme \'etat d'h\'elicit\'e correspondant \`a $\lambda =0$. Nous poserons $\epsilon^{(0)}_3 = z_3$, $\epsilon^{(0)}_4 = z_4$. Le couplage d'h\'elicit\'e utilis\'e est celui de la voie $s$.

\beq
\fbox{\rule[-0.7cm]{0cm}{6cm}~$ \begin{array}{c} ~\\
 p_1 = \di{\sqrt{s} \over 2} \left[ T+ Z\right],~~p_2 = \di{\sqrt{s} \over 2} \left[ T- Z\right] \\~\\

 \epsilon_1 = \di{1\over \sqrt{2}} \left[ - \lambda_1 X -i Y\right], ~~\epsilon_2 = \di{1\over \sqrt{2}} \left[  \lambda_2 X -i Y\right],~~\epsilon_1 \cdot \epsilon_2 = \di{1\over 2} \left[ 1+ \lambda_1 \lambda_2 \right] = \delta_{\lambda_2, \lambda_1} \\~\\

 p_3 = \di{\sqrt{s} \over 2} \left[ T+\beta Z' \right],~~ p_4 = \di{\sqrt{s} \over 2} \left[ T-\beta Z' \right],~~Z' = Z\cos \theta  + X \sin \theta \\~\\

{\rm Pour}~\lambda_{3,4} = \pm 1~ : ~ \epsilon_3 = \di{1\over \sqrt{2}} \left[ - \lambda_3 X' -i Y\right], ~~\epsilon_4 = \di{1\over \sqrt{2}} \left[  \lambda_4 X' -i Y\right], \\~\\

X' = X \cos \theta - Z \sin \theta  \\~\\

 \epsilon_1 \cdot p_3 = - \epsilon_1 \cdot p_4 = \di{\lambda_1 \over{2 \sqrt{2}}} \,\beta \sqrt{s} \sin \theta,~~ \epsilon_2 \cdot p_3 = - \epsilon_2 \cdot p_4 = -\di{\lambda_2 \over{2 \sqrt{2}}} \,\beta \sqrt{s} \sin \theta  \\~\\

 p_1 \cdot \epsilon^\star_3 = - p_2 \cdot \epsilon^\star_3 = - \di{\lambda_3 \over{2 \sqrt{2}}} \, \sqrt{s} \sin \theta, ~~p_1 \cdot \epsilon^\star_4 = - p_2 \cdot \epsilon^\star_4 = \di{\lambda_4 \over{2 \sqrt{2}}} \, \sqrt{s} \sin \theta \\~\\

 \epsilon_1 \cdot \epsilon^\star_3 = - \di{1\over 2} \left[ 1 + \lambda_1 \lambda_3 \cos \theta \right] ,~~\epsilon_1 \cdot \epsilon^\star_4 = - \di{1\over 2} \left[ 1 - \lambda_1 \lambda_4 \cos \theta \right] \\~\\

 \epsilon_2 \cdot \epsilon^\star_3 = - \di{1\over 2} \left[ 1 - \lambda_2 \lambda_3 \cos \theta \right] ,~~\epsilon_2 \cdot \epsilon^\star_4 = - \di{1\over 2} \left[ 1 + \lambda_2 \lambda_4 \cos \theta \right] \\~\\

{\rm Pour}~ \lambda_{3,4} = 0~ :~  \epsilon^{(0)}_3 = z_3 = \di{\sqrt{s}\over{2m}} \left[ \beta T + Z' \right],~~\epsilon^{(0)}_4 = z_4 = \di{\sqrt{s}\over{2m}} \left[   \beta T -Z' \right] \\~\\

z_3 \cdot z_4 = \di{s \over{4m^2}} ( 1+ \beta^2) \\~\\

 p_1 \cdot z_3 = p_2 \cdot z_4 = \di{s \over{4m}} \left[ \beta - \cos \theta \right],~~ p_1 \cdot z_4 = p_2 \cdot z_3 = \di{s \over{4m}} \left[ \beta + \cos \theta \right] \\~\\

 \epsilon_1 \cdot z_3 = -\epsilon_1 \cdot z_4 =   \di{\lambda_1 \over \sqrt{2}} \,\di{\sqrt{s}\over{2m}}\, \sin \theta,~~\epsilon_2 \cdot z_3 = -\epsilon_2 \cdot z_4 =   - \di{\lambda_2 \over \sqrt{2}} \,\di{\sqrt{s}\over{2m}}\, \sin \theta \\~\\
a = m^2 - (p_1 - p_3)^2 = 2 (p_1 \cdot p_3) = 2 (p_2 \cdot p_4) =\di{s \over 2} \left[ 1 - \beta \cos \theta \right] \\~\\ 
b = m^2 - (p_1 - p_4)^2 = 2 (p_1 \cdot p_4) = 2 (p_2 \cdot p_3)= \di{s \over 2} \left[ 1 + \beta \cos \theta \right] \\~
\end{array} 
$~}  \label{formulesWW}
\enq

\newpage
\nin Bien avant la d\'ecouverte exp\'erimentale du $W$ en 1983, de nombreux th\'eoriciens avaient d\'ej\`a \'etudi\'e certains processus offrant la possibilit\'e de mettre en \'evidence cette particule, notamment des r\'eactions o\`u apparaissent, en tant que sous-processus, l'effet Compton $\gs + W^{\pm}  \rightarrow \gs + W^{\pm}$ ou m\^eme $\gs + \gs \rightarrow W^- + W^+$.   

\vv \nin Cependant, il semblerait que les amplitudes d'h\'elicit\'e de $\gs + \gs \rightarrow W^- + W^+$ aient \'et\'e calcul\'ees pour la premi\`ere fois par J-F Loiseau dans sa th\`ese, soutenue \`a Paris en 1973\footnote{Plus pr\'ecis\'ement, ce sont plut\^ot les amplitudes avec des polarisations rectilignes que l'on trouve dans cette th\`ese.}. Dans ce travail, l'auteur a d\'eduit l'amplitude g\'en\'erique en appliquant, selon l'usage, le principe de couplage \'electromagn\'etique minimum au Lagrangien libre du boson charg\'e $W$, tout en envisageant pour cette particule un moment magn\'etique anomal $\kappa$, ce qui conduit \`a l'amplitude ({\ref{agWW}) \'ecrite plus haut. Il a montr\'e en particulier que la valeur $\kappa =1$, qui est en fait celle assign\'ee par la th\'eorie \'electro-faible actuelle, pr\'emunit la section efficace de toute divergence \`a tr\`es haute \'energie\footnote{Pour $s \rightarrow \infty$, la section efficace correspondant \`a $\kappa=1$ tend vers la constante $\di{{128 \pi \alpha^2}\over m^2}$.}. C'est cette valeur de $\kappa$ que nous retenons dans la suite. Outre l'int\'er\^et physique certain que pr\'esente ledit processus\footnote{Voir M. Baillargeon, G. B\'elanger, F, Boudjema, loc. cit.}, le calcul de ses amplitudes d'h\'elicit\'e est un excellent exercice de manipulation de 4-vecteurs de polarisation et de leurs produits scalaires ! Une liste de d\'efinitions et de formules utiles au calcul est donn\'ee dans (\ref{formulesWW}).  

\vv \nin Le lecteur s'assurera que le tenseur $ T_{\rho, \sigma ; \mu, \nu} $ est compatible avec l'invariance de jauge, c'est-\`a-dire qu'il v\'erifie bien les \'equations $p^\mu_1\,T_{\rho, \sigma ; \mu, \nu} =0$, $p^\nu_2\,T_{\rho, \sigma ; \mu, \nu} =0$, mais \`a la condition qu'on le projette pr\'ealablement sur les polarisations des bosons $W^-$ et $W^+$.  

\vv \nin La premi\`ere \'etape du calcul peut para\^itre r\'ebarbative mais s'av\`ere rentable. Elle consiste \`a effectuer les d\'eveloppements des produits tensoriels ${A_1}_{\mu \rho \omega} a^{\omega \gs} A_{2\, \nu \sigma \gs} $ et 
${B_1}_{\mu \sigma \omega} b^{\omega \gs} B_{2\, \nu \rho \gs} $, tout en \'eliminant dans le r\'esultat les termes proportionnels \`a $p_{1 \mu}$, $p_{2 \nu}$, $p_{3 \rho}$ ou encore  $p_{4 \sigma}$, car ils ont des projections nulles sur les polarisations des particules vectorielles, celles-ci \'etant associ\'ees aux indices $\mu$, $\nu$, $\rho$ et $\sigma$, respectivement\footnote{En interm\'ediaire, il est utile de poser $p= p_1 - p_3= p_4 - p_2$, $p'=p_1 - p_4=p_3 -p_2$.} . Dans cette op\'eration, les tenseurs $g_{\rho \mu}\, g_{\sigma \nu}$ et $g_{\rho \nu}\, g_{\sigma \mu}$ du ``terme de contact" $C_{\rho, \sigma ; \mu, \nu}$ disparaissent, et les termes proportionnels \`a $1/m^2$ provenant des propagateurs de $W$ disparaissent aussi. Tenant compte de $p_4 \cdot \epsilon_{1,2} = - p_3\cdot \epsilon_{1,2}$, on aboutit ainsi \`a une forme ``simplifi\'ee" de l'amplitude g\'en\'erique   :

\beq \begin{array}{c} 
 T(\lambda_3, \lambda_4\,;\, \lambda_1, \lambda_2) =  4 \left( \di{1\over a} + \di{1\over b} \right) \left\llbracket  \di{{}\over{}} - (p_3 \cdot \epsilon_1)(p_3\cdot \epsilon_2) (\epsilon^\star_3 \cdot \epsilon^\star_4)  +(p_3 \cdot \epsilon_1) \left[\hskip -0.1 cm \di{{}\over{}}  (p_2 \cdot \epsilon^\star_4) (\epsilon_2 \cdot \epsilon^\star_3) \right. \right. \\~\\ 
\left. \left. \left.\di{{}\over{}}  - (p_2 \cdot \epsilon^\star_3) (\epsilon_2 \cdot \epsilon^\star_4 )\right]\right.  + (p_3 \cdot \epsilon_2) \left[\hskip -0.1 cm \di{{}\over{}}  (p_1 \cdot \epsilon^\star_4) (\epsilon_1 \cdot \epsilon^\star_3) - (p_1 \cdot \epsilon^\star_3) (\epsilon_1 \cdot \epsilon^\star_4)   \right]\, \right\rrbracket \\~\\
+ \di{4\over a} \left\llbracket \hskip -0.05cm \di{{}\over{}}  (p_1 \cdot \epsilon^\star_3) (p_2 \cdot \epsilon^\star_4) (\epsilon_1 \cdot \epsilon_2) + (p_1 \cdot p_2) (\epsilon_1 \cdot \epsilon^\star_3) (\epsilon_2 \cdot \epsilon^\star_4) \,\right\rrbracket \\~\\
 + \di{4 \over b} \left\llbracket \hskip -0.05cm\di{ {} \over {} } (p_1 \cdot \epsilon^\star_4) (p_2 \cdot \epsilon^\star_3) (\epsilon_1 \cdot \epsilon_2) + (p_1 \cdot p_2) (\epsilon_1 \cdot \epsilon^\star_4) (\epsilon_2 \cdot \epsilon^\star_3) \,\right\rrbracket \\~\\
 - 2 (\epsilon^\star_3 \cdot \epsilon^\star_4)(\epsilon_1 \cdot \epsilon_2 ) 
\end{array}
\enq

\vv \nin plus adapt\'ee aux \'etapes suivantes du calcul.

\vv
\vv \nin $\bullet$ \und{\bf Amplitudes $T(0, 0\,;\, \lambda_1, \lambda_2)$} 
\vv \nin On trouve : 

$$
E = - (p_3 \cdot \epsilon_1)(p_3\cdot \epsilon_2) (\epsilon^\star_3 \cdot \epsilon^\star_4)  +(p_3 \cdot \epsilon_1) \left[\hskip -0.1 cm \di{{}\over{}}  (p_2 \cdot \epsilon^\star_4) (\epsilon_2 \cdot \epsilon^\star_3) 
 \left.\di{{}\over{}}  - (p_2 \cdot \epsilon^\star_3) (\epsilon_2 \cdot \epsilon^\star_4 )\right]\right. $$
$$ + (p_3 \cdot \epsilon_2) \left[\hskip -0.1 cm \di{{}\over{}}  (p_1 \cdot \epsilon^\star_4) (\epsilon_1 \cdot \epsilon^\star_3) - (p_1 \cdot \epsilon^\star_3) (\epsilon_1 \cdot \epsilon^\star_4)   \right] $$
$$= \di{{ \lambda_1 \lambda_2 \beta^2 s^2 \sin^2\theta  } \over{32 m^2}} \left[- 4 + (1 + \beta^2) \right]
$$

$$ F = (p_1 \cdot \epsilon^\star_3) (p_2 \cdot \epsilon^\star_4) (\epsilon_1 \cdot \epsilon_2) + (p_1 \cdot p_2) (\epsilon_1 \cdot \epsilon^\star_3) (\epsilon_2 \cdot \epsilon^\star_4) $$
$$ = \di{ s^2 \over{16 m^2}} \left[ \delta_{\lambda_1 ,\lambda_2} (\beta - \cos \theta)^2 + \lambda_1 \lambda_2 \sin^2 \theta \right] $$

$$ G = (p_1 \cdot \epsilon^\star_4) (p_2 \cdot \epsilon^\star_3) (\epsilon_1 \cdot \epsilon_2) + (p_1 \cdot p_2) (\epsilon_1 \cdot \epsilon^\star_4) (\epsilon_2 \cdot \epsilon^\star_3) $$
$$ = \di{ s^2 \over{16 m^2}} \left[ \delta_{\lambda_1 ,\lambda_2} (\beta + \cos \theta)^2 + \lambda_1 \lambda_2 \sin^2 \theta \right] $$

$$ H =  - 2 (\epsilon^\star_3 \cdot \epsilon^\star_4)(\epsilon_1 \cdot \epsilon_2 ) = - 2\, \delta_{\lambda_1, \lambda_2} \di{s \over{4 m^2}} (1 + \beta^2 ) $$

\vv \nin D'o\`u 

$$T(0, 0\,;\, \lambda_1, \lambda_2) = \di{s \over{2 m^2 (1 - \beta^2 \cos^2 \theta)}} \left\llbracket \di{{}\over{}} \lambda_1 \lambda_2 \,\beta^2  \sin^2\theta \left[ -4 + (1 + \beta^2) \right] \right. $$
$$\left. \left. + (1 + \beta \cos \theta ) \left[ \delta_{\lambda_1, \lambda_2} (\beta - \cos \theta)^2 + \lambda_1 \lambda_2 \sin^2 \theta \right] + (1 - \beta \cos \theta ) \left[ \delta_{\lambda_1, \lambda_2} (\beta + \cos \theta)^2 + \lambda_1 \lambda_2 \sin^2 \theta \right] \right.\right.$$
$$\left. \di{{}\over{}}  - \,\delta_{\lambda_1, \lambda_2} (1 + \beta^2)(1 -\beta^2 \cos^2 \theta) \right\rrbracket = \di{2 \over{1 - \beta^2 \cos^2 \theta}} \left\llbracket \hskip -0.1cm \di{{}\over{}} \right.  \lambda_1 \lambda_2 \sin^2 \theta (2 - \beta^2) +  \delta_{\lambda_1, \lambda_2} \times$$
$$\left.\di{{}\over{}} \times \left[ 1- \beta^2 - \sin^2 \theta (2 - \beta^2) \right] \right\rrbracket $$

\vv \nin Notant que $\lambda_1 \lambda_2 = \delta_{\lambda_1, \lambda_2} - \delta_{\lambda_1, -\lambda_2} $ (puisque $\lambda_{1,2} = \pm 1$), on aboutit \`a 

\beq 
\fbox{\fbox{\rule[-0.7cm]{0cm}{1.5cm}~$ 
T(0, 0\,;\, \lambda_1, \lambda_2) = \di{2 \over{1 - \beta^2 \cos^2 \theta}} \left\llbracket \hskip -0.05cm \di{{}\over{}} \delta_{\lambda_1, \lambda_2} (1 - \beta^2) - \delta_{\lambda_1, -\lambda_2} \sin^2 \theta (2 - \beta^2) \right\rrbracket 
$~}} 
\enq

\vv
\vv \nin $\bullet$ \und{\bf Amplitudes $T(0, \lambda_4\,;\, \lambda_1, \lambda_2)$, avec $\lambda_4 = \pm 1$} 

\vv \nin On a maintenant $\epsilon_3 \cdot \epsilon_4 =0$, et donc $H=0$. Il vient 

$$ E = \di{{s \sqrt{s} \beta \sin \theta}\over{16 m \sqrt{2}}} \left[\di{{}\over{}} \lambda_1 (\beta + \cos \theta) (1 +\lambda_2 \lambda_4 \cos \theta) -\lambda_2 (\beta - \cos \theta) (1 -\lambda_1 \lambda_4 \cos \theta) \right]   $$
$$F = - \di{{s \sqrt{s} \sin \theta}\over{8 m \sqrt{2}}} \left[ \hskip -0.05cm\di{{}\over{}} \lambda_4 \delta_{\lambda_1, \lambda_2} ( \beta - \cos \theta) + \lambda_1 ( 1+ \lambda_2 \lambda_4 \cos \theta )\, \right]   $$

$$G =  \di{{s \sqrt{s} \sin \theta}\over{8 m \sqrt{2}}} \left[ \hskip -0.05cm\di{{}\over{}} \lambda_4 \delta_{\lambda_1, \lambda_2} ( \beta + \cos \theta) + \lambda_2 ( 1- \lambda_1 \lambda_4 \cos \theta )\, \right]   $$

\vv \nin D'o\`u 

$$T(0, \lambda_4\,;\, \lambda_1, \lambda_2) = \di{{\sqrt{s} \sin \theta }\over{ m \sqrt{2} (1 - \beta^2 \cos^2 \theta)}} \left\llbracket \hskip -0.05cm \di{{}\over{}} \lambda_1 \left\{ \beta(\beta + \cos \theta) -(1 + \beta \cos \theta) \right\}  \right. $$
$$ + \lambda_2 (1 - \lambda_1 \lambda_4 \cos \theta) \left\{ 1- \beta \cos \theta - \beta( \beta - \cos \theta)\right\} + \lambda_4 \delta_{\lambda_1, \lambda_2} \left\{ (\beta + \cos \theta)(1- \beta \cos \theta) \right.$$
$$\left.\left. \di{{}\over{}}  - (\beta - \cos \theta )(1 + \beta \cos \theta) \right\} \right\rrbracket = \di{{2 m \sqrt{2} \sin \theta }\over{ \sqrt{s} (1 - \beta^2 \cos^2 \theta)}} \left\llbracket \hskip-0.05cm \di{{}\over{}} \lambda_2 ( 1- \lambda_1 \lambda_4 \cos \theta) - \lambda_1( 1+ \lambda_2 \lambda_4 \cos \theta) \right.$$
$$ \left. \di{{}\over{}} + 2 \lambda_4 \delta_{\lambda_1,\lambda_2} \cos \theta\, \right\rrbracket $$

\vv \nin Or, ~~$\lambda_2 ( 1- \lambda_1 \lambda_4 \cos \theta) - \lambda_1( 1+ \lambda_2 \lambda_4 \cos \theta) + 2 \lambda_4 \delta_{\lambda_1,\lambda_2} \cos \theta = \lambda_2 - \lambda_1 + 2 \lambda_4 \cos \theta ( \delta_{\lambda_1, \lambda_2} - \lambda_1 \lambda_2 ) $ ~et~ 

$$ \lambda_2 - \lambda_1 = - 2 \lambda_1 \delta_{\lambda_1, - \lambda_2},~~~\delta_{\lambda_1, \lambda_2} - \lambda_1 \lambda_2 = \delta_{\lambda_1, - \lambda_2} $$

\nin d'o\`u il ressort que 

$$ \lambda_2 - \lambda_1 + 2 \lambda_4 \cos \theta ( \delta_{\lambda_1, \lambda_2} - \lambda_1 \lambda_2 ) = 2 \delta_{\lambda_1, - \lambda_2} \left[ - \lambda_1 + \lambda_4 \cos \theta \right] = - 2 \lambda_1 \delta_{\lambda_1, - \lambda_2} \left[ 1 - \lambda_1 \lambda_4 \cos \theta \right] $$
$$ \equiv (\lambda_2 - \lambda_1) ( 1 - \lambda_1 \lambda_4 \cos \theta ) \equiv  (\lambda_2 - \lambda_1) ( 1 + \lambda_2 \lambda_4 \cos \theta ) $$

\nin et finalement 

\beq 
\fbox{\fbox{\rule[-0.7cm]{0cm}{1.5cm}~$ 
 T(0, \lambda_4\,;\, \lambda_1, \lambda_2) =\di{{2 m \sqrt{2} \sin \theta }\over{ \sqrt{s} (1 - \beta^2 \cos^2 \theta)}} \left[ \lambda_2 - \lambda_1 \right]  \left[ 1 - \lambda_1 \lambda_4 \cos \theta \right]  
$~}} 
\enq

\vv
\vv \nin $\bullet$ \und{\bf Amplitudes $T(\lambda_3, 0\,;\, \lambda_1, \lambda_2)$, avec $\lambda_3 = \pm 1$} 

\vv \nin Le lecteur v\'erifiera que 

\beq 
\fbox{\fbox{\rule[-0.7cm]{0cm}{1.5cm}~$ 
 T(\lambda_3, 0\,;\, \lambda_1, \lambda_2) =\di{{2 m \sqrt{2} \sin \theta }\over{ \sqrt{s} (1 - \beta^2 \cos^2 \theta)}} \left[ \lambda_1 - \lambda_2 \right]  \left[ 1 + \lambda_1 \lambda_3 \cos \theta \right]  
$~}} 
\enq

\vv
\vv \nin $\bullet$ \und{\bf Amplitudes $T(\lambda_3, \lambda_4\,;\, \lambda_1, \lambda_2)$, avec $\lambda_{3,4} = \pm 1$} 

\vv \nin On a cette fois 

$$ E= \di{{ s \beta \sin^2 \theta}\over{16}} \left[\hskip -0.1cm \di{{}\over{}}  2 \beta \lambda_1 \lambda_2 \delta_{\lambda_3, \lambda_4} +(\lambda_1 + \lambda_2)(\lambda_3 + \lambda_4) \right] $$

$$F= \di{s \over 8} \left[\hskip -0.1cm \di{{}\over{}} \lambda_3 \lambda_4 \delta_{\lambda_1, \lambda_2} \sin^2 \theta +(1 + \lambda_1 \lambda_3 \cos \theta)(1+ \lambda_2 \lambda_4 \cos \theta) \right] $$

$$G = \di{s \over 8} \left[\hskip -0.1cm \di{{}\over{}} \lambda_3 \lambda_4 \delta_{\lambda_1, \lambda_2} \sin^2 \theta +(1 - \lambda_1 \lambda_4 \cos \theta)(1- \lambda_2 \lambda_3 \cos \theta) \right] $$ 

$$H = -2\, \delta_{\lambda_3, \lambda_4}\, \delta_{\lambda_1, \lambda_2} $$

\nin d'o\`u 

$$ T(\lambda_3, \lambda_4\,;\, \lambda_1, \lambda_2) = \di{1\over{1 - \beta^2 \cos^2 \theta}} \left[\hskip -0.1cm \di{{}\over{}}  2 \beta^2 \sin^2 \theta \lambda_1 \lambda_2 \delta_{\lambda_3, \lambda_4}   + \beta \sin^2 \theta (\lambda_1 + \lambda_2)(\lambda_3 + \lambda_4)  \right.$$
$$+2\, \lambda_3 \lambda_4 \sin^2 \theta \delta_{\lambda_1, \lambda_2}  + (1 + \beta \cos \theta) (1 + \lambda_1 \lambda_3 \cos \theta)(1+ \lambda_2 \lambda_4 \cos \theta)  $$
$$\left. \di{{}\over{}}  + (1 -\beta \cos \theta) (1 - \lambda_1 \lambda_4 \cos \theta)(1- \lambda_2 \lambda_3 \cos \theta) -2\,\delta_{\lambda_3, \lambda_4}  \delta_{\lambda_1, \lambda_2}(1 - \beta^2 \cos^2 \theta ) \right]$$
\newpage

$$ = \di{1\over{1 - \beta^2 \cos^2 \theta}} \left[\hskip -0.1cm \di{{}\over{}} 2 \beta^2 \sin^2 \theta \lambda_1 \lambda_2 \delta_{\lambda_3, \lambda_4} +2\, \lambda_3 \lambda_4 \sin^2 \theta \delta_{\lambda_1, \lambda_2} +2 + 2 \lambda_1 \lambda_2 \lambda_3 \lambda_4 \cos^2 \theta  \right. $$
$$  \left. \di{{}\over{}} - 2\,\delta_{\lambda_3, \lambda_4}  \delta_{\lambda_1, \lambda_2}(1 - \beta^2 \cos^2 \theta ) +\beta (\lambda_1 + \lambda_2)(\lambda_3 + \lambda_4) +(\lambda_1 - \lambda_2)(\lambda_3 - \lambda_4) \cos \theta \right]$$ 

\nin En faisant usage des relations 
\vskip -0.3cm
$$ \delta_{\lambda_1, \lambda_2} + \delta_{\lambda_1, - \lambda_2} = 1,~~\lambda_1 \lambda_2 =\delta_{\lambda_1, \lambda_2} + \delta_{\lambda_1, -\lambda_2},~~ \lambda_1 + \lambda_2 = 2\lambda_1 \delta_{\lambda_1, \lambda_2},~~ \lambda_1 - \lambda_2 = 2\lambda_1 \delta_{\lambda_1,- \lambda_2}$$
$$ 1 + x^2 + 2 \lambda_1 \lambda_3 x = \left[1 + \lambda_1 \lambda_3 x \right]^2 ~~({\rm car}~|\lambda_i| = 1)$$

\vv\nin on aboutit \`a la formule\footnote{A titre d'exercice, nous proposons au lecteur de comparer les formules des amplitudes \'etablies ici \`a celles donn\'ees dans l'appendice A de l'article de Baillargeon et. al., loc. cit.} 
\vskip 0.2cm
\beq 
\fbox{\fbox{\rule[-0.8cm]{0.cm}{2.cm}~$ \begin{array}{c} 
 T(\lambda_3, \lambda_4\,;\, \lambda_1, \lambda_2) =\di{2\over{ 1 - \beta^2 \cos^2 \theta}} \left\llbracket \hskip -0.05cm \di{{}\over{}} \,\delta_{\lambda_1, \lambda_2}\delta_{\lambda_3, \lambda_4}   \left[ \hskip -0.1cm \di{{}\over{}} 1 + \lambda_1 \lambda_3 \,\beta \right]^2  \right.\\~\\
\left. + \delta_{\lambda_1, - \lambda_2} \delta_{\lambda_3, \lambda_4} (1 - \beta^2) \sin^2 \theta 
+ \delta_{\lambda_1, - \lambda_2} \delta_{\lambda_3, -\lambda_4} \left[ \hskip -0.1cm \di{{}\over{}} 1 +\lambda_1 \lambda_3 \cos \theta \right]^2 \,\right\rrbracket 
\end{array} 
$~}} 
\enq
\vskip 0.35cm

\nin On remarque que l'amplitude correspondant \`a $\lambda_1 = \lambda_2,~\lambda_3 = - \lambda_4$ est nulle, que celle correspondant \`a $\lambda_1 = - \lambda_2,~ \lambda_3 = \lambda_4$ est d\'efavoris\'ee \`a tr\`es haute \'energie ($\beta \rightarrow 1$) par le facteur $(1-\beta^2) \sin^2 \theta$, et que celle correspondant \`a $\lambda_1 = \lambda_2 = - \lambda_3 = - \lambda_4$ est elle aussi d\'efavoris\'ee dans ce domaine par le facteur $(1 - \beta)^2$.

\newpage


\section{Amplitudes d'h\'elicit\'e de \,$\gs + \gs \rightarrow e^{\star -} + e^+$ \label{eexcite}}

\vv \nin L'\'eventuelle existence de quarks ou de leptons ``excit\'es" a fait l'objet de nombreuses \'etudes th\'eoriques, portant \`a la fois sur leur spectroscopie vis-\`a-vis du groupe $SU(2) \times U(1)$ de la th\'eorie \'electro-faible, et sur les possibilit\'es de les mettre en \'evidence dans certaines r\'eactions\footnote{F.M. Renard, Phys. Lett. 139B, 449 (1982) ; N. Cabibbo, L. Maiani, Y. Srivastava, Phys. Lett. 139B, 459 (1984) ;  A. De Rujula, L. Maiani, R. Petronzio, Phys. Lett. 140B, 253 (1984) ; G. Pancheri, Y. Srivastava, Phys. Lett. 146B, 87 (1984) ;  J. K\"uhn, P. Zerwas, Phys. Lett. 147B, 189 (1984) ; J. K\"uhn, H.D. Tholl, P.M. Zerwas, Phys. Lett. 158B, 270 (1985) ; K. Hagiwara, S. Komamiya, D. Zeppenfeld, Z. Phys. C29, 115 (1985) ; F. Boudjema, A. Djouadi, Phys. Lett. B240, 485 (1990) ; U. Baur, M. Spira, P.M. Zerwas, Phys. Rev. D42, 815 (1990) ; I.F. Ginzburg, D.Yu. Ivanov, Phys. Lett. B276, 214 (1992) ; Y.A. Coutinho, J.A. Martins Simo\~es, C.M. Porto, P.P. Queiroz Filho, Phys. Rev. D57, 6975 (1998) ; O.J.P. \'Eboli, S.M. Lietti, Prakash Mathews, Phys. Rev. D65 (2002) 075003.}. Nous consid\'erons ici la production d'une paire \{\'electron excit\'e ($e^{\star -}$) - positron ($e^+$)\} par collision de deux photons r\'eels de suffisamment haute \'energie. Dans l'\'etat actuel des techniques, une telle r\'eaction n'est pas r\'ealisable avec deux photons strictement r\'eels, mais seulement avec des photons ``quasi r\'eels" se trouvant, par exemple, dans le champ \'electromagn\'etique intense d'ions lourds ultra-relativistes\footnote{Voir \`a ce sujet : C. Carimalo et al. : ``Nuclei as Generators of Quasi-real Photons", Phys. Rev. D10, 1561 (1974) ; G. Baur, C.A. Bertulani, Nucl. Phys. A505, 835 (1989) ; N. Baron, G. Baur, Phys. Rev. C, 1999 (1993) ; K. Hencken, D. Trautmann, G. Baur, Z. Phys. C, 473 (1995) ; G. Baur, K. Hencken, D. Trautmann, Prog. Part. Nucl. Phys. 42, 357 (1999).}. 

\vv \nin Nous utiliserons un mod\`ele, couramment sugg\'er\'e, pour lequel le couplage $\gs$\,-\,$e^\star$\hskip -0.05 cm-\,$e$ est d\'ecrit par le Lagrangien 

\beq L_1 = \sqrt{4 \pi \alpha}~\,f_{\gs e^\star e} \left\{ \ov{e^\star} \, \sigma_{\mu \nu} (1 - \gs_5) e\, \partial^\mu A^\nu + {\rm h.c.} \right\}\enq   

\vv \nin $f_{\gs e^\star e} $ \'etant une constante de couplage\footnote{h.c. = conjugu\'e hermitique, $\alpha = 1/137$, $A^\mu$ est le champ du photon...}, $\sigma_{\mu \nu} = \di{i \over 4} \left[ \gs_\mu, \gs_\nu \right]$, tandis que le couplage $\gs$\,-\,$e^\star$\hskip -0.05 cm-\,$e^\star$ est d\'ecrit par le Lagrangien 
  
\beq L_2 = -\sqrt{4 \pi \alpha} ~\ov{e^\star} \left\{ \gs_\mu A^\mu + \di{\kappa \over M} \sigma_{\mu \nu} \partial^\mu A^\nu \right\} e^\star \enq

\vv \nin $\kappa$ \'etant un moment magn\'etique, $M$ la masse du lepton excit\'e. Cette masse, ainsi que  l'\'energie totale de la r\'eaction dans le r\'ef\'erentiel de son centre de masse sont suppos\'ees tr\`es grandes devant la masse du positron, et pour cette raison, nous n\'egligerons cette derni\`ere.  

\vv \nin Les constantes de couplage \'etant mises \`a part, l'amplitude tensorielle du processus, invariante de jauge, est 

$$ T_{\mu \nu} = -i \ov{U_4} \left\llbracket \,\sigma_{\mu \rho} q^\rho_1 (1 - \gs_5) \di{{\gs(q_2 - p_3)}\over{2 q_2 \cdot p_3}} \gs_\nu + \sigma_{\nu \omega} q^\omega_2 (1 - \gs_5) \di{{\gs(q_1 - p_3)}\over{2 q_1 \cdot p_3}} \gs_\mu \right.  $$
\beq  + \left. \left[ \gs_\mu + i \di{\kappa \over M} \sigma_{\mu \rho} q^\rho_1 \right] \di{{M + \gs(p_4 - q_1)}\over{M^2 - (p_4 - q_1)^2}} \sigma_{\nu \omega} q^\omega_2 (1 - \gs_5) \right. \label{ampexcit} \enq
$$ \left.+ \left[ \gs_\nu + i \di{\kappa \over M} \sigma_{\nu \omega} q^\omega_2 \right] \di{{M + \gs(p_4 - q_2)}\over{M^2 - (p_4 - q_2)^2}} \sigma_{\mu \rho} q^\rho_1 (1 - \gs_5)   \right\rrbracket  V_3 $$

\vv \nin Les notations sont les suivantes : $(q_1, \mu)$, $(q_2, \nu)$ sont les 4-impulsions et les indices de Lorentz respectifs des photons initiaux $1$ et $2$ ; $(p_3 ,  V_3 = \gs_5 U_3)$, $(p_4, U_4)$ sont les 4-impulsions et les spineurs respectifs du positron $3$ et de l'\'electron excit\'e $4$. Les deux premiers termes correspondent aux diagrammes (a) de la figure \ref{fig:excit} o\`u un \'electron virtuel est \'echang\'e, tandis que les deux derniers termes correspondent aux diagrammes (b) de cette figure, o\`u un \'electron excit\'e virtuel est \'echang\'e. Pour ce calcul, nous utiliserons encore le couplage d'h\'elicit\'e de la voie $s$. Notant $s = (q_1 + q_2)^2 = 2 q_1 \cdot q_2$, $T = \di{{q_1 + q_2}\over \sqrt{s}}$, $Z = \di{{q_1 - q_2}\over \sqrt{s}}$, on a $q_1 = \di{\sqrt{s} \over 2}(T+Z)$, $q_2 = \di{\sqrt{s} \over 2}(T-Z)$. Dans le r\'ef\'erentiel du centre de masse, l'axe $Y$ est encore pris selon $\Vec{q_1} \wedge \Vec{p_3}$, et l'axe $X$ perpendiculaire 

\vskip 0.2cm \nin \`a $\Vec{q_1}$, dans le plan $(\Vec{q_1}, \Vec{p_3})$ : $Y_\mu \,\propto \, \epsilon_{\mu \nu \rho \omega}\,q^\nu_1\, q^2_\rho\, p^\omega_3$, $X_\mu = \epsilon_{\mu \nu \rho \omega}\,T^\nu\, Y^\rho\, Z^\omega$.

\vvv
\begin{figure}[hbt]
\centering
\includegraphics[scale=0.3, width=14cm, height=3.5cm]{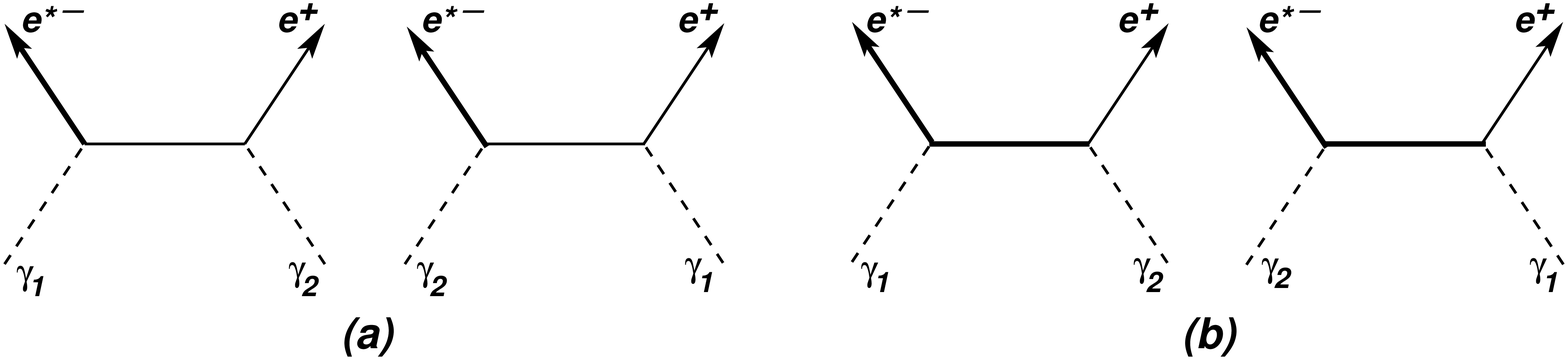}
\vskip 0.25cm

\caption{Diagrammes de Feynman pour $\gs + \gs \rightarrow e^{\star -} \hskip -0.1cm + e^+$} \label{fig:excit}
\end{figure}

\vv \nin Les vecteurs de polarisation des deux photons (1) et (2) sont respectivement d\'efinis par 

$$\epsilon^{(\lambda)}_1 = - \di{1\over \sqrt{2}} \left[  \lambda \,X + i Y \right],~~~\epsilon^{(\lambda)}_2= \epsilon^{(-\lambda)}_1$$   

\vv \nin Par rapport aux calculs pr\'ec\'edents, celui envisag\'e ici pr\'esente les diff\'erences suivantes. Tout d'abord, les spineurs $V_3$ et $U_4$ sont associ\'es \`a des masses tr\`es diff\'erentes, dont l'une est en outre prise \'egale \`a z\'ero. Ensuite, apr\`es avoir fait dispara\^itre, par des anticommutations appropri\'ees, la masse de l'\'electron excit\'e du num\'erateur des propagateurs, la matrice sandwich\'ee apparaissant finalement dans l'amplitude (\ref{ampexcit}) est une somme de produits de matrices $\gs$, certains comptant un nombre pair de ces matrices, d'autres en comptant un nombre impair. Enfin, on note la pr\'esence du projecteur de chiralit\'e $(1- \gs_5)/2$.  

\vv \nin Consid\'erons tout d'abord des spineurs associ\'es \`a des masses non nulles et normalis\'es selon $\ov{U} U = 2$, et posons $\epsilon = \pm 1$. A partir des formules \'etablies au paragraphe 3.2.3, on d\'eduit  :
\vvv

$$ \ov{U_4}_{\sigma'} \left[1 + \epsilon \gs_5 \right] V_{3 \sigma} = - 2 \delta_{\sigma' , -\sigma} \left[ \delta_{\epsilon, 2 \sigma} \,e^{\chi^+_{34}} - \delta_{\epsilon, - 2 \sigma} \,e^{- \chi^+_{34}} \right]$$

$$\ov{U_4}_{\sigma'} \,\gs_\mu \left[1 + \epsilon \gs_5 \right] V_{3 \sigma} = 2 \delta_{\epsilon, 2 \sigma} \left\llbracket  \delta_{\sigma', \sigma}\, (2 \sigma) \,e^{\chi^+_{34}} \left[ X' + i (2 \sigma) Y \right]_\mu - \delta_{\sigma', - \sigma}\,e^{\chi^-_{34}} [T+Z']_\mu \right\rrbracket $$
$$ + 2 \delta_{\epsilon, - 2 \sigma} \left\llbracket \delta_{\sigma', \sigma} (2 \sigma) e^{- \chi^+_{34}} \left[X' +i(2 \sigma)Y \right]_\mu + \delta_{\sigma', - \sigma} e^{- \chi^-_{34}} [T - Z']_\mu \right\rrbracket $$

$$\ov{U_4}_{\sigma'} \,\sigma_{\mu \nu} \left[1 + \epsilon \gs_5 \right] V_{3 \sigma} = \delta_{\epsilon, 2 \sigma} \left\llbracket \hskip -0.05cm \di{{}\over{}} i\,\delta_{\sigma', \sigma} (2 \sigma) e^{\chi^-_{34}}\,  \left[ \hskip -0.05cm \di{{}\over{}} [T+Z']_\mu [X' + i(2\sigma) Y]_\nu \right. \right. $$
$$ \left. \left. \left. \di{{}\over{}} - [T+Z']_\nu [X' + i(2 \sigma) Y]_\mu \right] - \delta_{\sigma', - \sigma} \,e^{\chi^+_{34}} \left[\hskip -0.05cm \di{{}\over{}}  i\, [ T_\mu Z'_\nu - T_\nu Z'_\mu \right] + (2 \sigma) \left[ X'_\mu Y_\nu - X'_\nu Y_\mu \right] \right] \right\rrbracket$$
$$ + \delta_{\epsilon, - 2 \sigma'} \left\llbracket  \hskip -0.05cm \di{{}\over{}} i\,\delta_{\sigma', \sigma} (2 \sigma) e^{-\chi^-_{34}}\,  \left[ \hskip -0.05cm \di{{}\over{}} [T-Z']_\mu [X' + i(2\sigma) Y]_\nu - [T-Z']_\nu [X' + i(2\sigma) Y]_\mu \right] \right. $$           
$$\left. \left. - \delta_{\sigma', - \sigma} \,e^{-\chi^+_{34}} \left[\hskip -0.05cm \di{{}\over{}}  i\, [ T_\mu Z'_\nu - T_\nu Z'_\mu \right] - (2 \sigma) \left[ X'_\mu Y_\nu - X'_\nu Y_\mu \right] \right] \right\rrbracket$$

\vv \nin Des d\'efinitions $\cosh \chi_3 = \di{E_3 \over m_3}$, $\cosh \chi_4 = \di{E_4 \over m_4}$, on tire ($m_3 = m_{e^+}$, $m_4 = M$)

$$ \cosh 2 \chi^+_{34} = \cosh (\chi_3 + \chi_4) = \di{1 \over{m_3\, m_4}} [ E_3 E_4 + p_3 p_4 ] = \di{{p_3 \cdot p_4}\over{m_3\, m_4}} $$

\vv \nin Lorsque $m_3 \rightarrow 0$, 

$$ e^{\chi_3/2} = \sqrt{e^{\chi_3}} = \sqrt{ \cosh \chi_3 + \sinh \chi_3} \simeq \sqrt{\di{{2 E_3}\over m_3}},~~{\rm donc}$$
$$ \sqrt{m_3}\, e^{\chi_3/2} \simeq \sqrt{2 E_3} ,~~ \sqrt{m_3}\, e^{- \chi_3/2}\simeq \di{m_3 \over \sqrt{2 E_3}} \rightarrow 0 $$

\vv \nin et, revenant aux spineurs normalis\'es selon $\ov{U} U = 2 m$, les relations pr\'ec\'edentes conduisent  
finalement \`a 

\beq \fbox{\fbox{\rule[-0.4cm]{0cm}{1cm}~$\begin{array}{c} ~\\
\ov{U_4}_{\sigma'} \left[1 + \epsilon \gs_5 \right] V_{3 \sigma} = - 2 \delta_{\sigma' , -\sigma} \, \delta_{2 \sigma, \epsilon} \,\sqrt{2 p_3 \cdot p_4} 
\\~\\  \hline\\
\ov{U_4}_{\sigma'} \,\gs_\mu \left[1 + \epsilon \gs_5 \right] V_{3 \sigma} = 2 \delta_{2 \sigma, \epsilon} \left\llbracket  \delta_{\sigma', \sigma}\, (2 \sigma) \,\sqrt{2 p_3 \cdot p_4} \left[ X' + i (2 \sigma) Y \right]_\mu \right. 
\\~\\
\left. - \delta_{\sigma', - \sigma}\,\di{{2 M}\over{\sqrt{2 p_3 \cdot p_4}}} \, p_{3 \mu} \right\rrbracket 
\\~\\ \hline \\
\ov{U_4}_{\sigma'} \,\sigma_{\mu \nu} \left[1 + \epsilon \gs_5 \right] V_{3 \sigma} = \delta_{2 \sigma, \epsilon} \left\llbracket \hskip -0.05cm \di{{}\over{}} i\,\delta_{\sigma', \sigma} (2 \sigma) \di{{2M}\over{\sqrt{2 p_3 \cdot p_4}}} \,  \left[ \hskip -0.05cm \di{{}\over{}} p_{3 \mu} [X' + i(2\sigma) Y]_\nu \right. \right. 
\\~\\
\left. \left. \left. \di{{}\over{}} - p_{3 \nu} [X' + i(2 \sigma) Y]_\mu \right] - \delta_{\sigma', - \sigma} \,\sqrt{2 p_3 \cdot p_4} \left[\hskip -0.05cm \di{{}\over{}}  i\, [ T_\mu Z'_\nu - T_\nu Z'_\mu \right] \right. \right.
\\~\\
\left. \left.\di{{}\over{}}  + (2 \sigma) \left[ X'_\mu Y_\nu - X'_\nu Y_\mu \right] \,\right] \,\right\rrbracket
\\~\\
{\rm avec}~~~~T_\mu Z'_\nu - T_\nu Z'_\mu = - \di{{p_{3 \mu} p_{4 \nu} - p_{3 \nu} p_{4 \mu}} \over{p_3 \cdot p_4}} 
\\~
\end{array} $~}}  \enq

\vv \nin En utilisant syst\'ematiquement la relation $\gs_5 = i \gs(X) \gs(Y) \gs(Z) \gs(T)$, valable pour toute base $T,X,Y,Z$ orthonorm\'ee et d'orientation directe, on montre les relations (tr\`es) utiles suivantes. 

\vvv
\beq \fbox{\fbox{\rule[-0.4cm]{0cm}{1cm}~$\gs(\lambda X + i Y) \, \gs(T + \eta Z) (1 + \epsilon \gs_5) \left[ 1- \epsilon \eta \lambda \right] = 0  $~}}  \label{R1} \enq

\vv \nin o\`u $\lambda, \eta, \epsilon = \pm1$, d'o\`u l'on d\'eduit notamment 

\beq \fbox{\fbox{\rule[-0.4cm]{0cm}{1cm}~$\begin{array}{c} ~\\
\gs(\epsilon^{(\pm)}) \gs(q_1) (1 \mp \gs_5) = 0,~~~
\gs(\epsilon^{(\pm)}) \gs(q_2) (1 \pm \gs_5) = 0 
\\~
\end{array} 
 $~}} \label{R2}  \enq

\vv \nin Notant que 

\beq
\begin{array}{c}  
\gs(\epsilon_1) \gs(\epsilon_2) = \delta_{\lambda_1, \lambda_2} \left[ 1 +  \lambda_1 \gs(Z) \gs(T) \gs_5 \right],~~~ \gs(\epsilon_2) \gs(\epsilon_1) = \delta_{\lambda_1, \lambda_2} \left[ 1 - \lambda_1 \gs(Z) \gs(T)\gs_5 \right] \\~\\
 \gs(q_1) \gs(q_2) = \di{s\over 2} \left[1 + \gs(Z) \gs(T) \right],~~~\gs(q_2) \gs(q_1) =\di{ s\over 2} \left[ 1 - \gs(Z) \gs(T) \right] 
\end{array} 
\label{R3} \enq

\vv \nin on d\'eduit aussi

\beq \fbox{\fbox{\rule[-0.4cm]{0cm}{1cm}~$\begin{array}{c} ~\\
\gs(\epsilon_1) \gs(\epsilon_2) \gs(T + \eta Z)\left[1+ \epsilon \gs_5 \right]  = \delta_{\lambda_2, \lambda_1} (1 - \lambda_1 \eta \epsilon) \gs(T + \eta Z) \left[1+ \epsilon \gs_5 \right]
\\~\\
= 2\,\delta_{\lambda_2, \lambda_1} \delta_{\lambda_1 \eta \epsilon, -1} \,\gs(T + \eta Z) \left[1 + \epsilon \gs_5 \right] 
\\~\\ 
\gs(\epsilon_2) \gs(\epsilon_1) \gs(T + \eta Z)\left[1+ \epsilon \gs_5 \right]  = \delta_{\lambda_2, \lambda_1} (1 + \lambda_1 \eta \epsilon) \gs(T + \eta Z) \left[1+ \epsilon \gs_5 \right]
\\~\\
= 2\,\delta_{\lambda_2, \lambda_1} \delta_{\lambda_1 \eta \epsilon, 1}\, \gs(T + \eta Z) \left[1 + \epsilon \gs_5 \right] 
\\~\\ 
\gs(\epsilon_1) \gs(\epsilon_2) \gs(q_1) \gs(q_2) \left[1 + \epsilon \gs_5 \right] =  s \delta_{\lambda_1, \lambda_2} \delta_{\lambda_1, \epsilon} \left[ 1+ \gs(Z) \gs(T) \right] \left[1 + \epsilon \gs_5 \right] 
\\~\\
  \gs(\epsilon_2) \gs(\epsilon_1) \gs(q_2) \gs(q_1) \left[1 + \epsilon \gs_5 \right] =  s \delta_{\lambda_1, \lambda_2} \delta_{\lambda_1, \epsilon} \left[ 1- \gs(Z) \gs(T) \right] \left[1 + \epsilon \gs_5 \right] 
\\~\\
\gs(\lambda X + iY) \gs(q_1) \gs(q_2) \left[1+ \epsilon \gs_5 \right] = s \delta_{\lambda, - \epsilon} \gs(\lambda X + i Y)\left[1+ \epsilon \gs_5 \right] 
\\~\\
\gs(\lambda X + iY) \gs(q_2) \gs(q_1) \left[1+ \epsilon \gs_5 \right] = s \delta_{\lambda,  \epsilon} \gs(\lambda X + i Y)\left[1+ \epsilon \gs_5 \right] 
\\~
\end{array} 
 $~}} \label{R4} \enq

\vv \nin Rappelons enfin que dans le cas $m_3 =0$, on a $\gs_5\,V_3 = (2 \sigma_3) \,V_3$. Il en r\'esulte que, dans cette approximation, la pr\'esence de la matrice $(1 - \gs_5)$ fait que les huit amplitudes pour lesquelles $\sigma_3 = +1/2$ {\it sont nulles}.   

\vv \nin Les relations (\ref{R2}) montrent aussi que les amplitudes pour lesquelles $\epsilon_1 = \epsilon^{(+)}$ et $\epsilon_2 = \epsilon^{(+)}_2 = \epsilon^{(-)}$ sont \'egalement {\it nulles}. En effet, compte tenu de $q_1 \cdot \epsilon^{(+)} =0$, on a par exemple $\sigma_{\mu \rho} \epsilon^\mu_1 q^\rho_1 (1 - \gs_5) = \di{i \over 2} \gs(\epsilon^{(+)}) \gs(q_1) (1 - \gs_5) = 0$. Au final, il ne reste donc que six amplitudes \`a calculer.

\vv \nin On a 

$$ T(\sigma_4, \sigma_3 ; \lambda_1, \lambda_2) = \di{1\over 2}\, \ov{U_4}_{\sigma_4} \left\llbracket \,\gs(\epsilon_1) \gs(q_1) (1 - \gs_5) \di{{\gs(q_2 - p_3)}\over{2 q_2 \cdot p_3}} \gs(\epsilon_2) \right. $$
$$ + \gs(\epsilon_2) \gs(q_2) (1 - \gs_5) \di{{\gs(q_1 - p_3)}\over{2 q_1 \cdot p_3}} \gs(\epsilon_1) $$
$$+ \left. \left[ \gs(\epsilon_1) - \di{\kappa \over{2 M}} \gs(\epsilon_1) \gs(q_1) \right] \di{{M + \gs(p_4 - q_1)}\over{M^2 - (p_4 - q_1)^2}} \gs(\epsilon_2)\gs(q_2)  (1 - \gs_5) \right. $$
$$ \left.+ \left[ \gs(\epsilon_2) - \di{\kappa \over{2 M}} \gs(\epsilon_2)\gs(q_2) \right] \di{{M + \gs(p_4 - q_2)}\over{M^2 - (p_4 - q_2)^2}} \gs(\epsilon_1) \gs(q_1) (1 - \gs_5)   \right\rrbracket { V_3}_{\sigma_3} $$

$$ = \di{1\over 2}\, \ov{U_4}_{\sigma_4} \left\llbracket \,\left[ \di{1\over{2 q_2 \cdot p_3} } + \di{{1 + \kappa}\over{2 q_1 \cdot p_4} } \right] \gs(\epsilon_1) \gs(\epsilon_2) \gs(q_1) \gs(q_2) (1 - \gs_5)  \right. $$

$$ + \left[ \di{1\over{2 q_1 \cdot p_3}} +  \di{{1+ \kappa}\over{2 q_2 \cdot p_4}} \right]\gs(\epsilon_2) \gs(\epsilon_1)\gs(q_2)\gs(q_1)  (1 - \gs_5) $$

$$ +\left[ \di{{\epsilon_2 \cdot p_4}\over{q_2 \cdot p_4}}- \di{{\epsilon_2 \cdot p_3}\over{q_2 \cdot p_3}} \right]\gs(\epsilon_1)\gs(q_1) (1-\gs_5) + \left[\di{{\epsilon_1 \cdot p_4}\over{q_1 \cdot p_4}}  - \di{{\epsilon_1 \cdot p_3}\over{q_1 \cdot p_3}}\right] \gs(\epsilon_2) \gs(q_2) (1-\gs_5)$$

$$  - \di{\kappa \over{2 M}} \left[  \di{{\epsilon_1 \cdot p_4} \over{ q_1 \cdot p_4}}\gs(\epsilon_2) \gs(q_1) \gs(q_2) (1-\gs_5)+ \di{{\epsilon_2 \cdot p_4} \over{ q_2 \cdot p_4}}\gs(\epsilon_1) \gs(q_2) \gs(q_1) (1-\gs_5)  \right.  $$
$$ \left.\left. \di{{}\over{}} + \gs(\epsilon_1)\gs(\epsilon_2)\gs(q_2)(1-\gs_5) + \gs(\epsilon_2) \gs(\epsilon_1) \gs(q_1)(1-\gs_5)  \right] \,\right\rrbracket {V_3}_{\sigma_3} $$

\vv \nin Calculons : 
\vv \nin \ding{172} ~~~$\ov{U_4}_{\sigma_4} \gs(\epsilon_1) \gs(\epsilon_2) \gs(q_1) \gs(q_2) (1 - \gs_5) {V_3}_{\sigma_3} =  s \,\delta_{\lambda_2, \lambda_1} \delta_{\lambda_1, -1}\,\ov{U_4}_{\sigma_4} \left[1 + \gs(Z) \gs(T) \right] \left[1 - \gs_5 \right]  {V_3}_{\sigma_3} $ 

\vv \nin Comme $T\cdot Z=0$, on peut \'ecrire 

$$ \ov{U_4}_{\sigma_4} \,\sigma_{\mu \nu} Z^\mu T^\nu\,\left[1 -  \gs_5 \right] {V_3}_{\sigma_3} = \di{i \over 2} \,\ov{U_4}_{\sigma_4} \gs(Z) \gs(T) \,\left[1 -  \gs_5 \right] {V_3}_{\sigma_3} $$
$$ = \delta_{2 \sigma, -1} \,\left\llbracket -i\, \delta_{\sigma_4, \sigma_3} \di{{2M}\over{\sqrt{2 p_3\cdot p_4}}} \left[ -p_3 \cdot T\, X'\cdot Z  \right] - \delta_{\sigma_4, - \sigma_3} \sqrt{2 p_3 \cdot p_4} \left[-i Z' \cdot Z \right] \right\rrbracket $$
$$ = i \delta_{2 \sigma, -1} \left\llbracket \, \delta_{\sigma_4, \sigma_3}  \di{{2M E_3}\over{\sqrt{2 p_3\cdot p_4}}} \sin \theta- \delta_{\sigma_4, - \sigma_3}\sqrt{2 p_3 \cdot p_4}  \cos \theta\right\rrbracket $$

\vv \nin o\`u $\theta$ est l'angle de diffusion du positron dans le r\'ef\'erentiel du centre de masse de la r\'eaction. Compte tenu de la relation $2 \sqrt{s} E_3 = 2 p_3 \cdot (p_3 + p_4 ) = 2 p_3 \cdot p_4 = s - M^2$, il vient 

$$ \ov{U_4}_{\sigma_4} \gs(Z) \gs(T) \,\left[1 -  \gs_5 \right] {V_3}_{\sigma_3} = 2 \delta_{2\sigma_3 , -1} \sqrt{2 p_3 \cdot p_4} \left[ \delta_{\sigma_4, \sigma_3} \di{M\over \sqrt{s}} \sin \theta - \delta_{\sigma_4, - \sigma_3} \, \cos \theta \right] $$

\vv \nin Par suite, 

$$\ov{U_4}_{\sigma_4} \gs(\epsilon_1) \gs(\epsilon_2) \gs(q_1) \gs(q_2) (1 - \gs_5) {V_3}_{\sigma_3}   = $$
$$  s \,\delta_{\lambda_2, \lambda_1} \delta_{\lambda_1, -1} \delta_{2 \sigma_3, -1} \sqrt{s-M^2} \left[ \delta_{\sigma_4, \sigma_3} \di{M\over \sqrt{s}} \sin \theta - \delta_{\sigma_4, - \sigma_3} \left[1 + \cos \theta \right] \right]$$
\vvv \nin \ding{173} ~~~$\ov{U_4}_{\sigma_4} \gs(\epsilon_2) \gs(\epsilon_1) \gs(q_2) \gs(q_1) (1 - \gs_5) {V_3}_{\sigma_3} =  s \delta_{\lambda_2, \lambda_1} \delta_{\lambda_1, -1}\,\ov{U_4}_{\sigma_4} \left[1 - \gs(Z) \gs(T) \right] \left[1 - \gs_5 \right]  {V_3}_{\sigma_3} $ 
$$= - 2 s \,\delta_{\lambda_2, \lambda_1} \delta_{\lambda_1, -1} \delta_{2 \sigma_3, -1} \sqrt{s-M^2} \left[ \delta_{\sigma_4, \sigma_3} \di{M\over \sqrt{s}} \sin \theta + \delta_{\sigma_4, - \sigma_3} \left[1 - \cos \theta \right] \right]$$

\vvv \nin \ding{174} \hskip 2cm $\ov{U_4}_{\sigma_4} \gs(\epsilon_1) \gs(q_1) (1-\gs_5)  {V_3}_{\sigma_3} = -2 i\, \ov{U_4}_{\sigma_4} \sigma_{\mu \nu} \epsilon^\mu_1 q^\nu_1 (1-\gs_5)  {V_3}_{\sigma_3}  $
$$= -2i\, \delta_{2 \sigma_3, -1} \left\llbracket \, -i \delta_{\sigma_4, \sigma_3} \di{{2M}\over{\sqrt{2 p_3 \cdot p_4}}} \,  \left[ \hskip -0.05cm \di{{}\over{}} \epsilon_1 \cdot p_3\, q_1 \cdot X' - q_1 \cdot p_3\, \epsilon_1 \cdot (X'-iY)  \right] \right.$$
$$\left. \di{{}\over{}}  - \delta_{\sigma_4, - \sigma_3} \, \sqrt{2 p_3\cdot p_4}\,\left[ -i q_1 \cdot T\, \epsilon_1 \cdot Z' + q_1 \cdot X'\, \epsilon_1 \cdot Y \right] \, \right\rrbracket $$

\vv \nin Tenant compte de 

$$ q_1 \cdot X' = \di{\sqrt{s}\over 2} \sin \theta,~~q_1 \cdot p_3 = \di{{E_3 \sqrt{s}}\over 2} (1 - \cos \theta),~~\epsilon_1 \cdot Z' = \di{\lambda_1 \over \sqrt{2}} \sin \theta,~~\epsilon_1 \cdot X' = \di{\lambda_1 \over \sqrt{2}} \cos \theta,$$
$$ \epsilon_1 \cdot p_3 = \di{{E_3 \lambda_1}\over \sqrt{2}} \sin \theta,~~\epsilon_1 \cdot Y = \di{i \over \sqrt{2}}, ~~\lambda_1 - 1 = - 2 \delta_{\lambda_1, -1} ,$$

\vv \nin il vient

$$\ov{U_4}_{\sigma_4} \gs(\epsilon_1) \gs(q_1) (1-\gs_5)  {V_3}_{\sigma_3} =- \delta_{2 \sigma_3, -1} \delta_{\lambda_1, -1}\, \sqrt{2s} \sqrt{s-M^2} \left[ \delta_{\sigma_4, - \sigma_3} \sin \theta - \delta_{\sigma_4, \sigma_3} \di{M\over \sqrt{s}} [1-\cos \theta] \right]$$ 

\vvv \nin \ding{175} \hskip 5cm $\ov{U_4}_{\sigma_4} \gs(\epsilon_2) \gs(q_2) (1-\gs_5)  {V_3}_{\sigma_3} =$
$$ \delta_{2 \sigma_3, -1} \delta_{\lambda_2, -1}\, \sqrt{2s} \sqrt{s-M^2} \left[ \delta_{\sigma_4, - \sigma_3} \sin \theta + \delta_{\sigma_4, \sigma_3} \di{M\over \sqrt{s}} [1+\cos \theta] \right]$$
\vvv \nin \ding{176} ~De la relation 

\beq \gs(\epsilon_2) \gs(q_1) \gs(q_2) = \di{s\over 2}\, \gs(\epsilon_2) \left[1 + \lambda_2 \gs_5 \right] \enq

\vv \nin on d\'eduit 

$$\gs(\epsilon_2) \gs(q_1) \gs(q_2) \left[ 1 - \gs_5 \right] =   s \,\delta_{\lambda_2, -1} \, \gs(\epsilon_2) \left[1 - \gs_5 \right] $$ 

\vv \nin d'o\`u \hskip 1.5cm $\ov{U_4}_{\sigma_4} \gs(\epsilon_2) \gs(q_1) \gs(q_2) (1-\gs_5)  {V_3}_{\sigma_3} =  s \,\delta_{\lambda_2, -1} \ov{U_4}_{\sigma_4} \gs(\epsilon_2) \left[1-\gs_5 \right]  {V_3}_{\sigma_3}$
$$ =  2 s \,\delta_{\lambda_2, -1} \delta_{2 \sigma_3, -1}\,\left\llbracket - \delta_{\sigma_4, \sigma_3}\,\sqrt{2 p_3 \cdot p_4} \,\epsilon_2 \cdot (X' - i Y) - \delta_{\sigma_4, - \sigma_3}\di{{2M}\over \sqrt{2 p_3\cdot p_4}} \epsilon_2 \cdot p_3 \right\rrbracket  $$
$$ =-  s \sqrt{2} \sqrt{s-M^2} \, \delta_{2 \sigma_3, -1}\, \delta_{\lambda_2, -1}\, \left[ \delta_{\sigma_4, \sigma_3}\,[1 + \cos \theta ] + \delta_{\sigma_4, - \sigma_3} \di{M \over \sqrt{s}} \sin \theta \right] $$

\vvv \nin \ding{177} ~De m\^eme, la relation  

\beq \gs(\epsilon_1) \gs(q_2) \gs(q_1) = \di{s\over 2}\, \gs(\epsilon_1) \left[1 + \lambda_1 \gs_5 \right] \enq

\vv \nin conduit \`a 

$$\ov{U_4}_{\sigma_4} \gs(\epsilon_1) \gs(q_2) \gs(q_1) (1-\gs_5)  {V_3}_{\sigma_3} =  s \,\delta_{\lambda_1, -1} \ov{U_4}_{\sigma_4} \gs(\epsilon_1) \left[1-\gs_5 \right]  {V_3}_{\sigma_3}$$
$$ = - s \sqrt{2} \sqrt{s-M^2} \, \delta_{2 \sigma_3, -1}\, \delta_{\lambda_1, -1}\, \left[ \delta_{\sigma_4, \sigma_3}\,[1 - \cos \theta ] - \delta_{\sigma_4, - \sigma_3} \di{M \over \sqrt{s}} \sin \theta \right] $$

\vvv \nin \ding{178} \hskip 1.5 cm $\ov{U_4}_{\sigma_4} \gs(\epsilon_1) \gs(\epsilon_2) \gs(q_2) (1-\gs_5)  {V_3}_{\sigma_3} =2 \delta_{\lambda_2, \lambda_1}\, \delta_{\lambda_1, -1}\, \ov{U_4}_{\sigma_4} \gs(q_2) (1 - \gs_5) {V_3}_{\sigma_3} $
$$ = 4 \,\delta_{\lambda_2, \lambda_1}\, \delta_{\lambda_1, -1}\,\delta_{2 \sigma_3,-1} \left[ - \delta_{\sigma_4, \sigma_3} \sqrt{2 p_3\cdot p_4} \, q_2 \cdot X' - \delta_{\sigma_4, - \sigma_3} \,\di{{2M}\over \sqrt{2p_3\cdot p_4 }}\,q_2\cdot p_3  \right]$$
$$ = 2 \sqrt{s} \sqrt{s-M^2} \,\delta_{\lambda_2, \lambda_1}\, \delta_{\lambda_1, -1}\,\delta_{2 \sigma_3,-1} \left[ \delta_{\sigma_4, \sigma_3}\,\sin \theta - \delta_{\sigma_4, - \sigma_3} \di{M \over \sqrt{s}} [1+ \cos \theta ] \right] $$

\vvv \nin \ding{179} \hskip 1.5 cm $\ov{U_4}_{\sigma_4} \gs(\epsilon_2) \gs(\epsilon_1) \gs(q_1) (1-\gs_5)  {V_3}_{\sigma_3} =2 \delta_{\lambda_2, \lambda_1}\, \delta_{\lambda_1, -1}\,\ov{U_4}_{\sigma_4} \gs(q_1) (1 - \gs_5) {V_3}_{\sigma_3} $

$$ = - 2 \sqrt{s} \sqrt{s-M^2} \,\delta_{\lambda_2, \lambda_1}\, \delta_{\lambda_1, -1}\,\delta_{2 \sigma_3,-1} \left[ \delta_{\sigma_4, \sigma_3}\,\sin \theta + \delta_{\sigma_4, - \sigma_3} \di{M \over \sqrt{s}} [1- \cos \theta ] \right] $$

\vvv \nin On a $E_4 = \sqrt{s} - E_3 = \di{{s+M^2}\over{2 \sqrt{s}}}$. Nous poserons $\beta^\star = \di{E_3 \over E_4} = \di{{s-M^2}\over{s+M^2}}$. On a 

$$ 2 q_{1,2} \cdot p_4 = 2 \sqrt{s} E_4 ( 1\pm \beta^\star \cos \theta) = 2 \sqrt{s} E_3 (1 \pm \beta^\star \cos \theta)/\beta^\star$$ 

$$ \epsilon_1 \cdot p_3 = - \epsilon_1 \cdot p_4 = \di{{\lambda_1 E_3} \over \sqrt{2}} \sin \theta,~~\epsilon_2 \cdot p_3 = - \epsilon_2 \cdot p_4 = -\di{{\lambda_2 E_3} \over \sqrt{2}} \sin \theta $$

\vv \nin Rassemblant toutes ces expressions, on aboutit aux formules des six amplitudes d'h\'elicit\'e non nulles $T^{\sigma_4, \sigma_3}_{\lambda_1, \lambda_2}$ list\'ees dans le tableau VI et dont nous laissons la v\'erification au lecteur.  

\vv
\beq \fbox{\fbox{\rule[-0.4cm]{0cm}{1cm}~$\begin{array}{c} 
~\\
T^{\uparrow, \downarrow}_{+,-} = - \sqrt{s-M^2} \left[ 1 + \cos \theta \right] \left\{ 1 + \beta^\star (1 + \di{\kappa \over 2}) \di{{1 - \cos \theta}\over{1 + \beta^\star \cos \theta}}  \right\} 
\\ ~\\ 
\hline \\
T^{\downarrow, \downarrow}_{+,-} =- \di{{M \sqrt{s-M^2}}\over \sqrt{s}}\, \di{{\sin \theta \left[1+ \cos \theta\right]}\over{1 - \cos \theta}} \left\{ 1 + \beta^\star \left( 1 + \di{{\kappa s}\over{2 M^2}} \right) \di{{1 - \cos \theta}\over{1 + \beta^\star \cos \theta}} \right\}
\\ ~\\ 
\hline \\
T^{\uparrow, \downarrow}_{-,+} = - \sqrt{s-M^2} \left[ 1 - \cos \theta \right] \left\{ 1 + \beta^\star (1 + \di{\kappa \over 2}) \di{{1 + \cos \theta}\over{1 - \beta^\star \cos \theta}}  \right\} 
\\ ~\\ 
\hline \\
T^{\downarrow, \downarrow}_{-,+} =  \di{{M \sqrt{s-M^2}}\over \sqrt{s}}\, \di{{\sin \theta \left[1- \cos \theta\right]}\over{1 + \cos \theta}} \left\{ 1 + \beta^\star \left( 1 + \di{{\kappa s}\over{2 M^2}} \right) \di{{1 + \cos \theta}\over{1 - \beta^\star \cos \theta}} \right\}
\\ ~\\ 
\hline \\
T^{\uparrow, \downarrow}_{-,-} = - \sqrt{s - M^2}\, \di{{1+\beta^\star}\over \beta^\star} \left [ 2 + \beta^\star \kappa \, \di{{1- \beta^\star \cos^2 \theta}\over{1 - \beta^{\star 2} \cos^2 \theta}}\right]
\\ ~\\ 
\hline \\
T^{\downarrow, \downarrow}_{-,-} = - \di{M \over \sqrt{s}} \sqrt{s-M^2}\, \cot \theta \, \left [ 2 \di{{1-\beta^\star}\over \beta^\star} +\di{{ \beta^\star \sin^2 \theta \left[ 4 \beta^\star + \kappa (1 + \beta^\star)\right]}\over {1 - \beta^{\star 2} \cos^2 \theta}} \right] 
\\ ~
\end{array} $~}}  \label{amplitudes-excit} \enq

\vvv
\centerline{\bf Tableau VI - Amplitudes d'h\'elicit\'e non nulles pour $\gs + \gs \rightarrow e^{\star -} + e^+$ ($m_{e^+} =0$)}
\newpage

\section{Amplitudes d'h\'elicit\'e de $e^- + \,e^+  \rightarrow W^- +\, W^+$ via l'\'echange de leptons excit\'es}

\vv \nin Dans le cadre du mod\`ele standard \'electro-faible, cette r\'eaction s'effectue, \`a l'ordre le plus bas suivant les constantes de couplage, selon l'un ou l'autre des deux processus d\'ecrits par les diagrammes de la figure \ref{fig:eewwst}. Le premier, (a), proc\`ede de l'\'echange dans la voie $t$ d'un neutrino \'electronique, et le second, (b), de l'\'echange dans la voie $s$ d'un photon ou d'un $Z$.

\vvv
\begin{figure}[hbt]
\centering
\includegraphics[scale=0.3, width=10cm, height=3.5cm]{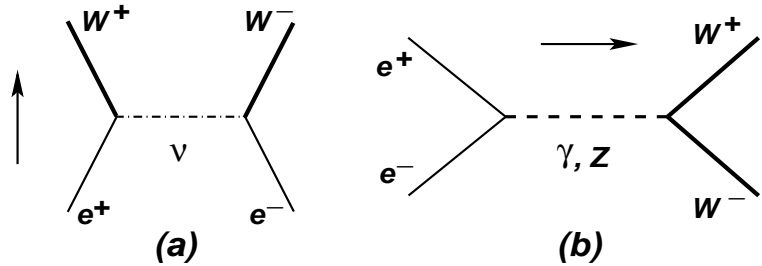}
\vskip 0.25cm

\caption{Diagrammes de Feynman du mod\`ele standard pour $e^-  + e^+  \rightarrow W^- + W^+$} \label{fig:eewwst}
\end{figure}

\nin Cependant, si l'on imagine que les leptons actuellement observ\'es sont en r\'ealit\'e des \'etats li\'es de particules plus fondamentales, appel\'ees ``pr\'eons" dans les mod\`eles de ``compositeness", et qui restent \`a d\'ecouvrir, on peut envisager, \`a l'instar de ce qui est observ\'e dans le secteur hadronique, des multiplets d'isospin faible de leptons excit\'es ; et si l'on suppose en outre que ces leptons excit\'es peuvent \^etre coupl\'es aux leptons usuels par les particules de jauge $\gs, W, Z$, d'autres voies de r\'ealisation de r\'eactions telles que celle consid\'er\'ee ici peuvent \^etre imagin\'ees\footnote{Voir par exemple les r\'ecentes publications suivantes, dans lesquelles le lecteur trouvera de nombreuses r\'ef\'erences : S. Biondini, Frascati Phys, Ser. 55 (2012) 7-12 ; S. Biondini, O. Panella, G. Pancheri, Y. Srivastava, Phys. Rev. D85 (2012) 095018 ; Nuovo Cim. C037, (2014) 02, 91-97 ; S. Biondini, O. Panella, Phys. Rev. D92 (2015) 015023 ; S. Biondini, O. Panella, Frascati Phys. Ser. 61 (2016) 141-146.}. Ainsi, dans le secteur d'isospin et d'hypercharge $I_W = \frac{3}{2}, Y=-1$ on peut avoir quatre \'electrons excit\'es $(E^+, E^0, E^-, E^{--})$ (de spin 1/2), dont un neutre et un doublement charg\'e n\'egativement, ces derniers \'etant susceptibles d'intervenir dans la production d'une paire $W^- W^+$ par collisions $e^- e^+$, via les processus d\'ecrits par les diagrammes de la figure \ref{fig:eewwexo}, qui proc\`edent par \'echange dans la voie $t$ d'un $E^{--}$, diagramme (a), ou d'un $E^0$, diagramme (b).      

\vvv
\begin{figure}[hbt]
\centering
\includegraphics[scale=0.3, width=7cm, height=4cm]{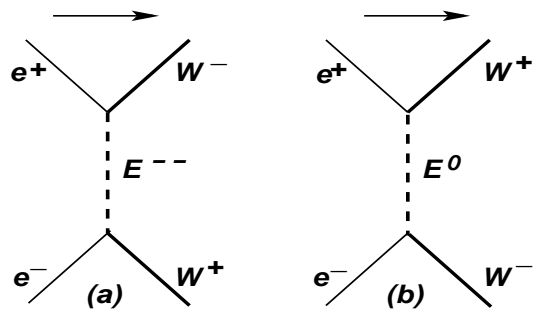}
\vskip 0.25cm

\caption{Diagrammes de Feynman pour $e^-  + e^+  \rightarrow W^- + W^+$ via l'\'echange de leptons excit\'es} \label{fig:eewwexo}
\end{figure}

\nin Le couplage qui nous int\'eresse ici entre ces \'electrons excit\'es et l'\'electron usuel est suppos\'e de type magn\'etique\footnote{Ce qui garantit l'invariance de jauge.} et d\'ecrit par le Lagrangien effectif suivant 

$$  {\cal L} = C \left\{\hskip -0.05cm \di{{}\over{}}  \ov{e} \,\sigma_{\mu \nu} \partial^\mu W^{(+) \nu} \left[1 + \gs_5 \right] E^{--} + \ov{E^{--}} \, \sigma_{\mu \nu} \partial^\mu W^{(-) \nu} \left[ 1- \gs_5 \right] \, e \right.  $$
\beq  \left. \di{1 \over \sqrt{3}} \,\ov{e} \,\sigma_{\mu \nu} \partial^\mu W^{(-) \nu} \left[1 + \gs_5 \right] E^{0} +  \di{1 \over \sqrt{3}} \, \ov{E^{0}} \, \sigma_{\mu \nu} \partial^\mu W^{(+) \nu} \left[ 1- \gs_5 \right] \, e\right\} \enq

\vv \nin o\`u $C$ est une constante de couplage dont la valeur est ici sans importance et o\`u, ici encore, $\sigma_{\mu \nu} = \di{i\over 4} \left[ \gs_\mu, \gs_\nu \right]$. Avec ce Lagrangien, les amplitudes correspondant aux diagrammes (a) et (b) de la figure \ref{fig:eewwexo} sont respectivement 

$$ {\cal A}_a =  \di{1\over D_a} \ov{V_2} \, \sigma_{\mu \nu}\, p^\mu_3 \epsilon^{\star \nu}_3 \,\left[1 + \gs_5 \right] \left[ M_a + \gs(Q_a) \right]\, \sigma_{\rho \omega}\, p^\rho_4 \epsilon^{\star \omega}_4 \left[ 1- \gs_5 \right]\, U_1 $$
$$ {\rm avec} ~~~Q_a = p_1 - p_4,~~D_a = M^2_a - Q^2_a, $$
$$ {\cal A}_b =  \di{1\over{ 3 D_b}} \ov{V_2} \, \sigma_{\mu \nu}\, p^\mu_4 \epsilon^{\star \nu}_4 \,\left[1 + \gs_5 \right] \left[ M_b + \gs(Q_b) \right]\, \sigma_{\rho \omega}\, p^\rho_3 \epsilon^{\star \omega}_3 \left[ 1- \gs_5 \right]\, U_1 $$
$$ {\rm avec} ~~~Q_b = p_1 - p_3,~~D_b = M^2_b - Q^2_b $$

\vv \nin apr\`es division par $C^2$. Les notations sont les suivantes : $(U_1, p_1)$ et $(V_2, p_2)$ sont les spineurs et 4-impulsions respectives de l'\'electron et du positron incidents ; $(\epsilon_3, p_3)$ et $(\epsilon_4, p_4)$ sont les vecteurs de polarisation et les 4-impulsions respectives du $W^-$ et du $W^+$ finals ; $M_a$ et $M_b$ sont les masses respectives de $E^{--}$ et de $E^0$, que l'on peut supposer \'egales, conform\'ement \`a la sym\'etrie suppos\'ee. Ces masses sont suppos\'ees tr\`es grandes devant la masse de l'\'electron, de sorte que nous n\'egligerons cette derni\`ere\footnote{Compte tenu du fait qu'on a aussi $M_W \gg m_e$ ...}.  Dans le r\'ef\'erentiel du centre de masse de la r\'eaction, l'\'electron incident se propage selon l'axe $Z$ et le positron en sens inverse, tandis que le $W^-$ sortant est \'emis avec l'angle $\theta$ par rapport \`a $Z$. Dans ce r\'ef\'erentiel, l'axe $Y$ est choisi selon $\Vec{p_1} \wedge \Vec{p_3}$ et l'on a 

$$ p_1 = \di{\sqrt{s} \over 2} \left[ T+ Z \right],~~~p_2 = \di{\sqrt{s} \over 2} \left[ T- Z \right],~~{\rm avec}~~~s = (p_1 +p_2)^2, $$   
$$ p_3 = \di{\sqrt{s} \over 2} \left[ T+ \beta Z' \right],~~~ p_4 = \di{\sqrt{s} \over 2} \left[ T -\beta Z' \right],~~{\rm avec}~~~\beta = \sqrt{1 - \di{{4 M^2_W}\over s}}, $$
$$ Z' = Z\, \cos \theta + X\, \sin \theta, ~~X' = X \,\cos \theta - Z\, \sin \theta $$
$$ \epsilon^{(\lambda)}_3 = - \di{1\over \sqrt{2}} \left[ \lambda X' + i Y \right],~~~\epsilon^{(\lambda)}_4 = \epsilon^{(-\lambda)}_3~~~{\rm pour}~~~\lambda = \pm 1,$$
$$ \epsilon^{(0)}_3 = \di{\sqrt{s}\over{2 M_W}} \left[ Z' + \beta T \right],~~~\epsilon^{(0)}_4 = \di{\sqrt{s}\over{2 M_W}} \left[ -Z' + \beta T \right]$$

\vv \nin De par la forme mag\'etique du couplage et la pr\'esence des matrices $(1 + \gs_5)$ et $(1 - \gs_5)$, les amplitudes ``r\'eduites"  $T_a =  4 D_a {\cal A}_a$, $T_b =  12 D_b {\cal A}_b$ s'expriment comme  

$$ T_a = - \ov{V_2} \, \left[1+ \gs_5 \right]\,\gs(p_3) \gs(\epsilon^\star_3) \gs(Q_a) \gs(p_4) \gs(\epsilon^\star_4) \left[1 - \gs_5 \right]\,U_1 ~~{\rm et}$$
\beq T_b = - \ov{V_2} \, \left[1+ \gs_5 \right]\,\gs(p_4) \gs(\epsilon^\star_4) \gs(Q_b) \gs(p_3) \gs(\epsilon^\star_3) \left[1 - \gs_5 \right]\,U_1 \enq

\vv \nin La masse de l'\'electron \'etant n\'eglig\'ee, on a\footnote{A d\'emontrer !} 

\beq \fbox{\fbox{\rule[-0.4cm]{0cm}{1cm}~$\begin{array}{c}~\\
{U_1}_{\sigma_1} \ov{V_2}_{\sigma_2} = \di{1 \over{ 2 \sqrt{s}}} \delta_{\sigma_2,- \sigma_1} \gs(p_1) \gs(p_2) \left[ 1 + (2 \sigma_1) \gs_5 \right]
\\~\\
+ (2 \sigma_1) \di{ \sqrt{s} \over 4} \delta_{\sigma_2, \sigma_1} \gs(X + i (2 \sigma_1) Y) \left[1 -(2 \sigma_1 \gs_5 \right] 
\\~
\end{array} $~}} \label{formule1} \enq

\vv \nin et  

\beq \fbox{\rule[-0.4cm]{0cm}{1cm}~$\begin{array}{c}~\\
 \left[1 - \gs_5 \right] \,{U_1}_{\sigma_1} \ov{V_2}_{\sigma_2} \, \left[1+ \gs_5 \right] = - \sqrt{s} \,  \delta_{\sigma_2, \sigma_1} \, \delta_{2 \sigma_1, -1} \gs(X - i Y) \left[ 1+ \gs_5 \right] \\~
\end{array}$~}
\label{formule2} \enq

\vv \nin Comme on pouvait le pr\'evoir, seules les h\'elicit\'es $\sigma_2 = \sigma_1 = -1/2$ contribuent. On en d\'eduit 
 
$$ T_a = \sqrt{s}\,\delta_{\sigma_2, \sigma_1} \, \delta_{2 \sigma_1, -1}\,X_a~~~{\rm avec} $$
$$ X_a =  {\rm Tr} \, \gs(p_3) \gs(\epsilon^\star_3) \gs(Q_a) \gs(p_4) \gs(\epsilon^\star_4) \gs(X- i Y) \left[1 + \gs_5 \right] $$

\vv \nin De fa\c{c}on \'evidente, $T_b$ se d\'eduit de $T_a$ en \'echangeant les r\^oles de $W^-$ et $W^+$. Il suffit donc de d\'eterminer uniquement les amplitudes $T_a$. 

\vv \nin \leftpointright~Commen\c{c}ons par l'\'evaluation de la trace $X_a$ pour des polarisations uniquement circulaires. On a

$$ X_a(\lambda_1, \lambda_2) = {\rm Tr} \, \gs(p_3) \gs(\epsilon^{(- \lambda_1)}) \gs(Q_a) \gs(p_4) \gs(\epsilon^{(\lambda_2)}) \gs(X- i Y) \left[1 + \gs_5 \right] $$
$$ = \delta_{\lambda_2, \lambda_1} \,  {\rm Tr} \, \gs(p_3) \gs(\epsilon^{(- \lambda_1)}) \gs(Q_a) \gs(p_4) \gs(\epsilon^{(\lambda_1)}) \gs(X- i Y) \left[1 + \gs_5 \right] $$
$$ + \delta_{\lambda_2, -\lambda_1} \,  {\rm Tr} \, \gs(p_3) \gs(\epsilon^{(- \lambda_1)}) \gs(Q_a) \gs(p_4) \gs(\epsilon^{(-\lambda_1)}) \gs(X- i Y) \left[1 + \gs_5 \right]  $$
\centerline {et~~ $ Q_a = \di{\sqrt{s} \over 2} \left[ Z + \beta Z' \right],~~~Z = Z' \cos \theta - \di{{\sin \theta}\over \sqrt{2}} \left[ \epsilon^{(-)} - \epsilon^{(+)} \right] $}

\vv \nin \ding{172} ~Comme les vecteurs de polarisation circulaire sont isotropes, il vient 

$$ {\rm Tr} \, \gs(p_3) \gs(\epsilon^{(- \lambda_1)}) \gs(Q_a) \gs(p_4) \gs(\epsilon^{(\lambda_1)}) \gs(X- i Y) \left[1 + \gs_5 \right] $$
$$ = \di{\sqrt{s}\over 2} \,\left[ \cos \theta + \beta \right]  {\rm Tr} \, \gs(p_3) \gs(\epsilon^{(- \lambda_1)}) \gs(Z') \gs(p_4) \gs(\epsilon^{(\lambda_1)}) \gs(X- i Y) \left[1 + \gs_5 \right] $$
$$ = \di{{s \sqrt{s}} \over 8} \left[ \cos \theta + \beta \right] \, {\rm Tr}\, \gs(T + \beta Z')  \gs(\epsilon^{(- \lambda_1)}) \gs(Z') \gs(T - \beta Z')\gs(\epsilon^{(\lambda_1)})  \gs(X- i Y) \left[1 + \gs_5 \right] $$
$$ = \di{{s\sqrt{s}} \over 8} \left[ \cos \theta + \beta \right] \, {\rm Tr}\,\gs(T + \beta Z') \gs(T + \beta Z') \gs(\epsilon^{(- \lambda_1)}) \gs(Z')  \gs(\epsilon^{(\lambda_1)}) \gs(X- i Y) \left[1 + \gs_5 \right] $$ 
$$ = M^2_W \,\di{\sqrt{s} \over 2}\,\left[ \cos \theta + \beta \right] \, {\rm Tr}\,\gs(\epsilon^{(-\lambda_1)}) \gs(Z')   \gs(\epsilon^{( \lambda_1)})\gs(X- i Y) \left[1 + \gs_5 \right] $$

\nin Mais 

$$ \gs(\epsilon^{(-\lambda_1)}) \gs(\epsilon^{(\lambda_1)}) = 1 - i \lambda_1 \gs(X') \gs(Y) = 1 - \lambda_1 \gs(Z') \gs(T) \gs_5,~~~{\rm et} $$  
$$ {\rm Tr}\,\gs(\epsilon^{(-\lambda_1)}) \gs(Z')   \gs(\epsilon^{( \lambda_1)})\gs(X- i Y) \left[1 + \gs_5 \right]  
= - {\rm Tr}\, \left[ \gs(Z') - \lambda_1 \gs(T) \gs_5 \right]\gs(X- i Y) \left[1 + \gs_5 \right] $$
$$ = - 4 Z' \cdot X = 4 \sin \theta $$

\nin Ainsi, 

$$ {\rm Tr} \, \gs(p_3) \gs(\epsilon^{(- \lambda_1)}) \gs(Q_a) \gs(p_4) \gs(\epsilon^{(\lambda_1)}) \gs(X- i Y) \left[1 + \gs_5 \right] = 2 M^2_W \sqrt{s}\,\sin \theta \,\left[ \cos \theta + \beta \right]  $$

\vvv \nin \ding{173} ~~$ {\rm Tr} \, \gs(p_3) \gs(\epsilon^{(- \lambda_1)}) \gs(Q_a) \gs(p_4) \gs(\epsilon^{(-\lambda_1)}) \gs(X- i Y) \left[1 + \gs_5 \right]  $
$$= - \di{\sqrt{s} \over 2}  {\rm Tr} \, \gs(p_3) \gs(\epsilon^{(- \lambda_1)}) \gs(Z)  \gs(\epsilon^{(-\lambda_1)}) \gs(p_4) \gs(X- i Y) \left[1 + \gs_5 \right] $$
$$= -\sqrt{s}\, Z \cdot  \epsilon^{(- \lambda_1)}\, {\rm Tr} \, \gs(p_3) \gs(\epsilon^{(- \lambda_1)})  \gs(p_4) \gs(X- i Y) \left[1 + \gs_5 \right] $$

\nin Or, 

\beq \fbox{\rule[-0.4cm]{0cm}{1cm}~$ \gs(p_3) \gs(\epsilon^{(- \lambda_1)})  \gs(p_4) = - \di{s \over 4} \gs(\epsilon^{(- \lambda_1)})  \,\left[ 1 - \lambda_1 \beta \gs_5 \right]^2$~}  \label{formule3} \enq

\nin donc 

$$ {\rm Tr} \, \gs(p_3) \gs(\epsilon^{(- \lambda_1)})  \gs(p_4) \gs(X- i Y) \left[1 + \gs_5 \right] = -\di{s \over 4}\, (1 + \lambda_1 \beta)^2\, {\rm Tr}\, \gs(\epsilon^{(- \lambda_1)}) \gs(X- i Y) \left[1 + \gs_5 \right]$$
$$ = -s (1 + \lambda_1 \beta)^2\,\epsilon^{(- \lambda_1)} \cdot (X- i Y) = -\di{s \over \sqrt{2}} (1 + \lambda_1 \beta)^2\,\left[ 1 - \lambda_1 \cos \theta \right] $$

\nin Comme ~$ Z \cdot  \epsilon^{(- \lambda_1)} = \di{\lambda_1 \over \sqrt{2}} \sin \theta$, on aboutit \`a 

$$ {\rm Tr} \, \gs(p_3) \gs(\epsilon^{(- \lambda_1)}) \gs(Q_a) \gs(p_4) \gs(\epsilon^{(-\lambda_1)}) \gs(X- i Y) \left[1 + \gs_5 \right]  $$
$$ =  \lambda_1 \, \di{{s \sqrt{s}}\over 2} \,(1 + \lambda_1 \beta )^2 \sin \theta \left[ 1 - \lambda_1 \cos \theta \right] $$

\vv \nin Finalement, pour $\left|\lambda_1\right| = \left|\lambda_2\right| = 1$, on obtient 

$$ X_a(\lambda_1, \lambda_2) = \di{{s \sqrt{s}}\over 2} \, \sin \theta\,\left[ \hskip -0.05cm\di{{}\over{}} \delta_{\lambda_1, \lambda_2} (1 - \beta^2) \left[\cos \theta + \beta\right] + \delta_{\lambda_2, - \lambda_1} \lambda_1\,(1 + \lambda_1 \beta)^2 \left[1 - \lambda_1 \cos \theta \right] \right]$$

\beq \fbox{\fbox{\rule[-0.4cm]{0cm}{1cm}~$\begin{array}{c}~\\
 X_a(\lambda_1, \lambda_2) = \di{{s \sqrt{s}}\over 2} \, \sin \theta\,\left\llbracket \hskip -0.05cm\di{{}\over{}} \delta_{\lambda_1, \lambda_2} (1 - \beta^2) \left[\cos \theta + \beta\right] \right.
\\~\\
\left. \di{{}\over{}} + \delta_{\lambda_2, - \lambda_1} \lambda_1\,(1 + \lambda_1 \beta)^2 \left[1 - \lambda_1 \cos \theta \right] \right\rrbracket ,~~~\left|\lambda_1\right| = \left|\lambda_2 \right|=1 \\~
\end{array} $~}}  \enq


\vvv
\vvv \nin \leftpointright~Prenons maintenant $\lambda_1 = 0$, $\left|\lambda_2 \right|=1$. Ici, $\epsilon^{(0)}_3 \equiv z_3$ et, dans le couplage d'h\'elicit\'e en voie $s$, $\gs(p_3) \gs(\epsilon^{(0)}_3) = M_W \gs(t_3) \gs(z_3) = M_W \gs(T) \gs(Z')$. Donc\footnote{Rappelons que $\epsilon^{\star}_2 = \left[\epsilon^{(- \lambda_2)}_1 \right]^\star = - \epsilon^{(\lambda_2)}$.}  

$$ X_a(0, \lambda_2) = - M_W \di{s \over 4} \,{\rm Tr} \, \gs(T) \gs(Z') \gs(Z+ \beta Z') \gs(T - \beta Z') \gs(\epsilon^{(\lambda_2)}) \gs(X- i Y) \left[1 + \gs_5 \right] $$
$$ =  M_W \di{s \over 4} \,{\rm Tr} \, \gs(T) \left[ \cos \theta + \beta + \sin \theta \gs(Z') \gs(X') \right] \gs(T - \beta Z') \gs(\epsilon^{(\lambda_2)}) \gs(X- i Y) \left[1 + \gs_5 \right] $$
$$ =  M_W \di{s \over 4} \,{\rm Tr} \,\left[ \cos \theta + \beta + \sin \theta \gs(Z') \gs(X') \right] \left[ 1+ \beta \gs(Z') \gs(T) \right] \gs(\epsilon^{(\lambda_2)}) \gs(X- i Y) \left[1+ \gs_5 \right] $$

\newpage

\nin Or, d'une part, $\gs(Z') \gs(X') = \gs(Z) \gs(X)$, et, d'autre part, 

\beq \fbox{\rule[-0.4cm]{0cm}{1cm}~$\left[ 1+ \beta \gs(Z') \gs(T) \right] \gs(\epsilon^{(\lambda_2)}) = \gs(\epsilon^{(\lambda_2)}) \left[ 1 - \lambda_2 \beta \gs_5 \right] $~}  \label{formule4} \enq

\vv \nin On a ainsi 

$$ X_a(0, \lambda_2) =  M_W \di{s \over 4} \left[ 1+ \lambda_2 \beta \right] {\rm Tr} \, \gs(X - iY) \left[ \cos \theta + \beta + \sin \theta \gs(Z) \gs(X) \right] \gs(\epsilon^{(\lambda_2)}) \left[1 - \gs_5 \right]$$
$$ =   M_W \di{s \over 4} \left[ 1+ \lambda_2 \beta \right] {\rm Tr} \left\{ \hskip -0.1cm \di{{}\over{}} \left[ \cos \theta + \beta \right] \gs(X - i Y)  + \sin \theta \gs(Z) + \sin \theta \gs(T) \gs_5 \right\}  \gs(\epsilon^{(\lambda_2)}) \left[1 - \gs_5 \right]$$
$$ =  M_W s  \left[ 1+ \lambda_2 \beta \right] \left\{ \hskip -0.1cm \di{{}\over{}} \left[ \cos \theta + \beta \right] (X - i Y) + \sin \theta Z \right\} \cdot \epsilon^{(\lambda_2)}$$

\nin soit, tous calculs effectu\'es, 
\vv
\beq \fbox{\fbox{\rule[-0.4cm]{0cm}{1cm}~$\begin{array}{c}~\\
 X_a(0, \lambda_2) = \di{{M_W s}\over \sqrt{2}} \left[1 + \lambda_2 \beta \right] \left[ 1 + \lambda_2 \cos \theta \right] \left[ \beta - \lambda_2 + 2 \cos \theta \right], 
\\~\\
\left|\lambda_2 \right| = 1 \\~
\end{array} $~}}  \enq


\vvv 
\vvv \nin \leftpointright Le lecteur v\'erifiera que l'on obtient de fa\c{c}on similaire : 
\vv
\beq \fbox{\fbox{\rule[-0.4cm]{0cm}{1cm}~$\begin{array}{c}~\\
 X_a(\lambda_1, 0) = -\di{{M_W s}\over \sqrt{2}} \left[1 - \lambda_1 \beta \right] \left[ 1 - \lambda_1 \cos \theta \right] \left[ \beta + \lambda_1 + 2 \cos \theta \right], 
\\~\\
\left|\lambda_1 \right| = 1 \\~
\end{array} $~}}  \enq


\vvv
\vvv \nin \leftpointright~ Prenons maintenant $\lambda_1 = \lambda_2 =0$. On a alors $\gs(p_3) \gs(\epsilon^{(0)}_3) = - \gs(p_4) \gs(\epsilon^{(0)}_4) = M_W\, \gs(T) \gs(Z')$, et  

$$ X_a(0, 0) = - M^2_W \, \di{\sqrt{s}\over 2}\, {\rm Tr}\, \gs(T) \gs(Z') \gs(Z + \beta Z') \gs(T) \gs(Z') \gs(X -i Y) \left[1 + \gs_5\right] $$
$$ =  M^2_W \, \di{\sqrt{s}\over 2}\, {\rm Tr} \left\{\hskip -0.1cm \di{{}\over{}} \left[ \cos \theta + \beta \right] \gs(Z') + \sin \theta \gs(X') \right\} \gs(X - iY) \left[ 1 + \gs_5 \right] $$
$$ = 2 M^2_W \sqrt{s} \left[ (\cos \theta + \beta ) Z' \cdot X + \sin \theta X'\cdot X \right],~~{\rm soit} $$

\vv
\beq \fbox{\fbox{\rule[-0.4cm]{0cm}{1cm}~$\begin{array}{c}~\\
 X_a(0, 0) = -2 M^2_W \sqrt{s} \sin \theta \left[ \beta + \cos \theta \right]
\\~
\end{array} $~}}  \enq

\newpage
\section{Amplitudes d'h\'elicit\'e de \,$ g + g \rightarrow g + g$ \label{4gluons}}

\vv \nin La diffusion de deux gluons, $g + g \rightarrow g + g$, est un processus de pure Chromodynamique Quantique. A l'ordre le plus bas selon la constante de couplage forte $ \sqrt{4 \pi \alpha_s}$, il est d\'ecrit par les diagrammes de la figure (\ref{fig:4g})

\vvv
\begin{figure}[hbt]
\centering
\includegraphics[scale=0.3, width=12cm, height=3.5cm]{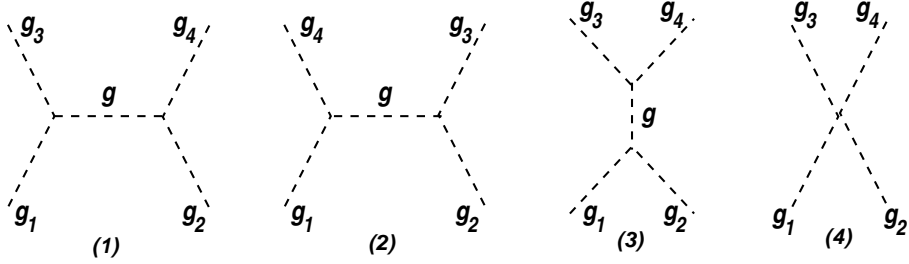}
\vskip 0.25cm

\caption{Diagrammes de Feynman pour $g+ g \rightarrow g+g$} \label{fig:4g}
\end{figure}

\nin correspondant respectivement aux amplitudes tensorielles suivantes (obtenues apr\`es division par $4 \pi \alpha_s$) :

$$ A_1 = - \di{1\over{(p_1-p_3)^2}} \,f_{\alpha \gs a} \, f_{\beta \delta a} \left\{ \hskip -0.07cm \di{{}\over{}} g_{\mu \rho} (p_1 + p_3)_\omega +g_{\rho \omega} (p_1 - 2 p_3)_\mu - g_{\mu \omega} ( 2 p_1 - p_3)_\rho \right\} g^{\omega \pi} \times$$
$$ \left\{ \hskip -0.07cm \di{{}\over{}} g_{\nu \sigma} (p_4 + p_2)_\pi +g_{\nu \pi} (p_4 - 2 p_2)_\sigma - g_{\sigma \pi}( 2p_4 - p_2)_\nu \right\} $$

$$ A_2 = -\di{1\over{(p_1-p_4)^2}}\, f_{\alpha \delta a} \, f_{\beta \gs a} \left\{ \hskip -0.07cm \di{{}\over{}} g_{\mu \sigma} (p_1 + p_4)_\omega +g_{\sigma \omega} (p_1 - 2 p_4)_\mu - g_{\mu \omega} ( 2 p_1 - p_4)_\sigma \right\} g^{\omega \pi} \times$$
$$ \left\{ \hskip -0.07cm \di{{}\over{}} g_{\nu \rho} (p_3 + p_2)_\pi +g_{\nu \pi} (p_3 - 2 p_2)_\rho - g_{\rho \pi}( 2p_3 - p_2)_\nu \right\} $$

$$ A_3 = \di{1 \over{(p_1 + p_2)^2}} \, f_{\alpha \beta a} \,f_{\gs \delta a} \left\{ \hskip -0.07cm \di{{}\over{}} g_{\mu \nu} (p_1 - p_2)_\omega +g_{\nu \omega} (p_1 + 2 p_2)_\mu - g_{\mu \omega} ( 2 p_1 + p_2)_\nu \right\} g^{\omega \pi} \times$$
$$ \left\{ \hskip -0.07cm \di{{}\over{}} g_{\rho \sigma} (p_3 - p_4)_\pi +g_{\sigma \pi} (p_3 + 2 p_4)_\rho - g_{\rho \pi}( 2p_3 + p_4)_\sigma \right\} $$

$$ A_4 =  f_{\alpha \beta a} \, f_{\gs \delta a} \left\{ \hskip -0.07cm \di{{}\over{}} g_{\mu \sigma}\, g_{\nu \rho} - g_{\mu \rho}\,g_{\nu \sigma} \right\} + f_{\alpha \gs a} \, f_{\delta \beta a} \left\{ \hskip -0.07cm \di{{}\over{}} g_{\mu \nu}\, g_{\rho \sigma} - g_{\mu \sigma}\,g_{\rho \nu} \right\} $$
$$+ f_{\gs \beta a} \, f_{\alpha \delta a} \left\{ \hskip -0.07cm \di{{}\over{}} g_{\rho \sigma}\, g_{\nu \mu} - g_{\rho \mu}\,g_{\nu \sigma} \right\}  $$

\vv \nin Dans ces expressions, les grandeurs $f_{\alpha \beta a}$, etc, sont les constantes de structure de l'alg\`ebre de Lie su(3), et une sommation sur l'indice $a$ y est implicite. Ces constantes ne sont pas ind\'ependantes car elles satisfont l'identit\'e de Jacobi 

\beq  f_{\alpha \beta a}\, f_{a \gs \delta} + f_{\beta \gs a}\, f_{a \alpha \delta} + f_{\gs \alpha a} \, f_{a \beta \delta} =0 \label{jacobi1} \enq

\vv \nin (avec sommation sur $a$), soit 

\beq  f_{\alpha \beta a} \, f_{\gs \delta a} = f_{\alpha \gs a} \, f_{\beta \delta a} - f_{\alpha \delta a}\, f_{\beta \gs a} \label{jacobi2} \enq 

\vv \nin Cette derni\`ere relation permet des regroupements de termes dans l'amplitude totale. Mais avant de les effectuer, pr\'ecisons les notations. Le r\'ef\'erentiel du centre de masse de la r\'eaction a pour base $(T, X,Y,Z)$. Les gluons sont des particules vectorielles de masse nulle et n'ont donc que des polarisations circulaires. Leurs 4-impulsions, indices de Lorentz, indices de couleur, 4-vecteurs de polarisation sont les suivants :

\vv \nin  \ding{51}~$(p_1, \mu, \alpha, \epsilon_1)$ et $(p_2, \nu, \beta, \epsilon_2)$ pour les gluons initiaux, avec $p_1 = \di{\sqrt{s} \over 2} \left[ T+ Z\right]$, $p_2 = \di{\sqrt{s} \over 2} \left[ T- Z\right]$, $\epsilon_1 = - \di{1\over \sqrt{2}} \left[ \lambda_1 X + i Y \right]$, $\epsilon_2 = - \di{1\over \sqrt{2}} \left[ -\lambda_2 X + i Y \right]$ ; 

\nin \ding{51}~$(p_3, \rho, \gs, \epsilon_3)$ et $(p_4, \sigma, \delta, \epsilon_4)$ pour les gluons finals, avec $p_3 = \di{\sqrt{s} \over 2} \left[ T+ Z'\right]$, $p_4 = \di{\sqrt{s} \over 2} \left[ T- Z'\right]$, $\epsilon_3 = - \di{1\over \sqrt{2}} \left[ \lambda_3 X' + i Y \right]$, $\epsilon_4 = - \di{1\over \sqrt{2}} \left[ -\lambda_4 X' + i Y \right]$, $Z' = Z \cos \theta + X \sin \theta$, $X' = X \cos \theta - Z \sin \theta$ ; 

\vv \nin \ding{51}~enfin : $s = 2 p_1\cdot p_2 = 2 p_3 \cdot p_4$, $t = 2 p_1\cdot p_3 = 2 p_2 \cdot p_4$, $u = 2 p_1 \cdot p_4 = 2 p_2 \cdot p_3$, et l'on a $s = t+u$. 

\vv
\vv \nin En couplage d'h\'elicit\'e dans la voie $s$, les 4-impulsions $p_1$ et $p_2$ sont toutes deux orthogonales \`a la fois \`a $\epsilon_1$ et $\epsilon_2$ ; de m\^eme, les 4-impulsions $p_3$ et $p_4$ sont toutes deux orthogonales \`a la fois \`a $\epsilon_3$ et $\epsilon_4$. En cons\'equence, nous \'eliminerons dans l'amplitude tensorielle les termes proportionnels \`a $p_{1 \mu},\, p_{1 \nu}, \,p_{2 \mu}, \,p_{2 \nu}, \,p_{3 \rho}, \,p_{3 \sigma}, \, p_{4 \rho}, \,p_{4 \sigma}$, dont les contributions sont finalement nulles\footnote{Notons ici que l'amplitude tensorielle est invariante de jauge relativement \`a chacun des indices $\mu$, $\nu$, $\rho$ et $\sigma$, \`a la condition que les autres indices soient contract\'es avec les vecteurs de polarisation qui leur sont respectivement attach\'es.}. Tenant compte aussi du fait que $(p_1 + p_2)_{\mu\, {\rm ou}\, \nu} = (p_3 + p_4)_{\mu\,{\rm ou}\, \nu} \equiv 0$ et $(p_3 + p_4)_{\rho\, {\rm ou}\, \sigma} = (p_1 + p_2)_{\rho \,{\rm ou}\, \sigma} \equiv 0$, on obtient ainsi l'amplitude tensorielle ``effective"  
suivante

$$ T = 2\,f_{\alpha \gs a}\, f_{\beta \delta a} \,T_1 +2\, f_{\alpha \delta a}\, f_{\beta \gs a} \,T_2,~~{\rm o\grave{u}}$$
$$ T_1 =  g_{\mu \sigma}\, g_{\nu \rho} + \di{{u}\over t} \, g_{\mu \rho}\, g_{\nu \sigma} - \di{{u}\over s} \, g_{\mu \nu}\, g_{\rho \sigma} + \di{2 \over t} \left\{ \hskip -0.1cm \di{{}\over{}}  p_{1 \rho}\,\left[ p_{3 \mu}\, g_{\nu \sigma}  - p_{3 \nu}\, g_{\mu \sigma} \right]  \right. $$
\beq \left. \di{{}\over{}}  +p_{1 \sigma}\, \left[ p_{3 \nu}\, g_{\mu \rho} -p_{3 \mu}\, g_{\nu \rho} \right] - p_{1 \rho} \,p_{1 \sigma}\, g_{\mu \nu} - p_{3 \mu}\, p_{3 \nu}  \, g_{\rho \sigma}   \right\}  \enq
$$ T_2 = g_{\mu \rho}\, g_{\nu \sigma} + \di{{t}\over u} \, g_{\mu \sigma}\, g_{\nu \rho} - \di{{t}\over s} \, g_{\mu \nu}\, g_{\rho \sigma} + \di{2 \over u} \left\{ \hskip -0.1cm \di{{}\over{}}  - p_{1 \sigma}\,\left[ p_{3 \mu}\, g_{\nu \rho}  - p_{3 \nu}\, g_{\mu \rho} \right]  \right. $$
$$\left. \di{{}\over{}}  -p_{1 \rho}\, \left[ p_{3 \nu}\, g_{\mu \sigma} -p_{3 \mu}\, g_{\nu \sigma} \right] - p_{1 \rho} \,p_{1 \sigma}\, g_{\mu \nu} - p_{3 \mu}\, p_{3 \nu}  \, g_{\rho \sigma}   \right\}  $$

\nin Clairement, on obtient $T_2$ \`a partir de $T_1$ en \'echangeant entre eux soit les gluons finals, soit les gluons initiaux\footnote{Notons aussi que les facteurs de couleur associ\'es respectivement \`a $T_1$ et \`a $T_2$ sont ind\'ependants et que de ce fait, $T_1$ et $T_2$ sont s\'epar\'ement invariants de jauge, sous la condition mentionn\'ee dans la note pr\'ec\'edente.}. Pour le calcul des amplitudes d'h\'elicit\'e, qui est ici encore un excellent exercice de manipulation des 4-vecteurs de polarisation et de leurs indices de spin, les formules suivantes sont utiles : 

$$ \epsilon_1 \cdot \epsilon_2 = \delta_{\lambda_2, \lambda_1} = \di{1\over 2} (1 + \lambda_1 \lambda_2),~~\epsilon_3 \cdot \epsilon_4 = \delta_{\lambda_3, \lambda_4} = \di{1\over 2} (1 + \lambda_3 \lambda_4), $$
$$ \epsilon_1 \cdot \epsilon^\star_3 = - \di{1\over 2} \left[ 1+ \lambda_1 \lambda_3 \cos \theta \right],~~\epsilon_1 \cdot \epsilon^\star_4 = - \di{1\over 2} \left[ 1 - \lambda_1 \lambda_4 \cos \theta \right], $$
$$ \epsilon_2 \cdot \epsilon^\star_3 = - \di{1\over 2} \left[ 1- \lambda_2 \lambda_3 \cos \theta \right],~~\epsilon_2 \cdot \epsilon^\star_4 = - \di{1\over 2} \left[ 1 + \lambda_2 \lambda_4 \cos \theta \right], $$
$$ p_1 \cdot \epsilon^\star_3 = -\lambda_3 \,\di{\sqrt{s} \over{2 \sqrt{2}}} \,\sin \theta,~~p_1 \cdot \epsilon^\star_4 = \lambda_4 \,\di{\sqrt{s} \over{2 \sqrt{2}}} \,\sin \theta, $$
$$ p_3 \cdot \epsilon_1 = \lambda_1 \,\di{\sqrt{s} \over{2 \sqrt{2}}} \,\sin \theta,~~p_3 \cdot \epsilon_2 = -\lambda_2 \,\di{\sqrt{s} \over{2 \sqrt{2}}} \,\sin \theta $$
$$ \cos \theta = 1 - \di{{2t}\over s} = \di{{2u}\over s} -1,~~\sin^2 \theta = \di{{4 ut}\over s^2} $$

\vv \nin \leftpointright~Pour simplifier l'\'ecriture, posons $c = \cos \theta$ puis calculons : 

$$ {T_1}_{\lambda_1\, \lambda_2\,\lambda_3\,\lambda_4} = (\epsilon_1 \cdot \epsilon^\star_4)(\epsilon_2 \cdot \epsilon^\star_3) + \di{u \over t} (\epsilon_1 \cdot \epsilon^\star_3)(\epsilon_2 \cdot \epsilon^\star_4) - \di{u \over s} (\epsilon_1 \cdot \epsilon_2)(\epsilon^\star_3 \cdot \epsilon^\star_4) $$
$$ + \di{2 \over t} \left\llbracket  \hskip -0.1cm \di{{}\over {}}  (p_1 \cdot \epsilon^\star_3) \left[ \hskip -0.1cm \di{{}\over{}} (p_3 \cdot \epsilon_1)(\epsilon_2\cdot \epsilon^\star_4) - (p_3 \cdot \epsilon_2)(\epsilon_1\cdot \epsilon^\star_4) \right]  \right. $$ $$ +  (p_1 \cdot \epsilon^\star_4) \left[ \hskip -0.1cm \di{{}\over{}} (p_3 \cdot \epsilon_2)(\epsilon_1\cdot \epsilon^\star_3) - (p_3 \cdot \epsilon_1)(\epsilon_2\cdot \epsilon^\star_3) \right]$$
$$\left. \di{{}\over{}}  - (p_1 \cdot \epsilon^\star_3)(p_1\cdot \epsilon^\star_4)(\epsilon_1\cdot \epsilon_2) - (p_3 \cdot \epsilon_1 )(p_3 \cdot \epsilon_2)(\epsilon^\star_3 \cdot \epsilon^\star_4) \right\rrbracket $$

$$ = \di{1\over 4} \left[1- \lambda_1 \lambda_4 c \right] \left[1- \lambda_2 \lambda_3 c \right] + \di{u \over{4t}} \left[1 +\lambda_1 \lambda_3 c \right] \left[1+ \lambda_2 \lambda_4 c \right] - \di{u\over s} \delta_{\lambda_2, \lambda_1} \delta_{\lambda_4, \lambda_3} $$
$$ + \di{1 \over {4t}} \left[ \di{\sqrt{s}\over{2 \sqrt{2}}} \sin \theta\right]^2 \left\{  \di{\lambda_3 \over 2} \left[ \hskip -0.1cm \di{{}\over{}} \lambda_1[ 1 + \lambda_2 \lambda_4 c ] + \lambda_2 [ 1 - \lambda_1 \lambda_3 c]  \right] \right. $$
$$\left. + \di{\lambda_4 \over 2} \left[ \hskip -0.1cm \di{{}\over{}}  \lambda_2 [ 1 + \lambda_1 \lambda_3 c ] + \lambda_1 [ 1 - \lambda_2 \lambda_3 c ] \right] + \lambda_3 \lambda_4 \delta_{\lambda_2, \lambda_1} + \lambda_1 \lambda_2 \delta_{\lambda_4, \lambda_3} \right\}$$
$$ = \di{1\over 4} \left[1- \lambda_1 \lambda_4 c \right] \left[1- \lambda_2 \lambda_3 c \right] + \di{{s-t} \over{4t}} \left[1 +\lambda_1 \lambda_3 c \right] \left[1+ \lambda_2 \lambda_4 c \right] $$
$$ \left.+ \di{u \over {4s}} \left\{ \hskip -0.2cm \di{{}\over{}} \right. - 4 \delta_{\lambda_2, \lambda_1} \delta_{\lambda_4, \lambda_3} + \di{1\over 2} [\lambda_1 + \lambda_2][\lambda_3 + \lambda_4 ] + \lambda_3 \lambda_4 \delta_{\lambda_2, \lambda_1} + \lambda_1 \lambda_2 \delta_{\lambda_4, \lambda_3} \right\}$$
$$= - \di{1\over 4} [\lambda_1 + \lambda_2 ] [\lambda_3 + \lambda_4] [ \di{{2u} \over s} - 1] + \di{s \over{4t}} \left[\hskip -0.1 cm \di{{}\over{}}  1 + [ \lambda_1 \lambda_3 + \lambda_2 \lambda_4 ] [ 1 - \di{{2t}\over s}] \right.$$
$$\left. + \lambda_1 \lambda_2 \lambda_3 \lambda_4 [ 1 - \di{{4 ut}\over s^2} ] \right] + \di{u \over{4s}} \left\{\cdots \right\}$$

\vv \nin Rempla\c{c}ant ensuite $4 \delta_{\lambda, \lambda'}$ par $1 + \lambda \lambda' $, on trouve finalement 
$$  {T_1}_{\lambda_1\, \lambda_2\,\lambda_3\,\lambda_4} = \di{s \over{4 t} } [ 1+ \lambda_1 \lambda_3 ][ 1 + \lambda_2 \lambda_4 ] - \di{u \over{4 s}} [ 1- \lambda_1 \lambda_2 ][1- \lambda_3 \lambda_4] - \di{1\over 4}[\lambda_1 - \lambda_2][ \lambda_3 - \lambda_4 ] $$

\nin puis 
$${T_2}_{\lambda_1\, \lambda_2\,\lambda_3\,\lambda_4} = \di{s \over{4 u} } [ 1+ \lambda_1 \lambda_4 ][ 1 + \lambda_2 \lambda_3 ] - \di{t \over{4 s}} [ 1- \lambda_1 \lambda_2 ][1- \lambda_3 \lambda_4] - \di{1\over 4}[\lambda_1 - \lambda_2][ \lambda_4 - \lambda_3 ] $$

\vv \nin L'amplitude g\'en\'erique totale s'\'ecrit donc (sommation implicite sur $a$) 

\beq \fbox{\fbox{\rule[-0.4cm]{0cm}{1cm}~$\begin{array}{c}
~\\
T^{\alpha \beta \gs \delta}_{\lambda_1\, \lambda_2\,\lambda_3\,\lambda_4} = \di{1\over 2}  \left[ \hskip -0.07 cm \di{{}\over{}} f_{\alpha \gs a} \, f_{\beta \delta a}\,X_1 + f_{\alpha \delta a}\, f_{\beta \gs a} 
\,X_2 \right]~~~{\rm avec} 
\\~\\
X_1 = \di{s \over t }\, [ 1+ \lambda_1 \lambda_3 ][ 1 + \lambda_2 \lambda_4 ] 
- \di{u \over s}\, [ 1- \lambda_1 \lambda_2 ][1- \lambda_3 \lambda_4] - [\lambda_1 - \lambda_2][ \lambda_3 - \lambda_4 ],
\\~\\
X_2 = \di{s \over u }\, [ 1+ \lambda_1 \lambda_4 ][ 1 + \lambda_2 \lambda_3 ] 
- \di{t \over s}\, [ 1- \lambda_1 \lambda_2 ][1- \lambda_3 \lambda_4] - [\lambda_1 - \lambda_2][ \lambda_4 - \lambda_3 ]  
\\~
\end{array} $~}}  \enq

\vv \nin Dix amplitudes sont nulles ind\'ependamment des facteurs de couleur $f_{\alpha \gs a} \, f_{\beta \delta a}$ et $f_{\alpha \delta a}\, f_{\beta \gs a}$ : 

$$ T^{\cdots}_{+++-} = T^{\cdots}_{++-+} = T^{\cdots}_{+-+++} = T^{\cdots}_{-+++} = T^{\cdots}_{---+} = T^{\cdots}_{--+-}=T^{\cdots}_{-+--} = T^{\cdots}_{+---} =0 $$ 
$$ T^{\cdots}_{++--} = T^{\cdots}_{--++} =0 $$

\vv \nin et les six restantes, nulles ou non selon les valeurs des facteurs de couleur, sont list\'ees dans le tableau VII : 

\beq \fbox{\fbox{\rule[-0.4cm]{0cm}{1cm}~$\begin{array}{c} ~\\
T^{\alpha \beta \gs \delta}_{++++} = T^{\alpha \beta \gs \delta}_{----} = 2 \left[ \hskip -0.1 cm \di{{}\over{}}  f_{\alpha \gs a} \, f_{\beta \delta a}\,\di{ s\over t} + f_{\alpha \delta a}\, f_{\beta \gs a} \,\di{s \over u} \right]
\\~\\
T^{\alpha \beta \gs \delta}_{+-+-} = T^{\alpha \beta \gs \delta}_{-+-+} = 2 \left[ \hskip -0.1 cm \di{{}\over{}}  f_{\alpha \gs a} \, f_{\beta \delta a}\,\di{ u^2\over{s t}} + f_{\alpha \delta a}\, f_{\beta \gs a} \,\di{u \over s} \right]
\\~\\
T^{\alpha \beta \gs \delta}_{+--+} = T^{\alpha \beta \gs \delta}_{-++-} = 2 \left[ \hskip -0.1 cm \di{{}\over{}}  f_{\alpha \gs a} \, f_{\beta \delta a}\,\di{ t\over s} + f_{\alpha \delta a}\, f_{\beta \gs a} \,\di{t^2 \over{su}} \right]
\\~
\end{array} $~}}  \enq

\vv \vv
\centerline{\bf Tableau VII - Amplitudes d'h\'elicit\'e de $g+g \rightarrow g+g$ }

\vvv
\vv \nin Utilisant les relations 

$$ \di{\sum_{\alpha \gs}} \, f_{\alpha \gs a}\, f_{\alpha \gs b} = 3 \delta_{a b},~~~\di{\sum_{\alpha a \delta}}  \, f_{\alpha \gs a} \, f_{\beta \delta a}\, f_{\alpha \delta b} = - \di{\sum_{\alpha a \delta}}  \,f_{\delta \beta a}\,
f_{a \gs \alpha} \, f_{\alpha b \delta} =  \di{3\over 2} \, f_{\beta \gs b},$$

\vv \nin on d\'eduit ensuite le taux d'interaction\footnote{B.L. Combridge, J. Kripfganz, J. Ranft, Phys. Lett. 70B (1977) 234. Attention, dans cette r\'ef\'erence $t$ et $u$ sont les variables de Mandelstam, oppos\'ees \`a $t$ et $u$ d\'efinis ici, lesquels sont positifs.} 

\beq  \di{\sum} \, \left|T\right|^2 = 128 \times 9\times \left[ 3 + \di{{st}\over u^2} + \di{{su} \over t^2} - \di{{ut}\over s^2} \right] \label{taux4g} \enq


\newpage

\section{Compl\'ement I :\, Obtention du taux d'interaction de \\ $\gs^\star + \gs^\star \rightarrow e^{-} +\, e^{+} $ par un calcul de trace ; amplitudes $F_{m\, \ov{m} \,;\, n \,\ov{n}}$ } 

\vv \nin Pour bien comprendre un calcul, il est toujours bon de l'aborder de diff\'erentes fa\c{c}ons afin de cerner ses difficult\'es et de rechercher des astuces permettant de contourner celles-ci. Ainsi, \`a titre de comparaison avec le calcul d'amplitudes d'h\'elicit\'e et pour compl\'eter l'\'etude du processus virtuel  $\gs^\star + \gs^\star \rightarrow e^{-} +\, e^{+} $, nous pr\'esentons le calcul du tenseur 

$$ X_{\mu \rho ; \nu \sigma} = \di{\sum_{\rm pol.}}\, T_{\mu \nu}\, T^\star_{\nu \sigma} ,~~{\rm avec} ~~T_{\mu \nu} = \ov{U_3} \,X_{\mu \nu} \,V_4,~~T^\star_{\rho \sigma} =  \ov{V_4} \,\ov{X}_{\rho \sigma} \,U_3, $$
\beq  X_{\mu \nu} =2 \left[ \di{p_{3 \mu} \over a} - \di{p_{4 \mu} \over b} \right] \gs_\nu  - \di{1\over a}\gs_\mu \gs(p_1) \gs_\nu   + \di{1\over b}  \gs_\nu \gs(p_1) \gs_\mu,   \label{TX1} \enq
$$ \ov{X}_{\rho \sigma} =2 \left[ \di{p_{3 \rho} \over a} - \di{p_{4 \rho} \over b} \right] \gs_\sigma  - \di{1\over a}\gs_\rho \gs(p_1) \gs_\sigma   + \di{1\over b}  \gs_\sigma \gs(p_1) \gs_\rho,   $$

\vv \nin o\`u $\di{\sum_{\rm pol.}}$ est la somme sur tous les \'etats de spin de l'\'electron et du positron, que nous effectuerons en calculant la trace     

$$ X_{\mu \rho ; \nu \sigma} =  {\rm Tr} \left[\gs(p_3) + m \right] X_{\mu \nu} \left[ \gs(p_4) - m \right] \ov{X}_{\rho \sigma} $$

\vv \nin Pour s'\'epargner le fastidieux calcul d'une trace d'un produit de huit matrices $\gs$, nous utiliserons pr\'ealablement la d\'ecomposition suivante d'un produit de trois de ces matrices\footnote{Voir ITL, Eq, 7.59.} : 

$$ \gs_\mu \gs_\alpha \gs_\nu = g_{\mu \alpha}\, \gs_\nu - g_{\mu \nu} \gs_\alpha + g_{\alpha \nu}\gs_\mu - i \epsilon_{\mu \alpha \nu \omega} \gs^\omega \gs_5 $$

\vv \nin pour \'ecrire les tenseurs $X_{\mu \nu}$ et $\ov{X}_{\rho \sigma}$ sous la forme 

$$ X_{\mu \nu} = A_{\mu \nu \alpha} \gs^\alpha  + i B_{\mu \nu \alpha} \gs^\alpha \gs_5, ~~~\ov{X}_{\rho \sigma} = A_{\rho \sigma \beta} \gs^\beta  - i B_{\rho \sigma \beta} \gs^\beta \gs_5,~~{\rm avec} $$ 
$$ A_{\mu \nu \alpha} = 2 \left[ \di{p_{3 \mu} \over a} - \di{p_{4 \mu} \over b} \right] g_{\nu \alpha} +\left( \di{1\over b} - \di{1\over a} \right) \left[\hskip -0.05cm \di{{}\over{}}  p_{1 \mu} g_{\nu \alpha}  - p_{1 \alpha} g_{\mu \nu} + p_{1 \nu} g_{\mu \alpha} \right] $$
$$ B_{\mu \nu \alpha} = \epsilon_{\mu \nu \alpha \omega}\, p^\omega_1 \left( \di{1 \over a} + \di{1\over b} \right) $$

\vv \nin et obtenir ``simplement" : 

$$ \di{1\over 4} X_{\mu \nu ; \rho \sigma} = \left\{\hskip -0.1 cm \di{{}\over {}} A_{\mu \nu \alpha} A_{\rho \sigma \beta} + B_{\mu \nu \alpha} B_{\rho \sigma \beta} \right\} L^{\alpha \beta} $$
$$ + g^{\alpha \beta}  \left\{\hskip -0.1 cm \di{{}\over {}} \left( m^2 - p_3 \cdot p_4\right) B_{\mu \nu \alpha} B_{\rho \sigma \beta} - \left( m^2 + p_3 \cdot p_4 \right) A_{\mu \nu \alpha} A_{\rho \sigma \beta} \right\} $$
 \beq + \left\{\hskip -0.1 cm \di{{}\over {}} B_{\mu \nu \alpha} A_{\rho \sigma \beta} - A_{\mu \nu \alpha} B_{\rho \sigma \beta} \right\} \epsilon^{\alpha \beta \omega \pi}\, p_{3 \omega} p_{4 \pi}, ~~{\rm o\grave{u}}  \label{TX2} \enq
$$ L^{\alpha \beta} = p^\alpha_3 p^\beta_4 + p^\alpha_4 p^\beta_3  $$

\vv \nin L'\'etape suivante, la plus longue, consiste \`a effectuer les contractions des tenseurs dans (\ref{TX2}). Observons tout d'abord que la sym\'etrie entre les deux photons virtuels a \'et\'e \'egar\'ee lorsque le tenseur $X_{\mu \nu}$ a \'et\'e pr\'esent\'e sous la forme (\ref{TX1}), r\'esultant d'un souci  naturel de simplification. Pour tenter de la retrouver et exploiter au mieux l'expression (\ref{TX2}), nous projetterons cette derni\`ere sur deux tenseurs $F^{\mu \rho}$ et $G^{\nu \sigma}$, le premier orthogonal \`a $p_1$ suivant ses deux indices, le second orthogonal \`a $p_2$ suivant ses deux indices. Nous noterons $p_{i \mu} p_{j \rho} F^{\mu \rho} = F(p_i, p_j) = F_{ij}$, et utiliserons le m\^eme type de notation pour les contractions des tenseurs  $G^{\nu \sigma}$ et $L^{\alpha \beta}$.

\newpage 

\vv \nin \ding{172} Calcul de ~$Q_1 =  A_{\mu \nu \alpha} A_{\rho \sigma \beta} L^{\alpha \beta} F^{\mu \rho} G^{\nu \sigma} $    

$$ \bullet ~~ Q_{11} = 4 \left[ \di{p_{3 \mu}\over a} - \di{p_{4 \mu}\over b} \right]  \left[ \di{p_{3 \rho}\over a} - \di{p_{4 \rho}\over b} \right] F^{\mu \rho} G_{\alpha \beta} L^{\alpha \beta} = 4 \left[ \di{F_{33} \over a^2} + \di{F_{44} \over b^2} - \di{\left[ F_{34} + F_{43} \right] \over {ab}} \right] \left[ G_{34} + G_{43} \right] $$  

$$ \bullet~~Q_{12} = \di{2 \over a} \,p_{3 \mu} \left( \di{1\over b} - \di{1\over a} \right) g_{\nu \alpha} \left[ p_{1 \sigma} g_{\rho \beta} - p_{1 \beta} g_{\rho \sigma} \right] F^{\mu \rho} G^{\nu \sigma} L^{\alpha \beta} = \di{2\over a} \left( \di{1\over b} - \di{1\over a} \right) \left[ F_{33} G_{41} + F_{34} G_{3 1} \right] $$
$$ - \di{2\over a} \left( \di{1\over b} - \di{1\over a} \right) \left[ p_1 \cdot p_4 \,H_{33} + p_1 \cdot p_3 \,H_{34} \right],~~ {\rm o\grave{u} }~~ H^{\mu \nu} = g_{\rho \sigma} F^{\mu \rho} G^{ \nu \sigma} $$

$$ \bullet~~Q_{13} = -\di{2 \over b} \,p_{4 \mu} \left( \di{1\over b} - \di{1\over a} \right) g_{\nu \alpha} \left[ p_{1 \sigma} g_{\rho \beta} - p_{1 \beta} g_{\rho \sigma} \right] F^{\mu \rho} G^{\nu \sigma} L^{\alpha \beta} = -\di{2\over b} \left( \di{1\over b} - \di{1\over a} \right) \left[ F_{43} G_{41} + F_{44} G_{3 1} \right] $$
$$ + \di{2\over b} \left( \di{1\over b} - \di{1\over a} \right) \left[ p_1 \cdot p_4 \,H_{43} + p_1 \cdot p_3 \,H_{44} \right] $$

$$ \bullet~~Q_{14} = \di{2 \over a} \,p_{3 \rho} \left( \di{1\over b} - \di{1\over a} \right) g_{\sigma \beta} \left[ p_{1 \nu} g_{\mu \alpha} - p_{1 \alpha} g_{\mu \nu} \right] F^{\mu \rho} G^{\nu \sigma} L^{\alpha \beta} = \di{2\over a} \left( \di{1\over b} - \di{1\over a} \right) \left[ F_{33} G_{14} + F_{4 3} G_{1 3} \right] $$
$$ - \di{2\over a} \left( \di{1\over b} - \di{1\over a} \right) \left[ p_1 \cdot p_4 \,K_{33} + p_1 \cdot p_3 \,K_{34} \right],~~ {\rm o\grave{u} }~~ K^{\rho \sigma} = g_{\mu \nu} F^{\mu \rho} G^{ \nu \sigma} $$

$$ \bullet~~Q_{15} = -\di{2 \over b} \,p_{4 \rho} \left( \di{1\over b} - \di{1\over a} \right) g_{\sigma \beta} \left[ p_{1 \nu} g_{\mu \alpha} - p_{1 \alpha} g_{\mu \nu} \right] F^{\mu \rho} G^{\nu \sigma} L^{\alpha \beta} = -\di{2\over b} \left( \di{1\over b} - \di{1\over a} \right) \left[ F_{34} G_{14} + F_{4 4} G_{1 3} \right] $$
$$ + \di{2\over b} \left( \di{1\over b} - \di{1\over a} \right) \left[ p_1 \cdot p_4 \,K_{43} + p_1 \cdot p_3 \,K_{44} \right] $$

$$ \bullet~~Q_{16} = \left( \di{1\over b} - \di{1\over a} \right)^2 \left[ p_{1 \nu} g_{\mu \alpha} - p_{1 \alpha} g_{\mu \nu} \right] \left[ p_{1 \sigma} g_{\rho \beta} - p_{1 \beta} g_{\rho \sigma} \right] F^{\mu \rho} G^{\nu \sigma} L^{\alpha \beta} $$
$$ = \left( \di{1\over b} - \di{1\over a} \right)^2 \left\{ \hskip -0.1cm \di{{}\over {}} \left[F_{34} + F_{43} \right] G_{11} - p_1 \cdot p_3\, H_{41} - p_1 \cdot p_4 \,H_{31} \right. $$
$$\left. \hskip -0.1cm \di{{}\over {}}  - p_1 \cdot p_3\, K_{41} - p_1 \cdot p_4 \,K_{31}  + 2 (p_1\cdot p_3)(p_1 \cdot p_4) H \right\},~~{\rm o\grave{u}}~~ H = g_{\mu \nu} H^{\mu \nu} $$ 

\vvv
\vv \nin \ding{173} Calcul de ~$Q_2 =  B_{\mu \nu \alpha} B_{\rho \sigma \beta} L^{\alpha \beta} F^{\mu \rho} G^{\nu \sigma} $

\vv \nin Le produit $\epsilon_{\mu \nu \alpha r}\, \epsilon_{\rho \sigma \beta s}$ ne permet pas d'effectuer imm\'ediatement des produits scalaires. On peut lui trouver une forme adapt\'ee en utilisant la formule (2.183) du chapitre 2, reposant sur la d\'efinition du tenseur de Levi-Civita. Compte tenu de la m\'etrique de Minkowski, ledit produit s'exprime selon la longue formule (\ref{eps-eps}) ci-dessous (d\'eveloppement d'un d\'eterminant $4 \times 4$).  

$$ \bullet~~Q_{21} = t_1 \left( \di{1\over b} + \di{1\over a} \right)^2 \left[\hskip -0.07cm \di{{}\over {}} F\,G\,L  + M_{34} + M_{43}  + N_{34} + N_{43} - F \left[ G_{34} + G_{43} \right] - G \left[ F_{34} + F_{43} \right] - M L \right] $$ 

$$ {\rm o\grave{u}}~~~F = g_{\mu \rho} F^{\mu \rho},~~G = g_{\nu \sigma} G^{\nu \sigma},~~M^{\rho \nu} = g_{\mu \sigma} F^{\mu \rho} G^{\nu \sigma},~~N^{\mu \sigma} = g_{\nu \rho} F^{\mu \rho} G^{\nu \sigma},  $$
$$ L = g_{\alpha \beta} L^{\alpha \beta} = 2\, p_3 \cdot p_4,~~M= g_{\rho \nu} M^{\rho \nu} = N = g_{\mu \sigma} N^{\mu \sigma} $$

\newpage

\beq \fbox{\fbox{\rule[-0.4cm]{0cm}{1cm}~$ \begin{array}{c} ~\\ 
\epsilon_{\mu \nu \alpha r}\, \epsilon_{\rho \sigma \beta s} = -g_{rs} \left[\hskip -0.07cm \di{{}\over {}}  g_{\mu \rho}\, g_{\nu \sigma} \,g_{\alpha \beta} + g_{\mu \sigma}\, g_{\nu \beta} \,g_{\alpha \rho} +g_{\mu \beta}\, g_{\nu \rho} \,g_{\alpha \sigma} -g_{\mu \rho}\, g_{\nu \beta} \,g_{\alpha \sigma}  \right.
\\~\\
\left.  \di{{}\over {}} - g_{\mu \beta}\, g_{\nu \sigma} \,g_{\alpha \rho}  - g_{\mu \sigma}\, g_{\nu \rho} \,g_{\alpha \beta}     \right]  + g_{r \beta} \left[\hskip -0.07cm \di{{}\over {}}  g_{\mu s}\, g_{\nu \rho} \,g_{\alpha \sigma} + g_{\mu \rho}\, g_{\nu \sigma} \,g_{\alpha s} +g_{\mu \sigma}\, g_{\nu s} \,g_{\alpha \rho} \right.
\\~\\
 \left.  \di{{}\over {}} -g_{\mu s}\, g_{\nu \sigma} \,g_{\alpha \rho}  - g_{\mu \sigma}\, g_{\nu \rho} \,g_{\alpha s}  - g_{\mu \rho}\, g_{\nu s} \,g_{\alpha \sigma}   \right]  -  g_{r \sigma} \left[\hskip -0.07cm \di{{}\over {}}  g_{\mu \beta}\, g_{\nu s} \,g_{\alpha \rho} + g_{\mu s}\, g_{\nu \rho} \,g_{\alpha \beta} \right. 
\\~\\
\left.  \di{{}\over {}} + g_{\mu \rho}\, g_{\nu \beta} \,g_{\alpha s}  - g_{\mu \beta}\, g_{\nu \rho} \,g_{\alpha s} - g_{\mu \rho}\, g_{\nu s} \,g_{\alpha \beta} - g_{\mu s}\, g_{\nu \beta} \,g_{\alpha \rho} \right] + g_{r \rho} \left[\hskip -0.07cm \di{{}\over {}} g_{\mu \sigma}\, g_{\nu \beta} \,g_{\alpha s}  \right.
\\~\\
\left.  \di{{}\over {}}  + g_{\mu \beta}\, g_{\nu s} \,g_{\alpha \sigma} +  g_{\mu s}\, g_{\nu \sigma} \,g_{\alpha \beta} - g_{\mu \sigma}\, g_{\nu s} \,g_{\alpha \beta}  -g_{\mu s}\, g_{\nu \beta} \,g_{\alpha \sigma} -g_{\mu \beta}\, g_{\nu \sigma} \,g_{\alpha s} \right] 
\\~\\
= g_{rs} \,\epsilon_{\lambda \mu \nu \alpha} \,\epsilon_{\lambda \rho \sigma \beta} - g_{r \beta} \,\epsilon_{\lambda \mu \nu \alpha}\, \epsilon_{\lambda s \rho \sigma} + g_{r \sigma} \,\epsilon_{\lambda \mu \nu \alpha} \,\epsilon_{\lambda \beta s \rho} - g_{r \rho} \,\epsilon_{\lambda \mu \nu \alpha} \,\epsilon_{\lambda \sigma \beta s} 
\\~
\end{array}
$~}}  \label{eps-eps}  \enq

\vvv
\vvv
$$ \bullet~~Q_{22} =  \left( \di{1\over b} + \di{1\over a} \right)^2 \left[\hskip -0.07cm \di{{}\over {}}  F\, G\, L_{11} +p_1 \cdot p_4 \,M_{3 1} + p_1 \cdot p_3\, M_{4 1} - M L_{11} -F \left[ p_1 \cdot p_4\, G_{1 3} + p_1 \cdot p_3 \,G_{1 4} \right] \right] $$

$$ \bullet~~Q_{23} = -\left( \di{1\over b} + \di{1\over a} \right)^2  \left[\hskip -0.07cm \di{{}\over {}} G_{11} \left[ F_{34} + F_{43} \right]+ F \left[ p_1 \cdot p_3\, G_{4 1} + p_1 \cdot p_4\, G_{3 1} \right] \right.$$
$$ \left. \hskip -0.07cm \di{{}\over {}} - p_1 \cdot p_3 \,N_{4 1 } -p_1 \cdot p_4\, N_{31}   -F\, L \,G_{11} \right] $$

\vvv

\vv \nin \ding{174} Calcul de ~$Q_3 = (m^2 - p_3 \cdot p_4 ) B_{\mu \nu \alpha} B_{\rho \sigma \beta} \,g^{\alpha \beta} F^{\mu \rho} G^{\nu \sigma} $

$$ \bullet~~Q_3 = - (m^2 - p_3 \cdot p_4 )\left( \di{1\over b} + \di{1\over a} \right)^2 p^r_1 p^s_1\, F^{\mu \rho} G^{\nu \sigma}  \left[\hskip -0.07cm \di{{}\over {}} g_{\mu \rho} \, g_{\nu \sigma}\, g_{rs} + g_{\mu \sigma}\, g_{\nu s} \, g_{r \rho} + g_{\mu s}\, g_{\nu \rho} \, g_{r \sigma}   \right.$$
$$ \left.\hskip -0.07cm \di{{}\over {}} - g_{\mu \rho} \, g_{\nu s}\, g_{r \sigma} - g_{\mu s}\, g_{\nu \sigma} \, g_{r \rho} - g_{\mu \sigma}\, g_{\nu \rho} \, g_{r s} \right] =  - (m^2 - p_3 \cdot p_4 )\left( \di{1\over b} + \di{1\over a} \right)^2 \left[\hskip -0.07cm \di{{}\over {}} -t_1 F G    \right.$$
$$ \left.\hskip -0.07cm \di{{}\over {}} - F \,G_{11} + t_1 M \right]$$

\vvv

\vv \nin \ding{175} Calcul de ~$Q_4 = -(m^2 + p_3 \cdot p_4 ) A_{\mu \nu \alpha} A_{\rho \sigma \beta} \,g^{\alpha \beta} F^{\mu \rho} G^{\nu \sigma} $

$$ Q_4 = -(m^2 + p_3 \cdot p_4 ) \left\{\hskip -0.07cm \di{{}\over {}} 4 G\, \left[ \di{F_{33} \over a^2} + \di{F_{44} \over b^2} - \di{{[F_{34} + F_{43} ]}\over {ab}} \right] + 2\left( \di{1\over b} - \di{1\over a} \right)\left[ \di{{ H_{41} + K_{41} } \over b} \right.  \right. $$
$$\left. \left. -\di{{ M_{41} + N_{41} } \over b} - \di{{ H_{31} + K_{31}} \over a} + \di{{ M_{31} + N_{31}} \over a} \right]+ \left( \di{1\over b} - \di{1\over a} \right)^2\left[ -t_1 \,H + F\, G_{11} \right] \right\} $$

\newpage

\vv \nin \ding{176} Calcul de ~$Q_5 =  B_{\mu \nu \alpha} \,A_{\rho \sigma \beta} \,\epsilon^{\alpha \beta \omega \pi}\, p_{3 \omega} p_{4 \pi}\,F^{\mu \rho}\,G^{\nu \sigma}$

$$ \bullet~~Q_5 = -\left( \di{1\over b} + \di{1\over a} \right)F^{\mu \rho} G^{\nu \sigma} \left[\hskip -0.07cm \di{{}\over {}} \delta^\beta_\mu\, p_{3 \nu} \,p_1 \cdot p_4 + p_{3 \mu}\,p_{4 \nu}\, p^\beta_1 + p_{4 \mu}\,\delta^\beta_\nu \, p_1 \cdot p_3 - \delta^\beta_\mu\, p_{4 \nu}\, p_1 \cdot p_3 \right.$$
$$ \left. \hskip -0.07cm \di{{}\over {}} - p_{4 \mu}\,p_{3 \nu}\, p^\beta_1 - p_{3 \mu}\, \delta^\beta_\nu\,p_1 \cdot p_4   \right]  \left\{ 2 \left[ \di{ p_{3 \rho}\over a} - \di{p_{4 \rho} \over b} \right] g_{\sigma \beta} + \left( \di{1\over b} - \di{1 \over a} \right) \left[ p_{1 \sigma} \, g_{\rho \beta} - p_{1 \beta}\,g_{\rho \sigma} \right] \right\}$$
$$ = -\left( \di{1\over b} + \di{1\over a} \right) \left\llbracket  \,\di{2 \over a} \left[\, p_1 \cdot p_4\, M_{33} + G_{41} \,F_{33} + p_1\cdot p_3\,G\, F_{43} - p_1\cdot p_3\,M_{34} - G_{31}\,F_{43} - p_1 \cdot p_4\,G\,F_{33}  \right] \right.$$
$$\left.\hskip -0.07cm \di{{}\over {}}   - \di{2 \over b}  \left[\,  p_1 \cdot p_4 \, M_{43} +G_{41}\,F_{34} + p_1 \cdot p_3\, G\,F_{44} -p_1 \cdot p_3\, M_{44} - G_{31}\,F_{44} - p_1 \cdot p_4 \,G\,F_{34}\right] \right.$$
$$ + \left( \di{1\over b} - \di{1\over a} \right)  \left[\hskip -0.07cm \di{{}\over {}} p_1\cdot p_4\,F\,G_{31} +t_1 \,H_{34} +p_1\cdot p_3\, N_{41} -p_1\cdot p_3\, H_{41}   -p_1\cdot p_3\,F\,G_{41}\right. $$
$$\left. \left.\hskip -0.07cm \di{{}\over {}}-t_1\,H_{43} -p_1\cdot p_4\,N_{31} +p_1\cdot p_4\,H_{31} \right] \,\right\rrbracket$$

\vvv

\vv \nin \ding{177} Calcul de ~$Q_6 = - A_{\mu \nu \alpha} \,B_{\rho \sigma \beta} \,\epsilon^{\alpha \beta \omega \pi}\, p_{3 \omega} p_{4 \pi}\,F^{\mu \rho}\,G^{\nu \sigma}$

$$Q_6  = -\left( \di{1\over b} + \di{1\over a} \right) \left\llbracket  \,\di{2 \over a} \left[\, p_1 \cdot p_4\, N_{33} + G_{14} \,F_{33} + p_1\cdot p_3\,G\, F_{34} - p_1\cdot p_3\,N_{34} - G_{13}\,F_{34}   \right. \right.$$
$$ \left.- p_1 \cdot p_4\,G\,F_{33} \right] \left.\hskip -0.07cm \di{{}\over {}}   - \di{2 \over b}  \left[\,  p_1 \cdot p_4 \, N_{43} + G_{14}\,F_{43} + p_1 \cdot p_3\, G\,F_{44}  - p_1 \cdot p_3\, N_{44} \right. \right.$$
$$ \left. - G_{13}\,F_{44} - p_1 \cdot p_4 \,G\,F_{43}\right] + \left( \di{1\over b} - \di{1\over a} \right)  \left[\hskip -0.07cm \di{{}\over {}} p_1\cdot p_4\,F\,G_{13} +t_1 \,K_{34} +p_1\cdot p_3\, M_{41} \right. $$
$$\left. \left.\hskip -0.07cm \di{{}\over {}} -p_1\cdot p_3\, K_{41}   -p_1\cdot p_3\,F\,G_{14} -t_1\,K_{43} -p_1\cdot p_4\,M_{31} +p_1\cdot p_4\,K_{31} \right] \,\right\rrbracket$$

\vv \nin Nous ordonnerons ensuite les termes de la fa\c{c}on suivante : facteur de $F\,G$, facteur de $F$, facteur de $G$, facteurs des $F_{ij} G_{k \ell}$, puis le reste. 

\vvv 
\vv \nin \ding{172} Facteur de $F\,G$ 

$$ R_1 = \left( \di{1\over b} + \di{1\over a} \right)^2 \left[ \hskip -0.07cm \di{{}\over {}} t_1 \, L + L_{11} -t_1 (p_3 \cdot p_4 - m^2) \right],~~{\rm et} $$
$$ t_1\,L + L_{11} -t_1 ( p_3 \cdot p_4 -m^2) = \di{1\over 2} t_1 \,W^2 + \di{1\over 2} (a+t_1)(b+t_1) ~~{\rm o\grave{u}} $$
$$ W^2 = (p_3 + p_4)^2 = 2\, m^2 + 2 \,p_3\cdot p_4 = (p_1 + p_2)^2 = a +b - t_1 - t_2,~~{\rm d'o\grave{u}}  $$
$$R_1 = \di{1\over 2} \left( \di{1\over b} + \di{1\over a} \right)^2 \left( a\,b -t_1\,t_2\right) $$

\vv \nin Pour \'eviter des confusions, le carr\'e de la masse invariante de la paire $e^- e^+$ est ici not\'e $W^2$ plut\^ot que $s$ comme dans les formules de la section 4.4, car le processus {\it virtuel} $\gs^\star + \gs^\star \rightarrow e^- + e^+$ est un sous-processus d'un processus global dont le carr\'e de la masse invariante est \mbox{g\'en\'eralement not\'e $s$.}  

\vvv
\vv \nin \ding{173} Facteur de $F$ 

$$ R_2 = \left( \di{1\over b} + \di{1\over a} \right)^2 \left\{\hskip -0.15cm \di{{}\over {}} -t_1\, [ G_{34} + G_{43} ] -p_1 \cdot p_4\,\left[ G_{13} +G_{31} \right] -p_1\cdot p_3\,\left[  G_{14} +G_{41} \right] + L\, G_{11} \right. $$
$$ \left. \hskip -0.07cm \di{{}\over {}}- G_{11} (p_3 \cdot p_4 - m^2) \right\} 
-(p_3 \cdot p_4 + m^2)  \left( \di{1\over b} - \di{1\over a} \right)^2 \,G_{11} $$
$$ + \left( \di{1\over a^2} - \di{1\over b^2} \right) \left\{ \hskip -0.07cm \di{{}\over {}}p_1 \cdot p_4 \left[ G_{31} + G_{13} \right] - p_1 \cdot p_3\, \left[ G_{14} + G_{41} \right] \right\}$$
$$ = \di{2 \over a^2}\, t_1\,G_{44} + \di{2\over b^2}\,t_1\, G_{33} - \di{2 \over{ab}} \left[\hskip -0.07cm \di{{}\over {}} t_1 \left[ G_{34} + G_{43} \right] +t_2\, G_{11} \right]$$

\vv \nin Pour arriver \`a ce r\'esultat, nous avons effectu\'e, d'une part, les simplifications qui r\'esultent de $G^{\nu \sigma} p_{2 \nu} =  G^{\nu \sigma} p_{2 \sigma} = 0$, et qui permettent d'\'ecrire, par exemple, $G_{3 1}=  G_{33} + G_{34}$, puisque $p_1 = p_3 + p_4 - p_2$, et, d'autre part, tenu compte de la relation $W^2 = 2 ( p_3\cdot p_4 + m^2) = a + b -t_1 -t_2$. 

\vvv
\vv \nin \ding{174} Facteur de $G$ 

\vv \nin En \'ecrivant $F_{22} = F_{33} + F_{44} + F_{34} + F_{43}$, on trouve 

$$R_3 = \di{2 \over a^2}\, t_2\,F_{33} + \di{2\over b^2}\,t_2\, F_{44} - \di{2 \over{ab}} \left[\hskip -0.07cm \di{{}\over {}} t_2 \left[ F_{34} + F_{43} \right] +t_1\, F_{22} \right]$$

\vvv
\vv \nin \ding{174} Facteur de $H$ 

$$ R_4 =  \di{1\over 2} \left( \di{1\over b} - \di{1\over a} \right)^2\,(ab - t_1 t_2) $$

\vvv
\vv \nin \ding{174} Facteur de $M$ 

$$ R_5 = - \di{1\over 2} \left( \di{1\over b} + \di{1\over a} \right)^2\,(ab - t_1 t_2) $$

\vvv
\vv \nin \ding{175} Termes du type $F_{ij} G_{k \ell}$ 

$$ R_6 = - \di{2 \over a^2}  \left\{ \hskip -0.07cm \di{{}\over {}}  4 \,F_{33}\,G_{44} + \left[F_{34} - F_{43}\right]\left[G_{34} - G_{43} \right] \right\} - \di{2 \over b^2}  \left\{ \hskip -0.07cm \di{{}\over {}}  4 \,F_{44}\,G_{33} + \left[F_{34} - F_{43}\right]\left[G_{34} - G_{43} \right] \right\} $$
$$ - \di{4\over{ab}} \left[F_{34} + F_{43} \right]\left[G_{34} + G_{43} \right] $$

\vvv
\vv \nin  \ding{176} Terme proportionnel \`a $H_{34} + K_{34}$ : $R_7 = \di{4\over a} \left[ H_{34} + K_{34} \right]$.

\vvv
\vv \nin  \ding{177} Terme proportionnel \`a $H_{43} + K_{43}$ : $R_7 = \di{4\over b} \left[ H_{43} + K_{43} \right]$.

\vvv
\vv \nin  \ding{178} Termes restants : 

$$ R_8 = - \Delta_{34} \, \left[ \di{{2(t_1 + t_2)}\over ab} + \di{W^2\over a}\left(\di{1\over a} + \di{1\over b} \right) \right] - \Delta_{43} \, \left[ \di{{2(t_1 + t_2)}\over ab} + \di{W^2\over b}\left(\di{1\over a} + \di{1\over b} \right) \right] $$

\vv \nin o\`u l'on a pos\'e $\Delta_{34} = H_{34} + K_{34} - M_{34} - N_{34}$, $\Delta_{43} = H_{43} + K_{43} - M_{43} - N_{43}$.

\vvv
\vv \nin Finalement, on aboutit \`a la formule (\ref{XFG}) suivante, o\`u l'on retrouve toutes les sym\'etries souhait\'ees (entre les photons initiaux d'une part, et entre les leptons finals d'autre part) :

\newpage 

\beq \fbox{\fbox{\rule[-0.5cm]{0cm}{1.2cm}~$\begin{array}{c}
~\\
X_{\mu \rho ; \nu \sigma}\,F^{\mu \rho}\, G^{\nu \sigma} = X_{F G} = 
\\~\\
2\,F\,G\,\left( \di{1\over b} + \di{1\over a} \right)^2 \left( a\,b -t_1\,t_2\right) 
\\~\\
+\, 8\,F\, \left\{ \di{t_1 \over a^2}\,G_{44} + \di{t_1\over b^2}\, G_{33} - \di{1 \over{ab}} \left[\hskip -0.07cm \di{{}\over {}} t_1 \left[ G_{34} + G_{43} \right] +t_2\, G_{11} \right]\right\} 
\\~\\
+\, 8\,G\, \left\{ \di{t_2 \over b^2}\,F_{44} + \di{t_2\over a^2}\, F_{33} - \di{1 \over{ab}} \left[\hskip -0.07cm \di{{}\over {}} t_2 \left[ F_{34} + F_{43} \right] +t_1\, F_{22} \right]\right\} 
\\~\\
+ 2 (ab - t_1 t_2) \left\{ H\left( \di{1\over b} - \di{1\over a} \right)^2 - M \left( \di{1\over b} + \di{1\over a} \right)^2 \right\} 
\\~\\
- \di{8 \over a^2}  \left\{ \hskip -0.07cm \di{{}\over {}}  4 \,F_{33}\,G_{44} + \left[F_{34} - F_{43}\right]\left[G_{34} - G_{43} \right] \right\} 
\\~\\
- \di{8 \over b^2}  \left\{ \hskip -0.07cm \di{{}\over {}}  4 \,F_{44}\,G_{33} + \left[F_{34} - F_{43}\right]\left[G_{34} - G_{43} \right] \right\} 
\\~\\
- \di{16\over{ab}} \left[F_{34} + F_{43} \right]\left[G_{34} + G_{43} \right] 
\\~\\
+ \di{16\over a} \left[ H_{34} + K_{34} \right] + \di{16\over b} \left[ H_{43} + K_{43} \right] 
\\~\\
- 4\,\Delta_{34} \, \left[ \di{{2(t_1 + t_2)}\over ab} + \di{W^2\over a}\left(\di{1\over a} + \di{1\over b} \right) \right] - 4\,\Delta_{43} \, \left[ \di{{2(t_1 + t_2)}\over ab} + \di{W^2\over b}\left(\di{1\over a} + \di{1\over b} \right) \right] 
\\~
\end{array}
$~}}  \label{XFG} \enq

\vv \nin Dans les cas courants, les tenseurs $F^{\mu \nu}$ et $G^{\nu \sigma}$ utilis\'es sont sym\'etriques et l'expression (\ref{XFG}) se simplifie consid\'erablement, puisqu'alors $H_{\omega \pi} = K_{\omega \pi}= M_{\omega \pi} = N_{\omega \pi}$, $\Delta_{34} = \Delta_{43} = 0$ :   

\vvv \vvv
\beq \fbox{\fbox{\rule[-0.5cm]{0cm}{1.2cm}~$\begin{array}{c}
~\\
X_{F G} = 
2\,F\,G\,\left( \di{1\over b} + \di{1\over a} \right)^2 \left( a\,b -t_1\,t_2\right) - 8\,H\,\di{{(ab -t_1 t_2)}\over{ab}} 
\\~\\
+\, 8\,F\, \left\{ \di{t_1 \over a^2}\,G_{44} + \di{t_1\over b^2}\, G_{33} - \di{1 \over{ab}} \left[\hskip -0.07cm \di{{}\over {}} 2\,t_1 G_{34} +t_2\, G_{11} \right]\right\} 
\\~\\
+\, 8\,G\, \left\{ \di{t_2 \over b^2}\,F_{44} + \di{t_2\over a^2}\, F_{33} - \di{1 \over{ab}} \left[\hskip -0.07cm \di{{}\over {}} 2\,t_2 \,F_{34} +t_1\, F_{22} \right]\right\} 
\\~\\
- \di{32 \over a^2}   F_{33}\,G_{44}  
- \di{32 \over b^2} F_{44}\,G_{33}
- \di{64\over{ab}} F_{34} G_{34} 
+ \di{32\over a}  H_{34} + \di{32\over b} H_{43}  
\\~
\end{array}
$~}}  \label{XFG-sym} \enq

\newpage
\vv \nin Prenons par exemple $F^{\mu \rho} = g^{\mu \rho} + \di{{p^\mu_1 p^\rho_1}\over t_1}$. On obtient\footnote{Dans le calcul menant de (\ref{XFG-sym}) \`a (\ref{XG}), tous les termes en $1/t_1$ ou en $1/t^2_1$, qui pourraient \^etre cause de divergence lorsque $t_1 \rightarrow 0$, disparaissent. C'est, en fait, une cons\'equence de l'invariance de jauge du tenseur complet $X_{\mu \rho ; \nu \sigma}$. Cette remarque s'applique aussi au calcul de (\ref{XR}).}

\beq \fbox{\rule[-0.5cm]{0cm}{1.2cm}~$\begin{array}{c}~\\
 X_G = 4\,G \left\{ 2 m^2 t_2 \left(\di{1\over a} + \di{1 \over b} \right)^2 + (ab - t_1 t_2 ) \left( \di{1 \over a^2} + \di{1 \over b^2} \right)  -\di{{2 t_2 W^2}\over {ab}}  \right\} 
\\~\\
 + 16 (t_1 - 2 m^2 ) \left\{ \di{G_{33} \over b^2} + \di{G_{44} \over a^2} - \di{{2 G_{34}}\over {ab}} \right\} - \di{{16 t_2} \over{a b}} \left[ G_{33} + G_{44} \right] \\~
\end{array} $~} \label{XG} \enq

\vv \nin et si, de plus, $G^{\nu \sigma} = g^{\nu \sigma} + \di{{p^\nu_2 p^\sigma_2} \over t_2}$, cette expression devient

\beq \fbox{\rule[-0.5cm]{0cm}{1.2cm}~$\begin{array}{c}~\\
 X_T = 8\, \left( ab - t_1 t_2 \right) \left( \di{1\over a^2} + \di{1 \over b^2} \right)  -\di{{16(t_1 + t_2) W^2} \over {ab}} -32\, m^4 \left( \di{1\over a} + \di{1 \over b} \right)^2 
\\~\\
 +16 m^2 \left[ (t_1 + t_2) \left( \di{1\over a} + \di{1 \over b} \right)^2 + \di{{2 W^2}\over{ab}} \right] \\~
\end{array} $~} \label{XT} \enq

\vv \nin Puis, prenant $F^{\mu \rho} = q^\mu q^\rho$ avec $q = r + p_1 \di{{p_1 \cdot r}\over t_1}$ o\`u $r$ est un 4-vecteur quelconque, (\ref{XFG}) prend la forme   

\beq \fbox{\rule[-0.5cm]{0cm}{1.2cm}~$\begin{array}{c}~\\
X_r = G \left\{ 2 r^2 \left(\di{1\over a} + \di{1 \over b}\right)^2 (ab - t_1 t_2) \right. 
\\~\\
\left. + 8 t_2 \left[ \di{1 \over b} \left( r\cdot p_4 - \di{{r \cdot p_1} \over 2} \right) - \di{1 \over a} \left( r\cdot p_3 - \di{{r \cdot p_1} \over 2} \right) \right]^2 \right. 
\\~\\
 \left. - 2 t_2 (r \cdot p_1)^2 \left(\di{1\over a} + \di{1 \over b}\right)^2 - \di{8 \over {ab}} \left[ t_1 (r \cdot p_2)^2 + r\cdot p_1 (a+b) \right] \right\} -8\, G_{rr} \di{{ab -t_1 t_2}\over{ab}} 
\\~\\
+ \di{{16 G_{3r}}\over {ab}} \left[ \hskip -0.07cm \di{{}\over{}} a \left[2 r\cdot p_4 - r\cdot p_1 \right] + t_2 r\cdot p_1 \right]  + \di{{16 G_{4r}}\over {ab}} \left[ \hskip -0.07cm \di{{}\over{}} b\left[2 r\cdot p_3 - r\cdot p_1 \right] + t_2 r\cdot p_1 \right] 
\\~\\
 + \di{{8\, G_{33}}\over b^2}  \left \{\hskip -0.07cm \di{{}\over{}} 4(r \cdot p_4) \left[ r\cdot p_1 - r\cdot p_4 \right] + r^2  t_1 - r^2t_2 \di{b \over a } \right\} 
\\~\\
+ \di{{8\, G_{44}}\over a^2}  \left \{\hskip -0.07cm \di{{}\over{}} 4(r \cdot p_3) \left[ r\cdot p_1 - r\cdot p_3 \right] + r^2  t_1 -r^2 t_2 \di{a \over b } \right\} 
\\~\\
- \di{{16\, G_{34}}\over {ab}}  \left \{\hskip -0.07cm \di{{}\over{}} (2 r \cdot p_3 - r\cdot p_1) (2 r \cdot p_4 - r\cdot p_1)  + (r\cdot p_1)^2 + r^2 \left[ t_1 + t_2 \right] \right\} \\~
\end{array} $~} \label{XR} \enq

\newpage 
\vv \nin Ci-dessous sont pr\'esent\'ees certaines combinaisons d'amplitudes 

$$F_{m \ov{m}\, ;\, n \ov{n}} = \left[ \epsilon^{m}_1\right]^\mu  \left[ \epsilon^{\bar{m}}_1\right]^{\star \rho} X_{\mu \rho\,;\, \nu \sigma} \left[ \epsilon^{n}_2\right]^\nu  \left[ \epsilon^{\bar{n}}_1\right]^{\star \sigma} = \di{\sum_{\rm pol}} ~T_{ m n} T^\star_{\bar{m} \,\bar{n}} $$ 

\nin intervenant dans le taux d'interaction de la production d'une paire $e^- e^+$ par \'echange de deux photons virtuels, selon le processus d\'ecrit au paragraphe 1.5.3 du chapitre 1\footnote{Voir, par exemple, N. Arteaga, C. Carimalo, P. Kessler, S. Ong, O. Panella, Phys. Rev. 52, 4920 (1995).}. On peut les obtenir en sommant des amplitudes d'h\'elicit\'e calcul\'ees dans la section 4.4 (avec le changement $s \rightarrow W^2$), ou bien en utilisant directement la formule (\ref{XFG-sym}).  

\vskip -0.2cm
\beq \fbox{\fbox{\rule[-0.4cm]{0cm}{1cm}~$\begin{array}{c}  
~\\
F_{++\,;\, ++} + F_{++\,;\, --} = \di{4 \over{a^2 b^2}} \left\{ \hskip -0.07cm \di{{}\over {}} \left[ab -t_1 t_2\right]\left[ Z^2 - 2 ab \right] - 4 \eta^4 Z^2  \right. 
\\~\\
\left. \hskip -0.07cm \di{{}\over {}}+ 2 \eta^2 Z \left[ 2ab - Z(t_1 + t_2) \right] \right\} 
\\~\\
 F_{++\,;\, +-} = \di{{4 Z \eta^2} \over{a^2 b^2}} \left\{ \hskip -0.07cm \di{{}\over {}} 2
\left[Z \eta^2 -ab \right] + Z t_1 \right\} 
\\~\\
 F_{+-\,;\, ++} = \di{{4 Z \eta^2} \over{a^2 b^2}} \left\{ \hskip -0.07cm \di{{}\over {}} 2
\left[Z \eta^2 -ab \right] + Z t_2 \right\}
\\~\\
F_{+-\,;\, +-} = \di{8 \over{a^2 b^2}} \left\{ \hskip -0.07cm \di{{}\over {}} ab\, t_1 t_2 -\left[ ab -Z \eta^2\right]^2 \right\} ,~~F_{+-\,;\,-+} = - \di{{8 Z^2 \eta^4}\over{a^2 b^2}} 
\\~\\
F_{++\,;\, 0-} - F_{++\,;\,+0} = \di{8 \over{a^2 b^2}} \sqrt{ \di{{2t_2} \over W^2}}\, \eta \,\zeta\, X_1  \left\{ \hskip -0.07cm \di{{}\over {}} ab - Z \left[2 \eta^2 + t_1 \right] \right\}
\\~\\
 F_{+0\,;\, ++} - F_{0-\,;\,++} = \di{8 \over{a^2 b^2}} \sqrt{ \di{{2t_1} \over W^2}}\, \eta \,\zeta\, X_2  \left\{ \hskip -0.07cm \di{{}\over {}} ab - Z \left[2 \eta^2 + t_2 \right] \right\} 
\\~\\
F_{+-\,;\,0-} - F_{+-\,;\,+0} = \di{16 \over{a^2 b^2}} \sqrt{ \di{{2t_2} \over W^2}}\, \eta \,\zeta\, X_1  \left\{ \hskip -0.07cm \di{{}\over {}} Z \eta^2 -ab \right\} 
\\~\\
 F_{+0\,;\,+-} - F_{0-\,;\,+-} = \di{16 \over{a^2 b^2}} \sqrt{ \di{{2t_1} \over W^2}}\, \eta \,\zeta\, X_2  \left\{ \hskip -0.07cm \di{{}\over {}} Z \eta^2 -ab \right\} 
\\~\\
F_{0+\,;\,+-} - F_{-0\,;\,+-} = \di{16 \over{a^2 b^2}} \sqrt{ \di{{2t_1} \over W^2}}\, \eta^3 \,\zeta\, X_2 Z
\\~\\
F_{++ \,;\, 00} = \di{{2 t_2} \over W^2} \left\{ \hskip -0.07cm \di{{}\over {}} \left[2 \eta^2 + t_1 \right] \left[ W^2 Z^2 - 4 X^2_1 \zeta^2 \right] -4 t_1 W^2 a b \right\} 
\\~\\
 F_{00 \,;\, ++} = \di{{2 t_1} \over W^2} \left\{ \hskip -0.07cm \di{{}\over {}} \left[2 \eta^2 + t_2 \right] \left[ W^2 Z^2 - 4 X^2_2 \zeta^2 \right] -4 t_2 W^2 a b \right\} 
\\~\\
F_{+-\,;\, 00} = \di{{4 t_2 \eta^2}\over{a^2 b^2 W^2}} \left\{ \hskip -0.07cm \di{{}\over {}} 4 X^2_1 \zeta^2 - Z^2 W^2 \right\} 
\\~\\
 F_{+0\,;\, 00} = \di{{8 t_2}\over {a^2 b^2}} \sqrt{\di{{2 t_1 }\over W^2}} \eta \zeta \left\{ \hskip -0.07cm \di{{}\over {}} Z W^2 - 4 X_1 \zeta^2 \right\},~~~F_{00\,;\,00} = \di{{32 t_1 t_2}\over {a^2 b^2}} \zeta^2 \left[ W^2 - 4 \zeta^2 \right] 
~\\~\\
\end{array}
$~}}  \enq

\newpage

\vv \nin avec les notations suivantes : 

$$ Z = W^2 + t_1 + t_2 = a+b ,~~~X_1 = W^2-t_1 + t_2,~~~X_2 = W^2 +t_1 - t_2, $$
$$ \eta = \di{W\over 2} \beta \sin \theta,~~~\zeta = \di{W\over 2} \beta \cos \theta,~~~\beta = \sqrt{ 1 - \di{{4 m^2}\over W^2}} $$ 

\vv \nin Dans les formules (\ref{XFG}) \`a (\ref{XR}), le taux d'interaction est exprim\'e \`a l'aide des invariants de la r\'eaction et les expressions correspondantes sont directement utilisables pour obtenir, par un prolongement analytique appropri\'e, le taux d'interaction de la diffusion Compton virtuelle, avec des photons virtuels du genre espace ou du genre temps.

\newpage

\section{Compl\'ement II : La conjugaison de charge}

\vv \nin Du point de vue du groupe de Lorentz, la conjugaison de charge est l'op\'eration qui transforme l'\'etat d'une particule correspondant \`a une 4-impulsion $p$ et une composante de spin $\sigma$ donn\'ees en l'\'etat de son anti-particule correspondant \`a ces m\^emes grandeurs. Pour ce m\^eme \'etat $|\,[\,p\,], \sigma>$, notons $a_\sigma( [\,p\,])$ et $b_\sigma([\,p\,])$ les op\'erateurs d'annihilation de la particule et de son anti-particule, respectivement. Agissant sur ces op\'erateurs, la conjugaison de charge est repr\'esent\'ee par un op\'erateur $U(C)$, {\it lin\'eaire} et {\it unitaire}\footnote{Par conjugaison de charge, le nombre d'anti-particules finales doit \^etre \'egal au nombre de particules initiales.} , tel que  

\beq U(C) \,a_\sigma( [\,p\,]) \,U(C)^{-1} = \eta_c\,b_\sigma([\,p\,]),~~~ U(C) \,a^\dagger_\sigma( [\,p\,]) \,U(C)^{-1} = \eta^\star_c\,b^\dagger_\sigma([\,p\,])\enq
$$ {\rm avec}~~~\eta_c\, \eta^\star_c = 1 $$

\vv \nin On doit avoir sym\'etriquement 

\beq U(C) \,b_\sigma( [\,p\,]) \,U(C)^{-1} = \eta^\prime_c\,a_\sigma([\,p\,]),~~~ {\rm avec}~~~\eta^\prime_c\, \eta^{\prime \star}_c = 1 \enq

\vv \nin ce qui implique que 

\beq {U(C)}^2 = \eta_c \,\eta^\prime_c \times {\rm id.} \enq 

\vv \nin o\`u ${\rm id.}$ est l'op\'erateur identit\'e dans l'espace de Fock. Consid\'erons le champ de spineurs 

\beq  \Psi(x) = \di{\int} d\rho(p) \di{\sum_\sigma} \left[ \, e^{-i p \cdot x}\,U_\sigma([\,p\,])\, a_\sigma([\,p\,])     + e^{i p \cdot x} \,W_\sigma([\,p\,])\, b^\dagger_\sigma([\,p\,]) \right] \label{psi} \enq

\vv \nin o\`u $d\rho(p) = \di{{d^3 p}\over{(2 \pi)^3 2 E}}$\, avec $E = \sqrt{m^2 + \Vec{\,p\,}^2}$. L'analogue de ce champ qui correspondrait \`a un univers o\`u anti-particule et particule auraient \'echang\'e leurs r\^oles s'\'ecrit  

\beq  \Psi^c(x) = \di{\int} d\rho(p) \di{\sum_\sigma} \left[ \, e^{-i p \cdot x}\,U_\sigma([\,p\,])\, b_\sigma([\,p\,])     + e^{i p \cdot x} \,W_\sigma([\,p\,])\, a^\dagger_\sigma([\,p\,]) \right] \label{psic} \enq

\vv \nin et ladite conjugaison de charge doit permettre de passer de l'un \`a l'autre :  

\beq   U(C)\, \Psi(x)\, U(C)^{-1} = \eta\, \Psi^c(x) ~~~{\rm o\grave{u}}~~~|\eta|=1 \label{psic-2}\enq 

\vv \nin Or, 

\beq U(C)\, \Psi(x)\, U(C)^{-1} = \di{\int} d\rho(p) \di{\sum_\sigma} \left[ \, \eta_c\, e^{-i p \cdot x}\,U_\sigma([\,p\,])\, b_\sigma([\,p\,])  + \eta^{\prime \star}_c\,e^{i p \cdot x} \,W_\sigma([\,p\,])\, a^\dagger_\sigma([\,p\,]) \right] \label{psic} \enq

\vv \nin et l'on est donc amen\'e \`a poser $\eta^{\prime \star}_c = \eta_c = \eta$. Cependant, le champ $\Psi(x)$ \'etant suppos\'e \^etre l'unique grandeur d\'ecrivant \`a la fois les particules et les anti-particules de l'esp\`ece consid\'er\'ee, son transform\'e $\Psi^c(x)$ doit pouvoir \^etre exprim\'e en fonction de celui-ci. Comparant (\ref{psic}) et (\ref{psi}), on voit que cela n\'ecessite une conjugaison hermitique et une transposition des spineurs :   

 $$ ~^t\Psi^\dagger(x) =\di{\int} d\rho(p) \di{\sum_\sigma} \left[ \,  e^{- i p \cdot x} \,^tW^\dagger_\sigma([\,p\,])\, b_\sigma([\,p\,]) + e^{i p \cdot x}\,^tU^\dagger_\sigma([\,p\,])\, a^\dagger_\sigma([\,p\,])  \right]  $$

\vv \nin puis une transformation lin\'eaire permettant d'exprimer $^tU^\dagger_\sigma([\,p\,])$ et $^tW^\dagger_\sigma([\,p\,])$ en fonction de $W_\sigma([\,p\,])$ et $U_\sigma([\,p\,])$, c'est-\`a-dire, telle que 

\beq  L_c\, ^tW^\dagger_\sigma([\,p\,]) = \eta\,\eta_c \, U_\sigma([\,p\,])~~{\rm et}~~L_c \, ^tU^\dagger_\sigma([\,p\,]) = \eta\,\eta^{\prime \star}_c\, W_\sigma([\,p\,]) \label{defW} \enq

\vv \nin afin d'obtenir 

\beq L_c \,^t\Psi^\dagger(x) = \eta\,\Psi^c(x) \label{psic-3} \enq

\vv \nin Du point de vue des fonctions d'onde spinorielles et des facteurs $e^{\pm i p\cdot x}$, le passage de $\Psi(x)$ \`a $\Psi^c(x)$ appara\^it donc comme une op\'eration {\it anti-lin\'eaire}. En fait, on peut consid\'erer que (\ref{defW}) est la d\'efinition appropri\'ee du spineur $W$ conduisant \`a (\ref{psic-3}). Pour la suite, nous renvoyons le lecteur \`a la section 6.9 et au paragraphe 7.4.3 de notre cours ITL. Nous y montrons  
que le spineur \`a \'energie n\'egative 
\vv
\beq \fbox{\rule[-0.4cm]{0cm}{1cm}~$ W_\sigma([\,p\,]) = (-1)^{1/2- \sigma}\, V_{-\sigma}([\,p\,])~~~~(V = \gs_5 U) $~}  \enq

\vv \nin est tel que 

\beq \fbox{\rule[0.35cm]{0cm}{1cm}~$\begin{array}{c}~\\
 U_\sigma([\,p\,]) = {\cal C}\, ^t\ov{W}_\sigma([\,p\,]),~~~W_\sigma([\,p\,]) = {\cal C}\, ^t\ov{U}_\sigma([\,p\,])
\\~\\
{\rm o\grave{u}}~~~{\cal C} = i \,\gs^2\, \gs_0 
\\~\\
\end{array}
$~} \enq

\vv \nin Prenant cette d\'efinition du spineur $W$, on obtient alors 

\beq U(C)\, \Psi(x)\, U(C)^{-1} = \Psi^c(x) = {\cal C} \,^t\ov{\Psi}(x) \enq 

\vv \nin Notons en passant que les relations simples   

$$ W^\uparrow = V^\downarrow,~~~W^\downarrow = - V^\uparrow  $$

\vv \nin justifient l'utilisation syst\'ematique des spineurs $V$ pour repr\'esenter les \'etats spinoriels d'une  anti-particule de spin 1/2. La matrice ${\cal C}$ a  les propri\'et\'es suivantes : 

\beq  {\cal C}^\star = {\cal C} =- {\cal C}^{-1},~~~{\cal C}\, \gs_\mu \,{\cal C}^{-1} = - \,^t \gs_\mu,~~~{\cal C}\, \gs_5\, {\cal C}^{-1} = \gs_5 \enq 

\vv \nin Consid\'erons alors la forme $X(\sigma', \sigma) = \ov{U}^\prime_{\sigma'}\, \Gamma\, U_\sigma$ o\`u $\Gamma$ est le produit de $n$ matrices $\gs_\mu$, ne comportant {\it aucune} matrice $\gs_5$ : $\Gamma = \gs_1 \cdots \gs_n$. Consid\'erons son conjugu\'e complexe : 

$$ X^\star(\sigma', \sigma) = \ov{U}_\sigma\, \gs_n \cdots \gs_1\, U^\prime_{\sigma'} = \,^tU^\prime_{\sigma'}\,^t\gs_1 \cdots\, ^t\gs_n\,^t\ov{U}_\sigma  = \ov{W}^\prime_{\sigma'}\, {\cal C}^{-1} \,^t\gs_1 \cdots\, ^t\gs_n\, {\cal C}^{-1}\, W_\sigma$$
\beq = - (-1)^n\, \ov{W}^\prime_{\sigma'}\, \gs_1 \cdots \gs_n\, W_\sigma = (-1)^n \,(-1)^{1 - \sigma' - \sigma} \,\ov{U}^\prime_{- \sigma'}\, \gs_5\, \gs_1 \cdots \gs_n\, \gs_5\, U_{- \sigma}  \enq
$$ = (-1)^{1 - \sigma' - \sigma}\, X(-\sigma', - \sigma) = (2 \sigma)(2\sigma') \, X(-\sigma', - \sigma) $$

\vv \nin Si $\Gamma$ comporte une matrice $\gs_5$\footnote{Le cas d'un nombre impair sup\'erieur \`a 1 de matrices $\gs_5$ pr\'esentes dans le produit $\Gamma$ se ram\`ene, par anti-commutations successives, \`a celui d'une seule matrice $\gs_5$ pr\'esente.}, du fait que $\gs_0\, ^t\gs_5\, \gs_0 = \gs_0\, \gs_5\,\gs_0 = - \gs_5$, un signe ``$-$" suppl\'emen-taire s'introduit et l'on a dans ce cas 

\beq X^\star(\sigma', \sigma) =(-1)^{\sigma' + \sigma}\, X_{-\sigma', - \sigma} = - (2 \sigma)(2\sigma')\, X_{-\sigma', - \sigma} \enq

\newpage

\nin Envisageons alors des amplitudes d'h\'elicit\'e relatives \`a des processus relevant de l'Electrodynamique Quantique ou de la Chromodynamique Quantique, impliquant des leptons ou des quarks. Prenons comme premier exemple la diffusion M\o ller. Son amplitude g\'en\'erique comporte des formes du type $X(\sigma', \sigma)$ o\`u chacune des matrices $\Gamma$ ``sandwich\'ees" ne comporte aucune matrice $\gs_5$. Les spineurs impliqu\'es dans ce processus \'etant au nombre de 4, on en d\'eduit la propri\'et\'e suivante des amplitudes d'h\'elicit\'e : 

$$T^\star_{\sigma_3 , \sigma_4\,;\, \sigma_1, \sigma_2} = (-1)^{\,\sum \sigma_i} \,T_{- \sigma_3, - \sigma_4\,;\, -\sigma_1, -\sigma_2 } $$  

\vv \nin ce qui se v\'erifie ais\'ement sur les formules de la section \ref{secmoller}. Consid\'erons ensuite le processus $\gs^\star + \gs^\star \rightarrow e^- + e^+$. Le positron \'etant d\'ecrit par un spineur $V$, une matrice $\gs_5$ est pr\'esente dans l'amplitude g\'en\'erique de type $X(\sigma', \sigma)$. Les fonctions d'onde des photons sont des vecteurs de polarisation v\'erifiant $\epsilon^\star_\lambda = (-1)^\lambda\,\epsilon_{- \lambda}$. Pour ce processus, on a donc 

$$T^\star_{\lambda_1, \lambda_2\,;\, \sigma_3, \sigma_4} = (-1)^{\lambda_1 + \lambda_2 + \sigma_3 + \sigma_4}\, T_{-\lambda_1, -\lambda_2\,;\, -\sigma_3, - \sigma_4} $$

\vv \nin ce qui se v\'erifie tout aussi ais\'ement sur les formules de la section \ref{secggee}.

\end{document}